\documentclass[11pt, a4paper, logo, copyright, nonumbering]{map}
\usepackage{CJKutf8}
\usepackage{xargs}  
\usepackage{todonotes}
\usepackage{amsmath}
\usepackage{dsfont}

\usepackage{longtable}

\usepackage{svg}
\usepackage{fontawesome}
\usepackage{array}
\usepackage{tabularx}
\usepackage{latexsym}
\usepackage{graphicx}
\usepackage[utf8]{inputenc} %
\usepackage[utf8]{inputenc} %
\usepackage{amssymb} %
\usepackage{url}            %
\usepackage{amsfonts}       %
\usepackage{nicefrac}       %
\usepackage{microtype}      %
\usepackage{xspace}

\usepackage{listings}
\usepackage{makecell}
\usepackage{inconsolata}
\usepackage{multicol}
\usepackage{amstext}
\usepackage{tcolorbox}

\definecolor{darkblue}{RGB}{84, 112, 198}

\tcbset{
  aibox/.style={
    width=\linewidth,
    top=8pt,
    bottom=4pt,
    colback=blue!6!white,        %
    colframe=white,              %
    colbacktitle=darkblue,           %
    fonttitle=\bfseries\color{white},  %
    enhanced,
    center,
    attach boxed title to top left={yshift=-0.1in,xshift=0.15in},
    boxed title style={
      boxrule=0pt,               %
      colframe=white,            %
      colback=blue,              %
    },
  }
}
\newtcolorbox{AIbox}[2][]{aibox,title=#2,#1}

\definecolor{lightblue}{rgb}{0.85, 0.95, 1.0}    %
\definecolor{lightgreen}{rgb}{0.90, 1.0, 0.90}    %
\definecolor{lightorange}{rgb}{1.0, 0.95, 0.85}   %
\definecolor{lightpurple}{rgb}{0.95, 0.90, 1.0}   %
\definecolor{lightgray}{rgb}{0.97, 0.97, 0.97}    %

\usepackage{tikz}
\usetikzlibrary{calc}
\definecolor{battery-empty}{rgb}{0.9, 0.9, 0.9}
\newcommand{\difficultybar}[1]{%
  \begin{tikzpicture}[baseline, scale=0.5, every node/.style={scale=0.8}]
    \foreach \i in {1,2,3,4,5} {
      \ifnum\i>#1
        \draw[fill=battery-empty] (\i*0.5-0.5, 0) rectangle (\i*0.5, 0.25);
      \else
        \pgfmathsetmacro{\colorlevel}{80 - 12*(\i)} %
        \edef\x{\noexpand\draw[fill=blue!\colorlevel!white, opacity=0.9] (\i*0.5-0.5, 0) rectangle (\i*0.5, 0.25);}
        \x
        \draw[blue!50!black] (\i*0.5-0.5, 0) rectangle (\i*0.5, 0.25);
      \fi
    }
    \fill[battery-empty!70] (2.5, 0.08) rectangle (2.6, 0.17);
    \draw[battery-empty!70!black] (2.5, 0.08) rectangle (2.6, 0.17);
  \end{tikzpicture}%
}
\renewcommand{\arraystretch}{0.96}

\definecolor{hidden-draw}{RGB}{20,68,106}
\definecolor{hidden-pink}{RGB}{255,245,247}
\usepackage[edges]{forest}

\usepackage{natbib}
\usepackage{CJKutf8}
\usepackage{xargs}
\usepackage{todonotes}
\usepackage{multirow}
\usepackage{cleveref}
\usepackage{amsmath}
\usepackage{dsfont}
\usepackage{subcaption}
\usepackage{url}
\usepackage{svg}
\usepackage{fontawesome}
\usepackage{xcolor}
\usepackage{float}

\usepackage[utf8]{inputenc}
\usepackage{booktabs}
\usepackage{graphicx}
\usepackage{array}
\usepackage{tabularx}
\usepackage{xcolor,colortbl}  %
\definecolor{boxcolor}{HTML}{d92523} %
\definecolor{bulbcolor}{HTML}{e3b87f} %
\usepackage{url}
\usepackage{ragged2e}   %
\usepackage{makecell} 
\DeclareRobustCommand{\cmark}{\textcolor{green!60!black}{\ding{51}}} %
\DeclareRobustCommand{\xmark}{\textcolor{red!70!black}{\ding{55}}}   %

\usepackage{hyperref}       %
\usepackage{url}            %
\usepackage{booktabs}       %
\usepackage{amsfonts}       %
\usepackage{nicefrac}       %
\usepackage{microtype}      %
\usepackage{xcolor}         %
\usepackage{graphicx}
\usepackage{longtable}
\usepackage{array}
\usepackage{pbox}
\usepackage{amsmath}
\usepackage{amssymb}
\usepackage{tabularx}
\usepackage{multirow}
\usepackage{wrapfig}
\usepackage{pifont}
\usepackage{algorithm}
\usepackage{algorithmic}
\usepackage{lipsum}
\usepackage{subcaption}
\usepackage{tcolorbox}
\usepackage{enumitem}
\usepackage{tikz}
\usepackage{xstring}

\usepackage{color-edits}

\usepackage{threeparttable} %

\addauthor[Zhejian]{zzj}{blue}

\newcommand{\mcell}[1]{\makecell[c]{#1}}

\hypersetup{
    colorlinks=true,            %
    linkcolor=blue,             %
    filecolor=magenta,          %
    urlcolor=cyan,              %
    citecolor=purple,             %
    pdftitle={Overleaf Example},
    pdfpagemode=FullScreen,
    }

\renewcommand{\today}{}
\newcommandx{\info}[2][1=]{\todo[linecolor=red,backgroundcolor=red!25,bordercolor=red,#1]{#2}}

\title{\vspace{-0.2in}
\centering \fontsize{15pt}{16pt}\selectfont From Code Foundation Models to Agents and Applications: A Comprehensive Survey and Practical Guide to Code Intelligence
\vspace{-0.2in}
}

\author{
BUAA-SKLCCSE, Alibaba, ByteDance, M-A-P, BJTU, OPPO, HKUST (GZ), BUPT, TeleAI, Shanghai AI Lab, Manchester, StepFun, UoS, SCU, CASIA, NJU, Kuaishou, HIT, Huawei Cloud, Tencent, Monash/CSIRO, NTU, ZJU, BIT, Ubiquant, NUS, HNU, PKU, CSU
\vspace{-5pt}
}

\begin{abstract}

\vspace{-0.2in}

Large language models (LLMs) have fundamentally transformed automated software development by enabling direct translation of natural language descriptions into functional code, driving commercial adoption through tools like GitHub Copilot (Microsoft), Cursor (Anysphere), Trae (ByteDance), and Claude Code (Anthropic). While the field has evolved dramatically from rule-based systems to Transformer-based architectures, achieving performance improvements from single-digit to over 95\% success rates on benchmarks like HumanEval. In this work, we provide a comprehensive synthesis and practical guide (a series of analytic and probing experiments) about code LLMs, systematically examining the complete model life cycle from data curation to post-training through advanced prompting paradigms, code pre-training, supervised fine-tuning, reinforcement learning, and autonomous coding agents. We analyze the code capability of the general LLMs (GPT-4, Claude, LLaMA) and code-specialized LLMs (StarCoder, Code LLaMA, DeepSeek-Coder, and QwenCoder), critically examining the techniques, design decisions, and trade-offs. Further, we articulate the research-practice gap between academic research (e.g., benchmarks and tasks) and real-world deployment (e.g., software-related code tasks), including code correctness, security, contextual awareness of large codebases, and integration with development workflows, and map promising research directions to practical needs. Last, we conduct a series of experiments to provide a comprehensive analysis of code pre-training, supervised fine-tuning, and reinforcement learning, covering scaling law, framework selection, hyperparameter sensitivity, model architectures, and dataset comparisons.
\end{abstract}

\begin{document}

    \maketitle

\let\oldthefootnote\thefootnote

\begin{figure}[H]
    \vspace{-5mm}
    \centering
    \includegraphics[width=0.90\textwidth]{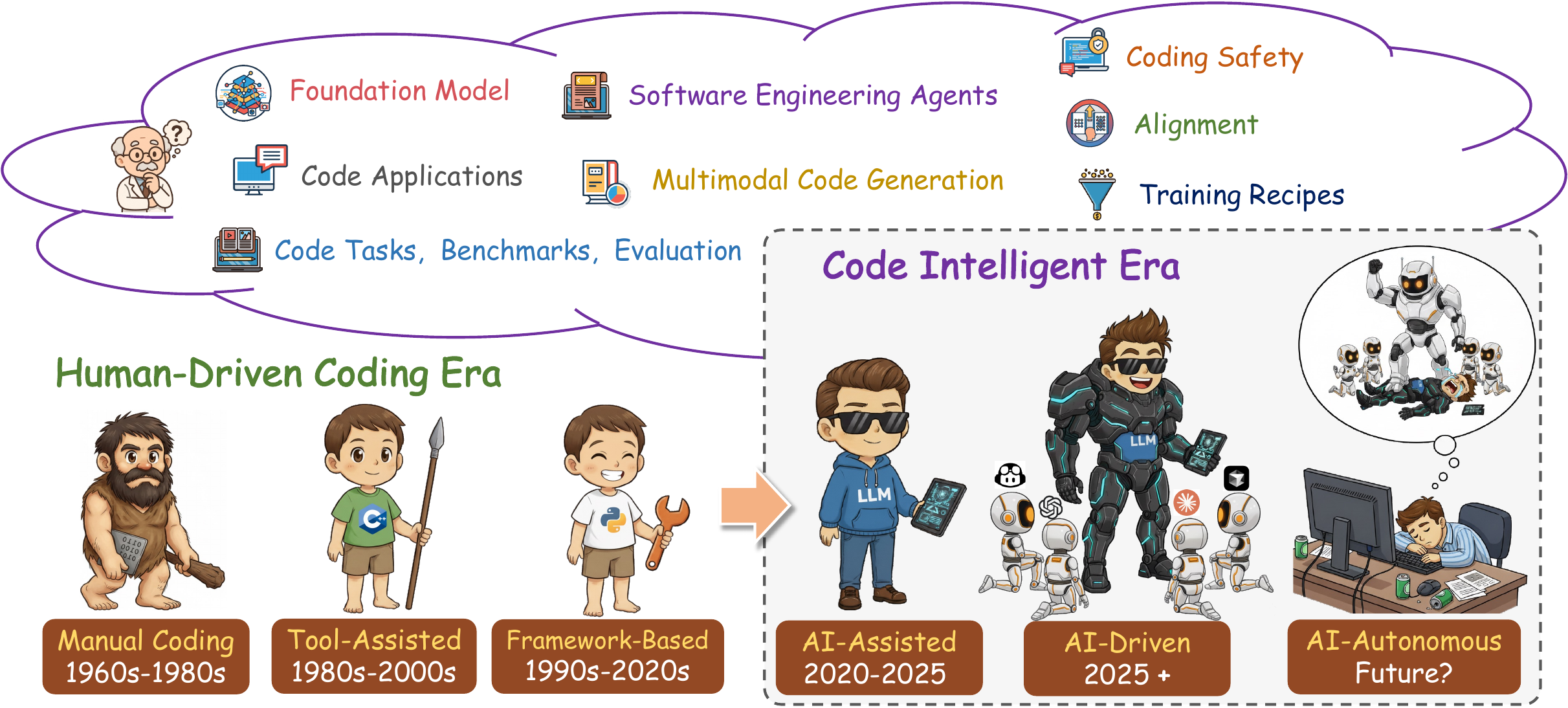}
    \vspace{-10pt}
    \caption{Evolution of programming development and research landscapes in AI-powered code generation. The upper section highlights the key research areas covered in this work. The timeline below illustrates the six-stage evolution from the human-driven coding era to the emerging code intelligence era.}
    \label{fig:intro}
\end{figure}

\newpage

\tableofcontents

\newpage
\section{Introduction}
The emergence of large language models (LLMs)~\cite{claude3_7,claude4,claude45,openai2024gpt4o,qwen25coder,openai2025gpt5codex,Huang2024OpenCoderTO,gpt5,openai2025gpt5} has catalyzed a paradigm shift in automated software development, fundamentally reconceptualizing the relationship between human intent and executable code~\cite{zhuo2025nlp+}. Modern LLMs have achieved remarkable capabilities across a wide range of code-related tasks, including code completion~\cite{bavarian2022efficient}, translation~\cite{CodeTransOcean}, repair~\cite{debugbench,mdeval}, and generation~\citep{chen2021codex,MultiPL-E}. These LLMs effectively distill years of accumulated programming expertise into accessible, instruction-following tools that can be deployed by developers at any skill level using code from sources such as GitHub, Stack Overflow and other code-related websites. 
Among LLM-related tasks, code generation stands as one of the most transformative, enabling the direct translation of natural language descriptions into functional source code, thereby dissolving traditional barriers between domain knowledge and technical implementation. This capability has transcended academic curiosity to become a commercial reality through a series of commercial and open-source tools, including (1) GitHub Copilot (Microsoft)~\cite{github_copilot}, which provides intelligent code completion within development environments; (2) Cursor (Anysphere)~\citep{cursor_ai_2025}, an AI-first code editor that enables conversational programming; (3) CodeGeeX (Zhipu AI)~\cite{codegeex}, which offers multilingual code generation; (4) CodeWhisperer (Amazon)~\cite{aws2024codewhisperer}, which integrates seamlessly with AWS services; (5) Claude Code (Anthropic)~\cite{claude_code_repo}/Gemini CLI (Google)~\cite{gemini_cli_docs}, which are both command-line tools that allows developers to delegate coding tasks directly to Claude or Gemini~\cite{claude45,geminiteam2025gemini} from their terminal for agentic coding workflows. These applications reshape software development workflows, challenge conventional assumptions about programming productivity, and redefine the boundary between human creativity and machine assistance.

In \autoref{fig:intro}, the evolutionary trajectory of code generation reveals a compelling narrative of technological maturation and paradigm shifts. Early approaches, constrained by heuristic rules and probabilistic grammar-based frameworks~\cite{mine_from_code,inducing_tree_substitution,question_selection_for_interactive}, were inherently brittle—optimized for narrow domains and resistant to generalization across the vast diversity of programming contexts. The advent of transformer-based architectures~\citep{feng2020codebertpretrainedmodelprogramming,graph_code_bert} represented not merely an incremental improvement but a fundamental reconceptualization of the problem space, leveraging attention mechanisms~\citep{vaswani2017attention} and scale to capture the intricate relationships between natural language intent and code structure. More remarkably, these models exhibit emergent instruction-following capabilities that were neither explicitly programmed nor directly optimized for, suggesting that the capacity to translate high-level goals into executable implementations may be a natural consequence of learning rich representations at scale. This democratization~\cite{vibe_coding_agentic_coding,code_world_model} of coding, enabling non-experts to generate sophisticated programs through natural language, carries profound implications for workforce development, innovation pace, and the very essence of computational literacy in the 21st century~\cite{how_to_write_code1,how_to_write_code2}.

The contemporary landscape of code LLMs reveals a strategic bifurcation between generalist and specialist approaches, each with distinct advantages and trade-offs. General-purpose models like the GPT~\cite{gpt4Code,gpt5,openai2024gpt4o}, Claude~\cite{claude3_7,claude4,claude45}, and LLaMA~\cite{touvron2023llama,touvron2023llama2,meta2024llama3_2,meta2025llama4} series offer remarkable breadth, leveraging vast corpora of natural language alongside code to develop a nuanced understanding of context, intent, and domain knowledge. Conversely, specialized code LLMs such as StarCoder~\cite{li2023starcoder}, Code LLaMA~\citep{roziere2024codellama}, DeepSeek-Coder~\cite{deepseek2024coder}, CodeGemma~\cite{codegemma2024}, and QwenCoder~\cite{qwen25coder,qwen3coder} achieve superior performance on code-specific benchmarks through focused pre-training on programming-centric data and task-specific architectural optimizations. Dramatic performance improvements from single digits to 95\%+ success rates on standardized benchmarks like HumanEval~\cite{chen2021codex} reflect both algorithmic innovations and deeper insights. While code is highly formalized, it shares core characteristics with natural language, particularly in compositional semantics and contextual dependencies.

Despite vigorous research activity and rapid commercial adoption, a critical gap persists between the breadth of innovation and the depth of systematic analysis in the literature. Existing surveys have largely adopted panoramic approaches, surveying broad categories of code-related tasks, or focusing on earlier generations of models, leaving contemporary advances inadequately synthesized. 
Crucially underexplored are the sophisticated data curation strategies of state-of-the-art systems, which balance quantity with quality instruction tuning methods to align model behavior with developer intent.
Such alignment techniques involve incorporating human feedback to refine outputs, advanced prompting paradigms including chain-of-thought reasoning and few-shot learning, the emergence of autonomous coding agents capable of multi-step problem decomposition, retrieval-augmented generation (RAG) approaches that ground outputs in authoritative references, and novel evaluation frameworks that move beyond simple binary correctness to assess code quality, efficiency, and maintainability. 

\begin{figure*}[t]
\begin{center}
    \includegraphics[width=1.0\textwidth]{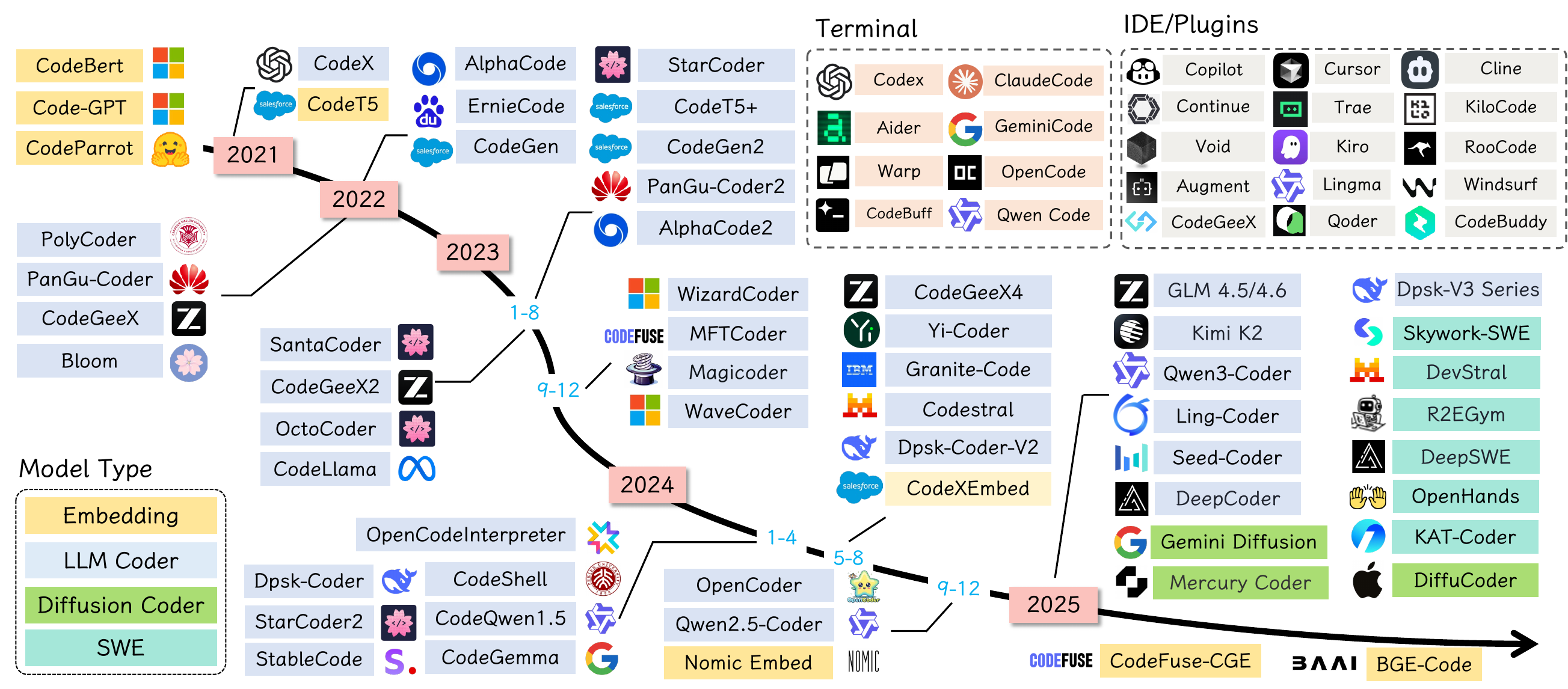}
    \caption{
    Overview of the evolution of code large language models (Code-LLMs) and related ecosystems from 2021 to 2025. The landscape begins with early models and quickly expands into a diverse set of LLM coders across 2022–2024. From 2025 onward, research focus shifts toward reinforcement learning (RL)-based training, software engineering (SWE) agents, and novel architectures such as diffusion-based code models. In parallel, a rich ecosystem of terminal tools, IDE integrations, and plugins emerges, highlighting the transition from pure modeling to practical developer-oriented applications.
    }
    \label{fig:coder_overview}
    \vspace{-15pt}
\end{center}
\end{figure*}

In \autoref{fig:coder_overview}, recent LLMs like Kimi-K2~\cite{k2}, GLM-4.5/4.6~\cite{glm2025glm4_5,glm46}, Qwen3Coder~\cite{qwen3coder}, Kimi-Dev~\cite{kimi_dev}, Claude~\cite{claude45}, Deepseek-V3.2-Exp~\cite{deepseekai2024deepseekv32}, and GPT-5~\cite{gpt5} embody these innovations, yet their contributions remain scattered across disparate publications without cohesive integration. \autoref{tab:work_compare} compares various surveys related to code intelligence or LLM, evaluating them across eight dimensions: domain, whether focus on Code, LLM usage, pretraining, supervised fine-tuning (SFT), reinforcement Learning (RL), Training Recipes for code LLM, and applications. These surveys cover diverse areas, including general code generation, software engineering using GenAI, code summarization, and LLM-based agents. Most surveys focus on code and applications, but vary significantly in their coverage of technical aspects. While some address LLMs and pretraining, very few cover reinforcement learning methods. This survey offers a comprehensive and contemporary synthesis of research literature on large language models (LLMs) for code intelligence, providing a systematic examination of the entire model life cycle. It explores critical phases—from initial data curation and instruction tuning to advanced code applications and the development of autonomous coding agents.

To provide a comprehensive and practical study from code foundation models to agents and applications, we present a detail guide that bridges theoretical foundations with implementations in modern code generation systems, as shown in \autoref{tab:work_compare}. Our work makes several key contributions: (1) We provide a unified taxonomy of contemporary code LLMs, tracing their evolution from early transformer-based models to the latest generation of instruction-tuned systems with emergent reasoning capabilities; (2) We systematically analyze the complete technical pipeline from data curation and preprocessing strategies, through pretraining objectives and architectural innovations, to advanced fine-tuning methodologies including supervised instruction tuning and reinforcement learning; (3) We examine cutting-edge paradigms that define state-of-the-art performance, including prompting techniques (e.g., chain-of-thought~\cite{yang2024chain}), retrieval-augmented generation approaches, and autonomous coding agents capable of complex multi-step problem solving; (4) We critically evaluate the landscape of benchmarks and evaluation methodologies, discussing their strengths, limitations, and the ongoing challenge of assessing not merely functional correctness but code quality, maintainability, and efficiency; (5) We synthesize insights from recent breakthrough models (e.g., GPT-5, Claude 4.5 among others) to identify emerging trends and open challenges that will shape the next generation of code generation systems. This survey aims to serve as both a comprehensive reference for researchers entering the field and a strategic roadmap for practitioners seeking to leverage these technologies in production environments. (6) We perform extensive experiments to comprehensively examine code pre-training, supervised fine-tuning, and reinforcement learning across multiple dimensions including scaling laws, frameworks, hyperparameters, architectures, and datasets.

\begin{table*}[h!]
\centering
\caption{Comparison between our study and existing works.}
\label{tab:work_compare}
\renewcommand{\arraystretch}{1.25}
\resizebox{\textwidth}{!}{
\begin{tabular}{
    @{}p{4.8cm}  %
    p{4.0cm}     %
    >{\centering\arraybackslash}p{1.5cm}  %
    >{\centering\arraybackslash}p{1.0cm}  %
    >{\centering\arraybackslash}p{1.5cm}  %
    >{\centering\arraybackslash}p{0.8cm}  %
    >{\centering\arraybackslash}p{0.8cm}  %
    >{\centering\arraybackslash}p{1.5cm}  %
    >{\centering\arraybackslash}p{1.5cm}@{} %
}
\toprule
\textbf{Survey} & \textbf{Scope} & \makecell[c]{\textbf{Focus}\\\textbf{on Code}} & \textbf{LLM} & \makecell[c]{\textbf{Pretrain}} & \textbf{SFT} & \textbf{RL} & \makecell[c]{\textbf{Appli-}\\\textbf{cation}} & \makecell[c]{\textbf{Training}\\\textbf{Recipes}} \\
\midrule
\rowcolor{gray!5}
\raggedright A Survey on Language Models for Code~\citep{code_survey_ant} 
& \raggedright All & \cmark & \cmark & \cmark & \cmark & \xmark & \cmark & \xmark \\

\raggedright Deep Learning for Code Generation: A Survey~\citep{zhang2023deep} 
& \raggedright Deep Learning, Code Generation, Automated SE & \cmark & \xmark & \xmark & \xmark & \xmark & \cmark & \cmark \\

\rowcolor{gray!5}
\raggedright Code to Think, Think to Code~\citep{yang2025codethinkthinkcode} 
& \raggedright Code reasoning, planning, debugging & \cmark & \xmark & \xmark & \xmark & \xmark & \cmark & \xmark \\

\raggedright A Survey on LLMs for Code Generation~\citep{jiang2024surveylargelanguagemodels} 
& \raggedright Code Generation, Data Process & \cmark & \cmark & \cmark & \xmark & \xmark & \cmark & \xmark \\

\rowcolor{gray!5}
\raggedright A Survey of ML for Big Code and Naturalness~\citep{allamanis2018surveymachinelearningbig}
& \raggedright Code patterns, model design & \xmark & \xmark & \xmark & \xmark & \xmark & \cmark & \xmark \\

\raggedright A Survey on Code Generation with LLM-based Agents~\citep{wang2023survey}
& \raggedright Code Gen, LLM Agents, Multi-agent Systems & \cmark & \cmark & \cmark & \cmark & \cmark & \cmark & \xmark \\

\rowcolor{gray!5}
\raggedright A Survey of Automatic Source Code Summarization~\citep{liu2022survey}
& \raggedright Code Summarization, Program Analysis, NMT & \cmark & \xmark & \xmark & \xmark & \xmark & \cmark & \cmark \\

\raggedright A Review of Automatic Source Code Summarization~\citep{feng2024review}
& \raggedright Code Summarization, Program Analysis, NMT & \cmark & \xmark & \xmark & \xmark & \xmark & \cmark & \xmark \\

\rowcolor{gray!5}
\raggedright Survey on NN-based Automatic Source Code Summarization~\citep{gao2023survey}
& \raggedright Intelligent SE, Code Summarization, Deep Learning & \cmark & \xmark & \xmark & \xmark & \xmark & \cmark & \xmark \\

\raggedright A Survey of Large Language Models~\citep{zhao2025surveylargelanguagemodels}
& \raggedright General LLM & \xmark & \cmark & \cmark & \cmark & \xmark & \cmark & \xmark \\

\rowcolor{gray!5}
\raggedright Source code data augmentation for deep learning: A survey~\citep{zhuo2023source}
& \raggedright Code Data Augmentation, Program Analysis, Deep Learning & \cmark & \cmark & \xmark & \cmark & \xmark & \cmark & \xmark \\

\raggedright A Survey of Vibe Coding with LLMs~\citep{vibe_coding_survey}
& \raggedright Vibe Coding & \xmark & \cmark & \cmark & \cmark & \xmark & \cmark & \xmark \\

\midrule
\textbf{Ours} & \textbf{All} & \cmark & \cmark & \cmark & \cmark & \cmark & \cmark & \cmark \\
\bottomrule
\end{tabular}
}
\end{table*}

\section{Code Foundation Models}

\subsection{General Large Language Models}
\label{subsec:general_llms}

\subsubsection{The Rise of General LLMs}

\begin{figure*}[t]
\begin{center}
    \includegraphics[width=0.9\textwidth]{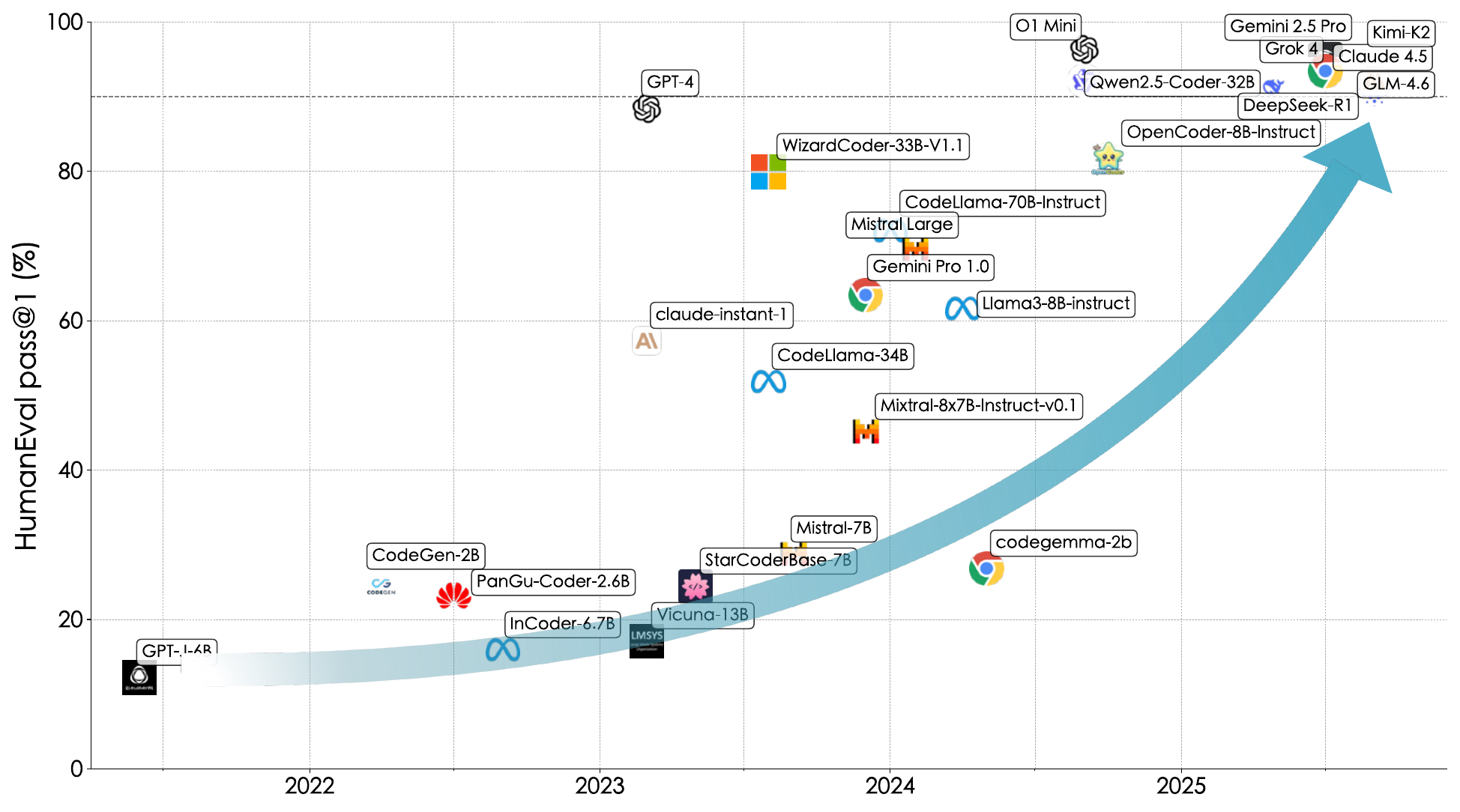}
    \caption{
    The timeline of code language models’ progress on HumanEval. The dashed line represents a score of 90. The vertical axis does not indicate actual scores but signifies that model scores exceed 90 points.
    }
    \label{fig:timeline_humaneval}
    \vspace{-15pt}
\end{center}
\end{figure*}

The advent of LLMs built on the transformer architecture \cite{vaswani2017transformer} marked a decisive shift in AI. Before transformers, progress was fragmented across specialized systems, including sequence-to-sequence models for translation \cite{sutskever2014seq2seq, bahdanau2015nmt, wu2016googlesneuralmachinetranslation}, handcrafted pipelines for dialogue \cite{young2013pomdp, williams2013dstc, xu2000agenda}, and domain-specific engines for program synthesis \cite{gulwani2011flashfill, polozov2015flashmeta, alur2013sygus}. Transformer-based pretraining and knowledge transfer unified these strands into a single, scalable framework that could be adapted across tasks and modalities \cite{devlin2019bert, raffel2020t5, brown2020gpt3}. Scaling laws show predictable gains with more model parameters, data, and compute \cite{kaplan2020scalinglaws}, while reports of \emph{emergent} abilities, defined as capabilities that appear only at larger scales, suggest LLMs generalize beyond their training distribution \cite{wei2022emergent}. Yet recent work argues some emergence may stem from metric choice rather than true leaps in capability, offering a more nuanced view of the benefits of scale \cite{schaeffer2023mirage}. Two classes of abilities are especially salient: coding and agentic behavior. First, general-purpose LLMs revealed surprising coding competence, catalyzing the development of models explicitly trained on code. OpenAI's Codex demonstrated functional code generation from natural-language prompts and introduced standardized evaluation like HumanEval~\cite{chen2021codex}.
LLMs have achieved outstanding performance on HumanEval, as illustrated in \autoref{fig:timeline_humaneval}. In parallel, DeepMind's AlphaCode~\cite{li2022alphacode} showed that large-scale sampling and filtering could reach competitive-programming proficiency at roughly the median human level under simulated Codeforces settings. These results established that linguistic modeling and code synthesis share exploitable structure, making LLMs immediately useful for tasks from boilerplate generation to algorithmic problem solving \cite{hendrycks2021apps, austin2021mbpp, li2023starcoder, roziere2024codellama, jimenez2024swebench}.

\begin{figure*}[t]
\begin{center}
    \includegraphics[width=1.0\textwidth]{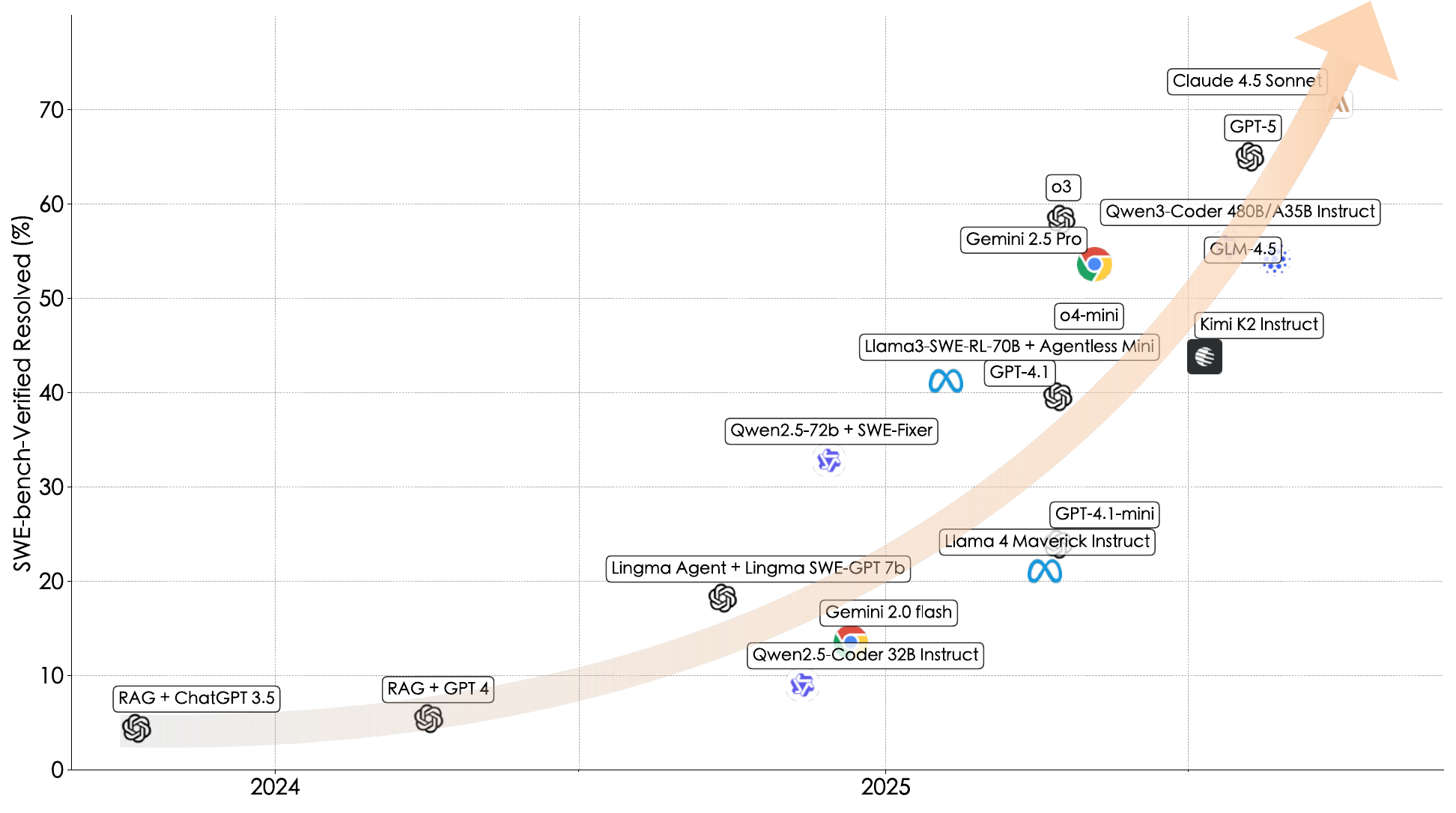}
    \caption{
     The timeline of code language models’ progress on SWE-bench-Verified. All models without scaffold annotations uniformly use mini-SWE-agent.
    }
    \label{fig:timeline_swe}
    \vspace{-15pt}
\end{center}
\end{figure*}

Second, when paired with external tools, memory, and closed-loop reasoning, LLMs begin to look like decision-making agents rather than static predictors. Methods such as ReAct~\cite{yao2023react} interleave reasoning traces with environment actions to plan, gather information, and correct course \cite{yao2023react}. Complementary approaches such as Toolformer~\cite{toolformer} show that models can learn \emph{when} and \emph{how} to call APIs in a self-supervised way, improving reliability on tasks that benefit from calculators, search, or retrieval \cite{schick2023toolformer,nakano2022webgpt, yao2023tot, shinn2023reflexion, liu2024agentbench, deng2023mind2web}. Among them, the most representative software engineering (SWE) agents have made remarkable progress, as shown in \autoref{fig:timeline_swe}.

Taken together, these developments mark a clean break from narrow, task-specific systems to general coding system, which provides a unified substrate for language, programming, and tool-mediated reasoning. At the same time, their breadth exposes limits in accuracy, security, and system-level reliability in professional software settings~\cite{swe_agent_2024,swe-bench-live,swebenchmultilingual}, which in turn motivate the specialized coding models and agents represented in the rest of this work.

\subsubsection{Model Architectures}

Alongside tremendous growth in scale and data~\cite{code_scaling_law,chinchilla_scaling_law}, innovations in model architecture have been a central pillar of the rapid progress of LLMs. This architectural evolution is primarily defined by a shift away from dense models, where every parameter is engaged in every computation, and toward sparser, more specialized designs that optimize the trade-offs between efficiency, scalability, and performance.

\paragraph{Dense Models}
The transformer model \cite{vaswani2017transformer} remains the foundation of modern LLMs, leveraging dense architectures where every parameter is involved in processing each token. This design, built on stacks of attention and feed-forward layers, has enabled remarkable progress in capturing long-range dependencies and driving breakthroughs across NLP tasks. Building on this, models like LLaMA \cite{touvron2023llama, touvron2023llama2, grattafiori2024llama3} and its successors have shown that high-quality open models can rival proprietary systems, scaling from 7B to 70B parameters. The GLM series \cite{du2022glm, glm2024chatglm} extended dense architectures into bilingual and multilingual domains, while the Qwen family \cite{bai2023qwen, qwen2024qwen1_5, yang2024qwen2, qwen2025qwen2_5} emphasized strong performance in both understanding and generation with scalable dense models. Meanwhile, Mistral \cite{jiang2023mistral7b} highlighted how careful engineering, such as grouped query attention (GQA), can deliver competitive results with fewer parameters. Collectively, these dense models illustrate a consistent trend: while computationally demanding, they continue to evolve toward greater efficiency and versatility, cementing their central role in modern NLP research and applications.

\paragraph{Mixture-of-Experts (MoE)}
MoE expands model capacity through conditional computation without proportionally increasing activated compute: each token is routed to only a small number of experts, typically the top-$k$ experts, for forward computation, thereby trading sparse activation for higher effective capacity \cite{lepikhin2020gshard, fedus2022switchtransformers, du2022glam}.
In the open-source community, the Mixtral series made two-expert routing a de facto engineering standard: 8$\times$7B demonstrated that activating fewer parameters can outperform larger dense baselines, and the subsequent 8$\times$22B further pushed the limits of capability and throughput in open-source models \cite{jiang2024mixtral}.
The Qwen series introduced MoE variants across its 1.5/2.5/3 versions \cite{qwen2024qwen1_5, qwen2025qwen2_5, yang2025qwen3}.
DeepSeek~\cite{deepseek} systematized efficient co-design of sparse experts and Multi-head Latent Attention (MLA) in its V2/V3 series. V2 has 236B total parameters with about 21B activated, while V3 has 671B total parameters with about 37B activated. These models offered replicable open paradigms balancing cost and stability \cite{deepseekai2024deepseekv2, deepseekai2025deepseekv3}.
DeepSeek R1 further built on V3-Base with reinforcement learning to significantly enhance chain-of-thought reasoning \cite{deepseekai2025deepseekr1}.
GLM-4.5 employed large-scale MoE, integrating hybrid reasoning modes into a unified model for coding, reasoning, and agent applications \cite{glm2025glm4_5}.
In addition, the entire LLaMA-4 series also adopts the MoE architecture \cite{meta2025llama4}.
Overall, MoE has become one of the mainstream architectures for optimizing the effective capacity ratio, and in practice it works synergistically with long-context handling, KV cache compression, and multi-token prediction, forming an efficient paradigm for large-scale production environments.

\paragraph{Recurrent Models}
Recurrent-style architectures revisit sequence modeling to cut memory and latency while preserving parallel training.
RWKV~\cite{peng2023rwkv, peng2024eaglefinchrwkv,minmax_m1} blends transformer-like parallelizable training with recurrent inference, activating a constant-size state at each step so that decoding scales linearly and can approach transformer quality at similar sizes.
Retentive Networks (RetNet)  \cite{sun2023retnet} replace attention with a retention operator that supports fully parallel training and either recurrent or chunkwise-recurrent inference, yielding linear-time long-sequence processing with strong language-modeling results.
Mamba \cite{gu2024mamba} introduces selective state-space models whose parameters are input-dependent, enabling linear-time decoding and competitive performance on language while maintaining high throughput; a follow-up theoretical line frames transformers and SSMs under a shared state-space duality with efficient algorithms \cite{dao2024transformersaressms}.
Closely related long-range operators such as Hyena \cite{poli2023hyena} use implicitly parameterized long convolutions with gating to match attention quality at subquadratic cost, pushing feasible context lengths far beyond standard attention regimes and complementing recurrent approaches in practice.
Additionally, DeltaNet~\cite{yang2024deltanet} introduces a hardware-efficient way to parallelize linear transformers with the delta rule (a state update mechanism), which improves associative retrieval and enables scaling to standard language-modeling settings. Gated DeltaNet \cite{yang2024gateddeltanet} combines gating with the delta update to better control memory and consistently surpasses Mamba-2 and DeltaNet on long-context and retrieval benchmarks.

\paragraph{Diffusion-based Models}
Diffusion-based language models replace left-to-right decoding with iterative denoising steps that refine a noisy sequence into fluent text, enabling strong global control over attributes and structure.
Foundational work on discrete diffusion formalized corruption/denoising processes directly in token space (D3PM~\cite{austin2023d3pm}), establishing principled transition kernels for categorical data such as text.
Building on this, Diffusion-LM \cite{li2022diffusionlm} operates in a continuous embedding space and leverages gradient-based guidance for fine-grained controllability while remaining non-autoregressive.
For conditional generation, DiffuSeq \cite{gong2023diffuseq} adapts diffusion to sequence-to-sequence tasks and reports performance that is competitive with strong autoregressive baselines.
To better align diffusion with token vocabularies and practical decoding, SSD-LM \cite{han2023ssdlm} performs simplex-based diffusion over the discrete vocabulary and generates text in blocks, enabling modular classifier guidance that matches or surpasses GPT-style models.
AR-Diffusion \cite{wu2023ardiffusion} introduces an explicit autoregressive ordering within diffusion to reconcile sequential dependencies with iterative refinement.
Lately, several larger efforts have pushed diffusion LMs beyond small-scale prototypes: LLaDA \cite{nie2025llada} trains diffusion models for language from scratch via a masking schedule and reverse denoising with a vanilla transformer, reporting competitiveness with similarly sized autoregressive baselines.
On the commercial side, Mercury Coder \cite{labs2025mercury} frames coding as parallel multi-token denoising and markets substantial speed/throughput gains relative to autoregressive (AR) models. Gemini Diffusion \cite{deepmind2025geminidiffusion} is another research model exploring diffusion for text generation, signaling continued interest in non-autoregressive decoding at production scale.
While diffusion LMs offer controllability and parallelizable training objectives, they typically require many sampling steps, motivating research on faster samplers and hybrid AR–diffusion decoders.

\paragraph{Hybrid Architectures}
Hybrid architectures interleave complementary sequence operators, typically combining transformer attention with state-space or recurrent blocks, often in addition to MoE feed-forwards to trade off quality, context length, and throughput in one stack.
Jamba \cite{lieber2024jamba} is a canonical example: it interleaves transformer and Mamba layers with MoE, achieving high throughput at long contexts while retaining strong performance.
In the Qwen line, Qwen3-Next \cite{qwen2025qwen3next} adopts a hybrid attention design that mixes gated DeltaNet-style linear operators with gated attention and sparse-activation MoE, targeting 256K+ (more than 256K tokens) contexts with low active parameters per token. 
The DeepSeek family also fuses multiple ideas: V3 introduced MLA with DeepSeek-MoE for efficient training/inference \cite{deepseekai2025deepseekv3}, and the recent V3.2-Exp \cite{deepseekai2024deepseekv32} adds an experimental DeepSeek Sparse Attention (DSA) mechanism as an intermediate step toward its next-generation hybrid architecture, emphasizing longer-context efficiency across diverse hardware.

In summary, model architecture has diversified from a one-size-fits-all dense transformer to a toolkit of sparsity, recurrence/state-space, diffusion, hybrids, and efficient attention. These choices let practitioners trade off capacity, latency, and context length, providing the capabilities that underpin both general LLMs and the specialized coding systems discussed later.

\subsubsection{Multimodality}
Code LLMs need to process visual information like diagrams, screenshots, and UI elements to understand and generate code in real-world scenarios~\cite{li2024mmcode,yang2024chartmimic,webevolver,wu2023visualchatgpt,koh2024visualwebarena}. These capabilities form the foundation for code-oriented workflows. Modalities such as audio or speech are outside the present scope.

\subsubsection{Limitations of General LLMs}
The progress highlights the breadth and versatility of general-purpose LLMs, spanning dense and sparse architectures, recurrent and hybrid designs, as well as emerging multimodal capabilities. These developments underscore how far the field has advanced from narrow task-specific systems toward unified substrates for language, coding, and perception–action reasoning. Yet, this very breadth also exposes their limitations: general LLMs, while impressive in scope, often lack the depth, robustness, and domain alignment required for professional software engineering. We therefore turn next to a closer examination of their key shortcomings.

\textbf{Specialization and Accuracy}
Despite their breadth, general-purpose LLMs often lack the depth required for professional software engineering. They may produce functionally-looking code that superficially appears correct but fails to satisfy domain constraints such as subtle API contracts, security policies, and they struggle to maintain invariants across large systems. Evidence from repository-scale evaluations further indicates that real-world issue resolution remains challenging even for strong models and agentic toolchains~\cite{jimenez2024swebench}.

\textbf{Security and Reliability}~~~A growing body of empirical studies shows that \emph{functionally correct} code from general LLMs can still be \emph{insecure}. Large-scale evaluations involving more than one hundred models across eighty tasks report that about 45\% of generations contain known vulnerabilities, with little improvement from newer or larger models. Smaller focused studies likewise find that ChatGPT and similar LLMs often emit code that is not robust to attacks \cite{khoury2023howsecureiscode}, and recent outcome-driven benchmarks that evaluate both functionality and security confirm substantial rates of works-but-insecure solutions \cite{peng2025cweval, tihanyi2024howsecureisaigeneratedcode}.

\textbf{Repository-Level Understanding}~~~Even with expanded context windows, general LLMs do not robustly exploit very long inputs: performance degrades when pertinent information lies in the \emph{middle} of the context rather than near its ends \cite{liu2024lostinthemiddle}, and repository-level benchmarks covering tasks such as multi-file completion, retrieval, and editing reveal persistent difficulties in cross-file dependency tracking and global reasoning.

\textbf{Multimodal Friction}~~~General multimodal models provide useful perception for screenshots, documents, and diagrams, but fine-grained UI hierarchy and interaction semantics remain weak points. Recent analyses in GUI understanding note that existing systems often specialize in narrow sub-tasks rather than achieving holistic and consistent screen comprehension, which in turn limits stable perception-to-action transitions in real applications.

\textbf{Agentic Constraints}~~~For tool-augmented settings, benchmarked agents still fail due to brittle long-horizon reasoning, decision-making, and instruction following. Systematic evaluations highlight sizeable gaps across interactive environments and domains \cite{liu2024agentbench}, and new diagnostics document \emph{tool hallucinations} such as wrong tool choice, incorrect timing, or fabricated tool outcomes. These studies further propose reliability alignment to mitigate such issues, underscoring that robust planning and faithful tool use remain open challenges for general LLMs \cite{zhang2024toolbehonest, xu2025reducingtoolhallucination}.

Overall, breadth without domain alignment leads to gaps in depth, reliability, and system-level coherence. Addressing these limitations motivates \emph{coding-specialized} pretraining, data curation, safety alignment, and evaluation, with models optimized to act as expert programmers rather than generalists.

\subsection{Code Large Language Models}
\subsubsection{Closed-source Code Large Language Models}
\begin{figure}[h]
\begin{center}
    \includegraphics[width=0.75\textwidth]{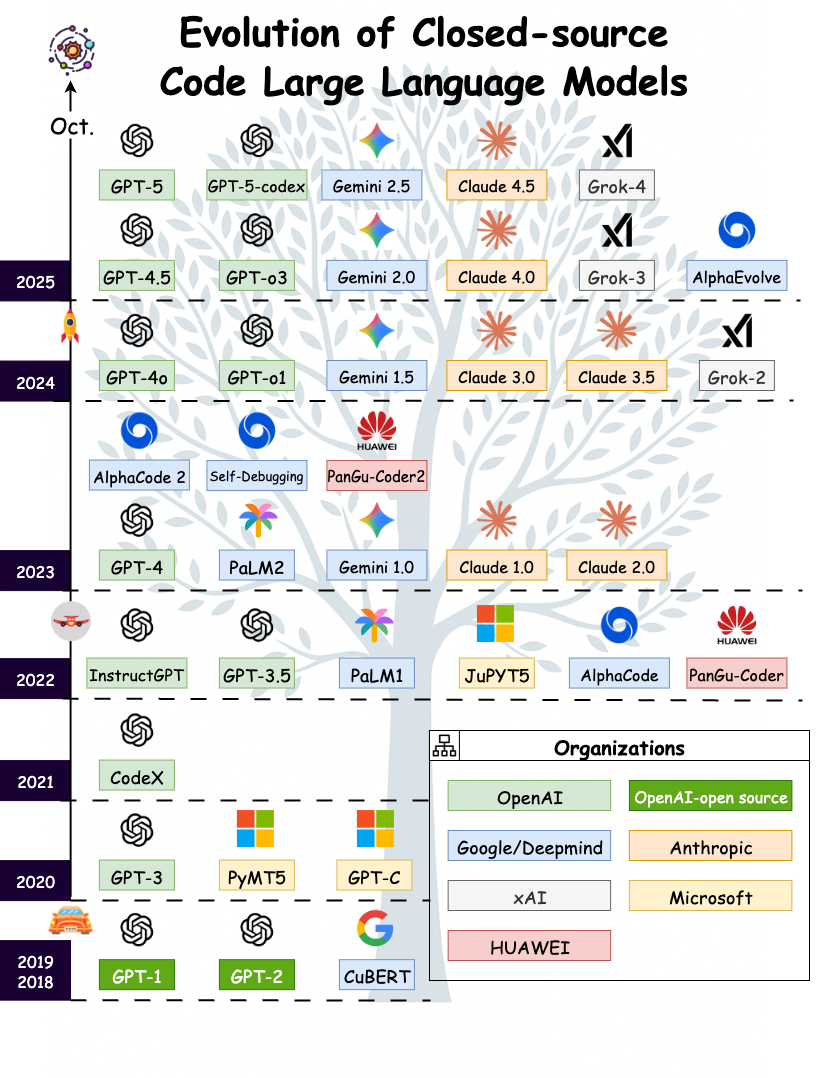}
    \caption{Evolution of closed-source large language models from 2018 to 2025. This figure depicts the chronological development of major proprietary LLMs released by leading research organizations, illustrating key milestones in the progression of model capabilities and architectures across systems such as GPT, Gemini, Claude, and Grok.}
    \label{fig:closed_source_llm}
    \vspace{-15pt}
\end{center}
\end{figure}

In \autoref{fig:closed_source_llm}, closed-source code LLMs have evolved from basic generation to agentic systems with repository-level capabilities. The GPT series~\cite{openai2024gpt4technicalreport,gpt5,openai2025gpt5codex} from OpenAI and Claude~\cite{claude3_7,claude4,claude45} from Anthropic achieve state-of-the-art results on SWE-Bench through reasoning and RL on engineering tasks.

\paragraph{GPT Series} The \textbf{GPT series} from OpenAI has strongly shaped code intelligence. Early open-weight GPT-1/2 validated generative pre-training. Proprietary successors—GPT-3, Codex, GPT-4, and the reasoning-focused \textit{o}-series—expanded from in-context learning and code synthesis to multimodal use and repository-level repair. GPT-OSS\cite{openai2025gptoss120bgptoss20bmodel} reintroduced open weights via mixture-of-experts. Most recently, GPT-5 and GPT-5-Codex set leading results on SWE-Bench and Aider Polyglot, pushing from passive generation toward agentic, feedback-driven software engineering. Overall, the family charts a path from general language modeling to systems optimized for end-to-end coding.

\begin{itemize}

\item \textbf{GPT-3}~\cite{DBLP:journals/corr/abs-2005-14165} scaled autoregressive pre-training on diverse web and curated text, and \textit{in-context learning} showed models can adapt from a few examples without gradient updates. It delivered strong zero-/few-shot results across language, reasoning, and code tasks, cementing large-scale pre-training as a foundation for code synthesis and program understanding.

\item \textbf{Codex}~\cite{chen2021codex} continued GPT-3 training on large GitHub corpora across many languages under an autoregressive decoder. It performed well on code generation and completion benchmarks (e.g., HumanEval, APPS) and powered GitHub Copilot. Conditioned on natural language, Codex synthesized code, translated between languages, and generated docstrings—an early large-scale alignment of LLMs to programming.

\item \textbf{InstructGPT}~\cite{ouyang2022traininglanguagemodelsfollow} aligned models with reinforcement learning from human feedback via supervised demonstrations, preference-based reward modeling, and PPO optimization. The resulting models were preferred by human raters, with fewer hallucinations and safer behavior; notably, a smaller aligned model surpassed a much larger base GPT-3 in preference evaluations and showed preliminary transfer to non-English and code instructions.

\item \textbf{ChatGPT}~\cite{openai2022chatgpt} (GPT-3.5) built on InstructGPT with additional instruction tuning and RLHF, stabilizing multi-turn dialogue and adding safety and refusal behaviors. Despite undisclosed details, it is broadly viewed as an extension of the GPT-3 line. As the first widely deployed conversational LLM with robust coding ability, it generated, explained, and debugged code in IDE workflows, paving the way for GPT-4.

\item \textbf{GPT-4}~\cite{openai2024gpt4technicalreport, openai2024gpt4o, openai2024gpt4omini, openai2025gpt41} advanced reasoning and code synthesis over GPT-3. GPT-4 Turbo improved efficiency for production use; GPT-4o integrated text, vision, and audio while keeping strong code performance; GPT-4o mini emphasized cost efficiency. GPT-4.1 expanded context and code-editing capabilities, enabling repository-level software engineering within the series.

\item \textbf{\textit{o}-series} targets \emph{reasoning-centered} modeling for complex problem solving with coding as a core focus. Early o1 and o1-mini introduced step-by-step internal deliberation, with o1-mini noted for software tasks~\cite{openai2024o1preview}. Successors o3 and o3-mini~\cite{openai2025o3} scaled context and optimized for repository-level editing and automated repair. On SWE-Bench Verified, the series outperformed prior GPT-4 models, establishing state-of-the-art proprietary performance in program repair and maintenance.

\item \textbf{GPT-5} was introduced as OpenAI’s most capable coding model to date, with leading results on \emph{SWE-Bench Verified} and \emph{Aider Polyglot}~\cite{openai2025gpt5developers}. \textbf{GPT-5-Codex}~\cite{openai2025gpt5codex} specializes in agentic coding via RL on real engineering tasks, sandboxed execution, and controlled tool use, deployed across CLI, IDEs, and cloud. External commentary suggests strong gains over baseline GPT-5 on synthesis tasks, though estimates remain provisional. Together they combine stronger benchmark results with interactive, feedback-driven development workflows.

\end{itemize}

\paragraph{PaLM–Gemini Series}\mbox{} Google’s \textbf{PaLM–Gemini} lineage evolves from dense, decoder-only Pathways scaling with SwiGLU and parallelized attention/FFN~\cite{chowdhery2022palmscalinglanguagemodeling} through an efficiency-oriented redesign with multilingual pre-training and UL2-style denoising~\cite{anil2023palm2technicalreport}, to native multimodality with sparse expert routing and memory-efficient long-context attention~\cite{geminiteam2025geminifamilyhighlycapable,geminiteam2024gemini15unlockingmultimodal}. Across generations, the series consolidates code intelligence for program synthesis, multilingual editing, and repository-level reasoning via scaled sequence modeling and integrated tool use.

\begin{itemize}

\item \textbf{PaLM}~\cite{chowdhery2022palmscalinglanguagemodeling} is a large decoder-only transformer using SwiGLU and parallelized attention/FFN to improve scaling. Trained on mixed natural language and substantial code, it transfers effectively to programming tasks; the finetuned \emph{PaLM-Coder} further strengthens generation, repair, and translation, showing general models adapt well to coding workloads.

\item \textbf{PaLM 2}~\cite{anil2023palm2technicalreport} refines the scaling/data balance with multilingual pre-training and UL2-style denoising, delivering stronger results at more compute-efficient sizes. Its code-specialized variant \textbf{PaLM 2-S*}—trained on multilingual code—shows competitive performance on HumanEval, MBPP, ARCADE, and BabelCode, highlighting robust cross-lingual synthesis and understanding.

\item \textbf{Gemini 1 \& 1.5}~\cite{geminiteam2025geminifamilyhighlycapable,geminiteam2024gemini15unlockingmultimodal} introduce native multimodality (text/code–vision–audio) under Pathways. Gemini 1.5 adds sparse MoE, efficiency improvements, and million-scale context, enabling repository-level comprehension and more reliable long-range code reasoning, with consistent gains over Gemini 1 on coding benchmarks (e.g., HumanEval, Natural2Code).

\item \textbf{Gemini 2  \& 2.5}~\cite{gemini2_flash_modelcard,comanici2025gemini25pushingfrontier} emphasizes efficiency, reasoning, and code intelligence. 2.0 Flash optimizes attention and memory for long contexts while retaining multimodality; 2.5 extends context length, parallelism, and agentic capabilities (tool use, iterative reasoning). Trained on mixed natural language and code and finetuned for repair, translation, and synthesis, the series reports strong results on Natural2Code, Bird-SQL, LiveCodeBench, Aider Polyglot, and SWE-Bench Verified.

\end{itemize}

\paragraph{Anthropic Claude Series} Anthropic’s \textbf{Claude} line evolves from RLHF/Constitutional-AI–aligned, decoder-only LLMs to long-context, tool-augmented agentic coders. \textbf{Claude~1$\rightarrow$2} adds long-context and safer instruction following, boosting standardized code synthesis and editing~\cite{anthropic2023introducingclaude,anthropic2023claude2,anthropic2023claude2modelcard}. \textbf{Claude~3/3.5} introduces native multimodality and function calling with documented gains on HumanEval and multi-file repository edits under sandboxed evaluation~\cite{anthropic2024claude3modelcard,anthropic2024introducingclaude3,anthropic2024claude35modelcard,anthropic2024claude35sonnet}. \textbf{Claude~4/4.5} integrates deliberative reasoning and a computer-use stack (terminal, editor, package manager, browser) with policy-controlled tool use and parallel test-time compute, showing strong results on repository-level program repair and terminal-coding suites~\cite{anthropic2025introducingclaude4,anthropic2025claude4systemcard,anthropic2025claude45}.

\begin{itemize}

\item The \textbf{Claude} family comprises proprietary decoder-only LLMs aligned via RLHF and Constitutional AI, with successive generations emphasizing longer context, safer instruction following, and robustness for structured outputs (JSON/XML and code)~\cite{anthropic2023introducingclaude,anthropic2023claude2}. \textbf{Claude~2} expands context and introduces training/service refinements for multistep reasoning and tool-friendly formatting, aiding repository comprehension, refactoring, and test-driven edits. Under standardized evaluation (e.g., HumanEval), Claude~2 shows clear gains in program synthesis~\cite{anthropic2023claude2modelcard}, translating to stronger generation, explanation, debugging, and cross-language editing in closed-source models.

\item The \textbf{Claude~3} family (Opus/Sonnet/Haiku) are proprietary, multimodal decoder-only LLMs with native tool use and vision inputs, with reported improvements in coding reliability over prior generations~\cite{anthropic2024claude3modelcard,anthropic2024introducingclaude3}. On HumanEval, Claude~3 demonstrates strong unit-style synthesis~\cite{anthropic2024claude3modelcard}. \textbf{Claude~3.5 Sonnet} further improves code performance and shows gains on repository-style multi-file editing in offline, sandboxed evaluations~\cite{anthropic2024claude35sonnet,anthropic2024claude35modelcard}. Long-context retrieval is also strengthened, supporting large-codebase comprehension~\cite{anthropic2024claude35modelcard}. Overall, the 3 $\rightarrow$ 3.5 transition centers on multimodal, tool-augmented modeling with improved synthesis and repository-level editing under controlled tests.

\item The \textbf{Claude~4} family integrates hybrid (deliberative) reasoning with first-class agentic coding and a computer-use toolchain (sandboxed shell, editor, package manager, browser), trained and aligned via RLHF and Constitutional AI~\cite{anthropic2025introducingclaude4,anthropic2025claude4systemcard}. The system card details coding-specific safeguards and safety instrumentation for tool use, alongside dedicated evaluations for agentic coding and terminal workflows~\cite{anthropic2025claude4systemcard}. On SWE-bench Verified, Claude~4 reports strong program-repair accuracy, further improved by parallel test-time compute. \textbf{Claude~4.5 (Sonnet)} advances repository-level repair and shows gains on terminal-coding and tool-use suites~\cite{anthropic2025claude4systemcard,anthropic2025claude45}. Collectively, Claude~4/4.5 shift toward long-horizon, tool-augmented coding agents that deliberate, invoke tools under policy controls, and iteratively validate patches, yielding measurable improvements in repair and structured editing.

\end{itemize}

\paragraph{Others}

\begin{itemize}

\item \textbf{Grok Series}~ xAI’s \textbf{Grok} evolves from a proprietary, instruction-following decoder-only LLM into an agentic, code-oriented family with longer context and specialized coding variants. \textbf{Grok-1} shipped with Chat and later released as open weights, enabling public inspection and downstream use~\cite{xai2023announcinggrok,xai2024grok15}. \textbf{Grok-1.5} introduced a 128k-token window with stronger math/coding and long-context reasoning for repository-scale comprehension/editing~\cite{xai2024grok15}. \textbf{Grok-2} reported gains on standardized coding evaluations such as HumanEval~\cite{xai2024grok2}. The \textbf{Grok-4} generation emphasizes native tool use and “think” modes with real-time search, plus a code-specialized endpoint (\texttt{grok-code-fast-1}) for synthesis, refactoring, and repair loops~\cite{xai2025grok4,xai2025grok4fast,xai2025apidocs}. Overall, Grok integrates longer context, tool-grounded reasoning, and a code-optimized serving path aligned with developer workflows.

\item \textbf{PanGu-Coder}~\cite{christopoulou2022pangucoderprogramsynthesisfunctionlevel} uses a decoder-only transformer (PanGu-$\alpha$) for function-level synthesis, translating between docstrings, signatures, and method bodies. Training follows large-scale causal pre-training on mixed language/code, then task adaptation on docstring--function pairs with code-focused losses (e.g., CODE-CLM). Emphasizing code tokens during fine-tuning outperforms docstring-side denoising, and the released 317M model is competitive on HumanEval under multi-sample evaluation. \textbf{PanGu-Coder2}~\cite{shen2023pangucoder2boostinglargelanguage} scales to 15B with longer context and introduces ranking-feedback alignment (RRTF): offline solutions are ranked by unit-test signals and teacher preferences, then optimized with a ranking loss. With expanded, leakage-screened instructions, it reports consistent gains on HumanEval and broader suites, showing that execution-aware ranking improves code generation without heavy online RL.

\item \textbf{PyMT5}~\cite{clement-etal-2020-pymt5} casts method-level NL$\leftrightarrow$Python generation as a T5-style seq2seq multi-mode translation problem, where a single encoder--decoder model reconstructs any missing method feature (signature, docstring, body) through span-masking and feature-filling denoising. This unified formulation enforces cross-feature consistency and preserves syntactic structure, enabling controlled, feature-conditioned generation of Python methods.\textbf{JuPyT5}~\cite{chandel2022trainingevaluatingjupyternotebook} extends this paradigm to Jupyter notebooks via a cell-infilling objective that predicts each cell from its surrounding context, modulated by cell-type control codes. This notebook-aware seq2seq scheme models cross-cell dependencies and executable semantics, framing notebook code generation as structured infilling under test-driven supervision.

\item \textbf{AlphaCode}~\cite{li2022competition} treats competitive programming as sequence-to-sequence translation from long natural-language statements to full programs, coupling large encoder--decoder transformers (pretrained on multilingual GitHub code and fine-tuned on the CodeContests dataset~\cite{li2022competition}) with massive stochastic sampling, execution-based filtering on public tests, and behavioural clustering over model-generated test inputs to select a small, diverse set of candidate solutions. \textbf{AlphaCode~2}~\cite{alphacode2_technical_report} refines this pipeline with Gemini-based policies and a learned scoring model, applying two-stage fine-tuning on an updated \emph{CodeContests v2} dataset and a curated higher-quality problem set~\cite{alphacode2_technical_report}, while aggressive execution filtering, behavioural clustering, and reranking concentrate the submission budget on high-likelihood, semantically diverse candidates. \textbf{AlphaEvolve}~\cite{novikov2025alphaevolvecodingagentscientific} instead casts program synthesis as evolutionary search in code-edit space, maintaining a population of programs and iteratively applying LLM-generated diff-style patches, compiling and executing under task-specific tests, and selecting high-scoring descendants, thereby exploiting structured edits and test-time compute for scientific and algorithmic discovery.

\end{itemize}

\subsubsection{Open-source Code Large Language Models}

\begin{table}[!h]
\centering
\caption{Open-source code-specialized LLMs.}
\label{tab:opensource_code_llms}
\resizebox{1.0\columnwidth}{!}{%
\begin{tabular}{lllllll}
\toprule
\textbf{Model} & \textbf{Layers} & \textbf{Hidden Size} & \textbf{Intermediate Size} & \textbf{Attention Method} & \textbf{Max Context} & \textbf{Extra} \\
\midrule
StarCoder 15B & 40 & 6144 & 24576 & MQA & 8192 & \makecell[l]{Multi Query Attention} \\
\midrule
StarCoder2-3B & 30 & 3072 & 12288 & GQA & \makecell[l]{16384\\(sliding 4096)} & \makecell[l]{BigCode consortium} \\
StarCoder2-7B & 32 & 4608 & 18432 & GQA & \makecell[l]{16384\\(sliding 4096)} & \makecell[l]{Multiple data sources} \\
StarCoder2-15B & 40 & 6144 & 24576 & GQA & \makecell[l]{16384\\(sliding 4096)} & \makecell[l]{Largest variant} \\
\midrule
Code Llama-7B & 32 & 4096 & 11008 & GQA & \makecell[l]{16k training\\(supports 100k)} & \makecell[l]{Based on Llama2 architecture} \\
Code Llama-13B & 40 & 5120 & 13824 & GQA & 16k & \makecell[l]{Python specialization} \\
Code Llama-34B & 48 & 8192 & 22016 & GQA & 16k & Larger version \\
\midrule
Qwen2.5-Coder-7B & 28 & 3584 & 18944 & GQA & \makecell[l]{131072\\(with YaRN)} & \makecell[l]{Base context 32768} \\
Qwen2.5-Coder-32B & 64 & 5120 & 27648 & GQA & 131072 & State-of-the-art \\
Qwen3-Coder-30B-A3B & 48 & 5120 & 25600 & GQA+MoE & \makecell[l]{262144\\(1M w/ YaRN)} & \makecell[l]{MoE: 30B total ~3.3B active\\(128 experts, 8 activated)} \\
Qwen3-Coder-480B-A35B & 62 & 6144 & 8192 & GQA+MoE & \makecell[l]{262144\\(1M w/ YaRN)} & \makecell[l]{MoE: 480B total ~35B active\\(160 experts, 8 activated)} \\
\midrule
IBM Granite Code-3B & 28 & 2560 & 10240 & GQA & 8192 & \makecell[l]{116 languages} \\
IBM Granite Code-8B & 36 & 4096 & 16384 & GQA & 8192 & \makecell[l]{Enterprise focused} \\
IBM Granite Code-20B & 52 & 6144 & 24576 & GQA & 8192 & \makecell[l]{High performance} \\
IBM Granite Code-34B & 88 & 6144 & 24576 & GQA & 8192 & \makecell[l]{Depth upscaling} \\
\midrule
DeepSeek-Coder V2-Lite & 27 & 2048 & — & MLA + MoE & 128k & \makecell[l]{MoE: 16B total ~2.4B active\\(2 shared + 64 routed)} \\
DeepSeek-Coder V2-236B & 60 & 5120 & 12288 & MLA + MoE & 128k & \makecell[l]{MoE: 236B total ~21B active\\338 languages} \\
\midrule
Codestral-22B & 56 & 6144 & 16384 & GQA & 32k & \makecell[l]{80+ languages FIM support} \\
Codestral 25.01 & - & - & - & - & 256k & \makecell[l]{Latest version\\2$\times$ faster\\80+ languages\\API only} \\
Codestral Mamba-7B & 64 & 4096 & — & Mamba2 SSM & 256k & \makecell[l]{State Space\\Model\\7.3B params} \\
\midrule
Microsoft Phi-4 & 40 & 5120 & 17920 & \makecell[l]{Full\\Attention} & \makecell[l]{16k\\(ext. from 4k)} & \makecell[l]{Strong math reasoning} \\
\midrule
Replit Code v1-3B & 32 & 2560 & 10240 & MQA & 4096 & \makecell[l]{Trained on Stack Dedup} \\
StableCode-3B & 32 & 2560 & 10240 & MQA & 16384 & \makecell[l]{Fill-in-the Middle} \\
WizardCoder-15B & 40 & 6144 & 24576 & MQA & 8192 & \makecell[l]{Fine-tuned StarCoder} \\
Magicoder-6.7B & 32 & 4096 & 11008 & GQA & 16384 & \makecell[l]{Based on Code Llama} \\
CodeGeeX2-6B & 28 & 4096 & 13696 & MHA & 8192 & \makecell[l]{Based on ChatGLM2} \\
CodeGeeX4-ALL-9B & 40 & 4096 & 14336 & GQA & 131072 & \makecell[l]{Multi-language} \\
OctoCoder-15.5B & 40 & 6144 & 24576 & MQA & 8192 & \makecell[l]{Fine-tuned StarCoder} \\
Yi-Coder-1.5B & 28 & 2048 & 8192 & GQA & 131072 & \makecell[l]{52 languages} \\
OpenCoder-1.5B & 24 & 2240 & 6144 & GQA & 4096 & \makecell[l]{Fully open-source} \\
OpenCoder-8B & 32 & 4096 & 14336 & GQA & 8192 & \makecell[l]{2.5T tokens training} \\
\bottomrule
\end{tabular}
}
\end{table}

\begin{figure}[h]
    \centering
    \includegraphics[width=1.0\textwidth]{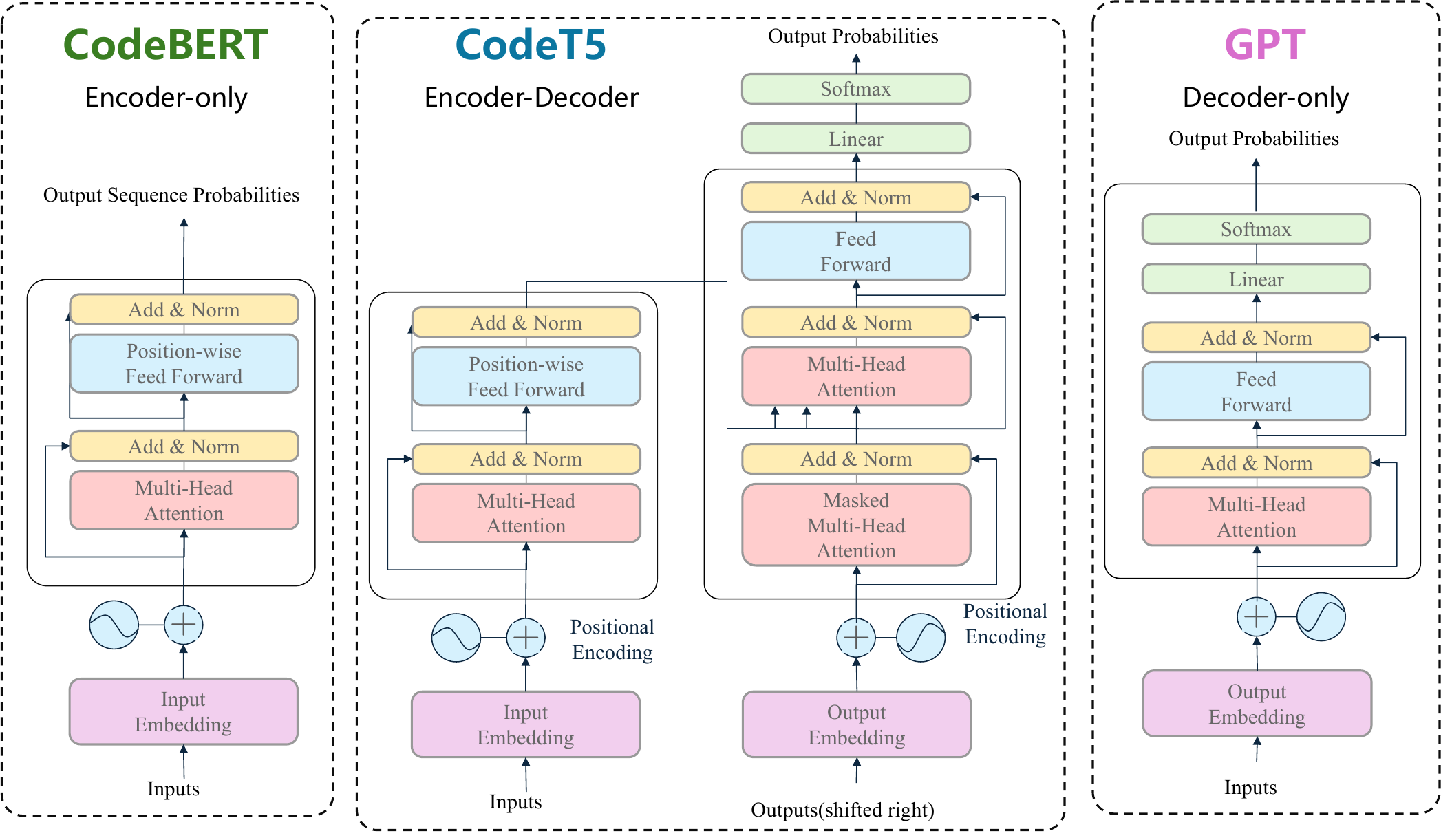} 
    \caption{Comparison of model architectures for CodeBERT, CodeT5, and GPT.}
    \label{fig:model_structure_revelution}
\end{figure}

As shown in \autoref{tab:opensource_code_llms}, This subsection concisely reviews classical encoder-centric NL (natural language) - PL (programming language) embedding methods, highlighting core architectures, denoising/contrastive pre-training over code–text corpora, and their primary uses in retrieval and code understanding that underpin modern open-source Code LLMs. \autoref{fig:model_structure_revelution} illustrates representative model structures including encoder-only, encoder–decoder, and decoder-only designs.

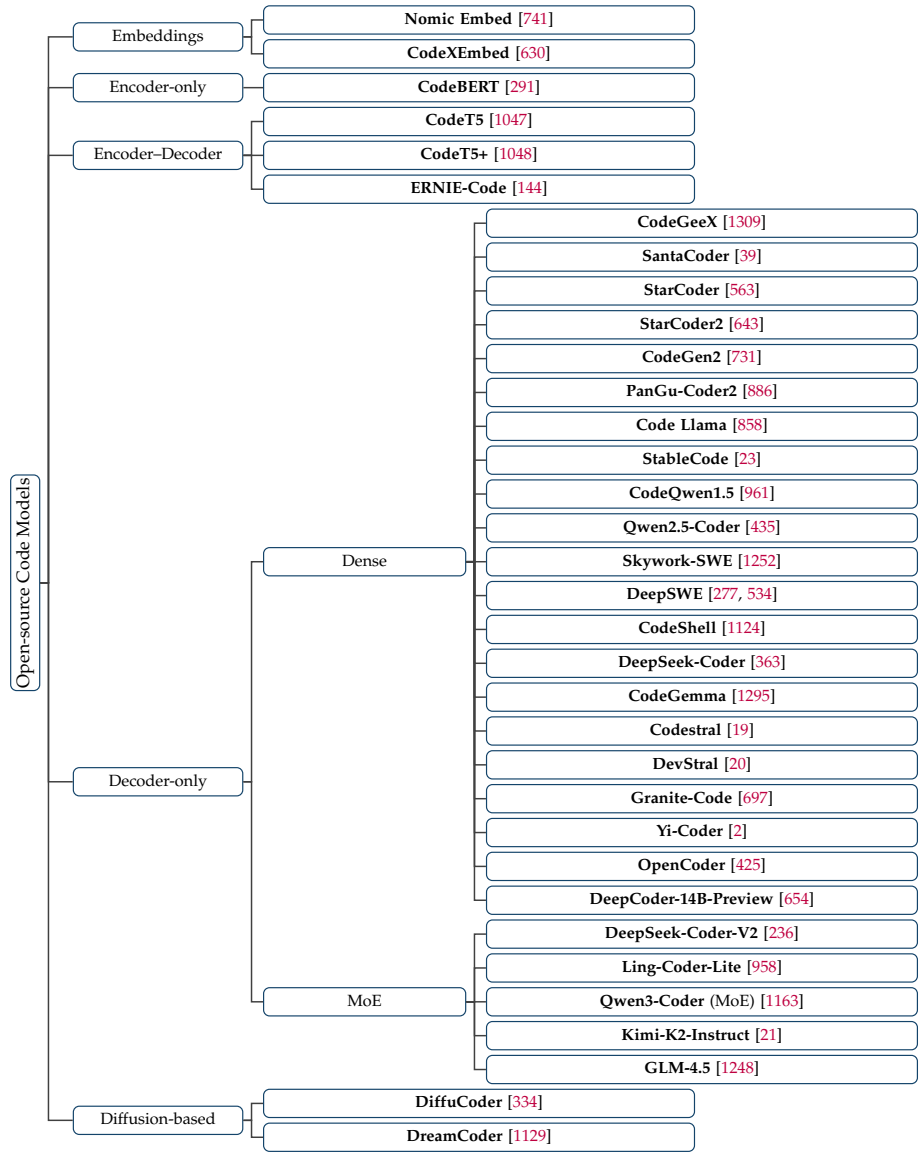
\begin{figure*}[h] 
    \centering
    \resizebox{0.75\textwidth}{!}{
    \begin{forest}
        forked edges,
        for tree={
                grow=east,
                reversed=true,
                anchor=base west,
                parent anchor=east,
                child anchor=west,
                base=center,
                font=\large,
                rectangle,
                draw=hidden-draw,
                rounded corners,
                align=left,
                text centered,
                minimum width=4em,
                edge+={darkgray, line width=1pt},
                s sep=3pt,
                inner xsep=2pt,
                inner ysep=3pt,
                line width=0.8pt,
                ver/.style={rotate=90, child anchor=north, parent anchor=south, anchor=center},
            },
            where level=1{text width=10em,font=\normalsize, }{},
            where level=2{text width=12em,font=\normalsize}{} ,
            where level=3{text width=12em,font=\normalsize,}{} ,
        [Open-source Code Models, ver
            [Embeddings
                [\textbf{Nomic Embed}~\cite{nussbaum2025nomicembedtrainingreproducible}, text width=26em]
                [\textbf{CodeXEmbed}~\cite{liu2025codexembedgeneralistembeddingmodel}, text width=26em]
            ]
            [Encoder-only
                [\textbf{CodeBERT}~\cite{feng2020codebertpretrainedmodelprogramming}, text width=26em]
            ]
            [Encoder--Decoder
                [\textbf{CodeT5}~\cite{wang-etal-2021-codet5}, text width=26em]
                [\textbf{CodeT5+}~\cite{wang-etal-2023-codet5}, text width=26em]
                [\textbf{ERNIE-Code}~\cite{chai2023erniecodeenglishcentriccrosslingualpretraining}, text width=26em]
            ]
            [Decoder-only
                [Dense
                    [\textbf{CodeGeeX}~\cite{zheng2023codegeex}, text width=26em]
                    [\textbf{SantaCoder}~\cite{allal2023santacoder}, text width=26em]
                    [\textbf{StarCoder}~\cite{li2023starcoder}, text width=26em]
                    [\textbf{StarCoder2}~\cite{lozhkov2024starcoder2stackv2}, text width=26em]
                    [\textbf{CodeGen2}~\cite{nijkamp2023codegen2}, text width=26em]
                    [\textbf{PanGu-Coder2}~\cite{shen2023pangucoder2}, text width=26em]
                    [\textbf{Code Llama}~\cite{roziere2023code}, text width=26em]
                    [\textbf{StableCode}~\cite{stabilityai2023stablecode}, text width=26em]
                    [\textbf{CodeQwen1.5}~\cite{qwen2024codeqwen1.5}, text width=26em]
                    [\textbf{Qwen2.5-Coder}~\cite{qwen25coder}, text width=26em]
                    [\textbf{Skywork-SWE}~\cite{zeng2025skyworkswe}, text width=26em]
                    [\textbf{DeepSWE}~\cite{li2025towards,deepswe2025}, text width=26em]
                    [\textbf{CodeShell}~\cite{xie2024codeshell}, text width=26em]
                    [\textbf{DeepSeek-Coder}~\cite{guo2024deepseek}, text width=26em]
                    [\textbf{CodeGemma}~\cite{codegemma2024}, text width=26em]
                    [\textbf{Codestral}~\cite{mistral2024codestral}, text width=26em]
                    [\textbf{DevStral}~\cite{devstral2025}, text width=26em]
                    [\textbf{Granite-Code}~\cite{DBLP:journals/corr/abs-2405-04324}, text width=26em]
                    [\textbf{Yi-Coder}~\cite{yicoder2024}, text width=26em]
                    [\textbf{OpenCoder}~\cite{huang2025opencoder}, text width=26em]
                    [\textbf{DeepCoder-14B-Preview}~\cite{deepcoder2025}, text width=26em]
                ]
                [MoE
                    [\textbf{DeepSeek-Coder-V2}~\cite{zhu2024deepseekcoderv2}, text width=26em]
                    [\textbf{Ling-Coder-Lite}~\cite{team2025every}, text width=26em]
                    [\textbf{Qwen3-Coder} (MoE)~\cite{yang2025qwen3}, text width=26em]
                    [\textbf{Kimi-K2-Instruct}~\cite{moonshotai2025kimik2}, text width=26em]
                    [\textbf{GLM-4.5}~\cite{glm2025glm4_5}, text width=26em]
                ]
            ]
            [Diffusion-based
                [\textbf{DiffuCoder}~\cite{gong2025diffucoder}, text width=26em][\textbf{DreamCoder}~\cite{xie2025dream}, text width=26em]
            ]
        ]
    \end{forest}}
    \caption{Taxonomy of selected open-source code models grouped by architecture. }
    \label{fig:taxonomy-open-source-code}
\end{figure*}

\newcommand{\arcdec}{%
  \raisebox{-0.13em}{%
    \includegraphics[height=1em]{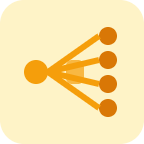}%
  }%
}
\newcommand{\arcenc}{%
  \raisebox{-0.13em}{%
    \includegraphics[height=1em]{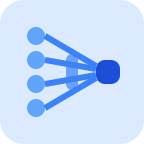}%
  }%
}
\newcommand{\arcencdec}{%
  \raisebox{-0.13em}{%
    \includegraphics[height=1em]{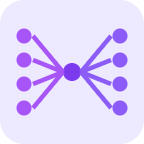}%
  }%
}
\newcommand{\arcdiff}{%
  \raisebox{-0.13em}{%
    \includegraphics[height=1em]{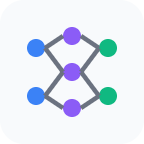}%
  }%
}
\newcommand{\arcmoe}{%
  \raisebox{-0.13em}{%
    \includegraphics[height=1em]{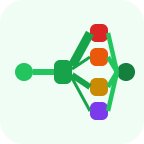}%
  }%
}
\newcommand{\funcembed}{%
  \raisebox{-0.13em}{%
    \includegraphics[height=1em]{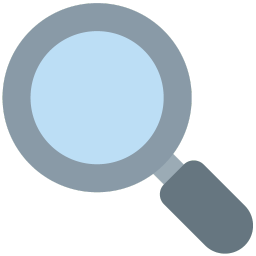}%
  }%
}
\newcommand{\funcgen}{%
  \raisebox{-0.13em}{%
    \includegraphics[height=1em]{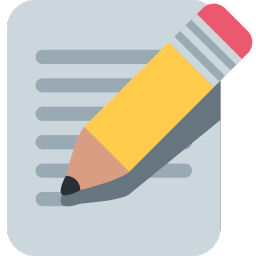}%
  }%
}
\newcommand{\funcunderstand}{%
  \raisebox{-0.13em}{%
    \includegraphics[height=1em]{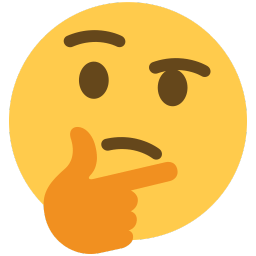}%
  }%
}
\newcommand{\funcfim}{%
  \raisebox{-0.13em}{%
    \includegraphics[height=1em]{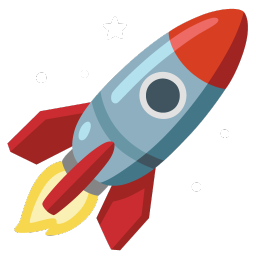}%
  }%
}
\newcommand{\funcswe}{%
  \raisebox{-0.13em}{%
    \includegraphics[height=1em]{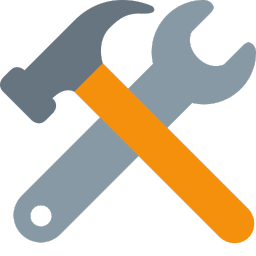}%
  }%
}

\subsubsection{Evolution of Open-Source Code Large Language Models}

The development of open-source code large language models can be systematically categorized into four distinct evolutionary stages based on their architectural innovations and functional capabilities. This taxonomy provides a comprehensive framework for understanding the technological progression in the open-source code intelligence community.

\paragraph{Icon legend.}
Throughout this subsection, we annotate models with small icons indicating their
architectures and primary capabilities.

\medskip

\noindent\emph{Architecture icons}
\begin{itemize}
  \item \arcenc{} — encoder-only
  \item \arcencdec{} — encoder–decoder
  \item \arcdec{} — decoder-only
  \item \arcdiff{} — diffusion-based
  \item \arcmoe{} — mixture-of-experts (MoE)
\end{itemize}

\noindent\emph{Functional icons}
\begin{itemize}
  \item \funcembed{} — retrieval / embedding
  \item \funcunderstand{} — code understanding
  \item \funcgen{} — code generation
  \item \funcfim{} — fill-in-the-middle / infilling
  \item \funcswe{} — software-engineering agents
\end{itemize}

\medskip

\textbf{Stage 1: Pre-trained Encoder Models.} The initial stage was dominated by encoder-based pre-trained models such as CodeBERT~\cite{graph_code_bert}, GraphCodeBERT~\cite{graph_code_bert}, and CodeT5~\cite{chen2023codet5}. These open-source models primarily focused on code understanding tasks, establishing fundamental code-text alignment through bidirectional attention mechanisms. Their core strengths lay in code classification, vulnerability detection, and semantic code search.

\begin{itemize}
\item \arcenc \textbf{CodeBERT}~(\funcunderstand)~\cite{feng2020codebertpretrainedmodelprogramming} is an encoder-only RoBERTa-style model pre-trained on paired natural language and source code using a hybrid objective (masked prediction on NL–PL pairs plus replaced-token detection) with data from CodeSearchNet. It is primarily used for representation tasks (retrieval, reranking, classification) and is typically combined with a decoder when applied to generation.

\item \arcencdec \textbf{ERNIE-Code}~(\funcgen\funcunderstand)~\cite{chai2023erniecodeenglishcentriccrosslingualpretraining} is a multilingual text–code encoder–decoder built on the T5 line with a single vocabulary covering many natural and programming languages and added tokens to capture code layout. Its pretraining mixes span-corruption on text and code with a pivot-based translation objective to promote cross-lingual and cross-modal alignment; fine-tuned ERNIE-Code shows strong transfer on summarization, text-to-code, translation and program repair.
\end{itemize}

\textbf{Stage 2: Generative Models.} The second stage witnessed the emergence of open-source generative models including CodeT5~\cite{chen2023codet5} and CodeGPT~\cite{codegpt}, which introduced encoder-decoder architectures capable of both code understanding and generation. These models demonstrated proficiency in code generation, cross-language translation, and automated code completion tasks.

\begin{itemize}
\item \arcdec\textbf{CodeParrot}~(\funcgen)~\cite{codeparrot} is a family of decoder-only GPT-2 models (110M, 1.5B parameters) specifically trained for Python code generation tasks. Built upon the cleaned CodeParrot dataset derived from GitHub dumps with aggressive deduplication and filtering heuristics, the model employs a GPT-2 tokenizer trained on code-specific vocabulary. The training methodology uses standard autoregressive language modeling with left$\to$right generatio. CodeParrot excels at Python code completion, docstring$\to$code generation, and unittest generation tasks, demonstrating strong performance on code synthesis despite being trained on significantly fewer tokens than larger models. The model architecture and training data are fully open-sourced, enabling reproducible research in neural code generation.
\item \arcdec \textbf{CodeGPT}~(\funcunderstand\funcgen)~\cite{codegpt} is a family of GPT-style transformer models (110M, 1.5B parameters) developed by Microsoft Research for Python code understanding and generation. Built upon large-scale filtered GitHub repositories with aggressive data cleaning and deduplication strategies, the model employs a combination of masked language modeling and next token prediction during pre-training. The training methodology incorporates multi-task learning across diverse code-related objectives including code completion, NL$\to$code generation, code$\to$NL summarization, and bug detection tasks. PyCodeGPT excels at syntactic and semantic code understanding, enabling applications in automated code review, documentation generation, and program repair. The model architecture and training approach contribute to Microsoft's broader research initiative in AI-assisted software development, demonstrating strong capabilities in code completion, comment generation, and educational programming assistance.
\item \arcencdec \textbf{T5 series}~(\funcgen\funcunderstand)~\cite{wang-etal-2021-codet5,wang-etal-2023-codet5} are T5-derived models for code understanding and generation that use a code-aware tokenizer and identifier-aware pretraining, alternating unimodal and bimodal data and employing a dual-generation stage to align NL and PL. The family spans encoder/decoder and seq2seq variants (from compact to large), applies a two-stage pretraining plus instruction tuning recipe, and is competitive across many code tasks.

\end{itemize}

\textbf{Stage 3: Large Language Models.} The third stage marked a paradigm shift with the advent of large-scale open-source language models such as StarCoder~\cite{li2023starcoder}, CodeLlama~\cite{roziere2024codellama}, DeepSeek-Coder~\cite{deepseek2024coder}, and CodeQwen~\cite{qwen2024codeqwen1.5}. These models exhibited remarkable capabilities in complex code generation, multi-turn conversational programming, and instruction following, demonstrating that open-source models could achieve competitive performance with proprietary counterparts.

\begin{itemize}
\item \arcdec\textbf{SantaCoder}~(\funcunderstand\funcgen)~\cite{allal2023santacoder} is an open-source decoder transformer from the BigCode project. It adopts Multi-Query Attention (MQA) for efficient serving and a Fill-in-the-Middle (FIM) objective to support both left-to-right generation and code infilling. Pretraining used permissively-licensed code (Python/Java/JavaScript) with strict data governance: opt-out honoring, PII redaction, aggressive near-deduplication, and documentation-aware filtering. A two-phase training recipe validated design choices before a final large-scale run. SantaCoder supports text-to-code, infilling, and multilingual synthesis, performs well on multi-language code benchmarks (e.g., MultiPL-E) compared to some larger models.
\item \arcdec\textbf{OctoCoder}~(\funcunderstand\funcgen)~\cite{humanevalpack} 
is an instruction-tuned Code LLM (StarCoder-16B base) trained on permissively licensed, commit-derived instructions mixed with natural-language targets. The work also releases \textbf{HumanEvalPack}, extending HumanEval to three tasks—code repair, explanation, and synthesis—across six languages (Python, JS, Java, Go, C++, Rust). OctoCoder attains the best average pass@1 among commercially usable (permissive) models on this suite—e.g., strong gains in bug-fixing from commit-style data and solid synthesis—while closed models like GPT-4 remain higher overall. The paper underscores practical deployability (permissive licensing, OpenAI-output–free data), multilingual generalization from pretraining, and the importance of mixing NL targets to avoid code-only bias.    
\item \arcdec\textbf{CodeGeeX}~(\funcunderstand\funcgen)~\cite{zheng2023codegeex} is a multilingual GPT-style decoder LLM aimed at generation and translation. It was pretrained on a large multilingual code corpus spanning many languages and introduces \textbf{HumanEval-X}, a multi-language suite of canonical problems for generation and cross-lingual translation evaluation. The work emphasizes deployability (INT8 quantized inference, integration with FasterTransformer) and developer tooling (IDE plugins), and shows that CodeGeeX is a top-performing open multilingual baseline, competitive with comparable models depending on language; a fine-tuned variant further improves translation. The release also reports substantial real-world usage and user-reported productivity gains.
\item \arcdec\textbf{StarCoder}~(\funcunderstand\funcgen)~\cite{li2023starcoder} and \textbf{StarCoderBase} are open-access decoder-only code models with long context and FIM training. StarCoderBase was trained on a large permissive-code collection (The Stack) and StarCoder is obtained after targeted fine-tuning on additional Python data. The project prioritizes data governance (near-deduplication, benchmark decontamination, PII redaction) and practical engineering (tokenizer sentinels for FIM, efficient attention/backends for long context). Evaluated across diverse code benchmarks, StarCoderBase/StarCoder lead among open multilingual code LLMs and compare favorably to some closed models. The release includes IDE demos and an OpenRAIL-M license that pairs permissive access with documented usage restrictions. 
\item \arcdec\textbf{CodeGen2}~(\funcunderstand\funcgen)~\cite{nijkamp2023codegen2} presents an open-source family of decoder-only code LLMs and a single, practical recipe that unifies architecture choices, sampling modes (left-to-right \& infill), and mixed NL/PL data. The study tests a Prefix-LM unification but finds \emph{no consistent gains} over a causal decoder; the final recipe therefore uses a decoder-only transformer with a mixed objective: with probability \(p=0.5\) train by next-token prediction (CLM), otherwise apply within-file span corruption (dynamic mask ratios/lengths) to enable infilling. Infill training is shown to trade off slightly with pure left-to-right performance, NL+PL mixing offers a robust compromise when one model must cover both modalities, and continued multi-epoch pretraining (the CodeGen2.5 variant) yields clear scaling benefits. Overall lessons: Prefix-LM is not superior for these tasks, infill is not free, a CLM+span-corruption mixture is effective, NL+PL mixing is promising under constrained compute, and multi-epoch training is important.
\item \arcdec\textbf{Code LLaMA}~(\funcunderstand\funcgen)~\cite{roziere2023code} is a LLaMA~2–based code family that emphasizes strong infilling, long-context support, and instruction-following for programming. It ships in foundation, Python-specialized, and instruction-tuned variants across multiple scales and is trained/finetuned for long sequences and repository-scale completion (RoPE base period increased from \(10^{4}\) to \(10^{6}\) during long-context fine-tuning). Training continues from LLaMA~2 on a code-heavy corpus; the Python and Instruct variants add focused token streams for language specialization and alignment. Ablations report modest trade-offs from infill training, clear gains from long-context fine-tuning for repository tasks, and consistent benefits from language specialization; safety-tuned instruct models improve truthfulness and reduce toxicity while preserving coding ability.
\item \arcdec\textbf{MFTCoder}~(\funcunderstand\funcgen)~\cite{MFTCoder} proposes a multi-task fine-tuning framework that trains a single decoder-only backbone to handle completion, text-to-code, commenting, translation, and unit-test generation concurrently. It addresses multi-task issues via a data-balanced, token-weighted loss, focal-style emphasis at sample and task levels, and a validation-driven dynamic reweighting that prioritizes slowest-converging tasks. Efficiency techniques—dynamic padding, packed SFT (concatenating samples with \texttt{eos}), and PEFT support (LoRA/QLoRA)—reduce padding and enable practical fine-tuning of large bases on modest hardware. Applied across multiple models, MFTCoder shows consistent gains over single-task SFT and mixed-data SFT and better generalization on unseen tasks.
\item \arcdec\textbf{DeepSeek-Coder}~(\funcunderstand\funcgen \funcfim)~\cite{guo2024deepseek} is an open-source code LLM family (1.3B–33B) trained from scratch on multi-language corpora. A key idea is repository-level pretraining that models cross-file dependencies, improving repo understanding and cross-file completion. It integrates a fill-in-the-middle objective with long context (up to 16{,}384 tokens), enhancing FIM infilling and long-range code reasoning. It reports strong results on HumanEval and MBPP, exceeding GPT-3.5 without proprietary data. Instruction-tuned variants show robust multi-turn problem solving, and the permissive license supports reproducibility and practical adoption.
\item \arcdec\textbf{StableCode}~(\funcunderstand\funcgen \funcfim)~\cite{stabilityai2023stablecode} is a 3B lightweight open model for code generation and understanding, trained on large GitHub corpora. It supports completion and natural language $\to$ code, with long-context handling (up to 16{,}384 tokens) for multi-file reasoning. Performance on HumanEval/MBPP is competitive among compact open models, trading peak accuracy for efficiency and easy deployment under limited compute.
\item \arcdec\textbf{StarCoder2}~(\funcunderstand\funcgen \funcfim)~\cite{lozhkov2024starcoder2stackv2} advances the BigCode line with \textbf{The Stack v2} (larger and more diverse, partnered with Software Heritage). Models at 3B/7B/15B are trained across hundreds of languages plus issues/PRs, docs, and math/logic data. Training uses two stages (4k $\to$ 16k) with repository-context formatting and FIM. On benchmarks, \textbf{StarCoder2-3B} surpasses similar-size peers, and \textbf{StarCoder2-15B} matches or exceeds larger models, with strong math and low-resource language performance. 
\item \arcdec\textbf{CodeShell}~(\funcunderstand\funcgen )~\cite{xie2024codeshell} is a 7B foundation model (8k context) extending GPT-2 with grouped-query attention and RoPE for efficient inference. Its hallmark is rigorous data governance: multi-stage filtering (deduplication, perplexity, structure rules, model-based scoring) to build a high-quality corpus. Despite modest scale, CodeShell outperforms comparable 7B models and shows competitive results on MultiPL-E and code completion, supporting the view that careful data curation can rival sheer scaling.
\item \arcdec\textbf{CodeGemma}~(\funcunderstand\funcgen\funcfim)~\cite{codegemma2024} adapts Gemma for coding via large-scale code-centric pretraining and instruction tuning. The suite includes a 2B model for low-latency completion/infilling and 7B variants (pretrained and instruction-tuned). Curated corpora employ deduplication, contamination removal, and multi-file packing guided by dependency graphs and tests; an improved FIM objective (high-rate) supports both prefix–suffix–middle and suffix–prefix–middle modes. The 7B-IT model performs strongly on HumanEval/MBPP and multilingual coding (e.g., BabelCode), with solid mathematical reasoning, while the 2B model offers competitive accuracy with fast inference for IDE use.
\item \arcdec \textbf{Granite-Code}~(\funcunderstand\funcgen\funcfim)~\cite{DBLP:journals/corr/abs-2405-04324} is a decoder-only open-source family from IBM (3B–34B) trained in two stages (large-scale code pretraining \(\rightarrow\) mixed code+NL enhancement). It integrates Fill-in-the-Middle (PSM/SPM) with causal LM, improving infilling and completion. The series shows solid results on coding and explanation/fix tasks while emphasizing enterprise-grade data transparency and Apache~2.0 licensing for practical deployment.
\item \arcdec \textbf{Codestral}~(\funcunderstand\funcgen\funcfim)~\cite{mistral2024codestral} is a 22B open-weight code model focused on instruction following and FIM across many languages, with a \(\sim\)32K context for repository-level reasoning. Public materials report strong long-range and FIM performance, including on RepoBench and Python-oriented evaluations. It is released for research under the Mistral AI Non-Production License with separate commercial options.

\arcdec\textbf{Yi-Coder}~(\funcunderstand\funcgen)~\cite{yicoder2024} targets high coding quality under compact sizes (1.5B/9B; base and chat) with up to 128K context over 52 languages. The models prioritize inference efficiency and interactive debugging, showing robust outcomes on HumanEval, MBPP, and LiveCodeBench relative to larger peers. Weights, code, and deployment guides are provided under Apache~2.0 for straightforward IDE and production integration.

\item \arcdec\textbf{CodeQwen1.5 \& Qwen2.5-Coder}~(\funcunderstand\funcgen\funcfim)~\cite{qwen2024codeqwen1.5} is a 7B decoder-only model trained on large-scale code corpora, covering many languages and long-context (up to 64K). It adopts GQA for efficient inference and extended context, demonstrating strong $\text{text}\!\rightarrow\!\text{SQL}$, bug fixing, and debugging. Qwen2.5-Coder~\cite{qwen25coder} expands to a family (0.5B–32B) trained on a balanced mix of code, natural language, and math. It combines file$\rightarrow$repository pretraining with FIM, and scales context to 128K via YARN. Instruction tuning blends multilingual synthesis and DPO with execution feedback, yielding solid results on MultiPL-E, RepoEval, and CrossCodeEval without relying on narrow prompt formats.

\arcdec\textbf{OpenCoder}~(\funcunderstand\funcgen)~\cite{huang2025opencoder} emphasizes full reproducibility: weights, inference code, curated RefineCode data, processing pipelines, and checkpoints are all released. Models (1.5B/8B) use a LLaMA-style transformer (RoPE, SwiGLU) and a two-stage instruction plan (general SFT \(\rightarrow\) code-specific SFT). The 8B variants report strong HumanEval/MBPP, multilingual (MultiPL-E), and debugging performance, surpassing StarCoder2-15B and CodeLlama-7B. The project serves both as a capable model and an open recipe for scientific reuse.
\end{itemize}

\textbf{Stage 4: Advanced Scaling and Agentic Models.} The current stage represents two major developments: massive parameter scaling through mixture-of-experts (MoE) architectures that maintain high inference efficiency while dramatically increasing model capacity, and the evolution toward agentic systems that integrate tool usage, multi-step reasoning, and environment interaction capabilities. Representative models like DeepSeek-Coder-V2~\cite{deepseek_coder_v2} demonstrate how MoE architectures enable unprecedented scaling while preserving computational efficiency. These models excel at complex software engineering tasks, including repository-level programming and systematic code maintenance as demonstrated in benchmarks like SWE-bench~\cite{swebench}.

\begin{figure}[h]
    \centering
    \includegraphics[width=1.0\textwidth]{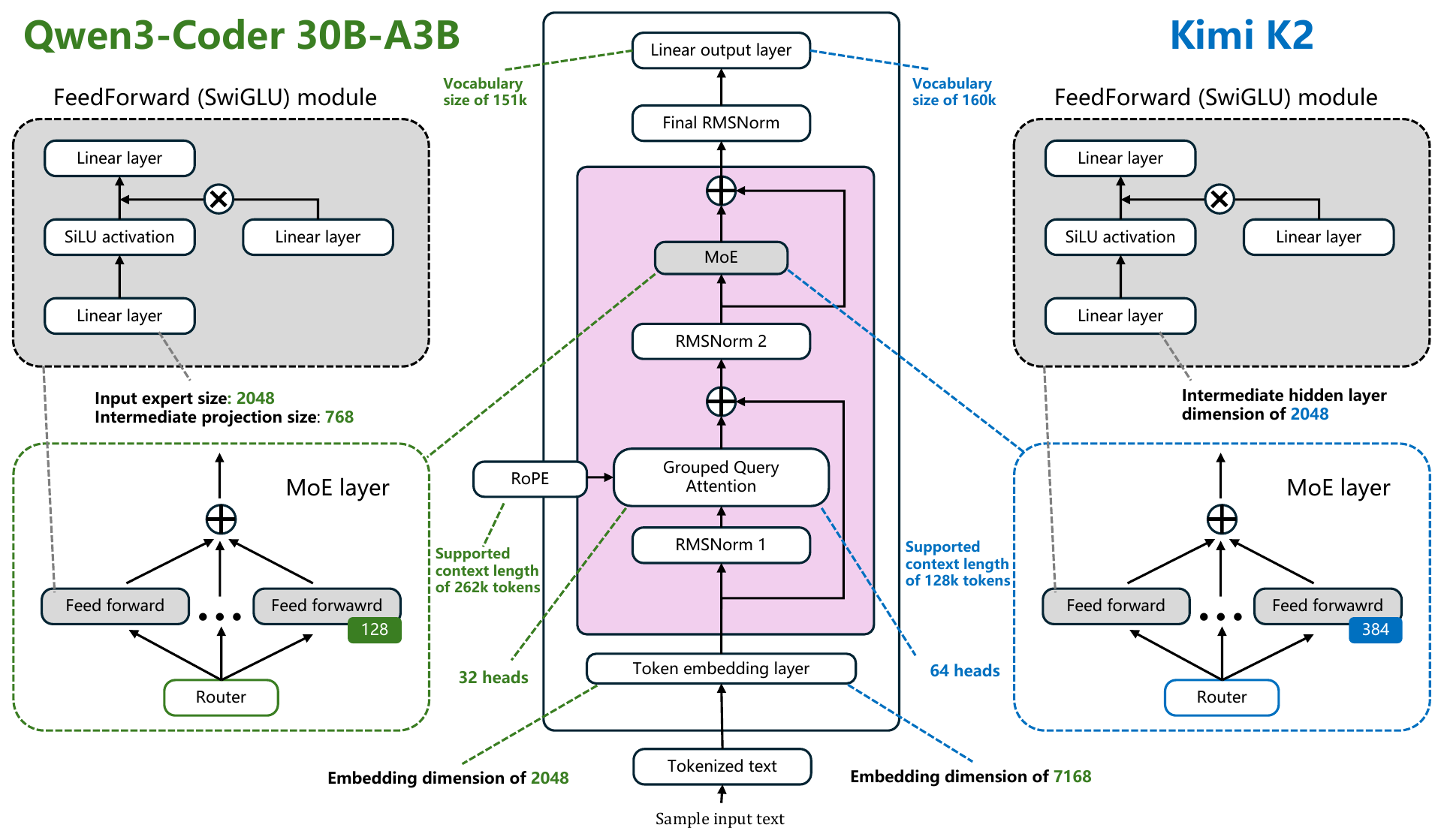} 
    \caption{Architectural comparison between Kimi-K2-Instruct and Qwen3-Coder.}
    \label{fig:qwen_vs_kimi}
\end{figure}

\begin{itemize}
\item \arcdec\arcmoe\textbf{DeepSeek-Coder-V2}~(\funcunderstand\funcgen\funcfim\funcswe)~\cite{zhu2024deepseekcoderv2} adopts a Mixture-of-Experts backbone (16B and 236B; small active experts per token) continued from DeepSeek-V2 with mixed code/math/NL data and extended context (up to 128K via YARN). It delivers strong results across synthesis, competitive programming, bug fixing, and math reasoning, with a lightweight variant offering compelling efficiency. Released under a permissive license, it narrows the gap with top closed models in both coding and reasoning.

\item \arcdec\arcmoe\textbf{Ling-Coder-Lite}~(\funcunderstand\funcgen\funcfim\funcswe)~\cite{team2025every} is an MoE code LLM (few active parameters per token) with shared+routed experts, top-\(6\) routing, and a refined NormHead design. Training proceeds via continued pretraining and instruction optimization (SFT \(\rightarrow\) DPO) over high-quality, execution-aware, repository-structured data. It shows competitive results on HumanEval, MBPP, LiveCodeBench, and BigCodeBench against similarly sized peers, achieving a favorable performance–efficiency trade-off for low-latency deployment.

\item \arcdec\textbf{Skywork-SWE}~(\funcunderstand\funcgen\funcswe)~\cite{zeng2025skyworkswe} presents an execution-aware curation pipeline plus an open 32B agent, revealing clear SWE data scaling laws with LLMs. It collects PR–issue pairs, builds per-instance Docker runtimes validated by tests, and filters multi-turn agent trajectories to retain only passing solutions. Fine-tuning within \textbf{OpenHands} on validated trajectories yields \textbf{Skywork-SWE-32B}, which improves over its base on SWE-bench-Verified and further benefits from test-time scaling. Ablations indicate log-linear gains with more trajectories and that execution-grounded data and framework quality matter more than parameter count. The work releases the checkpoint and practical guidance for leakage control and scalable evaluation.

\item \arcdec\textbf{DeepCoder}~(\funcunderstand\funcgen\funcswe)~\cite{deepcoder2025} is a fully open, RL-trained 14B code–reasoning model fine-tuned from \texttt{DeepSeek-R1-Distilled-Qwen-14B}. It targets repository-level coding with long-context editing (trained at 32K, inference-time scaled to 64K) and reaches competitive LiveCodeBench performance versus strong proprietary baselines. Training uses verifiable tasks and rewards (e.g., TACO-Verified, a verified subset of PrimeIntellect \texttt{SYNTHETIC-1}, and LiveCodeBench from 2023-05-01 \(\to\) 2024-07-31) enforced by unit tests. The release includes the RL pipeline, datasets, evaluation logs, and traces for reproducible study.

\item \arcdec\textbf{DeepSWE}~(\funcunderstand\funcgen\funcswe)~\cite{deepswe2025} is an open-source \emph{RL-only} coding agent on \textbf{Qwen3-32B} with a thinking mode. A compact RL recipe rapidly lifts SWE-bench-Verified, and test-time scaling with a \emph{DeepSWE-Verifier} selects high-quality patches; combining execution-free and execution-based verifiers yields additional gains. The public write-up details the rLLM-based\footnote{https://github.com/agentica-project/rllm} setup on real repository-level tasks with stability tweaks for long-horizon, multi-file editing, showing that RL-only post-training \(+\) lightweight TTS narrows the gap to larger proprietary systems.

\item \arcdec\textbf{Devstral}~(\funcunderstand\funcgen\funcswe)~\cite{devstral2025} is an Apache~2.0 agentic code LLM co-developed by Mistral~AI and All Hands~AI. \emph{Devstral Small} (24B, 128K context) targets repository-scale SWE on accessible hardware, while \emph{Devstral Medium} provides stronger cost–performance via API. On SWE-bench-Verified, both achieve top-tier open-weight results and are designed as agent backbones emphasizing multi-file reasoning, long-context editing, and verifier-friendly test-time scaling.

\item \arcdec\arcmoe\textbf{Qwen3-Coder}~(\funcunderstand\funcgen\funcfim\funcswe)~\cite{qwen3coder} advances agentic capabilities (e.g., Qwen3-Coder-480B-A35B-Instruct), showing strong open-model performance on agentic coding, browser use, and foundational coding tasks, competitive with leading assistants. It offers native 256K context (extendable to 1M via Yarn), a structured function-call schema, and integration with Qwen Code/Cline. With permissive licensing, the series stands as a leading open-source code-LLM family.

\item \arcdec\arcmoe\textbf{GLM-4.5/4.6}~(\funcunderstand\funcgen\funcswe)\cite{glm2025glm4_5} is an open MoE foundation model for agentic, reasoning, and coding tasks, featuring hybrid “think” \(\leftrightarrow\) direct modes with \(\sim\)32B active parameters per token within a larger MoE design. Both GLM-4.5 and its GLM-4.6 successor adopt GQA, QK-Norm, and an MoE multi-token prediction head for speculative decoding; training spans diverse corpora with mid-training that upsamples repo-level code, synthetic reasoning, and long-context/agent trajectories, with context extended from \(4\text{K}\to32\text{K}\to128\text{K}\) in GLM-4.5 and further to \(200\text{K}\) tokens in GLM-4.6. Post-training blends expert models via SFT and unified self-distillation, with RL innovations (difficulty curricula, long-output RL, dynamic temperature, code-weighted losses) yielding consistent gains across TAU-Bench~\cite{yao2024tau}, AIME~\cite{aime24,aime25}, SWE-bench-Verified, and BrowseComp, while GLM-4.6 additionally improves coding, tool-augmented reasoning, and agentic performance across a broader suite of public benchmarks and enhances writing quality.

\item \arcdec\arcmoe\textbf{Kimi-K2-Instruct}~(\funcunderstand\funcgen\funcswe)~\cite{moonshotai2025kimik2} is the instruction-tuned variant of the Kimi-K2 Mixture-of-Experts (MoE) series developed by Moonshot AI. It employs a large-scale MoE design with $\sim 10^{12}$ total parameters and $3.2\times10^{10}$ active per forward pass, as shown in ~\autoref{fig:qwen_vs_kimi}. The model is pretrained with the MuonClip optimizer on trillions of tokens, followed by agentic data synthesis and reinforcement learning for improved instruction following and tool usage. On code-related benchmarks, Kimi-K2-Instruct shows strong performance across multiple evaluation sets. It also maintains robust reasoning on mathematical and logic tasks, reflecting its cross-domain capability. With native tool invocation and extremely long context support ($\ge 128\text{K}$ tokens), it serves as a versatile open-weight foundation for agentic code assistants and general-purpose reasoning systems.

\item \arcdec \textbf{KAT-Dev}~(\funcunderstand\funcgen\funcswe)~\cite{katcoder} is an open-weight code-centric model series from Kwaipilot, built on the Qwen3 architecture and released under Apache~2.0. The 32B variant reaches 62.4\% on SWE-Bench Verified, ranking among the strongest open models. Its training pipeline combines mid-training for tool-use and instruction following, supervised and reinforcement fine-tuning across diverse programming tasks, and large-scale agentic RL. With long-context support and native tool invocation, KAT-Dev serves as a versatile foundation for autonomous coding agents and general-purpose software reasoning.

\item \arcdec\arcmoe\textbf{DeepSeek-V3/V3.1/V3.2}~(\funcunderstand\funcgen\funcswe)\cite{deepseekai2024deepseekv3technicalreport} is an open Mixture-of-Experts LLM series for agentic reasoning and code generation, featuring hybrid “thinking” vs. direct modes and $\sim$37B active parameters per token (671B total) with a 128K context window. DeepSeek-V3 adopts Multi-Head Latent Attention and a multi-token prediction head for efficient long-context inference, and is pre-trained on 14.8T tokens with auxiliary-loss-free MoE load balancing, followed by SFT and RL fine-tuning. It achieves state-of-the-art open-source performance on coding benchmarks, rivaling closed models on code tasks. Its successor DeepSeek-V3.1 underwent extended training (an additional ~840B tokens to reach 32K then 128K context) and integrated the “DeepThink” chain-of-thought mode, which boosted multi-step tool use and coding-agent capabilities. Post-training optimizations made V3.1 significantly stronger on software engineering challenges (e.g. SWE-bench, Terminal-Bench), outperforming earlier V3 and R1 models in code generation and search-agent benchmarks. The experimental DeepSeek-V3.2 builds on V3.1-Terminus with a novel DeepSeek Sparse Attention mechanism that yields near-linear attention scaling, cutting inference cost by ~50\% for long inputs while maintaining output quality on par with V3.1. This improves efficiency in handling large code bases and retrieval-augmented coding tasks without degrading coding accuracy.

\item \arcdec \arcmoe \textbf{MiniMax-M1/M2} (\funcunderstand\funcgen\funcswe) is an open MoE model pair for long-context reasoning, coding, and agentic tasks. M1 introduced a hybrid Mixture-of-Experts architecture with a custom ``lightning'' attention mechanism, enabling a 1-million-token context window (8× DeepSeek-R1’s length) while maintaining high FLOP efficiency. Trained via large-scale reinforcement learning, M1 excels at complex multi-step reasoning, software engineering, and tool use, outperforming earlier open models on long-horizon tasks. Its successor M2 emphasizes deployment efficiency – using 230B total (10B active) parameters to deliver near-frontier performance in code generation and autonomous tool use with only ~200K context. Post-training alignment further boosts M2’s capabilities across end-to-end coding benchmarks and agent planning tasks (e.g. SWE-Bench\cite{swebench}, BrowseComp\cite{wei2025browsecomp}), making it one of the most capable and practical open LLMs for complex workflows.

\end{itemize}

\textbf{Alternative Architecture Explorations.} Beyond the mainstream autoregressive transformer paradigm, diffusion-based language models for code have recently begun to attract attention. On the proprietary side, models such as Gemini Diffusion~\cite{deepmind2025geminidiffusion} and Mercury Coder~\cite{labs2025mercury} illustrate that discrete text diffusion can achieve competitive code quality while substantially reducing generation latency compared to standard autoregressive decoders. In parallel, the open-source community is also exploring this design space: for example, DiffuCoder~\cite{gong2025diffucoder} investigates masked diffusion models for code generation and reports encouraging results on standard coding benchmarks, suggesting that diffusion LLMs are a viable alternative architecture for code synthesis tasks.

\begin{itemize}
    \item \arcdiff \textbf{DiffuCoder} (\funcgen) is an open 7B masked diffusion coder that serves as a canonical testbed for diffusion-native code generation and reinforcement learning. Trained on 130B effective tokens of code, DiffuCoder delivers performance competitive with strong autoregressive coders while enabling non-autoregressive, globally planned decoding over the entire sequence. Using this model, the authors introduce local and global “AR-ness’’ metrics to quantify how closely diffusion LMs follow left-to-right generation, and show that raising the sampling temperature diversifies not only token choices but also generation order, creating a rich rollout space for RL. Building on this insight, they propose \emph{coupled-GRPO}, a diffusion-native variant of GRPO that applies complementary mask noise to paired completions, reducing variance in likelihood estimates and better exploiting the non-AR search space. Coupled-GRPO training yields a +4.4\% improvement on EvalPlus with only $\sim$21K RL samples, further strengthening DiffuCoder-Instruct and establishing DiffuCoder as a strong open baseline for future diffusion-based coding assistants and RL research.

\end{itemize}

\textbf{Code Retrieval Embeddings.} Parallel to the main evolutionary trajectory, open-source code retrieval embedding models have undergone their own transformation. Early approaches relied on BERT-based encoder models such as CodeBERT and UniXcoder~\cite{unixcoder} for generating code representations. Recent developments have shifted toward open-source LLM-based embedding models, leveraging the rich semantic understanding of large language models to produce more sophisticated code embeddings for retrieval and similarity tasks.

\begin{itemize}
\item \arcdec\textbf{Nomic Embed Code}(\funcembed)~\cite{nussbaum2025nomicembedtrainingreproducible} is a 7B parameter code embedding model that achieves state-of-the-art performance on CodeSearchNet through high-quality contrastive training. Built upon the CoRNStack dataset—a large-scale curated corpus derived from deduplicated Stackv2 with dual-consistency filtering—the model converts code retrieval tasks into semantic similarity matching using cosine distance between pooled representations. The training methodology employs a novel curriculum-based hard negative mining strategy with softmax-based sampling to progressively introduce challenging examples during contrastive learning. Nomic Embed Code excels at NL$\to$code, code$\to$NL, and code$\to$code retrieval tasks across multiple programming languages while maintaining full open-source availability of training data, model weights.
\item \arcenc \arcdec\textbf{CodeXEmbed}(\funcembed)~\cite{liu2025codexembedgeneralistembeddingmodel} is an open family of multilingual and multi-task retrieval models spanning both encoder and decoder architectures. The 400M variant is a BERT-style bi-encoder trained from scratch for efficiency-oriented deployment, while the 2B and 7B variants are decoder-only LLMs adapted into dual-tower encoders for generalist retrieval. All variants map diverse text–code tasks into a unified query–document matching framework, where pooled embeddings are compared via cosine similarity. A two-stage LoRA contrastive training pipeline—first on large-scale text retrieval and then jointly on text–code pairs with hard negatives—produces models specialized for Text$\to$Code, Code$\to$Text, and Code$\to$Code retrieval, as well as repository-level RAG. The 7B model achieves state-of-the-art results on CoIR, while smaller models maintain strong BEIR performance with lower latency and cost.
\item \arcenc \textbf{CodeSage}(\funcembed)~\cite{codesage} is a family of bidirectional encoder models (130M, 356M, 1.3B) trained for large-scale code representation learning across nine programming languages. It employs a two-stage training scheme: first, a mix of identifier deobfuscation and masked language modeling (without the 80-10-10 corruption) for token-level denoising, and second, bimodal contrastive learning using text–code pairs with hard positives and hard negatives. This design promotes semantic alignment between natural and programming languages. Evaluated on NL$\to$Code, Code$\to$Code, and classification benchmarks, CodeSage consistently outperforms prior models such as UnixCoder, GraphCodeBERT, and OpenAI-Ada embeddings. Larger variants yield stronger cross-lingual and retrieval performance, while smaller ones balance speed and efficiency.
\item \arcdec \textbf{BGE-Code}(\funcembed)~\cite{li2025towards} is a generalist code-embedding bi-encoder (Qwen2.5-Coder~1.5B) trained with an \emph{Annealing} curriculum to transfer from text-only to code-centric retrieval. It relies on the synthetic \textbf{CodeR-Pile} built under DRU (Diversity, Reliability, Usability), spanning Text2Code, Code2Text, Code2Code, and Hybrid tasks across many languages. Data are synthesized via LLM brainstorming, instruction refinement, pair generation/annotation, and \emph{hard-negative} mining on real code. LoRA-based training with staged schedules and difficulty filtering produces strong results on CoIR/CoIR-filter/CodeRAG. Ablations support that broader task coverage, mined negatives, and the curriculum \(\gg\) single-shot mixed training.

\item \arcdec \textbf{CodeFuse-CGE}(\funcembed)~\cite{codefuseCGE} is an open decoder-only family for code retrieval that adapts causal LLMs into dual-tower encoders via a lightweight cross-attention embedding head. The Large model builds on CodeQwen1.5-7B-Chat and the Small model on Phi-3.5-mini-instruct; both are LoRA-tuned to project text and code into a shared vector space and scored by cosine similarity. CGE reframes NL$\to$Code search as query–document matching and targets repository-level workflows; it reports strong results on CodeSearchNet and AdvTest, and has been used as the semantic retriever in repo-level systems. Compared with encoder baselines, the decoder-based design captures richer cross-modal semantics while remaining practical to deploy.

\end{itemize}

\subsubsection{Model Pre-Training Tasks}
\paragraph{Next Token Prediction}
Next Token Prediction (NTP) is the most fundamental and widely used self-supervised training task, whose goal is to predict the next likely word or subword unit based on a given preceding context sequence. Essentially, this task is a form of Causal Language Modeling (CLM), where the model can only access information up to the current moment and cannot ``peek into'' future content. During the training process, the input sequence is slid token by token, and the label for each position is the token that immediately follows it. The model learns the conditional probability distribution $P(x_{t+1} \mid x_1, x_2, \ldots, x_t)$ by minimizing the cross-entropy loss. When training, given a text sequence of length $T$, the model sequentially predicts the $t+1$-th token for each position $t \in [1, T-1]$. This process enables the models to capture the statistical laws of language, grammatical structures, semantic correlations, and world knowledge, thereby establishing robust capabilities in language understanding and generation.

\begin{figure}[h]
    \centering
    
    \includegraphics[width = 1.0\textwidth]{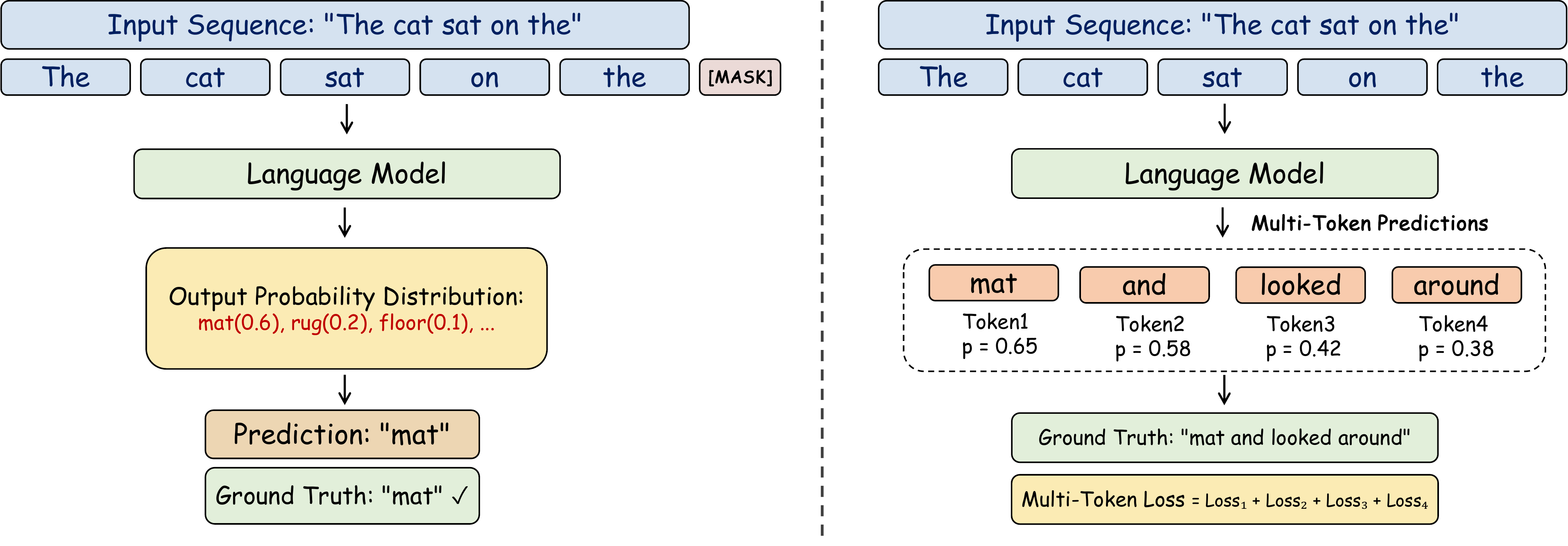} 
    \caption{Comparison between next token prediction (NTP) and multi-token prediction (MTP) training objectives in large language models. In NTP (left), the model predicts the next single token (``mat'') given the preceding context. In contrast, MTP (right) extends this to predict multiple future tokens (``mat and looked around'') simultaneously, optimizing a combined loss over all predicted tokens. This parallel formulation can improve training efficiency and better capture longer-range dependencies in sequence modeling.}
    \label{ntp_mtp}
\end{figure}

\paragraph{Multi-Token Prediction}
In \autoref{ntp_mtp}, we provide a comparison between next token prediction (NTP) and multi-token prediction (MTP) training objectives in large language models.
Multi-Token Prediction (MTP) is an extended task built on the foundation of Next Token Prediction. Its objective is to enable the model to predict multiple consecutive tokens at once based on the preceding context sequence, thereby improving the model's text generation efficiency and coherence.

\begin{figure}[h]
    \centering
    \includegraphics[width=1.0\textwidth]{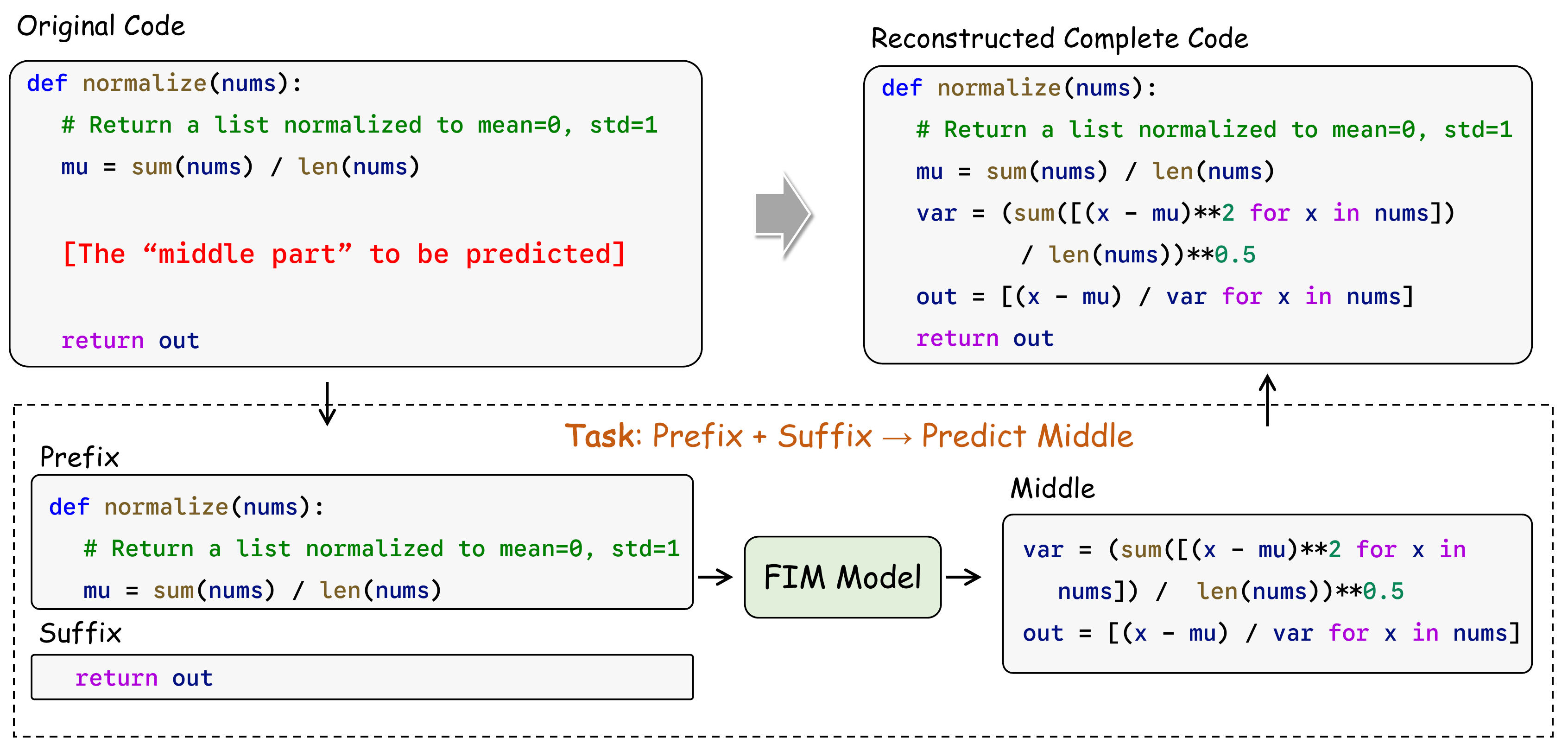} 
    \caption{Illustration of the Fill-in-the-Middle (FIM) training and inference process for code completion. The model receives both a prefix and a suffix of a function and learns to generate the missing middle segment. In this example, given the beginning and end of the \texttt{normalize()} function, the FIM model predicts the intermediate computation steps for variance and normalization, reconstructing the complete code.}
    \label{fig:fim_task}
\end{figure}

\paragraph{Fill-in-the-Middle}
In \autoref{fig:fim_task}, fill-in-the-Middle (FIM) is a non-autoregressive language modeling task. Its core objective is to enable the model to predict the missing token segment in the middle of a text sequence based on the given prefix and suffix of the sequence, thereby enhancing the model’s ability to understand the global semantics of text and its sequence completion capability. The execution logic of this task differs from autoregressive sequential prediction: first, the model inputs both the prefix and suffix sequences into the network simultaneously, and uses the bidirectional attention mechanism of the transformer to jointly encode the semantics of the prefix and suffix, capturing the semantic association between them; then, the model generates possible token sequences for the missing middle region, and optimizes its parameters by calculating the loss between the generated sequence and the real middle sequence. Formally, the input format is often expressed as: \texttt{<prefix>...<suffix> $\rightarrow$ <mask>}, and the model’s output is the original text with the masked part restored. The Code LLaMA~\cite{roziere2023code} models all adopt the FIM strategy for training. To enhance diversity and robustness, half of the segmentations use the prefix-suffix-middle (PSM) format, while the other half use the compatible suffix-prefix-middle (SPM) format. To support this new task, Code LLaMA extends the original tokenizer of Llama 2 with four special tokens: \texttt{<fim\_prefix>}, \texttt{<fim\_suffix>}, \texttt{<fim\_middle>}, and \texttt{<fim\_pad>}, which are used to clearly identify the boundaries of each part in the input sequence. By jointly optimizing FIM as a multi-task objective parallel to standard autoregressive prediction, the Code LLaMA model can directly implement intelligent completion for cursor positions or arbitrary code blocks in environments such as IDEs during inference without additional fine-tuning, significantly improving its practicality in real-world programming scenarios.

\begin{figure}[h]
    \centering
    \includegraphics[width=0.50\textwidth]{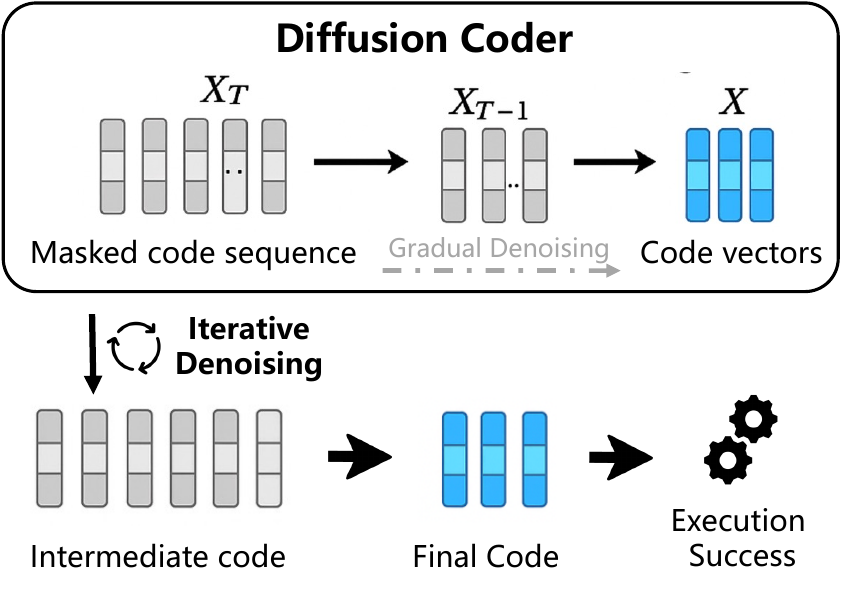} 
    \caption{Overall architecture of Diffusion Coder.}
    \label{fig:diffusion_coder}
\end{figure}

\paragraph{Diffusion Coder Training Task}
Diffusion Coder is a language model designed based on the principles of diffusion models. Its core objective is to enable the model to generate text token sequences that comply with semantic logic from random noise sequences through a ``gradual denoising'' process, primarily for enhancing the diversity and quality of text generation, as illustrated in \autoref{fig:diffusion_coder}. The execution logic of this task is divided into two phases: the first is the ``forward diffusion phase'', in which random noise is gradually added to the real text token sequence in accordance with a preset noise scheduling strategy, causing the sequence to gradually approximate pure noise; the second is the ``reverse denoising phase'', where the model needs to learn to start from the noisy sequence, gradually remove noise based on the noise level, and restore the real text sequence. During the training process, the model optimizes its parameters by calculating the loss between the ``predicted noise'' and the ``actually added noise'', and ultimately gains the ability to generate high-quality text from noise. Compared with autoregressive models, Diffusion Coder has the advantages of higher diversity in generated text and the ability to improve generation efficiency through parallel computing.

\begin{figure}[h]
    \centering
    \includegraphics[width=1.0\textwidth]{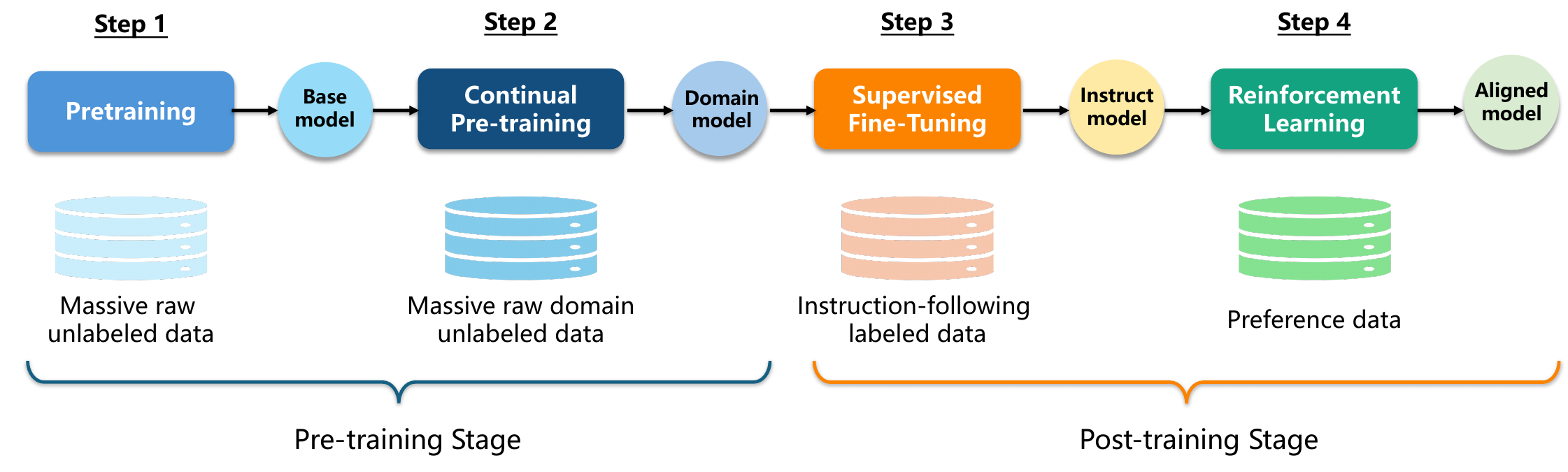} 
    \caption{Overview of the model training stages.}
    \label{fig:training_stages}
\end{figure}

\subsubsection{Model Training Stages}
The training stages of large language models are a critical process through which models progress from learning general linguistic knowledge to adapting to specific tasks and generating content that meets human needs. This process determines the model's capability boundaries and practical value. It primarily consists of two main parts: \textbf{Pre-training} and \textbf{Post-training}, as shown in \autoref{fig:training_stages}. Pre-training establishes the model's foundation in general language capabilities, while Post-training enables the model to operate more accurately and efficiently in specific tasks and interactive scenarios.

\newcommand{\na}{\textemdash}
\newcolumntype{L}[1]{>{\raggedright\arraybackslash}p{#1}}
\newcolumntype{Y}{>{\centering\arraybackslash}p{1.15cm}}
\newcolumntype{Z}{>{\centering\arraybackslash}p{1.8cm}}

\begin{table}[h!]
  \centering
  \caption{Training phases of different Code LLMs.}
  \label{tab:train-phases}
  \rowcolors{2}{gray!6}{white}
  \resizebox{0.85\columnwidth}{!}{%
  \renewcommand{\arraystretch}{1.15}
  \setlength{\tabcolsep}{4pt}
  \begin{tabular} {L{4.6cm} *{3}{Y} *{1}{Z} *{2}{Y} *{1}{Z}}
    \toprule
    & \multicolumn{4}{c}{\textbf{Pre-training}} & \multicolumn{3}{c}{\textbf{Post-training}} \\
    \cmidrule(lr){2-5} \cmidrule(lr){6-8}
    \textbf{Model} & \textbf{PT} & \textbf{CPT} & \textbf{ANN} & \textbf{Repo PT} & \textbf{SFT} & \textbf{RL} & \textbf{Repo SFT} \\
    \midrule
    SantaCoder~\cite{allal2023santacoder}              & \cmark &        &       &        & \cmark &       &       \\
    OctoCoder~\cite{muennighoff2023octopack}           &        & \cmark &       &        & \cmark &       &       \\
    CodeLlama~\cite{roziere2024codellama}              &        & \cmark &       &        & \cmark &       &       \\
    StarCoder~\cite{li2023starcoder}                    & \cmark &        &       &        & \cmark &       &       \\
    CodeT5+~\cite{wang2023codet5plus}                  &        & \cmark &       &        & \cmark &       &       \\
    PanGu\mbox{-}Coder2~\cite{shen2023pangucoder2}     &        & \cmark &       &        & \cmark & \cmark &       \\
    AlphaCode2~\cite{alphacode2_technical_report}      &        & \cmark &       &        & \cmark &       &       \\
    WizardCoder~\cite{luo2023wizardcoder}              &        & \cmark &       &        & \cmark &       &       \\
    MFTCoder~\cite{MFTCoder}                            &        & \cmark &       &        & \cmark &       &       \\
    WaveCoder~\cite{yu2024WaveCoderWidespreadVersatile} &        & \cmark &       &        & \cmark &       &       \\
    DeepSeek\mbox{-}Coder~\cite{deepseek2024coder}     & \cmark &        &       & \cmark & \cmark &       &       \\
    StarCoder2~\cite{starcoder2energy2024}             & \cmark &        &       &        & \cmark &       &       \\
    CodeGemma~\cite{codegemma2024}                     &        & \cmark &       &        & \cmark &       &       \\
    Yi\mbox{-}Coder~\cite{yicoder2024}                 &        & \cmark &       &        & \cmark &       &       \\
    Granite\mbox{-}Code~\cite{mishra2024granitecodemodelsfamily} & \cmark &        &       &        & \cmark &       &       \\
    Qwen2.5\mbox{-}Coder~\cite{qwen2025qwen2_5}        &        & \cmark &       & \cmark & \cmark &       &       \\
    DeepSeek\mbox{-}Coder\mbox{-}V2~\cite{deepseekai2024deepseekv2} &        & \cmark &       &        & \cmark & \cmark &       \\
    Opencoder~\cite{huang2025opencoder}                & \cmark &        & \cmark &       & \cmark &       &       \\
    DeepSeek\mbox{-}V3~\cite{deepseekai2025deepseekv3} &        & \cmark &       &        & \cmark & \cmark &       \\
    DeepSeek\mbox{-}R1~\cite{deepseekai2025deepseekv3} &        & \cmark &       &        & \cmark & \cmark &       \\
    Ling\mbox{-}Coder~\cite{team2025every}             &        & \cmark & \cmark &       & \cmark &       &       \\
    Seed\mbox{-}coder~\cite{seedcoder}                 & \cmark &        &       &        & \cmark & \cmark &       \\
    Qwen3\mbox{-}Coder~\cite{qwen3coder}               & -    & \cmark & \cmark & \cmark & \cmark & \cmark & \cmark \\
    \bottomrule
  \end{tabular}
  }%
  \vspace{4pt}
  \begin{tablenotes}\footnotesize
    \item \textbf{PT}: Pre-training. \textbf{CPT}: Continued pre-training. \textbf{ANN}: Annealing training. 
    \textbf{Repo PT}: Repository-level pre-training. 
    \textbf{SFT}: Supervised fine-tuning. 
    \textbf{RL}: Reinforcement learning. 
    \textbf{Repo SFT}: Repository-level supervised fine-tuning.
  \end{tablenotes}
\end{table}

\paragraph{Pre-Training Stages}
Pre-training is the core phase where large language models learn general language patterns and accumulate world knowledge. By learning from massive amounts of unlabeled data, the model acquires basic grammar, semantic understanding, and generation capabilities.
\begin{itemize}
\item \textbf{Data Collection and Processing}
Data collection and processing form the foundational step of pre-training. Data collection requires broadly sourcing texts spanning multiple domains and types, such as general web text, books, academic literature, and specialized domain materials. Data processing involves cleaning and deduplicating the collected raw text, filtering out low-quality and sensitive content, tokenization (splitting text into model-recognizable tokens), format standardization, unified encoding, and adding special markers, thereby transforming it into an input form that the model can directly learn from. Data collection and processing typically utilize methods like web crawlers and integration of public datasets, employing NLP tools and custom scripts to complete the cleaning, tokenization, and other processing workflows.
\item \textbf{PT (Pre-training Stage)} Pretraining refers to the process of training a model starting from randomly initialized parameters, utilizing processed large-scale unlabeled corpora through self-supervised learning objectives. These objectives include autoregressive language modeling, which involves predicting the next token based on preceding context, or masked language modeling, which entails predicting randomly masked tokens within the input sequence. During the training process, the model progressively learns grammatical structures, lexical relationships, and extensive general world knowledge by continuously optimizing its parameters.
\item \textbf{CPT (Continual Pre-training Stage)} CPT refers to the pre-training optimization task where, based on an existing pre-trained model, additional pre-training is performed using domain-specific or newly added corpora to continuously enrich the model's knowledge and enhance its performance. It is not training from scratch but rather a continuation of the pre-training process, often used for vertical domain adaptation or knowledge updates. The primary execution methods include freezing part of the lower-layer parameters or performing full fine-tuning, and continuing to perform autoregressive or masked language modeling tasks on the new corpus to learn new contextual patterns.
\item \textbf{Annealing Strategy} The annealing strategy primarily involves the dynamic adjustment of training hyperparameters such as the learning rate. Typically, a larger learning rate is used in the early stages of training to accelerate convergence, followed by a gradual reduction in the learning rate for fine-tuning the model parameters. This helps prevent the model from struggling to converge to an optimal solution in the later stages due to an excessively large learning rate. 
\end{itemize}

\paragraph{Post-training Stage} Post-training is the crucial step following pre-training that enables the model to adapt to specific tasks and align with human interaction needs. It transforms the model from ``understanding general language''~\cite{feng2020codebertpretrainedmodelprogramming} to ``accurately completing specific tasks and generating content satisfactory to humans''~\cite{codearena}.
\begin{itemize}
\item \textbf{Supervised Fine-Tuning (SFT)} Supervised Fine-Tuning is a post-training method that uses manually annotated ``input-output'' data pairs to perform supervised training on the pre-trained model, enabling it to quickly adapt to specific tasks. This involves collecting annotated data for specific tasks (e.g., dialogue generation, text summarization, code generation), inputting this data into the pre-trained model, and adjusting the model parameters by optimizing the loss function to minimize the prediction error on the task data. This allows the pre-trained model to learn task-specific mapping relationships from this data, making its output more aligned with task requirements.
\item \textbf{Reinforcement Learning (RL)} Reinforcement Learning is a post-training method that further optimizes the model's output by incorporating human feedback or reward mechanisms, guiding the model to generate content that better aligns with human preferences. By introducing a reward signal, the model adjusts its behavior through interaction with specific task scenarios to maximize this signal, thereby producing higher-quality content. A common implementation is Reinforcement Learning from Human Feedback (RLHF). The process typically involves first training a Reward Model (RM) using annotated data to quantify human preferences, then using reinforcement learning algorithms to optimize the generative model's output under the guidance of the Reward Model.
\end{itemize}

\subsection{Open-source Code Pre-training Data}
The capabilities of open-source code LLMs are fundamentally shaped by their pre-training data. The evolution of these datasets shows a clear trend: a shift from prioritizing sheer volume to a rigorous focus on data quality, licensing, and governance. This maturation has been critical for the open-source community to build models that are not only powerful but also safer and more reliable. \autoref{tab:code_datasets} shows a few key datasets have defined this landscape.
\definecolor{stackcolor}{RGB}{230,240,250}   %
\definecolor{derivedcolor}{RGB}{250,240,230} %
\definecolor{othercolor}{RGB}{240,250,230}   %

\begin{table}[H]
\centering
\caption{Open-source pre-training code datasets.}
\label{tab:code_datasets}
\small
\setlength{\tabcolsep}{6pt}
\renewcommand{\arraystretch}{1.1}
\resizebox{1.0\textwidth}{!}{
\begin{tabular}{lllp{3.2cm}p{5cm}}
\toprule
\textbf{Dataset} & \textbf{Size} & \textbf{Languages} & \textbf{Source} & \textbf{Key Features} \\
\midrule
\rowcolor{stackcolor}
\multicolumn{5}{c}{\textbf{The Github Dataset}} \\
\addlinespace[0.3em]
The Stack v1~\cite{kocetkov2022stack} &  2.9 TB &  358 &  \makecell[l]{GitHub repositories} &  \makecell[l]{Permissively licensed; two-stage\\ deduplication (exact-match +\\ MinHash LSH)} \\
\addlinespace[0.1em]
The Stack v2~\cite{lozano2024starcoder2} &  32.1 TB &  600+ &  \makecell[l]{Software Heritage,\\GitHub PRs/issues,\\Jupyter notebooks} &  \makecell[l]{4$\times$ larger than v1; formal opt-out\\ process; industry standard} \\
\addlinespace[0.1em]
OpenCoder~\cite{huang2025opencoder} &  3.3 TB &  13 &  \makecell[l]{GitHub,skypike\\cc,FineWeb\\AutoMathText} &  \makecell[l]{Complete Open Source\\Reproducible} \\
\rowcolor{derivedcolor}
\multicolumn{5}{c}{\textbf{Derived Datasets}} \\
\addlinespace[0.3em]
StarCoderData~\cite{li2023starcoder} &  783 GB &  86 &  The Stack v1.2 &  \makecell[l]{Additional decontamination;\\ includes GitHub issues and \\commits; contamination-free} \\
\rowcolor{othercolor}
\multicolumn{5}{c}{\textbf{Other Foundational Datasets}} \\
\addlinespace[0.3em]
The Pile~\cite{gao2020pile} &  825 GB &  30+ &  \makecell[l]{GitHub,\\ Stack Exchange} &  \makecell[l]{EleutherAI; pioneering open \\reproducible dataset; foundation \\for GPT-Neo and Pythia} \\
\addlinespace[0.1em]
RedPajama~\cite{together2023redpajama} &  120 GB &  50+ &  GitHub &  \makecell[l]{Permissive licenses only (MIT, \\BSD, Apache 2.0); LLaMA data replication} \\
\addlinespace[0.1em]
CodeParrot~\cite{codeparrot} &  50 GB &  1 (Python) &  GitHub & \makecell[l]{Python-only; $\sim$70\% data\\ removed via deduplication} \\
\bottomrule
\end{tabular}
}
\end{table}

\subsubsection{The Github Datasets}
The Stack dataset family, from the BigCode collaboration, is the most influential resource\footnote{\url{https://huggingface.co/datasets/bigcode/the-stack}} for pre-training code LLMs.

\textbf{The Stack v1}~\cite{kocetkov2022stack}. In the first version, The Stack v1 was a landmark, providing 3.1 TB of permissively licensed source code across 358 programming languages after extensive deduplication. Its creation set a vital precedent by sourcing exclusively from GitHub repositories with licenses permitting redistribution, addressing key legal concerns in open-source development. The data pipeline was notable for its two-stage deduplication strategy, combining exact-match hashing with a MinHash LSH algorithm for near-deduplication, a technique proven to significantly boost model performance.
    
\textbf{The Stack v2}~\cite{lozano2024starcoder2}. Building on this, The Stack v2 represents the current industry standard. It expands the data scale fourfold to approximately 900 billion tokens. Its sources are more diverse, drawing primarily from the Software Heritage archive and supplemented with GitHub pull requests, issues, and Jupyter notebooks. The Stack v2 also increased coverage to over 600 languages and introduced a crucial governance mechanism: a formal opt-out process, allowing developers to have their code removed and ensuring the dataset is periodically updated to respect these requests.

\subsubsection{StarCoderData}
\textbf{StarCoderData}~\cite{li2023starcoder}. Derived from The Stack, the StarCoderData dataset  was curated specifically for training the StarCoder model series. While originating from The Stack v1.2, it underwent an additional, cleaning and decontamination phase to filter against common evaluation benchmarks, preventing data leakage and ensuring fair model assessment. Totaling 783 GB, it provides a rich mix of source code across 86 languages, augmented with contextual data from GitHub issues and commits, making it a benchmark for high-quality, contamination-free pre-training.
\subsubsection{Others}
Beyond these large-scale multilingual datasets, several others have been instrumental.

\textbf{The Pile}~\cite{gao2020pile}. The Pile, an 825 GB collection from EleutherAI, is a dedicated effort that included significant code from GitHub and Stack Exchange. It is one of the first large-scale, open, and reproducible datasets, serving as the foundation for influential early models like GPT-Neo and Pythia.

\textbf{RedPajama}~\cite{together2023redpajama}. Similarly, the code portion of RedPajama-Data-1T is a vital contribution, created as an open-source replication of the data used to train the original LLaMA models. Its 59-billion-token code slice was carefully filtered to include only projects with highly permissive licenses (MIT, BSD, Apache 2.0), making it a valuable resource for training models in the LLaMA architectural family without legal ambiguity.
    
\textbf{CodeParrot}~\cite{codeparrot}. Language-specific datasets have also proven highly effective. CodeParrot, for instance, is a high-quality dataset focused exclusively on Python. Its processing revealed the high degree of duplication in public code repositories, with its pipeline removing nearly 70\% of the raw data, underscoring the critical need for this step in creating efficient training corpora.

Collectively, the trajectory of these open-source datasets highlights a maturation in data engineering. The community has moved from raw data collection to implementing sophisticated, transparent, and reproducible pipelines for filtering, deduplication, PII redaction, and license verification. These high-quality, well-governed datasets have become the bedrock upon which the modern ecosystem of open-source Code LLMs is built.

\subsection{Future Trends}

Based on the comprehensive evolution of code foundation models above, we identify three key trends that will likely shape the future landscape of code intelligence:

\paragraph{From General to Specialized Code Intelligence.} The trajectory from general-purpose LLMs to dedicated code specialists represents a fundamental shift in model development philosophy. While early models like GPT-3~\cite{dettmers2022gpt3} demonstrated surprising code competence as an emergent capability, the success of specialized systems like GPT-5-Codex~\cite{openai2025gpt5codex}, and Claude-4~\cite{claude4}'s coding variants illustrates the substantial gains achievable through domain-specific optimization.  We anticipate continued differentiation between general conversational AI and purpose-built coding assistants, with the latter achieving superior performance on repository-level tasks, complex debugging scenarios, and multi-step software engineering processes.

\paragraph{Agentic Training and Complex Scenario Mastery.} The emergence of agentic code models represents a paradigm shift from passive code generation to active software engineering. Future developments will likely emphasize training methodologies that enable models to operate autonomously across complex, multi-step programming scenarios. This includes reinforcement learning from execution feedback, curriculum learning on progressively complex repository-level tasks, and integration with external tools and environments. Models will be trained not just to write code, but to understand project contexts, navigate codebases, execute iterative debugging cycles, and collaborate with human developers through extended interactions. The success of systems like SWE-Bench\cite{swebench} agents suggests that future code LLMs will increasingly function as autonomous software engineers capable of end-to-end problem solving rather than mere code completion tools.

\paragraph{Scaling Laws and Scientific Model Development.} The application of scaling laws to code model development will drive more systematic and efficient training strategies. Unlike the early era of empirical model scaling, future code LLMs will leverage principled understanding of how model performance scales with parameters, training data, and compute resources specifically for coding tasks~\cite{code_scaling_law}. This scientific approach will inform optimal resource allocation between model scale, data curation quality, and training duration. Additionally, scaling laws will guide the development of mixture-of-experts architectures optimized for code tasks, enabling models to achieve superior performance while maintaining computational efficiency. We expect future code models to be developed through data-driven optimization of the scaling trade-offs unique to programming domains, leading to more capable and cost-effective systems.

These trends collectively point toward a future where code foundation models evolve into sophisticated, specialized systems that combine deep programming expertise with autonomous problem-solving capabilities, developed through scientifically principled scaling strategies that maximize both capability and efficiency.

\begin{figure*}[!h]
    \centering
    \resizebox{\textwidth}{!}{
    \begin{forest}
        forked edges,
        for tree={
                grow=east,
                reversed=true,
                anchor=base west,
                parent anchor=east,
                child anchor=west,
                base=center,
                font=\large,
                rectangle,
                draw=hidden-draw,
                rounded corners,
                align=left,
                text centered,
                minimum width=4em,
                edge+={darkgray, line width=1pt},
                s sep=3pt,
                inner xsep=2pt,
                inner ysep=3pt,
                line width=0.8pt,
                ver/.style={rotate=90, child anchor=north, parent anchor=south, anchor=center},
            },
            where level=1{text width=10em,font=\normalsize, }{},
            where level=2{text width=15em,font=\normalsize}{},
            where level=3{text width=14em,font=\normalsize,}{},
        [Coding Tasks and Benchmarks, ver
            [{Statement{,} Function \\and Class-Level}
                [Completion and FIM
                    [ %
                        {CodeXGLUE~\cite{CodeXGLUE}{,}HumanEval-Infill~\cite{bavarian2022efficient}{,}BigCodeBench~\cite{zhuo2024bigcodebench}{,}MultiPL-E~\cite{MultiPL-E}{,}\\          ClassEval~\cite{du2023classeval}{,}ClassEval-T~\cite{xue2024classevalT}{,}OOP~\cite{oop}{,}Qwen2.5 FIM~\cite{qwen25coder}{,}CCCI~\cite{jin2025cccicodecompletioncontextual}{,}\\
                        SAFIM~\cite{gong2024evaluationllmssyntaxawarecode}{,}AST-FIM~\cite{gong2025structureawarefillinthemiddlepretrainingcode}{,}SAFIM-LLM~\cite{zhang2025comparativeanalysislargelanguage}{,}SynthCoder~\cite{synthcoder}}
                    , text width=45em, align=left, text centered=false 
                    ]
                ]
                [Code Generation
                    [ %
                    {CodeXGLUE~\cite{CodeXGLUE}{,}HumanEval~\cite{chen2021codex}{,}MBPP~\cite{austin2021mbpp}{,}MBXP~\cite{mbxp}{,}HumanEval+~\cite{evalplus}{,}MBPP+~\cite{evalplus}{,}\\CodeFuseEval~\cite{CodeFuseEval}{,}Multilingual HumanEval~\cite{multilingualhumaneval}{,}Human Eval Pro~\cite{humanevalpro}{,}MBPP Pro~\cite{humanevalpro}{,}\\BigCodeBench-LitePro~\cite{humanevalpro}{,}MBUPP~\cite{mbupp}{,}HumanEval-XL~\cite{humanevalxl}{,}PythonSaga~\cite{pythonsaga}{,}\\
                    CodeScope~\cite{CodeScope}{,}BigCodeBench~\cite{zhuo2024bigcodebench}{,}AutoCodeArena~\cite{zhuo2025bigcodearena}{,}McEval~\cite{mceval}{,}MERA Code~\cite{MERACode}{,}YABLoCo~\cite{yabloco}{,}\\
                    ClassEval~\cite{du2023classeval}{,}
                    ClassEval-T~\cite{xue2024classevalT}{,}OOP~\cite{oop}{,}AutoCodeBench~\cite{autocodebench}{,}CodeNet~\cite{codenet}{,}APPS~\cite{apps}{,}\\
                    MultiPL-E~\cite{MultiPL-E}{,}CruxEval(-O/-I)~\cite{gu2024cruxeval}{,}LiveCodeBench~\cite{jain2024livecodebench}{,}LiveCodeBenchPro~\cite{livecodebenchpro}{,}\\
                    CodeElo~\cite{codeelo}{,}ProBench~\cite{probench}{,}ICPC-Eval~\cite{icpceval}{,}OJBench~\cite{ojbench}{,}HLCE~\cite{hlce}{,}MathQA-X~\cite{mathqax}{,}\\
                    SciCode~\cite{scicode}{,}CodeInsight~\cite{codeinsight}{,}CodeRAG-Bench~\cite{CodeRAG-Bench}{,}FullStackBench~\cite{liu2024fullstackbench}{,}Mercury~\cite{mercury2024}{,}\\
                    EffiBench~\cite{huang2024effibench}{,}EffiBench-X~\cite{effibenchx2025}{,}Deep-Bench~\cite{deepbench}{,}KernelBench~\cite{kernelbench}{,}TritonBench~\cite{tritonbench}{,}\\
                    OSS-Bench~\cite{ossbench}{,}COFFE~\cite{coffe}{,}BigO(Bench)~\cite{bigobench}{,}DynaCode~\cite{dynacode}{,}FVAPPS~\cite{fvapps}{,}FPBench~\cite{fpbench}}
                    , text width=45em, align=left, text centered=false
                    ]
                ]
                [Edit and Bug Fix
                    [ 
                       {Megadiff~\cite{megadiff}{,}TSSB-3M~\cite{wen2022tssb}{,}TFix~\cite{tfix}{,}FixJS~\cite{fixjs}{,}TypeBugs~\cite{wei2022pyter}{,}RunBugRun~\cite{runbugrun}{,}\\xCodeEval~\cite{ziyao2023xcodeeval}{,}DebugBench~\cite{debugbench}{,}HumanEvalPack~\cite{humanevalpack}{,}MdEval~\cite{mdeval}{,}SWT-Bench~\cite{jimenez2024swtbench}{,}\\
                       FeedbackEval~\cite{dai2025feedbackeval}{,}DebugEval~\cite{debugeval}{,}RepoFixEval~\cite{sun2024repofixevaluation}{,}CodeEditorBench~\cite{codeeditorbench}}
                    , text width=45em, align=left, text centered=false
                    ]
                ]
                [Code Efficiency
                    [
                       {EffiBench~\cite{huang2024effibench}{,}Mercury~\cite{mercury2024}{,}EffiBench-X~\cite{effibenchx2025}{,}BigO(Bench)~\cite{bigobench}{,}\\
                       DynaCode~\cite{dynacode}{,}COFFE~\cite{coffe}{,}
                       EvalPerf~\cite{liu2024evaluatinglanguagemodelsefficient}{,}ECCO~\cite{ecco2024}{,}\\
                       \textit{AI-Powered, But Power-Hungry?}~\cite{energy2025}{,}\textit{Generating Energy-efficient Code with LLMs}~\cite{Cappendijk_2025}}
                    , text width=45em, align=left, text centered=false
                    ]
                ]
                [Code Preference
                    [
                    {CodeArena~\cite{du-etal-2025-codearena}{,}Long CodeArena \cite{bogomolov2024long}{,}CODEPREFBENCH~\cite{liu2024codefavor}}
                    , text width=45em, align=left, text centered=false
                    ]
                ]
                [{Reasoning and QA}
                    [
                       {CodeQA~\cite{liu2021codeqa}{,}CS1QA~\cite{lee2022cs1qa}{,}CodeMMLU~\cite{manh2024codemmlu}{,}CodeSense~\cite{roy2025codesense}{,}RepoQA~\cite{liu2024repoqa}{,}\\CodeRepoQA~\cite{hu2024coderepoqa}{,}CoreQA~\cite{chen2025coreqa}{,}LONGCODEU~\cite{li2025longcodeu}{,}SpecEval~\cite{ma2025specevalevaluatingcodecomprehension}{,}CRUXEval~\cite{gu2024cruxeval}{,}\\CRUXEval-X~\cite{xu2024cruxeval}{,}EquiBench~\cite{wei2025equibench}{,}CORE~\cite{xie2025core}{,}ScratchEval~\cite{fu2024scratchevalgpt4osmarterchild}}
                    , text width=45em, align=left, text centered=false
                    ]
                ]
                [Code Translation
                    [{Neural code translation evaluation~\cite{jiao2023evaluationneuralcodetranslation}{,}MuST~\cite{zhu2022multilingual}{,}\citet{Pan_2024}{,}\citet{yin2024rectifier0}}, text width=45em, align=left, text centered=false
                    ]
                ]
                [Test-Case Generation
                    [
                      {SWT-Bench~\cite{mundler_swt-bench_2025}{,}TestGenEval~\cite{jaintestgeneval}{,}TestBench~\cite{zhang_testbench_2024}{,}CLOVER~\cite{xu_clover_2025}{,}\\TestEval~\cite{wang_testeval_2025}{,}TCGBench~\cite{ma2025rethinkingverificationllmcode}{,}TestCase-Eval~\cite{yang_can_2025}}
                    , text width=45em, align=left, text centered=false
                    ]
                ]
            ]
            [Repository-Level
                [{Generation and Completion}
                    [
                       {RepoBench~\cite{liu2023repobench}{,}RepoEval (RepoCoder)~\cite{zhang2023repocoderrepositorylevelcodecompletion}{,}Execrepobench~\cite{yang2024execrepobenchmultilevelexecutablecode}{,}CoderEval~\cite{yu2024codereval}{,}\\CrossCodeEval~\cite{ding2023crosscodeevaldiversemultilingualbenchmark}{,}M2rc-Eval~\cite{liu2024m2rcevalmassivelymultilingualrepositorylevel}{,}Codev-Bench~\cite{pan2024codevbenchllmsunderstanddevelopercentric}{,}RepoCod~\cite{liang2025languagemodelsreplaceprogrammers}{,}DI-Bench~\cite{zhang2025dibenchbenchmarkinglargelanguage}{,}\\
                       DependEval~\cite{du2025dependevalbenchmarkingllmsrepository}{,}REPOST~\cite{xie2025repostscalablerepositorylevelcoding}{,}SecRepoBench~\cite{dilgren2025secrepobenchbenchmarkingllmssecure}{,}DevEval~\cite{li2024devevalmanuallyannotatedcodegeneration}}
                    , text width=45em, align=left, text centered=false
                    ]
                ]
                [{Domain-Specific\\and Complex Code}
                    [ %
                       {BioCoder~\cite{tang2024biocoderbenchmarkbioinformaticscode}{,}PaperBench~\cite{starace2025paperbenchevaluatingaisability}{,}Commit0~\cite{zhao2024commit0}{,}\\HackerRank-ASTRA~\cite{xing2025hackerrankastraevaluatingcorrectness}{,}ProjectEval~\cite{liu2025projectevalbenchmarkprogrammingagents}{,}DA-Code~\cite{huang2024dacodeagentdatascience}}
                    , text width=45em, align=left, text centered=false %
                    ]
                ]
                [{Code Editing{,} Refactoring{,}\\and Agent Collaboration}
                    [ %
                       {Aider’s code editing benchmark~\cite{aider-code-edit}{,}Aider’s refactoring benchmark~\cite{aider-refactoring-leaderboard}{,}\\Aider's Polyglot benchmark~\cite{polyglot-benchmark}{,}
                       RES-Q~\cite{labash2024resqevaluatingcodeeditinglarge}{,}
                       \\LiveRepoReflection~\cite{zhang2025LiveRepoReflection}{,}HumanEvo~\cite{zheng2025humanevoevolutionawarebenchmarkrealistic}{,}RepoExec~\cite{hai2025impactscontextsrepositorylevelcode}}
                    , text width=45em, align=left, text centered=false
                    ]
                ]
                [Commit Message Generation
                    [ %
                      {CommitBench~\cite{schall2024commitbench}{,}MCMD~\cite{tao2021evaluationcommitmessagegeneration}}
                    , text width=45em, align=left, text centered=false
                    ]
                ]
                [SWE Task Resolution
                    [ %
                       {SWE-bench~\cite{jimenez2024swtbench}{,}JavaBench~\cite{cao2024javabenchbenchmarkobjectorientedcode}{,}SWE-bench Lite~\cite{jimenez2024swtbench}{,}SWE-bench Multilingual~\cite{yang2025swesmithscalingdatasoftware}{,}\\SWE-bench Multimodal~\cite{yang2025swebench}{,}SWE-bench Verified~\cite{swebenchverified}{,}SWE-bench Live~\cite{zhang2025swebenchgoeslive}{,}\\SWE-Perf~\cite{he2025sweperflanguagemodelsoptimize}{,}SWE-rebench~\cite{badertdinov2025swerebenchautomatedpipelinetask}{,}SWE-Dev~\cite{du2025swedevevaluatingtrainingautonomous}{,}SWE-PolyBench~\cite{rashid2025swepolybenchmultilanguagebenchmarkrepository}{,}Multi-SWE-bench~\cite{zan2025multiswebenchmultilingualbenchmarkissue}{,}\\
                       SWE-bench+~\cite{aleithan2024swebenchenhancedcodingbenchmark}{,}SWE-bench M~\cite{yang2024swebenchmultimodalaisystems}{,}SWE-bench-java~\cite{zan2024swebenchjavagithubissueresolving}{,}SWE-Lancer~\cite{SWE-Lancer}{,}FAUN-Eval~\cite{hu2024realworldbenchmarkevaluatingfinegrained}{,}\\FEA-Bench~\cite{li2025fea}{,}SwingArena~\cite{xu2025swingarenacompetitiveprogrammingarena}{,}CoreCodeBench~\cite{fu2025corecodebenchconfigurablemultiscenariorepositorylevel}{,}AgentIssue-Bench~\cite{rahardja2025agentsfixagentissues}}
                    , text width=45em, align=left, text centered=false
                    ]
                ]
                [Comprehensive Software\\Development
                    [ %
                      {Aider’s code editing benchmark~\cite{aider-code-edit}{,}Aider’s refactoring benchmark~\cite{aider-refactoring-leaderboard}{,}\\
                      Aider's Polyglot benchmark~\cite{polyglot-benchmark}{,}RES-Q~\cite{labash2024resqevaluatingcodeeditinglarge}{,}LiveRepoReflection~\cite{zhang2025LiveRepoReflection}{,}\\HumanEvo~\cite{zheng2025humanevoevolutionawarebenchmarkrealistic}{,}RepoExec~\cite{hai2025impactscontextsrepositorylevelcode}{,}CodePlan~\cite{bairi2023codeplanrepositorylevelcodingusing} }
                    , text width=45em, align=left, text centered=false
                    ]
                ]
            ]
            [Agentic Systems
                [Agent Tool Use
                    [ %
                      { API-Bank~\cite{li2023api}{,}ToolBench~\cite{qin2023toolllm}{,}BFCL~\cite{patil2025bfcl}{,}Tau Bench~\cite{yao2024tau}}
                    , text width=45em, align=left, text centered=false
                    ]
                ]
                [Deep Research Benchmarks
                    [ %
                      {GAIA~\cite{mialon2023gaia}{,}xbench~\cite{chen2025xbench}{,}DeepResearch Bench~\cite{du2025deepresearch}}
                    , text width=45em, align=left, text centered=false
                    ]
                ]
                [Web Research Benchmarks
                    [ %
                      {BrowseComp~\cite{wei2025browsecomp}{,}BrowseComp-ZH~\cite{chen2025browsecomp}{,}BrowseComp-Plus~\cite{zhou2025browsecomp}{,}\\WebWalkerQA~\cite{wu2025webwalker}{,}Widesearch~\cite{wong2025widesearch}}
                    , text width=45em, align=left, text centered=false
                    ]
                ]
                [Benchmarking Agents for\\ Graphical User Interface
                    [ %
                      {WebShop~\cite{yao2022webshop}{,}Mind2Web~\cite{deng2023mind2web}{,}OmniACT~\cite{kapoor2024omniactdatasetbenchmarkenabling}{,}WebChoreArena~\cite{miyai2025webchorearena}{,}PersonalWAB~\cite{cai2025personalwab}{,}\\Sphinx~\cite{shi2024sphinx}{,}NovelScreenSpot~\cite{fan2025gui-bee}{,}Design2Code~\cite{si2024design2code}{,}WebCode2M~\cite{gui2025webcode2m}{,}Sketch2Code~\cite{li2024sketch2code}{,}\\Interaction2Code~\cite{xiao2024interaction2code}{,}WebGen-Bench~\cite{lu2025webgen}{,}Web-Bench~\cite{xu2025web} }
                    , text width=45em, align=left, text centered=false
                    ]
                ]
                [Terminal Use
                    [ %
                      {Terminal-Bench \cite{tbench_2025}}
                    , text width=45em, align=left, text centered=false
                    ]
                ]
            ]
        ]
    \end{forest}}
    \caption{A Taxonomy of coding tasks and benchmarks.}
    \label{fig:task_taxonomy}
\end{figure*}
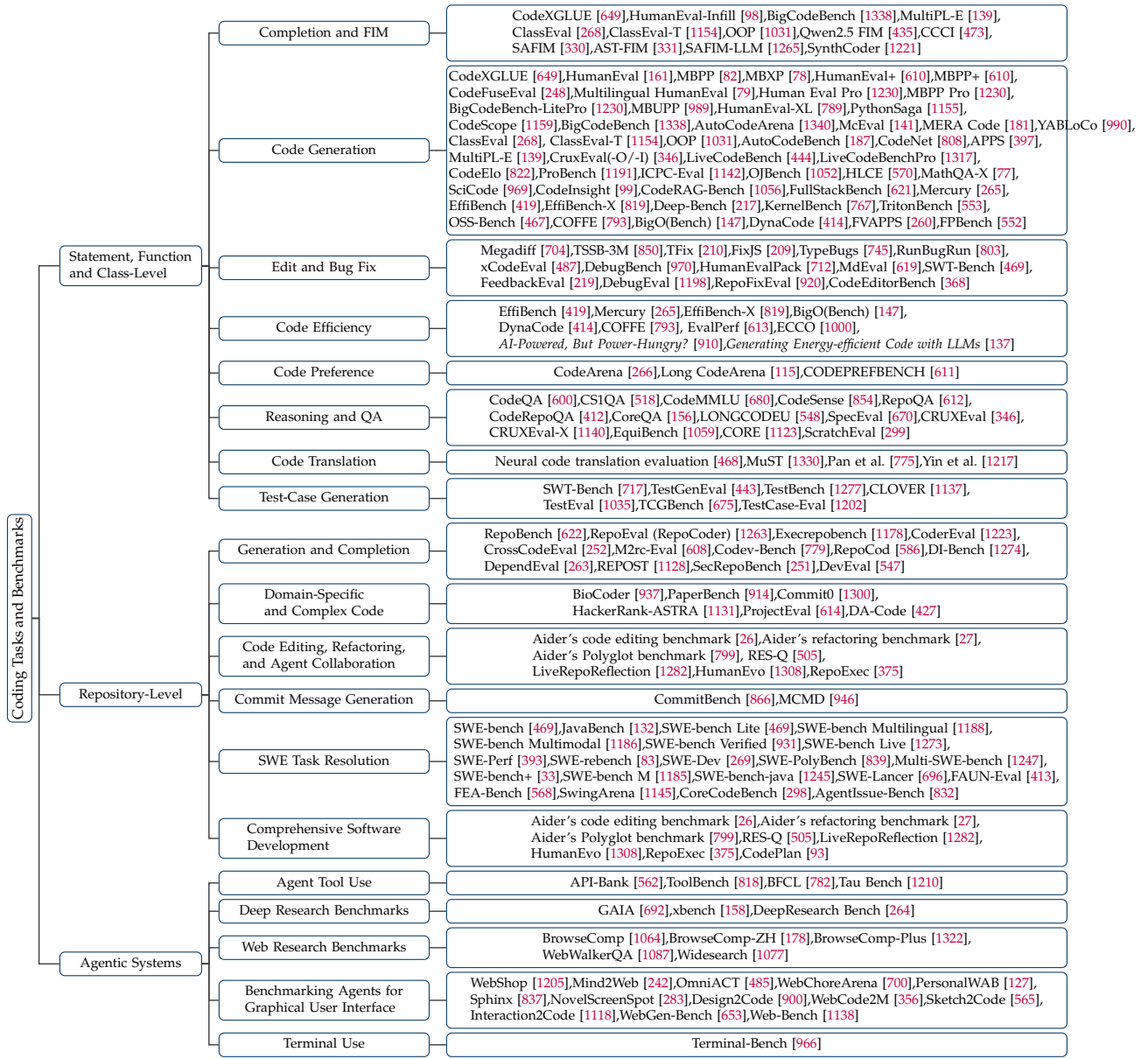

\section{Code Tasks, Benchmarks, and Evaluation}\label{sec:Code Task & Benchmark & Evalution}

\autoref{fig:task_taxonomy} presents a hierarchical taxonomy of coding tasks and benchmarks. The tree organizes the landscape along three major granularities: (i) statement, function, and class-level tasks; (ii) repository-level tasks; and (iii) agentic systems. At the statement/function/class level, the taxonomy covers completion and fill-in-the-middle, code generation, editing and bug fixing, code efficiency, code preference, reasoning and question answering, code translation, and test-case generation, with representative benchmarks listed at the leaves. The repository-level branch groups benchmarks for generation and completion in larger codebases, domain-specific and complex code, code editing/refactoring and agent collaboration, commit message generation, software engineering (SWE) task resolution, and comprehensive software development workflows. The agentic systems branch highlights benchmarks for tool use, deep and web research, graphical user interface interaction, and terminal use. Together, the leaves enumerate both widely used datasets and recent additions, providing a consolidated map of current evaluation resources across capabilities and scales. \autoref{sec:evaluation_metrics} first introduces the evaluation metrics for code LLMs and then \autoref{sec:statement_function_class-level} describes code tasks across three major granularities across diverse code llm tasks like code generation, code completion, code edit, code efficiency and code preference, code translation, and test case generation, while \autoref{sec:Repo-Level} introduces repo-level code benchmarks. Finally, \autoref{sec:Agentic systems} presents advanced code benchmarks, such as agent tool use, deep research, web search and terminal use.

\subsection{Evaluation Metrics}
\label{sec:evaluation_metrics}
\autoref{tab:code_metrics} lists the evolution from the string-matching methods to the execution-based methods. CodeBLEU~\cite{ren2020codebleumethodautomaticevaluation} measures syntax-level similarity using $n$-gram matching and abstract tree parsing but fails to assess functional correctness. Pass@$k$ then emerges as an execution-based metric, testing whether code passes predefined test cases across $k$ attempts. Most recently, LLM-based code judgment~\cite{codearena} leverages LLMs for comprehensive quality assessment, including correctness, efficiency, and readability beyond simple pass/fail outcomes.
\subsubsection{Extensions Based on Traditional Metrics}

\begin{itemize}

\item CodeBLEU~\cite{ren2020codebleumethodautomaticevaluation} borrows BLEU~\cite{wmt2021_nmt,papineni2002bleu} from machine translation, rewarding $n$-gram overlap between the model output and a single reference. That helps track superficial similarity but struggles with code with small lexical changes. And there are many correct implementations that share few tokens. It extends BLEU~\cite{papineni2002bleu} by incorporating $n$-gram matching, abstract syntax tree (AST) matching, and data flow semantic matching to provide a more comprehensive code quality assessment. 

\item CodeBERTScore~\cite{zhou2023codebertscoreevaluatingcodegeneration} is a pre-training model-based code generation evaluation metric that assesses generation quality by encoding natural language context and code snippets to calculate semantic similarity.
Round-trip correctness (RTC)~\cite{allamanis2024unsupervisedevaluationcodellms} is an unsupervised evaluation method for LLMs that verifies semantic consistency through the model’s own bidirectional tasks (e.g., code translation), avoiding reliance on manually annotated data. $RTC_{sim}(x) $ in \autoref{tab:code_metrics} measures the semantic similarity between original code $x$ and round-trip generated code $\hat{x}$. 

\item Pass@k~\cite{chen2021codex,SPoC} in \autoref{tab:code_metrics} sample $k$ completions for each task and score the chance that at least one passes the unit tests. pass@k directly measures functional correctness, captures the benefit of sampling, handles multiple valid solutions, and tends to correlate better with developer utility, making it the de facto standard for modern LLM code generation benchmarks.

\end{itemize}

\subsubsection{LLM-as-a-Judge Paradigm}

\begin{itemize}

\item ICE-Score~\cite{zhuo2024ice} is the first evaluation framework that instructs LLMs to rate code quality using structured, criteria-driven scoring rubrics. In contrast to test-based evaluation, ICE-Score leverages rubric-guided LLM judgments to assess dimensions such as \textit{usefulness} and \textit{functional correctness}. In the formula presented in \autoref{tab:code_metrics}, the evaluation function $f$ takes as input the task description $p$ and the generated code $r$, and outputs a discrete score $S \in \{0,1,2,3,4\}$ based on predefined evaluation criteria. Higher scores indicate better semantic alignment with the intended functionality. In a continuous adaptation, this score may be normalized as $S/4$, where values closer to 1 denote stronger correctness.

\item CodeJudge~\cite{tong2024codejudgeevaluatingcodegeneration} is a code evaluation framework built upon ICE-Score, enhancing evaluation reliability through a slow-thinking mechanism. In the formula presented in \autoref{tab:code_metrics}, $f$ denotes the evaluation function, which outputs either a binary judgment (correct/incorrect) or a deviation score ranging from 0 to 1 points. This function in the continuous setting, lower deviation scores indicate smaller semantic errors (i.e., better alignment with the reference).

\item LLM-as-a-Judge~\cite{he2025code} is a method that uses LLMs to directly evaluate code quality. In the formula presented in \autoref{tab:code_metrics}, \textit{src} represents the task source instruction, \textit{target} denotes the code/text to be evaluated, and the LLM directly outputs a score. The primary advantage of this approach is its ability to leverage LLMs for analyzing deep semantic aspects of code functionality. However, its limitation lies in unstable performance, attributed to LLMs' propensity for generating verbose explanations.

\item CodeJudgeBench~\cite{jiang2025codejudgebenchbenchmarkingllmasajudgecoding} adopts the LLM-as-a-Judge paradigm, where an LLM directly evaluates the functional correctness of code responses without executing the code. The core formula of the framework, presented in \autoref{tab:code_metrics}, involves concatenating \textit{p} (the programming task description), \textit{r} (the response to be evaluated), and \textit{q} (the model’s judging instruction), after which the LLM outputs J to represent the judgment result. 

\item BigCodeReward~\cite{zhuo2025bigcodearena}
is the first benchmark designed for evaluating reward models in practical coding scenarios. Unlike prior reward-model benchmarks that focus on general domains, BigCodeReward targets code-specific evaluation by incorporating real-world execution feedback, including textual logs, webpage screenshots, interactive application states, and plots. In the formulation shown in \autoref{tab:code_metrics}, the evaluation function $f$ takes as input a coding prompt $p$, two candidate responses $(r_A, r_B)$, and optional execution feedback $e$, and the model outputs a preference label $J \in \{\text{A}, \text{B}, \text{Tie}\}$. BigCodeReward operates entirely within the LLM-as-a-Judge paradigm, requiring models to directly judge which response better satisfies user intent, both with and without execution results. This design enables assessment of whether reward models can reliably leverage multimodal execution signals that extend beyond pure text.

\end{itemize}

\definecolor{cat1}{RGB}{230,240,250}   %
\definecolor{cat2}{RGB}{250,240,230}   %
\definecolor{cat3}{RGB}{240,250,230}   %
\definecolor{cat4}{RGB}{250,230,240}   %
\definecolor{cat5}{RGB}{240,240,250}   %
\definecolor{cat6}{RGB}{230,250,240}   %

\begin{table}[H]
\centering
\caption{Summarized metrics for code evaluation.}
\label{tab:code_metrics}
\small %

\setlength{\tabcolsep}{6pt} 
\renewcommand{\arraystretch}{1.1} %
\resizebox{0.95\textwidth}{!}{%
\begin{tabular}{lll}
\toprule
\textbf{Metrics} & \textbf{How to calculate} & \textbf{Explanation} \\
\rowcolor{cat1}
\multicolumn{3}{c}{\textbf{Extensions Based on Traditional Metrics}} \\
\addlinespace[0.8em]

Biased pass@k~\cite{chen2021codex} &
  $\displaystyle 1 - \frac{\binom{n-c}{k}}{\binom{n}{k}}$ &
  \makecell[l]{$n$ = total generation times, $c$ = correct generation times, \\ $k$ = Number of samples from all $n$ generation without replacement.} \\
\midrule
Unbiased pass@k~\cite{SPoC} &
  $\displaystyle 1 - \bigl(\tfrac{n-c}{n}\bigr)^k$ &
  \makecell[l]{$n$ = total generation times, $c$ = correct generation times, \\ $k$ = Number of samples from all $n$ generation with replacement.} \\
\midrule
CodeBERTScore~\cite{zhou2023codebertscoreevaluatingcodegeneration} &
  \makecell[l]{%
    $\displaystyle P=\frac{1}{|\hat y|}\sum_j\max_i\mathrm{sim}(y^*_i,\hat y_j)$\\
    $\displaystyle R=\frac{1}{|y^*|}\sum_i\max_j\mathrm{sim}(y^*_i,\hat y_j)$\\
    $\displaystyle F_1=\frac{2PR}{P+R},\quad F_3=\frac{10PR}{9P+R}$
  } &
  \makecell[l]{%
    $\hat y$ = generated tokens, $y^*$ = reference tokens,\\
    $\mathrm{sim}(\cdot,\cdot)$ = embedding similarity,\\
    $P,R$ = precision, recall; $F_1,F_3$ = harmonic f‐scores%
  } \\[1em]
\midrule
RTC~\cite{allamanis2024unsupervisedevaluationcodellms} &
  \makecell[l]{$\displaystyle 
    RTC_{sim}(x)\approx
    \frac{1}{N_f N_b}$ \\ \quad $
    \sum_{y\sim M(x)}\sum_{\hat x\sim M^{-1}(y)}
    \mathrm{sim}(\hat x,x)
  $} &
  \makecell[l]{%
    $N_f,N_b$ = \# forward/backward samples,\\
    $M,M^{-1}$ = round‐trip models, $\mathrm{sim}$ denotes code similarity%
  } \\

\rowcolor{cat2}
\multicolumn{3}{c}{\textbf{LLM-as-a-Judge Paradigm}} \\
\addlinespace[0.3em]

ICE-Score~\cite{zhuo2024ice} &
  $f(\mathrm{Score},\mathrm{Task})$ &
  \makecell[l]{%
    $f$ = scoring function,\\
    $\mathrm{Score}$ = numeric rating,
    $\mathrm{Task}$ = problem description%
  } \\
\midrule
CodeJudgeBench~\cite{jiang2025codejudgebenchbenchmarkingllmasajudgecoding} &
  $J \leftarrow \mathrm{LLM}(p \oplus r \oplus q)$ &
  \makecell[l]{%
    $J$ = judgment score,
    $p$ = prompt template,\\
    $r$ = reference solution,
    $q$ = model output,\\
    $\oplus$ = text concatenation%
  } \\
  \midrule
BigCodeReward~\cite{zhuo2025bigcodearena} &
  $J \leftarrow f(p, r_A, r_B, e)$ &
  \makecell[l]{%
    $J \in \{\mathrm{A},\mathrm{B},\mathrm{Tie}\}$ = preference judgment,\\
    $p$ = task description,\\
    $r_A, r_B$ = candidate responses,\\
    $e$ = (optional) execution feedback,\\
    $f$ = LLM-as-a-Judge evaluator%
  } \\

\rowcolor{cat3}
\multicolumn{3}{c}{\textbf{Execution-Based Metrics}} \\
\addlinespace[0.3em]
ProbeGen~\cite{allamanis2025disprovingprogramequivalencellms} &
  $\exists p_k: p_k(f_i)\neq p_k(f_j)\Rightarrow f_i\not\equiv f_j$ &
  \makecell[l]{%
    $p_k$ = test probe input, $f_i,f_j$ = two candidate functions,\\
    $p_k(f)$ = function output on probe%
  } \\
\midrule
REFUTE~\cite{sinha2025languagemodelsfalsifyevaluating} &
  $H(x^*)\cap\neg P(x^*)$ &
  \makecell[l]{%
    $x^*$ denotes input, $H$ represents the first presupposed hypothesis, \\ and $P$ represents the second hypothesis.} \\
\midrule
EvaLooop~\cite{fang2025evaloopassessingllmrobustness} &
  $\displaystyle ASL=\frac{\sum_{i=1}^M n_i\,i^2}{M\cdot N}$ &
  \makecell[l]{%
    $n_i$ = \# tests with $i$ iterations,
    $i$ = loop‐iteration count,\\
    $M$ = \# distinct $i$ values,
    $N$ = total test cases%
  } \\
\rowcolor{cat4}
\multicolumn{3}{c}{\textbf{Multi-Agent \& Advanced Reasoning Framework}} \\
\addlinespace[0.3em]
MCTS-Judge~\cite{wang2025mctsjudgetesttimescalingllmasajudge} &
  $\displaystyle Reward=\varepsilon\ \text{if}\ f(t,g)=h(x)$ &
  \makecell[l]{%
    $\varepsilon$ = small reward constant, $h(x)$ = expected output for input $x$ \\
    $f(t,g)$ = model’s predicted output given trace $t$ and goal $g$,\\
  } \\

\rowcolor{cat5}
\multicolumn{3}{c}{\textbf{Statistical \& Consistency Analysis Metrics}} \\
\addlinespace[0.3em]
Incoherence~\cite{valentin2025estimatingcorrectnessoraclesllmbased} &
  \makecell[l]{$\displaystyle I_{\mathrm{Gen}}(d)=P\bigl([\Pi_1^d]^1(X)$ \\ \quad $\neq[\Pi_2^d]^1(X)\bigr)$} &
  \makecell[l]{%
    $d$ = decoding depth, $P$ = probability,\\
    $\Pi_1^d,\Pi_2^d$ = two runs sampling depth $d$,\\
    $[\cdot]^1(X)$ = first token on input $X$,\\
  } \\
\midrule
MAD~\cite{moon2025dontjudgecodecover} &
  $\displaystyle \frac{1}{K}\sum_{k=1}^K\bigl|Acc_{\mathrm{org}}^{(k)}-Acc_{\mathrm{bias}}^{(k)}\bigr|$ &
  \makecell[l]{%
    $K$ = \# of trials, $Acc_{\mathrm{bias}}^{(k)}$ = accuracy with biased prompt \\
    $Acc_{\mathrm{org}}^{(k)}$ = accuracy on trial $k$ with original prompt} \\

\rowcolor{cat6}
\multicolumn{3}{c}{\textbf{Other Unique Paradigms}} \\
\addlinespace[0.3em]
SBC~\cite{ponnusamy2025bridgingllmgeneratedcoderequirements} &
  \makecell[l]{$(0.7\times \text{semantic}) + (0.1\times \text{BLEU})$ \\ \quad  $+ (0.2\times \text{completeness})$} &
  \makecell[l]{%
    semantic = semantic‐similarity score,
    BLEU = n-gram precision,\\
    completeness = test‐coverage measure%
  } \\
\midrule
Copilot Arena~\cite{chi2025copilotarenaplatformcode} &
  $\text{Rank}=\mathrm{BradleyTerry}(\text{User Votes})$ &
  \makecell[l]{%
    Rank = model’s overall rank,\\
    User Votes = pairwise preferences aggregated via Bradley–Terry%
  } \\
\midrule
BigCodeArena~\cite{zhuo2025bigcodearena} &
  $\text{Rank}=\mathrm{BradleyTerry}(\text{User Votes})$ &
  \makecell[l]{%
    Rank = model’s overall rank,\\
    User Votes = pairwise preferences aggregated via Bradley–Terry%
  } \\
\midrule
CodeCriticBench~\cite{zhang2025codecriticbenchholisticcodecritique} &
  $\displaystyle MSE=\frac{1}{N}\sum_{i=1}^N(\hat{y}_i-y_i)^2$ &
  \makecell[l]{%
    $N$ = number of samples,
    $\hat y_i$ = predicted rating,
    $y_i$ = true rating%
  } \\
\bottomrule
\end{tabular}
}
\end{table}

\subsubsection{Execution-Based Metrics} 

\begin{itemize}

\item ProbeGen~\cite{allamanis2025disprovingprogramequivalencellms}  leverages LLMs to generate targeted test cases (probes) that verify code semantic equivalence through execution feedback. Its key insight is captured by the formula in \autoref{tab:code_metrics}. when a probe $p_k$
causes different implementations, $f_i$ and $f_j$, to produce divergent outputs, their functional inequivalence is immediately established.
REFUTE~\cite{sinha2025languagemodelsfalsifyevaluating} is an evaluation framework specifically designed to assess the ability of large language models (LLMs) to find counterexamples. Its core concept involves prompting models to generate test cases that cause buggy code to fail. Its core formula is presented in \autoref{tab:code_metrics}, where $x^*$
denotes the counterexample input, $H$ represents the constraint, and $P$ stands for the expected correctness of the buggy code.

\item EvaLooop~\cite{fang2025evaloopassessingllmrobustness} introduces a self-contained feedback-loop evaluation framework to assess the robustness of LLMs in programming. The framework leverages two dualities to repeatedly transform the model’s outputs back into new prompts until functional correctness fails, including (a) code generation$\leftrightarrow$code summarization and (b) cross-language code translation. Its core metric, average sustainable loops (ASL), is defined in \autoref{tab:code_metrics}, where \(M\) denotes the maximum loop count, \(n_i\) the number of tasks still passing on loop \(i\), and \(N\) the total number of tasks. A higher ASL indicates stronger semantic consistency and robustness across iterations.

\end{itemize}

\subsubsection{Multi-Agent \& Advanced Reasoning Framework}
\begin{itemize}

\item MCTS-Judge~\cite{wang2025mctsjudgetesttimescalingllmasajudge} is a test-time computation framework for code correctness evaluation that enables LLMs to conduct multi-perspective reasoning through monte carlo tree search (MCTS). Each node in the search process represents a distinct reasoning perspective, such as boundary condition analysis, exception handling, or specification compliance. In \autoref{tab:code_metrics}, $f(t,g)$ denotes the aggregated prediction along the reasoning trajectory, $h(x)$ indicates the verification result from simulated test execution, and $\epsilon$ represents the reward assigned when the prediction aligns with the verification outcome.

\item CodeVisionary~\cite{wang2025codevisionaryagentbasedframeworkevaluating} constructs a two-stage evaluation framework. The first stage collects comprehensive information through multi-source knowledge analysis, while the second stage employs a multi-LLM collaborative scoring mechanism. In \autoref{tab:code_metrics}, $S_i$ represents the consensus score from multi-LLM collaboration and $n$ denotes the number of LLMs involved.

\end{itemize}

\subsubsection{Statistical \& Consistency Analysis Metrics}  

\begin{itemize}

\item Incoherence~\cite{valentin2025estimatingcorrectnessoraclesllmbased} quantifies the uncertainty of LLMs in code generation tasks and measures the probability that two independently generated programs for the same problem produce different outputs on identical inputs. Higher incoherence values reflect increased stochasticity and reduced consistency in model reasoning, whereas lower values imply more deterministic and consistent behavior. In \autoref{tab:code_metrics}, computes the divergence rate across program pairs $\left[ \Pi_1^d \right]$ and $\left[ \Pi_2^d \right]$ sampled for task $d$ over input distribution $\mathrm{Gen}(d)$.

\item Mean absolute deviation (MAD)~\cite{moon2025dontjudgecodecover} evaluates the robustness of LLM-based evaluators under superficial perturbations and measures the deviation between the original evaluation accuracy ($Acc_{\text{org}}$) and the accuracy after introducing $K$ systematic biases ($Acc_{\text{bias}}$), such as variable renaming, comment injection, or code formatting changes. A smaller MAD indicates higher evaluator consistency and less sensitivity to presentation-level noise.

\end{itemize}

\subsubsection{Other Unique Paradigms}
\begin{itemize}
\item The core innovation of the SBC~\cite{ponnusamy2025bridgingllmgeneratedcoderequirements} (\textbf{S}emantic similarity score: measuring meaning-level alignment between requirements, \textbf{B}LEU score: capturing lexical overlap, and \textbf{C}ompleteness score: identifying missing and extra elements) metric lies in its reverse generation technique: instead of directly evaluating code, it leverages LLMs to reverse-engineer requirements from the generated code and then compares these reverse-engineered requirements with the original ones. Its core calculation formula is presented in \autoref{tab:code_metrics}, where \textit{Semantic} refers to cosine similarity, and 
\textit{Completeness} is based on penalizing missing keywords and redundancy extracted from nouns and verbs.

\item Copilot Arena~\cite{chi2025copilotarenaplatformcode} and BigCodeArena~\cite{zhuo2025bigcodearena} are platforms that collect user preference votes for LLM-generated code in real developer environments and computes the relative win rates of models using the Bradley-Terry model (a probability model that estimates the likelihood of one item being preferred over another in a series of pairwise comparisons).

\item CodeCriticBench~\cite{zhang2025codecriticbenchholisticcodecritique} is a comprehensive code critique benchmark covering two main tasks (code generation and code QA) and structured into two dimensions: basic evaluation and advanced evaluation. The basic evaluation focuses on binary classification of correctness (ACC metric), while the advanced evaluation uses multi-dimensional fine-grained checklists for scoring with MSE metric (mean squared error).

\end{itemize}

\subsection{Statement, Function, and Class-Level Tasks and Benchmarks}
\label{sec:statement_function_class-level}
This section presents statements, function, and class-level code benchmarks across different tasks, such as code completion, FIM, generation, editing, and bug-fixing.

\subsubsection{Code Completion and Code FIM}

This section reviews benchmarks for code completion and Fill-in-the-Middle (FIM) tasks, which assess a model’s ability to predict correct code fragments in partially observed contexts. Unlike function-level generation from prompts or docstrings, code completion focuses on continuing or repairing existing code, often using local context. FIM further generalizes this by requiring the prediction of missing segments within a sequence, reflecting real-world scenarios such as IDE autocompletion, refactoring, and bug fixing. \autoref{tab:code_completion_benchmarks} list some code completion benchmarks.

\begin{table}[ht]
\centering
\caption{Code Completion and Fill-in-the-Middle Benchmarks Overview.}
\label{tab:code_completion_benchmarks}
\small
\setlength{\tabcolsep}{4pt}
\renewcommand{\arraystretch}{1.15}
\begin{tabular}{lccccc}
\toprule
\textbf{Benchmark} & \textbf{Year} & \textbf{Granularity} & \textbf{Languages} & \textbf{Size (K)} & \textbf{Evaluation} \\
\midrule
\multicolumn{6}{c}{\cellcolor{lightblue}\textbf{Code Completion Benchmarks}} \\
\midrule
CodeXGLUE~\cite{CodeXGLUE} & 2021 & Statement & 3+ & 104K & Exact Match \\
HumanEval-Infill~\cite{bavarian2022efficient} & 2022 & Function & Python & 164 & Unit Tests \\
ExecRepoBench~\cite{execrepobench} & 2024 & Statement & Python & 1.2K & Tests+Accuracy \\
BigCodeBench~\cite{zhuo2024bigcodebench} & 2024 & Function & 3+ & 1.1k & Functional Tests \\
MultiPL-E~\cite{MultiPL-E} & 2022 & Function & 3+ & 12.7K & Diversity \\
ClassEval~\cite{du2023classeval} & 2024 & Class & Python & 100/412 & Unit Tests \\
ClassEval-T~\cite{xue2024classevalT} & 2024 & Class & Java, C++ & 94 & Unit Tests \\
OOP~\cite{oop} & 2024 & Class & Python & 431 & OOP Patterns \\
\midrule
\multicolumn{6}{c}{\cellcolor{lightorange}\textbf{Fill-in-the-Middle (FIM) Benchmarks}} \\
\midrule
CCCI~\cite{jin2025cccicodecompletioncontextual} & 2025 & Statement & Java & 289 & Contextual Quality \\
SAFIM~\cite{gong2024evaluationllmssyntaxawarecode} & 2024 & Syntax-aware & 3+ & 17,720 & Syntax Correctness \\
AST-FIM~\cite{gong2025structureawarefillinthemiddlepretrainingcode} & 2025 & AST-based & 12 & 30k+ & Structural Accuracy \\
\bottomrule
\end{tabular}
\end{table}

\paragraph{Code Completion} 
\begin{itemize}

\item \textbf{Statement-level} completion evaluates the prediction of individual lines or expressions based on preceding context. CodeXGLUE~\cite{CodeXGLUE} includes a next-line completion task across six programming languages, using exact match accuracy as the primary metric. However, it lacks functional validation. More rigorous benchmarks like HumanEval-Infill~\cite{bavarian2022efficient} adapt function-level datasets by treating docstring-conditional body generation as a completion task, with correctness determined via unit test execution. Qwencoder2.5~\cite{qwen25coder} further refines this with a dedicated base completion benchmark that tests Python statement prediction within functions, measuring both lexical accuracy and test-based correctness under varying context scopes. 

\item \textbf{Function-level} completion involves generating full implementations from signatures or partial bodies. BigCodeBench~\cite{zhuo2024bigcodebench} includes function recovery tasks where models complete stubs extracted from real repositories, evaluated on functional correctness and complexity preservation. MultiPL-E~\cite{MultiPL-E} extends this by requiring multiple valid solutions per prompt, probing a model’s generative diversity—a key aspect of realistic completion behavior. Class-level completion introduces challenges in modeling stateful interactions and method dependencies. ClassEval~\cite{du2023classeval} requires models to implement individual methods within a partially defined class structure, assessing coherence with existing logic through unit tests. Results show significant performance drops compared to function-level tasks, especially for smaller models. Its multilingual extension, ClassEval-T~\cite{xue2024classevalT}, adds Java and C++ variants and identifies common errors such as incorrect field initialization and broken inter-method dependencies. OOP~\cite{oop} complements these by evaluating object-oriented design patterns during method completion, including proper use of inheritance and encapsulation.

\end{itemize}

\paragraph{Fill-in-the-Middle (FIM)}  
FIM specifically targets infilling missing code segments given both left and right context, simulating real editing and refactoring scenarios. While originally used as a pretraining objective (e.g., in StarCoder~\cite{lozhkov2024starcoder2stackv2}), systematic FIM evaluation has only recently emerged. CCCI~\cite{jin2025cccicodecompletioncontextual} improves code completion by incorporating contextual information like database relationships and object models into LLMs, achieving higher functional correctness and code quality. SAFIM~\cite{gong2024evaluationllmssyntaxawarecode} introduces a syntax-aware FIM benchmark across multiple programming languages, demonstrating that pretraining strategies and data quality significantly affect FIM performance. AST-FIM~\cite{gong2025structureawarefillinthemiddlepretrainingcode} leverages abstract syntax trees to mask syntactic structures during pretraining, improving model performance on real-world FIM tasks.

\subsubsection{Code Generation}
In \autoref{tab:codegen_benchmarks}, this section surveys code generation benchmarks, organized along different granularities (function-level, class-level, and domain-specific tasks) and extended dimensions such as efficiency and verification.

{\small
\begin{longtable}{>{\raggedright\arraybackslash}p{4.8cm}
                   >{\raggedright\arraybackslash}p{0.8cm}
                   >{\raggedright\arraybackslash}p{0.8cm}
                   >{\raggedright\arraybackslash}p{2.9cm}
                   >{\raggedright\arraybackslash}p{2.4cm}
                   >{\centering\arraybackslash}p{1.8cm}@{}}
\caption{Code generation Benchmarks Overview.} 
\label{tab:codegen_benchmarks} \\
\toprule
\textbf{Benchmark} & \textbf{Year} & \textbf{Size} & \textbf{Languages} & \textbf{Domain} & \textbf{Difficulty} \\

\endfirsthead

\multicolumn{6}{c}%
{{\bfseries Table \thetable\ continued from previous page}} \\
\toprule
\textbf{Benchmark} & \textbf{Year} & \textbf{Size} & \textbf{Languages} & \textbf{Domain} & \textbf{Difficulty} \\
\midrule
\endhead

\midrule \multicolumn{6}{r}{{Continued on next page}} \\
\endfoot

\bottomrule
\endlastfoot

\arrayrulecolor{black}
\cline{1-6}
\rowcolor{lightblue}\multicolumn{6}{c}{\textbf{Function-Level Benchmarks}} \\
\cline{1-6}
\noalign{\vskip 0.4em}   
\arrayrulecolor{black}
CodeXGLUE~\cite{CodeXGLUE} & 2021 & 104k & Java & General & \difficultybar{1} \\
HumanEval~\cite{chen2021codex} & 2021 & 164 & Python & Algorithmic & \difficultybar{2} \\
MBPP~\cite{austin2021mbpp} & 2021 & 974 & Python & Intro Coding & \difficultybar{2} \\
MBXP~\cite{mbxp} & 2022 & 12.4k & Multi-PLs & Intro Coding & \difficultybar{2} \\
HumanEval+~\cite{evalplus} & 2023 & 164 & Python & Algorithmic & \difficultybar{3} \\
MBPP+~\cite{evalplus} & 2023 & 378 & Python & Algorithmic & \difficultybar{3} \\
CodeFuseEval~\cite{CodeFuseEval} & 2023 & 820 & Multi-PLs & General & \difficultybar{2} \\
Multilingual HumanEval~\cite{multilingualhumaneval} & 2023 & 1.9K & Python,Java,JS & Algorithmic & \difficultybar{2} \\
HumanEval Pro~\cite{humanevalpro} & 2024 & 164 & Python & Algorithmic & \difficultybar{4} \\
MBPP Pro~\cite{humanevalpro} & 2024 & 378 & Python & Algorithmic & \difficultybar{4} \\
BigCodeBench-LitePro~\cite{humanevalpro} & 2024 & 57 & Multi-PLs & General & \difficultybar{4} \\
MBUPP~\cite{mbupp} & 2024 & 466 & Python & Algorithmic & \difficultybar{4} \\
HumanEval-XL~\cite{humanevalxl} & 2024 & 22k & Multi-PLs\&NLs & Cross-lingual & \difficultybar{5} \\
PythonSaga~\cite{pythonsaga} & 2024 & 185 & Python & Library/API & \difficultybar{2} \\
CodeScope~\cite{CodeScope} & 2024 & 400 & Multi-PLs & Control Flow & \difficultybar{3} \\
BigCodeBench~\cite{zhuo2024bigcodebench} & 2024 & 1.1k & Python & General & \difficultybar{5} \\
McEval~\cite{mceval} & 2025 & 12k & Python & Mathematical & \difficultybar{5} \\
MERA Code~\cite{MERACode} & 2025 & 1.3k & Python, Java & Systems & \difficultybar{3} \\
YABLoCo~\cite{yabloco} & 2025 & 208 & C, C++ & Systems & \difficultybar{4} \\
IFEvalCode~\cite{yang2025ifevalcode} & 2025 & 1.6K & Multi-PLs  & Systems & \difficultybar{4} \\
\cline{1-6}
\rowcolor{lightgreen}\multicolumn{6}{c}{\textbf{Class-Level Benchmarks}} \\
\cline{1-6}
\noalign{\vskip 0.4em}
ClassEval~\cite{du2023classeval} & 2024 & 100 & Python & OOP & \difficultybar{3} \\
ClassEval-T~\cite{xue2024classevalT} & 2024 & 94 & Python, Java, C++ & OOP & \difficultybar{4} \\
OOP~\cite{oop} & 2024 & 431 & Python & OOP & \difficultybar{3} \\
utoCodeBench~\cite{autocodebench} & 2025 & 3.9k & Multi-PLs & General & \difficultybar{4} \\

\cline{1-6}
\rowcolor{lightorange}\multicolumn{6}{c}{\textbf{Competition Benchmarks}} \\
\cline{1-6}
\noalign{\vskip 0.4em}
CodeNet~\cite{codenet} & 2021 & 14M+ & Multi-PLs & Cross-lingual & \difficultybar{3} \\
APPS~\cite{apps} & 2021 & 10k+ & Python & Algorithmic & \difficultybar{4} \\
Multip-E~\cite{MultiPL-E} & 2022 & 12.7k & Multi-PLs & Cross-lingual & \difficultybar{4} \\
CruxEval(-O / -I)~\cite{gu2024cruxeval} & 2024 & 800 & Multi-PLs & Algorithmic & \difficultybar{4} \\
LiveCodeBench~\cite{jain2024livecodebench} & 2024 & 121 & Multi-PLs & Dynamic & \difficultybar{4} \\
LiveCodeBenchPro~\cite{livecodebenchpro} & 2024 & 584 & C++ & Dynamic & \difficultybar{4} \\
CodeElo~\cite{codeelo} & 2025 & 387 & C++,Python & Algorithmic & \difficultybar{4} \\
ProBench~\cite{probench} & 2025 & 790 & C++, Java, Python & Cross-lingual & \difficultybar{4} \\
ICPC-Eval~\cite{icpceval} & 2025 & 118 & C++ & Algorithmic & \difficultybar{4} \\
OJBench~\cite{ojbench} & 2025 & 232 & C++, Python & Algorithmic & \difficultybar{3} \\
HLCE~\cite{hlce} & 2025 & 235 & C++, Python & Algorithmic & \difficultybar{3} \\

\cline{1-6}
\rowcolor{lightpurple}\multicolumn{6}{c}{\textbf{Other Domain-Specific Benchmarks}} \\
\cline{1-6}
\noalign{\vskip 0.4em}
MathQA-X~\cite{mathqax} & 2022 & 5.6k & Python, Java, JS & Mathematical & \difficultybar{5} \\
SciCode~\cite{scicode} & 2024 & 338 & Python & Scientific & \difficultybar{3} \\
CodeInsight~\cite{codeinsight} & 2024 & 3.4k & Python & Hybrid & \difficultybar{4} \\
CodeRAG-Bench~\cite{CodeRAG-Bench} & 2024 & 2.1k & Python & RAG & \difficultybar{3} \\
FullStackBench~\cite{liu2024fullstackbench} & 2024 & 3K & Multi-PLs & Web & \difficultybar{4} \\
Mercury~\cite{mercury2024} & 2024 & 1.9k & Multi-PLs & Efficiency & \difficultybar{3} \\
EffiBench~\cite{huang2024effibench} & 2024 & 1k & Multi-PLs & Efficiency & \difficultybar{3} \\
EffiBench-X~\cite{effibenchx2025} & 2025 & 623 & Multi-PLs & Efficiency & \difficultybar{3} \\
Deep-Bench~\cite{deepbench} & 2025 & 520 & Python & Deep Learning & \difficultybar{3} \\
KernelBench~\cite{kernelbench} & 2025 & 250 & Python & GPU kernel & \difficultybar{4} \\
TritonBench~\cite{tritonbench} & 2025 & 184 & Triton & GPU kernel & \difficultybar{4} \\
AutoCodeArena~\cite{zhuo2025bigcodearena}  & 2025 & 600 & Multi-PLs & Hybrid & \difficultybar{4} \\
OSS-Bench~\cite{ossbench} & 2025 & 17.9k & Php,Sql & Software & \difficultybar{2} \\
COFFE~\cite{coffe} & 2025 & 756 & Python & Efficiency & \difficultybar{3} \\
BigO(Bench)~\cite{bigobench} & 2025 & 3.1k & Python & Complexity & \difficultybar{4} \\
DynaCode~\cite{dynacode} & 2025 & 405 & Python & Complexity & \difficultybar{4} \\
FVAPPS~\cite{fvapps} & 2025 & 4.7k & Python & Verification & \difficultybar{4} \\
FPBench~\cite{fpbench} & 2025 & 1.8k & Python & Verification & \difficultybar{4} \\
\end{longtable}}

\paragraph{Function-Level Benchmarks}
Early benchmark suites such as CodeXGLUE~\cite{CodeXGLUE} includes code generation tasks (e.g., NL2Code, code-to-code translation), which are primarily formulated at the function or statement level. Function-level evaluation was further developed by HumanEval~\cite{chen2021codex} and MBPP~\cite{austin2021mbpp}, which target Python code synthesis from docstrings using pass/fail outcomes on unit tests as correctness criteria. To address limited test coverage, EvalPlus~\cite{evalplus} introduces strengthened variants—HumanEval+ and MBPP+—that incorporate adversarially generated and hidden test cases, and automatically synthesize more comprehensive test suites with robust metrics. Building on this line, HumanEval Pro, MBPP Pro, BigCodeBench-Lite Pro~\cite{humanevalpro}, and MBUPP~\cite{mbupp} extend classical settings with self-invoking code execution, larger problem pools, and one-to-many evaluation. Multi-lingual evaluation expands the scope beyond Python. MBXP~\cite{mbxp} and Multilingual HumanEval~\cite{multilingualhumaneval} provide function-level coding problems across multiple programming languages, HumanEval-XL~\cite{humanevalxl} scales this to 22,080 problems in 23 natural languages and 12 programming languages, and PythonSaga~\cite{pythonsaga} offers a curated Python set to assess generation quality under diverse problem types. CodeScope~\cite{CodeScope} adds execution-based feedback for multilingual code generation, while MERA Code~\cite{MERACode} introduces a systematic, cross-granularity benchmarking framework spanning function-level, algorithmic, and system-level tasks. McEval~\cite{mceval} focuses on method-level code completion in large codebases, emphasizing contextual coherence. BigCodeBench~\cite{zhuo2024bigcodebench} samples from real-world repositories to assess functional correctness and code complexity. CodeFuseEval~\cite{CodeFuseEval} further evaluates models on multi-task scenarios including code completion, cross-language translation, and code generation from Chinese commands.

\paragraph{Class-Level Benchmarks}
Beyond single-function tasks, ClassEval~\cite{du2023classeval} introduces a manually curated benchmark of 100 Python classes (412 methods) that require modeling inter-method dependencies and stateful behavior. Results across 11 LLMs show significant performance drops compared to function-level tasks. Its multilingual extension, ClassEval-T~\cite{xue2024classevalT}, adds Java and C++ variants and provides a fine-grained failure taxonomy, including dependency and initialization errors. OOP~\cite{oop} further targets object-oriented programming in Python, complementing ClassEval benchmarks by evaluating class-level design, method interactions, and stateful behavior.

\paragraph{Domain-Specific Benchmarks}
To better reflect domain-specific requirements, specialized benchmarks have emerged. For competitive programming, CodeNet~\cite{codenet} provides a large-scale, multilingual dataset with performance metadata, supporting diverse evaluation scenarios. APPS~\cite{apps} is a widely used benchmark featuring programming competition problems and human-written solutions. MultiPL-E~\cite{MultiPL-E} assesses solution diversity by requiring models to generate multiple distinct correct implementations per problem, thereby probing deeper algorithmic reasoning rather than mere pattern matching. CruxEval~\cite{gu2024cruxeval}, comprising CRUXEval-I and CRUXEval-O, curates problems from online judges that emphasize reasoning and execution, enabling evaluation of both input and output prediction to assess code understanding and runtime behavior. LiveCodeBench~\cite{jain2024livecodebench} evaluates performance on LeetCode problems with temporal alignment, capturing the evolution of model capabilities over time and in relation to human solution trends. Several follow-up benchmarks extend this paradigm: CodeElo~\cite{codeelo} introduces Elo-style ranking for competition-level problems, while ProBench~\cite{probench}, ICPC-Eval~\cite{icpceval}, OJBench~\cite{ojbench}, LiveCodeBench Pro~\cite{livecodebenchpro}, and HLCE~\cite{hlce} probe increasingly difficult Olympiad- and ICPC-level challenges with human baselines, highlighting the gap between LLMs and expert programmers. In \textbf{other specialized domains}, benchmarks target diverse tasks: Deep-Bench~\cite{deepbench} focuses on code generation for deep learning frameworks; MathQA-X~\cite{mathqax} provides a domain-specific evaluation for mathematical coding problems; FullStackBench~\cite{liu2024fullstackbench} is a multilingual full-stack programming benchmark covering multiple domains and difficulty levels with reference solutions and automated correctness tests; KernelBench~\cite{kernelbench} and TritonBench~\cite{tritonbench} probe low-level GPU kernel and operator generation; YABLoCo~\cite{yabloco} stresses long-context code generation beyond typical function-level inputs; SciCode~\cite{scicode} evaluates scientific coding tasks curated by researchers; CodeInsight~\cite{codeinsight} collects practical coding solutions from Stack Overflow; AutoCodeBench~\cite{autocodebench} and OSS-Bench~\cite{ossbench} propose meta-benchmarking frameworks that automatically construct new evaluation datasets; and retrieval-augmented generation benchmarks such as CodeRAG-Bench~\cite{CodeRAG-Bench} highlight the importance of contextual grounding for realistic evaluation.

\subsubsection{Code Edit and Bug Fix}
Code edit and bug fix are the process of modifying the source code of a software program to correct errors, malfunctions, or unexpected behaviors, known as ``bugs''. This is a fundamental and continuous activity in the software development lifecycle, aimed at improving the stability, reliability, and correctness of an application. \autoref{tab:code_editing_benchmarks} lists most popular bug fixing benchmarks.

\begin{table}[ht]
\centering
\caption{Code Editing and Bug Fixing Benchmarks Overview.}
\label{tab:code_editing_benchmarks}
\small
\setlength{\tabcolsep}{4pt}
\renewcommand{\arraystretch}{1.15}
\begin{tabular}{lccccl}
\toprule
\textbf{Benchmark} & \textbf{Year} & \textbf{Language} & \textbf{Size} & \textbf{Source} & \textbf{Key Feature} \\
\midrule
\multicolumn{6}{c}{\cellcolor{lightblue}\textbf{Statement-Level Bug Fixing}} \\
\midrule
NL2SQL-BUGs~\cite{10.1145/3711896.3737427} & 2025 & SQL & 2K & 
\begin{tabular}[t]{@{}l@{}}
Curated\\ (expert-annotated)
\end{tabular}
  & \begin{tabular}[t]{@{}l@{}}
Semantic error detection\\ for Text-to-SQL\\
(9 main + 31 subcategories)
\end{tabular}
 \\
Megadiff~\cite{megadiff} & 2021 & Java & 663K & VCS Commits & Real-world changes \\
TSSB-3M~\cite{wen2022tssb} & 2022 & Python & 3M & Synthetic & Mutation-based \\
FixJS~\cite{fixjs} & 2022 & JavaScript & 324K & GitHub & Large-scale patches \\
PyTer~\cite{wei2022pyter} & 2022 & Python & 93 & Curated & Type errors \\
RunBugRun~\cite{runbugrun} & 2023 & Multi-PLs & 450K & GitHub & Executable pairs \\
xCodeEval~\cite{ziyao2023xcodeeval} & 2023 & Multi-PLs & 4.7M & Multilingual & Cross-lingual repair \\
DebugBench~\cite{debugbench} & 2024 & C++/Java/Python & 4.2K & Curated & Controlled debugging \\
HumanEvalPack~\cite{humanevalpack} & 2023 & Multi-PLs & 984 & HumanEval & Multilingual debug \\
MdEval~\cite{mdeval} & 2024 & Multi-PLs & 3.5K & Instruction & 18 languages \\
\midrule
\multicolumn{6}{c}{\cellcolor{lightorange}\textbf{Interactive and Feedback-Based Repair}} \\
\midrule
SWT-Bench~\cite{jimenez2024swtbench} & 2024 & Python & 1.9K & GitHub & Test-driven repair \\
FeedbackEval~\cite{dai2025feedbackeval} & 2025 & Python & 394 & Synthetic & Iterative refinement \\
DebugEval~\cite{debugeval} & 2024 & Python & 4253 & Curated & Self-debugging \\
CodeEditorBench~\cite{codeeditorbench} & 2024 & Multi-PLs & 1216 & IDE simulation & Incremental editing \\
\bottomrule
\end{tabular}
\end{table}

\paragraph{Statement-Level Benchmarks}
These benchmarks evaluate LLMs on localized bug fixes, typically at the statement or function level. Early datasets are primarily derived from version control commit histories: Megadiff~\cite{megadiff} aggregates 663K Java changes mined from real-world repositories, while TSSB-3M~\cite{wen2022tssb} provides 3M synthetic single-statement bugs in Python, systematically generated by mutation operators.  
Beyond commit logs, several resources emphasize specific bug types or languages: TFix~\cite{tfix} applies sequence-to-sequence repair to 105K JavaScript fixes paired with unit tests; FixJS~\cite{fixjs} expands this direction with 324K JavaScript patches from GitHub, and PyTer~\cite{wei2022pyter} targets Python type errors, offering controlled benchmarks for static typing issues.  
Later efforts improve ``scale, executability, and multilinguality'': RunBugRun~\cite{runbugrun} contributes 450K executable bug-fix pairs across eight languages (e.g., Java, Python, C++), enabling end-to-end validation. xCodeEval~\cite{ziyao2023xcodeeval} scales to 4.7M multilingual samples (10+ languages) spanning understanding, generation, and repair. DebugBench~\cite{debugbench} provides 4,253 curated debugging tasks in C++, Java, and Python with accompanying tests, emphasizing controlled evaluation. More recent datasets integrate debugging into instruction tuning: HumanEvalPack~\cite{humanevalpack} extends HumanEval to six languages with debugging variants, and MdEval~\cite{mdeval} introduces 3,513 debugging tasks across 18 languages, highlighting multilingual generalization.

\paragraph{Interactive and Feedback-Based Benchmarks}
Recent benchmarks move beyond one-shot fixes toward multi-step, context-sensitive repair with test feedback, self-refinement, or IDE-like interactions. SWT-Bench~\cite{jimenez2024swtbench} collects 1,900 real GitHub bugs with executable test suites, showing how model-generated tests can validate candidate patches. FeedbackEval~\cite{dai2025feedbackeval} systematically studies iterative repair with structured external hints, finding that feedback improves performance but with diminishing returns over multiple refinement cycles. DebugEval~\cite{debugeval} explores self-debugging, where models generate internal diagnostic signals to guide multi-step patching. Finally, CodeEditorBench~\cite{codeeditorbench} simulates IDE-like incremental editing, assessing behaviors such as local modification, error propagation, and interactive change application. Together, these benchmarks emphasize repair as an iterative process requiring state tracking, execution feedback, and realistic editing workflows, moving beyond isolated bug-fix pairs. Debug-gym~\cite{debuggym} develops a gym environment that equips agents with debugger tools (like pdb) to teach LLMs to use stateful tools through SFT/RL, addressing their scarcity in pre-training data.

\subsubsection{Code Efficiency}

Code efficiency task is dedicated to evaluating and optimizing the performance of large language model generated code across multi dimensional efficiency metrics, including runtime, memory usage, algorithmic complexity, and energy consumption, aiming to generate code that is not only functionally correct but also resource-efficient.

\paragraph{Performance and Complexity Benchmarks}
Recent work has moved beyond functional correctness to systematically examine the runtime, memory, and energy efficiency of LLM-generated code. EffiBench~\cite{huang2024effibench} establishes a benchmark over 1,000 algorithmic tasks, showing that GPT-4 solutions can be up to 3$\times$ slower and use 14$\sim$44$\times$ more memory than optimized human baselines. Mercury~\cite{mercury2024} extends this evaluation with a multi-dimensional framework that measures execution time, memory footprint, and best/worst-case complexity across diverse programming problems. EffiBench-X~\cite{effibenchx2025} further generalizes efficiency benchmarking to multiple programming languages, enabling cross-lingual comparisons of computational resource usage. Several benchmarks focus on algorithmic complexity. BigO (Bench)~\cite{bigobench} evaluates whether generated code achieves the correct asymptotic complexity, while DynaCode~\cite{dynacode} introduces input scaling to expose inefficiencies that are not visible under small test cases. COFFE~\cite{coffe} emphasizes practical runtime evaluation under realistic execution loads, providing a complementary system-level perspective. EvalPerf~\cite{liu2024evaluatinglanguagemodelsefficient} introduces a Differential Performance Evaluation framework to systematically assess the efficiency of LLM-generated code on performance-challenging tasks.

\paragraph{Energy Consumption and Efficiency Optimization}
Energy consumption has also become an important evaluation axis. ECCO~\cite{ecco2024} examines whether efficiency improvements can be obtained through code-level transformations without affecting correctness. \citet{energy2025} quantifies the energy overhead of LLM-generated solutions compared to human-written code. \citet{Cappendijk_2025} explores fine-tuning strategies to reduce power consumption, and a study on \textit{StarCoder2}~\cite{starcoder2energy2024} analyzes the impact of quantization on inference cost and downstream code efficiency. Several approaches aim to directly incorporate efficiency into modeling. \textit{EffiCoder}~\cite{efficoder2024} introduces fine-tuning methods that integrate runtime and memory signals into the training objective. \textit{ACECode}~\cite{acecode2024} applies reinforcement learning to jointly optimize correctness and efficiency. \textit{Rethinking Code Refinement}~\cite{refine2024} trains models to detect and improve inefficient code through iterative refinement, offering a post-generation strategy for efficiency enhancement.

\subsubsection{Code Preference}

Code preference is a task that evaluates whether code language models can make preference judgments between different code solutions that align with human developer preferences across dimensions like correctness, efficiency, security, and readability.

\paragraph{Holistic and Composite Scoring Benchmarks}
CodeArena~\cite{du-etal-2025-codearena} calculates a dynamic score from two components. The correctness component is weighted by a problem's overall pass rate, thus giving more credit for solving difficult problems. The efficiency component is determined by a solution's runtime ranking among all correct submissions, with faster solutions scoring higher. The final score is a composite of these two. Moving beyond algorithm-level problems, Long CodeArena \cite{bogomolov2024long} assesses the understanding of LLMs at the project level, requiring a holistic comprehension of an entire codebase rather than a small part of code or one single file, such as library-based code generation, CI build repair, and commit message generation.

\paragraph{Aligning with Human Preferences}
CodePrefBench~\cite{liu2024codefavor} focuses on evaluating whether a code LLM's preferences align with human developers. It tests the judgment of LLMs by presenting them with pairs of code solutions that differ across dimensions like correctness, efficiency, security, and human preference and requiring LLMs to select the superior option. Similarly, other arena-style evaluations employ an LLM-as-a-judge to systematically compare two models. In this setup, the judge selects the better of two model-generated responses, providing a scalable method for assessing which model's output is more aligned with human preferences.
CodeArena (Yang)~\cite{codearena} employs an LLM-as-a-judge to systematically compare two LLMs. The judge selects the better of two LLM-generated responses, providing a scalable method for assessing which LLM's output is more aligned with human preferences (if LLM judgment can reflect the human preference). AutoCodeArena~\cite{zhuo2025bigcodearena} extends this arena-style evaluation paradigm by replacing costly human preference collection with automatic judgments from a strong LLM. Motivated by the high resource demands of BigCodeArena~\cite{zhuo2025bigcodearena}, AutoCodeArena leverages an LLM-as-a-judge to compare each model's output against a fixed baseline system, using the Bradley-Terry~\cite {sun2024rethinking} model to aggregate pairwise comparisons into final preference scores. To approximate real-world usage, the benchmark selects 600 representative prompts sampled across six programming topics, executes model-generated code using a local Docker-based sandbox, and provides execution outputs to the judge model. This enables a scalable and efficient automatic arena for tracking LLM coding capability while maintaining alignment with human preferences.

\subsubsection{Code Reasoning and Question Answering}

Code reasoning and question answering is a task category that evaluates language models' ability to understand, analyze, and reason about code semantics through question-answer formats, deep semantic reasoning challenges, and specialized tasks like code execution prediction, program equivalence checking, and static analysis.

\paragraph{Evaluation Based on Question-Answer Pairs}
Some benchmarks evaluate models through question-and-answer formats. For example, CodeQA~\cite{liu2021codeqa} converts code comments into question-answer pairs through syntactic and semantic analysis, creating a dataset with over 100,000 entries. On the other hand, CS1QA~\cite{lee2022cs1qa} collects question-answer pairs from chat logs of an introductory Python programming course. To more comprehensively evaluate a model's code reasoning ability, CodeMMLU~\cite{manh2024codemmlu} designed a variety of tasks, including code repair, execution inference, and fill-in-the-blank challenges. Similarly, CodeSense~\cite{roy2025codesense} proposes a benchmark covering fine-grained code semantic reasoning tasks in real-world software projects, aiming to address the inadequacy of existing benchmarks that largely rely on synthetic data or educational programming problems.

\paragraph{Deep Code Semantic Reasoning}
There are also benchmarks that measure models' deep code reasoning capabilities through more specific tasks. SpecEval~\cite{ma2025specevalevaluatingcodecomprehension} innovatively uses formal program specifications to evaluate models' code understanding ability, designing four tasks including specification judgment, selection, completion, and generation. CRUXEval~\cite{gu2024cruxeval} and CRUXEval-X~\cite{xu2024cruxeval} design two sub-tasks, output prediction and input prediction, to measure models' code reasoning and understanding capabilities. CRUXEval-X extends the benchmark to 19 different programming languages, providing a more comprehensive evaluation. EquiBench~\cite{wei2025equibench} evaluates models' understanding of program semantics through program equivalence checking, i.e., determining whether two functionally identical but syntactically different pieces of code are equivalent. This task directly tests the model's understanding of code execution semantics. CORE~\cite{xie2025core} evaluates LLMs' code reasoning capabilities through fundamental static analysis tasks such as data dependency, control dependency, and information flow. With the development of multimodal models, new evaluation dimensions have been introduced. ScratchEval~\cite{fu2024scratchevalgpt4osmarterchild} utilizes Scratch, a block-based visual programming language for children, to evaluate the reasoning abilities of large multimodal models in integrating visual information and programming logic.

\subsubsection{Code Translation}

The automated translation of code from one programming language to another is a long-standing challenge, aimed at modernizing legacy systems, improving performance, and unifying disparate codebases. The field has evolved from early structure-aware models to the current paradigm dominated by Large Language Models (LLMs) enhanced with sophisticated verification and reasoning techniques.

\paragraph{Foundational Approaches: From Syntax to Sequence}
Early research focused on capturing the rigid structure of programming languages. Methods like Tree-to-tree Neural Networks~\cite{chen2018treetotreeneuralnetworksprogram} and Grammar-Driven models~\cite{drissi2018programlanguagetranslationusing} treated translation as a transformation between Abstract Syntax Trees (ASTs), ensuring syntactic correctness by design. This structural approach was later advanced by sequence-to-sequence models like TransCoder~\cite{lachaux2020unsupervisedtranslationprogramminglanguages}, which demonstrated the power of unsupervised pre-training on vast monolingual code corpora. These models set the stage for modern LLMs by showing that deep semantic patterns could be learned without requiring parallel, line-by-line translation examples.

\paragraph{Ensuring Correctness through Execution and Feedback}
A key challenge in code translation is that syntactic validity does not guarantee functional equivalence. A translated program might compile but produce incorrect results. To address this, a significant trend has emerged: leveraging program execution as a feedback signal. This test-and-repair paradigm involves using automated unit tests to filter incorrect translations~\cite{roziere2022leveragingautomatedunittests}, employing compiler feedback within a reinforcement learning loop to guide the model toward valid programs~\cite{Jana_2024}, and using dynamic analysis to compare runtime states and pinpoint semantic errors~\cite{xinye2025enhancingllmslongcode}. This approach, which integrates generation with verification, has become critical for producing reliable translations.

\paragraph{Prompt Engineering and Reasoning for Large Language Models}
While powerful, general-purpose LLMs require specialized strategies to excel at the nuanced task of code translation. A primary method is the prompt engineering and reasoning. Techniques like the ``Explain-then-Translate'' method~\cite{tang2023explainthentranslateanalysisimprovingprogram} improve reasoning by forcing the model to first create a natural language summary. Others use intermediate languages and planning algorithms to decompose complex translations into manageable steps~\cite{macedo2024intertrans}. Another major direction is the development of agentic and automated workflows, which move beyond single-shot generation, employing multiple LLM agents that collaborate to repair syntax and semantics~\cite{guo2024transagenttransfervisionlanguagefoundation} or creating closed-loop, self-correcting frameworks that automatically use compiler errors to fix their own output until the code is executable~\cite{dearing2025lassillmbasedautomatedselfcorrecting}.

\paragraph{Specialization: Safety-Critical Translation}
A particularly high-stakes application of code translation is the migration from memory-unsafe languages (e.g., C/C++) to memory-safe languages like Rust. The goal here is not just correctness but also the elimination of entire classes of security vulnerabilities. This has spurred the development of specialized tools that combine LLMs with formal methods and rigorous testing. These frameworks use techniques like differential fuzzing to verify equivalence~\cite{eniser2025translatingrealworldcodellms}, generate reference oracle programs for validation~\cite{yang2024vert}, and leverage static analysis to guide the LLM in producing safer, more idiomatic Rust code~\cite{nitin2025c2saferrusttransformingcprojects,zhou2025llmdrivenmultisteptranslationc}.

\paragraph{Overcoming Data Limitations and Improving Generalization}
The performance of any model is dependent on its training data, and high-quality parallel code corpora are rare. Researchers have developed several techniques to overcome this bottleneck. These include training on aligned code snippets from multiple languages to improve low-resource performance (e.g., MuST~\cite{zhu2022multilingual}), using back-translation to generate synthetic parallel data (e.g., BabelTower~\cite{wen2022babeltower}), and employing federated learning to train models across organizations without sharing private code (e.g., FedCoder~\cite{kumar2025icantsharecode}). These methods are crucial for building models that can generalize across a wide range of languages and domains.

\paragraph{Error Analysis}
As translation models become more capable, evaluating them becomes more complex. Simple metrics like BLEU are often insufficient. New evaluation frameworks propose a multi-level taxonomy of translation complexity, from simple token replacement to complex algorithmic rewriting, providing a more nuanced assessment of model capabilities~\cite{jiao2023evaluationneuralcodetranslation}. Furthermore, significant research has focused on taxonomizing common LLM translation errors~\cite{Pan_2024} and developing lightweight, post-hoc models designed specifically to recognize and rectify these errors, thereby improving the reliability of any underlying translation model~\cite{yin2024rectifier0}.

\subsubsection{Test‑Case Generation}

In \autoref{tab:tcb}, test case generation is the task of automatically creating input-output test cases that can effectively evaluate and distinguish between correct and incorrect code implementations, serving as a critical component for assessing program correctness in both software engineering and competitive programming domains.

\begin{table}[ht]
\centering
\caption{Test case benchmarks overview. \textbf{Solutions per problem (SPP)} indicates the average number of solutions that are used to evaluate generated test cases for each problem in the benchmark.}
\label{tab:tcb} 
\begin{tabular}{lccccc}
\toprule
 \textbf{Benchmark} & \textbf{Year} & \textbf{Source} & \textbf{Language} & \textbf{\#Problems} & \textbf{SPP} \\
\midrule
\multicolumn{6}{c}{\cellcolor{green!15} \textbf{Software Engineer}} \\
\midrule
SWT-Bench~\cite{jimenez2024swtbench} & 2024 & SWE-Bench~\cite{swebench} & Python & 1900+ & 1 \\
\midrule
TestGenEval~\cite{jaintestgeneval} & 2024 & SWE-Bench~\cite{swebench} & Python & 1210 & 1 \\
\midrule
TestBench~\cite{zhang_testbench_2024} & 2024 & Github & Java & 108 & 1 \\
\midrule
CLOVER~\cite{xu_clover_2025} & 2025 & Github & Python & 845 & 1 \\
\midrule
\multicolumn{6}{c}{\cellcolor{blue!15} \textbf{Algorithm Competition}} \\
\midrule
TestEval~\cite{wang_testeval_2025} & 2025 & Leetcode & Python & 210 & 1 \\
\midrule
CodeForce-SAGA~\cite{ma2025rethinkingverificationllmcode}& 2025 & \mcell{Atcoder \\ Codeforces \\ Nowcoder} & Python & 1840 & 36.66 \\
\midrule
TestCase-Eval~\cite{yang_can_2025} & 2025& Codeforces & \makecell{C++ \\ Python \\ Java} & 500 & 200 \\
\midrule
TCGBench~\cite{cao_can_2025}& 2025 & \makecell{NOIP \\ Luogu} & C++ & 208 & 5 \\
\bottomrule
\end{tabular}
\end{table}

\paragraph{Test Case Generation for Software Engineering}
The correctness of code is often evaluated through test cases, making the generation of these test cases a new, core, and critical problem. 
Influenced by traditional software testing, early evaluations of test cases primarily focused on engineering practicality, such as code coverage and the ability to distinguish between historical (buggy) and current (correct) code. SWT-Bench~\cite{mundler_swt-bench_2025} and TestGenEval~\cite{jaintestgeneval} transform tasks from SWE-Bench, providing buggy code and its corresponding fix patch, requiring test cases to fail on the buggy code while passing on the fixed version. As these benchmarks are specific to Python, TestBench~\cite{zhang_testbench_2024} extends this approach to Java. CLOVER~\cite{xu_clover_2025} utilizes the data from GitHub and supplements the benchmark to address the lack of long-form evaluation.

\paragraph{Test Case Generation for Competitive Programming}
Beyond the field of software engineering, there is also a significant demand for test case generation in competitive programming. This is due to the inaccessibility of private test cases, necessitating the generation of a sufficient number of test cases to determine the correctness of a generated solution. TestEval~\cite{wang_testeval_2025} collects 210 problems from LeetCode; however, its evaluation is still limited to coverage metrics. 
In contrast, CodeForce-SAGA, TCGBench, TestCase-Eval and TCGBench~\cite{ma2025rethinkingverificationllmcode, yang_can_2025, cao_can_2025} collects a large number of wrong and correct code submissions to conduct an end-to-end evaluation of the proportion of test cases that could reject the wrong code while passing the correct code.
This progression highlights the critical shift from coverage-based metrics to more robust, functionality-driven benchmarks that assess the ability of test cases to precisely discriminate between correct and incorrect program behaviors.

\subsection{Repository-Level Tasks}\label{sec:Repo-Level}

This section presents repository-level code benchmarks, using various sources like repository-based commit messages, issues, and PRs to evaluate the performance of multiple large language models (LLMs). We will introduce their contributions and methods.

\begin{figure}[h]
    \centering
    \includegraphics[width=0.65\textwidth]{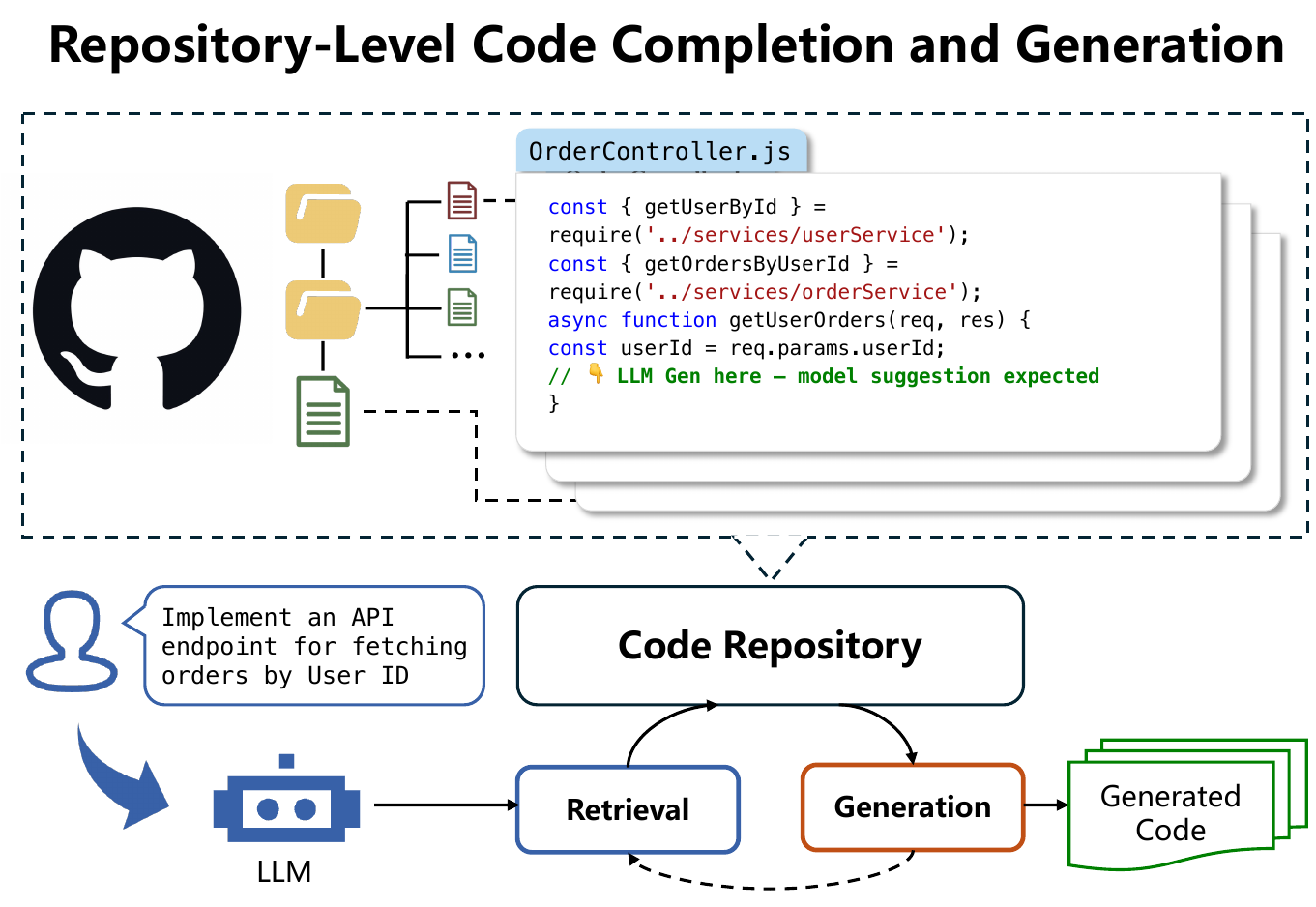} 
    \caption{Illustration of the repository-based code generation and completion task.}
    \label{fig:code_generation_and_completion_task}
\end{figure}

\subsubsection{Code Generation and Completion}
In \autoref{fig:code_generation_and_completion_task}, repository-level code completion and generation refers to AI-powered techniques that leverage the entire codebase context (including multiple files, project structure, dependencies, and cross-file relationships) to predict, complete, or generate code segments that are contextually aware of the broader software repository rather than just the immediate file or function. \textbf{RepoBench}~\cite{liu2023repobench} is a repository-level code completion benchmark using pre-2022 GitHub repos for training and post-2023 for testing to ensure temporal validity, covering Python and Java. It evaluates models (Codex, StarCoder variants) on metrics and tasks (Retrieval, Completion), showing better Python performance. \textbf{RepoEval}~\cite{zhang2023repocoderrepositorylevelcodecompletion} evaluates code completion from 14 real repositories using unit tests; its RepoCoder agent, leveraging full repository context, outperforms others in 90\% of tests. \textbf{Execrepobench}~\cite{yang2024execrepobenchmultilevelexecutablecode} assesses repository-based completion of LLM by regenerating multi-granularity masked spans (expressions, statements, functions) and validating via repository test files and fine-tune base Qwen2.5-Coder on it to enhance completion performance. \textbf{CoderEval}~\cite{yu2024codereval} is a practical benchmark with 460 Java/Python problems from open-source repositories, focusing on non-standalone functions; models perform significantly worse on these than standalone ones. \textbf{CrossCodeEval}~\cite{ding2023crosscodeevaldiversemultilingualbenchmark} is a multilingual benchmark (10k examples) testing cross-file context needs via static analysis. \textbf{M2rc-Eval}~\cite{liu2024m2rcevalmassivelymultilingualrepositorylevel} is a large-scale multilingual repository-level benchmark with AST-based annotations. \textbf{Codev-Bench}~\cite{pan2024codevbenchllmsunderstanddevelopercentric} uses industrial data to evaluate repository-level completion; specialized code LLMs outperform general ones, but all struggle with incomplete suffix scenarios. \textbf{RepoCod}~\cite{liang2025languagemodelsreplaceprogrammers} is a Python benchmark (980 tasks, 50\%+ needing repo context) from 11 projects. \textbf{DI-Bench}~\cite{zhang2025dibenchbenchmarkinglargelanguage} evaluates dependency reasoning across 4 languages (581 testable repos); even top models achieve low pass rates. \textbf{DependEval}~\cite{du2025dependevalbenchmarkingllmsrepository} hierarchically assesses repository-level dependency understanding across 8 languages (15,576 repos); significant performance gaps exist, with advanced models struggling. \textbf{REPOST}~\cite{xie2025repostscalablerepositorylevelcoding} builds repository-level environments via sandbox testing; models trained on its REPOST-TRAIN dataset show modest gains on HumanEval/RepoEval, but perform poorly on REPOST-EVAL. \textbf{SecRepoBench}~\cite{dilgren2025secrepobenchbenchmarkingllmssecure} evaluates secure code generation at the repository level (318 tasks, 15 CWE types in C/C++). \textbf{DevEval}~\cite{li2024devevalmanuallyannotatedcodegeneration} is a manually annotated benchmark (1,874 samples, 117 repos); current LLMs perform poorly, confirming its difficulty.

\begin{figure}[h]
    \centering
    \includegraphics[width=0.65\textwidth]{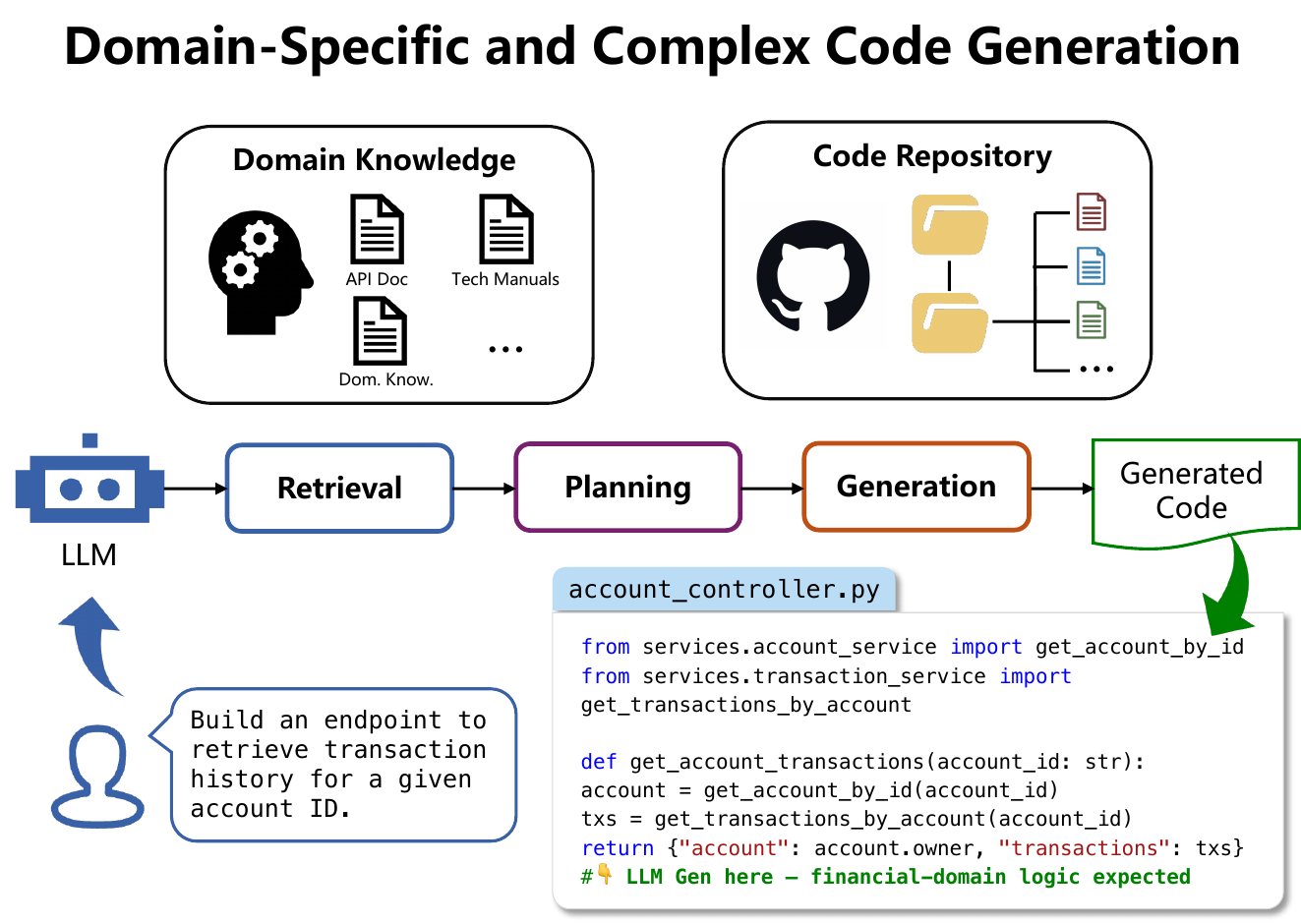} 
    \caption{Illustration of the domain-specific and complex code generation task.}
    \label{fig:domain_code_generation_task}
\end{figure}

\subsubsection{Domain-Specific and Complex Code Generation}
In \autoref{fig:domain_code_generation_task}, the domain-specific and complex code generation task refers to the automated creation of source code that requires specialized knowledge in a particular field and involves intricate logic, multiple dependencies, or sophisticated algorithmic implementations. \textbf{BioCoder}~\cite{tang2024biocoderbenchmarkbioinformaticscode} is a code generation benchmark for bioinformatics. It contains 2,269 high-quality coding problems from 1,700 bioinformatics repositories. 
\textbf{PaperBench}~\cite{starace2025paperbenchevaluatingaisability} uses the reproduction of 20 ICML 2024 papers as its benchmark standard, encompassing understanding core contributions, implementing code, and conducting experiments. A rubric tree decomposes the overall task into subtasks; every leaf node is an independently scored unit that cannot be split further. PaperBench contains 8,316 individually gradable tasks. Rubrics are co-developed with the authors of each ICML paper for accuracy and realism. To enable scalable evaluation, PaperBench also develops an LLM-based judge to automatically grade replication attempts against rubrics and assess our judge's performance by creating a separate benchmark for judges. PaperBench evaluates several frontier models in experiments.
\textbf{Commit0}~\cite{zhao2024commit0} is a benchmark that tests AI agents' ability to write software libraries from scratch. It includes 54 Python libraries where LLMs are given specification documents and empty function bodies to complete. The task is to implement the functions and pass all unit tests. The results show that current agents can pass some unit tests but cannot reproduce entire libraries. Interactive feedback is very helpful for models to generate code that passes more tests.
\textbf{HackerRank-ASTRA}~\cite{xing2025hackerrankastraevaluatingcorrectness} is a benchmark that evaluates the correctness of LLMs on cross-domain multi-file project problems. This benchmark assesses LLMs by creating multi-file project problems, with each problem run 32 times for evaluation. 
\textbf{ProjectEval}~\cite{liu2025projectevalbenchmarkprogrammingagents} is a new benchmark that uses simulated user interaction to test how well LLM agents can generate code at the repository level. It is built with LLMs and human review, has 284 test cases, and offers three levels of input, including \textbf{Level 1} natural language prompt (NL
Prompt), \textbf{Level 2} Natural Language Checklist (NL Checklist), and \textbf{Level 3} Skeleton. The results show that current agents perform poorly, revealing that creating systematic project code and understanding the whole project are key challenges for LLM agents.
\textbf{DA-Code}~\cite{huang2024dacodeagentdatascience} is a code generation benchmark designed to evaluate LLMs on agent-based data science tasks. It includes 500 complex data science tasks covering various aspects and a developed DA-Agent baseline, which shows that even advanced LLMs perform poorly on DA-Code, indicating that current models still face significant challenges with complex data science tasks.

\begin{figure}[h]
    \centering
    \includegraphics[width=0.65\textwidth]{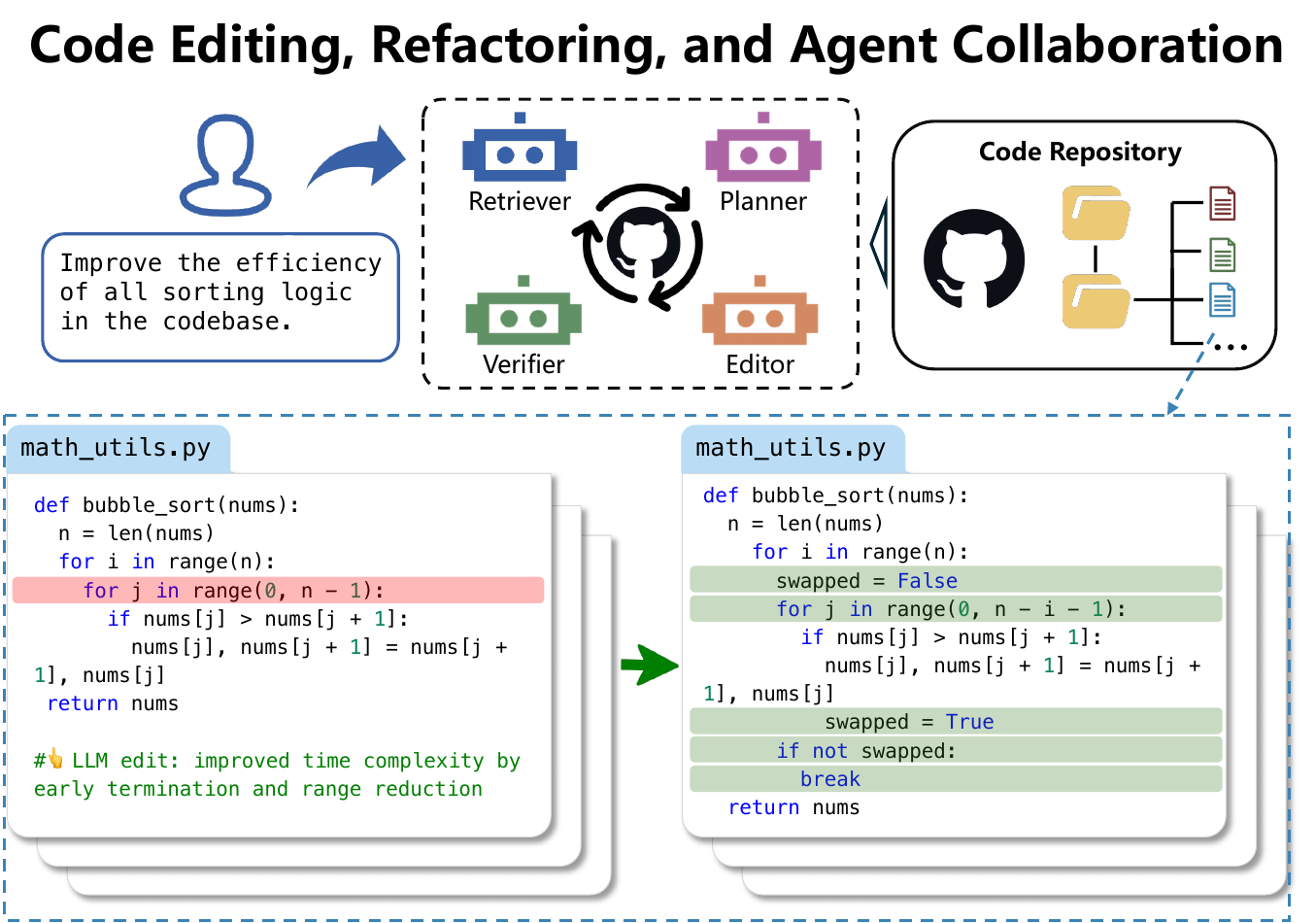} 
    \caption{Illustration of the code editing, refactoring, and agent collaboration task.}
    \label{fig:code_editing_task}
\end{figure}

\subsubsection{Code Editing, Refactoring, and Agent Collaboration}
In \autoref{fig:code_editing_task}, code editing, refactoring, and agent collaboration tasks are a type of tasks where code is modified, restructured for improvement, and multiple AI agents work together to complete programming objectives. \textbf{Aider’s code editing benchmark}~\cite{aider-code-edit} asks the LLM to edit Python source files to complete 133 small coding exercises from Exercism\footnote{\url{https://github.com/exercism}}. This measures the LLM’s coding ability and whether it can write new code that integrates into existing code. The model also has to successfully apply all its changes to the source file without human intervention. \textbf{Aider’s refactoring benchmark}~\cite{aider-refactoring-leaderboard} asks the LLM to refactor 89 large methods from large Python classes. This is a more challenging benchmark, which tests the model’s ability to output long chunks of code without skipping sections or making mistakes. It was developed to provoke and measure GPT-4 Turbo’s ``lazy coding'' habit (refers to an LLM tendency to avoid fully writing long, exact code—skipping sections, hand-waving with placeholders or summaries, and introducing mistakes instead of producing a complete, faithful implementation). The refactoring benchmark requires a large context window to work with large source files. Therefore, results are available for fewer models. \textbf{Aider's Polyglot benchmark}~\cite{polyglot-benchmark} is based on Exercism coding exercises like Aider’s original code editing benchmark. The new polyglot benchmark contains coding problems in C++, Go, Java, JavaScript, Python, and Rust. The old benchmark was solely based on Python exercises. It focuses on the most difficult 225 exercises out of the 697 that Exercism provides for those languages. The old benchmark simply included all 133 Python exercises, regardless of difficulty. \textbf{RES-Q}~\cite{labash2024resqevaluatingcodeeditinglarge} is a benchmark based on GitHub commits, containing 100 handcrafted repository-level editing tasks. It is used to evaluate LLMs in software development tasks, focusing on their ability to follow human instructions and then make code changes. The results show that RES-Q can effectively distinguish the performance of different LLMs. \textbf{LiveRepoReflection}~\cite{zhang2025LiveRepoReflection} establishes a rigorous benchmark for evaluating code comprehension and generation capabilities in multi-file repository contexts. LiveRepoReflection contains 6 programming languages to ensure diversity and increase complexity, and demonstrates significantly greater difficulty than Aider Polyglot Benchmark in experimental evaluations. Additionally, it provides the RepoReflection-Instruct dataset and fine-tunes the RepoReflectionCoder based on Qwen2.5-Coder-32B. \textbf{HumanEvo}~\cite{zheng2025humanevoevolutionawarebenchmarkrealistic} is an evolution-aware repository-level code generation benchmark designed to address the issue of overestimated performance in LLMs caused by ignoring project dynamics. It constructs a benchmark of 400 task instances with evolution-aware settings, combining dependency level categorization and automated evaluation tools to compare code generation capabilities across multiple mainstream LLMs. The results show that neglecting project evolution leads to performance overestimation ranging from 10.0\% to 61.1\%, validating the critical importance of dynamic context for realistic evaluation. \textbf{RepoExec}~\cite{hai2025impactscontextsrepositorylevelcode} is a repository-level code generation benchmark. It constructs executable environments to analyse how context affects the quality of generated code. The benchmark designs three context modes and compares 18 LLMs using the pass@k metric and the Dependency Invocation Rate(DIR). The results show that complete dependencies significantly improve model performance, and instruction-tuned models are better at using dependencies than pre-trained models. \textbf{CodePlan}~\cite{bairi2023codeplanrepositorylevelcodingusing} is a framework for repository-level coding tasks. It automates complex tasks that require extensive edits across the entire repository. CodePlan uses incremental dependency analysis and other algorithms to create a multi-step chain of edits. The results show that CodePlan performs better than baseline methods on 2 repository-level tasks. Most repositories pass the validity checks when using CodePlan.

\begin{figure}[h]
    \centering
    \includegraphics[width=0.65\textwidth]{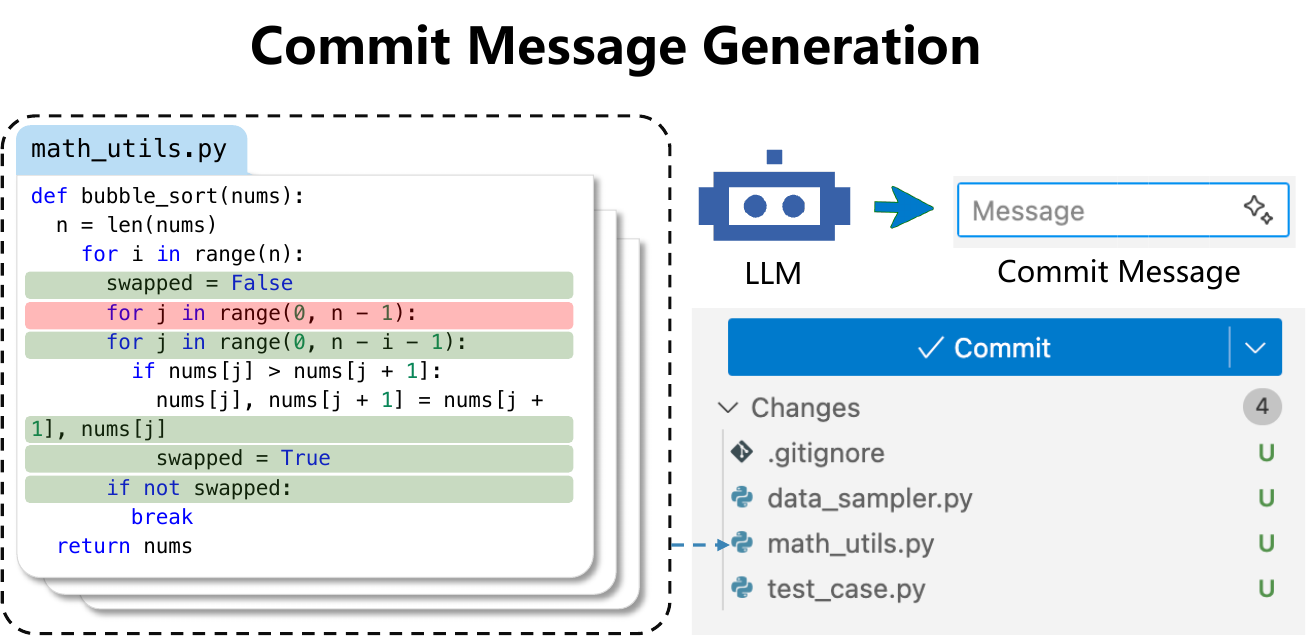} 
    \caption{Illustration of the commit message generation task.}
    \label{fig:commit_message_generation_task}
\end{figure}

\subsubsection{Commit Message Generation}
In \autoref{fig:commit_message_generation_task}, the commit message generation task is the process of automatically creating concise, informative textual descriptions that summarize the changes made in a code commit based on the modified code differences. Commit message generation aims to automatically summarize source code changes into concise natural-language descriptions. Early works treat this as a supervised translation problem from code diffs to text. \citet{jiang2017automaticgenerationshortsummaries} apply naive bayes classification to predict verbs and objects from diffs, but struggles with full-sentence generation due to limited semantic modeling. \citet{loyola2017neuralarchitecturegeneratingnatural} introduce an attention-based encoder–decoder for diff-to-text generation, while \citet{jiang2017automaticallygeneratingcommitmessages} add a verb–direct-object (VDO) structural filter to improve syntactic coherence and BLEU scores.

Later studies focus on representation and retrieval. \citet{liu2018neural} propose \textit{NMTGen}, a nearest-neighbor retrieval model that reuses messages from similar diffs, achieving higher performance and faster inference efficiency than neural baselines. \citet{vanhal2019generatingcommitmessagesgit} revisited NMT methods with stricter preprocessing, showing that performance gains often stemmed from memorization rather than semantic understanding. \citet{xu2019commit} developed \textit{CoDiSum}, which integrates code structure and a copy mechanism for out-of-vocabulary (OOV) words, improving BLEU, METEOR~\cite{banerjee2005meteor}, and Recall. \citet{liu2020atom} propose \textit{ATOM} combines abstract syntax trees (ASTs) with retrieval and CNN-based ranking to boost accuracy.

With the rise of pre-trained models, transformer-based and context-aware approaches became dominant. \citet{jung2021commitbert} introduces \textit{CommitBERT}, jointly encoding added and deleted code with a code-pretrained BERT model, outperforming earlier NMT systems. \citet{wang2021context} presented \textit{CoRec}, a retrieval–generation hybrid mitigating exposure bias and handling rare vocabulary via decay sampling. \citet{wang2023delving}’s \textit{ExGroFi} incorporate linked issue descriptions, enhancing the rationale and conciseness of messages, while \citet{eliseeva2023commitmessagegenerationhistoryaware}’s \textit{CommitChronicle} leveraged commit history to improve temporal and stylistic coherence.

For standardized evaluation, CommitBench~\cite{schall2024commitbench} compiles 1.6M commit–diff pairs from 72K GitHub repositories, enabling large-scale comparison of transformer models such as CodeTrans~\cite{codetrans} and T5~\cite{t5}. Using ROUGE-L~\cite{rouge_l} and BLEU~\cite{papineni2002bleu}, code-pretrained transformers consistently outperforms general-purpose LMs. The MCMD framework~\cite{tao2021evaluationcommitmessagegeneration} proposes the BLEU variant B-Norm, which applies smoothing and case-insensitivity and was shown to correlate more strongly with human judgments than traditional BLEU measurements..

Overall, research has evolved from rule-based and RNN systems to pre-trained, retrieval-augmented transformers that incorporate structure and history. Remaining challenges include aligning messages with developer intent, handling multilingual repositories, and balancing informativeness with brevity.

\subsubsection{Software Engineering Tasks}
Software engineering task is the process of analyzing, implementing, and completing specific software development tasks, including bug fixes, feature implementations, and technical problem-solving, from identification through to verified completion.

The SWE-bench family of benchmarks~\cite{jimenez2024swtbench} has become the foundation for evaluating large language models on real-world software engineering tasks. The original SWE-bench contains 2,294 Python issue–PR pairs, where models must resolve GitHub issues verified via ''fail-to-pass'' unit tests. \textbf{JavaBench}~\cite{cao2024javabenchbenchmarkobjectorientedcode} extends this evaluation to Java, focusing on object-oriented reasoning across four open-source projects, where models consistently lag behind trained human developers. To reduce computational cost, \textbf{SWE-bench Lite} offers a 300-task subset that preserves the overall distribution of the full benchmark. \textbf{SWE-bench Multilingual}~\cite{yang2025swesmithscalingdatasoftware} broadens coverage to nine programming languages (42 repositories, 300 manually verified tasks), addressing the Python-centric limitation of earlier versions. Meanwhile, \textbf{SWE-bench Multimodal}~\cite{yang2025swebench} introduces 517 image-based issues, testing whether models can interpret visual cues such as screenshots or GUI errors during debugging. \textbf{SWE-bench Verified}~\cite{swebenchverified} contributes 500 human-curated examples with Docker-based evaluation, ensuring consistent reproducibility across models and environments. 

Recent extensions aim to enhance temporal and practical realism. \textbf{SWE-bench Live}~\cite{zhang2025swebenchgoeslive} continuously incorporates new GitHub issues (1,565 tasks across 164 repositories), making the benchmark more dynamic and substantially harder than static datasets. Beyond bug fixing, \textbf{SWE-Perf}~\cite{he2025sweperflanguagemodelsoptimize} evaluates performance optimization through 140 repository-level pull requests, revealing that models still fall short of expert programmers in efficiency-related tasks. \textbf{SWE-rebench}~\cite{badertdinov2025swerebenchautomatedpipelinetask} automates large-scale issue collection (>21,000 tasks), enabling longitudinal evaluation of model degradation and temporal generalization. \textbf{SWE-Dev}~\cite{du2025swedevevaluatingtrainingautonomous} shifts the focus from bug fixing to feature implementation, introducing 14K training and 500 evaluation examples—fine-tuned 7B models approach GPT-4-class performance. BugPilot~\cite{bugpilot} generates more realistic synthetic bugs by having LLMs add new features rather than directly inserting bugs, achieving SOTA results on SWE-bench-Verified with Qwen3 models. Gistify~\cite{gistify} requires agents to distill a repository into minimal single-file code that replicates its runtime behavior, testing deeper codebase understanding beyond SWE-bench's localization-focused tasks.
Several newer datasets extend this paradigm in specific directions. \textbf{SWE-PolyBench}~\cite{rashid2025swepolybenchmultilanguagebenchmarkrepository} evaluates multi-language repositories across 21 projects (2,110 tasks), revealing substantial variance in cross-language reasoning ability. \textbf{Multi-SWE-bench}~\cite{zan2025multiswebenchmultilingualbenchmarkissue} similarly covers seven languages (1,632 verified issues), showing strongest results in Python and weakest in C/C++/Rust. \textbf{SWE-bench+}~\cite{aleithan2024swebenchenhancedcodingbenchmark} mitigates data leakage by including post-cutoff repositories, demonstrating significant performance drops compared to pre-2023 datasets. Finally, \textbf{SWE-bench M}~\cite{yang2024swebenchmultimodalaisystems} evaluates visual debugging across 17 JavaScript projects (619 image-grounded issues), where even top-tier AI systems struggle with multimodal reasoning. 

Beyond the SWE-bench family, \textbf{SWE-Lancer}~\cite{SWE-Lancer} reframes evaluation as freelance-style project execution using 1,488 real UpWork tasks (valued at roughly \$1M), probing economic value creation rather than accuracy alone. \textbf{FAUN-Eval}~\cite{hu2024realworldbenchmarkevaluatingfinegrained} benchmarks fine-grained GitHub issue resolution (300 manually curated entries) and finds that open- and closed-source models excel on different categories. \textbf{FEA-Bench}~\cite{li2025fea} measures repository-level feature implementation across 83 repositories, with the best models solving only around 10\% of cases. \textbf{SwingArena}~\cite{xu2025swingarenacompetitiveprogrammingarena} and \textbf{CoreCodeBench}~\cite{fu2025corecodebenchconfigurablemultiscenariorepositorylevel} extend evaluation toward CI-integrated and composite tasks, respectively. Finally, \textbf{AgentIssue-Bench}~\cite{rahardja2025agentsfixagentissues} examines the self-maintenance abilities of software agents (50 tasks derived from 201 issues), revealing persistent challenges in long-term memory and LLM compatibility.

\subsubsection{Comprehensive Software Development}

Recent benchmarks are expanding the scope of evaluation from isolated code snippets to the broader and more complex ecosystem of software development. This trend reflects a growing understanding that a model's value lies not just in writing code, but in its ability to comprehend, document, and interact with the entire development lifecycle.
This shift is evident in benchmarks targeting different facets of this lifecycle. README Eval~\cite{readme-eval} moves beyond code to assess high-level project understanding, tasking models with generating repository documentation from contextual metadata like issues and commits. Pushing into the collaborative process, OmniGIRL~\cite{guo2025omnigirlmultilingualmultimodalbenchmark} evaluates a model's ability to resolve GitHub issues, introducing multilingual and, critically, multimodal challenges by including images within bug reports—a task where current LLMs significantly struggle.
Furthermore, new benchmarks are beginning to evaluate the use of essential developer tools. GitGoodBench~\cite{lindenbauer2025gitgoodbenchnovelbenchmarkevaluating} is the first to test AI agents on their mastery of version control systems, revealing that even sophisticated agents fail at complex but common tasks like resolving merge conflicts. To ensure these evaluations remain relevant, EvoCodeBench~\cite{li2024evocodebenchevolvingcodegeneration} introduces a dynamic paradigm, with tasks derived from evolving, real-world repositories to better reflect the moving target of ongoing software development. Underpinning all these advanced capabilities is the foundational importance of data quality, with projects like Stack-Repo~\cite{kocetkov2022stack3tbpermissively} demonstrating that curating massive, deduplicated source code datasets is critical for improving model performance across all these real-world tasks.

\subsubsection{Repository-Level and Long Context Understanding}
\begin{figure*}[t]
\begin{center}
    \includegraphics[width=1.0\textwidth]{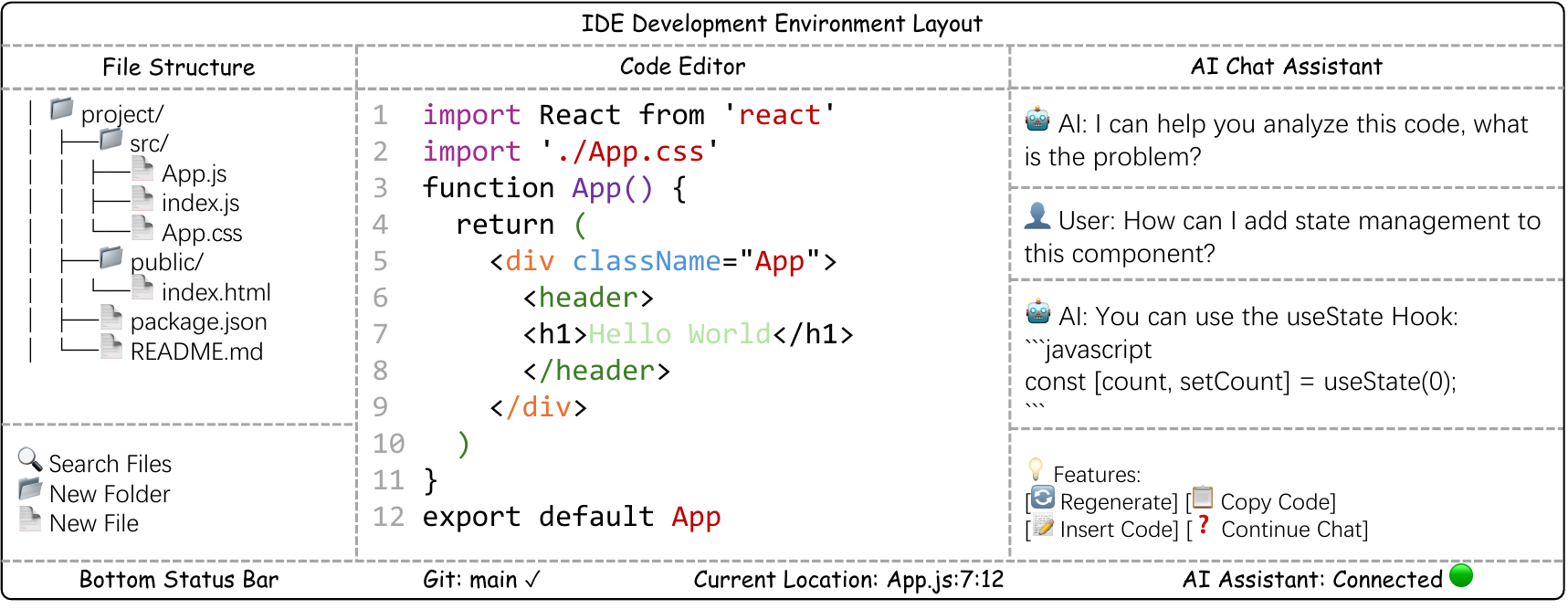}
    \caption{The core task of RepoQA~\cite{liu2024repoqa} is searching needle function (SNF), asking a model to find and reproduce a target function from an entire repository using only a natural language description of what the function does.}
    \label{fig:cursor-repoqa}
\end{center}
\end{figure*}

As shown in \autoref{fig:cursor-repoqa} of IDE, the repository-level and long context understanding refers to the task of comprehending and reasoning across entire codebases or extensive documents that span multiple files and require maintaining context over thousands or millions of tokens

As models' ability to process longer contexts improves, evaluating repository-level code understanding becomes particularly important. Benchmarks such as RepoQA~\cite{liu2024repoqa} focus on the repository-level question-answering task, where a system must answer natural-language queries by grounding its response in the contents of a code repository. The task requires retrieval and reasoning over heterogeneous project artifacts (e.g., source files, README documents, test files, issues/PRs) distributed across multiple files and directories, and producing an answer that is both correct and supported by one or more evidence items from the repository (for example, file paths, code snippets, or documentation excerpts). Formally: given a repository \(R\) composed of artifacts \(A = \{a_{1}, a_{2}, \dots, a_{n}\}\) and a natural-language question \(q\), the goal is to produce an answer \(a\) based on the repository’s contents.

RepoQA~\cite{liu2024repoqa} and CodeRepoQA~\cite{hu2024coderepoqa} both focus on evaluating the long-context code understanding capabilities of large language models. Specifically, RepoQA requires models to find functions based on natural language descriptions, while CodeRepoQA assesses their repository-level question-answering ability in the field of software engineering. Similarly, CoreQA~\cite{chen2025coreqa} also targets code repository-level question-answering tasks, building its dataset by collecting questions and comments from real GitHub repositories to reflect the complexity of real-world software development. Furthermore, LongCodeU~\cite{li2025longcodeu} provides a more comprehensive and challenging set of tasks, evaluating models' long code understanding ability across four key aspects: code unit awareness, intra-code unit understanding, inter-code unit relationship understanding, and long document understanding.

To reduce the context length, LongCodeZip~\cite{longcodezip} proposes a novel, training-free framework designed to efficiently compress long code contexts for large language models (LLMs). Traditional context pruning and retrieval-based methods struggle with the structural and semantic complexity of code. To address this, LongCodeZip introduces a two-stage hierarchical compression strategy: (1) coarse-grained compression that ranks and selects function-level chunks based on conditional perplexity (approximated mutual information with respect to the instruction), and (2) fine-grained compression that segments retained functions into blocks via perplexity-based boundary detection, followed by adaptive token-budget optimization using a knapsack algorithm. Experiments across code completion, summarization, and question answering tasks demonstrate that LongCodeZip achieves up to 5.6× compression without degrading performance, outperforming baselines. It also generalizes well across models (from 0.5B to 8B parameters), significantly reduces API costs and latency, and represents the first dedicated framework for long-context code compression in LLMs.

\subsection{Agentic Systems}
\label{sec:Agentic systems}

\subsubsection{Agent Tool Use}

A fundamental capability of any agent is its ability to interact with the digital world through tools like APIs and functions. Early benchmarks like API-Bank~\cite{li2023api} and ToolBench~\cite{qin2023toolllm} established the foundation for this evaluation, testing an agent's ability to correctly select and invoke the right tool for a given task. However, the complexity of these evaluations is quickly advancing. BFCL~\cite{patil2025bfcl} exemplifies this trend with its iterative versions, progressing from single function calls to complex, multi-turn, and multi-step scenarios that better mimic real-world interactions. Furthermore, benchmarks are moving from generic tool-use tasks to domain-specific applications, with frameworks like Tau Bench~\cite{yao2024tau} assessing agent performance in realistic business workflows like customer service.

\subsubsection{Deep Research Benchmarks}

The true measure of an advanced agent lies in its ability to go beyond simple information retrieval and perform deep, human-like reasoning. Benchmarks in this category are designed to be trivially easy for humans but expose profound limitations in current AI systems. GAIA~\cite{mialon2023gaia} pioneered this approach with real-world questions that require a combination of reasoning, web browsing, and tool use, revealing a stark performance gap where leading models fail on tasks humans solve with over 90\% accuracy. This challenge is being pushed further into specialized, high-stakes domains. xbench~\cite{chen2025xbench} evaluates agents on professional workflows like talent sourcing and marketing, while DeepResearch Bench~\cite{du2025deepresearch} raises the bar to an academic standard, requiring agents to synthesize analyst-grade reports on PhD-level topics, testing the limits of autonomous research and synthesis.

\subsubsection{Web Search Benchmarks}
The open web presents perhaps the most unconstrained and unforgiving environment for autonomous agents. 
Unlike closed benchmarks with predefined states, real-world web search requires persistence, adaptability, and the ability to filter noisy or contradictory information. 
Recent benchmarks have begun to capture this challenge more faithfully. 
For example, BrowseComp~\cite{wei2025browsecomp} frames information retrieval as a compositional reasoning task, requiring agents to aggregate fragmented evidence distributed across multiple websites. 
Its extensions, including BrowseComp-ZH~\cite{chen2025browsecomp} for multilingual search and BrowseComp-Plus~\cite{zhou2025browsecomp}, which decouples reasoning from retrieval, push evaluation toward more realistic and interpretable agent behavior. 
Complementary efforts such as WebWalkerQA~\cite{wu2025webwalker} emphasize systematic exploration and question answering across hyperlinked environments, while Widesearch~\cite{wong2025widesearch} focuses on large-scale, open-domain information synthesis. 
The results reported so far remain sobering: even top-performing models achieve near-zero success on end-to-end tasks, underscoring that robust web-scale reasoning remains a major unsolved problem.

\subsubsection{Benchmarking Agents for Graphical User Interfaces}
The development of autonomous agents capable of understanding and interacting with graphical user interfaces (GUIs)—across web, desktop, and mobile platforms—represents a significant milestone toward general-purpose artificial intelligence. Early work in this area typically focused on narrow tasks, but the integration of visual and linguistic reasoning in Large Language Models (LLMs) has enabled agents with far greater adaptability. Nonetheless, grounding natural language commands into concrete UI actions, maintaining coherent multi-step plans, and generalizing across heterogeneous interfaces remain open challenges. These difficulties have motivated the creation of increasingly sophisticated benchmarks designed to systematically evaluate and advance GUI-agent capabilities. Broadly, these benchmarks target two complementary dimensions of capability: interface navigation (acting as a user) and interface development (acting as a creator).

\paragraph{Benchmarks for Frontend Navigation}
A core challenge for GUI agents is navigating complex, interactive environments. Foundational benchmarks such as WebShop~\cite{yao2022webshop} and Mind2Web~\cite{deng2023mind2web} established this paradigm by providing thousands of goal-oriented tasks that require agents to operate within real or simulated websites. These evaluations revealed that even powerful models struggle with long-horizon reasoning and precise UI grounding. Building on this foundation, recent benchmarks have expanded along several directions.  
\textbf{Cross-Platform Generalization:} OmniACT~\cite{kapoor2024omniactdatasetbenchmarkenabling} introduces unified benchmarks spanning web, desktop, and mobile settings, exposing limitations in cross-domain generalization.  
\textbf{Higher-Level Cognitive Reasoning:} WebChoreArena~\cite{miyai2025webchorearena} challenges agents with compound “chore” tasks requiring long-term memory and computation, while PersonalWAB~\cite{cai2025personalwab} introduces personalization through user histories and preferences.  
\textbf{Fine-Grained Capability Diagnosis:} Benchmarks such as Sphinx~\cite{shi2024sphinx} and NovelScreenSpot~\cite{fan2025gui-bee} decompose GUI interaction into subskills, such as goal understanding, UI grounding, and planning.

\paragraph{Benchmarks for Frontend Development}
The second major direction focuses on automating GUI creation, translating human intent into executable frontend code. This research area has evolved from single-page generation to complex multi-file workflows that simulate real development pipelines. Early efforts such as Design2Code~\cite{si2024design2code} and WebCode2M~\cite{gui2025webcode2m} provide large-scale paired datasets linking design inputs with corresponding code implementations. Increasingly sophisticated tasks have since emerged: Sketch2Code~\cite{li2024sketch2code} examines generation from informal sketches, while Interaction2Code~\cite{xiao2024interaction2code} evaluates dynamic, interactive webpage generation.  
Recent benchmarks reflect a broader ambition to develop autonomous agents capable of full-stack web creation. WebGen-Bench~\cite{lu2025webgen} requires the generation of entire multi-file websites from scratch, while Web-Bench~\cite{xu2025web} models realistic software engineering workflows with sequentially dependent tasks, shifting evaluation from single-turn code generation toward continuous, project-level reasoning.

\subsubsection{Terminal Use}
Terminal-Bench \cite{tbench_2025} is an emerging benchmark that evaluates a code agent's ability to autonomously operate in a terminal environment on real tasks. 
It transcends the traditional paradigm of code generation or fixing within a defined environment, requiring the code agent to have system-level development capabilities. 
This includes being able to explore the system and execute shell commands and various command-line tools to complete complex, system-level tasks, such as compiling and booting a complete Linux kernel from source, or deploying a functional server from scratch.
Unlike SWE-bench~\cite{swebench}, which has clear problem boundaries (a repo/folder), in Terminal-Bench, the entire environment is the problem space to be solved. The agent's objective is to deliver a successfully running or configured system by performing system configuration, dependency management, and executing a complete workflow.

\section{Alignment}
Alignment in code LLMs refers to the process of adapting pre-trained LLMs to follow human instructions and perform coding tasks effectively. \autoref{subsec:sft} introduce the progress about supervised fine-tuning (SFT) for code LLMs, which learn from labeled instruction-following datasets covering tasks like code generation, repair, and translation. \autoref{sec:code_rl}
and Reinforcement Learning (RL), which uses reward signals to further refine model behavior. A particularly powerful variant is Reinforcement Learning with Verifiable Rewards (RLVR), where models receive deterministic pass/fail feedback from test cases or compilers, enabling them to develop structured reasoning, self-verification, and error-correction capabilities. Together, these alignment methods transform general pre-trained models into specialized coding assistants capable of understanding requirements, generating correct solutions, and handling complex real-world software development scenarios. 

\subsection{Supervised Fine-tuning (SFT)}
\label{subsec:sft}
Supervised fine-tuning (SFT) is the process of training a pre-trained language model on a labeled dataset. A common and powerful way to format this data is through instruction tuning, where the model learns to follow natural language commands. This method not only enables the model with the ability to understand and execute instructions but also significantly enhances its performance on targeted tasks~\citep{zhang2024instructiontuninglargelanguage,xcot}. In the training of code LLMs, SFT is a key strategy for improving model performance.

The scope of code instruction tuning data is broad, covering a diverse range of programming-related tasks, including but not limited to code generation, code repair, and code translation. A multi-task dataset design aims to cultivate comprehensive capabilities in LLMs, allowing them to respond flexibly to human instructions and perform various code-related tasks~\citep{MFTCoder}. The emergence of high-quality code instruction datasets has significantly enhanced the generality and adaptability of code LLMs, bringing them closer to meeting the demands of real-world scenarios.
Early code instruction tuning data were extracted from various developer communities and platforms~\citep{muennighoffOctoPackInstructionTuning2024}, such as code-comment pairs from GitHub repositories~\citep{husain2020CodeSearchNet} and user question-answer data from StackExchange~\citep{codecontests}. Data of this type are written by humans and conform to real-world data distributions; therefore, fine-tuning techniques based on such data are also known as {Natural-Instruct}~\citep{mishra2022CrossTask}. However, because each data point is provided by different users from diverse sources, the quality is often inconsistent and difficult to filter. Moreover, the original data were not specifically created for instruction tuning and may not conform to the required format. For instance, code in a repository may not have a corresponding natural language comment. These factors limit the quality and extraction efficiency of Natural-Instruct data.

\begin{figure}[H]
    \centering
    \includegraphics[width=0.8\textwidth]{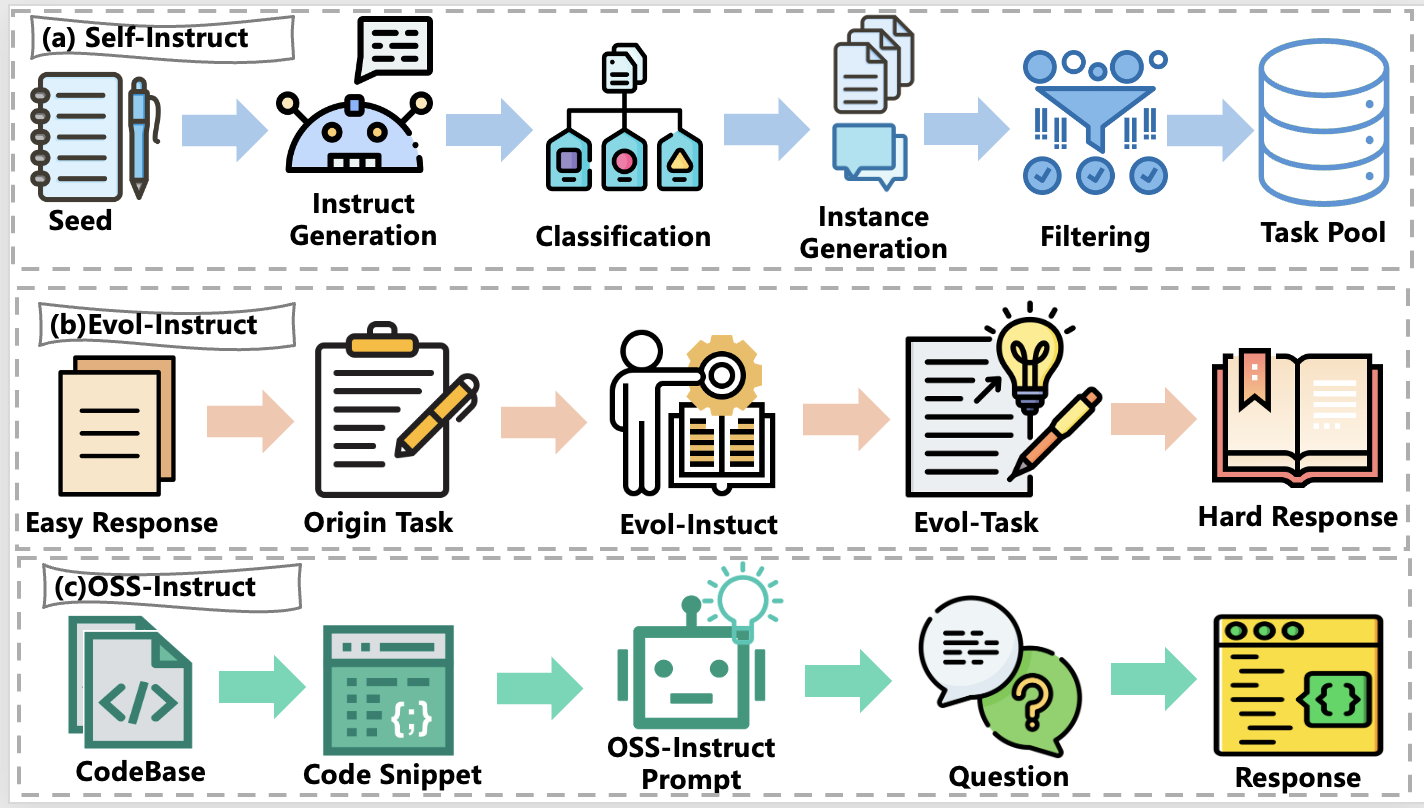}
    \vspace{-2mm}
    \caption{Three typical methods for synthesizing code alignment data.}
    \label{fig:sft_synthesis}
\end{figure}

In contrast, self-instruct~\cite{luo2023wizardcoder}, an approach to enhance the instruction-following abilities of pretrained language models by iteratively learning from their own generated outputs, for leveraging more powerful LLMs to generate higher-quality and more standardized data through high-quality demonstrations and in-context learning. The Alpaca dataset~\citep{alpaca} in the field of LLMs is a typical example based on self-instruct technology. Following this work, Code-Alpaca~\citep{codealpaca} replaced the general-purpose seed examples with code-related ones, thereby constructing the first Self-Instruct-based dataset in the code domain. Inspired by this, subsequent works have emerged, optimizing the self-Instruct method for code-related tasks. \autoref{fig:sft_synthesis} summarzies the three typical methods for synthesizing code alignment data.

\subsubsection{Single-Turn Supervised Fine-tuning}
\paragraph{Complexity} The code in the CodeAlpaca dataset mostly consists of basic operations such as object creation and arithmetic, lacking advanced algorithms and complex logic. To address this, \citet{luo2023wizardcoder} proposed the Evol-Instruct method, which guides the model to generate more complex augmented data based on the original data using a series of manually crafted heuristic rules. The augmentation process can be iterated for multiple rounds to continuously increase data complexity.

\paragraph{Diversity} As the volume of generated data increases, the likelihood of the Self-Instruct method producing repetitive or similar data also increases, limiting the quality of the dataset. An effective solution is to introduce human-written code data from Natural-Instruct during the generation process, leveraging its rich diversity to avoid repetition. To this end, \citet{luoSemiInstructBridgingNaturalInstruct2024} proposed the Semi-Instruct method, which uses diverse data from Natural-Instruct as a foundation and employs the Self-Instruct method to rewrite it, thereby enhancing data standardization. Similarly, the OSS-Instruct method~\citep{weiMagicoderEmpoweringCode2024} adopts a similar construction process, with the difference that the raw data consists of code snippets rather than complete code, providing greater flexibility in the subsequent rewriting phase. Furthermore, \citet{yu2024WaveCoderWidespreadVersatile} constructed the Code Ocean dataset by first selecting diverse representative data through heuristic rules and embedding-based semantic similarity, and then using a CoT-like approach to verify the correctness of the selected data with a large model, thus obtaining correct and diverse results. \citet{wu2024inversecoder} explored using the inverse generation capability of the model to be fine-tuned, without relying on a larger model. It first cleans the outputs from Evol-Instruct, then uses the model itself to generate multiple instructions, and finally, the model itself determines whether the generated instructions are correct.

\paragraph{Sensitivity} \citet{luo2025successdetailsevaluateenhance} propose a new sensitive dimension, which means the ability to capture detailed changes in instructions. It uses the CTF-Instruct framework, which augments counterfactuals to improve sensitivity. Such as changing use bullet points touse numbered lists or tweaking a length constraint. To measure this, they generate minimal‐edit instruction pairs and check whether the model’s outputs differ in exactly the prescribed way, using both automated metrics (e.g., edit‐distance, style classifiers) and targeted human judgments.

\subsubsection{Multi-Turn Supervised Fine-tuning}

\paragraph{Execution Feedback} A key feature that distinguishes code from natural language is its executability. Therefore, generated code can be verified using compilers and interpreters, providing low-cost and high-efficiency feedback signals without human intervention to guide improvements after errors occur. 

\paragraph{Multi Agent}
AIEV-Instruct (Instruction Tuning with Agent-Interaction and Execution-Verified)~\cite{leiAutoCoderEnhancingCode2024} sets up a questioner agent and a programmer agent based on the same large model. When the code generated by the programmer agent fails verification, the questioner agent intervenes, asking questions based on the error to facilitate improvement. Data constructed with feedback are inherently multi-turn, whereas users in practical applications generally prefer single-turn interactions whenever possible. To bridge this gap between training and inference, \citet{renReflectionCoderLearningReflection2024} proposed a self-distillation method. It adds a summary-like turn at the end of the multi-turn data, which directly generates the multi-turn optimized code based on the user's original question. The training process adopts a method similar to scheduled sampling~\citep{bengio2015scheduled}, masking all but the last turn with a certain probability, and gradually increasing the masking probability as training progresses. This allows the model to fully learn from the knowledge in multi-turn interactions during training while ultimately adapting to the single-turn generation requirement at the inference stage. \citet{yuan2024advancing} introduce multi-turn interaction from the verification level to the problem-solving level using the feedback at each step.

\subsubsection{SFT for Repository Tasks}

While traditional code instruction tuning has focused on isolated functions or file-level tasks, the complexity of real-world software development demands models that can understand and navigate entire repository structures. Repository-level SFT datasets have emerged as a critical resource for training code LLMs to handle cross-file dependencies, multi-file edits, and long-context reasoning that characterize practical software engineering scenarios.

\paragraph{Software Engineering Task Datasets}
The SWE family of datasets has become the cornerstone for training autonomous coding agents. SWE-smith~\citep{yang2025swesmith} introduces a scalable pipeline that generates a large collection of task instances from GitHub repositories, representing an order of magnitude increase over previous works. SWE-Synth~\citep{pham2025swesynth} complements this by synthesizing verifiable bug-fix data through LLM-driven debugging workflows, producing not only bug-fix pairs but also test cases and structured repair trajectories. SWE-Gym~\citep{pan2024swegym} provides Python task instances with executable runtime environments, enabling reinforcement learning approaches. SWE-Dev~\citep{du2025swedev} specifically targets feature-driven development, addressing a task type that comprises a significant portion of real-world development efforts. Skywork-SWE~\citep{zeng2025skyworkswe} tries to demonstrates clear data scaling laws across thousands of task instances from multiple repositories.

\paragraph{Offensive Cybersecurity Task Datasets}
On the offensive cybersecurity side, recent work has introduced CTF-style benchmarks that target vulnerability discovery and exploitation. Cyber-Zero~\citep{zhuo2025cyber} proposes a runtime-free framework that synthesizes high-quality agent trajectories from public CTF writeups, using persona-driven LLM agents to reverse-engineer plausible environment behaviors and generate long-horizon interaction sequences capturing both successful exploits and realistic failed attempts. The resulting corpus spans thousands of challenges across diverse CTF categories and enables open-weight models to approach the performance of frontier proprietary systems on standard CTF benchmarks. Complementing this trajectory-focused perspective, CTF-Dojo~\citep{zhuo2025training} provides a large-scale, execution-ready environment containing hundreds of fully functional CTF challenges packaged in secure Docker containers. Built on top of pwn.college artifacts via the CTF-FORGE pipeline, which automatically constructs and validates runtime environments with over 98\% success rate, CTF-Dojo supports systematic trajectory collection from multiple LLMs and reveals key factors for building effective cybersecurity agents, including the importance of writeup-guided interaction, runtime environment augmentation, and teacher-model diversity.

\paragraph{Code Completion and Repository Navigation}
Several datasets focus on repository-level code completion and editing capabilities. RepoBench~\citep{liu2023repobench} evaluates auto-completion through three interconnected tasks across 10,345 Python and 14,956 Java repositories, specifically measuring cross-file context understanding. CoEdPilot~\citep{liu2024coedpilot} addresses incremental code edits, collecting over 180,000 commits from 471 projects across 5 programming languages to evaluate edit location prediction. RepoST~\citep{xie2025repost} introduces sandbox testing to isolate target functions for execution, containing 7,415 functions from 832 repositories. For hardware design, RTL-Repo~\citep{allam2024rtlrepo} extends repository-level training to Verilog, a hardware description language (HDL)~\cite{yang2025large}, with over 4,000 samples including full repository context ranging from 2K to 128K tokens.

Repository-level SFT data presents unique challenges compared to traditional code datasets. The computational cost of maintaining execution environments is substantial, with some datasets requiring terabytes of storage for Docker images. Ensuring data quality while scaling remains difficult, as automated generation methods may introduce subtle errors that are hard to detect without comprehensive testing. Additionally, the diversity of repository structures, dependency management systems, and coding conventions across projects complicates unified training frameworks. 

Despite these challenges, the consistent improvements across benchmarks and the clear data scaling laws indicate that investing in larger, more diverse repository-level datasets is a promising path for advancing autonomous programming systems.

\subsubsection{Reasoning-based Methods}
\paragraph{Paradigm Shift Towards Reasoning}
Reasoning-based LLMs~\cite{deepseekai2025deepseekr1} represent a significant advancement over traditional instruction-tuned models by incorporating explicit chain-of-thought (CoT) processes during inference, enabling them to decompose complex problems into intermediate steps and perform more systematic analysis. While instruction-tuned models are trained to directly map inputs to outputs through supervised fine-tuning on task-specific demonstrations, reasoning models employ techniques such as reinforcement learning from human feedback and process supervision to develop metacognitive capabilities that allow them to verify their own logic, explore multiple solution pathways, and self-correct errors before producing final answers. Specifically, CoT reasoning prompts models to generate a step-by-step thinking process, transforming complex requests into sequences of simpler, manageable steps that allow for more computational focus on intricate problems~\citep{Wei}.
This approach has proven highly effective in code generation, enabling models to build upon their own intermediate conclusions and verify logical consistency at each step~\citep{jiang2025makes}. By making the reasoning process explicit, CoT provides a more robust framework for tackling complex coding challenges in real-world software development (e.g., repository-based tasks)~\citep{zhang2025unveiling} that require deep semantic understanding rather than just syntactic fluency.

\paragraph{Mechanism Interpretation}
The mechanism behind CoT's success is tied to how it structures the problem-solving process for the model. Generating an explicit reasoning chain serves as a computational ``scratchpad'', allowing the model to offload intermediate steps and conclusions into its context window~\citep{Wei}. This process helps the model organize relevant information and focus its attention, which is crucial for complex tasks. Interestingly, the effectiveness of this process seems to depend more on the structure of the reasoning than the factual correctness of its content. Studies have shown that models can learn effective reasoning behaviors even when fine-tuned on CoT demonstrations with flawed logic or incorrect final answers \citep{li2025llms,wang2022towards}. This suggests that the primary role of the CoT during fine-tuning is to teach the model a structured, deliberative approach to problem-solving. The model learns the pattern of breaking down a problem, exploring alternatives, and verifying steps, a process that is more critical for success than learning the specific logical rules from any single example~\citep{li2025llms}.

\paragraph{Supervised Fine-Tuning for Reasoning}
SFT on datasets of reasoning traces has become the primary method for instilling this capability in code LLMs~\citep{zhang2025unveiling}. Research has shown that fine-tuning on a relatively small number of high-quality reasoning demonstrations can lead to significant performance gains on challenging benchmarks \citep{li2025llms}. The quality of these demonstrations is crucial and selecting for difficult problems that require diverse reasoning strategies, such as exploration, backtracking, and self-verification, is more sample-efficient than simply scaling up the quantity of data~\citep{li2025naturalthoughts, jiang2025makes}. Furthermore, techniques that prune redundant or irrelevant tokens from the reasoning chain during generation can improve both performance and efficiency, encouraging the model to maintain a clear and focused thought process~\citep{choi2025think}.

\paragraph{Rejection Sampling Fine-Tuning and Reinforcement Learning}
This SFT-based approach to teaching reasoning has a symbiotic relationship with Reinforcement Learning with Verifiable Rewards (RLVR) in \autoref{sec:code_rl}, the other major paradigm for training advanced reasoning models. SFT provides the foundational reasoning structures and patterns, essentially teaching the model how to think. An important enhancement to standard SFT is rejection sampling fine-Tuning (RFT), which bridges supervised learning and reward-based optimization by generating multiple candidate solutions, filtering them using verifiable rewards, and fine-tuning exclusively on correct outputs~\citep{rft}. This selective approach substantially improves data quality and reasoning diversity compared to basic SFT, while remaining more computationally efficient than full RL. RLVR then builds upon this foundation, using rewards to further refine the model's policy and amplify the generation of correct solutions~\citep{deng2025decomposing}. However, RLVR is fundamentally constrained by the reasoning patterns established during its initial training phases, including both SFT and RFT. It primarily sharpens the model's existing capabilities, concentrating probability mass on known, high-reward reasoning paths rather than discovering entirely new ones~\citep{wu2025invisible}. This limitation exists because RLVR gains are driven almost entirely by optimizing the model's policy at a small fraction of high-entropy ``forking'' tokens, which represent critical decision points in the reasoning process~\citep{wang_beyond_2025, yan2025reformreducinghuman}. Therefore, high-quality initial training, encompassing both diverse SFT and selective RFT, is a crucial prerequisite for developing powerful and robust code LLMs.

\subsubsection{Training Strategies}

After data expansion, some work has focused on how to train more effectively. One aspect is quality filtering of large datasets. \citet{tsai2024codelessalignmore} first clusters all data and retains a different number of data points from each cluster based on the density of the cluster center, thereby achieving data de-duplication. \citet{wang-etal-2024-code} first focuses on data difficulty and correctness before considering diversity, training a complexity scorer to select the most difficult data and a test generator to compute the test case pass rate for each data point to select the most accurate ones. Besides data quality, multi-task balance is also important. \citet{dolphcoder} treats code generation and code evaluation as a two-stage training task, significantly improving model performance. \citet{MFTCoder} integrates multiple loss functions and proposes a multi-task fine-tuning framework that can fine-tune tasks in parallel.
\citet{wang-etal-2024-make} introduces a denoising strategy that transforms some output tokens into random noise while keeping the LLMs predicting the following correct tokens based on them. This not only keeps the instruction following ability, but also adds denoising ability to accelerate inference.

\subsubsection{Challenges}
Although various methods and datasets have emerged in code instruction tuning, several challenges still remain.

\paragraph{Data Leakage} Current instruction tuning datasets suffer from potential data leakage, where training data may contain information from test or benchmark sets. This leakage not only leads to inflated performance but may also conceal the model's deficiencies in real-world scenarios~\citep{matton2024leakagecodegenerationevaluation}. Therefore, there is an urgent need to develop more rigorous and effective data filtering methods to ensure the purity of instruction datasets~\citep{riddell-etal-2024-quantifying, li2023starcoder, yang2023rethinkingbenchmarkcontaminationlanguage}.

\paragraph{Data Bias} Existing instruction tuning datasets often exhibit a significant task complexity bias, with an excessive focus on simple programming tasks, such as programs that can be completed in just a few lines of code. This bias leads to poor model performance on complex tasks (e.g., long code generation, complex algorithm implementation, and system design), resulting in more pronounced performance disparities across tasks of varying difficulty~\citep{jain2024livecodebenchholisticcontaminationfree}. Consequently, it is crucial to construct more balanced datasets that cover a wider range of task complexities.

\paragraph{Insufficient Multilingual Support} Current code instruction tuning datasets are predominantly based on a narrow set of popular programming languages, such as Python or JavaScript. This focus limits the applicability and effectiveness of models for developers working in other critical, high-demand languages (e.g., C++, Rust, Go, or Swift). The lack of broad programming language support not only restricts the model's utility across the diverse software development ecosystem but may also lead to significant performance degradation for tasks in less-represented languages. To address this issue, it is necessary to construct instruction datasets that cover a wider spectrum of programming languages. 
This requires not just a simple ``translation'' of problems into different syntaxes. Languages vary fundamentally in both surface syntax (variable declaration syntax, type annotations, comment formats, file extensions) and deeper semantics (idioms, standard libraries, memory management, and design patterns).

\begin{figure}[H]
    \centering
    \includegraphics[width=0.95\textwidth]{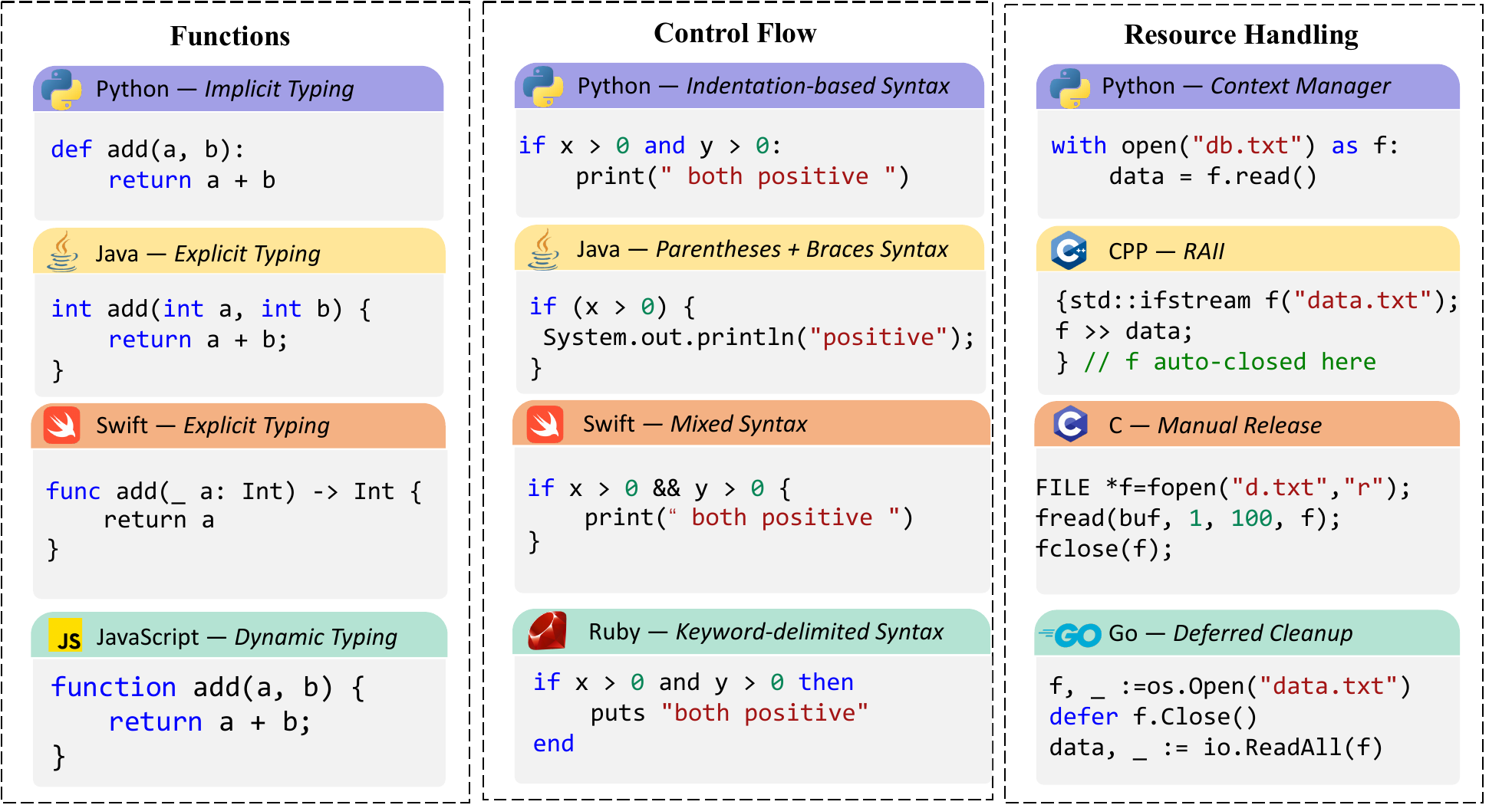}
    \vspace{-2mm}
    \caption{A comparison of programming language syntax across ``Functions'', ``Control Flow'', and ``Resource Handling'', highlighting key differences in their design philosophies for typing, block structure, and memory management.}
    \label{fig:multilingual_support}
\end{figure}

\subsection{Cold-start / Distill Reasoning SFT data for Code LLMs}
\label{sec:sft_data}
\subsubsection{Data Sourcing}
The objective of data sourcing is to acquire the raw problems and materials that form the foundational layer of a dataset. Across the research, a consistent practice is to begin with large-scale existing resources, which include online forums like Art of Problem Solving and Math StackExchange, academic systematicallycompetitions such as AIME and AMC, as well as extensive synthetic or distilled datasets.
A variety of data sourcing methods are currently employed. 
OpenThoughts3~\cite{guha2025openthoughts} systematically evaluated numerous sources and found that mixing a small number of high-quality sources, such as StackExchange, CodeGolf, and OpenCodeReasoning~\cite{ahmad2025opencodereasoning}, produced the best results. The rationale is that quality of the source data is more beneficial than sheer diversity from a large number of sources.
LIMO~\cite{ye2025limo} and s1~\cite{muennighoff2025s1} focused on extracting a minimal yet highly effective set of samples from their extensive corpora. This approach is motivated by the hypothesis that a small number of carefully chosen examples are sufficient to elicit complex reasoning in models that already have a rich knowledge base from pre-training.
DeepMath-103K~\cite{he2025deepmath} and OpenMathReasoning~\cite{moshkov2025aimo} concentrated on extracting raw content from diverse but less structured sources, such as Math StackExchange and Art of Problem Solving community forums, and converting this content into a structured format. This method is designed to find novel and unique problems that are not present in more common, well-formatted datasets.
AceReason-Nemotron~\cite{chen2025acereason} and Skywork-OR1~\cite{he2025skywork} sourced data from a combination of distilled datasets (e.g., DeepScaler~\cite{meng2023deepscaler}, NuminaMath~\cite{li2024numinamath}) and competitive programming platforms to gather problems with high-quality, verifiable answers and test cases. This strategy is essential for their reinforcement learning pipelines, which rely on precise, rule-based reward signals.

In conclusion, while all projects begin with sourcing from large datasets, their strategies diverge based on their specific goals. Some prioritize finding a few high-quality samples to prove a hypothesis about data efficiency, while others aim for a massive scale to push the boundaries of model performance. The choice of source materials is closely tied to the subsequent data curation and filtering pipelines, which are tailored to the desired training paradigm, whether it's standard SFT, or more advanced methods like RL with verifiable rewards.

\subsubsection{Data Cleaning and Decontamination}
The primary purpose of data cleaning and decontamination is to enhance the integrity and reliability of the training data. This process is crucial to prevent models from overfitting to specific examples and to ensure that evaluation results are not skewed by contamination from benchmark datasets. By removing low-quality, incomplete, or duplicate samples, researchers can create a more robust foundation for training powerful reasoning models.
Several distinct strategies for data cleaning and decontamination have been developed.
To address the issue of subtle overlap with evaluation benchmarks, DeepMath-103K~\cite{he2025deepmath} implemented a highly rigorous process using an LLM-Judge (specifically, Llama-3.3-70B-Instruct) to perform semantic comparisons. This method identifies not only exact duplicates but also paraphrased problems, ensuring a truly clean dataset. OpenMathReasoning~\cite{moshkov2025aimo} and OpenThoughts3~\cite{guha2025openthoughts} employed similar LLM-based comparisons or n-gram matching to remove similar questions from their datasets, providing a strong defense against data leakage. This level of scrutiny is an advantage as it goes beyond simple string matching to catch more nuanced forms of contamination.
For models trained with reinforcement learning or tool-integrated reasoning, data quality is defined by more than just uniqueness. Skywork-OR1~\cite{he2025skywork} and AceReason-Nemotron~\cite{chen2025acereason} focused on filtering out problems that were unsuitable for rule-based verification, such as proofs, multiple-choice questions, or those with insufficient test cases. DeepMath-103K~\cite{he2025deepmath} took this a step further by filtering out problems that were not only verifiable but also produced consistent answers across multiple solutions generated by a teacher model. This ensures that the reward signals used in RL are unambiguous and reliable, a crucial aspect for stable training.
The s1~\cite{muennighoff2025s1} project's methodology included filtering out questions that were either too easy or too difficult for the target model (Qwen2.5-32B-Instruct) to answer. This approach ensures that the final dataset contains only a concise set of high-leverage examples, which is a key advantage for projects aiming for sample efficiency.

In summary, current practices for data cleaning and decontamination go far beyond basic deduplication. They now incorporate sophisticated techniques like semantic analysis with LLMs to identify hidden contamination, verifiability checks to ensure data is suitable for advanced training paradigms like RL, and model-aware filtering to create highly-efficient, targeted datasets. These rigorous steps are fundamental to building robust and generalizable reasoning models.

\subsubsection{Question Filtering and Quality/Difficulty Assessment}
The objective of question filtering and quality assessment is to select a high-leverage subset of problems from a larger pool, ensuring that these problems are challenging enough to stimulate complex reasoning without being unsolvable. This process is critical because training on every available data point is often computationally prohibitive, and a carefully curated subset can be significantly more effective for developing robust model capabilities.
The research employs several distinct methodologies for question filtering and quality assessment.
A common approach is to use model performance as a proxy for problem difficulty. The LIMO~\cite{ye2025limo} and s1~\cite{muennighoff2025s1} projects adopt a ``Goldilocks'' strategy by retaining problems of intermediate difficulty. LIMO~\cite{ye2025limo}, for example, filters out problems that a weak model could easily solve and keeps only those a strong model (DeepSeek-R1-Distill-Qwen-32B) could solve in a narrow range of attempts (1-3 out of 32). This method is highly advantageous for creating a minimal, high-signal dataset that is exceptionally sample-efficient. In contrast, DeepMath-103K~\cite{he2025deepmath} and AceReason-Nemotron~\cite{chen2025acereason} deliberately seek out highly difficult problems. DeepMath-103K~\cite{he2025deepmath} uses GPT-4o to rate problems on a 1-10 scale and retains only those at level 5 or higher. AceReason-Nemotron~\cite{chen2025acereason} uses the pass rate of a powerful model (DeepSeek-R1-671B) over multiple rollouts to assign a difficulty score, ultimately filtering problems that the model consistently fails to solve. The benefit of this strategy is to push the model's reasoning boundaries and enable it to solve problems that were previously unsolvable.
Researchers have increasingly leveraged LLMs for qualitative filtering tasks. OpenThoughts3~\cite{guha2025openthoughts} found that for math and science problems, filtering for questions that elicit longer LLM-generated responses was an effective strategy. This is based on the intuition that a longer reasoning trace indicates a more complex problem. Similarly, the Skywork-OR1~\cite{he2025skywork} project used LLMs to classify problems (e.g., as proof-based, multiple-choice, or valid) and to assess data quality based on criteria like clarity, completeness, and formatting. This approach ensures that problems are not only difficult but also well-formed and suitable for the intended training task.
For datasets designed for reinforcement learning with verifiable rewards, filtering for verifiability is paramount. AceReason-Nemotron~\cite{chen2025acereason} and Skywork-OR1~\cite{he2025skywork} explicitly filter out problems such as proofs, interactive problems, or questions lacking comprehensive test cases, as these cannot provide reliable reward signals. This step is a key advantage for ensuring training stability by eliminating sources of noisy or ambiguous reward signals.

In conclusion, question filtering is a multi-faceted process that has evolved from simple data statistics to sophisticated, model-aware strategies. The optimal approach is highly dependent on the research goal: for maximizing sample efficiency, a ``Goldilocks'' strategy that targets intermediate-difficulty problems is effective; whereas for pushing the performance ceiling with RL, actively curating a dataset of high-difficulty and verifiable problems is crucial. The pervasive use of LLMs in this phase has enabled more nuanced filtering that assesses not just difficulty, but also problem quality, diversity, and suitability for specific training paradigms.

\subsubsection{Reasoning Chain Generation}
The primary purpose of reasoning chain generation is to create detailed, step-by-step solutions, such as CoT or tool-integrated reasoning, for each problem. These explicit reasoning traces are crucial for teaching a model the logical process required to solve complex problems, as opposed to simply memorizing the final answer. By providing a ``cognitive template'', these chains enable a model to apply and structure its pre-trained knowledge effectively at inference time.
Current methods for reasoning chain generation vary significantly in their approach and complexity.
The most common method involves using one or more large, powerful ``teacher'' models to generate solutions. These teachers, such as DeepSeek-R1, QwQ-32B, and Gemini Thinking Experimental, are often state-of-the-art models that can produce high-quality reasoning traces. The advantage of this approach is its relative simplicity and scalability, as demonstrated by projects like OpenThoughts3~\cite{guha2025openthoughts} and s1~\cite{muennighoff2025s1} which use this method to generate millions or thousands of high-quality samples, respectively. DeepMath-103K~\cite{he2025deepmath} takes this a step further by generating three distinct solutions per problem to support diverse training paradigms.
For more specialized tasks, a multi-stage approach is often required. OpenMathReasoning~\cite{moshkov2025aimo} developed a complex pipeline specifically for building TIR data, which integrates natural language reasoning with Python code execution. This process involves an iterative cycle of generating, filtering, and re-training models to produce increasingly high-quality TIR solutions. This method's advantage is its ability to create datasets that teach models to use external tools effectively, a capability that standard reasoning models often lack.
The generated solutions are also tailored to the downstream training method. For example, AceReason-Nemotron~\cite{chen2025acereason} and Skywork-OR1~\cite{he2025skywork} create solutions that are formatted for rule-based verification, a necessary component for their reinforcement learning pipelines. This ensures that the generated data provides reliable and unambiguous reward signals, which is critical for stable and effective RL training.

In summary, the generation of reasoning chains has become a sophisticated process that goes beyond simple prompting. While distillation from powerful models remains a cornerstone, projects are increasingly employing more specialized techniques, such as iterative tool-integrated generation or targeted formatting for RL, to create datasets that are not only large but are also precisely engineered for their specific training goals. The ultimate goal is to craft reasoning traces that act as effective ``cognitive template'', allowing models to generalize and solve new, unseen problems.

\subsubsection{Solution Filtering and Refinement}
The main purpose of solution filtering and refinement is to ensure that generated solutions are not just correct, but also of high quality and in an appropriate format. This process aims to create ideal ``cognitive templates'' for the model to learn from, which are characterized by logical coherence, clarity, and efficiency. By refining these solutions, researchers can improve the model's ability to structure its own reasoning process, leading to better generalization and performance on unseen problems.
Several distinct methodologies for filtering and refining solutions have been developed, each with specific advantages.
To identify the most effective reasoning chains, LIMO~\cite{ye2025limo} employed a unique rule-based scoring system. This system evaluated solutions based on ``meta-reasonin'' features such as elaborated reasoning, self-verification, an exploratory approach, and adaptive granularity. The advantage of this approach is that it allows for the precise selection of a minimal set of high-quality examples, which is crucial for achieving high performance with few training samples.
For specialized tasks like tool-integrated reasoning, filtering ensures that generated solutions provide meaningful value. OpenMathReasoning~\cite{moshkov2025aimo} applied a strict filtering process to its TIR data to retain only solutions where code execution was ``novel and significant''. This involved using an LLM to classify whether a code block performed a new calculation or merely verified a previous step. The advantage is that it prevents the model from learning to use tools for trivial or redundant tasks, thereby reinforcing effective and purposeful tool usage.
DeepMath-103K~\cite{he2025deepmath} implement a rigorous answer verification process to ensure the robustness of its data. After generating three distinct solutions for each problem with a powerful teacher model, it retained only problems where all three solutions produced an identical, verifiable final answer. This strategy's main advantage is that it creates a highly reliable dataset with unambiguous ground truths, which is essential for stable reinforcement learning with verifiable rewards.
Interestingly, OpenThoughts3~\cite{guha2025openthoughts} conducted extensive experiments on various answer filtering techniques and found that none of them provided a significant performance improvement over simply training on all generated answers. As such, their final pipeline omits this step. This suggests that for some training paradigms, the noise introduced by lower-quality or redundant answers may be negligible and not worth the computational cost of filtering.

In conclusion, solution filtering and refinement has become a sophisticated, goal-oriented process. While some projects, like LIMO~\cite{ye2025limo} and OpenMathReasoning~\cite{moshkov2025aimo} , use complex filtering to curate highly specific datasets for sample-efficient or tool-integrated learning, others, like OpenThoughts3~\cite{guha2025openthoughts}, find that simpler pipelines can be just as effective. The optimal strategy depends on the trade-off between data quality, data volume, and the specific requirements of the downstream training paradigm.

\subsubsection{Final Dataset Construction}

The final step of assembling the training dataset is to compile all processed problems and their refined solutions into a cohesive format ready for model fine-tuning. The primary goal is to create a dataset that aligns with the specific research objectives, whether that's to demonstrate data efficiency or to push the boundaries of large-scale model performance. This stage represents the culmination of all prior steps, from sourcing and cleaning to filtering and refining, ensuring the final product is optimized for its intended use.
Current approaches to constructing the final dataset vary primarily in their scale and the strategic rationale behind that scale.
Some projects, like LIMO~\cite{ye2025limo} and s1~\cite{muennighoff2025s1}, prioritize creating extremely small yet high-quality datasets to prove that extensive data isn't always necessary for achieving strong reasoning capabilities. The final LIMO~\cite{ye2025limo} dataset, for instance, contains just 800 examples, while s1~\cite{muennighoff2025s1} is built on only 1,000 samples. This strategy's main advantage is its ability to demonstrate the power of data curation over data quantity, showing that carefully selected ``cognitive templates'' can be highly effective for alignment and generalization with minimal computational resources.
Other projects, aiming for state-of-the-art performance and broad generalization, construct much larger datasets. OpenThoughts3~\cite{guha2025openthoughts} scaled its final dataset to 1.2 million samples to achieve its performance gains. Similarly, DeepMath-103K~\cite{he2025deepmath} and OpenMathReasoning~\cite{moshkov2025aimo} built datasets with over 100,000 and 500,000 unique problems, respectively, to create a robust foundation for training powerful reasoning models across multiple domains. The advantage of this approach is that a large volume of diverse, high-quality data can push models to higher performance ceilings and enhance their ability to handle a wider variety of complex problems.
Certain datasets are designed to serve multiple purposes. OpenMathReasoning~\cite{moshkov2025aimo} compiled a final dataset of 5.5 million samples from three different tasks: CoT solution generation, tool-integrated reasoning, and generative solution selection. This multi-task approach allows a single model to learn different inference modes, such as providing a CoT solution, using a code interpreter, or selecting the best answer from multiple candidates. This creates a versatile and powerful model capable of adapting to various reasoning challenges.

In summary, final dataset construction is a strategic decision that reflects the core hypothesis of a research project. The choice between a minimalist or large-scale approach dictates the kind of model that can be trained, with some researchers prioritizing data efficiency and others focusing on maximizing performance and versatility. Regardless of scale, the quality and intentional design of the final dataset remain the most critical factors for success.

\begin{table*}[!h]
\centering
\caption{Summary of datasets and filtering strategies.}
\label{tab:dataset_summary}
\resizebox{\textwidth}{!}{%
\begin{tabular}{llllp{6cm}}
\toprule
\textbf{Dataset} & \textbf{Scale} & \textbf{Source} & \textbf{Model} & \textbf{Filtering Strategy} \\
\midrule
OpenThoughts3~\cite{guha2025openthoughts} & \makecell[l]{1.2M\\(850k math,\\250k code,\\100k science)} & \makecell[l]{OpenMath-2-Math\\StackExchange-CodeGolf\\OpenCodeReasoning\\StackExchange-Physics} & QwQ-32B & \makecell[l]{$\leq$2 top sources per domain\\Code: difficulty-based filter\\Math/science: answer-length filter\\Exact deduplication (math/science only)\\16× answer sampling for all\\No answer filtering} \\
\midrule
LIMO~\cite{ye2025limo} & 800 & \makecell[l]{NuminaMath~\cite{li2024numinamath}-CoT\\DeepScaler~\cite{meng2023deepscaler}\\AIME, MATH} & \makecell[l]{DeepSeek R1\\DeepSeek-R1-Distill-Qwen-32B\\QwQ-32B} & \makecell[l]{Multi-stage difficulty filtering:\\- Remove problems solvable by weak model\\- Keep those strong model solved 1-3/32 attempts\\N-gram deduplication\\Rule-based scoring on elaborateness} \\
\midrule
Skywork-OR1~\cite{he2025skywork} & \makecell[l]{105k math\\13.7k code} & \makecell[l]{Math: NuminaMath~\cite{li2024numinamath}-1.5\\Code: LeetCode, TACO} & \makecell[l]{DeepSeek-R1-Distill\\-Qwen-7B/32B} & \makecell[l]{Keep verifiable, correct, challenging problems\\Model pass-rate difficulty estimation\\Human + LLM-as-Judge for quality check} \\
\midrule
DeepMath-103K~\cite{he2025deepmath} & 103k & \makecell[l]{Math StackExchange\\MMIQC\\WebInstructSub} & DeepSeek-R1 & \makecell[l]{LLM-Judge semantic decontamination\\GPT-40 difficulty rating ($\geq$ level 5)\\Retain problems with 3 consistent solutions} \\
\midrule
AceReason-Nemotron~\cite{chen2025acereason} & \makecell[l]{49k math\\8.5k code} & \makecell[l]{Math: DeepScaler~\cite{meng2023deepscaler},\\NuminaMath~\cite{li2024numinamath}\\Code: competitive platforms} & DeepSeek-R1-671B & \makecell[l]{Remove benchmark contamination\\Model pass-rate scoring (8 attempts)\\Rule-based verification:\\- Math: \texttt{sympy}\\- Code: sandbox + test cases} \\
\midrule
OpenMathReasoning~\cite{moshkov2025aimo} & \makecell[l]{540k problems\\3.2M CoT solutions} & AoPS forums & \makecell[l]{DeepSeek-R1\\QwQ-32B} & \makecell[l]{LLM extracts and classifies problems\\LLM-based decontamination\\Filter CoTs not reaching expected answer\\Iterative TIR pipeline} \\
\midrule
s1~\cite{muennighoff2025s1} & 1k & \makecell[l]{NuminaMath~\cite{li2024numinamath}\\OlympicArena\\AGIEval\\s1~\cite{muennighoff2025s1}-prob} & \makecell[l]{Gemini Thinking\\Experimental\\DeepSeek-r1} & \makecell[l]{Remove API errors and poor formatting\\Exclude problems easily solved\\Classify by domain for diversity\\Favor longer reasoning traces} \\
\bottomrule
\end{tabular}
}
\end{table*}

\subsection{Multilingual Code Understanding and Generation}
\label{subsec:multilingual_sft}
\subsubsection{Multilingual Code LLMs}
\begin{table*}[htb]
\centering
\caption{Representative multilingual code models.}
\footnotesize
\resizebox{1.0\columnwidth}{!}{
\begin{tabular}{l|c|l|l|l|l}
\toprule
\textbf{Model} & \textbf{Year} & \textbf{Task Type} & \textbf{Scale} & \textbf{Language} & \textbf{Corpus Type} \\
\midrule
\multicolumn{6}{c}{\cellcolor{blue!15}Single Programming Languages} \\
\midrule
JavaBERT~\cite{javabert} & 2020 & Java code understanding & 125M & Java & Java GitHub repositories \\
C-BERT~\cite{cbert} & 2021 & C code understanding & 125M & C & Linux kernel code \\
JSCoder~\cite{jsbert} & 2021 & JavaScript generation & 350M & JavaScript & npm packages \\
GPT-Neo-Python~\cite{gpt_neox_20b} & 2021 & Python generation & 2.7B & Python & Python code corpus \\
PyMT5~\cite{clement-etal-2020-pymt5} & 2021 & Python code generation & 220M & Python & GitHub Python corpus \\
InCoder-Python~\cite{fried2023incoder} & 2022 & Python code infilling & 6.7B & Python & GitHub Python \\
RustCoder~\cite{rustcoder} & 2023 & Rust code generation & 2.7B & Rust & Rust GitHub repositories \\
WizardCoder-Python~\cite{luo2023wizardcoder} & 2023 & Python code generation & 7B-34B & Python & Evol-Instruct + Code Llama \\
VeriGen~\cite{thakur2023verigen} & 2023 & Verilog code generation & 16B & Verilog & GitHub, Verilog textbooks \\
VerilogEval~\cite{verilog_eval} & 2023 & Verilog evaluation benchmark & - & Verilog & HDLBits (156 problems) \\
ReasoningV~\cite{reasoningv} & 2025 & Verilog code generation & 7B & Verilog & ReasoningV-5K, PyraNet \\
\midrule
\multicolumn{6}{c}{\cellcolor{green!15}Multiple Programming Languages} \\
\midrule
Codex~\cite{chen2021codex} & 2021 & Code generation & 12B & Python, JavaScript, Go, Java, PHP, Ruby & GitHub \\
CodeBERT~\cite{feng2020codebertpretrainedmodelprogramming} & 2020 & Code search, generation & 125M & Python, Java, JavaScript, PHP, Ruby, Go & GitHub, CodeSearchNet \\
GraphCodeBERT~\cite{graph_code_bert} & 2021 & Code understanding, generation & 125M & Python, Java, JavaScript, PHP, Ruby, Go & GitHub repositories \\
CodeT5~\cite{chen2023codet5} & 2021 & Code generation, summarization, translation & 220M & Java, Python, JavaScript, PHP, Ruby, Go, C, C\# & GitHub, CodeSearchNet \\
CodeT5+~\cite{wang2023codet5plus} & 2023 & Code understanding, generation & 220M-16B & Python, Java, JavaScript, C, C++, Go, etc. & GitHub, The Stack \\
PolyCoder~\cite{polycoder} & 2022 & Code generation & 2.7B & 12 languages (C, C++, Python, Java, etc.) & GitHub (245GB) \\
AlphaCode~\cite{li2022alphacode} & 2022 & Competitive programming & 41B & C++, Python, Java & Codeforces, GitHub \\
CodeGen~\cite{nijkamp2023codegen} & 2022 & Code generation & 350M-16B & Python, Java, JavaScript, C, C++, Go & The Pile, BigQuery, BigPython \\
PanGu-Coder~\cite{shen2023pangucoder2boostinglargelanguage} & 2022 & Code generation & 317M-2.6B & Python, Java, JavaScript, C, C++ & GitHub \\
PanGu-Coder2~\cite{shen2023pangucoder2} & 2023 & Code generation & 15B & Python, Java, JavaScript, C++, Go & GitHub, internal data \\
SantaCoder~\cite{allal2023santacoder} & 2023 & Code completion & 1.1B & Python, Java, JavaScript & The Stack \\
StarCoder~\cite{liu2025rstarcoder} & 2023 & Code completion, generation & 15.5B & 80+ languages & The Stack (GitHub) \\
CodeGen2~\cite{nijkamp2023codegen2} & 2023 & Code generation, infilling & 1B-16B & Python, Java, JavaScript, C, C++ & The Stack, StarCoder data \\
Code Llama~\cite{roziere2024codellama} & 2023 & Code generation, infilling & 7B-34B & Python, C++, Java, PHP, C\#, TypeScript, Bash & GitHub, Stack Overflow \\
WizardCoder~\cite{luo2023wizardcoder} & 2023 & Code generation & 15B-34B & Python, Java, JavaScript, C++, C\#, PHP & Evol-Instruct + StarCoder \\
phi-1~\cite{phi} & 2023 & Code generation & 1.3B & Python, JavaScript, Java & Synthetic textbooks, StackOverflow \\
phi-1.5~\cite{phi15} & 2023 & Code generation & 1.3B & Python, Java, JavaScript, C++ & Synthetic data, web corpus \\
DeepSeek-Coder~\cite{deepseek2024coder} & 2023 & Code generation & 1.3B-33B & 87 languages & GitHub (2TB) \\
DeepSeek-Coder-V2~\cite{deepseek_coder_v2} & 2024 & Code generation & 16B-236B & 338 languages & GitHub (10TB+) \\
Phind-CodeLlama & 2023 & Code generation & 34B & Python, JavaScript, C++, Java, TypeScript & Code Llama + instruction tuning \\
CodeFuse~\cite{CodeFuseEval} & 2023 & Code generation & 13B-15B & Python, Java, JavaScript, C++, SQL & GitHub, internal data \\
Magicoder~\cite{chen2024magicoder} & 2023 & Code generation & 7B & Python, Java, JavaScript, C++, TypeScript & OSS-Instruct synthetic data \\
OctoCoder~\cite{muennighoff2023octopack} & 2023 & Code generation & 15.5B & Python, Java, JavaScript, C++, TypeScript & StarCoder + Git commits \\
Yi-Coder~\cite{yicoder2024} & 2024 & Code generation & 1.5B-9B & 52 languages & GitHub, StackOverflow \\
OpenCodeInterpreter~\cite{opencodeinterpreter} & 2024 & Code execution, generation & 6.7B-33B & Python, JavaScript, Java, C++, SQL & Code execution data \\
Qwen2.5-Coder~\cite{qwen25coder} & 2024 & Code generation & 0.5B-32B & 92 languages & GitHub, synthetic data (5.5TB) \\
CodeStral~\cite{mistral2024codestral} & 2024 & Code generation & 22B & 80+ languages & Multilingual code corpus \\
TransCoder~\cite{transcoder} & 2020 & Unsupervised code translation & - & C++, Java, Python & GitHub (monolingual) \\
DOBF~\cite{dobf} & 2021 & Code translation & - & Java, C\# & Parallel corpus \\
TreeBERT~\cite{ni2025tree} & 2021 & Code understanding & 125M & Java, Python, JavaScript, Ruby, Go, PHP & Abstract syntax trees \\
SPT-Code~\cite{spt_code} & 2022 & Code translation & 220M & C++, Java, Python, C\#, JavaScript & Syntax-preserved translation \\
UniXcoder~\cite{unixcoder} & 2022 & Cross-lingual code understanding & 125M & Python, Java, JavaScript, Ruby, Go, PHP, C, C\# & GitHub, CodeSearchNet \\
CodeRL~\cite{code_rl} & 2022 & Code generation with RL & 770M & Python, Java, JavaScript, C++, Go, Rust & CodeNet, GitHub \\
CodeTransOcean~\cite{CodeTransOcean} & 2023 & Low-resource language translation & - & Java, C\#, Python, JavaScript, C, C++ & Parallel corpus with back-translation \\
CodeExecutor~\cite{code_executor} & 2023 & Execution-based translation & - & Python, Java, C++, JavaScript, Ruby, PHP & Execution traces \\
REEF~\cite{reef} & 2024 & Low-resource code generation & - & Fortran, COBOL, Ada, Lisp & Legacy code corpus \\
Qwen2.5-xCoder~\cite{xcoder} & 2025 & code generation & - & Fortran, COBOL, Ada, Lisp & Legacy code corpus \\
\midrule
\multicolumn{6}{c}{\cellcolor{purple!15}Massively Multiple Programming Languages} \\
\midrule
CodeGeeX~\cite{codegeex} & 2023 & Multilingual code generation & 13B & 23 languages (Python, Java, C++, etc.) & The Pile, CodeParrot, GitHub \\
CodeGeeX4~\cite{zheng2024codegeex4} & 2023 & Multilingual code generation & 6B & 100+ languages & The Stack, GitHub \\
CodeShell~\cite{xie2024codeshell} & 2023 & Code generation & 7B & Python, Java, C++, JavaScript, Go, Rust & GitHub, internal Chinese data \\
StarCoder2~\cite{lozano2024starcoder2} & 2024 & Code generation & 3B-15B & 600+ languages & The Stack v2 (4TB+) \\
CodeQwen1.5~\cite{qwen2024codeqwen1.5} & 2024 & Code generation & 7B & 92 languages & GitHub, internal data \\
Granite-Code~\cite{mishra2024granitecodemodelsfamily} & 2024 & Code generation & 3B-34B & 116 languages & The Stack, GitHub \\
CodeGemma~\cite{codegemma2024} & 2024 & Code generation & 2B-7B & 20+ languages & GitHub, web corpus \\
PolyglotCode~\cite{polyglotcode} & 2024 & Cross-family code translation & - & 25+ languages across paradigms & Parallel multilingual corpus \\
\bottomrule
\end{tabular}
}
\label{multilingual_codellm_table}
\end{table*}
\autoref{multilingual_codellm_table} comprehensively illustrates the development trajectory of code large language models from 2020 to 2025, revealing significant evolutionary progress in model scale, language coverage, task capabilities, and training data.

\paragraph{Development Phases and Scale Evolution} Early models (2020-2021) such as JavaBERT and C-BERT primarily focused on single programming languages with relatively small parameter scales (125M-350M) and training data from limited sources (e.g., specific GitHub repositories or Linux kernel code). The intermediate period (2022-2023) witnessed breakthrough progress, with multilingual models like Codex and AlphaCode expanding to 12B-41B parameters and supporting 6-12 mainstream programming languages. The latest generation of models (2024-2025) achieved a leap: DeepSeek-Coder-V2 reached 236B parameters supporting 338 languages, while StarCoder2 supports 600+ languages trained on 4TB+ data, marking the entry of code intelligence into the era of ``ultra-large-scale multilingualism~\cite{lozano2024starcoder2}''

\paragraph{Data and Architecture Innovation} Training corpora have evolved from single GitHub repositories (hundreds of MB) to ultra-large-scale multi-source fusion (10TB+), with data sources including GitHub, Stack Overflow, CodeSearchNet, The Stack series, The Pile, synthetic textbook data, execution traces, and parallel translation corpora. Architecturally, the field exhibits diversity: BERT-based encoder models (CodeBERT, UniXcoder), T5-based encoder-decoder models (CodeT5 series, PyMT5), GPT-based decoder models (Codex, Code Llama), and graph neural network approaches integrating code structure (GraphCodeBERT, TreeBERT).

\paragraph{Ecosystem and Application Trends} \autoref{multilingual_codellm_table} reflects a complete industrial ecosystem: open-source foundation models (StarCoder, DeepSeek-Coder), commercial closed-source models (Codex, AlphaCode), instruction-tuned variants (WizardCoder, Phind-CodeLlama), domain-specific models (CodeFuse, CodeShell, incorporating Chinese data), and evaluation benchmarks (VerilogEval). Latest trends include: (1) small efficient models (phi series 1.3B, Yi-Coder 1.5B) pursuing the balance between performance and cost; (2) cross-lingual translation emerging as an independent research direction (TransCoder, CodeTransOcean, PolyglotCode); (3) multimodal fusion (code-natural language-execution results); (4) specialized optimization for low-resource and legacy languages, demonstrating that code intelligence is transitioning from generalization to refinement, and from high-resource languages to long-tail languages along a mature development path.

\subsubsection{Multilingual Code Evaluation}

\autoref{tab:multilingual_code_benchmarks} chronicles the evolution of code evaluation benchmarks from 2021 to 2025, showing a clear progression from single-language Python-focused datasets to comprehensive multilingual frameworks. Early benchmarks like HumanEval (164 problems) and MBPP (974 tasks) establish foundational evaluation paradigms for code generation, which were subsequently enhanced with expanded test coverage and specialized domains including data science (DS-1000), algorithmic challenges (APPS), and real-world software engineering tasks (SWE-bench series). The field evolved dramatically toward multilingual evaluation starting in 2022, with benchmarks like MultiPL-E (18 languages), HumanEval-X (5 languages), and eventually McEval (40 languages) addressing the need for cross-lingual code generation and understanding. Recent benchmarks (2023-2025) demonstrate increasing sophistication, incorporating diverse tasks such as code reasoning (CRUXEval-X across 19 languages), multilingual debugging (MdEval with 18 languages), full-stack programming (FullStackBench with 16 languages), and cross-lingual natural language generalization (HumanEval-XL with 23 natural languages × 12 programming languages). The most recent entries reflect emerging trends toward multimodal evaluation (SWE-bench Multimodal) and practical repository-level tasks (Multi-SWE-bench), indicating a maturation from simple code generation benchmarks to comprehensive, real-world software engineering evaluation frameworks that span multiple programming languages, natural languages, domains, and modalities.

\paragraph{Specialized Single-Language Tasks}
The evolution of single-language benchmarks diversified into specialized domains. DS-1000 (2022) focused specifically on data science code generation across 7 Python libraries, while APPS (2021) tackled algorithmic challenges with 10,000 problems. More sophisticated benchmarks emerged to evaluate code reasoning and execution, such as CRUXEval (2023), which introduced input/output prediction tasks. The field further matured with real-world engineering challenges through the SWE-bench series (2024), which brought authentic GitHub issues into the evaluation paradigm, offering variants like Verified and Lite versions for different use cases.

\paragraph{The Multilingual Expansion}
Starting in 2021-2022, the field witnessed a paradigm shift toward multilingual benchmarks. Early efforts like mCoNaLa (2021) and MBXP (2022) began exploring code generation across multiple programming languages. MultiPL-E (2022) made significant strides by translating HumanEval to 18 different languages, while HumanEval-X (2023) provided hand-written multilingual tasks across 5 major languages. These benchmarks recognized that modern developers work across diverse programming ecosystems and that models need cross-lingual capabilities.

\paragraph{Emerging Trends and Future Directions}
The most recent benchmarks (2024-2025) reflect increasing sophistication and practical orientation. HumanEval-XL (2024) introduced an ambitious cross-product of 23 natural languages and 12 programming languages, creating over 22,000 prompts to evaluate true cross-lingual generalization. The field is also expanding beyond pure code generation: SWE-bench Multimodal (2025) incorporates visual elements into programming tasks, while Multi-SWE-bench (2025) tackles multilingual issue resolution, reflecting the complex, multimodal nature of modern software development. This progression demonstrates a clear trajectory from simple, single-language code generation toward comprehensive, real-world software engineering evaluation across languages, modalities, and domains.

\begin{table*}[htbp]
\centering
\caption{Multilingual code benchmarks and evaluation datasets.}
\label{tab:multilingual_code_benchmarks}
\footnotesize
\resizebox{1.0\textwidth}{!}{
\begin{tabular}{l|c|l|l|l}
\toprule
\textbf{Benchmark} & \textbf{Year} & \textbf{Task Type} & \textbf{Languages} & \textbf{Description} \\
\midrule
\multicolumn{5}{c}{\cellcolor{blue!15} Single Language} \\
\midrule
HumanEval~\cite{chen2021codex} & 2021 & Code generation & Python & 164 hand-written problems \\
MBPP~\cite{austin2021mbpp} & 2021 & Basic programming problems & Python & 974 entry-level tasks \\
HumanEval+~\cite{evalplus} & 2023 & Enhanced test coverage & Python & HumanEval with 80$\times$ more tests \\
MBPP+~\cite{evalplus} & 2023 & Enhanced test coverage & Python & MBPP with 35$\times$ more tests \\
APPS~\cite{apps} & 2021 & Algorithmic challenges & Python & 10,000 coding problems \\
DS-1000~\cite{ds1000} & 2022 & Data science code generation & Python (7 libraries) & 1,000 StackOverflow problems \\
CRUXEval~\cite{gu2024cruxeval} & 2023 & Code reasoning \& execution & Python & 800 functions, input/output prediction \\
BigCodeBench~\cite{zhuo2024bigcodebench} & 2024 & Practical programming & Python & 1,140 challenging real-world tasks \\
SWE-bench~\cite{swebench} & 2024 & Software engineering & Python & 2,294 real GitHub issues \\
SWE-bench Verified~\cite{swebench} & 2024 & Human-filtered issues & Python & 500 curated problems \\
SWE-bench Lite~\cite{swebench} & 2024~ & Cost-effective subset & Python & 300 selected issues \\
EvalPlus~\cite{evalplus} & 2024 & Enhanced test suite framework & Multiple languages & Augmented HumanEval/MBPP \\
\midrule
\midrule
\multicolumn{5}{c}{\cellcolor{green!15} Multiple Languages} \\
\midrule
mCoNaLa~\cite{mconala} & 2021 & Multilingual code generation from NL & Python, Java, JavaScript & StackOverflow, CoNaLa \\
MBXP~\cite{mbxp} & 2022 & Multi-lingual execution-based & 13 languages (Python, Java, Go, etc.) & MBPP transpiled to multiple PLs \\
ODEX~\cite{odex} & 2022 & Open-domain execution & Python (79 libraries), 4 NLs & 945 StackOverflow NL-code pairs \\
XLCoST~\cite{xlcost} & 2022 & Cross-lingual code translation & C++, Java, Python, C\#, JavaScript, PHP, C & Parallel code snippets \\
MultiPL-E~\cite{MultiPL-E} & 2022 & Multilingual evaluation & 18 languages (Lua, Racket, Scala, etc.) & HumanEval translations \\
CodeContests~\cite{codecontests} & 2022 & Competitive programming & Multiple languages & AlphaCode training data \\
HumanEval-X~\cite{codegeex} & 2023 & Cross-lingual generation & 5 languages (Python, C++, Java, JS, Go) & 820 hand-written multilingual tasks \\
xCodeEval~\cite{ziyao2023xcodeeval} & 2023 & Cross-lingual multitask & Python, Java, JavaScript, C++, Go, Rust, etc. & GitHub + synthetic data \\
XCodeSearchNet~\cite{xcodesearchnet} & 2023 & Multilingual code search & Python, Java, JavaScript, Go, PHP, Ruby, C, C\# & Extended CodeSearchNet \\
CodeScore~\cite{codescore} & 2023 & Cross-lingual evaluation & 20+ languages including Dart, Kotlin, Swift & Execution-based metrics \\
CrossCodeEval~\cite{ding2023crosscodeevaldiversemultilingualbenchmark} & 2023 & Cross-lingual understanding & 11 languages (Scala, Swift, Kotlin, Rust, etc.) & Parallel benchmark dataset \\
ClassEval~\cite{du2023classeval} & 2023 & Class-level generation & Python & 100 classes with 410 methods \\
RepoBench & 2023 & Repository-level completion & Python, Java & Cross-file code completion \\
CRUXEval-X~\cite{xu2024cruxeval} & 2024 & Multilingual code reasoning & 19 languages (C++, Java, Rust, etc.) & 12,660 subjects, 19K tests, I/O prediction \\
HumanEval-XL~\cite{humanevalxl} & 2024 & Cross-lingual NL generalization & 23 NLs $\times$ 12 PLs & 22,080 prompts, parallel data \\
McEval~\cite{mceval} & 2024 & Massively multilingual & 40 languages (16K samples) & Code generation, completion, understanding \\
MdEval~\cite{mdeval} & 2024 & Multilingual debugging & 18 languages (3.6K samples) & APR, code review, bug identification \\
FullStackBench~\cite{liu2024fullstackbench} & 2024 & Full-stack programming & 16 languages (3,374 problems) & 11 domains, real-world scenarios \\
SWE-bench M~\cite{yang2024swebenchmultimodalaisystems} & 2025 & Issues with visual elements & Python & 517 multimodal tasks \\
Multi-SWE-bench~\cite{zan2025multiswebenchmultilingualbenchmarkissue} & 2025 & Multilingual issue resolving & Multiple languages & Multilingual benchmark for issue resolving \\
\bottomrule
\end{tabular}
}
\end{table*}

\subsection{Multimodal Code Understanding and Generation}
\label{subsec:multimodal_sft}

\subsubsection{Vision-Language Foundation Models for Code}
\label{sec:vlm-foundation}

Modern multimodal LLMs trace back to contrastive vision–language pretraining like CLIP \cite{radford2021clip} that established robust zero-shot perception by aligning image and text embeddings at scale.
Subsequent ``connector'' designs couple frozen encoders with LLMs. BLIP \cite{li2022blip} and BLIP-2 \cite{li2023blip2} demonstrate that lightweight adapters can efficiently bridge modalities while retaining strong generation and understanding.
Architectures that ingest interleaved image–text sequences push few-shot generalization: Flamingo \cite{alayrac2022flamingo} conditions a language backbone on visual tokens for strong in-context transfer, and instruction-tuned stacks like LLaVA \cite{liu2023llava, liu2023llava1_5,li2025llavaonevision} demonstrate that synthetic visual dialogue corpora can elicit broad perception–reasoning skills with modest compute. 

Large open families extend coverage and resolution handling: the Qwen-VL line \cite{qwen2023qwenvl, qwen2024qwen2vl, qwen2025qwen2_5vl} brings grounding, OCR and dynamic-resolution processing.
The InternVL Series \cite{chen2024internvl, chen2025internvl2_5, zhu2025internvl3, wang2025internvl3_5} evolve toward competitive open performance via model, data and test-time scaling.
DeepSeek-VL \cite{lu2024deepseekvl, wu2024deepseekvl2} target real-world screenshots, documents, and charts, adding Mixture-of-Experts variants for efficiency.

Foundation releases from major labs have made vision first-class: Meta's Llama-3.2-Vision \cite{meta2024llama3_2} adds image understanding to an open LLM family, Google's Gemma 3 brings a lightweight, open, multimodal stack derived from Gemini \cite{gemmateam2025gemma3}, while the Gemini series itself advances long-context multimodality \cite{geminiteam2025gemini, geminiteam2024gemini1_5} and newer models focus on agentic use \cite{comanici2025gemini2_5}.
OpenAI's GPT-4V \cite{openai2023gpt4v} established broad image-understanding with a detailed safety analysis, GPT-4o \cite{openai2024gpt4o} unified end-to-end text-vision-audio for low-latency interaction, and GPT-5 \cite{openai2025gpt5} continues the trend with updated capabilities.

Across these lines, we observe converging design motifs such as frozen or lightly trained visual front-ends, compact cross-modal adapters, and instruction or preference tuning at scale, all of which yield steady gains in perception-reasoning breadth. These foundation models provide the underlying capabilities for specialized code-related multimodal tasks discussed in the following sections. Meanwhile, there are many popular multimodal benchmarks in \autoref{tab:multimodal-benchmarks} to evaluate the performance of the multimodal code understanding and generation.

\begin{table*}[htbp]
\centering
\caption{Representative Multimodal Code Generation Benchmarks.}
\label{tab:multimodal-benchmarks}
\footnotesize
\resizebox{0.95\columnwidth}{!}{
\begin{tabular}{p{3.1cm}|c|p{3cm}|p{3cm}|p{3.5cm}|p{4.0cm}}
\toprule
\textbf{Benchmark} & \textbf{Year} & \textbf{Task Type} & \textbf{Scale} & \textbf{Key Metrics} & \textbf{Innovation} \\
\midrule
\multicolumn{6}{c}{\cellcolor{blue!15}\textbf{Frontend Interface Generation}} \\
\midrule
Design2Code~\cite{si2024design2code} & 2024 & Screenshot→HTML & 484 pages & TreeBLEU, DOM-ED & Real-world benchmark \\
UICoder~\cite{wu2024uicoder} & 2024 & UI→Code & 3M pairs & Compile-Render-CLIP & Auto feedback loop \\
Interaction2Code~\cite{xiao2024interaction2code} & 2024 & Interaction→Code & 374 interactions & Event accuracy & Dynamic interaction \\
Sketch2Code~\cite{li2024sketch2code} & 2024 & Sketch→Code & 731 sketches & User preference & Interactive eval \\
\midrule
\multicolumn{6}{c}{\cellcolor{green!15}\textbf{Data Visualization}} \\
\midrule
nvBench~\cite{nvBench2021} & 2021 & NL→Chart & Cross Domain & Cross Domain Database & First Large Scale Benchmark \\
Plot2Code~\cite{wu2024plot2code} & 2024 & Chart→Python & Scientific plots & Execution rate & Scientific chart focus \\
ChartMimic~\cite{yang2024chartmimic} & 2024 & Chart→Code & Research papers & Cross-modal reasoning & Reasoning eval \\
nvAgent~\cite{ouyang2025nvagent} & 2025 & NL→Chart & nvBench & SOTA performance & Multi-agent workflow \\
DeepVis~\cite{shuai2025deepvis} & 2025 & NL→Chart & nvBench & SOTA performance & Transparent Decision Refine \\
nvBench 2.0~\cite{nvbench2-2025luo} & 2025 & NL→Chart & Cross Domain & Ambiguity Identification & Disambiguiation Reasoning\\
ChartCoder~\cite{zhao2025chartcoder} & 2025 & Chart→Code & Large dataset & Code quality & Open-source enhance \\
\midrule
\multicolumn{6}{c}{\cellcolor{orange!15}\textbf{Web-Embodied Intelligence}} \\
\midrule
WebArena~\cite{zhou2023webarena} & 2024 & Web tasks & 800+ tasks & Task completion & Multi-domain platform \\
VisualWebArena~\cite{koh2024visualwebarena} & 2024 & Visual web tasks & 910 tasks & Success vs human & Visual + navigation \\
WebVoyager~\cite{webvoyager2023} & 2024 & End-to-end agent & 15 websites & GPT-4V eval & Screenshot-based \\
Agent-E~\cite{abuelsaad2024agent} & 2024 & Hierarchical agent & WebVoyager & Error recovery & Planner + Navigator \\
\midrule
Flow2Code~\cite{he2025flow2code} & 2025 & Flowchart→Code & 15 languages & Logic accuracy & Flowchart understanding \\
ArtifactsBench~\cite{zhang2025artifactsbench} & 2025 & Code$\to$Artifacts & Multi-type & Auto eval & Programmatic rendering \\
MMCode~\cite{li2024mmcode} & 2024 & Visual programming & Rich visuals & Programming ability & Multi-visual elements \\
HumanEval-V~\cite{zhang2024humaneval} & 2024 & Visual reasoning & Complex diagrams & High-level reasoning & Complex chart focus \\
MM-Coder~\cite{islam2024mapcoder} & 2025 & Design→Code & Multi-language & Code quality & UML + Flowchart \\
\bottomrule
\end{tabular}
}
\end{table*}

\begin{table}[ht]
\centering
\caption{Evolution of Evaluation Metrics in Multimodal Code Generation}
\label{tab:evaluation-evolution}
\footnotesize
\begin{tabular}{l|l|l}
\toprule
\textbf{Generation} & \textbf{Metrics} & \textbf{Focus} \\
\midrule
\textbf{Traditional} & BLEU~\cite{papineni2002bleu}, Code similarity~\cite{chen2024survey_code_evaluation} & Syntactic correctness \\
\textbf{Structural} & TreeBLEU~\cite{gui2025webcode2m}, DOM-ED~\cite{suri2024docedit} & Hierarchical structure \\
\textbf{Multimodal} & Visual fidelity~\cite{screencoder}, Rendering similarity~\cite{chen2025learning} & Cross-modal alignment \\
\textbf{Automated} & Code-in-the-Loop, MLLM judging~\cite{zhang2025artifactsbench} & End-to-end evaluation \\
\bottomrule
\end{tabular}
\end{table}

\begin{table}[ht]
\centering
\caption{Key Technical Trends in Multimodal Code Generation}
\label{tab:technical-trends}
\footnotesize
\resizebox{0.9\textwidth}{!}{
\begin{tabular}{l|l|l}
\toprule
\textbf{Trend} & \textbf{Evolution} & \textbf{Representative Works} \\
\midrule
\textbf{Agent Workflows} & Plan$\to$Execute$\to$Observe$\to$Reflect & \makecell[l]{nvAgent, Agent-E, \\ Frontend Diffusion} \\
\textbf{Self-Correction} & One-shot$\to$Iterative optimization & \makecell[l]{UICoder, DesignCoder, \\ ChartIR, ReLook} \\
\textbf{Hierarchical Generation} & Flat$\to$Multi-level decomposition & \makecell[l]{UICopilot, DesignCoder, \\ WebDreamer} \\
\textbf{Code-in-the-Loop} & Static evaluation$\to$Dynamic rendering & \makecell[l]{ArtifactsBench, UICoder} \\
\bottomrule
\end{tabular}}
\end{table}

\tikzstyle{my-box}=[
    rectangle,
    draw=hidden-draw,
    rounded corners,
    text opacity=1,
    minimum height=1.5em,
    minimum width=5em,
    inner sep=2pt,
    align=center,
    fill opacity=.5,
    line width=0.8pt,
]
\tikzstyle{leaf}=[my-box, minimum height=1.5em,
    fill=hidden-pink!80, text=black, align=left,font=\normalsize,
    inner xsep=2pt,
    inner ysep=4pt,
    line width=0.8pt,
]
\begin{figure*}[H]
	\centering
    \resizebox{\textwidth}{!}{
	\begin{forest}
        forked edges,
		for tree={
                grow=east,
                reversed=true,
                anchor=base west,
                parent anchor=east,
                child anchor=west,
                base=center,
                font=\large,
                rectangle,
                draw=hidden-draw,
                rounded corners,
                align=left,
                text centered,
                minimum width=4em,
                edge+={darkgray, line width=1pt},
                s sep=3pt,
                inner xsep=2pt,
                inner ysep=3pt,
                line width=0.8pt,
                ver/.style={rotate=90, child anchor=north, parent anchor=south, anchor=center},
            },
            where level=1{text width=8em,font=\normalsize, }{},
            where level=2{text width=12em,font=\normalsize}{},
            where level=3{text width=14em,font=\normalsize,}{},
            where level=4{text width=22em,font=\normalsize,}{},
	    [Multimodal Code \\Generation, ver
			[Visual Interface \\Code Generation
                [Screenshot-to-Code
                    [Foundational Works
                        [
                            pix2code~\cite{beltramelli2018pix2code}{, }Design2Code~\cite{si2024design2code}{, }\\Web2Code~\cite{yun2024web2code}{, }UICoder~\cite{wu2024uicoder}{, }\\Prototype2Code~\cite{xiao2024prototype2code}
                           , text width=26em
                        ]
                    ]
                ]
                [Sketch-to-Code
                    [Hand-drawn Interfaces
                        [
                            Sketch2Code\cite{li2024sketch2code}{, }Designing with Language\cite{feng2023designing}{, }\\ML Sketches and Visual Code Assistants\cite{gomes2024exploratory}
                           , text width=26em
                        ]
                    ]
                ]
                [Interactive \\Prototype-to-Code
                    [Dynamic UI Generation
                        [
                            Interaction2Code~\cite{xiao2024interaction2code}{, }SWE-bench Multimodal~\cite{yang2024swe}{, }\\CodeV~\cite{zhang2024codev}
                           , text width=26em
                        ]
                    ]
                ]
                [Hierarchical Code \\Generation
                    [Advanced Methods
                        [
                            UICopilot~\cite{gui2025uicopilot}{, }MLLM-Based UI2Code~\cite{wu2025mllm}{, }\\DesignCoder~\cite{chen2025designcoder}{, }Frontend Diffusion~\cite{ding2025frontend}{, }ReLook~\cite{li2025relook}
                           , text width=26em
                        ]
                    ]
                ]
			]
			[Data Visualization \\Code Generation
                [Scientific Plot \\Code Generation
                    [Chart Understanding
                        [
                            Plot2Code~\cite{wu2024plot2code}{, }ChartCoder~\cite{zhao2025chartcoder}{, }VisCoder~\cite{ni2025viscoder}{, }\\ChartIR~\cite{bouhlal2024assembly}{, }AskChart~\cite{yang2024askchart}{, }ChartInsight~\cite{wu2024chartinsight}
                           , text width=26em
                        ]
                    ]
                ]
                [Agent-based \\Visualization
                    [Multi-Agent Systems
                        [
                            nvAgent~\cite{ouyang2025nvagent}{, }MatPlotAgent~\cite{ouyang2025nvagent}{, } DeepVis~\cite{shuai2025deepvis}{, }\\Step-Text2Vis~\cite{nvbench2-2025luo}{, }METAL~\cite{ouyang2025nvagent}
                           , text width=26em
                        ]
                    ]
                ]
                [Cross-modal Chart \\Understanding
                    [Reasoning Evaluation
                        [
                            ChartMimic~\cite{yang2024chartmimic}
                           , text width=26em
                        ]
                    ]
                ]
            ]
            [Web Agent \\Interaction
                [Task-specific \\Environments
                    [Specialized Platforms
                        [
                            WebShop~\cite{yao2022webshop}{, }WebArena~\cite{zhou2023webarena}{, }\\Reinforcement Learning on Web Interfaces~\cite{liu2018reinforcement}
                           , text width=26em
                        ]
                    ]
                ]
                [End-to-End \\Navigation
                    [Multimodal Navigation
                        [
                            Multimodal Web Navigation~\cite{furuta2023multimodal}{, }WebLINX~\cite{furuta2023multimodal}{, }\\WebVoyager~\cite{he2024webvoyager}
                           , text width=26em
                        ]
                    ]
                ]
                [Agent Architectures
                    [Reasoning Frameworks
                        [
                            ReAct~\cite{yao2023react}{, }Agent-E~\cite{abuelsaad2024agent}{, }\\WebDreamer~\cite{gu2024your}{, }Toolformer~\cite{schick2023toolformer}
                           , text width=26em
                        ]
                    ]
                ]
                [Multi-Agent \\Systems
                    [Collaborative Agents
                        [
                            AgentVerse~\cite{chen2023agentverse}{, }Voyager~\cite{wang2023voyager}{, }\\Generative Agents~\cite{park2023generative}
                           , text width=26em
                        ]
                    ]
                ]
            ]
            [Software \\Engineering \\Artifacts
                [Flowcharts \& UML
                    [Diagram Generation
                        [
                            DiagrammerGPT~\cite{zala2023diagrammergpt}{, }Unified Modeling Language~\cite{bates2025unified}{, }\\Flow2Code~\cite{he2025flow2code}{, }Code-Vision~\cite{wang2025code}{, }\\Multilingual Multimodal Software Developer~\cite{chai2025multilingual}
                           , text width=26em
                        ]
                    ]
                ]
                [Scientific Diagram \\Reconstruction
                    [Visual Understanding
                        [
                            Draw with Thought~\cite{cui2025draw}{, }From Text to Visuals~\cite{lee2025text}
                           , text width=26em
                        ]
                    ]
                ]
                [Comprehensive \\Programming Evaluation
                    [Benchmark Development
                        [
                            MMCode~\cite{li2024mmcode}{, }HumanEval-V~\cite{zhang2024humaneval}
                           , text width=26em
                        ]
                    ]
                ]
            ]
            [Evaluation \\Methodology
                [Automated \\Evaluation
                    [Multi-modal Assessment
                        [
                            ArtifactsBench~\cite{zhang2025artifactsbench}
                           , text width=26em
                        ]
                    ]
                ]
                [Hierarchical \\Metrics
                    [Structural Evaluation
                        [
                            TreeBLEU\cite{si2024design2code, chen2025designcoder}{, }DOM-ED\cite{si2024design2code, wu2025mllm}{, }\\
                            Visual Fidelity Metrics\cite{lee2025text, yang2024chartmimic}
                           , text width=26em
                        ]
                    ]
                ]
                [Closed-loop \\Feedback
                    [Rendering-based \\Evaluation
                        [
                            Compile-Render-CLIP Pipeline~\cite{wu2024uicoder}{, }\\
                            Browser Screenshot Validation~\cite{chen2025designcoder, xiao2024prototype2code}
                           , text width=26em
                        ]
                    ]
                ]
            ]
		]
	\end{forest}}
	\caption{Taxonomy of Multimodal Code Generation. This comprehensive framework categorizes the field into four main domains: Visual Interface Code Generation (focusing on converting visual designs to executable code), Data Visualization Code Generation (emphasizing chart and plot understanding), Web Agent Interaction (enabling autonomous web navigation and task completion), and Software Engineering Diagram Generation (creating technical diagrams and documentation). The evaluation methodology spans automated assessment, hierarchical metrics, and closed-loop feedback systems, reflecting the field's evolution from static code generation to dynamic, intelligent agent-driven interactions.}
    \label{fig:taxonomy-multimodal-code}
\end{figure*} 
With the rapid development of vision-language multimodal large language models (MLLMs), code generation research is expanding from pure text input to a new paradigm that incorporates multiple modalities such as images, sketches, and interactive signals. Multimodal Code Generation (MCG) aims to enable models to understand visual design intentions and generate executable, renderable high-quality code, thereby bridging the gap between high-level abstract thinking and low-level code implementation.

\subsubsection{Core Challenges and Technical Positioning}

Multimodal code generation faces two unified challenges: \textbf{Fidelity} - how to achieve high-fidelity restoration of visual details, structural hierarchies, and functional semantics in cross-modal transformations; and \textbf{Executability} - how to ensure that generated code is syntactically correct, renders without errors, and is functionally complete. These challenges drive researchers to explore new model architectures, training strategies, and evaluation paradigms.

Based on application scenarios and technical focus, this section divides multimodal code generation into three core subfields: frontend interface generation, web-embodied intelligence, and software engineering artifact generation.

\subsubsection{Frontend Interface Generation}\label{sec:frontend-interface-generation}

\paragraph{Task Evolution Trajectory}

Frontend interface generation, as a pioneering field in multimodal code generation, has undergone an evolution from single modality to multimodality, and from static to dynamic.

Early screen-to-code systems such as pix2code \cite{beltramelli2017pix2code} demonstrated the feasibility of mapping GUI screenshots to platform-specific code, establishing the foundational paradigm for subsequent research. This pioneering work validated that deep learning approaches could bridge the gap between visual UI representations and executable code.
    
\begin{itemize}

\item \textbf{Image-to-Code} represents the starting point of this field. pix2code~\cite{beltramelli2018pix2code} pioneer the use of CNNs to extract GUI screenshot features combined with LSTMs to generate corresponding platform code.

\item \textbf{Design-to-Code} represents standardization efforts in this field. Design2Code~\cite{si2024design2code} built a large-scale benchmark containing 43,000 webpage screenshot-HTML pairs, systematically evaluate 9 MLLMs (multimodal large language models), and proposed hierarchical evaluation metrics such as TreeBLEU~\cite{gui2025webcode2m}, DOM-ED~\cite{suri2024docedit}. The study finds that even GPT-4V still has many label omissions in maintaining hierarchical structures, highlighting the severity of fidelity challenges. Prototype2Code~\cite{xiao2024prototype2code} further utilize Figma APIs to build a 9k real prototype$\to$React code dataset, proposing a pipeline consisting of hierarchical layout trees, component library retrieval, and incremental repair.

\item \textbf{Sketch-to-Code} explores more natural interaction methods. Sketch2Code~\cite{li2024sketch2code} released the first benchmark containing 731 high-quality hand-drawn sketches and designed two interactive evaluation modes: ``passive feedback acceptance'' and ``active questioning''. The researchers find that although current models perform poorly in active questioning, this mode is more favored by UI/UX experts. WireGen~\cite{feng2023designing} utilizes generative LLMs to automatically generate medium-fidelity wireframes from simple design intent descriptions. \cite{gomes2024exploratory} further explored the feasibility of integrating visual code assistants in IDEs, validating the practical potential of sketch-to-code generation. Sketches are quick, usually hand-drawn UI drawings used early in design to explore ideas and communicate layout concepts without worrying about visual polish.
Wireframes are more structured, low-/medium-fidelity blueprints that define the information hierarchy, layout, and interaction flows of a screen while intentionally omitting final styling and detailed content.

\item \textbf{Interaction-to-Code} extends tasks to the dynamic level. Interaction2Code~\cite{xiao2024interaction2code} defined a new task paradigm, releasing a large benchmark containing 127 webpages, 374 interactions, and 31 event types, systematically revealing four major failure modes of SOTA MLLMs in interaction generation: event omission, logical errors, detail confusion, and visual detail loss.

\end{itemize}

\paragraph{Key Methodological Innovations}
Hierarchical Generation and Layout Modeling has become the mainstream approach for handling complex UI structures. UICopilot~\cite{gui2025uicopilot} splits HTML generation into a two-stage process of coarse-grained hierarchical skeleton followed by fine-grained tags and CSS, significantly reducing MLLM context length. LayoutCoder~\cite{wu2025mllm} introduces element relationship graphs and layout tree prompts to better capture complex grid and float layouts. Further advancing this paradigm, DesignCoder~\cite{chen2025designcoder} proposes a UI Grouping Chain to automatically group mock-ups into nested hierarchies, which is combined with a divide-and-conquer decoding strategy to enhance performance on industrial datasets.

\begin{itemize}
\item  \textbf{Automated feedback and self-correction loops} significantly improve generation quality. UICoder~\cite{wu2024uicoder} pioneered compile-render-CLIP triple automatic feedback, filtering 3M LLM synthetic data for fine-tuning, enabling 7B-13B open-source models to achieve 12-18 point BLEU improvements on the Design2Code benchmark, approaching GPT-4V performance. DesignCoder's self-inspection module uses browser screenshots + A Vision Encoder to detect and fix missing styles and invalid logic. ChartIR~\cite{bouhlal2024assembly} transfers this idea to chart generation through a two-stage process of initial generation and optimization to gradually improve code quality. ReLook~\cite{li2025relook} introduces a reinforcement learning framework that closes the generate–diagnose–refine loop, leveraging a multimodal LLM as a vision-grounded critic and coach. By internalizing this self-correction capability during training, the model can perform low-latency self-editing at inference, even without relying on an external critic.

\item  \textbf{Agentic workflows} represent a shift toward intelligent workflows. Frontend Diffusion~\cite{ding2025frontend} proposes a Sketch-to-PRD (product requirements document)-to-Code three-stage agent chain, combining Pexels\footnote{\url{https://www.pexels.com/}} image retrieval with Claude-3.5, validating the feasibility of LLM-Agents in ``Code$\to$Rende$\to$Self-Assessmen$\to$
Revision'' closed loops. User studies reveal AI-human bidirectional alignment requirements and the authors propose prompt layering and visual iterative interfaces.
\end{itemize}

\paragraph{Evaluation System Development}
The rapid development of this field benefits from continuous improvement of high-quality benchmarks. Design2Code not only provides large-scale data but also breaks through limitations of traditional sequence metrics like BLEU, proposing hierarchical metrics such as TreeBLEU~\cite{gui2025webcode2m}, DOM-ED~\cite{suri2024docedit}. Web2Code~\cite{yun2024web2code} builds large-scale webpage-to-code datasets, proposing comprehensive evaluation frameworks including webpage understanding and code generation, with experiments proving that this dataset improves not only webpage task performance but also general vision tasks. To avoid data leakage, the Snap2Code dataset specifically distinguishes between seen and unseen sites to ensure evaluation fairness.

Recent evaluation efforts have broadened the scope of UI understanding tasks. WebUIBench \cite{lin2025webuibench} offers a comprehensive benchmark covering element classification, visual grounding, OCR, layout understanding, and code generation, providing a holistic assessment framework for WebUI-to-Code capabilities. Complementary directions explore grammar-guided layout generation, where LLMs act as high-level planners for UI composition \cite{lu2023uilayoutgenerationllms}, and HTML structure understanding for web automation \cite{gur2023understandinghtmlwithllms}. 

These developments suggest that general LLMs can increasingly bridge visual comprehension with front-end engineering. However, achieving robustness in fine-grained DOM (document object model) grounding and generating faithful, production-quality code remains a key challenge for the field.

\subsubsection{Web-Embodied Intelligence}\label{sec:web-arena}

\paragraph{Problem Definition and Challenges}

Web-embodied intelligence transcends the scope of static code generation, requiring AI agents to complete complex tasks in real web environments through ``observe-reason-act'' loops. These tasks not only test code generation capabilities but also require comprehensive reasoning, planning, and environmental interaction abilities.

\paragraph{Methodological Development Timeline}

\begin{itemize}

\item  \textbf{Stage 1: Foundational Frameworks} provide theoretical foundations for this field. ReAct~\cite{yao2023react} proposes the ``reasoning-acting'' interleaved paradigm, becoming the cornerstone of modern agent architectures. Models first perform Chain-of-Thought reasoning, then execute actions, observe results, and proceed to the next step of thinking, effectively reducing hallucination phenomena. Workflow-Guided Exploration~\cite{liu2018reinforcement} guides reinforcement learning exploration through human workflows, laying foundations for learning in complex web environments. Toolformer~\cite{schick2023toolformer} proposes methods for language models to self-learn tool usage, providing theoretical support for agent tool-use capabilities.

\item \textbf{Stage 2: Task-Specific Platforms} drive practical progress. WebShop~\cite{yao2022webshop} creates simulated online shopping website environments specifically for evaluating multimodal understanding and multi-step reasoning capabilities. WebArena~\cite{zhou2023webarena} provides more realistic and complex open platforms, including functionally complete environments replicated from real websites and over 800 long-term planning tasks, greatly promoting general web agent research.

\item \textbf{End-to-End Multimodal Navigation} achieves significant breakthroughs. WebGUM~\cite{furuta2023multimodal} innovatively combines transformer language models (T5) and vision models (ViT), capable of simultaneously processing webpage HTML text and screenshot information. WebVoyager~\cite{he2024webvoyager} is a milestone work in this field, first achieving end-to-end multimodal web agents that directly use screenshots as input to simulate human browsing behavior, and innovatively uses GPT-4V for automated evaluation. WebLINX~\cite{furuta2023multimodal} releases large-scale benchmarks supporting multi-turn dialogue, containing over 100,000 expert demonstrations recorded on 150+ real websites.

\item \textbf{Stage 3: Game Environments as Open-Ended Testbeds} provide rich, interactive settings for developing and evaluating multimodal agent capabilities. In Minecraft, video pre-training (VPT) \cite{baker2022videopretrainingvpt} learns to act from large-scale human videos using a small labeled set for inverse dynamics, demonstrating long-horizon control from native mouse–keyboard interfaces. MineDojo \cite{fan2022minedojo} contributes an internet-scale knowledge base plus a simulation suite, enabling language-conditioned agents and reward shaping from video–language models. Building on this ecosystem, Voyager \cite{wang2023voyager} uses an LLM to drive open-ended exploration and lifelong skill acquisition via an evolving library of executable programs. These results indicate that general LLMs, when augmented with perception and code execution, can acquire reusable competencies in complex environments. Complementing these interactive game-playing agents, V-GameGym~\cite{zhang2025vgamegym} addresses the inverse problem of visual game generation, providing a comprehensive benchmark of 2,219 Pygame samples with multimodal evaluation across code correctness, visual quality, and gameplay dynamics.

\item \textbf{Stage 4: Advanced Agent Architectures} improve complex task handling capabilities. Agent-E~\cite{abuelsaad2024agent} proposes hierarchical agent architectures, decomposing complex web tasks into ``planner'' and ``browser navigator'', enabling agents to more effectively handle long-term tasks, error recovery, and backtracking. WebDreamer~\cite{gu2024your} innovatively uses LLMs as internet world models, first simulating consequences of each possible action, then evaluating and selecting optimal paths, effectively reducing trial-and-error costs on real websites.

\item \textbf{Stage 5: Multi-Agent Collaboration} explores swarm intelligence. AgentVerse~\cite{chen2023agentverse} provides general multi-agent collaboration frameworks, improving problem-solving efficiency through dynamic task decomposition, agent assignment, and collaborative execution. Voyager~\cite{wang2023voyager} achieves lifelong learning in Minecraft, demonstrating possibilities for open-ended continuous learning. Generative Agents~\cite{park2023generative} creates virtual towns with 25 generative agents, simulating credible human behavior and pioneering large-scale social behavior simulation.

\item \textbf{Stage 6: Tool-Augmented Multimodal Reasoning} enables LLMs to leverage specialized visual modules for complex perception-action tasks. Systems such as Visual ChatGPT \cite{wu2023visualchatgpt} and MM-ReAct \cite{yang2023mmreact} orchestrate calls to specialist vision models under LLM control for multi-step tasks, demonstrating the effectiveness of modular architectures. Code-driven assemblers like ViperGPT \cite{suris2023vipergpt} compose VLM modules via generated Python, yielding explicit, verifiable pipelines for complex queries. For end-to-end interactive evaluation, benchmarks like Mind2Web \cite{deng2023mind2web} target general web instruction following, while AndroidWorld \cite{rawles2025androidworld} focuses on smartphone tasks. These studies highlight that long-horizon perception-to-action competence remains challenging even for top models, underscoring open problems in UI grounding, memory, safety, and reliability.
\end{itemize}

\paragraph{Evaluation Environment Evolution}

The progression of this field relies on a foundation of robust evaluation environments, which require constant iteration and improvement. From WebShop's~\cite{yao2022webshop} closed shopping scenarios to WebArena's~\cite{zhou2023webarena,zhou2023webarena2} open multi-domain websites, to WebVoyager's~\cite{webevolver} innovative evaluation paradigm using GPT-4V image scoring, evaluation dimensions have expanded from text correctness to comprehensive assessment of visual understanding, long-term planning, and interaction robustness.

\subsubsection{Software Engineering Artifact Generation}\label{sec:artifacts}

\paragraph{Task Value and Positioning}

Software engineering artifact generation aims to create supporting artifacts for code, including UML diagrams, data charts, flowcharts, architecture diagrams, and other visual software engineering documents. These tasks have important value throughout the software engineering lifecycle: business logic understanding in the requirements phase, architecture visualization in the design phase, supporting documentation generation in the development phase, and system understanding support in the maintenance phase.

\paragraph{Sub-task Classification and Progress}

\begin{itemize}
\item \textbf{Data Visualization Generation} is the most active branch in this field. nvAgent~\cite{ouyang2025nvagent} proposes a multi-agent collaborative natural language to visualization system, utilizing four roles - an insight miner, visualization recommender, code generator, and narrative generator - for collaboration, achieving SOTA performance on the nvBench~\cite{nvBench2021} benchmark. Similarly, DeepVis~\cite{shuai2025deepvis} propose an interactive visual interface that tightly integrates with the CoT reasoning process, allowing users to inspect reasoning steps, identify errors, and make targeted adjustments to improve visualization outcomes.
Plot2Code~\cite{wu2024plot2code} establishes a comprehensive benchmark for evaluating multimodal large language models in code generation from scientific plots. ChartCoder~\cite{zhao2025chartcoder} advances multimodal large language models for chart-to-code generation through instruction fine-tuning datasets. VisCoder~\cite{ni2025viscoder} focuses on fine-tuning LLMs for executable Python visualization code generation. MatPlotAgent~\cite{ouyang2025nvagent} explores single-agent data science visualization methods, while METAL~\cite{ouyang2025nvagent} investigates multi-agent frameworks for chart generation with test-time scaling. ChartMimic~\cite{yang2024chartmimic} emphasizes the cross-modal reasoning difficulty of Chart-to-Code, providing important evaluation benchmarks for this subfield.
Beyond recognition and captioning, chart understanding increasingly targets \emph{executable} reconstruction by producing plotting code that faithfully reproduces the input figure. ChartMimic \cite{yang2025chartmimic} frames this as a chart-to-code benchmark with multi-level automatic metrics, revealing substantial headroom even for strong proprietary models. Dedicated models like ChartCoder \cite{zhao2025chartcoder} pair code-centric backbones with large-scale chart-to-code corpora to boost executability and visual fidelity. Broader chart reasoning datasets such as ChartBench \cite{xu2024chartbench} probe complex visual–logical skills that underlie chart regeneration and editing. For code-centric multimodal pipelines, these tasks tie perception to verifiable outputs, offering precise failure signals that are valuable for iterative improvement.
Furthermore, nvBench 2.0~\cite{nvbench2-2025luo} extend the task to ambiguous scenarios, defining patterns of ambiguity in natural language queries for visualizations and resolving said ambiguity through step-wise reasoning.

\item \textbf{Software Diagram and Model Generation} covers broader software engineering scenarios. DiagrammerGPT~\cite{zala2023diagrammergpt} generates open-domain diagrams from natural language through plan and review dual loops. Draw with Thought~\cite{cui2025draw} uses chain-of-thought to reconstruct scientific diagrams into editable XML code. Flow2Code~\cite{he2025flow2code} evaluates flowchart-to-code mapping capabilities, covering 15 programming languages. Code-Vision~\cite{wang2025code} focuses on flowchart-to-program logic reasoning. Unified UML Generation~\cite{bates2025unified} explores automatic UML code generation from UML images. From Text to Visuals~\cite{lee2025text} converts mathematical problem text to SVG graphics, completing mathematical and scientific diagram sub-tasks. MM-Coder~\cite{chai2025multilingual} proposes a multilingual multimodal software developer for code generation that can jointly understand software design diagrams (UML, flowcharts) and textual descriptions.

\item \textbf{Comprehensive Multimodal Software Engineering Tasks} reflect systematic development in this field. SWE-bench Multimodal~\cite{yang2024swe} targets visual bug fixing in JavaScript software development, while CodeV~\cite{zhang2024codev} addresses issue resolving with visual data, particularly chart-related bug fixing methods. MMCode~\cite{li2024mmcode} benchmarks multimodal large language models for code generation with visually rich programming problems, and HumanEval-V~\cite{zhang2024humaneval} benchmarks high-level visual reasoning with complex diagrams in coding tasks. Both incorporate visual elements such as trees, graphs, charts, tables, and pseudocode into programming evaluation, building more comprehensive multimodal programming benchmarks.
\end{itemize}

\paragraph{Evaluation Paradigm Innovation}
ArtifactsBench~\cite{zhang2025artifactsbench} proposes a breakthrough automated multimodal evaluation paradigm, solving the pain point of traditional evaluation ignoring visual fidelity and interaction completeness. This framework programmatically renders visual artifacts, captures dynamic behavior, then uses MLLM judges to comprehensively score code, visuals, and interactions according to task checklists. The work focus on: (1) Automated closed loop - completing the full process from code generation to quality assessment without human intervention; (2) Multi-dimensional evaluation - considering code correctness, visual fidelity, and interaction completeness; (3) Good scalability - supporting unified evaluation of different types of software artifacts.

\subsubsection{Technical Trends and Future Outlook}

Through systematic analysis of three core areas in multimodal code generation, we identify four key technical trends:

\paragraph{Widespread Application of Agentic Workflows} \autoref{tab:technical-trends} shows that ``plan$\to$execute$\to$observe \\ $\to$reflect'' loop has become the mainstream paradigm for solving complex multimodal tasks. From nvAgent's multi-agent collaboration to Agent-E's hierarchical architecture to Frontend Diffusion's three-stage agent chain, all reflect the core value of the agent paradigm in handling multi-step, multi-modal information fusion tasks.

\paragraph{Maturation of Self-Correction and Iterative Optimization Mechanisms} The growing ability of models to autonomously detect and rectify errors has emerged as a pivotal technology for enhancing robustness, as evidenced by self-correction mechanisms like UICoder's ``compile–render–CLIP'' triple feedback~\cite{wu2024uicoder}, DesignCoder's browser screenshot self-inspection~\cite{chen2025designcoder}, ChartIR's structured instruction iteration~\cite{chartIR}, and ReLook's ``generate–diagnose–refine'' loop~\cite{li2025relook}. These approaches collectively underscore the critical role of iterative self-improvement, marking a significant paradigm shift from ``one-shot generation'' to ``iterative optimization.''

\paragraph{Standardization of Code-in-the-Loop Evaluation} Incorporating compilers and rendering engines as components of evaluation environments to achieve immediate, automated functional and visual feedback is becoming an important development direction in this field. ArtifactsBench's programmatic rendering evaluation and UICoder's rendering feedback mechanisms both reflect this trend, not only improving evaluation objectivity but also providing reliable feedback signals for model self-correction.

\paragraph{Widespread Adoption of Hierarchical Generation Strategies} Facing complex multimodal tasks, decomposing the generation process into multiple levels or stages has become the mainstream approach. UICopilot's two-stage generation, DesignCoder's hierarchical structure awareness, and WebDreamer's world model planning all effectively alleviate long sequence generation difficulties and improve generation quality and controllability.

Despite substantial progress, several fundamental challenges persist across the multimodal code generation landscape~\cite{li2024mmcode,viscodex,robocodex}. Fine-grained UI hierarchy understanding and interaction semantics remain difficult, particularly for complex DOM or canvas manipulations. Multi-step perception-to-action planning still suffers from robustness issues in real-world deployment. Analyses of multimodal hallucination and grounding errors continue to motivate the development of domain-aligned data, structured representations, and executable-by-design evaluations. The field is converging toward evaluation paradigms that emphasize verifiable outputs (such as chart or code regeneration) rather than purely perceptual metrics, particularly as applications move toward professional software engineering settings. This shift reflects a broader maturation from proof-of-concept demonstrations to production-ready systems.

Looking ahead, multimodal code generation will achieve deepened development in three dimensions: \textbf{Stronger Agent Capabilities} - including significant improvements in long-term planning, environmental adaptation, and error self-healing abilities; \textbf{More Complete Evaluation Systems} - building multi-dimensional, automated, and low-cost Code-in-the-Loop standards; \textbf{Richer Application Scenarios} - expanding from web development to multi-platform including mobile and desktop, and deep integration with human-computer collaboration.

Multimodal code generation is gradually converging toward the grand vision of ``digital world embodied intelligence''~\cite{code_world_model} - enabling AI to observe, think, program, and act, laying a solid foundation for the comprehensive release of next-generation software productivity. As technology continues to mature and application scenarios continue to expand, this field will become an important bridge connecting human creativity and machine execution power, driving fundamental transformation in software development paradigms.

\subsection{Task-based Overview of Reinforcement Learning in Code Intelligence}

\subsubsection{Reinforcement Learning (RL) Algorithms}
\label{sec:code_rl}
For code LLMs, RL algorithms~\cite{rl_survey} follow the specific pattern (compiles$\to$runs$\to$passes tests) by incorporating executable program-level rewards, such as unit-test pass rates, compiler/runtime errors, static-analysis flags, and verifcriticalitical feedback. This closes the loop between generation and external tools and directly optimizes correctness and reasoning depth. The proximal policy optimization (PPO) methods enable online exploration with KL-regularized policy updates that balance diversity against fidelity to a supervised prior, often coupled with execution-guided rewards and repair loops (generate$\to$run$\to$debug) to boost pass@k and reduce hallucinations \cite{ppo, zheng2024critic, yang2024learning, yang2024chain, li2025structured}. The direct Preference optimization (DPO) circumvents the need for online rollouts by learning from pairwise preferences distilled from execution-graded samples (success vs. failure), yielding stable alignment for code style and step-by-step reasoning without the variance of policy-gradient estimates \cite{rafailov2023direct, zuccotto2024reinforcement, ning2024survey}. Reinforcement Learning from AI Feedback (RLAIF) further leverages stronger verifier/critic LMs to supply dense, structured signals (critiques, edits, step-level checks), enabling self-verification and iterative refinement that particularly benefit long CoT traces in math and programming \cite{zheng2024critic, zuccotto2024reinforcement, gu2024review}. Large-scale post-training (e.g., o1 and R1) integrates these ingredients (execution-aware rewards, verifier guidance, and preference optimization) to elicit robust tool use, decomposition, and self-correction behaviors that translate into sizable accuracy gains on code-generation and reasoning benchmarks \cite{openai2024gpt4technicalreport, guo2025deepseek, yang2024chain}.

\paragraph{Proximal Policy Optimization Methods and Variants}
\begin{itemize}
\item \textbf{PPO}~\cite{ppo} is the standard policy-gradient algorithm for RL-based LLM fine-tuning. It balances policy improvement and stability by constraining each update via a clipped surrogate objective:

\begin{MiddleEquation}
\begin{equation}
\begin{aligned}
L_{\text{PPO}}(\theta) \;=\; \mathbb{E}_{t}\Big[ \min\!\Big(r_t(\theta)\,\hat{A}_t,\; \text{clip}\big(r_t(\theta),\,1-\epsilon,\,1+\epsilon\big)\,\hat{A}_t\Big)\Big],
\end{aligned}
\end{equation}
\end{MiddleEquation}where 
$
r_t(\theta) = \frac{\pi_\theta(a_t \mid s_t)}{\pi_{\theta_{\text{old}}}(a_t \mid s_t)}
$ is the probability ratio, and $\hat{A}_t$ the advantage estimate. Clipping with $\epsilon\!\in\![0.1,0.2]$ stabilizes training by preventing large updates. For advantage estimation, PPO typically uses Generalized Advantage Estimation (GAE)~\cite{schulman2015high}:

\begin{MiddleEquation}
\begin{equation}
\begin{aligned}
\hat{A}_t \;=\; \sum_{l=0}^{\infty} (\gamma\lambda)^l\,\delta_{t+l}, \qquad 
\delta_t = r_t + \gamma V(s_{t+1}) - V(s_t),
\end{aligned}
\end{equation}
\end{MiddleEquation}where $\lambda$ controls the bias–variance trade-off. In practice, PPO is combined with KL regularization~\cite{bai2022training,van2014renyi} to keep $\pi_\theta$ close to the reference model, preventing reward hacking (e.g., verbosity, repetition). This combination underpins RLHF pipelines such as InstructGPT~\cite{ouyang2022training}, which fine-tuned GPT-3~\cite{brown2020language} with a reward model trained from human labels, yielding significant alignment improvements. PPO has since been widely adopted (e.g., WebGPT~\cite{nakano2021webgpt}). However, vanilla PPO struggles on sparse-reward reasoning tasks due to \emph{value collapse} (when the value network becomes degenerate and outputs nearly constant predictions across different states, losing its ability to meaningfully distinguish state values and guide policy improvement), where critic training fails. For example, \citet{yuan2025s} observed near-zero accuracy on long-CoT math benchmarks. To overcome this, several PPO-inspired variants were proposed.
    
\item \textbf{Group Relative Policy Optimization (GRPO)}~\cite{shao2024deepseekmath} avoids value learning by sampling $G$ outputs for a prompt and computing relative advantages $A_i = r_i - \mu_{\text{group}}$. This yields stable updates without a critic and reduced memory overhead, proving effective in mathematical reasoning tasks. Deriving from similar ideas, there are akin research that try to improve the group baseline~\cite{li2024remax}, rollout efficiency~\cite{li2025treepo}, and group diversity~\cite{zheng2025fr3e}.

\item \textbf{Dr. GRPO}~\cite{liu2025understandingr1zeroliketrainingcritical} addresses a critical bias found in GRPO used in reinforcement learning for LLMs. Standard GRPO suffers from an optimization bias that artificially inflates response length during training, particularly for incorrect outputs, leading to inefficient token usage. Dr. GRPO corrects this bias by providing a more balanced optimization approach that maintains reasoning performance while significantly improving token efficiency. 

\item \textbf{Decoupled Clip and Dynamic Sampling Policy Optimization (DAPO)}~\cite{yu2025dapo} improves PPO/GRPO with engineering refinements: (i) \emph{Clip-Higher} to maintain entropy, (ii) \emph{Dynamic Sampling} to adapt $G$, (iii) \emph{Token-Level Loss} for long-horizon credit assignment, and (iv) \emph{Overlong Reward Shaping} to penalize verbosity. DAPO achieved $50\%$ accuracy on AIME 2024 with a 32B model, surpassing DeepSeek-R1 at half the training cost~\cite{tan2025gtpo}.

\item \textbf{Value-based Augmented PPO (VAPO)}~\cite{yue2025vapo,gao2024regressing} tackles three critical challenges that arise in value model based reinforcement learning. These challenges include value model bias over long sequences, heterogeneous sequence lengths during training, and sparse reward signals in verifier based tasks. To address these issues, VAPO integrates several key techniques including length adaptive Generalized Advantage Estimation, value pretraining to reduce initialization bias, and enhanced strategies for balancing exploration and exploitation during training.

\item \textbf{REINFORCE++}~\cite{hu2025reinforcestabilizingcriticfreepolicy}, a critic-free reinforcement learning framework, is designed to improve the efficiency and stability of reinforcement learning from human feedback (RLHF) for LLMs. While current methods like PPO require computationally expensive critic networks, and existing critic-free alternatives (e.g., GRPO~\cite{fu2025posteriorgrpo}) suffer from biased advantage estimation due to prompt-level normalization, REINFORCE++ addresses these issues through global advantage normalization by normalizing advantages across entire global batches rather than small, prompt-specific groups. REINFORCE++ is proposed for broad RLHF applications and complex reasoning tasks, demonstrating superior stability and performance compared to existing methods.

\end{itemize}

\paragraph{Direct Preference Optimization (DPO) and Veriants}

\begin{itemize}
\item  While PPO-based RLHF has been highly successful, it is resource-intensive, requiring online sampling, reward modeling, and delicate hyperparameter tuning. \textbf{DPO}~\cite{rafailov2023direct} simplifies this pipeline by formulating preference learning as a direct policy optimization problem without an explicit reward model or on-policy exploration. Given a dataset of preference pairs $(x, y^+, y^-)$, DPO trains the policy to assign higher probability to the preferred response, bypassing the intermediate reward modeling step.

Formally, for a prompt $x$ with preferred $y^+$ and dispreferred $y^-$, traditional RLHF would train a reward model $r(x,y)$ and then optimize via RL so that $r(x,y^+) > r(x,y^-)$. DPO collapses this into a direct adjustment of $\pi_\theta$ using a Bradley–Terry preference model. Under the assumption that the optimal policy $\pi^*$ induces the true reward up to a scaling $\beta$, one obtains the contrastive loss:

\begin{MiddleEquation}
\begin{equation}
\begin{aligned}
\mathcal{L}_{\text{DPO}}(\theta) \;=\; -\log \sigma\!\Big(\beta\big[\log \pi_\theta(y^+|x) - \log \pi_\theta(y^-|x)\big]\Big),
\end{aligned}
\end{equation}
\end{MiddleEquation}where $\sigma$ is the sigmoid and $\beta$ controls preference sharpness. Optimization is purely offline, resembling supervised fine-tuning with implicit labels. DPO avoids training a reward or value model, reduces memory overhead, and fully leverages offline preference data. This simplicity has driven strong interest in the open-source community~\cite{yao2023deepspeed, thapa2023humans, zhao2023can}. However, one work~\cite{ivison2024unpacking} finds PPO with a large reward model can still beat DPO in some scenarios. Likewise, another work~\cite{zhu2024starling} reports no gains of DPO over SFT for Starling-7B, whereas PPO with a tuned reward model yielded stronger results~\cite{ppo,trpo,fan2025online,wang2024conditional}. These findings suggest that the effectiveness of DPO depends on the base model, data distribution, and hyperparameter choices.

\item \textbf{DPO-VP}~\cite{tu2025enhancing} is a notable extension, which integrates verifiable signals (e.g., correctness in math/code). By treating correct outputs as preferred, DPO-VP directly optimizes against binary feedback. Compared to PPO baselines, a small LLM trained this way achieves better performance across five reasoning benchmarks, but with much lower compute. \citet{tu2025enhancing} further propose a multi-round framework where the policy generates candidates, verifiable signals or PRM/ORM labels, and DPO refines the model iteratively. This loop resembles highly off-policy RL but consistently uses the DPO loss. DPO occupies a middle ground between supervised fine-tuning and full RL, remaining close to SFT in implementation while still capturing reward-driven optimization effects. It incorporates human feedback without the inner loop of RL~\cite{ivison2024unpacking}. Current work explores hybrid strategies, where DPO provides a strong offline initialization, followed by PPO or other RL methods for further gains.

\item \textbf{CodeDPO} proposed by~\citet{miao2024aligningdpo} contribute a pipeline to collect preference pairs for model optimization. CodeDPO proposed by~\cite{zhang2024codedpo} integrates preference learning into code generation models to optimize both the correctness and execution efficiency of generated code. The approach uses a self-generation and validation mechanism where code snippets and test cases are simultaneously created and evaluated through a PageRank inspired algorithm that iteratively updates credibility scores, assuming that tests executable by multiple code snippets are more reliable and code passing more tests is more likely correct. 
\end{itemize}

\paragraph{Reinforcement Learning from AI Feedback (RLAIF)}
A central bottleneck of RLHF is the cost of collecting high-quality human feedback. \textbf{RLAIF} addresses this by replacing human supervision with AI-generated feedback—such as outputs from LLMs, symbolic verifiers, or hybrid systems~\cite{ahn2024tuning, lee2023rlaif}. The key motivation is that advanced LLMs can often evaluate responses comparably to humans, while many domains (e.g., code or math) admit automated correctness checks~\cite{tjuatja2024llms, chen2024humans, li2024llms, gu2024survey, tong2024codejudge, quan2025codeelo}. This substitution greatly improves scalability, especially where exhaustive human annotation is infeasible.

\citet{RLAIF} first proposed training reward models on preference labels produced by off-the-shelf LLMs rather than humans, coining the term RLAIF. Subsequent comparisons showed that strong AI labelers can yield performance comparable to RLHF. For instance, RLAIF-trained policies matched RLHF on summarization and dialogue~\cite{hoglund2023comparison, zhu2024factual}, and even same-model feedback improved policy quality~\cite{lee2023rlaif}. Notably, \citet{ivison2024unpacking} reported that querying an AI judge directly during PPO updates (skipping reward model training) outperformed two-step pipelines~\cite{ma2024highly, kumar2024training}, suggesting that learned reward models may introduce distortions.

RLAIF has been applied across reasoning and code domains. In mathematics, algebra solvers and verifier models provide correctness signals, underpinning recent reasoning systems such as DeepSeek-R1~\cite{deepseekmath, singh2023beyond, Math-Shepherd}. In code generation, compilers and unit tests enable automatic verifiable rewards. \citet{ma2023eureka} showed that even single-example RL with unit-test feedback can nearly double performance on math benchmarks. Beyond symbolic evaluators, LLM-based critics supply structured feedback through self-refinement, CoT review, or constitutional principles, as in Skywork-OR1~\cite{Skywork}, CRITIC~\cite{CRITIC}, and Constitutional AI~\cite{bai2022constitutional}. Together, these approaches establish RLAIF as a unifying paradigm encompassing symbolic signals, programmatic evaluators, and auxiliary LLM critics.

Despite its promise, RLAIF faces key challenges. Feedback quality remains a concern, since AI evaluators may propagate biases or blind spots, leading to reward hacking; mitigations include diverse evaluators, human oversight, or periodic evaluator refreshing. Computational cost is another limitation. Querying large models is expensive, motivating reward-model distillation and lightweight proxies~\cite{RLAIF, lee2023rlaif}. RLAIF demonstrates that scalable alignment is achievable by replacing or augmenting human supervision with automated feedback, but ensuring robustness and efficiency remains an open problem.

\begin{figure}[htbp]
    \centering
    \includegraphics[width=\textwidth]{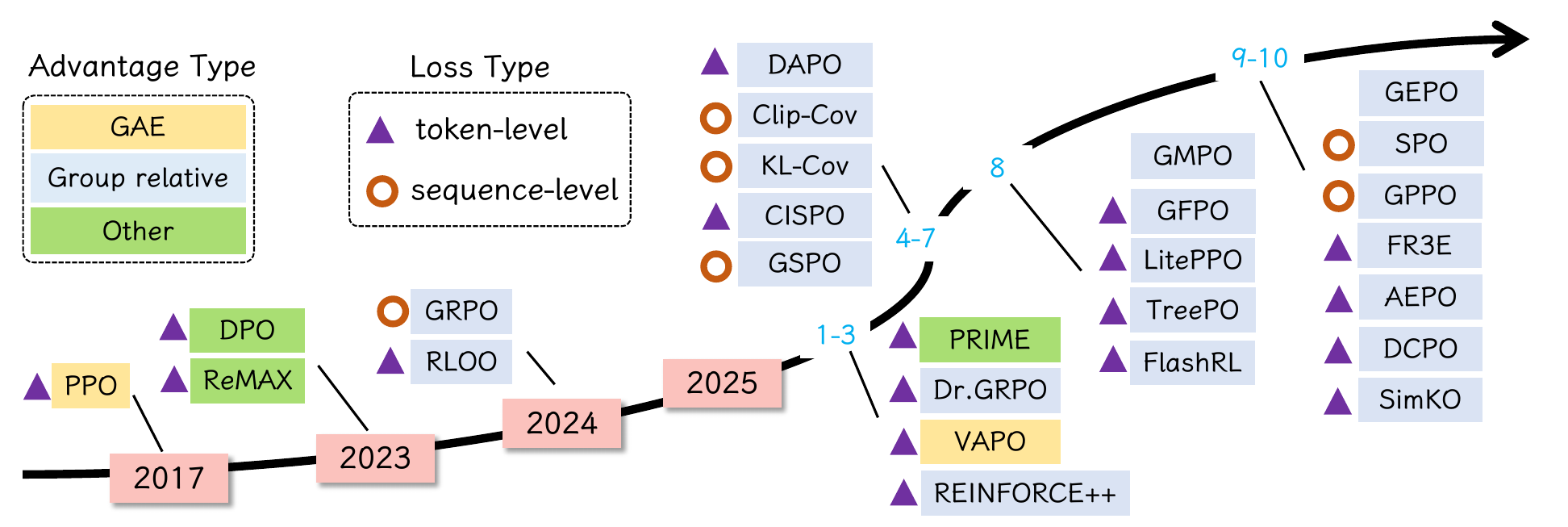}
    \caption{Overview of recent reinforcement learning algorithms for alignment.}
    \label{fig:rl_algorithms}
\end{figure}

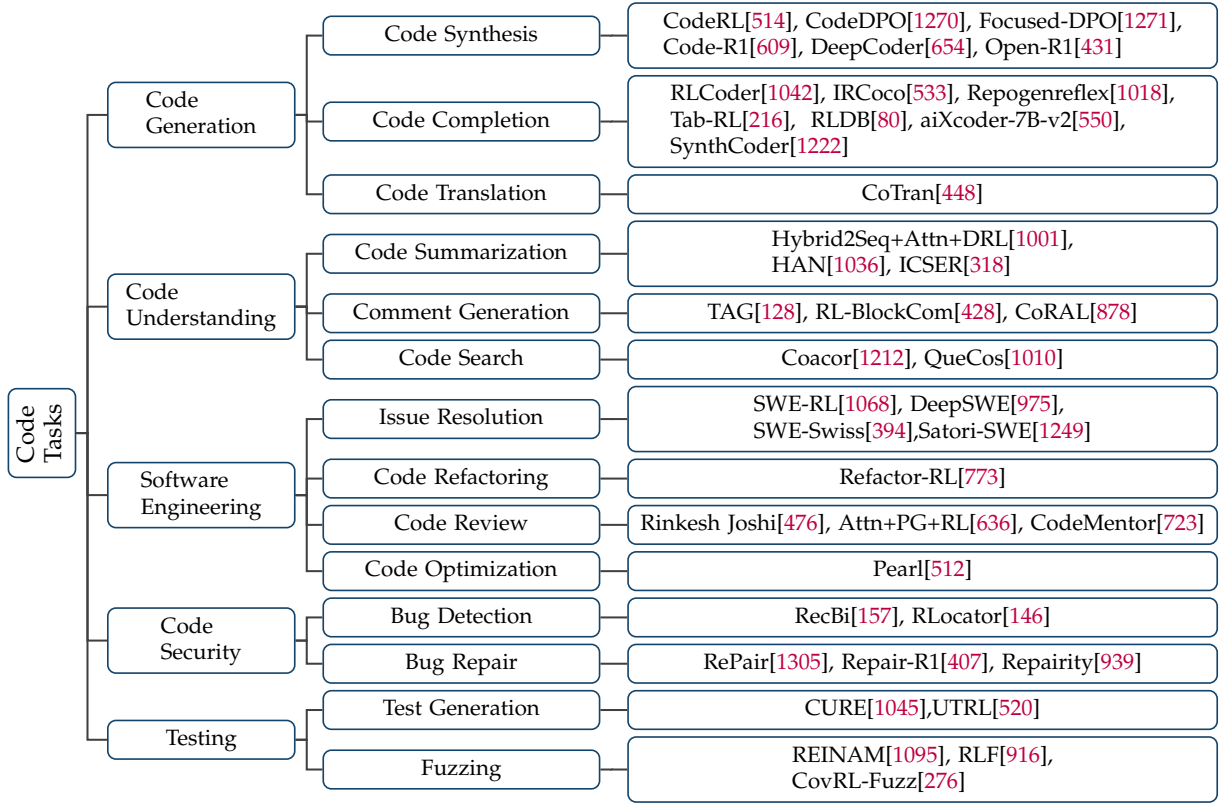
\begin{figure*}[!htbp]
    \centering
    \resizebox{\textwidth}{!}{
    \begin{forest}
        forked edges,
        for tree={
                grow=east,
                reversed=true,
                anchor=base west,
                parent anchor=east,
                child anchor=west,
                base=center,
                font=\large,
                rectangle,
                draw=hidden-draw,
                rounded corners,
                align=left,
                text centered,
                minimum width=4em,
                edge+={darkgray, line width=1pt},
                s sep=3pt,
                inner xsep=2pt,
                inner ysep=3pt,
                line width=0.8pt,
                ver/.style={rotate=90, child anchor=north, parent anchor=south, anchor=center},
            },
            where level=1{text width=8em,font=\normalsize, }{},
            where level=2{text width=12em,font=\normalsize}{},
            where level=3{text width=14em,font=\normalsize,}{},
        [Code \\Tasks, ver
            [Code \\Generation
                    [Code Synthesis
                        [
                            CodeRL\cite{le2022coderl}{, }CodeDPO\cite{zhang2024codedpo}{, }Focused-DPO\cite{zhang2025focuseddpo}{, }\\ Code-R1\cite{coder1}{, }DeepCoder\cite{deepcoder2025}{, }Open-R1\cite{openr1}
                           , text width=26em
                        ]
                    ]
                    [Code Completion
                        [
                            RLCoder\cite{wang2024rlcoder}{, }IRCoco\cite{li2024ircoco}{, }Repogenreflex\cite{wang2024repogenreflex}{, }\\Tab-RL\cite{tabrl}{, } RLDB\cite{rldb}{, }aiXcoder-7B-v2\cite{li2025aixcoder}{, }\\SynthCoder\cite{yu2025synthcoder}
                           , text width=26em
                        ]
                    ]
                    [Code Translation
                        [
                            CoTran\cite{jana2024cotran}
                           , text width=26em
                        ]
                    ]
            ]
            [Code \\Understanding
                    [Code Summarization
                        [
                            Hybrid2Seq+Attn+DRL\cite{wan2018improving}{, }\\HAN\cite{wang2020reinforcement}{, }ICSER\cite{geng2023interpretation}
                           , text width=26em
                        ]
                    ]
                    [Comment Generation
                        [
                            TAG\cite{cai2020tag}{, }RL-BlockCom\cite{huang2020towards}{, }CoRAL\cite{sghaier2025leveraging}
                           , text width=26em
                        ]
                    ]
                    [Code Search
                        [
                            Coacor\cite{yao2019coacor}{, }QueCos\cite{wang2022enriching}
                           , text width=26em
                        ]
                    ]
            ]
            [Software \\Engineering
                    [Issue Resolution 
                        [
                            SWE-RL\cite{wei2025swerl}{, }DeepSWE\cite{deepswe}{, }\\SWE-Swiss\cite{SWESwiss2025}{,}Satori-SWE\cite{zeng2025satori}
                           , text width=26em
                        ]
                    ]
                    [Code Refactoring
                        [
                            Refactor-RL\cite{palit2024generating}
                           , text width=26em
                        ]
                    ]
                    [Code Review
                        [
                            Rinkesh Joshi\cite{joshi2024comparative}{, }Attn+PG+RL\cite{liu2019automatic}{, }CodeMentor\cite{codementor}
                           , text width=26em
                        ]
                    ]
                    [Code Optimization
                        [
                            Pearl\cite{lamouri2025pearl}
                           , text width=26em
                        ]
                    ]
            ]
            [Code \\Security
                    [Bug Detection
                        [
                            RecBi\cite{chen2020enhanced}{, }RLocator\cite{chakraborty2024rlocator}
                           , text width=26em
                        ]
                    ]
                    [Bug Repair
                        [
                            RePair\cite{zhao2024repair}{, }Repair-R1\cite{hu2025repair}{, }Repairity\cite{tang2025boosting}
                           , text width=26em
                        ]
                    ]
            ]
            [Testing
                [Test Generation
                        [
                           CURE\cite{wang2025cure}{,}UTRL\cite{lee2025learning}
                           , text width=26em
                        ]
                ]
                [Fuzzing
                        [
                            REINAM\cite{wu2019reinam}{, }RLF\cite{su2022effectively}{, }\\CovRL-Fuzz\cite{eom2024fuzzing}
                           , text width=26em
                        ]
                ]
            ]
        ]
    \end{forest}}
    \caption{Taxonomy of coding tasks addressed by reinforcement learning methods}
    \label{fig:taxonomy_code_tasks}
\end{figure*}

\subsubsection{RL for Code Generation}
\autoref{fig:taxonomy_code_tasks} shows that reinforcement learning techniques are strategically adapted to address the distinct requirements of diverse code-related tasks. Code generation methods like CodeRL employ reward-based training for quality optimization, while completion tasks use incremental prediction approaches such as RLCoder. Understanding tasks integrate RL with attention mechanisms for summarization and documentation, whereas software engineering applications customize RL for multi-step reasoning in issue resolution and refactoring with task-specific rewards. Security and testing domains leverage RL for intelligent exploration, where fuzzing methods enable test input generation while bug repair systems use trial and error learning for defect localization and automated fixing.

\paragraph{Code Synthesis}
Code synthesis generates executable programs from natural language descriptions, particularly challenging in competitive programming tasks. CodeRL~\cite{le2022coderl} pioneered the application of PPO-based approaches by treating the code-generating LM as an actor network with a critic network providing dense feedback signals, achieving state-of-the-art results on APPS~\cite{apps} and MBPP~\cite{austin2021mbpp} benchmarks. The field then shifted towards Direct Preference Optimization (DPO) methods. CodeDPO~\cite{zhang2024codedpo} introduces a self-generation and validation mechanism with PageRank-inspired algorithms to create preference datasets optimizing both correctness and efficiency, while Focused-DPO~\cite{zhang2025focuseddpo} further refines this by targeting error-prone points through error point identification. More recently, RLVR has emerged as a breakthrough approach that substantially enhances code generation capabilities. Code-R1~\cite{coder1} demonstrates the importance of reliable reward signals, while the Open-R1 initiative~\cite{opencode,opencode2} releases validated datasets including Mixture-of-Thoughts with 350K reasoning trajectories and implements different reward functions with multiple sandbox providers. DeepCoder~\cite{deepcoder2025} showcases the effectiveness of these methods, achieving outstanding performance on LiveCodeBench~\cite{jain2024livecodebenchholisticcontaminationfree} through distributed RL training and matching o3-mini\cite{o3-o4} with just 14B parameters. The progression from PPO to RLVR represents a fundamental shift in how models learn to generate complex code, with RLVR's verifiable rewards significantly improving performance (a topic we will explore further in \autoref{sec:rlvr}).

\paragraph{Code Completion} Code completion has emerged as a prominent application of reinforcement learning in code generation, with recent advances addressing fundamental limitations of supervised learning approaches. RLCoder~\cite{wang2024rlcoder} introduces an RL framework for repository-level completion that learns to retrieve useful content without labeled data, achieving 12.2\% improvement on CrossCodeEval~\cite{ding2024crosscodeeval} and RepoEval benchmarks~\cite{repocoder2023}. IRCoCo~\cite{li2024ircoco} through immediate rewards-guided deep RL, providing instant feedback for dynamic context changes during continuous code editing. RepoGenReflex~\cite{wang2024repogenreflex} combines Retrieval-Augmented Generation (RAG) with Verbal Reinforcement Learning to dynamically optimize retrieval and generation processes. The aiXcoder-7B-v2~\cite{li2025aixcoder} addresses LLMs' tendency to ignore long-range contexts through explicit RL-based supervision signals, achieving up to 44\% improvement in exact match scores. SynthCoder~\cite{yu2025synthcoder} integrates multiple optimization techniques including Curriculum Learning and Direct Preference Optimization with rejection sampling, establishing state-of-the-art performance on Fill-in-the-Middle tasks. In production environments, Cursor Tab~\cite{tabrl} demonstrates the effectiveness of online RL by processing over 400 million daily requests and continuously updating models based on user feedback. While Augment Code's RLDB~\cite{rldb} achieves performance gains equivalent to doubling model size by learning directly from developer-IDE interactions. These RL-based approaches collectively demonstrate significant advantages over traditional supervised methods by addressing exposure bias, improving context utilization, and enabling real-time adaptation to developer behaviors.

\paragraph{Code Translation}
Code translation converts programs between languages while preserving functionality. Feedback-driven approaches leverage execution feedback to guide model training: CoTran~\cite{jana2024cotran} pioneers reinforcement learning with compiler and symbolic execution feedback for Java-Python translation, while error correction methods like Rectifier~\cite{yin2024rectifier0} utilize micro models to repair common translation errors across LLMs. kNN-ECD~\cite{xue2024an} employs k-nearest-neighbor search with error correction sub-datastores to enhance TransCoder-ST translations. Reasoning-enhanced methods improve translation through intermediate representations: Explain-then-Translate~\cite{tang-etal-2023-explain} demonstrates that self-generated natural language explanations as intermediate steps achieve 12\% average improvement in zero-shot scenarios across 19 languages. Retrieval-augmented approaches dynamically incorporate contextual examples: RAG-based few-shot learning~\cite{bhattarai2024enhancing0} retrieves relevant translation pairs to guide models, while task-specific embedding alignment~\cite{bhattarai2024enhancing1} optimizes retrieval quality for Fortran-C++ translation. Multi-agent systems decompose translation into specialized sub-tasks: TRANSAGENT~\cite{yuan2024semantic} employs four collaborative agents for initial translation, syntax fixing, code alignment, and semantic error correction, achieving superior performance through execution-based error localization. Repository-level translation addresses real-world complexity: AlphaTrans~\cite{ibrahimzada2024alphatrans0} introduces neuro-symbolic compositional translation with program transformation and multi-level validation for entire Java-to-Python projects. These diverse methodologies collectively advance code translation toward practical, scalable, and reliable automation.

\subsubsection{RL for Code Understanding}
\paragraph{Code Summarization}
Code summarization benefits from reinforcement learning to address the exposure bias issue inherent in traditional encoder-decoder frameworks. The Dual Model~\cite{wei2019code} incorporates abstract syntax tree structures into an actor-critic network, using the critic to evaluate reward values of possible extensions for global guidance. HAN~\cite{wang2020reinforcement} extends this approach with hierarchical attention mechanisms that integrate multiple code features including type-augmented ASTs and control flows, demonstrating substantial improvements in BLEU scores. Hybrid2Seq+Attn+DRL~\cite{wan2018improving} similarly employs deep reinforcement learning to mitigate the train-test discrepancy where models transition from ground-truth word training to full sequence generation during inference. ICSER~\cite{geng2023interpretation} introduces a two-stage paradigm that first identifies model focuses through interpretation techniques, then reinforces the model to generate improved summaries, significantly enhancing DeepCom's~\cite{DeepCom} performance. Collectively, these RL-based approaches demonstrate how actor-critic architectures and reward-guided training overcome the limitations of maximum likelihood training in capturing code semantics and structural information.

\paragraph{RL for Comment Generation}
Unlike code summarisation, comment generation targets developer-facing guidance tied to specific code spans and workflows as separate artifacts (e.g., review notes, API docs).
Comment generation leverages reinforcement learning to produce more accurate and useful code documentation. TAG~\cite{cai2020tag} introduces a Type Auxiliary Guiding framework that treats source code as an N-ary tree with type information, employing hierarchical reinforcement learning to handle the dependencies among type information for adaptive summarization. RL-BlockCom~\cite{huang2020towards} focuses specifically on block comments within methods, combining actor-critic algorithms with encoder-decoder architectures to achieve a better performance through statement-based AST traversal. CoRAL~\cite{sghaier2025leveraging} advances the field by designing reward mechanisms that consider both semantic similarity to expected comments and their utility as inputs for code refinement tasks, ensuring that generated comments are both meaningful and actionable. These approaches demonstrate how RL enables models to generate contextually relevant comments that better serve practical software development needs.

\paragraph{Code Search}
Code search leverages reinforcement learning to bridge the semantic gap between natural language queries and source code. CoaCor~\cite{yao2019coacor,xcodesearchnet,husain2020CodeSearchNet,code_search_survey} adopts a novel perspective by training code annotation models to generate descriptions that enhance code retrieval performance, using RL to explicitly encourage annotations that improve distinguishability of relevant code snippets. QueCos~\cite{wang2022enriching} addresses the semantic distance between user queries and code descriptions where code search performance serves as the reward signal for producing accurate query enrichment. Both approaches demonstrate that RL-guided semantic alignment substantially outperforms traditional matching-based methods in bridging the knowledge gap between natural language and code.

\subsubsection{RL for Software Engineering}

\paragraph{Issue Resolution}
Issue resolution in software repositories represents a complex challenge where RL has shown remarkable success in training models to autonomously fix real-world bugs. SWE-RL~\cite{wei2025swerl} advances LLM reasoning through reinforcement learning on open-source repository, demonstrating that RL can effectively guide models through the intricate process of understanding and resolving GitHub issues. DeepSWE~\cite{deepswe} scales RL training to create a fully open-sourced coding agent that achieves state-of-the-art performance on SWE-bench~\cite{swebench} through extensive reinforcement learning. SWE-Swiss~\cite{SWESwiss2025} employs a sophisticated two-phase training strategy that first embeds localization, repair, and unit testing capabilities via multi-task SFT, then sharpens repair skills through targeted RL with direct feedback from test environments, achieving strong performance on SWE-bench Verified with just a 32B model. Satori-SWE~\cite{zeng2025satori} introduces evolutionary test-time scaling (EvoScale) that treats generation as an evolutionary process, using RL to train models to self-evolve and improve their own generations across iterations, enabling the 32B model to match 100B+ parameter models while using fewer samples. These approaches demonstrate how RL transforms issue resolution from requiring human intervention to autonomous problem-solving at scale.

\paragraph{Code Refactoring}
Code refactoring~\cite{code_refactor_empirical_study} is the task of restructuring existing code to improve its internal structure, readability, and maintainability without altering its external behavior.
Refactor-RL~\cite{palit2024generating} addresses the limitations of traditional extract method refactoring approaches that require developers to manually identify refactoring boundaries and lack semantic understanding for meaningful method naming. The approach fine-tunes sequence-to-sequence models and aligns them using PPO, utilizing code compilation success and refactoring presence as reward signals for code-centric optimization. This RL-aligned approach achieves 11.96\% and 16.45\% improvements in BLEU and CodeBLEU scores respectively over supervised fine-tuning alone, while increasing successful unit test passes from 41 to 66, demonstrating the effectiveness of reinforcement learning in producing functionally correct refactorings that capture both syntactic and semantic aspects of code transformation.

\paragraph{Code Review}
Code review automation benefits from reinforcement learning in predicting pull request outcomes and generating review artifacts. \citet{joshi2024comparative} formalize PR outcome prediction as Markov Decision Processes with specialized reward functions to address data imbalance, achieving strong G-mean scores using PR features and even higher scores when focusing on PR discussions, effectively modeling both single-stage and multi-stage review processes. Attn+PG+RL~\cite{liu2019automatic} tackles the problem of missing PR descriptions by integrating pointer generator with direct ROUGE optimization through reinforcement learning, addressing out-of-vocabulary words in software artifacts while bridging the gap between training loss and evaluation metrics. CodeMentor~\cite{codementor} combines few-shot learning with RLHF, allowing domain experts to assess generated code and enhance model performance, achieving substantial improvements in code quality estimation, review generation, and bug report summarization. These approaches demonstrate how RL enables more effective collaboration in pull-based development by automating critical review processes.

\paragraph{Code Optimization}
Code optimization leverages reinforcement learning to automate complex compiler transformations. Pearl~\cite{lamouri2025pearl} introduces a deep RL framework that trains agents to select optimal sequences of polyhedral transformations, using a novel action space representation that determines both which optimization to apply and where in the loop nest to apply it. To address the data-intensive nature of compiler optimization where experiments typically require weeks, Pearl accelerates training through execution time memoization and actor-critic pre-training. Implemented in the Tiramisu compiler, Pearl achieves significant geometric mean speedups over state-of-the-art compilers while being the first RL-based system to support general tensor-manipulating loop nests with generalization to unseen programs, demonstrating how reinforcement learning can effectively navigate the complex optimization space previously requiring extensive manual tuning.

\subsubsection{RL for Code Security}
\paragraph{Bug Detection}
Bug detection~\cite{bugscope,zhong2024advancing_bug_detction} in compilers and software systems presents significant challenges due to limited debugging information and the vast search space for bugs. RecBi~\cite{chen2020enhanced} pioneers reinforcement learning for compiler bug isolation by augmenting traditional mutation operators with structural ones to transform bug-triggering test programs into passing variants, then using RL to intelligently guide the generation of diagnostic test programs that effectively isolate bugs through execution trace analysis. RLocator~\cite{chakraborty2024rlocator} formulates bug localization as a Markov Decision Process to directly optimize ranking metrics rather than using similarity-based heuristics, outperforming state-of-the-art tools FLIM and BugLocator. These RL-based approaches demonstrate how direct optimization of evaluation measures and intelligent search guidance substantially improve bug detection effectiveness compared to traditional static analysis methods.

\paragraph{RL for Bug Repair}
Automated program repair~\cite{debugbench,mdeval,foley2025apirl,lyu2025llm_differential_testing,zhang2023automatically_bug_reports} has been significantly advanced by RL, enabling smaller models to achieve competitive performance. RePair~\cite{zhao2024repair} introduces process-based supervision, using a reward model as a critic to iteratively optimize repair policies, allowing LLMs to approach the performance of larger commercial LLMs. Repair-R1~\cite{hu2025repair} shifts the paradigm by first generating discriminative test cases and then using RL to co-optimize both test generation and bug fixing, leading to substantial improvements in repair and test generation success rates. Repairity~\cite{tang2025boosting} bridges the performance gap via a three-stage methodology that extracts reasoning traces from closed-source models, transfers knowledge via supervised fine-tuning, and applies LLM-guided reinforcement learning, nearly closing the gap with a leading commercial model. Collectively, these approaches demonstrate how reinforcement learning facilitates effective program repair at a smaller scale through intelligent feedback and reasoning transfer.

\subsubsection{Code Testing}
\paragraph{Test Generation}
Test generation~\cite{gu2025llm_test_generation,promptpex} has emerged as a crucial component for training and evaluating code generation models, with reinforcement learning approaches enabling the creation of high-quality test suites that expose subtle bugs and edge cases. CURE~\cite{wang2025cure} proposes a co-evolution framework where coding and unit test generation capabilities improve through their interaction outcomes without ground-truth supervision, achieving significant improvement in code generation accuracy. UTRL~\cite{lee2025learning} trains test generators and code generators adversarially, where the test generator maximizes discrimination reward by exposing faults while the code generator maximizes code reward by passing tests, with Qwen3-4B trained via UTRL outperforming GPT-4.1~\cite{openai2025gpt41} in test quality.  These approaches collectively demonstrate how reinforcement learning transforms test generation from a manual bottleneck into an automated process that significantly enhances both model training and evaluation reliability.

\paragraph{Fuzzing}
Fuzzing~\cite{seq2seq_afl,fuzz_challenge_llm,understanding_fuzz_test_cases} is automated software-testing technique that feeds random or malformed inputs to programs to uncover bugs and security vulnerabilities. Fuzzing with reinforcement learning addresses the fundamental challenges of generating effective test inputs and navigating complex program behaviors. REINAM~\cite{wu2019reinam} pioneers RL-based input-grammar inference without seed inputs, formulating grammar generalization as an RL problem and using symbolic execution engines to iteratively generate inputs that achieve superior precision and recall over existing approaches. RLF~\cite{su2022effectively} tackles smart contract vulnerability detection by modeling fuzzing as a Markov Decision Process, designing reward mechanisms that consider both vulnerability detection and code coverage to guide generation of specific transaction sequences, detecting more vulnerabilities than state-of-the-art tools. CovRL-Fuzz~\cite{eom2024fuzzing} combines LLMs with coverage-guided reinforcement learning for JavaScript interpreter testing, integrating coverage feedback through TF-IDF weighted coverage maps to calculate fuzzing rewards, successfully identifying 58 real-world security bugs including 50 previously unknown bugs. These RL-based approaches collectively demonstrate how reinforcement learning transforms traditional random test generation into intelligent, adaptive exploration strategies that learn from program behaviors to discover vulnerabilities more effectively.

\newcommand{\faHuggingFace}{%
  \raisebox{-0.13em}{%
    \includegraphics[height=1em]{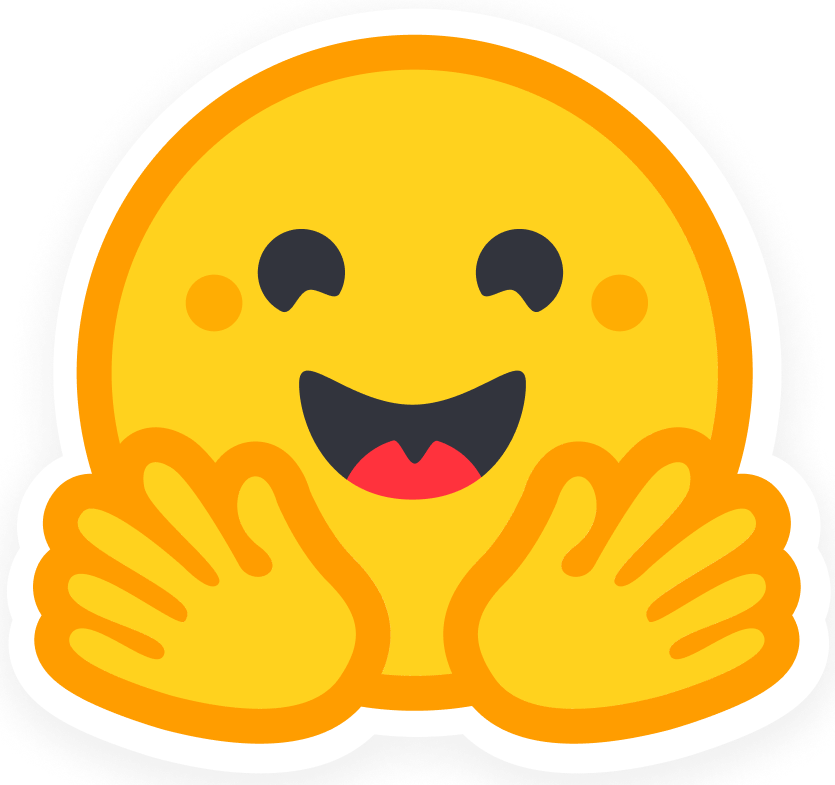}%
  }%
}

\subsection{Applying Reinforcement Learning with Verifiable Rewards}
\label{sec:rlvr}

Reinforcement Learning with Verifiable Rewards (RLVR) is an emerging paradigm designed to enhance reasoning capabilities in LLMs, particularly for domains such as mathematics and program synthesis.
Formally, RLVR can be viewed as a reinforcement learning framework defined by a tuple $
(\mathcal{S}, \mathcal{A}, \mathcal{R}_v, \pi_\theta)
$, where $\mathcal{S}$ represents the state space (e.g., the reasoning context or intermediate steps), $\mathcal{A}$ denotes the action space (i.e., token generation or chain-of-thought continuation), and $\pi_\theta$ is the model’s policy parameterized by $\theta$. 
The distinctive component is the verifiable reward function $\mathcal{R}_v$, which is derived from an external, deterministic verifier. Instead of relying on learned reward models or heuristic preference scores, $\mathcal{R}_v$ directly measures task correctness — for example, whether a generated mathematical proof reaches the correct answer or whether a piece of code passes all unit tests. 
The reward is typically binary (pass/fail), providing a clear and trustworthy signal that aligns precisely with task objectives.

The recent introduction of GRPO~\cite{deepseekmath} and its large-scale application in DeepSeek-R1~\cite{deepseekai2025deepseekr1} has driven RLVR to the forefront of research in reasoning LLMs. 
This family of algorithms and its application could be referred to a certain branch of RL algorithm (\autoref{sec:code_rl}), but focusing more the recently trending efficient PPO-variants, \emph{i.e.}, those designs pruning the critic model.
GRPO modifies the standard PPO framework by computing advantage estimates within a group of generated samples rather than training a separate value model. This ``group normalization'' approach eliminates the need for a learned critic network, significantly improving training efficiency and stability while maintaining alignment with the verifiable reward. Following DeepSeek-R1’s success in applying GRPO to mathematical reasoning — achieving high-quality long-chain reasoning purely from verifiable signals — a series of follow-up works have adopted the RLVR paradigm for both mathematical and code reasoning tasks, confirming its scalability and generality. These properties make RLVR particularly well-suited for reasoning-intensive domains such as algorithmic problem solving and code synthesis, where correctness can be explicitly checked by a computational oracle.

\subsubsection{RLVR-Suitable Datasets for Code Tasks}
\label{section:coderl_dataset}
A variety of specialized coding datasets have been developed to facilitate RLVR training. 
These datasets construct programming problems along with deterministic ``verifiers'' (e.g. input-output test cases or unit tests), ensuring the model outputs can be automatically checked for correctness. By providing clear pass/fail reward signals, they enable reliable reinforcement learning for code generation tasks. Notable representatives include the following datasets.

\paragraph{CodeContests} A large competitive-programming dataset ($\approx$ 13.3K problems) for training AlphaCode models~\cite{deepmind_code_contests}. It compiles coding challenges from platforms including AtCoder, CodeChef, Codeforces, and HackerEarth. Each problem is accompanied by multiple human reference solutions and paired input–output test cases. These comprehensive test suites allow verifiable evaluation of a model’s code correctness on each problem, making CodeContests an early cornerstone for RLVR in code generation.

\paragraph{TACO (Topics in Algorithmic COding)} An algorithmic code dataset introduced by BAAI in late 2023, consisting of 25,433 training problems and 1,000 test problems up to 1.55 million total solution codes~\cite{li2023taco}. TACO aggregates tasks from CodeContests, APPS, CodeChef, Codeforces, GeeksforGeeks, and HackerRank. Covering common competitive-programming topics including sorting, dynamic programming, and graph algorithms. Each problem is paired with a set of I/O examples that serve as rigorous test cases. By focusing on verifiable algorithmic tasks, TACO provides a rich resource for training and evaluating code LLMs with automatic reward signals.

\paragraph{Eurus-2-RL} A mixed-domain RL training set from the PRIME-RL project~\cite{cui2025processreinforcementimplicitrewards} containing 27K coding problems ($\approx$  450K math problems) all equipped with verifiers. The coding subset aggregates contest-style programming challenges sourced from APPS, CodeContests, TACO, Codeforces, and similar platforms, often translated into multiple programming languages. Each coding task is packaged with unit tests or I/O pairs that function as verifiable reward signals. Eurus-2-RL thus provides a large-scale, automatically-checkable corpus for jointly training models on code and math via RLVR.

\paragraph{IOI (International Olympiad in Informatics) Dataset} A collection of recent IOI competition problems (years 2020–2024) with their official input files, output files, and checker programs by using the \texttt{testlib} system\footnote{\url{https://github.com/MikeMirzayanov/testlib}}. Although relatively small (229 problems), these tasks are advanced algorithmic challenges used in the IOI, each with a thorough set of secret test cases and an official verifier. This dataset is valuable for RLVR because the high difficulty of IOI problems demands complex reasoning, and the provided test suites offer an authoritative pass/fail reward signal for each generated solution.

\paragraph{ACECode-87K} A synthetic coding dataset of roughly 87,000 problems created via automated test-case generation~\cite{zeng2502acecoder}. Starting from a seed corpus of coding questions, the curators used GPT-4o-mini to ``imagine'' ~16 diverse test cases per problem and filtered out any invalid or redundant cases. The result is a large collection of $(\text{problem}, {\text{tests}}, {\text{expected outputs}})$ triplets that are fully verifiable. ACECode-87K was primarily developed for training code LLMs with RL and the extensive test sets enable robust binary reward signals for code generation. This dataset demonstrated that synthetic but reliable test suites can substantially boost reward model training and subsequent RL fine-tuning of coder models.

\paragraph{SYNTHETIC-1} A massive open-source reasoning dataset comprising over 1.4 million tasks spanning mathematics, programming, software engineering, and more~\cite{2025synthetic1}. The coding subset alone contains approximately 144K program-synthesis and code-understanding problems, largely derived from public sources (e.g. APPS, CodeContests, Codeforces, and TACO). Crucially, each coding task includes a set of unit tests or I/O pairs as a verifier, often with the problem rephrased or ported to multiple languages for diversity. The broad coverage and automatic checkers in SYNTHETIC-1 make it suitable for both supervised fine-tuning and RL training of code-focused LLMs. Its diverse, verified problem set helps models learn generalizable reasoning strategies under the RLVR paradigm.

\paragraph{KodCode} The largest fully synthetic coding problem dataset to date, introduced in 2025, offering about 48.4K (question, solution, test) triplets~\cite{xu2025kodcode}. KodCode is composed of 12 distinct themed subsets, ranging from basic algorithmic puzzles and classic data-structure problems to complex domain-specific programming challenges. Every example in KodCode is equipped with a thorough set of unit tests or I/O checks, ensuring each generated solution can be definitively verified. By emphasizing a wide diversity of task types including unusual or specialized coding scenarios while retaining automatic verifiability, KodCode serves as a challenging benchmark for code LLMs and a rich training ground for RLVR and supervised learning alike.

\paragraph{HardTests} A competitive-programming dataset (about 47.1K problems) augmented with high-quality, synthesized test cases to reduce false positives in evaluation. Curated via an automated pipeline~\cite{he2025hardtests}, HardTests took coding problems from a broad array of online judges (Codeforces, AtCoder, Luogu, LeetCode, etc.) and generated extra thorough test inputs for each, including tricky corner cases. All proposed solutions were executed against these tests to validate correctness. The resulting dataset provides more stringent verifiers for each problem, making reward signals in RL training far more precise. By catching edge-case failures, HardTests helps ensure that an RL-trained model’s improvements reflect genuine problem-solving ability rather than exploitation of weak test suites.

\paragraph{Code-R1-12K} A curated RL training dataset consisting of 12k coding problem instances with associated tests, assembled as part of the open Code-R1 project~\cite{coder1}. It is built from 2k hand-picked LeetCode problems with reliable built-in unit tests and 10k additional verified problems filtered from the TACO corpus. The selection emphasizes low-noise, high-confidence reward signals and only problems with well-designed test cases and unambiguous correctness criteria are included. Code-R1-12K was used to train small code LLMs with the GRPO algorithm, showing that even a relatively compact, clean dataset can yield significant gains in code generation performance under RLVR.

\paragraph{LeetCodeDataset} A comprehensive Python coding dataset covering about 90\% of all LeetCode problems ($\approx$3.1K unique questions, as of mid-2024)~\cite{xia2025leetcodedatasettemporaldatasetrobust}. Each problem entry is enriched with metadata (difficulty level, topic tags, etc.) and paired with an extensive set of more than 100 test cases of varying difficulty. Notably, the dataset is temporally split into ``pre-July 2024'' and ``post-July 2024'' subsets to prevent data leakage between training and evaluation—ensuring that models can be trained on older problems and fairly tested on newer ones. This temporal split and the rich test suites make LeetCodeDataset ideal for verifiable-reward training for an RL agent can be rewarded for passing all tests on unseen problems, safe in the knowledge that it hasn’t memorized those solutions from training data.

\paragraph{CodeContests+} An enhanced version of the original CodeContests dataset, augmented by ByteDance’s SeedLab~\cite{wang2025codecontests+} with additional high-quality test cases generated by an LLM-based agent. For each of the ~11.4K competitive programming problems in CodeContests, the agent synthesized new challenging inputs and verified the correct outputs (through a sandbox executor or checker). These augmented test suites either supplement or replace the original ones, substantially increasing the true-positive rate of solution checking (i.e. reducing cases where an incorrect program might accidentally pass). Importantly, the underlying problem statements remain unchanged (only the evaluation criteria are strengthened). CodeContests+ thereby improves the accuracy and reliability of RL reward signals, without altering task content, enabling more trustworthy training and evaluation of code LLMs.

\paragraph{SYNTHETIC-2} The successor to SYNTHETIC-1, this dataset contains an even larger collection of automatically generated reasoning tasks (over 4 million verified problem-solving traces spanning coding, mathematics, logical puzzles, and more). Produced via a massive distributed inference pipeline (leveraging multiple LLMs in parallel), SYNTHETIC-2~\cite{primeintellect_synthetic} includes many problems beyond traditional competitive programming, such as code comprehension challenges, debugging tasks, and software engineering scenarios, all paired with known-correct answers or test-case validators. Every task is associated with a deterministic check (unit tests, proofs, or solutions) to serve as a reward signal. Although the coding subset ($\approx$ 61K) is only a portion of the full release, the sheer scale and variety of SYNTHETIC-2 make it one of the largest corpora for training LLMs in a verification-driven manner. It provides an unprecedented opportunity for both supervised fine-tuning and RL to push model reasoning capabilities to new heights.

\paragraph{Klear-CodeTest} A dataset of 27,965 competitive-programming problems curated by the Kwai Klear team~\cite{fu2025klearcodetestscalabletestcase}, each equipped with rigorously synthesized and validated test cases. Klear-CodeTest was constructed using a generator-and-checker framework. An LLM first proposes candidate test inputs for a given coding problem, then multiple solution attempts (and a reference solution) are run to check whether each test distinguishes correct code from incorrect. By iterating and filtering out redundant or ineffective cases, the pipeline produces a comprehensive test suite that covers the problem’s edge cases. The final dataset spans Codeforces, verified TACO/CodeContests problems, and other sources, with every example ready for automatic verification via either I/O comparison or a custom checker. Klear-CodeTest is intended for scalable RL training and robust evaluation of code LLMs, ensuring models learn to handle the full complexity of coding problems rather than overfitting to simplistic tests.

Collectively, the above datasets provide a foundation for verifiable-reward learning in code intelligence. They cover various programming challenges, each bundled with an oracle for correctness (test cases or checkers), so that a reinforcement learning agent can reliably gauge success. This abundance of automatically-checkable data enables large-scale RLVR training without human-in-the-loop labeling, and also supports rigorous evaluation of a model’s coding ability. By leveraging these resources, recent works have demonstrated substantial improvements in code generation accuracy and reasoning depth purely from self-play with verifiable rewards.

\begin{table}[!h]
\centering
\label{tab:coderl_dataset} 
\caption{Summary of representative open-source datasets for reinforcement learning with verifiable feedback in code intelligence tasks.} 
\resizebox{0.95\columnwidth}{!}{%
\begin{tabular}{llllllc}

\toprule
\textbf{Date} & \textbf{Name} & \textbf{Type} & \textbf{\#Sample} & \textbf{Test Format} & \textbf{Source} & \textbf{Link} \\
\midrule

2022.07 & CodeContests & \scalebox{0.8}{\makecell[l]{Algorithm}}  & 13.3k  & IO pairs   & \scalebox{0.8}{\makecell[l]{Aizu\\AtCoder\\CodeChef\\CodeForces\\HackerEarth}}  & \href{https://github.com/google-deepmind/code_contests}{\faGithub}\; \href{https://huggingface.co/datasets/deepmind/code_contests}{\faHuggingFace} \\
\midrule
2023.12 & TACO& \scalebox{0.8}{\makecell[l]{Algorithm}}   & 25.4k  & IO pairs    & \scalebox{0.8}{\makecell[l]{CodeContests\\APPS\\CodeChef\\CodeForces\\GeeksforGeeks\\HackerRank}}      & \href{https://github.com/FlagOpen/TACO}{\faGithub}\; \href{https://huggingface.co/datasets/BAAI/TACO}{\faHuggingFace} \\
\midrule
2025.01 & Eurus-2-RL& \scalebox{0.8}{\makecell[l]{Algorithm}}    & 26k  & IO pairs  & \scalebox{0.8}{\makecell[l]{APPS\\CodeContests\\ TACO\\Codeforces}}       & \href{https://github.com/PRIME-RL/PRIME}{\faGithub}\; \href{https://huggingface.co/datasets/PRIME-RL/Eurus-2-RL-Data}{\faHuggingFace} \\
\midrule
2025.02 & IOI& \scalebox{0.8}{\makecell[l]{Algorithm}}   & 229  & stdio and checker\footnote{https://github.com/MikeMirzayanov/testlib}    & \scalebox{0.8}{\makecell[l]{IOI 2020-2024}}        & \href{https://github.com/huggingface/ioi}{\faGithub}\; \href{https://huggingface.co/datasets/open-r1/ioi}{\faHuggingFace} \\
\midrule
2025.02 & ACECode& \scalebox{0.8}{\makecell[l]{Algorithm\\Code Understanding\\Debug}}  & 12.2k  & unit   & \scalebox{0.8}{\makecell[l]{Evol \\OSS\\Stack Python}}        & \href{https://github.com/TIGER-AI-Lab/AceCoder}{\faGithub}\; \href{https://huggingface.co/datasets/TIGER-Lab/AceCode-87K}{\faHuggingFace} \\
\midrule
2025.02 & SYNTHETIC-1& \scalebox{0.8}{\makecell[l]{Algorithm\\SWE\\Code Understanding}}    & 275k  & IO pairs  & \scalebox{0.8}{\makecell[l]{ Apps\\Codecontests\\Codeforces\\TACO}}       & \href{https://huggingface.co/datasets/PrimeIntellect/SYNTHETIC-1}{\faHuggingFace} \\
\midrule
2025.03 & KodCode& \scalebox{0.8}{\makecell[l]{Algorithm}}  & 48.4k  & pytest/IO pair    &   \scalebox{0.8}{\makecell[l]{Leetcode\\Codeforces\\Apps\\Taco\\Code Contests\\Evol \\ others}}    & \href{https://kodcode-ai.github.io/}{\faGithub}\; \href{https://huggingface.co/datasets/KodCode/KodCode-V1-SFT-R1}{\faHuggingFace} \\
\midrule
2025.03 & HardTests& \scalebox{0.8}{\makecell[l]{Algorithm}}   & 47.1k  & IO pair    &   \scalebox{0.8}{\makecell[l]{Luogu\\Codeforces\\AtCoder\\SPOJ\\CodeChef\\LeetCode\\GeekForGeeks\\ others}}  & \href{https://github.com/LeiLiLab/HardTestGen}{\faGithub}\; \href{https://huggingface.co/datasets/sigcp/hardtests_problems}{\faHuggingFace} \\
\midrule
2025.03 & Code-R1-12K & \scalebox{0.8}{\makecell[l]{Algorithm}}    & 12k  & IO pairs/unit   & \scalebox{0.8}{\makecell[l]{LeetCode\\TACO-verfied}}       & \href{https://github.com/ganler/code-r1}{\faGithub}\; \href{https://huggingface.co/datasets/ganler/code-r1-12k}{\faHuggingFace} \\
\midrule
2025.04 & LeetCodeDataset& \scalebox{0.8}{\makecell[l]{Algorithm}}     & 2.6k  & unit & \scalebox{0.8}{\makecell[l]{LeetCode}}   & \href{https://github.com/newfacade/LeetCodeDataset}{\faGithub}\; \href{https://huggingface.co/datasets/newfacade/LeetCodeDataset}{\faHuggingFace} \\
\midrule
2025.06 & CodeContests+& \scalebox{0.8}{\makecell[l]{Algorithm}}   & 11.4k  & stdio and checker  &  \scalebox{0.8}{CodeContests}   & \href{https://huggingface.co/datasets/ByteDance-Seed/Code-Contests-Plus}{\faHuggingFace} \\
\midrule
2025.07 & SYNTHETIC-2& \scalebox{0.8}{\makecell[l]{Algorithm\\Code Understanding}}    & 61k  & IO pairs  & \scalebox{0.8}{\makecell[l]{PRIME RL\\CodeForces}}  & \href{https://huggingface.co/datasets/PrimeIntellect/SYNTHETIC-2-RL}{\faHuggingFace} \\
\midrule
2025.08 & Klear-CodeTest& \scalebox{0.8}{\makecell[l]{Algorithm}}  & 28k  & IO Pairs/checker& \scalebox{0.8}{\makecell[l]{Codeforces\\TACO-verfied\\CodeContests}}       & \href{https://github.com/Kwai-Klear/CodeTest}{\faGithub}\; \href{https://huggingface.co/datasets/Jianlp/Klear-CodeTest}{\faHuggingFace} \\

\bottomrule
\end{tabular}
}
\end{table}

\subsubsection{Representative RLVR-Trained Open-Source Code LLMs}
\label{section:rl_coder}
Several open-source LLMs for code have recently been trained with reinforcement learning on verifiable tasks. Below, we highlight a number of such models, noting their base model, the scale and source of their training data, and the specific RL algorithms used. These examples illustrate the diversity of approaches in RLVR from standard PPO to GRPO and new techniques such as GPPO, as well as differences in whether a learned reward model or direct test-case feedback is employed. Despite varying in size and dataset scale, all these models report significant performance gains from verifiable reward optimization — often matching or surpassing much larger pretrained counterparts.

\begin{table}[ht]
\centering
\label{tab:rl_coder} 
\caption{Summary of representative open-source Code LLMs trained via reinforcement learning with verifiable rewards (RLVR). For each model we report the base model, datasets, RL algorithm, training length, and public links.} 
\resizebox{1.0\columnwidth}{!}{%
\begin{tabular}{llllccc}

\toprule
\textbf{Date} & \textbf{Model} & \textbf{Base Model} & \textbf{Data}& \textbf{Algorithm} & \textbf{Length} & \textbf{Link} \\
\midrule

2025.03 & AceCoder\cite{zeng2502acecoder}  & Qwen2.5-Coder-7B-Instruct & AceCode  & Reinforcement++   & 4K       & \href{https://github.com/TIGER-AI-Lab/AceCoder}{\faGithub}\; \href{https://huggingface.co/collections/TIGER-Lab/acecoder-67a16011a6c7d65cad529eba}{\faHuggingFace} \\
\midrule
2025.03 & Open-R1\cite{openr1}  & Qwen2.5-Math-7B  & \makecell[l]{IOI\\CodeForces}  & GRPO   & 32K & \href{https://github.com/huggingface/open-r1}{\faGithub}\; \href{https://huggingface.co/open-r1}{\faHuggingFace} \\
\midrule
2025.03 & Skywork-OR1\cite{Skywork} & DeepSeek-R1-Distill-Qwen-7B/DeepSeek-R1-Distill-Qwen-32B & \makecell[l]{LeetCodeDataset\\TACO}  & GRPO   & 16K$\xrightarrow{}$32K & \href{https://github.com/SkyworkAI/Skywork-OR1}{\faGithub}\; \href{https://huggingface.co/collections/Skywork/skywork-or1-67fa1bcb41b436ef2def76b9}{\faHuggingFace} \\
\midrule
2025.03 & Code-R1\cite{coder1} & Qwen2.5-7B  & Code-R1-12K   & GRPO   & 6K  & \href{https://github.com/ganler/code-r1}{\faGithub}\; \href{https://huggingface.co/ganler/CodeR1-Zero-Qwen2.5-7B-12k-832}{\faHuggingFace} \\
\midrule
2025.03 & DeepCoder\cite{deepcoder2025} & DeepSeek-R1-Distilled-Qwen-14B  & \makecell[l]{LeetCodeDataset\\TACO\\SYNTHETIC-1} & GRPO+   & 16K$\xrightarrow{}$32K    & \href{https://github.com/agentica-project/rllm}{\faGithub}\; \href{https://huggingface.co/agentica-org/DeepCoder-14B-Preview}{\faHuggingFace} \\
\midrule

2025.03 & Seed-Coder\cite{seedcoder} & Seed-Coder-8B-Base  & \makecell[l]{CodeContests\\ICPC problems\\CodeForces\\LiveCodeBench} & GRPO  & 16K$\xrightarrow{}$32K   & \href{https://github.com/ByteDance-Seed/Seed-Coder}{\faGithub}\; \href{https://huggingface.co/ByteDance-Seed/Seed-Coder-8B-Reasoning}{\faHuggingFace} \\
\midrule

2025.03 & AceReason \cite{chen2025acereason} & DeepSeek-R1-Distill-Qwen-7B/DeepSeek-R1-Distill-Qwen-14B  & \makecell[l]{AtCoder\\LeetCode\\Aizu} & GRPO  & 24K$\xrightarrow{}$32K   & \href{https://github.com/ByteDance-Seed/Seed-Coder}{\faGithub}\; \href{https://huggingface.co/ByteDance-Seed/Seed-Coder-8B-Reasoning}{\faHuggingFace} \\
\midrule

2025.03 & Klear-Reasoner\cite{klearreasoner} & Qwen3-8B-Base   & CodeSub-15K  & GPPO  & 32K  & \href{https://github.com/Kwai-Klear/KlearReasoner/}{\faGithub}\; \href{https://huggingface.co/Kwai-Klear/Klear-Reasoner-8B}{\faHuggingFace} \\

\bottomrule
\end{tabular}
}   
\end{table}

\paragraph{AceCoder \cite{zeng2024acecoder}} is a code-centric LLM based on models such as Llama-3.1-8B or Qwen2.5-7B trained via a novel automated RL pipeline with REINFORCE++ alogorithm. 
The model was trained on the AceCoder-87K synthetic dataset with GPT-generated tests, including a curated ``hard subset'' of ~22K challenging problems with ~16 test cases each. 
First, AceCoder constructs a reward model by generating preference pairs from test-case pass rates and training via Bradley–Terry loss. Then, the policy model is fine-tuned with either the learned reward model or direct binary test outcomes as the reward signal similar to DeepSeek-R1~\cite{deepseekai2025deepseekr1}’s strategy. AceCoder improved pass rates by approximately 10–25\% on HumanEval, MBPP, BigCodeBench, and LiveCodeBench compared to the supervised baseline. 
Notably, even with only 80  optimization steps on the base model, following the R1-style ``zero-shot'' RL setup, a ~7B AceCoder model achieved over a 25\% increase on HumanEval-plus and a 6\% increase on MBPP-plus.

\paragraph{Open-R1 \cite{openr1}} is an open reproduction of the DeepSeek-R1 methodology, applied to code reasoning tasks and trained on Qwen2.5 series with extended context, RoPE 300k as the starting point.
The training data comprises a collection of high-quality coding problems, including the IOI dataset, Codeforces tasks, and others, which is often supplemented with mathematical reasoning data to approximate the original R1 training distribution. 
The 7B-scale Open-R1 models, trained on a few thousand code problems with test oracles, demonstrated clear improvements over instruction-tuned baselines on coding benchmarks. 
Such an initiative showcases that R1-level performance improvements can be achieved and reproduced openly and group-based RLVR consistently enhances code generation accuracy without proprietary data or methods. 

\paragraph{Skywork-OR1~\cite{Skywork}} is a series of models trained via large-scale GRPO starting on weights distilled from DeepSeek-R1-Distill-Qwen models. The training corpus included substantial verified coding and math datasets (e.g., LeetCodeDataset, TACO) with an emphasis on long-chain tasks. 
Models underwent multi-stage RL, with context length increasing from 16K to 32K tokens. 
Their 32B model achieved ~63\% pass@1 on LiveCodeBench (Aug 2024--Feb 2025) and surpassed DeepSeek-R1 on AIME 2024/25~\cite{aime24,aime25}.

\paragraph{Code-R1 \cite{coder1}} demonstrates the efficacy of GRPO at a small scale. Fine-tuning a 7B Qwen2.5 model on just 12K high-quality problems yielded a 5-6\% pass@1 gain on HumanEval+ over its SFT base, proving RLVR's value even on a budget.

\paragraph{DeepCoder~\cite{deepcoder2025}} is a 14B model trained on 24K vetted problems using GRPO+, an enhanced algorithm removing entropy/KL terms. DeepCoder-14B achieved 60.6\% pass@1 on LiveCodeBench, matching OpenAI's much larger 34B ``o3-mini-2025'' model~\cite{o3-o4} and showing mid-sized open models can rival proprietary systems.

\paragraph{Seed-Coder \cite{seedcoder}} employs a hybrid, two-stage RL pipeline: first using DPO to create an ``Instruct'' model, then applying a PPO-style algorithm on verifiable tasks to create a ``Reasoning'' variant. This model-driven curation approach outperformed larger models on multi-step tasks.

\paragraph{AceReason \cite{chen2025acereason}} introduces a two-stage PPO curriculum. Stage 1 trained only on mathematical prompts, which unexpectedly boosted both math and coding reasoning. Stage 2 applied code-only RL. This math-first curriculum proved highly effective for unlocking general reasoning.

\paragraph{Klear-Reasoner \cite{klearreasoner}} is an 8B model trained on 45K expert-level tasks using gradient-preserving policy optimization (GPPO). This novel algorithm improved gradient utilization and mitigated premature entropy collapse, demonstrating the value of algorithmic innovation in RLVR.

In summary, these open-source models validate RLVR as a potent method for enhancing LLM reasoning. They showcase a diversity of successful strategies beyond the classic PPO, including value-free methods (GRPO) and gradient-preserving updates (GPPO), hybrid pipelines (DPO+PPO), and novel curricula (math-first-then-code). 
Two common themes emerge: the critical importance of high-quality, verified data (prioritized over quantity) and the use of strong, distilled base models to amplify RL gains. 
These openly-released models achieve SOTA results on benchmarks like LiveCodeBench, closing the gap with larger proprietary systems and providing a clear blueprint for advancing open-source reasoning AI.

\subsubsection{Reward Shaping in Code Post-training }\label{sec:reward_design}
\begin{figure}[h]
    \centering
    \includegraphics[width=1.0\textwidth]{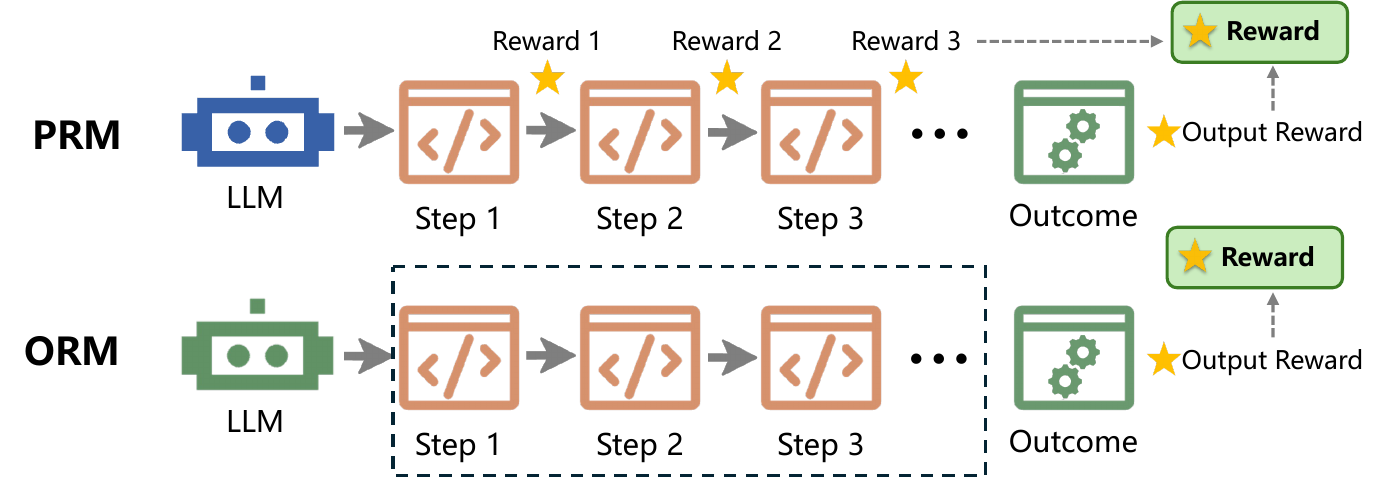} 
    \caption{Comparison between ORM and PRM.}
    \label{fig:orm_prm}
\end{figure}

In the post-training phase of code LLMs, reward design serves as one of the core mechanisms to align model outputs with functional requirements and nuanced developer expectations. 
Effective reward formulation incentivizes code that is not only functionally correct but also secure, maintainable, and contextually aligned with real-world software engineering standards. This section establishes four foundational dimensions, including \textit{correctness}, \textit{task alignment}, \textit{security}, and \textit{structural quality}. These approaches serve as foundational pillars for reward shaping, introducing three complementary paradigms: human preference modeling for capturing subjective quality attributes (e.g., readability and idiomatic style), outcome-supervised reward modeling (ORM) for evaluating the final generated code against correctness and quality criteria, and process reward modeling (PRM) for enabling real-time, stepwise code correction during generation, as shown in \autoref{fig:orm_prm}. Together, these paradigms enable LLMs to produce production-ready code that meets both functional requirements and quality standards.

\begin{figure}[h]
    \centering
    \includegraphics[width=1.0\textwidth]{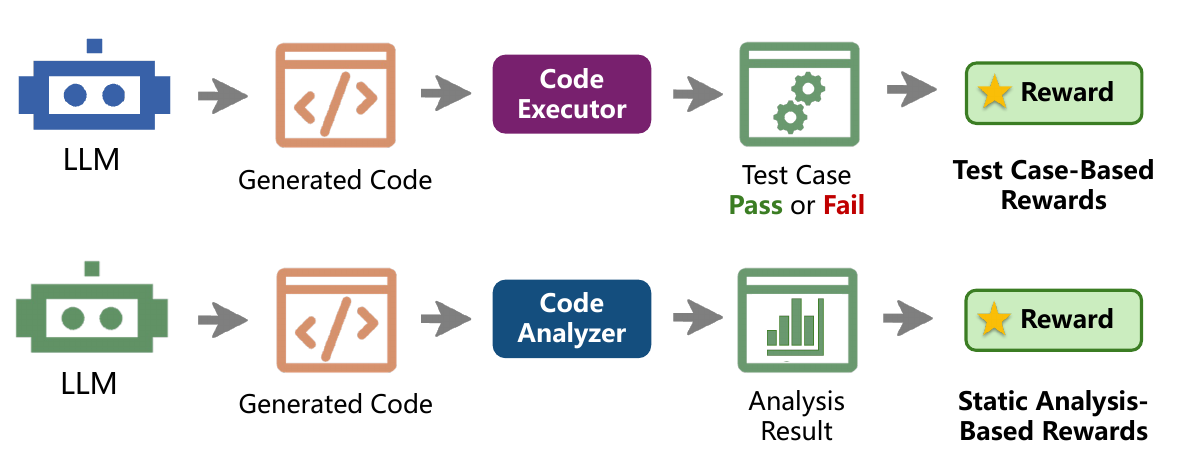} 
    \caption{Comparison of typical reward designs.}
    \label{fig:reawads_comparison}
\end{figure}

\paragraph{Correctness-Oriented Rewards}
Correctness is the most fundamental requirement for generated code. Correctness-oriented rewards ensure that code adheres to syntactic, semantic, and functional standards through static analysis and dynamic evaluation, as illustrated in \autoref{fig:reawads_comparison}.

\begin{itemize}[leftmargin=*]  
\item \textbf{Static Analysis-Based Rewards}: These are derived from static code analyzers that detect syntax errors, type mismatches, undefined variables, or style violations without executing the code. By integrating tools such as linters~\cite{kapadnis2025crscore++} or AST parsers~\cite{wu2025recode}, rewards can be assigned to syntactically valid and stylistically consistent code, providing immediate feedback during training.        
\item \textbf{Test Case-Based Rewards}: These evaluate functional correctness by executing the generated code against a set of input-output test cases~\cite{cui2025processreinforcementimplicitrewards,coder1}. A positive reward is assigned only when the code passes all or a sufficient number of tests. Automated frameworks can generate unit tests from function signatures or natural language descriptions, enabling scalable and systematic reward computation for complex programming tasks.
\end{itemize}

\subsubsection{Quality-Oriented Rewards}
Beyond functional correctness, high-quality code must meet broader software engineering standards for real-world deployment. Quality-oriented rewards target attributes such as task relevance, security, and structural maintainability.

\begin{itemize}[leftmargin=*]   
\item \textbf{Task Alignment}: This dimension assesses whether the generated code fulfills the intent specified in the prompt. For example, implementing a sorting function for a query asking to ``sort an array'' receives a positive reward, whereas incorrect logic (e.g., summing instead of multiplying) is penalized. Semantic similarity between natural language instructions and generated code can be used to quantify alignment accuracy.
\item \textbf{Security and Robustness}: These rewards encourage secure coding practices by discouraging common vulnerabilities:
\begin{itemize}
\item \textit{Vulnerability Avoidance}: Rewards are assigned for avoiding known security flaws such as SQL injection, cross-site scripting (XSS), or buffer overflows. Static analysis tools or taint tracking systems can detect such patterns and provide feedback signals.
\item \textit{Access Control Enforcement}: Additional rewards are granted when permission checks and authentication mechanisms are properly implemented (e.g., validating user roles before file deletion).    \end{itemize}    
\item \textbf{Structural Quality}: These rewards promote long-term usability and maintainability by encouraging sound software design:
\begin{itemize}        
\item \textit{Modularity and Reusability}: Code is rewarded if it is decomposed into well-defined functions or modules with clear responsibilities (e.g., separating input validation from business logic), enhancing readability and ease of maintenance.                
\item \textit{Interface Compatibility}: Positive reinforcement is provided when generated code adheres to existing APIs, class hierarchies, or interface contracts, enabling seamless integration into legacy systems or multi-component architectures.    
\end{itemize}
\end{itemize}

\paragraph{Rewards via Human Preference Modeling}\label{subsec:preference_modeling}

While correctness and quality-oriented rewards provide objective evaluation criteria, they often fail to capture nuanced aspects of code such as readability, clarity, and idiomatic style—qualities that are central to real-world developer preferences. Human preference modeling addresses this limitation by learning from explicit human judgments on code quality, enabling LLMs to align with subjective yet essential coding standards. This approach begins with the collection of high-quality preference data, where annotators rank or select among multiple code completions for the same prompt, typically comparing candidates in terms of efficiency, structure, or adherence to best practices. Specialized platforms like CodeRL and HumanEval-Preference facilitate systematic data gathering across diverse programming languages and task complexities. These preference pairs are then used to train a reward model via methods such as pairwise ranking loss or DPO, which learns to assign scalar rewards that reflect human judgment. To ensure robust generalization, recent models incorporate fine-grained attention mechanisms that focus on critical code segments, such as error-prone expressions or stylistic choices. Finally, the trained RM is integrated into post-training—serving as a reward signal in reinforcement learning (e.g., PPO or DPO)—to guide the LLM toward generating outputs that are not only functionally correct but also more readable and idiomatic, significantly improving alignment with human expectations.

\paragraph{Process Rewards Modeling}\label{subsec:prm}

Traditional reward mechanisms evaluate code only after full generation, which is inefficient for complex tasks where intermediate errors propagate irreversibly. Code Process Reward Modeling (PRM) addresses this limitation by providing granular, step-by-step feedback during the generation process, effectively turning the model into a ``guided coder'' that corrects itself in real time. PRM operates on the sequence of intermediate code states, assigning rewards at each token generation step based on contextual validity.

\begin{itemize}[leftmargin=*]
    \item \textbf{Design Principles}: PRM leverages two complementary signals: \textit{local validity} (e.g., syntactic correctness of the current token) and \textit{global trajectory coherence} (e.g., whether the partial code aligns with the problem's high-level structure). For instance, when generating a function, PRM rewards tokens that correctly extend the function signature (e.g., \texttt{def calculate\_sum(a: int, b: int) -> int:}) while penalizing tokens that introduce syntax errors or deviate from the expected API pattern.
    
    \item \textbf{Implementation Mechanisms}: PRM integrates with the LLM's decoding process through two primary architectures:
    \begin{itemize}
        \item \textit{Token-Level PRM}: A lightweight neural network (e.g., a small MLP) processes the current token and preceding context to predict a per-token reward. This is computationally efficient and directly integrates into autoregressive decoding.
        
        \item \textit{State-Tracking PRM}: For complex tasks, PRM maintains an internal state (e.g., AST or control flow graph) to evaluate whether the partial code satisfies structural constraints (e.g., ``all loops must have break conditions''). This requires symbolic reasoning modules but provides deeper contextual awareness.
    \end{itemize}
    
    \item \textbf{Advantages Over End-of-Sequence Rewards}: PRM reduces the exploration space by immediately correcting invalid paths, accelerating convergence. Empirical studies show PRM can reduce the failure rate of code generation by \textbf{TODO XX\%} compared to end-of-sequence reward models, particularly for tasks requiring multi-step reasoning (e.g., implementing a state machine or handling edge cases). It also enables progressive learning, where the model gradually builds complex structures from validated sub-components.
\end{itemize}

\medskip
In summary, reward shaping for code-focused LLMs in post-training combines explicit signals across multiple dimensions—\textit{correctness}, \textit{task alignment}, \textit{security}, and \textit{structural quality}—to ensure generated code is functionally sound and engineering-robust; furthermore, \textit{human preference modeling} and \textit{process reward modeling} enhance alignment with developer expectations and enable stepwise optimization, respectively, leading to more reliable, readable, and maintainable code generation.

\section{Software Engineering Agents}
\begin{figure*}[!htbp]
	\centering
    \resizebox{\textwidth}{!}{
	\begin{forest}
        forked edges,
		for tree={
                grow=east,
                reversed=true,
                anchor=base west,
                parent anchor=east,
                child anchor=west,
                base=center,
                font=\large,
                rectangle,
                draw=hidden-draw,
                rounded corners,
                align=left,
                text centered,
                minimum width=4em,
                edge+={darkgray, line width=1pt},
                s sep=3pt,
                inner xsep=2pt,
                inner ysep=3pt,
                line width=0.8pt,
                ver/.style={rotate=90, child anchor=north, parent anchor=south, anchor=center},
            },
            where level=1{text width=8em,font=\normalsize, }{},
            where level=2{text width=12em,font=\normalsize}{},
            where level=3{text width=14em,font=\normalsize,}{},
	    [SWE Agents, 
			[Requirements \\ Engineering
                    [Requirement Acquisition
                        [
                            Elicitron~\cite{Elicitron}
                           , text width=40em
                        ]
                    ]
                    [Requirement Examination
                        [
                           MARE~\cite{UserStory} MAD~\cite{MAD}
                           , text width=40em
                        ]
                    ]
                    [Requirement Modeling 
                        [
                        Progressive Prompting~\cite{Progressive} PrototypeFlow~\cite{PrototypeFlow} DCGen~\cite{DCGen}
                           , text width=40em
                        ]
                    ]
                    [Requirement Assurance
                        [
                            SimUser~\cite{SimUser} UXAgent~\cite{UXAgent}
                            , text width=40em
                        ]
                    ]
                    [End-to-End \\ Requirement Engineering
                        [
                            MARE~\cite{MARE} iReDev~\cite{iReDev}
                            , text width=40em
                        ]
                    ]
			]
			[Software \\ Development                   
                    [Program Synthesis
                        [
                            AlphaCodium~\cite{ridnik2024alphacodium} CodeCoT~\cite{huang2023codecot} Self-Refine~\cite{madaan2023selfrefine} PyCapsule~\cite{adnan2025pycapsule} \\MapCoder~\cite{islam2024mapcoder} ChatDev~\cite{qian2024chatdev} MetaGPT~\cite{hong2023metagpt} AgentCoder~\cite{pan2024agentcoder} \\CodeAgent~\cite{liao2024codeagent} Reflexion~\cite{shinn2023reflexion} AlphaCode~\cite{li2022alphacode} LargeLanguageMonkeys~\cite{brown2024largelanguagemonkeys} \\ MPSC~\cite{huang2024mpsc} SSTAR~\cite{li2025sstar}  Self-discover~\cite{zhou2024self} RepoCoder~\cite{repocoder2023} \\ CodeChain~\cite{codechain2023} HyperAgent~\cite{zhang2024hyperagent} OpenHands~\cite{openhands2024} CodePlan~\cite{codeplan2023} 
                            ,text width=40em
                        ]
                    ]
                    [Test to SQL
                        [
                            DAIL-SQL~\cite{Dial-sql} C3-SQL~\cite{C3-sql} DIN-SQL~\cite{Din-sql} MR-SQL~\cite{MR-SQL} \\MAC-SQL~\cite{MAC-SQL} MCS-SQL~\cite{MCS-SQL} TA-SQL~\cite{TA-SQL} CodeS~\cite{CodeS} \\CHESS~\cite{chess} MAG-SQL \cite{MAG-SQL} OpenSearch-SQL \cite{OpenSearch-SQL} \\MSc-SQL \cite{MSc-SQL} RSL-SQL~\cite{RSL-SQL} CHASE-SQL~\cite{CHASE-SQL} XiYan-SQL~\cite{XiYan-SQL} \\UCS-SQL~\cite{UCS-SQL} GenaSQL~\cite{GenaSQL} OmniSQL~\cite{OmniSQL} AskData~\cite{AskData}
                           , text width=40em
                        ]
                    ]
                    [Comment Generation
                        [
                            AutoComment~\cite{AutoComment} CloCom~\cite{CloCom} DeepCom~\cite{DeepCom} APIContext2Com~\cite{APIContext2Com}\\MESIA~\cite{MESIA} GTrans~\cite{GTrans} ByteGen~\cite{ByteGen} SCCLLM~\cite{SCCLLM} RAGcomment~\cite{RAGcomment}\\DeepCRCEval~\cite{DeepCRCEval}
                           , text width=40em
                        ]
                    ]
                    [Review Generation
                        [
                            AUGER~\cite{li2022auger} CodeReviewer~\cite{li2022codereviewer} \citet{froemmgen2024resovling} LLaMA-Reviewer~\cite{lu2023llama}\\CodeMentor~\cite{codementor} ~\citet{liu2025too}~\citet{jaoua2025combining} ~\citet{sghaier2025leveraging}\\CodeAgent~\cite{tang2024codeagent}
                           , text width=40em
                        ]
                    ]
                    [Fault Localization
                        [
                            Marko Vasic et al.~\cite{vasic2019neural} RESTORE~\cite{xu2020restore} FetaFix~\cite{louloudakis2025fetafix} InferFix~\cite{jin2023inferfix}\\~\citet{ji2025impact} AutoFL~\cite{kang2024quantitative} FlexFL~\cite{xu2025flexfl} AgentFL~\cite{qin2024agentfl} LLM4FL~\cite{rafi2024multi}\\SoapFL~\cite{qin2025s} RING~\cite{joshi2023repair}
                           , text width=40em
                        ]
                    ]
                    [Document Generation
                        [
                            CodeExp~\cite{CodeExp} \citet{KhanGPT} HotGPT~\cite{HotGPT} RepoAgent~\cite{RepoAgent}~\citet{Diggs}\\ DocAgent~\cite{DocAgent} METAMON~\cite{METAMON}
                           , text width=40em
                        ]
                    ]
                    [Patch Generation
                        [
                            RepairLLaMA~\cite{silva2025repairllama} AlphaRepair ~\cite{xia2022less} 
                            CIRCLE~\cite{yuan2022circle} ThinkRepair ~\cite{yin2024thinkrepair}\\FitRepair~\cite{xia2023revisiting} T5APR \cite{gharibi2024t5apr} MORepair~\cite{yang2024morepair}~\citet{ruiz2025artrepairoptimizingiterative}\\
                            LLM4CVE~\cite{fakih2025llm4cve} GAMMA~\cite{zhang2023gamma} 
                            KNOD~\cite{jiang2023knod} 
                            NSEdit~\cite{hu2022fix} 
                            SeqTrans~\cite{chi2022seqtrans}\\
                            SelfAPR~\cite{ye2022selfapr} 
                            NExT \cite{ni2024next} InferFix \cite{jin2023inferfix} RAP-Gen \cite{wang2023rap} SelRepair \cite{guo2025accelerating}\\MultiMend \cite{gharibi2025multimend} PredicateFix \cite{xiao2025predicatefix} DEAR \cite{li2022dear} GIANTREPAIR \cite{li2025hybrid}\\FLAMES \cite{le2024semantic} ESBMC-AI \cite{tihanyi2025new} Conversational APR \cite{xia2023conversational} ChatRepair \cite{xia2023keep}\\ContractTinker \cite{wang2024contracttinker} 
                            ITER \cite{ye2024iter} \citet{charalambous2024automated} RepairAgent \cite{bouzenia2403repairagent}\\AutoCodeRover \cite{zhang2024autocoderover} PatchPilot \cite{li2025patchpilot} MarsCode Agent \cite{liu2024marscode}\\EXPEREPAIR \cite{mu2025experepair} SpecRover \cite{ruan2024specrover} Repilot \cite{wei2023copiloting} SYNSHINE \cite{ahmed2022synshine}\\
                            GBPR \cite{silva2025gradient} 
                            RewardRepair \cite{ye2022neural} 
                            Recoder \cite{zhu2021syntax} 
                           , text width=40em
                        ]
                    ]
                    [Issue Resolving
                        [
                            SWE-Agent~\cite{yang2024swe} OpenHands~\cite{wang2024openhands} MAGIS~\cite{tao2024magis} CODER~\cite{chen2024coder} \\AutoCodeRover~\cite{zhang2024autocoderover} MarsCode Agent~\cite{liu2024marscode} AGENTLESS~\cite{xia2024agentless} SWE-Fixer~\cite{xie2025swe} \\Co-PatcheR~\cite{tang2025co} PatchPilot~\cite{li2025patchpilot} CODEXGRAPH~\cite{liu2024codexgraph} KGCompass~\cite{yang2025enhancing} \\CGM~\cite{tao2025code} LingmaAgent~\cite{ma2025alibaba} SemAgent~\cite{pabba2025semagent} SpecRover~\cite{ruan2024specrover} \\NEMOTRON-CORTEXA~\cite{sohrabizadehnemotron} SWE-Exp~\cite{chen2025swe} SWE-Debate~\cite{li2025swe} SE-Agent~\cite{lin2025se}
                           , text width=40em
                        ]
                    ]
            ]
            [Software \\ Testing
                    [Unit Test Generation
                        [
                            ChatTester~\cite{yuan2024manualtestsevaluatingimproving} 
                            ChatUniTest~\citep{chen2024chatunitestframeworkllmbasedtest} 
                            TestPilot~\cite{journals/corr/abs-2406-18181}                     CoverUp~\citep{pizzorno2025coverupeffectivehighcoverage} 
                            CodaMosa~\citep{Lemieux2023CODAMOSA} \\ 
                            MuTAP~\citep{dakhel2023effectivetestgenerationusing} 
                            HITS~\citep{wang2024hitshighcoveragellmbasedunit} 
                            TestART~\citep{gu2025testartimprovingllmbasedunit} 
                            SlipCover~\cite{Altmayer_Pizzorno_2023} 
                            TELPA~\cite{Yang2025EnhancingLT} 
                           , text width=40em
                        ]
                    ]
                    [Fuzz Testing
                        [                                                   AutoSafeCoder\cite{nunez2024autosafecodermultiagentframeworksecuring}  Mut4All \cite{wang2025mut4allfuzzingcompilersllmsynthesized} WhiteFox \cite{Yang_2024} 
                        CKGFuzzer \cite{xu2024ckgfuzzerllmbasedfuzzdriver} ToolFuzz \cite{milev2025toolfuzzautomatedagent}
                        , text width=40em
                        ]
                    ]
                    [Other Types of Testing
                        [
                            ACH2~\cite{foster2025mutationguidedllmbasedtestgeneration} 
                            AEGIS~\cite{wang2024aegis} 
                            MAdroid \cite{feng2025breaking}       EnvAgent~\cite{bouzenia2025you}         
                           , text width=40em
                        ]
                    ]
            ]
            [Software \\ Maintenance
                    [Log Analysis
                        [
                            DLog~\cite{DBLP:journals/tjs/LiMS18} 
                            FT-tree~\cite{DBLP:conf/iwqos/ZhangMBYLPXCDQS17} 
                            LPV~\cite{DBLP:conf/icdm/XiaoQW0L20} 
                            LTmatch~\cite{2021LTmatch} 
                            LogStamp~\cite{DBLP:journals/sigmetrics/TaoMCZLDHZWY22} 
                            LogBERT~\cite{DBLP:conf/ijcnn/GuoYW21} \\
                            LogPrompt~\cite{DBLP:conf/icse/0001TMYZY24} 
                            R-Log~\cite{liu2025r} 
                            AdaptiveLog~\cite{ma2025adaptivelog} 
                            ReAct-RCA~\cite{pei2025flow} 
                            LogRESP-Agent~\cite{lee2025logresp} \\
                            CyberSleuth~\cite{fumero2025cybersleuth} 
                            Audit-LLM~\cite{song2024audit}
                           , text width=40em
                        ]
                    ]
                    [Compiler Optimization
                        [
                            Iterative Compilation~\cite{1998Iterative} 
                            Cenetic search~\cite{DBLP:conf/lctrts/CooperSS99} 
                            Random exploration~\cite{DBLP:conf/cgo/AgakovBCFFOTTW06} \\
                            Greedy heuristics~\cite{DBLP:conf/cgo/PanE06} 
                            OpenTuner~\cite{DBLP:conf/IEEEpact/AnselKVRBOA14} 
                            CLTune~\cite{DBLP:conf/mcsoc/NugterenC15} 
                            MilepostGCC~\cite{2008MILEPOST} \\
                            DeepTune~\cite{2017End} 
                            AutoPhase~\cite{huang2019autophase} 
                            CompilerDream~\cite{deng2025compilerdream} 
                            \citet{DBLP:journals/corr/abs-2309-07062}
                           , text width=40em
                        ]
                    ]
                    [Decompilation
                        [
                            Neutron~\citep{DBLP:journals/cybersec/LiangCHC21} NeurDP~\citep{DBLP:conf/acsac/CaoL0H22} BTC~\citep{DBLP:journals/corr/abs-2212-08950} Slade~\citep{DBLP:journals/corr/abs-2305-12520} \\DecGPT~\citep{DBLP:journals/corr/abs-2310-06530} Nova+~\citep{DBLP:journals/corr/abs-2311-13721} LLM4Decompile~\citep{DBLP:conf/emnlp/TanL0Z24} 
                            CFADecLLM~\citep{liu2025control}
                           , text width=40em
                        ]
                    ]
                    [Deobfuscation
                        [
                            DeGuard~\citep{DBLP:conf/ccs/BichselRTV16} Autonym~\citep{DBLP:conf/sigsoft/VasilescuCD17} Debin~\citep{DBLP:conf/ccs/HeITRV18} JSNeat~\citep{DBLP:conf/icse/TranTNNN19} \\DIRE~\citep{DBLP:conf/kbse/LacomisYSAGNV19} VarBERT~\cite{DBLP:journals/corr/abs-2103-12801} DIRECT~\citep{nitin-etal-2021-direct} LmPa~\citep{DBLP:journals/corr/abs-2306-02546}
                            \\InFuncName~\citep{2019In} 
                            ALFREDO~\citep{nataliealfredo} 
                            Androidmeda~\citep{Androidmeda-Deobfuscate-android-app_2024} 
                            Patsakis~\citep{patsakis2024assessing}
                           , text width=40em
                        ]
                    ]
                    [DevOps and CI/CD
                        [
                            AutoDev~\cite{tufano2024autodev} 
                            GPT-Engineer~\cite{gptengineer} 
                            CodeAgent~\cite{tang2024codeagent}
                            , text width=40em 
                        ]
                    ]
            ]
            [End-to-End \\ Software Agents
                [Waterfall-based \\ full-cycle agents
                    [
                        ChatDev~\cite{qian2024chatdev} MetaGPT~\cite{hong2023metagpt} AISD ~\cite{zhang2024experimenting} CTC~\cite{du2024multiagent} CodePori~\cite{rasheed2024codepori}
                        , text width=40em
                    ]
                ]
                [Agile and iterative \\ frameworks
                    [
                        LCG \cite{lin2024whenllm} AgileCoder~\cite{nguyen2024agilecoder} CodeS \cite{zan2024codes} IER \cite{qian2024iterativeexperience} Croto \cite{du2024multiagent}
                        , text width=40em
                    ]
                ]
            ]
		]
	\end{forest}}
	\caption{SWE-Agents in software engineering lifecycles.}
    \label{fig:taxonomy-swe-agents}
\end{figure*}
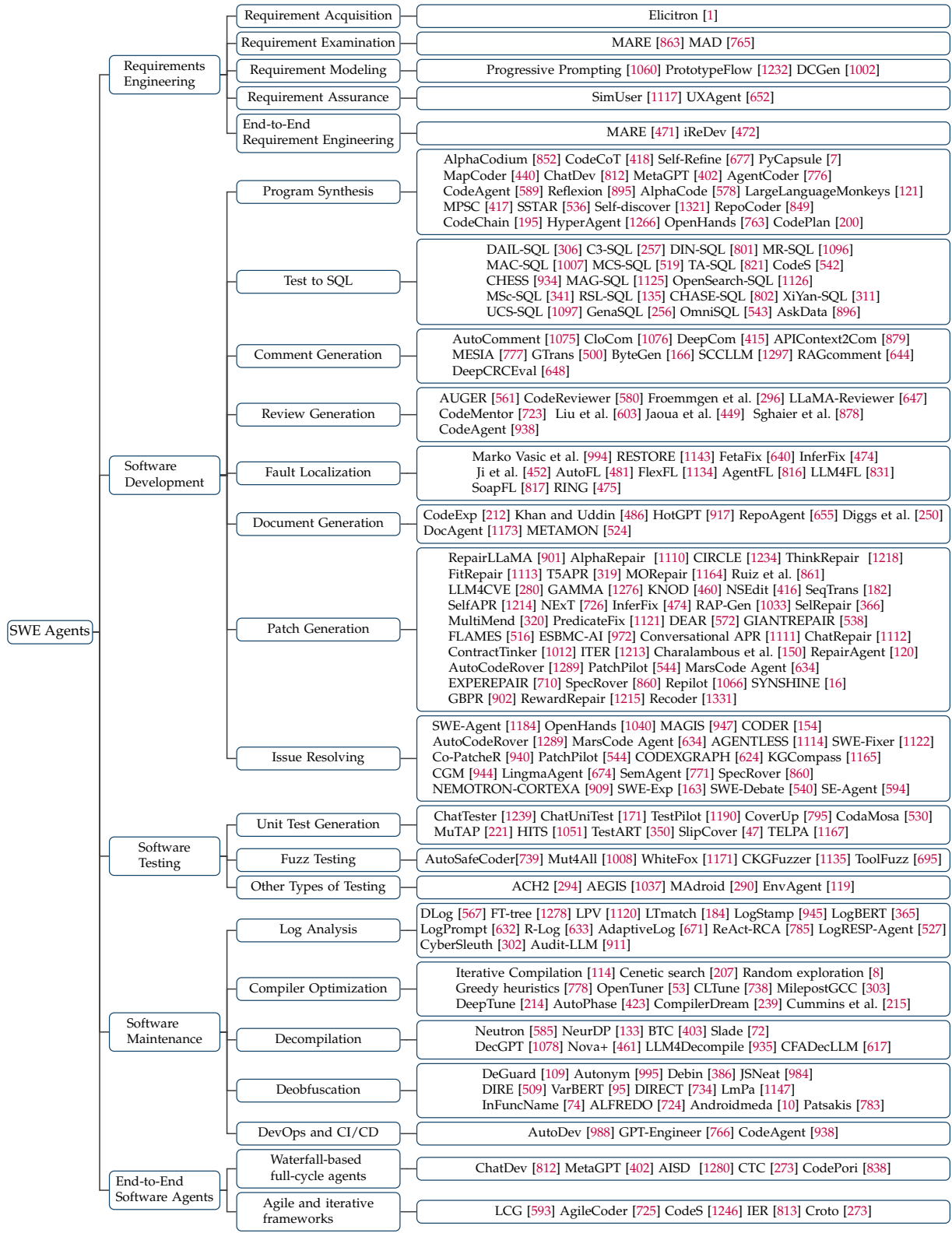

Recent developments in large language models (LLMs) have contributed to the emergence of SWE Agents, including autonomous or semi-autonomous systems designed to support, automate, or enhance traditional software engineering workflows. These agents demonstrate potential for efficiency, accuracy, and scalability throughout the software development lifecycle. To systematically examine their capabilities and implications, this section organizes SWE Agents by their downstream tasks within the classical phases of the waterfall model~\cite{royce1987managing} (to illustrate diverse agents in linear sequential software development), as shown in~\autoref{fig:taxonomy-swe-agents}: requirements analysis and design, software development, testing, and deployment. This organizational framework not only illustrates the range of agent applications but also reveals phase-specific challenges and innovations. Furthermore, in addition to introducing existing agent frameworks, this review analyzes the training paradigms, including data curation, fine-tuning, and reinforcement learning strategies, which enable effective SWE Agents. By connecting task-oriented categorization with training methodologies, this section aims to support future research in developing more robust, adaptable, and reliable agents for software engineering.

\subsection{SWE Agents Operate Across Lifecycles in Software Engineering}

\subsubsection{Requirements Engineering}
\label{subsec:requirements_engineering}
\begin{figure*}[!htbp]
	\centering
    \resizebox{\textwidth}{!}{
	\begin{forest}
        forked edges,
		for tree={
                grow=east,
                reversed=true,
                anchor=base west,
                parent anchor=east,
                child anchor=west,
                base=center,
                font=\large,
                rectangle,
                draw=hidden-draw,
                rounded corners,
                align=left,
                text centered,
                minimum width=4em,
                edge+={darkgray, line width=1pt},
                s sep=3pt,
                inner xsep=2pt,
                inner ysep=3pt,
                line width=0.8pt,
                ver/.style={rotate=90, child anchor=north, parent anchor=south, anchor=center},
            },
            where level=1{text width=8em,font=\normalsize, }{},
            where level=2{text width=12em,font=\normalsize}{},
            where level=3{text width=14em,font=\normalsize,}{},
	    [LLM-based Agents for \\Requirements Engineering, ver
			[Acquisition
                    [Simulated User Agents
                        [
                            Elicitron~\cite{Elicitron}
                           , text width=26em
                        ]
                    ]
			]
			[Examination \& \\Reconciliation
                    [Team Simulation
                        [
                            MARE~\cite{UserStory}
                           , text width=26em
                        ]
                    ]
                    [Multi-Agent Debate
                        [
                            MAD~\cite{MAD}
                           , text width=26em
                        ]
                    ]
            ]
            [Modeling \& \\Formalization
                    [Progressive Prompting
                        [
                            Progressive Prompting~\cite{Progressive}
                           , text width=26em
                        ]
                    ]
                    [UI Generation
                        [
                            PrototypeFlow~\cite{PrototypeFlow}{, }DCGen~\cite{DCGen}
                           , text width=26em
                        ]
                    ]
            ]
            [Assurance \& \\Confirmation
                    [Persona-based Simulation
                        [
                            SimUser~\cite{SimUser}
                           , text width=26em
                        ]
                    ]
                    [Multi-Persona Evaluation
                        [
                            UXAgent~\cite{UXAgent}
                           , text width=26em
                        ]
                    ]
            ]
            [End-to-End RE
                [End-to-End \\Automation
                        [
                            MARE~\cite{MARE}
                           , text width=26em
                        ]
                ]
                [Knowledge-driven \\Orchestration
                        [
                            iReDev~\cite{iReDev}
                           , text width=26em
                        ]
                ]
            ]
		]
	\end{forest}}
	\caption{Taxonomy of LLM-based Agents for Requirements Engineering.}
    \label{fig:taxonomy-re-agents}
\end{figure*}
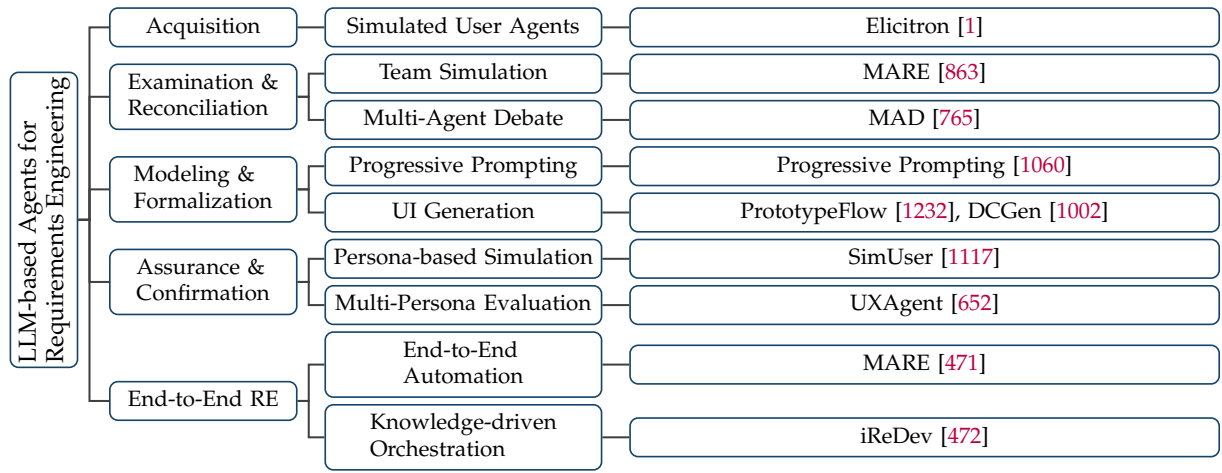

Requirements engineering (RE) constitutes the essential discipline in software engineering that bridges stakeholder demands with technical implementations, ensuring the delivered system fulfills user expectations. The RE process generally unfolds through several interlinked stages, including \textbf{acquisition}, \textbf{examination and reconciliation}, \textbf{modeling and formalization}, and \textbf{assurance and confirmation}, where each stage addresses distinct aspects of capturing, refining, representing, and validating requirements. Beyond these discrete phases, emerging \textbf{multi-stage frameworks} integrate multiple RE activities within unified, agentic pipelines. \autoref{subsec:requirements_engineering} explores representative agentic approaches across these stages, culminating in integrated multi-stage systems that holistically automate RE lifecycles, as shown in~\autoref{fig:taxonomy-re-agents}. 

\paragraph{Acquisition}
The acquisition phase focuses on identifying and extracting stakeholder expectations through diverse elicitation methods such as interviews, collaborative sessions, and observation. Agent-based user simulation has emerged as a powerful paradigm for this stage. \textbf{Elicitron}~\cite{Elicitron} generates diverse simulated user agents that autonomously interact with target products, logging their experiences and providing structured feedback. By interviewing and analyzing these simulated users, the system uncovers both explicit and latent requirements, significantly expanding coverage beyond conventional human-centered elicitation.

\paragraph{Examination and Reconciliation}
Following acquisition, the examination and reconciliation phase ensures that gathered requirements are coherent, feasible, and comprehensible. It often involves prioritization and conflict resolution among stakeholders. Multi-agent systems have proven effective in automating this analytical process. \citet{UserStory} simulate a virtual development team composed of agents with designated roles such as Product Owner and QA, collaboratively generating, evaluating, and ranking user stories. Meanwhile, \citet{MAD} introduce the \textbf{multi-agent debate (MAD)} mechanism, where opposing agents argue over requirement interpretations while a judge agent consolidates opinions into a refined and balanced specification. This adversarial collaboration enhances consistency and decision quality.

\paragraph{Modeling and Formalization}
In the modeling and formalization phase, requirements are translated into structured, machine-interpretable representations ranging from abstract UML models to concrete UI prototypes. \citet{Progressive} employ \textbf{progressive prompting} to iteratively map natural-language requirements to object-oriented code structures, facilitating traceability from text to implementation. For user-facing systems, \textbf{PrototypeFlow}~\cite{PrototypeFlow} coordinates multiple specialized design agents under a supervisory coordinator to transform textual requirements into coherent interface mockups. Similarly, \textbf{DCGen}~\cite{DCGen} utilizes multi-modal agent collectives that interpret UI screenshots and generate functional code, bridging the gap between visual and textual requirement representations.

\paragraph{Assurance and Confirmation}
This final stage validates that formalized requirements are both correctly specified and aligned with stakeholder intent to ensure that teams are building the system right and building the right system, where agent-based simulation plays a crucial role. \textbf{SimUser}~\cite{SimUser} models dual agents, where one emulates an application and another is a persona-based user to perform heuristic user experience (UX) validation. Extending this work, \textbf{UXAgent}~\cite{UXAgent} deploys thousands of diverse virtual users that autonomously navigate web interfaces and provide large-scale usability and satisfaction feedback, delivering an unprecedented level of automated assurance coverage.

\paragraph{End-to-End RE}
While the preceding sections address individual phases, the \textbf{end-to-end approach} integrates multiple RE processes into a continuous and feedback-driven pipeline. These frameworks manage an agent team, including collectors, analyzers, modelers, and validators, within a unified environment that allows outputs from one stage to dynamically inform the next. For instance, \textbf{MARE}~\cite{MARE} establishes an agentic workspace where acquisition agents capture stakeholder input, modeling agents translate it into formal structures, and validation agents continuously test requirement coherence, thereby closing the loop between discovery and verification. Similarly, \textbf{iReDev}~\cite{iReDev} introduces a knowledge-centric ecosystem of six domain-specific agents (including interviewers, analysts, and reviewers) that collaborate through shared event-based repositories. This design enables traceable transitions across stages, automated conflict detection, and consistent alignment with evolving stakeholder objectives. Collectively, multi-stage frameworks embody the agentic paradigm’s ultimate promise: end-to-end autonomy and adaptivity throughout the requirements lifecycle.

\subsubsection{Software Development}

\paragraph{5.1.2.1. Program Implementation}\mbox{}\\
\paragraph{Program Synthesis}
Program synthesis agents extend beyond static prompting by incorporating multi-step reasoning, test-based verification, and feedback-driven refinement loops. They aim to construct full programs from scratch with minimal human intervention, often by simulating aspects of a human programmer’s workflow (planning, coding, testing, debugging). 

\begin{figure}[h]
    \centering
    \includegraphics[width=1.0\textwidth]{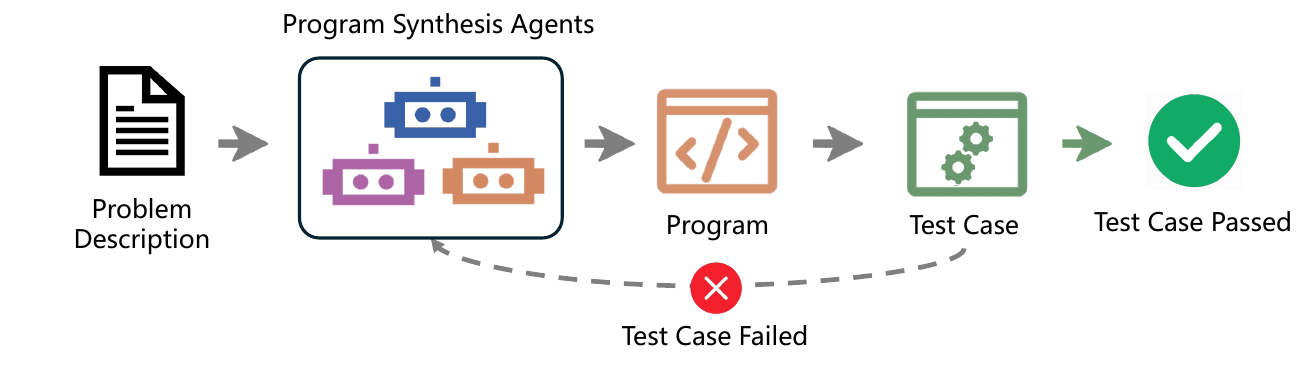} 
    \caption{Overview of the program synthesis.}
    \label{fig:program_synthesis}
\end{figure}

\subparagraph{(1) Problem Definition} Synthesize a program from a specification such that it passes a test suite. Formally, the input is a specification document (e.g. a natural language problem description, optional diagrams) and a set of test cases. The output is a program that satisfies the requirements and passes all tests, as illustrated in \autoref{fig:program_synthesis}. The agent must operate autonomously, without any ground-truth code, possibly through multiple internal reasoning and coding iterations.

\subparagraph{(2) Architectures of Program Synthesis Agents} Program synthesis agents can be organized in different ways depending on how they generate, test, and refine code. Recent research mainly explores two directions: single-agent systems, where one model iteratively improves its own outputs, and multi-agent systems, where several specialized agents collaborate to handle different tasks such as coding, testing, and debugging. The following discussion outlines these two main architectures.

\begin{itemize}
    \item \textbf{Single-Agent Iterative Systems } 
Early approaches to agentic program synthesis predominantly rely on a single-agent paradigm, in which one large language model iteratively improves its own outputs through cycles of reasoning, self-reflection, and re-prompting. Within this setting, a unified prompt loop governs planning, coding, testing, and debugging, allowing the model to coordinate the entire workflow without external controllers. AlphaCodium~\cite{ridnik2024alphacodium} exemplifies this methodology by prompting the model to draft implementations, analyze potential flaws in natural language, and regenerate improved solutions while using structured YAML outputs to maintain consistency. Intermediate reasoning steps further reduce hallucination and specification drift, and automated test execution provides feedback that enhances overall performance.
Similar strategies adopt the same self-improvement philosophy. CodeCoT~\cite{huang2023codecot} encourages the model to produce code and tests jointly to identify errors during generation, while few-shot self-refinement~\cite{madaan2023selfrefine} demonstrates that iterative correction can be achieved with minimal supervision. These methods effectively address syntactic and shallow logical issues, yet they struggle with deeper semantic reasoning and domain-intensive tasks. Overall, single-agent architectures benefit from simplicity, internal continuity of reasoning, and low operational overhead, but offer limited support for richer multi-perspective analysis and collaborative problem solving.

\item \textbf{Multi-Agent Pipelines }
Recent work extends beyond single-agent loops by introducing dual and multi-agent pipelines that strengthen modularity, reliability, and division of labor in program synthesis. PyCapsule~\cite{adnan2025pycapsule} exemplifies a dual-agent structure composed of a Programmer that generates and revises code and an Executor that runs the code and returns concrete feedback such as test failures or execution traces. This separation reduces hallucinated execution results and provides more grounded signals for iterative refinement. Later designs enhance the Executor with evaluative capabilities to better capture semantic issues that are not covered by basic tests.
Building on this foundation, multi-agent systems introduce additional roles to emulate collaborative software development. MapCoder~\cite{islam2024mapcoder} coordinates a retriever, a planner, a coder, and a debugger in a structured workflow that retrieves examples, drafts plans, implements solutions, and performs testing. Similar systems such as ChatDev~\cite{qian2024chatdev} and MetaGPT~\cite{hong2023metagpt} extend this idea by simulating multi-role organizations that reveal both the benefits of natural language coordination and the challenges of communication overhead and error propagation.
More recent approaches focus on streamlined collaboration with targeted specialization. AgentCoder~\cite{pan2024agentcoder} adopts a three-agent setup consisting of a coder, an independent test writer, and an executor that supports more rigorous self-evaluation. CodeAgent~\cite{liao2024codeagent} further integrates auxiliary agents and external tools (documentation readers, dependency analyzers, and compilers), to enable deeper contextual reasoning and repository-scale analysis.
\end{itemize}

\subparagraph{(3) Feedback as the Engine of Effective Code Search}
In agentic program synthesis, success depends not only on producing a single output but on how effectively the agent explores and refines the solution space. Feedback provides the mechanism that transforms generation into guided search. Through evaluation, critique, or self-testing, the model converts trial and error into informed iteration. Prior studies show that structured feedback significantly enlarges the effective action space of a language model, enabling correction of reasoning faults and gradual convergence toward functional correctness even with limited model capacity or compute~\citep{shinn2023reflexion}.
Across systems such as AlphaCode~\citep{li2022alphacode} and PyCapsule~\citep{adnan2025pycapsule}, feedback-driven exploration has become a defining characteristic of agentic intelligence in program synthesis. Four complementary strategies illustrate this paradigm: parallel sampling, iterative self-refinement, hybrid search, and consistency-based re-ranking. Together, these approaches provide distinct mechanisms through which agents probe, evaluate, and improve candidate programs.

\begin{itemize}
    \item \textbf{Parallel Sampling and Selection }
A common strategy for improving solution quality in code generation and program repair is to sample multiple candidate programs independently and select the most promising one, often referred to as best-of-N sampling. This approach leverages the stochasticity of large language models, as the probability of generating a correct solution increases with the number of samples. Prior work has characterized this trend through inference-time compute scaling, showing that broader sampling budgets systematically improve coverage of correct solutions~\citep{brown2024largelanguagemonkeys}.
In settings with reliable automatic verifiers, such as unit test driven coding benchmarks, expanded sampling frequently yields higher end-to-end success rates. However, naive sampling eventually encounters diminishing returns, particularly when verifiers are incomplete or when ranking depends on heuristic evaluators. Simple selection methods, including majority voting and reward model scoring, often plateau as sample sizes grow~\citep{brown2024largelanguagemonkeys}, indicating that unbounded sampling without principled selection is computationally inefficient.
Consequently, practical systems combine moderate sampling budgets with more structured filtering mechanisms. Typical approaches include execution based validation, LLM based scoring, or hybrid pipelines that integrate sampling with feedback driven refinement. Early systems such as AlphaCode~\citep{li2022alphacode} relied on large sample pools followed by clustering and execution checks, whereas more recent models attain similar or better performance with far fewer samples by coupling sampling with stronger verifiers or iterative improvement procedures. These developments underscore the value of pairing sampling with robust and cost effective selection strategies.
    \item \textbf{Iterative Refinement }
    A central strategy in agentic program synthesis is to generate an initial solution and iteratively improve it using feedback. This feedback may arise from executing public test cases, from static diagnostics such as compiler errors or lint warnings, or from the model’s own critique. Empirical studies show that one to three refinement rounds usually yield the largest performance gains, after which improvements plateau or may even regress due to noisy or misleading feedback. PyCapsule~\cite{adnan2025pycapsule}, for example, reports that a small number of self-debug attempts corrects many errors, while additional iterations often provide diminishing returns.
    Similar observations appear in interactive repair settings. On the Commit0 benchmark, \citet{zhao2024commit0} demonstrate that only a few repair rounds substantially increase the test pass rate, with later iterations offering limited benefit. Systems typically terminate once code passes available tests or after a fixed iteration budget. Complementary techniques encourage the model to articulate its reasoning before proposing fixes, as seen in CodeCoT~\cite{huang2023codecot} and self-refinement methods~\cite{madaan2023selfrefine}, which produce more targeted corrections. Some frameworks further cache previous mistakes to prevent oscillation and repeated ineffective edits~\cite{shinn2023reflexion, zhou2024self}.
    \item \textbf{Hybrid Search } A recent trend is to combine parallel generation with iterative refinement, capitalizing on both diversity and feedback. For example, \citet{li2025sstar} present a test-time scaling approach that first samples multiple candidates and then improves each one independently through a few rounds of self-debugging. After refinement, $S^*$ employs a selection step where it pits the refined solutions against each other using additional tests. Notably, these tests can be AI-generated: $S^*$ uses an LLM to synthesize adversarial inputs that differentiate between two candidate programs, then runs both programs to see which one produces correct or more robust outputs. This pairwise tournament of solutions yields a final winner that often inherits the strengths of many. By integrating both axes of search (breadth via sampling, depth via refinement), the hybrid strategy achieved near-SOTA results on benchmarks like LiveCodeBench\cite{jain2024livecodebench}, while using far less compute than an equivalent large model. The advantage is especially pronounced when using smaller models – techniques such as S narrowed the gap to proprietary models by making better use of compute at inference time. 
    A known drawback, however, is the added system complexity: hybrid search must coordinate multiple concurrent code-generation threads or agents and balance exploration depth against computational cost. Prior studies \cite{qian2024chatdev, madaan2023selfrefine, jain2024livecodebench} note that excessive iterations or candidate sampling can lead to diminishing returns and higher resource overheads, highlighting the need for adaptive stopping criteria. Nonetheless, this approach embodies the general philosophy of “generate, then iterate,” which underlies many agentic systems.
    \item \textbf{Consistency-based Re-ranking}
Consistency among multiple candidate outputs provides an implicit signal of correctness and offers an alternative to direct performance-driven selection. Inspired by self-consistency in chain-of-thought prompting, this line of work assumes that solutions independently derived in multiple coherent forms are more likely to be valid \cite{wang2022self}.
In program synthesis, Multi-Perspective Self-Consistency (MPSC)~\cite{huang2024mpsc} illustrates this idea by generating three complementary perspectives for each candidate: an implementation, a natural-language specification describing the intended behavior, and a set of unit tests. Consistency is assessed along two dimensions. The first dimension examines whether code execution aligns with the behavior implied by the specification. The second dimension evaluates whether execution results satisfy the expectations encoded in the tests. These relationships are modeled through a weighted graph in which edges reflect the strength of agreement, enabling candidate solutions to be ranked not only by external test results but also by internal coherence.
\end{itemize}

\subparagraph{(4) Scaling to Repository-Level Generation}
As code-generation research moves from single-function synthesis toward full-repository reasoning, the central challenge shifts from producing correct code snippets to managing long-context dependencies, iterative feedback, and large-scale integration. The Commit0 framework \cite{zhao2024commit0} catalyzed this transition by defining an interactive environment where agents must build complete libraries under test-driven development. While the benchmark itself remains difficult, it has spurred a new class of repository-level development agents that extend its core principles.

\begin{itemize}
    \item \textbf{Interactive Development Loops } The prototype agent SDE-I~\cite{zhao2024commit0} introduces a three-stage compile–test–debug cycle, including drafting modules in topological order, running static analysis, and iteratively patching unit-test errors. This feedback-centric design inspired follow-up agents such as OpenHands~\cite{openhands2024} and CodePlan~\cite{codeplan2023}, which extend the loop with automatic tool calls and explicit execution reasoning. OpenHands enhances the refinement process by diversifying repair strategies instead of repeating previous fixes, while CodePlan introduces structured planning to better coordinate compilation, execution, and debugging. Together, these systems demonstrate how integrating multi-round, verbalized feedback and execution reasoning can substantially improve the robustness of automated debugging workflows.
    \item \textbf{Dependency-Aware and Retrieval-Augmented Generation } Commit0~\cite{zhao2024commit0} highlights the difficulty of maintaining cross-file consistency, where agents must understand import graphs and API relations. Subsequent systems like RepoCoder \cite{repocoder2023} and CodeChain \cite{codechain2023} explicitly model repository dependency graphs, generating modules in a dependency-sorted manner and retrieving relevant snippets for each component, which shortens context length by orders of magnitude while retaining semantic coherence.
    \item \textbf{Tool-Integrated and Memory-Enhanced Agents } To cope with the long-horizon nature of repository tasks, recent frameworks embed agents within real development environments. OpenHands~\cite{openhands2024} and HyperAgent \cite{zhang2024hyperagent} employ multiple sub-agents for navigation, editing, and testing inside a sandboxed IDE. HyperAgent’s four-role team (planner, navigator, editor, executor) achieves state-of-the-art performance on RepoExec~\cite{le2024repoexec} by allocating different models to different functions (this idea derives from SDE-I’s separation of analysis and execution). These systems illustrate how scaling to repository level requires not only larger context windows but persistent memory and tool-aware interaction.

\end{itemize}

\paragraph{Text to SQL}

\begin{figure}[h]
    \centering
    \includegraphics[width=1.0\textwidth]{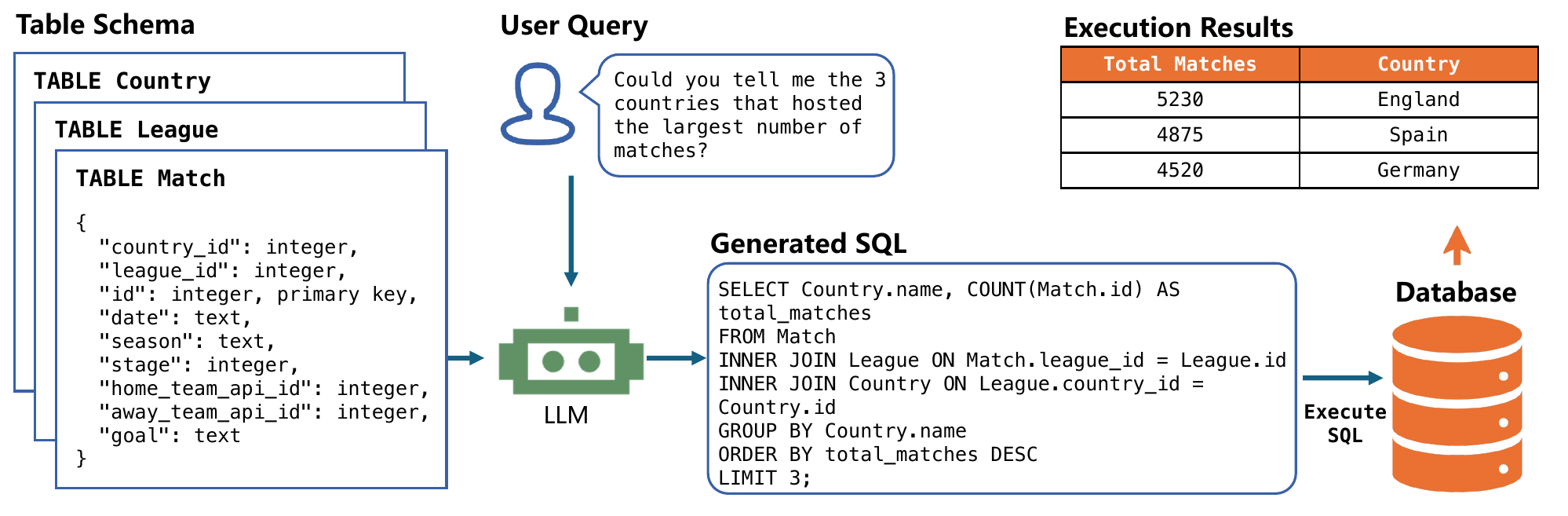} 
    \caption{Illustration of text to SQL.}
    \label{fig:nl2sql_task}
\end{figure}

Within the broader landscape of software engineering agents, Text-to-SQL plays a critical role during the development phase, where it enables intelligent systems to transform natural-language requirements into structured database queries, as shown in \autoref{fig:nl2sql_task}. By bridging human intent and data retrieval, it supports developers in automating data access, validating design assumptions, and constructing data-driven components without manual query crafting. As such, Text-to-SQL serves as both a reasoning benchmark and a practical development tool that unifies natural-language understanding, structured data querying, and autonomous decision-making. Over time, the field has evolved from early rule-based and template-guided systems to neural encoder–decoder models, and more recently to LLM-driven pipelines that incorporate schema linking, execution feedback, and multi-agent collaboration. This progression has shaped a rich research landscape that continues to expand in scope and sophistication.

\subparagraph{Pre-LLM Structure-aware Modeling}
Before the emergence of LLMs, systems paired BERT-style encoders with structure-aware decoders, such as abstract syntax trees~\cite{Xu,Guo,Wang} and predefined query sketches~\cite{He}. Encoder–decoder models also learned reusable NL{\textrightarrow}SQL patterns from supervised corpora~\cite{Hui,Li,Jinyang,Zheng,Dial-sql}.

\subparagraph{Prompting and In-context Learning}
Recently, some researchers have leveraged LLMs in text-to-SQL tasks to handle complex reasoning~\cite{Wei,yao2023react}. A critical aspect of this approach is the design and utilization of prompts, which directly influence the accuracy of SQL generation by guiding LLMs effectively. For example,\citet{Tai} improve the inference capabilities of LLMs using chain-of-thought prompting, including both the original chain-of-thought prompt and the least-to-most prompt. Further, Chang et al.~\cite{Chang} uncover the critical database knowledge and optimal representations for effective prompting through a comprehensive analysis. DAIL-SQL \cite{Dial-sql} considers both questions and SQL queries to select few-shot examples, adopts an example organization strategy to balance quality and quantity, and utilizes code representation prompting for question representation. Additionally, C3-SQL~\cite{C3-sql} and DIN-SQL~\cite{Din-sql} have introduced innovative frameworks for database simplification, query decomposition, and prompt engineering. Overall, C3 proves that well-crafted zero-shot prompts plus execution voting are enough to reach SOTA, whereas DIN-SQL introduces a modular, difficulty-aware pipeline and a built-in self-repair loop to raise the ceiling for complex, few-shot Text-to-SQL.

\subparagraph{Schema Grounding, Retrieval, and Agents}
To strengthen grounding and robustness, researchers have introduced schema linking to identify database tables and columns associated with natural language queries and have proposed complex and integrated prompting engineering methods. MR-SQL~\cite{MR-SQL} uses three independent retriever modules to pre-process the original information for accurate schema linking and to select few-shot examples with high reference significance. ~\citet{iswc2024} define a unified KGs-based schema for LLMs to generate SQL queries based on the evidence and retrieved information. MAC-SQL~\cite{MAC-SQL} and MCS-SQL~\cite{MCS-SQL}, centered on multi-agent collaboration, are designed to handle more intricate data scenarios and a broader range of error types for detection and correction. Beyond text-to-SQL generation, recent work has begun emphasizing the challenge of SQL debugging itself. BIRD-CRITIC~\cite{li2025swesqlilluminatingllmpathways} introduces a comprehensive benchmark and an associated training framework that highlight the limitations of current LLMs in diagnosing and refining complex SQL queries, while also demonstrating the potential of specialized open-source systems for strengthening this capability. TA-SQL~\cite{TA-SQL} and CodeS~\cite{CodeS} first pre-trains a code model to predict the SQL skeleton, then fine-tunes it to fill in tokens while using execution-checked data augmentation to multiply equivalent SQL variants, enabling accurate fine-tuning without any LLM prompting. CHESS~\cite{chess} enables more accurate schema linking by retrieving relevant information from database catalogs and database values. Another approach, MAG-SQL \cite{MAG-SQL} features a multi-agent generative approach with soft schema linking and iterative Sub-SQL refinement. OpenSearch-SQL \cite{OpenSearch-SQL} aligns the inputs and outputs of agents through the alignment module, reducing failures in instruction following and hallucination. MSc-SQL \cite{MSc-SQL} mitigates the performance gap of smaller open-source models by sampling and comparing multiple SQL query results. RSL-SQL \cite{RSL-SQL} robustly links the schema by first pruning it bidirectionally, then augments the slimmed schema with LLM-generated SQL components, asks the LLM to vote between full-schema and simplified-schema queries, and finally refines the chosen SQL through multi-turn self-correction until it executes successfully.
CHASE-SQL~\cite{CHASE-SQL} prompts multiple LLM agents to generate diverse SQL candidates via divide-and-conquer, query-plan reasoning and online synthetic examples, then trains a pairwise-selection agent to pick the best query. XiYan-SQL~\cite{XiYan-SQL} integrates the ICL approach to maximize the generation of high-quality and diverse SQL candidates. To better retrieve the correct identifiers from the schema and transform the linguistic structure, UCS-SQL~\cite{UCS-SQL} unites content and structure pipes to respectively extract key content and transform linguistic structure, thereby collaboratively enhancing SQL query generation. GenaSQL~\cite{GenaSQL} introduces ``N-rep'' consistency, which leverages multiple representations of the same schema input to mitigate weaknesses in any single representation, making the solution more robust and allowing the use of smaller and cheaper models. OmniSQL~\cite{OmniSQL} builds a scalable pipeline that bootstraps realistic relational databases from web tables, generates complexity-aware SQL, back-translates them into nine stylistically diverse natural-language questions, and synthesizes chain-of-thought solutions before fine-tuning open-source LLMs on the resulting quadruples to achieve state-of-the-art Text-to-SQL performance. AskData~\cite{AskData} explores the use of two standard and one newer metadata extraction techniques: profiling, query log analysis, and SQL-to-text generation using an LLM, which boosts BIRD~\cite{li2024can} benchmark accuracy by about 10 percentage points over the previous best.

\paragraph{5.1.2.2. Program Analysis}\mbox{}\\

\paragraph{Comment Generation}

\begin{table}[h!]
\centering
\resizebox{0.95\linewidth}{!}{%
\begin{threeparttable}
\caption{Summary of code comment generation methods and capabilities}
\label{tab:comment_summary}
\begin{tabular}{@{}lcccccc@{}}
\toprule
\textbf{Work} 
& \textbf{Mapping-Based} 
& \textbf{RAG} 
& \textbf{Structured Methods} 
& \textbf{IR-based} 
& \textbf{ICL} 
& \textbf{Eval} \\
\midrule
\rowcolor{gray!5}
AutoComment~\cite{AutoComment}       & \cmark & \xmark & \xmark & \xmark & \xmark & \xmark \\
CloCom~\cite{CloCom}                  & \cmark & \xmark & \cmark & \xmark & \xmark & \xmark \\
\rowcolor{gray!5}
DeepCom~\cite{DeepCom}                & \xmark & \xmark & \xmark & \xmark & \xmark & \xmark \\
APIContext2Com~\cite{APIContext2Com}  & \xmark & \xmark & \cmark & \xmark & \xmark & \xmark \\
\rowcolor{gray!5}
MESIA~\cite{MESIA}                    & \xmark & \xmark & \xmark & \xmark & \xmark & \cmark \\
GTrans~\cite{GTrans}                  & \xmark & \xmark & \cmark & \xmark & \xmark & \xmark \\
\rowcolor{gray!5}
ByteGen~\cite{ByteGen}                & \xmark & \xmark & \cmark & \cmark & \xmark & \xmark \\
SCCLLM~\cite{SCCLLM}                  & \xmark & \xmark & \xmark & \xmark & \cmark & \xmark \\
\rowcolor{gray!5}
RAGcomment~\cite{RAGcomment}          & \xmark & \cmark & \xmark & \xmark & \cmark & \xmark \\
DeepCRCEval~\cite{DeepCRCEval}        & \xmark & \xmark & \xmark & \xmark & \cmark & \cmark \\
\bottomrule
\end{tabular}
\begin{tablenotes}[flushleft]
\footnotesize
\item \textbf{Mapping-Based}: utilizes pre-existing code–comment alignments to guide generation.
\item \textbf{RAG}: employs retrieval-augmented generation to integrate external code or documentation knowledge.
\item \textbf{Structured Methods}: leverage program-structural signals (e.g., API context, ASTs) to enhance comment generation.
\item \textbf{IR-based}: operate on intermediate representations (e.g., bytecode, LLVM IR, Dex/Smali) for comment generation.
\item \textbf{ICL}: applies LLMs for comment generation via in-context learning (few-shot, tool-free adaptation).
\item \textbf{Eval}: designs dedicated benchmarks or metrics to assess the utility and quality of generated comments.
\end{tablenotes}
\end{threeparttable}
}
\end{table}

Comment generation~\cite{rag_comment_generation,rag_comment_generation2,developer_intent_comment_generation} refers to the automated creation of natural language annotations that describe the purpose, functionality, or behavior of code segments. Its primary significance lies in improving software readability, maintainability, and collaboration efficiency, particularly in large-scale or legacy codebases where manual comment writing is labor-intensive and error-prone. By bridging the gap between source code and human understanding, effective comment generation can accelerate onboarding for new developers, assist in code reviews, and support various downstream tasks such as bug localization or API usage comprehension.
    
\subparagraph{Mapping-Based Methods} Early research in this area predominantly adopted mapping-based methods, which established direct correspondences between code fragments and descriptive text. These systems typically relied on databases of code–description pairs to retrieve the most relevant comment for a given code segment. AutoComment~\citep{AutoComment} constructed such databases to retrieve suitable comments, while CloCom~\citep{CloCom} detected similar code segments across repositories and transferred corresponding annotations. While these methods prioritized accuracy and worked well within constrained settings, they suffered from limited semantic generalization and heavy dependence on pre-existing mappings.
\subparagraph{Neural Seq2Seq Models} The introduction of sequence-to-sequence (Seq2Seq) neural architectures marked a shift toward generative approaches, enabling models to learn comment patterns from large-scale annotated corpora. DeepCom~\citep{DeepCom} employed Seq2Seq networks for end-to-end code-to-comment generation, while Apicontext2com~\citep{APIContext2Com} incorporated predefined API context to improve the relevance of generated comments. MESIA~\citep{MESIA} introduced a framework for evaluating the supplementary information contained in generated comments, GTrans~\citep{GTrans} enhanced Transformer architectures with graph neural networks to capture richer structural properties of code, and ByteGen~\citep{ByteGen} extended comment generation to bytecode, bypassing the need for original source code. These neural approaches improved semantic fluency and adaptability but continued to require large annotated datasets and faced challenges in domain transfer.
\subparagraph{LLM- and Agent-Based Comment Generation}  
The emergence of LLMs and software engineering agents has transformed comment generation from a static text-to-text task into an interactive, reasoning-driven process. LLM-based methods such as SCCLLM~\citep{SCCLLM} and RAG-Comment~\citep{RAGcomment} leverage in-context learning and retrieval augmentation to produce more contextually grounded comments. Recent works integrate these capabilities into autonomous or collaborative agents that perform \textbf{self-reflection, multi-round reasoning, and tool-assisted documentation}. For instance, documentation-oriented agents in frameworks such as AutoDev~\citep{tufano2024autodev} continuously analyze repository histories, link code changes to their rationales, and generate or revise comments as part of an end-to-end development pipeline. Multi-agent paradigms further extend this process: a reader agent summarizes code intent, a review agent verifies factual correctness, and a refiner agent improves stylistic quality and consistency.

\paragraph{Review Generation}
Automated review generation~\cite{Wu2025Automated,reviewrl,automatically_review_generation_llm} refers to employing intelligent systems to automatically produce constructive, context-aware, and actionable comments for code changes. The input typically consists of diffs, code snippets, or function revisions, while the output is a coherent review comment identifying issues, suggesting improvements, or confirming correctness. Beyond accelerating human review and improving consistency, review generation agents also serve as autonomous participants in software quality assurance—capable of reasoning, critiquing, and coordinating with other agents within the development pipeline.

\subparagraph{Task-Specific Pretraining and Fine-Tuning}
Early works approach code review generation through task-specific model design. AUGER~\cite{li2022auger} pre-trains a T5 model on Java code–comment pairs using a masked language modeling objective. CodeReviewer~\cite{li2022codereviewer} further introduces diff tag prediction and denoising tasks, while the DIDACT framework~\cite{DIDACT} combines multi-task pretraining on billions of examples with specialized fine-tuning on real human comments. These models provide linguistic and semantic grounding for downstream reviewer agents.

\subparagraph{LLM-Based Code Review Models}
Recent research has increasingly adopted general-purpose large language models for automated review generation. LLaMA-Reviewer~\cite{lu2023llama} applies LoRA fine-tuning to adapt LLaMA for code diff analysis, while QLoRA-based models~\cite{haider2024prompting} and Carllm~\cite{yu2024fine} explore trade-offs between prompt-based adaptation and full fine-tuning. CodeMentor~\cite{codementor} employs RLHF to enhance contextual accuracy and reviewer tone. Collectively, these models demonstrate that lightweight adaptation of general-purpose LLMs can achieve competitive performance while maintaining scalability for downstream deployment within multi-agent systems.

\subparagraph{Data-Centric and Hybrid Quality Enhancement}
Beyond model improvements, a complementary line of work enhances data pipelines and hybrid reasoning. \citet{liu2025too} employ LLMs as classifiers to filter non-actionable review comments, improving dataset quality. \citet{jaoua2025combining} combine static analyzers and LLMs for hybrid data augmentation, while \citet{sghaier2025leveraging} explore comment reformulation via LLM rewriting. These efforts emphasize the synergistic integration of structured rules and generative flexibility.

\subparagraph{Agent-Based Review Frameworks}
The rise of software engineering agents has extended review generation from single-model prediction to multi-agent orchestration. In these systems, specialized reviewer agents simulate collaborative human review teams, coordinating through shared memory and role-based reasoning. 
Representative frameworks are summarized in ~\autoref{tab:agent-review-systems}. For instance, \textbf{Hydra-Reviewer}~\cite{ren2025hydra} introduces a parallel multi-agent architecture where each agent focuses on one review dimension (e.g., logic, readability, security), combining their assessments via a dimensional classification schema. \textbf{CodeAgent}~\cite{tang2024codeagent} models team collaboration with hierarchical roles (CEO, Coder, Reviewer, QA-Checker) to emulate human decision-making pipelines and prevent prompt drift during multi-turn discussions, achieving a 41\% higher vulnerability detection rate than GPT-4. \textbf{DeputyDev}~\cite{khare2025deputydev} scales this approach to industrial settings with expert-level agents specializing in security, maintainability, and performance, coordinated by a hybrid engine that merges and verifies outputs in near real-time. \textbf{iCodeReviewer}~\cite{peng2025icodereviewer} dynamically routes requests among prompt experts specialized for specific vulnerability categories, guided by code-aware routing algorithms that enhance issue identification.

 \newcommand{\iconsystem}{%
  \raisebox{-0.13em}{%
    \includegraphics[height=1em]{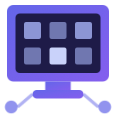}%
  }%
}
\newcommand{\iconcorearch}{%
  \raisebox{-0.13em}{%
    \includegraphics[height=1em]{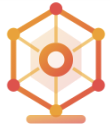}%
  }%
}
\newcommand{\iconagent}{%
  \raisebox{-0.13em}{%
    \includegraphics[height=1em]{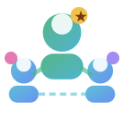}%
  }%
}
\newcommand{\iconkeyinnovation}{%
  \raisebox{-0.13em}{%
    \includegraphics[height=1em]{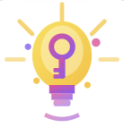}%
  }%
}

\begin{table}[h!]
\centering
\caption{Representative Agentic Review Generation Systems}
\label{tab:agent-review-systems}
\begin{tabular}{
  >{\raggedright\arraybackslash}p{3cm}
  >{\raggedright\arraybackslash}p{3.5cm}
  >{\raggedright\arraybackslash}p{4cm}
  >{\raggedright\arraybackslash}p{4cm}
}
\toprule
\iconsystem~\textbf{System} & \iconcorearch~\textbf{Core Architecture} & \iconagent~\textbf{Agent Roles} & \iconkeyinnovation~\textbf{Key Innovation} \\ 
\midrule
Hydra-Reviewer~\cite{ren2025hydra} &
Parallel multi-dimensional analysis &
Reviewers per dimension (logic, readability, etc.) &
Multi-dimensional classification taxonomy \\

CodeAgent~\cite{tang2024codeagent} &
Simulated human collaboration &
CEO, Coder, Reviewer, QA-Checker &
Prompt drift prevention via QA consistency \\

DeputyDev~\cite{khare2025deputydev} &
Industrial-scale expert orchestration &
Security, maintainability, performance agents &
Hybrid engine for output merging \\

iCode-Reviewer~\cite{peng2025icodereviewer} &
Security-focused dynamic expert mix &
Prompt specialists per vulnerability type &
Code-based dynamic routing algorithm \\ 
\bottomrule
\end{tabular}
\end{table}

\begin{figure}[H]
    \centering
    \includegraphics[width=0.75\textwidth]{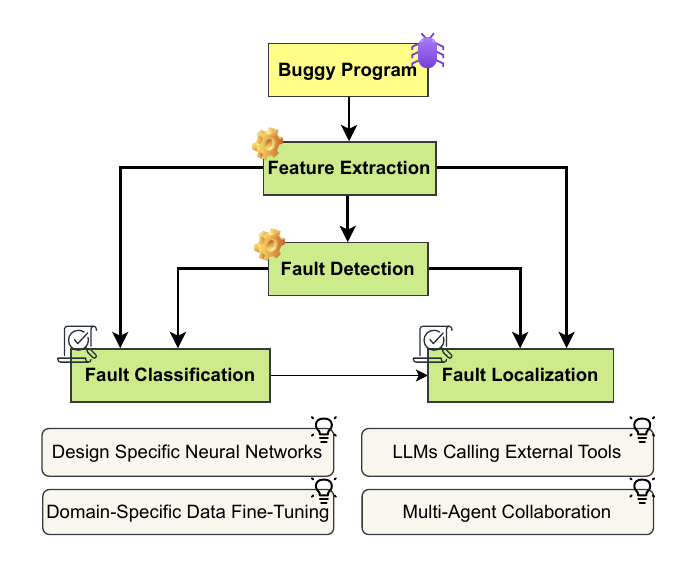}
    \vspace{-2mm}
    \caption{Taxonomy of End-to-end Fault Localization.}
    \label{fig:Taxonomy_of_fault_localization}
\end{figure}

\paragraph{Fault Localization}\label{sec:fault-localization}
Fault localization~\cite{swe_fault_localization,wu2308large_fault_localization, yang2024large_fault_localization, kang2024quantitative_fault_localization} is the process of automatically identifying the specific locations in source code that are responsible for causing a program failure, typically by analyzing program execution data such as test case outcomes.
As illustrated in~\autoref{fig:Taxonomy_of_fault_localization}, research on fault localization and automated repair has evolved into a rich and diverse area, spanning neural architectures, optimization-driven learning, external tool integration, and multi-agent collaboration. The overarching objective of these approaches is to enhance the precision, scalability, and adaptability of automated debugging, addressing varied scenarios such as variable misuse detection, large-scale code reasoning, and domain-specific model conversions.

\subparagraph{End-to-End Neural Methods}
This line of research focuses on unified neural architectures that jointly address fault detection, localization, and repair within a single learning pipeline. Early work explored pointer network based models capable of identifying faulty statements while simultaneously generating candidate repairs, illustrating the benefits of joint learning compared with enumerative approaches. Subsequent systems such as Restore~\cite{xu2020restore} incorporated feedback driven refinement, where failed validation attempts inform dynamic analysis and guide the search toward more effective patches. Domain-oriented frameworks like FetaFix~\cite{louloudakis2025fetafix} further extended neural repair to specialized settings, including the automated conversion of deep learning models, by identifying faults across representation frameworks such as ONNX and TVM and applying iterative correction.
Taken together, these efforts highlight the strengths of integrated neural reasoning for aligning localization and repair. By coupling end-to-end learning with structured feedback, these methods capture complex bug semantics while maintaining efficiency across diverse program domains.

\subparagraph{Optimization Methods}
Another research trajectory explores fine-tuning large language models with targeted and domain-specific datasets to enhance debugging performance.
Approaches such as DeepDebug~\cite{drain2021deepdebug,debug_like_human} synthesize artificial bugs from real-world commit histories to augment training data, thereby improving robustness in joint localization–repair tasks. InferFix~\cite{jin2023inferfix} integrates static analysis and retrieval-augmented learning, combining a contrastively trained retriever with a generative LLM to unify detection, classification, and correction under a single framework. Other works~\cite{ji2025impact} further demonstrates that line-aware fine-tuning can yield interpretable fault localization by training models to directly generate line-level predictions, circumventing traditional reliance on execution coverage. Overall, optimization-based methods highlight the importance of adaptive fine-tuning strategies that encode bug semantics and structural patterns into model representations, improving generalization across languages and repositories.

\subparagraph{External Tool Calling}
To overcome inherent LLM constraints such as limited context windows and incomplete code understanding, recent studies integrate external tools into the debugging pipeline.
AutoFL~\cite{kang2024quantitative} exemplifies this direction by enabling LLMs to interact with repositories via tool calls, thereby supporting explainable fault localization grounded in code navigation and repository traversal. Building upon this idea, FlexFL~\cite{xu2025flexfl} introduces a two-agent framework that combines static fault localization signals with LLM-based refinement: one agent prioritizes suspicious code regions, while another performs contextual inspection and re-ranking through interactive reasoning. These methods bridge the gap between reasoning and execution, allowing models to iteratively refine localization hypotheses through tool-assisted verification and repository-scale exploration.

\subparagraph{Multi-Agent Collaboration}
An emerging direction in fault localization employs multiple specialized agents that cooperate to decompose and coordinate debugging tasks. RING~\cite{joshi2023repair} exemplifies this trend by enabling language models to jointly conduct localization, ranking, and patch generation through inter-agent communication. LLM4FL~\cite{rafi2024multi} further develops a tri-agent architecture composed of context extraction, debugging, and reviewing agents, each responsible for subtasks such as reducing coverage data or reasoning over failure graphs. Reinforcement driven dialogue facilitates information exchange and consensus formation among agents.
Compared with single-agent approaches, these multi-agent paradigms introduce complementary reasoning perspectives and structured role separation, resulting in improvements in both interpretability and robustness. Collectively, they mark a shift toward collaborative fault localization, where distributed expertise and coordinated decision making enhance the autonomy and adaptability of the debugging pipeline.

\paragraph{Document Generation}
In software engineering, document generation~\cite{can_developers_prompt_code_documentation} refers to the automated production of comprehensive, structured, and often long-form textual documentation for software projects, APIs, or codebases. This task is essential for ensuring effective knowledge transfer, facilitating software maintenance, and enhancing usability for both internal and external stakeholders. High-quality documentation reduces the cognitive load on developers, accelerates onboarding, and supports compliance and quality assurance processes, making its automation a valuable goal in modern software development.

\subparagraph{Early Model-Based Approaches} Early work in this domain grappled with the inherent difficulty of producing long-form, contextually coherent explanations. CodeExp \citep{CodeExp} formalized the code explanation generation task and proposed a multi-stage fine-tuning strategy based on a retrieve-then-generate mechanism. Their methodology first trained the model to retrieve relevant contextual snippets from the codebase, and subsequently used these snippets to condition the final explanation generation process. This approach aimed to improve the relevance and completeness of the output. Despite this advancement, these initial efforts were limited by model capabilities and often fell short in handling complex, large-scale codebases.
\subparagraph{LLM-Driven Approaches} The emergence of powerful LLMs such as GPT and Codex rapidly advanced the field. \citet{KhanGPT} demonstrate Codex’s potential in generating comprehensive documentation, Hotgpt~\citep{HotGPT} investigate a general-purpose, one-model-fits-all solution for documentation tasks, and \citet{Dvivedi} provide a comparative analysis of different LLM-based documentation generation approaches. These works confirmed that LLMs excel in producing linguistically fluent, cross-domain documentation, though they also revealed weaknesses in technical accuracy and the ability to keep pace with evolving codebases.
\subparagraph{Agent-Based Approaches}
Recent work has increasingly explored agent-based architectures that integrate large language models into proactive and continuously evolving documentation pipelines. RepoAgent~\citep{RepoAgent} presents an open-source framework that automatically generates, maintains, and updates documentation as code evolves. \citet{Diggs} focus on bringing similar automation to legacy systems, while DocAgent~\citep{DocAgent} introduces a collaborative multi-agent framework composed of specialized roles including Reader, Searcher, Writer, Verifier, and Orchestrator. METAMON~\citep{METAMON} augments this paradigm by combining LLMs with metamorphic testing to identify inconsistencies between documentation and actual program behavior.
Collectively, these agent-based systems offer promising directions for addressing long-standing documentation challenges. Automating the maintenance cycle, as in RepoAgent, helps keep documentation accurate and sustainable over time. Role specialization and collaborative workflows, as demonstrated in DocAgent, improve scalability and enable coverage of large and complex codebases. The incorporation of active verification, exemplified by METAMON, further ensures that generated documentation aligns with real code semantics. Together, these techniques suggest that coordinated agentic workflows can substantially enhance both the quality and reliability of software documentation.

\paragraph{5.1.1.3. Program Editing}\mbox{}\\
\label{sec:program-editing}

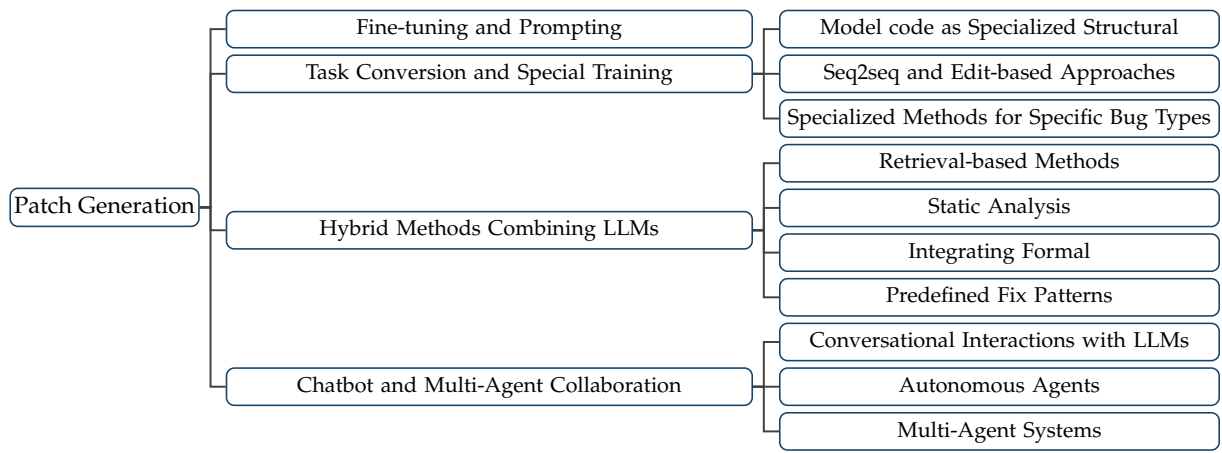
\begin{figure*}[!htbp]
	\centering
    \resizebox{\textwidth}{!}{
	\begin{forest}
        forked edges,
		for tree={
                grow=east,
                reversed=true,
                anchor=base west,
                parent anchor=east,
                child anchor=west,
                base=center,
                font=\large,
                rectangle,
                draw=hidden-draw,
                rounded corners,
                align=left,
                text centered,
                minimum width=6em,
                edge+={darkgray, line width=1pt},
                s sep=3pt,
                inner xsep=2pt,
                inner ysep=3pt,
                line width=0.8pt,
                ver/.style={rotate=90, child anchor=north, parent anchor=south, anchor=center},
            },
            where level=1{text width=24em,font=\normalsize, }{},
            where level=2{text width=20em,font=\normalsize}{},
            where level=3{text width=7em,font=\normalsize,}{},
	    [
                Patch Generation
			[
                    Fine-tuning and Prompting
                ]
			[
                    Task Conversion and Special Training
                    [
                        Model code as Specialized Structural
                    ]
                    [
                        Seq2seq and Edit-based Approaches
                    ]
                    [
                        Specialized Methods for Specific Bug Types
                    ]
                ]
                [
                    Hybrid Methods Combining LLMs
                    [
                        Retrieval-based Methods
                    ]
                    [
                        Static Analysis
                    ]
                    [
                        Integrating Formal
                    ]
                    [
                        Predefined Fix Patterns
                    ]
                ]
                [
                    Chatbot and Multi-Agent Collaboration
                    [
                        Conversational Interactions with LLMs
                    ]
                    [
                        Autonomous Agents
                    ]
                    [
                        Multi-Agent Systems
                    ]
                ]
		]
	\end{forest}}
	\caption{Taxonomy of End-to-end Patch Generation.}
    \label{fig:Taxonomy_of_patch_generation}
\end{figure*}

\paragraph{Patch Generation}\label{sec:patcg-generation}
Patch generation~\cite{wang2023rap,autopatch,bhandari2025generating,lin2024one} refers to the automated creation of code fragments that repair program vulnerabilities or logical errors. Its central objective is to automate the workflow traditionally executed by developers, spanning fault localization through patch synthesis, thereby improving software reliability and development efficiency. Recent years have witnessed substantial progress in this field, with a shift from template-based and example-driven repairs toward large language model-powered and agent-oriented frameworks.
Current approaches can be grouped into several paradigms, including fine-tuning and prompting techniques, task reformulation and training strategies, static analysis and rule-based repair, retrieval augmented generation, and multi-agent or conversational architectures. Each direction introduces distinct mechanisms for reasoning, adaptation, and verification, collectively advancing automated patch generation toward more autonomous, interpretable, and scalable systems.

\subparagraph{Fine-tuning and Prompting}
This line of work focuses on enhancing pre-trained language models for code through targeted fine-tuning or prompting to produce correct patches. Techniques such as RepairLLaMA~\cite{silva2025repairllama}, AlphaRepair~\cite{xia2022less}, and T5APR~\cite{gharibi2024t5apr} leverage parameter-efficient fine-tuning, masking strategies, and multitask learning to adapt general-purpose models for automated program repair. Others, like CIRCLE~\cite{yuan2022circle} and FitRepair~\cite{xia2023revisiting}, adopt continual learning and retrieval-based prompting to support multilingual and cross-domain repair. Emerging models, such as MORepair~\cite{yang2024morepair} and ThinkRepair~\cite{yin2024thinkrepair}, incorporate reasoning guidance, chain-of-thought prompting, and execution feedback, integrating interpretability into the repair process. Complementary work, including NExT~\cite{ni2024next} and LLM4CVE~\cite{fakih2025llm4cve}, further augments LLMs with execution traces and security-specific training, enabling them to reason over runtime behaviors and vulnerabilities. Collectively, these approaches highlight the growing capability of fine-tuned or prompted LLMs to internalize domain knowledge, generalize across languages, and perform repair through structured reasoning rather than surface pattern matching.

\subparagraph{Task Conversion and Specialized Training}
Several studies improve repair precision by reframing the patch generation task into other well-studied modeling problems, such as sequence transformation, grammar prediction, or optimization.
Recoder~\cite{zhu2021syntax} and NSEdit~\cite{hu2022fix} reformulate repair as structured code editing, employing grammar-constrained decoding and pointer-based edit modeling to preserve syntax. RewardRepair~\cite{ye2022neural} integrates execution feedback into training through reinforcement-style objectives, guiding models toward semantically correct fixes. SeqTrans~\cite{chi2022seqtrans} and KNOD~\cite{jiang2023knod} incorporate data-flow information and domain-specific rule distillation to enhance generalization and interpretability. Gradient-based program repair (GBPR)~\cite{silva2025gradient} introduces a differentiable optimization framework that maps program behavior to a continuous search space, allowing gradient descent to identify repairs. Collectively, these task conversion approaches push the boundary of APR from discrete token prediction toward learning structured, semantically grounded transformations.

\subparagraph{Static Analysis and Pattern-Guided Repair}
Another strand of research integrates symbolic reasoning, compiler diagnostics, or predefined fix patterns with LLM-based generation to guarantee correctness and robustness.
Dear~\cite{li2022dear} and Synshine~\cite{ahmed2022synshine} combine deep learning with static analysis and compiler feedback to identify multi-location or syntactic faults, enhancing precision through diagnostic-guided patching. Gamma~\cite{zhang2023gamma} and Hybrid~\cite{li2025hybrid} leverage predefined templates or structural patch skeletons to constrain the search space, improving both validity and efficiency. FLAMES~\cite{le2024semantic} bridges semantic search and LLM-driven repair, employing guided refinement based on test feedback. Similarly, hybrid systems such as ESBMC-AI~\cite{tihanyi2025new} and ContractTinker~\cite{wang2024contracttinker} integrate model checking and dependency analysis with iterative LLM reasoning to achieve verifiable and semantically consistent repairs. Overall, these methods demonstrate how formal reasoning and rule-driven heuristics can complement generative LLMs, yielding interpretable and verifiable patching pipelines.

\subparagraph{Retrieval-Augmented Approaches}
Retrieval-based methods enrich buggy code contexts with relevant bug–fix examples to inform generation.
InferFix~\cite{jin2023inferfix} exemplifies this trend by coupling static analysis with retrieval-augmented prompting, while RAP-Gen~\cite{wang2023rap} employs hybrid lexical and semantic retrievers to guide patch synthesis. SelRepair~\cite{guo2025accelerating} and MultiMend~\cite{gharibi2025multimend} propose dual retrieval and RAG selection mechanisms to enhance efficiency and multi-hunk repair capability. PredicateFix~\cite{xiao2025predicatefix} leverages static analysis predicates to retrieve targeted examples, reinforcing semantic alignment during patch validation. Collectively, retrieval-augmented methods strengthen contextual grounding, enabling models to reason from prior experience while maintaining generality across projects and programming languages.

\subparagraph{Conversational and Multi-Agent Frameworks}
Recent advances in patch generation increasingly adopt conversational paradigms and multi-agent collaboration to support interactive and adaptive repair. Frameworks such as Conversational APR~\cite{xia2023conversational} and ChatRepair~\cite{xia2023keep} employ iterative dialogue between patch generation and validation, enabling large language models to refine candidate patches through natural language reasoning and feedback from test execution. Systems including ITER~\cite{ye2024iter} and RepairAgent~\cite{bouzenia2403repairagent} further integrate fault localization, patch synthesis, and validation into continuous feedback loops, coordinating multiple agents or tool calls to support dynamic refinement.
At a broader scale, AutoCodeRover~\cite{zhang2024autocoderover}, PatchPilot~\cite{li2025patchpilot}, and MarsCode Agent~\cite{liu2024marscode} extend these ideas to repository-level repair, combining multi-agent planning, project-wide code search, and execution-based validation to emulate collaborative development workflows. More recent frameworks such as ExpeRepair~\cite{mu2025experepair} and SpecRover~\cite{ruan2024specrover} incorporate memory components, specification inference, and reviewer agents, supporting persistent learning and adaptive validation over extended interactions. Complementary approaches like Repilot~\cite{wei2023copiloting} explore coordination between language models and completion engines to enhance repair stability and patch consistency.
Collectively, these conversational and multi-agent systems represent a shift from single-pass patch generation to interactive, collaborative, and self-improving repair ecosystems. They illustrate how coordinated agentic behavior, iterative reasoning, and structured feedback can substantially enhance autonomy and reliability in automated software maintenance.

\begin{figure}[t]
    \vspace{-5mm}
    \centering
    \includegraphics[width=0.9\textwidth]{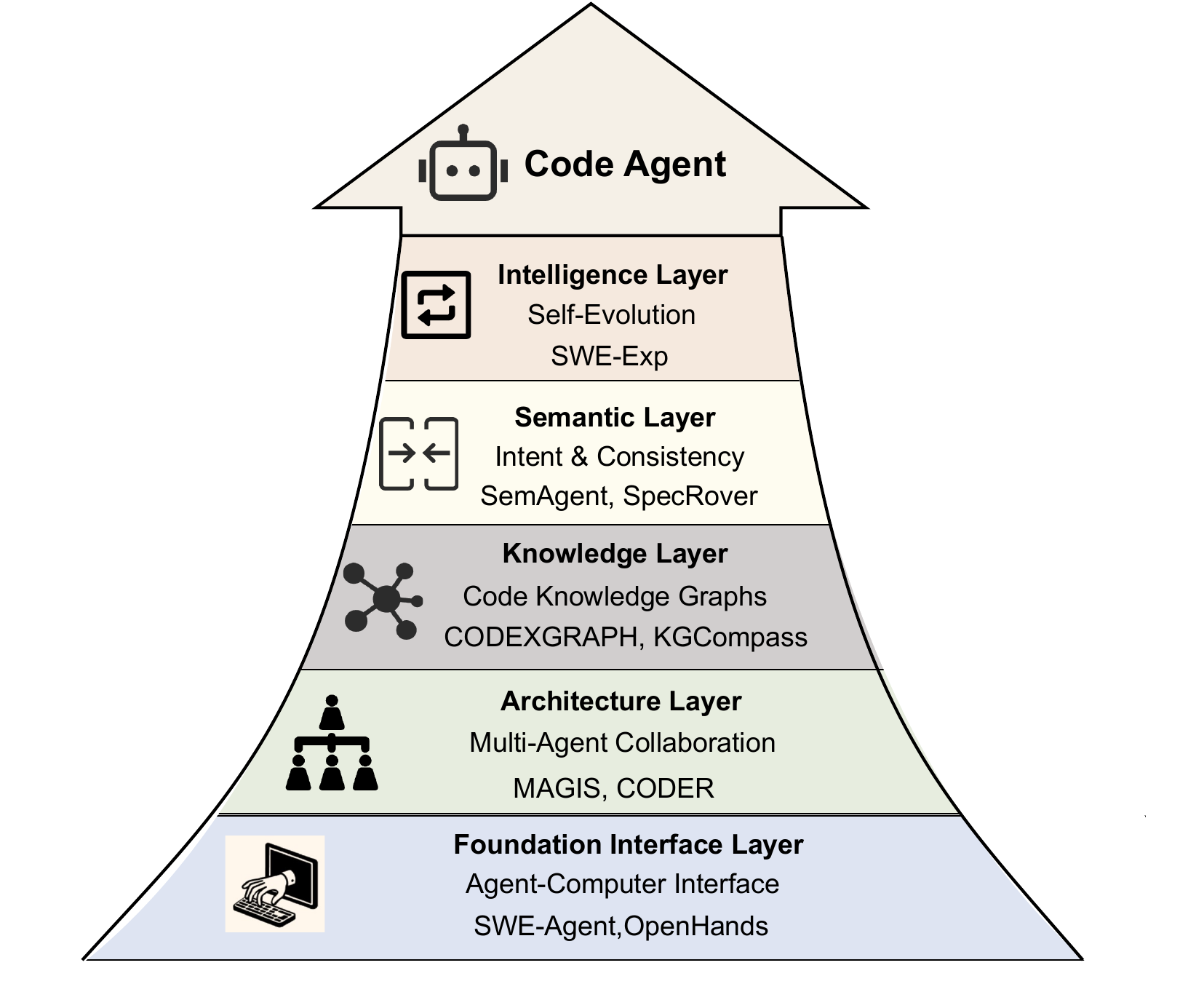}
    \vspace{-2mm}
    \caption{A hierarchical technical architecture for coding agents, illustrating the progression from the Foundation Interface Layer, through the architecture, knowledge, and semantic layers, to the culminating Intelligence Layer.}
    \label{fig:issue_resolving}
\end{figure}

\paragraph{Issue Resolving}
In recent years, the widespread adoption of large language models in software engineering has positioned agent-based automated code repair as a key direction for advancing intelligent software maintenance. Automated issue repair is inherently a complex decision-making process that spans the full workflow from understanding the problem to generating a patch~\autoref{sec:patcg-generation}, and typically involves several interconnected stages such as fault localization~\autoref{sec:fault-localization}, program editing~\autoref{sec:program-editing}, and validation~\autoref{sec:sofrware_test}.
From a technical perspective, the field has evolved across multiple layers, beginning with foundational interface designs and progressing toward more sophisticated decision-making mechanisms. These developments have led to substantial advances across the entire pipeline and have gradually shaped a hierarchical technical architecture for agent-based repair, as illustrated in~\autoref{fig:issue_resolving}.

\begin{figure}[H]
    \centering
    \includegraphics[width=0.8\textwidth]{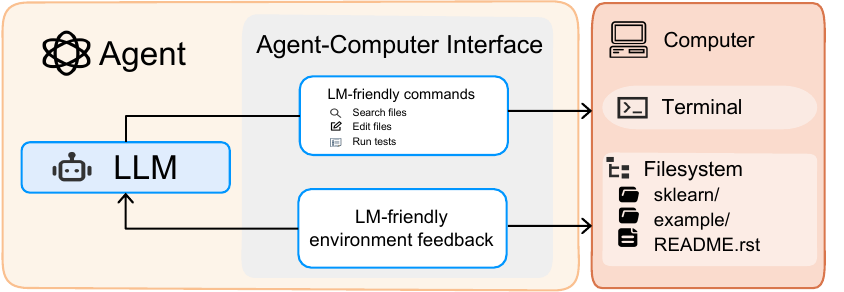}
    \vspace{-2mm}
    \caption{The paradigm of agent-environment interaction mechanisms for addressing issue resolution problems.}
    \label{fig:swe_agent_pdf}
\end{figure}

\subparagraph{Foundational interface layer} Effective automated issue repair first requires solving the interaction problem between LLMs and programming environments, since traditional text-only interfaces are insufficient for complex programming tasks. Foundational works such as SWE-Agent\cite{yang2024swe} and OpenHands \cite{wang2024openhands} were the first to introduce the language-model-as-agent concept, treating LLMs as a new kind of end user and designing dedicated agent–computer interfaces (ACIs) around their capabilities and limitations, as shown in~\autoref{fig:swe_agent_pdf}. ACIs simplify command structures, optimize environment feedback and history management, and integrate efficient editing and protective mechanisms (e.g., multi-line replacement, automatic rollback, dynamic history folding), substantially improving the operational efficiency and reliability of LLM agents in code creation, editing, and testing. This work provided critical technical foundations and design principles for subsequent agent-based automated repair systems, enabling LLMs to operate stably in complex programming environments.
    
\begin{figure}[H]
    \centering
    \includegraphics[width=0.8\textwidth]{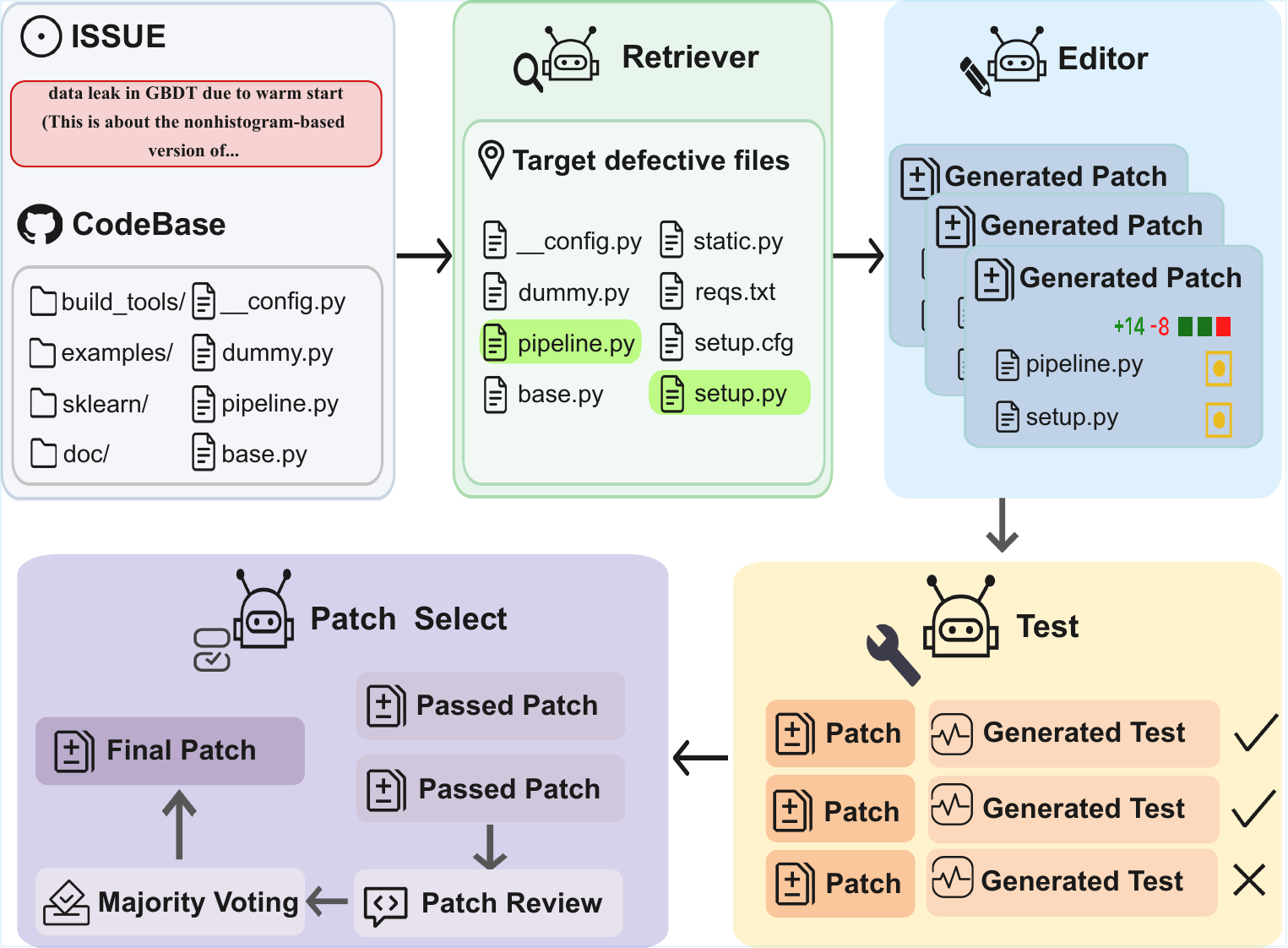}
    \caption{A typical workflow for addressing issue resolving problems.}
    \label{fig:swe_workflow}
\end{figure}

\subparagraph{Architectural Layer}
After basic interaction capabilities are established, the primary challenge becomes handling complex software issues while maintaining system efficiency and practical usability. Real-world faults often involve multiple files and layered dependencies, which are difficult for a single agent to manage. Modular collaborative architectures therefore decompose the repair workflow into specialized subtasks such as code understanding, dependency analysis, fault localization, patch generation, and validation, leading to notable improvements in scalability and performance.
Magis~\cite{tao2024magis} illustrates this idea by assigning responsibilities to four roles: manager, repository steward, developer, and quality assurance engineer, and by decomposing problems into role specific subtasks. With BM25 retrieval and memory mechanisms, Magis improves collaboration efficiency in complex multi file modification scenarios. Coder~\cite{chen2024coder} employs a predefined task graph to coordinate roles including manager, reproductor, fault localizer, editor, and verifier, enabling efficient repository level repair.
Systems such as AutoCodeRover~\cite{zhang2024autocoderover} and MarsCode Agent~\cite{liu2024marscode} adopt staged workflows for context retrieval and patch generation and use multi agent collaboration supported by code knowledge graphs, Language Server Protocol services\footnote{\url{https://microsoft.github.io/language-server-protocol/}}
, and AutoDiff tools\footnote{\url{https://www.autodiff.org/}}
 to achieve systematic repair.
At the same time, simplified workflow designs reduce system complexity, improve resource utilization, and enhance execution stability, making automated repair more practical. A representative workflow example is shown in~\autoref{fig:swe_workflow}. 
Agentless~\cite{xia2024agentless} proposes an agentless automation approach that uses a simple three-stage flow (localization, repair, patch validation) and simplifies decision-making and tooling via hierarchical localization, context windows, and majority-vote mechanisms, significantly lowering system complexity while retaining efficiency. SWE-Fixer~\cite{xie2025swe} focuses on efficiently addressing software engineering issues on GitHub through coarse-to-fine retrieval strategies and structured outputs, improving maintainability. Co-PatcheR~\cite{tang2025co} adopts a collaborative small-model system that assigns localization, generation, and verification tasks to specialized small reasoning models, greatly reducing computational cost. PatchPilot~\cite{li2025patchpilot} emphasizes stability, cost-effectiveness, and formal guarantees in the automated repair pipeline; its rule-based planning workflow uses self-reflection, hierarchical localization, and formal verification to ensure patch correctness and safety. This combination of modular collaboration and simplified workflows aims to improve execution efficiency through role specialization while preserving decision consistency, practicality, and deployability via process optimization.

\subparagraph{Knowledge Layer}
As system architectures advance, deeper understanding of code semantics and dependency relationships becomes essential for improving repair quality. Structured knowledge modeling, particularly representing a codebase as a graph, has enabled significant progress in semantic comprehension. CodexGraph~\cite{liu2024codexgraph} integrates large language models with repository structure through a graph database interface, supporting complex graph queries and multi task operations and improving the system’s ability to capture intricate dependencies among code entities. KGCompass~\cite{yang2025enhancing} constructs repository level knowledge graphs and guides repair processes by precisely linking issue descriptions to relevant code components, which narrows the search space and enhances both localization accuracy and the interpretability of generated patches.
CGM~\cite{tao2025code} incorporates semantic and structural properties of the codebase into model attention mechanisms and employs an agentless graph retrieval augmented generation framework to enable efficient subgraph retrieval and patch synthesis. LingmaAgent~\cite{ma2025alibaba} combines knowledge graphs with Monte Carlo tree search to achieve global repository understanding and dynamic patch generation. Together, these methods convert abstract semantic relationships within the code into computable graph structures, enabling structured reasoning that substantially improves problem understanding and solution generation.
\subparagraph{Semantic layer} On top of knowledge representation, accurately understanding developer intent and producing high-quality patches is a frontier in technical development. High-quality patch generation requires not only syntactic correctness but also semantic consistency and alignment with developer intentions. Methods based on deep semantic analysis fuse multi-dimensional semantic signals to significantly improve patch accuracy and applicability. SemAgent~\cite{pabba2025semagent} focuses on semantic-aware repair by performing semantic-driven fault localization and a two-stage repair architecture that combines problem semantics, code semantics, and execution semantics to produce complete and consistent patches. SpecRover~\cite{ruan2024specrover} centers on specification inference: integrating program structure, behavior, and tests to infer developer intent and generate high-quality patches, emphasizing multi-stage workflows and evidence generation mechanisms to provide interpretable guarantees of patch correctness. Nemotron-CORTEXA~\cite{sohrabizadehnemotron} develops specialized code-embedding models and localization agents, leveraging ASTs and LSP-derived information to construct code knowledge graphs, enabling multi-step reasoning and high-precision localization; it increases the diversity and accuracy of generated patches through varied context and ensemble learning. By deeply understanding the root causes and developer intent, these approaches avoid common overfitting seen in traditional methods and produce more generalizable and robust repair solutions.
\subparagraph{Intelligent Layer}
Once foundational repair capabilities are in place, a key objective becomes enabling systems to learn continuously and improve autonomously. Because software faults are diverse and highly dynamic, static repair strategies often struggle to generalize across evolving scenarios. Experience driven self evolution mechanisms address this limitation by learning from historical repair trajectories, refining decision strategies over time, and substantially enhancing system adaptability and robustness.
SWE Exp~\cite{chen2025swe} exemplifies this direction through multi dimensional experience repositories that capture reusable knowledge from both successful and failed repairs. It employs a dual agent architecture that separates strategic planning from tactical execution and augments decision making with Monte Carlo tree search to support continual and cross problem learning. SWE Debate~\cite{li2025swe} introduces a competitive multi agent debate framework that constructs fault propagation chains from code dependency graphs and organizes agents in staged debates to encourage diverse reasoning paths and more precise fault localization, addressing the limitations of single perspective analysis. SE Agent~\cite{lin2025se} applies self evolution principles by systematically expanding the exploration space and leveraging cross trajectory insights through the operations of revise, reorganize, and refine to iteratively optimize reasoning processes and enable dynamic decision making.
Together, these approaches transition software repair from a static and isolated workflow to a systematic, knowledge driven learning paradigm in which agents adapt, accumulate experience, and improve over time.

\subsubsection{Software Testing}
\label{sec:sofrware_test}

Software testing is a critical phase in the software development lifecycle that ensures the quality, reliability, and security of applications. As the complexity of software systems grows, traditional testing methods can be time-consuming and resource-intensive. The advent of LLMs has revolutionized this field, enabling the creation of autonomous agents that can automate and enhance various testing processes, from generating unit tests to performing security analyses and conducting fuzz testing. These AI-powered agents are transforming software testing from a manual, human-centric process into a more efficient and scalable system.

\paragraph{LLM-Driven Test Generation Frameworks}
In the field of software engineering, researchers have explored the use of LLM-based agents for automated test generation and dataset construction. 
Otter~\cite{ahmed2025otter} and SPICE~\cite{bhatia2025spice} respectively focus on two key aspects: automated test generation and large-scale dataset annotation. Specifically, Otter employs a locator–planner–generator architecture to automatically identify test targets and generate high-quality test cases, while SPICE constructs a scalable LLM-assisted annotation framework that integrates program analysis and human verification to build large, high-quality test datasets.
Automated test generation has further promoted the development of the test-driven development (TDD) field. 
SWE-Flow~\cite{zhang2025swe}, centered on TDD, achieves high-quality data synthesis by automatically inferring incremental development steps, constructing runtime dependency graphs (RDG), and generating structured development plans. 
This framework supports controllable data generation and automatic documentation creation, greatly enhancing dataset structure and usability. 
Experiments demonstrate that SWE-Flow–based data synthesis significantly improves LLM performance in realistic software development tasks.
Building upon these representative systems, subsequent studies have shifted attention from one-shot code generation to systematic test engineering — focusing not only on producing test cases, but also on improving their \textbf{quality, coverage, and executability}. This transition is most evident in the area of unit test generation.

\begin{figure}[h]
    \centering
    \includegraphics[width=1.0\textwidth]{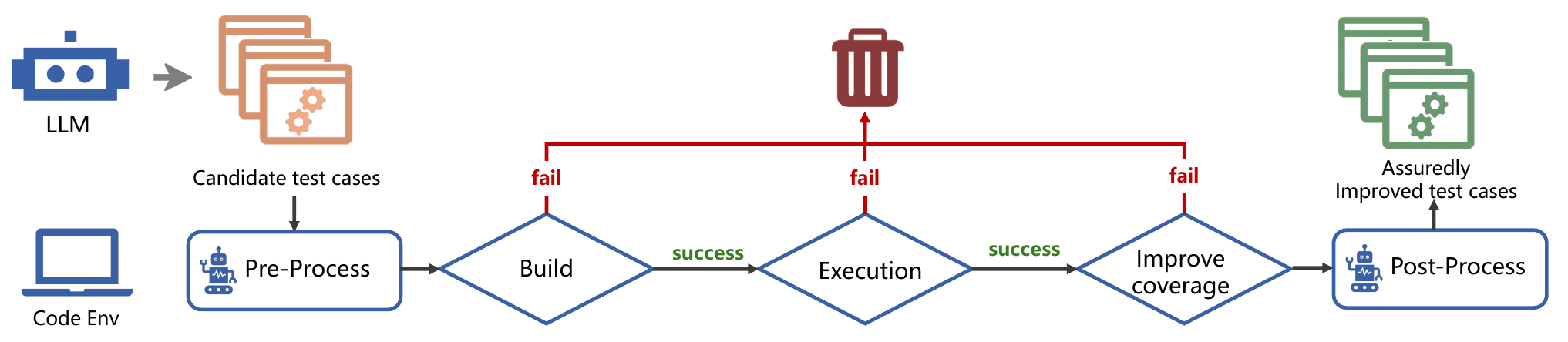} 
    \caption{Overview of unit test generation.}
    \label{fig:unit_test_generation}
\end{figure}

\paragraph{Unit Test Generation}
Unit testing validates the smallest executable parts of software such as functions or classes. Recent research on LLM-based unit test generation \cite{yuan2024manualtestsevaluatingimproving} has shifted from exploring \textbf{model capability} to building \textbf{systematic engineering} frameworks that ensure both correctness and meaningfulness of generated tests, as shown in \autoref{fig:unit_test_generation}.

\subparagraph{From Generation Capability to Generation Quality}
Early studies such as AthenaTest and A3Test focused primarily on whether large language models could generate syntactically correct test code~\cite{tufano2020unit,alagarsamy2024a3test}. With the emergence of more capable models, including ChatGPT and Codex~\cite{chatgpt,openai2025gpt5codex}, research has shifted from basic code generation toward ensuring correctness, coverage, and semantic relevance.
To improve semantic fidelity, ChatTester~\citep{yuan2024manualtestsevaluatingimproving} enhances assertion quality through intent inference and iterative refinement. TestPilot~\citep{journals/corr/abs-2406-18181} increases reliability by incorporating prompt refiner modules that leverage documentation and error guided regeneration. Coverage oriented methods such as CoverUp~\citep{pizzorno2025coverupeffectivehighcoverage} and CodaMosa~\citep{Lemieux2023CODAMOSA} integrate coverage feedback into the generation loop to improve line and branch coverage. MuTAP~\citep{dakhel2023effectivetestgenerationusing} further introduces mutation based feedback to strengthen test effectiveness and fault detection.
Collectively, these works illustrate a transition from static generation to adaptive and feedback driven refinement, highlighting that improvements in model capability must be matched by advances in test quality.

\subparagraph{From Single-Step Generation to Multi-Step Iterative Processes}
As research on unit test generation has progressed, it has become evident that a single generation step is insufficient for producing reliable tests. To address hallucinations and make more effective use of feedback, recent work has shifted from one shot generation to multi step iterative pipelines.
ChatUniTest~\citep{chen2024chatunitestframeworkllmbasedtest} enriches contextual prompting by incorporating detailed class and dependency information to support iterative generation and refinement. TestART~\citep{gu2025testartimprovingllmbasedunit} adopts template based repair strategies to iteratively correct errors in generated tests. HITS~\citep{wang2024hitshighcoveragellmbasedunit} introduces a structured multi step reasoning process that includes summarization, dependency extraction, slicing, and guided generation, coupled with self debugging feedback to improve coverage and robustness.
Subsequent frameworks such as SlipCover~\citep{Altmayer_Pizzorno_2023}, CoverUp~\citep{pizzorno2025coverupeffectivehighcoverage}, TestPilot~\citep{schäfer2023empiricalevaluationusinglarge}, and TELPA~\citep{Yang2025EnhancingLT} further strengthen this paradigm by integrating coverage feedback, orchestration mechanisms, and execution driven refinement.
Overall, recent systems no longer invoke large language models in a single step. Instead, they increasingly adopt semi autonomous pipelines that generate, evaluate, and repair tests iteratively, progressively improving coverage, correctness, and reliability.

\paragraph{Fuzz Testing}
While unit testing focuses on functional correctness, fuzz testing targets robustness and security by generating diverse, often malformed inputs. This direction extends the use of LLM agents from functional validation to vulnerability detection.  
\citet{nunez2024autosafecodermultiagentframeworksecuring} introduces a multi-agent system integrating coding, static analysis, and fuzzing agents in continuous feedback loops for autonomous vulnerability detection and repair.  
To overcome manual mutator design limitations, \citet{wang2025mut4allfuzzingcompilersllmsynthesized} presents a language-agnostic framework that synthesizes mutation operators using historical bug data.  
Similarly, \citet{Yang_2024} leverage LLMs for scalable white-box fuzzing of deep learning compilers, while \citet{xu2024ckgfuzzerllmbasedfuzzdriver} utilize code knowledge graphs for automatic fuzz driver generation.  
Further, \citet{milev2025toolfuzzautomatedagent} propose ToolFuzz, combining LLM-based query generation with fuzzing to test agentic tool reliability.  
Collectively, these studies demonstrate a gradual evolution from single-purpose fuzzing scripts to fully automated, cross-language LLM-driven fuzz testing ecosystems, capable of intelligent adaptation and learning.

\paragraph{Other Testing Agents}
Beyond unit and fuzz testing, large language model based agents have been extended to broader testing paradigms, including mutation testing, bug reproduction, interactive feature testing, and autonomous test execution, forming an increasingly comprehensive ecosystem for automated software testing.

\subparagraph{Mutation Testing}
Recent work applies LLMs to mutation testing by replacing handcrafted mutation operators with model generated mutants, as demonstrated by LLMorpheus~\citep{tip2025llmorpheus}. PRIMG~\citep{bouafif2025primg} introduces mutation informed prioritization to guide efficient test evaluation, while ACH2~\citep{foster2025mutationguidedllmbasedtestgeneration} utilizes mutation guided super bugs to stimulate high value test generation. These approaches strengthen the connection between test generation and fault detection and reinforce the broader trend toward feedback driven testing.

\subparagraph{Automated Bug Reproduction}
Agent based testing has also been extended to automated bug reproduction. AEGIS~\citep{wang2024aegis} combines structured context extraction with finite state optimization to reproduce software failures, and subsequent work~\citep{cheng2025agentic} adapts this framework to industrial settings. These methods demonstrate that agentic coordination can enhance not only test creation but also debugging and verification workflows.

\subparagraph{Multi-Agent Interactive Feature Testing}
Multi agent collaboration has further been applied to interactive feature testing. MAdroid~\citep{feng2025breaking} introduces an architecture composed of coordinator, operator, and observer agents that simulate realistic user interactions within graphical interfaces. This design enables automated evaluation of cross user workflows and expands the scope of agent based software testing.

\subparagraph{Autonomous Test Execution}
Despite progress in generating and repairing tests, executing them across heterogeneous environments remains a major challenge. ExecutionAgent~\citep{bouzenia2025you} addresses this by autonomously constructing execution environments and running test suites for complex software projects. Through the use of meta prompting, project artifacts such as configuration scripts, and iterative feedback for command execution and error recovery, the system demonstrates strong adaptability to diverse software stacks and highlights the feasibility of LLM driven automated test execution.

\subsubsection{Software Maintenance}
\begin{figure}[h]
    \centering
    \includegraphics[width=0.75\textwidth]{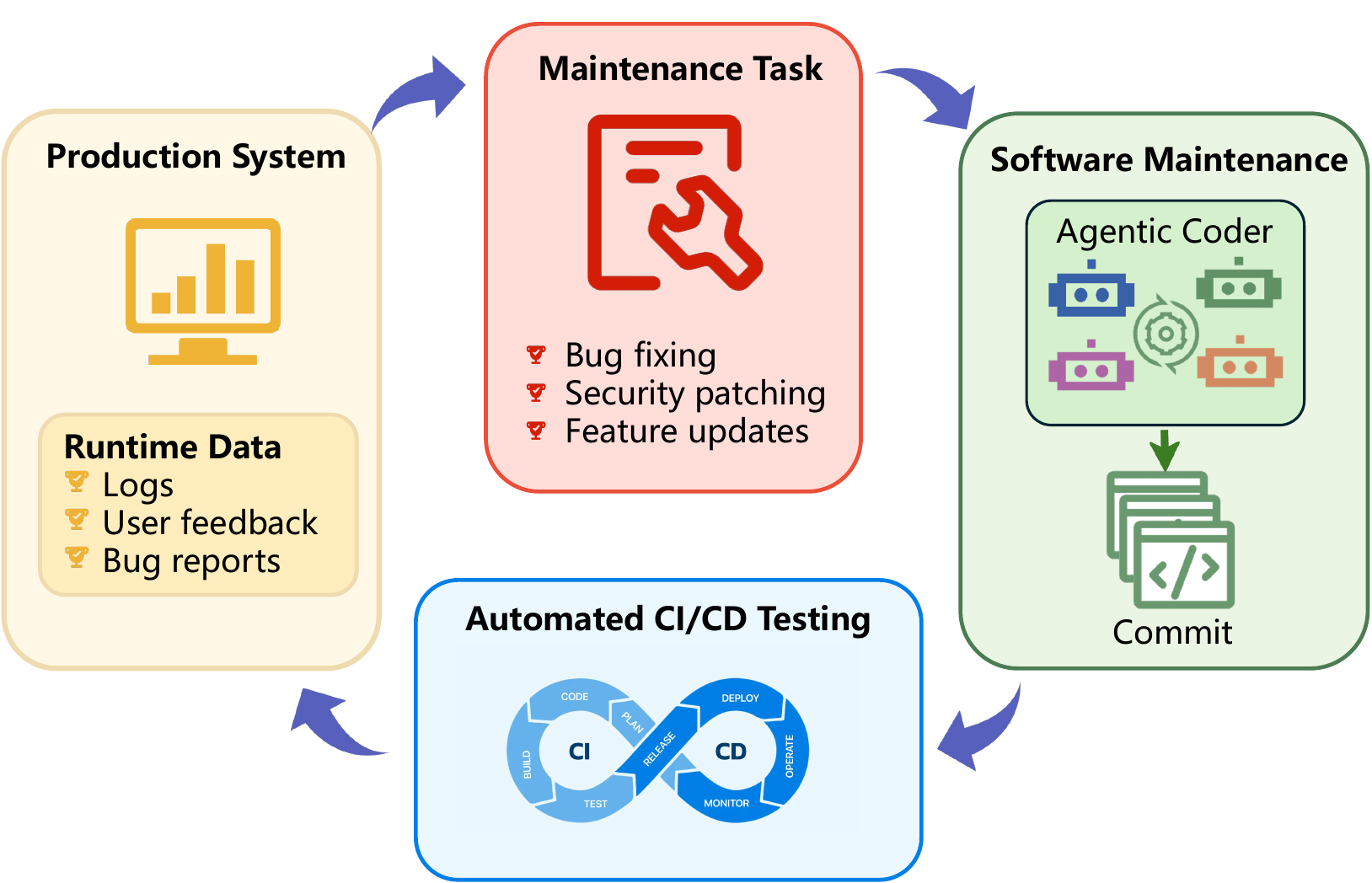} 
    \caption{Overview of software maintenance across the full software life-cycle.}
    \label{fig:software_maintenance}
\end{figure}

After software passes the testing stage, it is deployed into production, and often continuously maintained or updated~\citep{DBLP:journals/tmlr/ZhangCLLG0L024}, as illustrate in \autoref{fig:software_maintenance}. Therefore, some research has concentrated on addressing tasks within deployment and operations.

\paragraph{Log Analysis}
Log analysis is a core task in software operation and maintenance. It focuses on transforming raw, unstructured logs into actionable insights and identifying anomalies or faults in system behavior. Research in this area has evolved from traditional deterministic and statistical approaches to neural methods, and more recently to agent-based systems that emphasize autonomous reasoning and adaptive decision making. This progression can be organized into two main categories: traditional log parsing foundations and agent-based log analysis.

\subparagraph{Traditional Foundations for Log Parsing and Analysis}
Before the rise of neural and agent-based systems, research on log analysis was primarily grounded in deterministic and statistical techniques. These early foundations focused on identifying structural regularities within logs and developing parsers capable of handling large, heterogeneous datasets. Representative approaches can be grouped into two major lines of work, summarized as follows.

\begin{itemize}
    \item \textbf{Pattern Mining and Clustering.}
Early systems relied on deterministic pattern mining and clustering to extract structure from noisy log data. Methods such as DLog~\citep{DBLP:journals/tjs/LiMS18} and FT-tree~\citep{DBLP:conf/iwqos/ZhangMBYLPXCDQS17} identify invariant token sequences by constructing prefix or token-frequency trees. Clustering-based approaches, exemplified by LPV~\citep{DBLP:conf/icdm/XiaoQW0L20}, group logs with similar structures using hierarchical or density-based clustering, enabling unsupervised template discovery at scale.
    \item \textbf{Heuristic and Neural Parsing.}
Subsequent work introduced heuristic matching techniques such as longest-common-subsequence alignment in LTmatch~\citep{2021LTmatch}. Neural approaches emerged with models like LogStamp~\citep{DBLP:journals/sigmetrics/TaoMCZLDHZWY22}, which frame log parsing as a token classification task using pretrained encoders. Advanced neural systems such as LogBERT~\citep{DBLP:conf/ijcnn/GuoYW21} and LogPrompt~\citep{DBLP:conf/icse/0001TMYZY24} incorporate self-supervised learning and prompt tuning to improve adaptability across log formats. Although these methods significantly improved precision in structured environments, they lack dynamic reasoning and cross-context adaptability, motivating the emergence of more flexible agent-based paradigms.
\end{itemize}

\subparagraph{Agent-based Log Analysis}
Agent-based approaches redefine log analysis as an iterative reasoning task in which autonomous agents interpret system states, generate hypotheses, interact with external tools, and refine conclusions based on feedback. Compared with static parsers, these systems emphasize adaptability, cognitive reasoning, and coordinated decision making.

\begin{itemize}
    \item \textbf{Single-Agent Reasoning.}
Frameworks such as R-Log~\cite{liu2025r} reformulate log analysis from a static mapping problem to a reasoning-centric pipeline in which the agent produces intermediate explanations before synthesizing conclusions. By incorporating reinforcement learning in simulated O\&M environments, the system learns to optimize factual soundness and reasoning coherence across diverse log scenarios.
AdaptiveLog~\cite{ma2025adaptivelog} focuses on efficiency and accuracy by orchestrating a lightweight language model with a more capable one. Using Bayesian uncertainty estimation, the system selectively escalates challenging cases to the stronger model, achieving an effective balance between computational cost and analytical fidelity.
The ReAct-RCA framework~\cite{pei2025flow} employs a Thought--Action--Observation loop in which the agent iteratively interacts with domain knowledge bases and telemetry data. During root-cause analysis, the agent continuously refines intermediate hypotheses, enabling a more adaptive and process-oriented form of operational intelligence.

    \item \textbf{Multi-Agent Collaboration.}
Building on single-agent reasoning, recent frameworks employ multiple specialized agents to address heterogeneous log sources and complex operational contexts.
LogRESP-Agent~\cite{lee2025logresp} integrates recursive plan--act--observe reasoning with retrieval-augmented generation, enabling context-aware anomaly investigation and template-free parsing across varied log datasets.
CyberSleuth~\cite{fumero2025cybersleuth} organizes sub-agents dedicated to network-level and application-level forensics. By incorporating targeted web searches to ground reasoning in external vulnerability knowledge, the system minimizes hallucination and strengthens analytical robustness.
Audit-LLM~\cite{song2024audit} advances this paradigm through a structured decomposition of detection tasks into three agents: a \textit{Decomposer} that identifies subtasks, a \textit{Tool Builder} that configures external utilities, and an \textit{Executor} that performs the final analysis. These agents interact through an Evidence-Based Multi-Agent Debate mechanism, where candidate conclusions are iteratively challenged and validated to enhance decision fidelity.
\end{itemize}

Overall, the evolution of log analysis reflects a clear trajectory from deterministic parsing toward neural and agent-based reasoning. Traditional methods established the foundations for structuring and clustering logs, while modern agent-based systems introduce dynamic reasoning, tool interaction, and multi-agent coordination. Together, these approaches represent a shift toward more adaptive, interpretable, and context-aware operational intelligence.

\paragraph{Compiler Optimization}
Compiler optimization seeks to generate efficient executables by identifying transformation strategies that improve performance and portability across diverse architectures. Traditional compiler techniques laid the foundation for this task, and growing system complexity has gradually motivated a shift toward learning based and agent driven approaches.

\subparagraph{Traditional Foundations}
Before the development of intelligent or learning based optimizers, compiler optimization primarily relied on empirical exploration and hand crafted heuristics. Representative methods include the following:

\begin{itemize}
    \item \textbf{Iterative Compilation.}
    Early research explored compiler strategy spaces through repeated compilation, execution, and profiling, enabling systems to search for high quality optimization configurations~\citep{1998Iterative}. Search based variants, including genetic algorithms~\citep{DBLP:conf/lctrts/CooperSS99}, random exploration~\citep{DBLP:conf/cgo/AgakovBCFFOTTW06}, and greedy heuristics~\citep{DBLP:conf/cgo/PanE06}, further expanded this paradigm. Frameworks such as OpenTuner~\citep{DBLP:conf/IEEEpact/AnselKVRBOA14} and CLTune~\citep{DBLP:conf/mcsoc/NugterenC15} provided general infrastructures for empirical tuning, although these methods often lack adaptivity and incur substantial cost.

    \item \textbf{Machine Learning for Compilers.}
    To reduce search overhead, supervised learning techniques were introduced to predict optimization sequences or compiler flags directly from program features. Systems such as MilepostGCC~\citep{2008MILEPOST} and DeepTune~\citep{2017End} demonstrate that predictive models can guide optimization more efficiently than exhaustive exploration. However, these methods remain static and do not dynamically reason about optimization choices in context.
\end{itemize}

\subparagraph{Agent Based Compiler Optimization}
With the growth of intelligent decision making techniques, compiler optimization has increasingly been reframed as a sequential reasoning problem. Instead of relying on fixed heuristics, agent based systems explore, evaluate, and refine optimization strategies through interaction and feedback. Key directions include:

\begin{itemize}
    \item \textbf{LLM Augmented Optimization.}
    Recent work investigates large language models as meta optimizers capable of reasoning about code semantics and compiler configurations. \citet{DBLP:journals/corr/abs-2309-07062} demonstrate that language models can infer optimization flags from unoptimized assembly, revealing the potential of natural language guided optimization. These systems integrate in context reasoning and retrieval to dynamically adapt compiler strategies in a more interpretable manner.
    \item \textbf{Reinforcement Learning Agents.}
    Reinforcement learning based frameworks treat the compilation pipeline as a Markov Decision Process, where optimization passes correspond to actions and performance signals provide rewards. AutoPhase~\citep{huang2019autophase} shows that learned policies can outperform manually crafted pass orderings and better adapt to program characteristics.
    \item \textbf{World Model Driven Optimization.}
    Building on reinforcement learning, CompilerDream~\citep{deng2025compilerdream} introduces a world model that simulates compiler behavior. By training a predictive model to approximate compilation outcomes, the system enables broad strategy exploration in a simulated environment and supports generalization across languages and architectures through offline reinforcement learning.

\end{itemize}

Overall, compiler optimization has evolved from heuristic and search based exploration toward learning guided prediction and, more recently, agent driven adaptive reasoning. This progression reflects a broader trend toward compilers that continuously refine and generalize optimization strategies through interaction, feedback, and higher level semantic understanding.

\paragraph{Decompilation}
Decompilation aims to recover high-level source code from low-level representations such as assembly or bytecode. It is a fundamental task in reverse engineering, software maintenance, and security analysis. Traditional rule-based tools often struggle to reconstruct high-level semantics, especially when facing compiler optimizations. Recent advances in neural models and large language models have substantially reshaped this field by framing decompilation as a code translation and semantic reconstruction task.

\subparagraph{Neural and Learning-Based Decompilation}
Early learning-based systems adopt neural machine translation techniques to translate low-level programs into high-level languages. Neutron~\citep{DBLP:journals/cybersec/LiangCHC21} applies attention-based sequence modeling to generate human-readable source code while preserving functional behavior. To better handle compiler-optimized binaries, NeurDP~\citep{DBLP:conf/acsac/CaoL0H22} uses graph neural networks to map low-level representations into intermediate forms that bridge the semantic gap between binary and source code. Transformer-based approaches further demonstrate strong adaptability; BTC~\citep{DBLP:journals/corr/abs-2212-08950} treats assembly and source languages as plain text, enabling retargeting across different programming languages, while Slade~\citep{DBLP:journals/corr/abs-2305-12520} infers types and reconstructs coherent code structures using a sequence-to-sequence Transformer.

\subparagraph{LLM-Augmented Decompilation}
With the rise of large language models, decompilation has increasingly been approached as a semantic reasoning and code reconstruction task. DecGPT~\citep{DBLP:journals/corr/abs-2310-06530} proposes a hybrid workflow that uses pretrained models to repair syntax and memory issues in decompiled C programs, improving their correctness and executability. Nova+~\citep{DBLP:journals/corr/abs-2311-13721} enhances LLM training through hierarchical attention and contrastive objectives that better capture assembly-level semantics. LLM4Decompile~\citep{DBLP:conf/emnlp/TanL0Z24} introduces a family of open-source models trained specifically for decompilation, along with end-to-end variants capable of producing high-level code directly from binary input.

\subparagraph{Agent-Based Decompilation}
Recent work frames decompilation as an agentic reasoning task in which models interact with structured program information and external tools to improve reconstruction quality. CFADecLLM~\cite{liu2025control} exemplifies this direction by combining traditional disassembly outputs, such as control flow graphs, with natural language or structured token representations. Through a two-stage information fusion mechanism and role-guided prompting, the system integrates both instruction sequences and structural program analysis into a unified reasoning process. This approach leverages the strengths of program analysis techniques and the generative capabilities of language models, enabling more robust and semantically faithful recovery of high-level code.

\subparagraph{Summary}
Overall, the field has progressed from rule-based heuristics to neural translation models, and from general-purpose LLMs to agent-based systems that incorporate structured program analysis. This evolution reflects a broader trend toward decompilers that not only translate low-level code but also reason about program semantics, control flow, and high-level intent.

\paragraph{Deobfuscation}
Deobfuscation refers to the reverse process of renaming identifiers, where meaningful identifier names are recovered from obfuscated programs. DeGuard~\citep{DBLP:conf/ccs/BichselRTV16} is proposed for deobfuscating Android APKs based on probabilistic learning of large code bases. The key idea is to learn a probabilistic model over thousands of non-obfuscated Android applications and to use this probabilistic model to deobfuscate new, unseen Android APKs. Autonym~\citep{DBLP:conf/sigsoft/VasilescuCD17}, based on statistical machine translation (SMT), recovers some original names from JavaScript programs minified by the highly popular UglifyJS. Debin~\citep{DBLP:conf/ccs/HeITRV18} handles ELF binaries on three of the most popular architectures: x86, x64, and ARM through machine learning. JSNeat~\citep{DBLP:conf/icse/TranTNNN19} follows a data-driven approach to recover names by searching for them in a large corpus of open-source JS code. In addition to the aforementioned traditional approaches, recent studies have incorporated neural networks. DIRE~\citep{DBLP:conf/kbse/LacomisYSAGNV19} uses both lexical and structural information recovered by the decompiler for variable name recovery. \citet{2019In} investigate the problem of automatically naming pieces of assembly code. They justify training by assuming function names reflect semantics, using debug symbols as labels, and evaluate via standard NLP metrics against original names. VarBERT~\cite{DBLP:journals/corr/abs-2103-12801} is an early method to integrate a transformer-based model. It infers variable names in decompiled code based on Masked Language Modeling, Byte-Pair Encoding, and neural architectures such as Transformers and BERT. Similar to VarBERT, DIRECT~\citep{nitin-etal-2021-direct} is another transformer-based architecture customized specifically for decompiling binary executables to high-level code. More recent works apply LLM to handle the decompilation task. For example, LmPa~\citep{DBLP:journals/corr/abs-2306-02546} leverages the strengths of pre-trained generative models (CodeGemma-2B, CodeLlama-7B, and CodeLlama-34B) while mitigating model biases while mitigating model biases by aligning output distributions with developer symbol preferences and incorporating contextual information from caller/callee functions during both training and inference.

For multi-layer obfuscation and malicious code analysis scenarios, agents achieve adaptive deobfuscation through semantic recovery and tool collaboration. Traditional deobfuscation tools typically handle only specific types of obfuscation and struggle to cope with mixed and overlapping obfuscation techniques. Agents leverage the semantic understanding capabilities of large models, combined with tool assistance, to automatically identify and remove multiple obfuscation patterns: analyzing obfuscated code segments, identifying redundant patterns that do not affect business logic (such as useless loops or meaningless operations) and removing or simplifying them, while inferring more meaningful identifier names and adding comments based on context.
The ALFREDO framework~\cite{nataliealfredo} employs a classifier to identify obfuscation types, then invokes corresponding deobfuscation tools or strategies for different obfuscation types (such as calling Ghidra to extract low-level logic, or letting LLM attempt code reconstruction), gradually eliminating obfuscation components in the code through iterative loops. This multi-step reasoning and tool-using framework can process mixed and overlapping obfuscations in a single workflow.
Androidmeda~\cite{Androidmeda-Deobfuscate-android-app_2024} employs a context-based semantic reasoning approach, analyzing code context, method call relationships, and variable usage patterns, utilizing LLM's semantic understanding capabilities to infer more meaningful identifier names and generate comments. In malicious software analysis scenarios, agents combine pattern recognition and semantic recovery~\cite{patsakis2024assessing}: first identifying obfuscation patterns (such as control flow flattening, false branch insertion, etc.), then gradually restoring the original logic structure based on code semantics and contextual information. This class of Agentic frameworks adopts multi-step iterative methods: identifying obfuscation types, invoking corresponding deobfuscation tools or semantic reconstruction modules, and continuously optimizing deobfuscation results through feedback loops.

\paragraph{DevOps and CI/CD}
In modern DevOps, agents act as \emph{autonomous collaborators} that bring adaptivity and intelligence to continuous integration and deployment (CI/CD) pipelines.  
While traditional pipelines execute sequentially according to predefined scripts, agent-based DevOps frameworks dynamically adjust their workflows based on real-time context, execution feedback, and learned experience, transforming automation from static scripting into interactive decision-making.

\textbf{AutoDev}~\cite{tufano2024autodev} exemplifies this paradigm through a conversational multi-agent architecture. It integrates a \emph{Conversation Manager} for maintaining dialogue context, an \emph{Agent Scheduler} for task decomposition and delegation, and a unified tool library for interacting with IDEs, build systems, and test frameworks.  
The scheduler decomposes complex goals into subtasks—such as code editing, testing, and deployment—and dispatches them to specialized agents running in isolated Docker sandboxes. Execution results are continuously fed back into the conversation loop, allowing agents to iteratively plan, act, and refine their strategies. This architecture reframes DevOps automation as a problem of multi-agent coordination and dialogue-driven orchestration.

Within CI/CD pipelines, agents enhance control and decision-making through multiple mechanisms. They identify flaky tests via log analysis and selectively re-run or skip them; construct code–test dependency graphs to enable incremental test selection; and apply reinforcement learning or heuristic strategies to optimize concurrency and resource allocation. During deployment and monitoring, agents employ anomaly detection and priority assessment to trigger rollback actions or notify engineers of critical failures, integrating multimodal signals such as logs, metrics, and code changes for holistic evaluation.

\textbf{GPT-Engineer}~\cite{gptengineer} further extends DevOps automation toward development itself, enabling end-to-end workflows from natural-language requirements to executable code through iterative generation, testing, and self-improvement.  
\textbf{CodeAgent}~\cite{tang2024codeagent} advances this paradigm through the \emph{Code-as-Action} framework, where agents generate and execute Python code as dynamic actions to perform editing, analysis, and deployment tasks. In CI/CD contexts, CodeAgent integrates static analysis, dependency tracking, and test coverage estimation for intelligent test selection, while reinforcement learning models guide deployment adjustments and post-release optimization.

Collectively, these agentic DevOps frameworks transform software delivery from static, rule-driven automation into adaptive ecosystems capable of autonomous reasoning, coordinated task execution, and continuous self-optimization across distributed software infrastructures.

\subsubsection{End-to-End Software Agents}
Building on repository-level reasoning, a recent wave of research \cite{qian2024chatdev, hong2023metagpt, du2024multiagent, zhang2024experimenting, rasheed2024codepori, lin2024whenllm, nguyen2024agilecoder, zan2024codes, qian2024iterativeexperience, du2024multiagent, Elicitron} extends agentic coding beyond implementation toward the full software development life cycle (SDLC)—covering requirements elicitation, design, implementation, testing, and maintenance. These systems aim to realize end-to-end software development through coordinated multi-agent workflows that emulate human engineering teams.

\paragraph{Waterfall-based full-cycle agents}
Early end-to-end frameworks such as Self-Collaboration, ChatDev \cite{qian2024chatdev}, and MetaGPT \cite{hong2023metagpt} model a sequential, role-specialized workflow inspired by the waterfall process, where designated agents (e.g., CEO, CTO, Developer, Tester) collaborate through standardized operating procedures (SOPs) to complete projects from requirement analysis to final implementation. Subsequent systems including AISD \cite{zhang2024experimenting}, CTC \cite{du2024multiagent}, and CodePori \cite{rasheed2024codepori} extend this paradigm into iterative requirement–design–implementation–testing cycles, integrating execution feedback to refine upstream design decisions and improve code correctness. These works demonstrate that structured role assignment and hierarchical coordination can enable LLMs to manage complex multi-stage engineering tasks autonomously.

\paragraph{Agile and iterative frameworks}
Complementary to waterfall-style orchestration, other frameworks adopt agile or test-driven methodologies to enhance adaptability and incremental refinement. LCG \cite{lin2024whenllm}, AgileCoder \cite{nguyen2024agilecoder}, and CodeS \cite{zan2024codes} organize multiple agents around sprint-style development loops, combining planning, coding, and testing within each iteration. Systems such as Iterative Experience Refinement \cite{qian2024iterativeexperience} and Cross-Team Collaboration \cite{du2024multiagent} further incorporate experience reuse and inter-agent learning to stabilize long-horizon collaboration. By embedding feedback and shared memory, these agile frameworks approximate the iterative development patterns of human teams while retaining automation benefits.

\subsection{General Code Agents in Software Engineering}

As large language models continue to make breakthroughs in code generation, understanding, and reasoning capabilities, researchers have begun exploring how to build general code agents that can span all stages of software development. Unlike specialized agents that focus on only a single aspect (such as requirement analysis, testing, or debugging), these systems aim for full-process intelligent collaboration, seeking to take complete responsibility from task planning and code generation to debugging and deployment in real software engineering environments. Through multi-turn interactions, context tracking, and tool invocation, they achieve autonomous understanding and execution of complex engineering tasks, providing developers with continuous and unified intelligent support. The following will introduce several representative general code agent systems.

\begin{itemize}
\item \textbf{CodeAct}~\cite{wang2024executable}: CodeAct extends traditional text and JSON-formatted actions to executable python code, supporting dynamic code generation and multi-turn interaction through an integrated python interpreter. This framework leverages the control flow features of programming languages, enabling the agent to store intermediate results, combine multiple tools, and possess self-debugging capabilities.
\item \textbf{OpenHands}~\cite{wang2024openhands}: OpenHands 3adopts an event-stream architecture to manage agent-environment interactions and ensures code execution security through a Docker sandbox environment. It provides diverse execution environments, including a bash terminal, a Jupyter IPython server, and a playwright-based browser, and supports a multi-agent collaboration mechanism

\item \textbf{OpenCode}~\cite{opencode2024}: OpenCode aims to provide a controllable and extensible intelligent programming environment. Its core adopts a \textbf{multi-agent architecture}, consisting of a \textit{Plan Agent} responsible for analysis and planning, a \textit{Build Agent} for executing modifications, and a \textit{General Agent} for auxiliary queries. This design enables the \textbf{decoupling of reasoning and action} while ensuring safety and controllability.

\item \textbf{Aider}~\cite{aider2024}: Aider emphasizes human-AI collaboration rather than full autonomy. It achieves excellent performance on the SWE-Bench benchmark through precise static code analysis and an efficient LLM-based code editing mechanism. The framework incorporates multi-level linting and automated test repair logic, simulating the pair-programming workflow of actual developers.

\item \textbf{Augment}~\cite{augment2024}: {Augment} is a code agent designed for professional software engineering scenarios, targeting \textbf{end-to-end intelligent collaboration} from code comprehension to execution. It possesses proactive task execution and contextual management capabilities. Its core architecture incorporates a powerful \textbf{repository-level semantic understanding engine (context engine)} that indexes and retrieves functions, dependencies, and documentation within large codebases to maintain global consistency across multi-file and multi-module environments. In addition, Augment introduces \textbf{memory} and \textbf{rule} systems to learn developer habits and project conventions, enabling personalized and continuous collaboration. With its comprehensive codebase understanding, executable capabilities, and extensible design, it represents a robust paradigm for industrial-scale code agents.

\item \textbf{Trae Agent}~\cite{gao2025trae}: Trae Agent is an open-source, general-purpose code agent designed for software engineering workflows, notably capable of natural-language command execution, file editing, bash invocation, and large-codebase reasoning. Trae Agent formulates repository-level issue resolution as an optimal-search problem in a large ensemble space, and achieves strong results by combining generation, pruning, and selection modules.

\item \textbf{Refact Agent}~\cite{Refact2025}: Refact Agent is designed to achieve \textbf{full-process automation} from requirement parsing and code generation to debugging and deployment. The system can deeply analyze entire code repositories, synchronize with development environments in real time, and perform multi-step tasks through integration with terminal commands, version control systems (e.g., Git), and CI/CD pipelines. Its core features include support for \textbf{over 25 programming languages}, the adoption of \textbf{retrieval-augmented generation (RAG)} techniques for project-specific contextual understanding, and \textbf{self-hosted deployment} options to ensure data privacy and security.
\end{itemize}

\subsection{Training Techniques for SWE Agents}
\subsubsection{Fine-tuning SWE Agents}
SFT is a critical process for adapting pre-trained language models into specialized SWE agents capable of performing specific and nuanced tasks. In practice, LLM is fine-tuned only with successful trajectories by filtering failed samples (cannot pass the unit tests) for better performance, which is denoted as the rejection sampling fine-tuning (RFT)~\cite{rft}.
The common methodologies observed across the literature can be synthesized into three core areas: the strategic creation of training data, the design of sophisticated training objectives, and the adoption of scalable and adaptive training frameworks.

\paragraph{Strategic Curation and Synthesis of High-Fidelity Training Data}
The foundation of any effective fine-tuned agent lies in the quality and relevance of its training data. A prominent trend is the move beyond using raw datasets toward strategic data refinement and synthesis, which involves meticulously filtering noise, verifying correctness through execution, and creating complex, structured instances from the ground up. This data-centric approach ensures agents learn from clear, relevant, and representative examples. Key strategies include:

\begin{itemize}
\item \textbf{High-Fidelity Data Enhancement and Filtering} State-of-the-art methods actively enhance existing corpora by filtering noise and augmenting them with high-quality synthetic data. This includes using LLMs as classifiers to systematically remove non-actionable comments from code review datasets~\cite{liu2025too}, creating cleaner training signals. Concurrently, datasets are augmented by synthesizing artificial but realistic bugs from real-world commit histories, providing rich training pairs for fault localization and repair tasks~\cite{drain2021deepdebug}. Furthermore, hybrid augmentation methods enrich datasets by combining the formal rigor of static analyzers with the generative flexibility of LLMs, producing training examples that are both semantically diverse and structurally sound~\cite{jaoua2025combining}.

\item \textbf{Execution-Driven and Verifiable Data Augmentation} To guarantee the semantic correctness of training data, many frameworks incorporate an active verification loop. The CodeS framework, for instance, employs an execution-checked augmentation strategy that generates multiple equivalent SQL variants and verifies their functional correctness before adding them to the training set~\cite{CodeS}. A broader approach is seen in SPICE, which integrates program analysis with human verification to construct large-scale, high-quality test datasets~\cite{bhatia2025spice}. The principle of using execution feedback extends to shaping training objectives themselves; RewardRepair integrates execution outcomes into training via reinforcement-style objectives, effectively using runtime success as a verified signal to guide the model toward correct fixes~\cite{ye2022neural}.

\item \textbf{End-to-End Structured Data Synthesis} The most sophisticated approaches involve synthesizing entire complex datasets from scratch to mirror real-world software engineering processes. The OmniSQL pipeline is a prime example, as it bootstraps relational databases, generates complexity-aware SQL, back-translates it into diverse natural-language questions, and synthesizes chain-of-thought solutions to create comprehensive training quadruples~\cite{OmniSQL}. Similarly, SWE-Flow achieves high-quality data synthesis for test-driven development by automatically inferring incremental development steps and constructing runtime dependency graphs to generate structured plans and code~\cite{zhang2025swe}. This philosophy of building large, structured corpora is also seen in frameworks like DIDACT, which constructed a massive, billion-example dataset specifically for multi-task pretraining on code, providing a strong foundation for various downstream tasks~\cite{DIDACT}.
\end{itemize}

\paragraph{Refining Model Behavior through Structural Objectives}
The efficacy of fine-tuning is also determined by the specific algorithms and learning objectives used to guide the model's adaptation. Research has progressed from generic sequence-to-sequence modeling toward more sophisticated objectives that explicitly incorporate the structural and semantic properties of code, leading to more reliable and accurate agents.

\begin{itemize}
\item \textbf{Multi-Task and Curriculum-Based Learning} To build a strong semantic foundation, agents are often pre-trained or fine-tuned on a curriculum of related tasks. In review generation, for instance, models are trained on a combination of objectives, including masked language modeling, diff tag prediction, and code denoising, before being fine-tuned on human comments~\cite{li2022auger, li2022codereviewer, DIDACT}. This multi-task approach equips the model with a more robust and generalized understanding of code changes, which directly benefits the final review generation task. Similarly, in patch generation, multitask learning is used to adapt a single model to the varied demands of automated program repair~\cite{gharibi2024t5apr}.

\item \textbf{Incorporating Structural Constraints} A critical advancement is the design of training objectives that respect the inherent structure of code. To ensure syntactic validity in patch generation, the task is often reframed as a structured editing problem that uses grammar-constrained decoding or pointer-based models to perform precise, syntactically correct modifications~\cite{zhu2021syntax, hu2022fix}. In the text-to-SQL domain, models may first be trained to predict a high-level SQL ``skeleton'' before being fine-tuned to fill in the specific table and column names, decomposing the task in a structure-aware manner~\cite{CodeS}. Other methods explicitly incorporate data-flow information into the learning process of model or use graph neural networks to process code's structural representations~\cite{chi2022seqtrans, DBLP:conf/acsac/CaoL0H22}.

\item \textbf{Specialized Learning Formulations} Researchers have also explored novel learning formulations tailored to specific SWE challenges. In fault localization, contrastive learning is employed to train a retriever to distinguish between relevant and irrelevant code snippets for a given bug, which sharpens the model's ability to identify contextually important information for repair~\cite{jin2023inferfix}. A similar contrastive approach is used in decompilation to train models to learn the subtle semantic nuances of assembly optimizations, which is difficult with traditional loss functions~\cite{DBLP:journals/corr/abs-2311-13721}. These specialized objectives guide the model to learn more effectively from the unique characteristics of the problem domain.
\end{itemize}

\paragraph{Evolving Paradigms for Scalable and Continuous Adaptation}
The overarching strategies, or paradigms, for applying fine-tuning have also matured, with a growing emphasis on computational efficiency, adaptability to new contexts, and the capacity for continuous improvement over time.

\begin{itemize}
\item \textbf{Parameter-Efficient Fine-Tuning (PEFT)} The immense size of LLMs makes full fine-tuning computationally prohibitive. Consequently, PEFT methods such as low-rank adaptation (LoRA)~\cite{hu2022lora} and QLoRA~\cite{dettmers2024qlora} have become standard practice~\cite{zhuo2025parameter}. These techniques allow for the adaptation of very large models by training only a small fraction of their parameters. This approach is widely used across tasks like code review generation~\cite{lu2023llama, haider2024prompting} and patch generation~\cite{silva2025repairllama}, enabling the creation of highly specialized agents in a scalable and cost-effective manner.

\item \textbf{Multi-Stage and Continual Learning Frameworks} For complex tasks that require deep reasoning, a multi-stage training process is often more effective. In document generation, a "retrieve-then-generate" fine-tuning strategy first trains the model to identify relevant contextual information from a codebase before training it to synthesize the final explanation, mirroring a more effective human-like workflow~\cite{CodeExp}. To ensure agents remain effective over time, continual learning paradigms are employed, allowing a model to adapt to new programming languages, libraries, or evolving codebases without catastrophically forgetting previously learned knowledge, a key requirement for long-term multilingual program repair~\cite{yuan2022circle}.

\item \textbf{Direct Task-Specific Adaptation} While more complex paradigms are emerging, the foundational approach of directly fine-tuning a general model for a single, well-defined task remains highly effective. This is demonstrated across numerous domains where models like T5 or LLaMA are successfully adapted into specialized agents for code review~\cite{yu2024fine} or automated program repair~\cite{xia2022less}. The success of this direct approach is heavily contingent on the quality of the curated training dataset, reinforcing the central importance of the data-centric strategies discussed earlier.
\end{itemize}

\subsubsection{Reinforcement Learning for SWE Agents}
RL provides a distinct paradigm for training SWE agents, shifting the focus from mimicking static and expert-annotated datasets to learning optimal behaviors through direct interaction with an environment. Unlike SFT, which relies on pre-existing ground-truth data, RL allows agents to improve their decision-making policies by receiving feedback (in the form of rewards) for the outcomes of their actions. This approach is particularly well-suited for complex and multi-step tasks in software engineering where the notion of the correctness of a trajectory is ill-defined, but the quality of a final outcome (e.g., a passing test suite or an optimized program) is empirically measurable. Existing works highlight multiple key strategies for applying RL to enhance agent capabilities, centering on the design of reward functions, the formulation of the learning environment, and the algorithms used to drive policy improvement.

\paragraph{RL Algorithm Selection}
\begin{itemize}
\item \textbf{Policy Gradient Methods (e.g., REINFORCE~\cite{li2024remax,hu2025reinforceplusplus} and A2C/PPO~\citep{ppo})} Policy gradient methods directly optimize the decision-making policy and are well-suited for tasks with large, complex, or continuous action spaces where a direct mapping from state to action is needed. This makes them ideal for sequential decision-making problems common in software engineering. For instance, in compiler optimization, the task of selecting an optimal sequence of transformation passes from a vast library of options is a perfect use case. This is exemplified by AutoPhase~\citep{huang2019autophase}, which employs policy gradients to learn how to arrange and combine optimization passes for the LLVM compiler. The same algorithmic principle would be effective in learning multi-step debugging and repair strategies within agentic frameworks like ITER~\citep{ye2024iter} and RepairAgent~\citep{bouzenia2403repairagent}, where the agent must learn a policy to decide the next action (e.g., rerun tests, attempt a different patch, refine localization) based on the current state of the repair process.

\item \textbf{Offline Reinforcement Learning} Offline RL algorithms are designed to learn from a fixed, pre-collected dataset of interactions without actively exploring the environment. This approach is critical for SWE tasks where live interaction with the environment is prohibitively expensive, slow, or risky. The foremost example is compiler optimization, where compiling and benchmarking thousands of program variants is infeasible. CompilerDream~\citep{deng2025compilerdream} explicitly uses offline RL to learn from a world model trained on existing compilation data, thereby avoiding costly real-world interactions. This paradigm is also highly relevant for automated program repair, where running a full test suite after every minor change is a major bottleneck. An offline RL agent could learn from a large historical dataset of bug reports, code changes, and test outcomes from systems like RewardRepair~\citep{ye2022neural} or the iterative attempts logged by Reflexion~\citep{shinn2023reflexion} to derive an effective repair policy without needing extensive live testing during training.

\item \textbf{Value-based and Preference-based Methods (e.g., Q-Learning and RLHF)} This class of algorithms is optimal when the reward signal is not easily defined by a simple and objective metric but is instead based on qualitative or preferential feedback. Reinforcement learning from human feedback (RLHF) is the most prominent example, where a reward model is trained to reflect human preferences. This is essential for tasks involving human-computer interaction or subjective quality assessment, such as automated code review. CodeMentor~\citep{codementor} uses RLHF to align its generated reviews with human expectations for tone and contextual relevance. This approach is equally applicable to refining the quality of reasoning in other domains; for instance, the reward for generating high-quality, factual reasoning traces in R-Log~\citep{liu2025r} could be learned from human ratings. Similarly, in multi-agent debate frameworks like LLM4FL~\citep{rafi2024multi} or SWE-Debate~\citep{li2025swe}, a preference-based reward model could be trained to score the persuasiveness and correctness of agents' arguments, guiding them toward more effective collaborative problem-solving.
\end{itemize}

\paragraph{Optimizing for Task Success with Execution-Based Rewards}
A primary application of RL in SWE agents is to directly optimize for tangible, verifiable outcomes. In this setup, the environment is often a simulated or real software project, and the agent's actions (e.g., generating a patch, selecting a compiler flag) are evaluated based on their real-world effect. The reward signal is typically sparse but unambiguous, tied directly to the success or failure of the task. This aligns the agent’s learning objective with the ultimate goal of the SWE task itself.

\begin{itemize}
\item \textbf{Automated Program Repair} In this domain, the reward is directly correlated with the successful validation of a generated patch. Frameworks like RewardRepair~\citep{ye2022neural} integrate execution feedback from test suites into the training process using reinforcement objectives, rewarding the model for generating fixes that pass tests. This principle of learning from trial-and-error is central to agentic systems like Reflexion~\citep{shinn2023reflexion}, which use feedback to convert failed attempts into informed iterations and cache past mistakes to avoid repeated failures, effectively learning a policy to escape negative reward cycles. Similarly, agentic frameworks such as ITER~\citep{ye2024iter} and RepairAgent~\citep{bouzenia2403repairagent} operate on a continuous feedback loop of generation and validation, where the successful outcome of the validation phase serves as a positive reward signal that reinforces the agent's repair strategy.

\item \textbf{Compiler Optimization} RL agents learn to navigate the vast search space of compiler transformations by receiving rewards based on program performance. Systems like AutoPhase~\citep{huang2019autophase} model the selection of optimization passes as a Markov Decision Process and use policy gradient algorithms, where the agent is rewarded for sequences of passes that result in faster execution times. Further, CompilerDream~\citep{deng2025compilerdream} employs a world-model–based approach with offline reinforcement learning, allowing the agent to learn an effective optimization policy by exploring in a simulated environment, where rewards are based on predicted performance outcomes, thus drastically reducing the cost of real-world compilation and profiling. The foundational concept of iterative compilation~\citep{1998Iterative}, which relies on a cycle of compiling, executing, and profiling, establishes the feedback mechanism that these modern RL agents formalize and automate to learn their policies.

\item \textbf{Security and Vulnerability Management} The feedback loops inherent in security tasks provide clear reward signals for RL agents. The multi-agent system for securing code introduced by \citet{nunez2024autosafecodermultiagentframeworksecuring} uses continuous feedback loops between coding, analysis, and fuzzing agents; an RL framework can reward the system for autonomously detecting and repairing vulnerabilities identified by the fuzzer. In the forensic analysis system CyberSleuth~\citep{fumero2025cybersleuth}, an agent could be rewarded for correctly identifying threats or linking log data to known CVEs, thereby learning an efficient investigation policy. In multi-agent debate systems like Audit-LLM~\citep{song2024audit}, RL could improve performance by rewarding agents that successfully challenge opponents or contribute to accurate threat detection.
\end{itemize}

\paragraph{Refining Qualitative Behaviors and Multi-Agent Collaboration}
Beyond simple task success, RL is also employed to shape more nuanced and qualitative aspects of an agent's behavior, such as the clarity of its reasoning, the helpfulness of its feedback, and its ability to collaborate effectively with other agents. In these cases, the reward mechanism is often more complex, sometimes incorporating human feedback or rewarding intermediate reasoning steps to encourage more robust and interpretable decision-making processes.

\begin{itemize}
\item \textbf{Aligning with Human Preferences and Quality Standards} Reinforcement Learning from Human Feedback (RLHF) is a key technique for training agents to produce outputs that are not just technically correct but also useful and well-regarded by developers. CodeMentor~\citep{codementor}, for instance, employs RLHF to improve the contextual accuracy and tone of its automated code reviews, learning a reward model from human ratings of its suggestions. This same principle can be applied to data curation; for example, an agent could be trained using RLHF to filter non-actionable review comments, learning from human feedback which comments are considered valuable, a process explored in the work by \citet{liu2025too}. Similarly, in the context of reformulating comments for clarity, as studied by \citet{sghaier2025leveraging}, RLHF can provide the necessary reward signal to guide an agent toward producing more helpful and understandable text.

\item \textbf{Enhancing Reasoning and Interpretability} RL can be used to reward the cognitive process of an agent, not just its final output, leading to more transparent and reliable systems. The R-Log~\citep{liu2025r} framework exemplifies this by using reinforcement learning in a simulated environment to directly reward an agent based on the factual soundness and quality of its intermediate reasoning traces during log analysis. This focus on the thought process is also seen in the ReAct-RCA framework~\citep{pei2025flow}, where an agent's \textit{Thought–Action–Observation} loop could be optimized with RL to reward thought patterns that lead to more efficient root-cause analysis. This approach could also enhance models like ThinkRepair~\citep{yin2024thinkrepair}, which uses chain-of-thought prompting; an RL layer could be added to reward the generation of more logical and effective reasoning chains that result in successful program repairs.

\item \textbf{Optimizing Multi-Agent Dialogue and Collaboration} In systems with multiple interacting agents, RL provides a mechanism for learning effective communication and collaboration strategies. The LLM4FL~\citep{rafi2024multi} framework utilizes a reinforcement-based dialogue to enable agents to dynamically exchange feedback and achieve consensus during fault localization. This concept is extended in debate-based frameworks like Audit-LLM~\citep{song2024audit} and SWE-Debate~\citep{li2025swe}, where the competitive or collaborative debate provides a natural environment for RL. Agents can be rewarded for constructing persuasive arguments, successfully identifying flaws in others' reasoning, or contributing evidence that moves the group toward a correct and robust conclusion, thereby learning an optimal policy for collaborative problem-solving.
\end{itemize}

\paragraph{Learning Sequential Decision-Making in Dynamic Environments}
Many software engineering tasks are inherently sequential and take place in dynamic, stateful environments. RL is a natural fit for these problems, as it allows agents to learn complex, multi-step policies for navigating these environments. By modeling the task as a Markov decision process (MDP), agents can learn to make a sequence of decisions that maximizes a cumulative reward, enabling them to handle long-horizon tasks like interactive debugging, strategic planning, or managing CI/CD pipelines.

\begin{itemize}
\item \textbf{Strategic Exploration of Code and Solution Spaces}  RL, particularly in conjunction with search algorithms, enables agents to learn how to explore vast and complex solution spaces more efficiently. The SWE-Exp~\citep{chen2025swe} framework explicitly uses Monte Carlo tree search (MCTS), a method with strong ties to RL, to enhance decision-making and learn from both successful and failed repair trajectories, effectively building an experience-driven policy. This use of MCTS is also seen in LingmaAgent~\citep{ma2025alibaba}, which combines it with knowledge graphs to inform its search for dynamic patch synthesis. The self-evolution principles of SE-Agent~\citep{lin2025se}, which systematically expands its exploration and optimizes reasoning trajectories based on past insights, represent a high-level form of policy improvement that is central to the reinforcement learning paradigm.

\item \textbf{Interactive and State-Aware Repair Processes} Automated repair is not a one-shot task but an interactive process of hypothesis, testing, and refinement. Frameworks such as Reflexion~\citep{shinn2023reflexion} embody this by turning trial-and-error into informed iteration and using memory to avoid past mistakes, which is analogous to an RL agent learning a policy from environmental feedback (e.g., negative rewards on failed patch attempts). Conversational APR systems~\citep{xia2023conversational, xia2023keep} create an interactive loop where an RL agent could learn an optimal dialogue policy, deciding when to attempt a new patch versus when to ask for more information to maximize its long-term success rate. Likewise, the continuous feedback loop in RepairAgent~\citep{bouzenia2403repairagent}, which integrates localization, generation, and validation, defines a stateful environment where an RL agent could learn a policy to dynamically decide the next best action (e.g., re-run localization, attempt a different kind of patch) based on the history of its interactions.

\item \textbf{Navigating CI/CD and DevOps Pipelines} RL can be used to train agents that intelligently manage and optimize the long-horizon processes of continuous integration and deployment. CodeAgent~\citep{tang2024codeagent} applies reinforcement learning models to guide deployment adjustments and post-release optimizations, learning from real-time feedback to improve pipeline performance over time. The AutoDev~\citep{tufano2024autodev} framework, with its \emph{Agent Scheduler} for task decomposition, provides an ideal setting for an RL-trained meta-agent that learns the optimal policy for delegating subtasks to specialized agents based on the evolving state of the project. Similarly, the ExecutionAgent~\citep{bouzenia2025you}, which handles test execution through iterative feedback loops and error recovery, leverages RL to learn the most effective sequence of recovery commands to try when encountering different types of build or test failures.
\end{itemize}

\paragraph{Implementation Details and Synergies with Supervised Fine-Tuning}
The successful application of RL in SWE agents is rarely a standalone process. It is almost always preceded by and intertwined with SFT. This symbiotic relationship is critical for stabilizing training, improving sample efficiency, and ensuring that the exploration of agent is grounded in a solid foundation of domain knowledge. The key implementation considerations revolve around the necessity of SFT as a pre-training step, the strategic balance between SFT and RL training phases, and the potential for cyclical, data-driven improvement.
\begin{itemize}
\item \textbf{SFT as a Necessary Foundation for RL} Attempting to train an SWE agent with RL from a general-purpose, non-fine-tuned model is often intractable. The action space (e.g., the space of all possible code edits or review comments) is vast and sparsely rewarded. SFT serves as a critical bootstrapping phase that initializes the agent's policy with a strong prior based on high-quality, human-generated examples. This pre-training teaches the model the fundamental syntax, semantics, and common patterns of the target task, dramatically constraining the search space for the subsequent RL phase. For example, in automated program repair, a model is first fine-tuned on large datasets of bug-fix commits, as seen in approaches like AlphaRepair~\citep{xia2022less} and T5APR~\citep{gharibi2024t5apr}. Only after this SFT phase does it become feasible to apply an RL objective, like in RewardRepair~\citep{ye2022neural}, to optimize for the specific goal of passing a test suite. Without the initial SFT, the agent would generate syntactically invalid or semantically nonsensical code, making it nearly impossible to discover a rewarding trajectory.

\item \textbf{Balancing SFT and RL Training Ratios} The allocation of training resources (in terms of data volume and training epochs) between SFT and RL is a critical hyperparameter that balances knowledge acquisition with goal-oriented optimization. The standard and most effective methodology is a multi-stage approach: a comprehensive SFT phase followed by a more targeted RL phase. An extensive SFT phase, such as the one enabled by the massive, multi-task DIDACT~\citep{DIDACT} dataset for code review or the structured data synthesis in OmniSQL~\citep{OmniSQL} for Text-to-SQL, builds a robust and generalized base model. The subsequent RL phase (e.g., using RLHF as in CodeMentor~\citep{codementor}) can then be shorter, as its primary role is to refine the model's behavior and align it with specific success metrics rather than teaching it the task from scratch. Over-relying on SFT risks creating a rigid policy that mimics the training data too closely and struggles to explore novel solutions, while insufficient SFT leads to an inefficient and unstable RL process. The optimal ratio ensures the agent begins exploration from a high-quality starting point without being overly constrained by it.

\item \textbf{Iterative Cycles of RL and SFT for Continual Improvement} The most advanced training paradigms treat the relationship between SFT and RL not as a linear and one-time sequence, but as a continuous and iterative cycle. In this paradigm, the high-quality and successful trajectories discovered by the RL agent are used to create new and high-fidelity data for a subsequent round of SFT. This process distills the knowledge gained through environmental interaction and exploration back into the model's parameters, creating a self-improving loop. For instance, the experience repository in SWE-Exp~\citep{chen2025swe}, which stores successful and failed repair trajectories, provides a perfect source of data for this cycle. The effective patches discovered through its exploration can be added to an SFT dataset to improve the next iteration of the base model. Similarly, the self-evolutionary process of SE-Agent~\citep{lin2025se}, which revises and optimizes reasoning trajectories, generates improved problem-solving strategies that can be used as high-quality examples for fine-tuning. This hybrid approach leverages RL for discovery and SFT for knowledge consolidation, enabling the agent to learn and adapt over time in a scalable and data-efficient manner.
\end{itemize}

\subsection{Future Trends: Towards Integrated and Autonomous Software Engineering Ecosystems}
The development of SWE Agents, as described in this section, marks a clear technological trajectory from task-specific automation toward more autonomous and integrated systems that span the entire software development lifecycle. While current research has made significant progress in various domains such as requirements engineering, code generation, and testing, there remains ample opportunity for further integration and improvement. The following trends represent potential directions that may influence the next generation of SWE Agents, as they evolve from specialized tools toward more comprehensive and capable SWE partners.

\paragraph{From Specialized Agents to Full-Lifecycle Orchestration}
Current SWE agents often operate in independent stages, optimized for discrete phases of the software lifecycle. For instance, some agents excel at requirements acquisition, while others like Otter focus on test generation~\cite{ahmed2025otter}. A significant future direction will be the development of integrated, full-lifecycle agentic frameworks. These systems will orchestrate workflows that seamlessly transition from one phase to the next: an agent could take a high-level user need, engage in a multi-agent process to refine requirements, generate the corresponding code and documentation, create robust test suites for validation~\cite{yuan2024manualtestsevaluatingimproving}, and finally, manage deployment and maintenance by analyzing runtime logs~\cite{DBLP:conf/icse/0001TMYZY24}. End-to-end frameworks such as ChatDev~\cite{qian2024chatdev}, MetaGPT~\cite{hong2023metagpt}, and AgileCoder~\cite{nguyen2024agilecoder} have already begun to model the complete software development process, providing a practical blueprint for this vision of cross-lifecycle integration.

\paragraph{Deep Contextual Understanding and Long-Term Memory via Structured Knowledge}
A persistent challenge for SWE agents is the limited context window of LLMs, which hinders their ability to reason about large, complex codebases. While techniques like repository-aware indexing and dependency graph analysis have provided initial solutions, the future lies in agents that can build and maintain persistent, dynamic knowledge graphs of software projects. Moving beyond the static representations seen in systems like CodexGraph~\cite{liu2024codexgraph}, these agents will develop a dynamic ``mental model'' of a repository, encompassing not only its static structure but also its evolutionary history, runtime behavior, and implicit design principles. For instance, KGCompass~\cite{yang2025enhancing} accurately links issue descriptions to code entities via a knowledge graph, while CGM~\cite{tao2025code} combines graph retrieval with generation. This enables agents to perform complex, repository-level tasks with a level of understanding that may eventually rival, and perhaps exceed, that of human developers.

\paragraph{From Collaboration to Self-Evolution: The Rise of Multi-Agent Ecosystems}
To address the increasing complexity of software engineering tasks, future research is shifting from single-agent systems to collaborative multi-agent ecosystems, emphasizing role specialization and dynamic interaction to enhance problem-solving capabilities. In these ecosystems, different agents assume specific roles (e.g., planner, developer, tester), mimicking the collaborative dynamics of human teams. Frameworks such as CodeAgent~\cite{tang2024codeagent} and MAGIS~\cite{tao2024magis} efficiently handle complex, multi-file tasks through clear role division. Looking further, agents will gain the ability to self-evolve. Inspired by the \emph{intelligent layer} of agent design, frameworks like SWE-Exp~\cite{chen2025swe}, SWE-Debate~\cite{li2025swe}, and SE-Agent~\cite{lin2025se} explore mechanisms for learning from historical experience and internal debates, allowing agents to continuously refine their strategies. This evolution from collaboration to self-iteration signals a shift from passive executors to intelligent ecosystems capable of autonomous learning and adaptation.

\paragraph{Synergistic Human-Agent Collaboration: Building Trustworthy AI Pair Programmers}
While the pursuit of full autonomy is a driving force in agent research, the most practical and impactful future will likely involve synergistic human-agent collaboration. Rather than replacing developers, agents will evolve into proactive, intelligent pair programmers. Frameworks like Aider~\cite{aider2024}, which emphasize human-AI collaboration over complete autonomy, are at the forefront of this trend. Future systems will feature mixed-initiative interaction models, where the agent handles the laborious and repetitive aspects of coding, testing, and debugging, while the human developer provides high-level strategic direction, resolves ambiguities, and validates critical decisions. Conversational frameworks such as ChatRepair~\cite{xia2023keep} and AutoDev~\cite{tufano2024autodev} create interactive feedback loops that allow agents to iterate under human guidance, promising to enhance developer productivity and creativity without sacrificing control or oversight.

\paragraph{Trust, Security, and Verifiability by Design}
As SWE agents gain more autonomy and are granted greater access to production systems, ensuring their actions are safe, secure, and verifiable becomes a paramount concern. While the current use of sandboxed environments for code execution is a necessary first step, future work must integrate security and verification as first-class citizens in the agent's reasoning process. This includes the development of agents that can proactively identify and mitigate security vulnerabilities during code generation, as explored in multi-agent fuzz testing systems~\cite{nunez2024autosafecodermultiagentframeworksecuring}. Furthermore, future agents will draw inspiration from hybrid systems like ESBMC-AI~\cite{tihanyi2025new} and ContractTinker~\cite{wang2024contracttinker}, which combine formal methods like model checking with the generative capabilities of LLMs to provide verifiable correctness guarantees. This ``security-by-design'' philosophy will be essential for deploying autonomous systems in safety-critical environments.

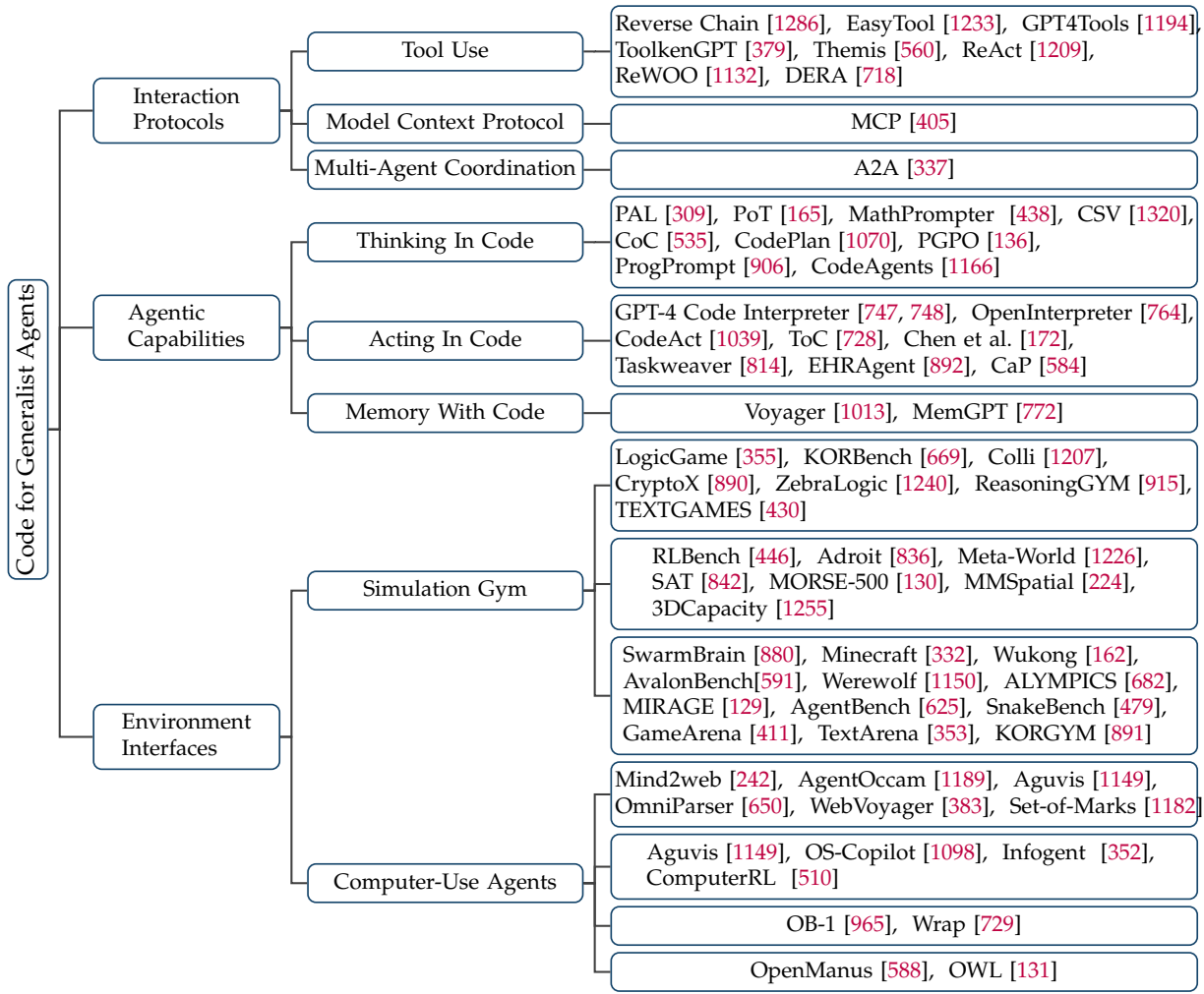
\begin{figure}[H]
	\centering
    \resizebox{\textwidth}{!}{
	\begin{forest}
        forked edges,
		for tree={
                grow=east,
                reversed=true,
                anchor=base west,
                parent anchor=east,
                child anchor=west,
                base=center,
                font=\large,
                rectangle,
                draw=hidden-draw,
                rounded corners,
                align=left,
                text centered,
                minimum width=4em,
                edge+={darkgray, line width=1pt},
                s sep=3pt,
                inner xsep=2pt,
                inner ysep=3pt,
                line width=0.8pt,
                ver/.style={rotate=90, child anchor=north, parent anchor=south, anchor=center},
            },
            where level=1{text width=8em,font=\normalsize, }{},
            where level=2{text width=12em,font=\normalsize}{},
            where level=3{text width=14em,font=\normalsize,}{},
	    [Code for Generalist Agents, ver
			[Interaction \\Protocols
                    [Tool Use
                        [
                            Reverse Chain~\citep{zhang2024reversechaingenericrulellms}{, }
                            EasyTool~\citep{yuan2024easytoolenhancingllmbasedagents}{, }
                            GPT4Tools~\citep{yang2023gpt4toolsteachinglargelanguage}{, }
                            \\
                            ToolkenGPT~\citep{hao2024toolkengptaugmentingfrozenlanguage}{, }
                            Themis~\citep{li2024toolaugmentedrewardmodeling}{, }
                            ReAct~\citep{yao2023react}{, }
                            \\ReWOO~\citep{xu2023rewoo}{, }
                            DERA~\citep{nair2023dera}
                           , text width=26em
                        ]
                    ]
                    [Model Context Protocol 
                        [
                            MCP~\citep{mcp_survey}
                           , text width=26em
                        ]
                    ]
                    [Multi-Agent Coordination
                        [
                            A2A~\citep{a2a}
                           , text width=26em
                        ]
                    ]
			]
			[Agentic \\Capabilities
                    [Thinking In Code
                        [
                            PAL~\citep{gao2023pal}{, }
                            PoT~\citep{chen2022program}{, }
                            MathPrompter ~\citep{imani2023mathprompter}{, }
                            CSV~\citep{zhou2023solving}{, }
                            \\CoC~\citep{li2023chain}{, }
                            CodePlan~\citep{wen2024unlocking}{, }
                            PGPO~\citep{cao2025pgpo}{, }
                            \\ ProgPrompt~\citep{singh2022progprompt}{, }
                            CodeAgents~\citep{yang2025codeagents}
                           , text width=26em
                        ]
                    ]
                    [Acting In Code
                        [
                            GPT-4 Code Interpreter~\citep{gptCodePlugin, gpt4Code}{, }
                            OpenInterpreter~\citep{openInterpreter}{, }
                            \\CodeAct~\citep{wang2024executable}{, }
                            ToC~\citep{ni2025tree}{, }
                            \citet{chen2024steering}{, }
                            \\Taskweaver~\citep{qiao2023taskweaver}{, }
                            EHRAgent~\citep{shi2024ehragent}{, }
                            CaP~\citep{liang2022code}
                           , text width=26em
                        ]
                    ]
                    [Memory With Code
                        [
                            Voyager~\citep{voyager}{, }
                            MemGPT~\citep{packer2023memgpt}
                           , text width=26em
                        ]
                    ]
            ]
            [Environment \\Interfaces
                    [Simulation Gym
                        [
                            LogicGame~\citep{LogicGame}{, }
                            KORBench~\citep{KORBench}{, }
                            Colli~\citep{Colli}{, }
                            \\CryptoX~\citep{CryptoX}{, }
                            ZebraLogic~\citep{ZebraLogic}{, }
                            ReasoningGYM~\citep{REASONINGGYM}{, }
                            \\TEXTGAMES~\citep{TextGames}
                           , text width=26em
                        ]
                        [
                            RLBench~\citep{RLBench}{, }
                            Adroit~\citep{Adroit}{, }
                            Meta-World~\citep{MetaWorld}{, }
                            \\
                            SAT~\citep{SAT}{, }
                            MORSE-500~\citep{MORSE500}{, }
                            MMSpatial~\citep{MMSpatial}{, }
                            \\
                            3DCapacity~\citep{3DCapacity}
                           , text width=26em
                        ]
                        [
                            SwarmBrain~\citep{SwarmBrain}{, }
                            Minecraft~\citep{MindAgent}{, }
                            Wukong~\citep{Wukong}{, }
                            \\AvalonBench\citep{AvalonBench}{, }
                            Werewolf~\citep{Werewolf}{, }
                            ALYMPICS~\citep{ALYMPICS}{, }
                            \\MIRAGE~\citep{MIRAGE}{, }
                            AgentBench~\citep{AgentBench}{, }
                            SnakeBench~\citep{snake_bench_2025}{, }
                            \\GameArena~\citep{GameArena}{, }
                            TextArena~\citep{TextArena}{, }
                            KORGYM~\citep{KORGYM}
                           , text width=26em
                        ]
                    ]
                    [Computer-Use Agents
                        [
                            Mind2web~\citep{deng2023mind2web}{, }
                            AgentOccam~\citep{yang2024agentoccam}{, }
                            Aguvis~\citep{xu2024aguvis}{, }\\
                            OmniParser~\citep{lu2024omniparser}{, }
                            WebVoyager \citep{he2024webvoyager}{, }
                            Set-of-Marks~\citep{yang2023set}
                           , text width=26em
                        ]
                        [
                            Aguvis~\citep{xu2024aguvis}{, }
                            OS-Copilot~\citep{wu2024copilot}{, }
                            Infogent ~\citep{guan2024intelligent}{, }
                            \\ComputerRL ~\citep{lai2025computerrl}
                           , text width=26em
                        ]
                        [
                            OB-1~\citep{ob_1}{, }
                            Wrap~\citep{wrap}
                           , text width=26em
                        ]
                        [
                            OpenManus~\citep{openmanus2025}{, }
                            OWL~\citep{owl2025}
                           , text width=26em
                        ]
                    ]
            ]
		]
	\end{forest}}
	\caption{Taxonomy of Code for Generalist Agents.}
    \label{fig:code_for_generalist_agents}
\end{figure}

\section{Code for Generalist Agents}
Using code as a universal medium allows AI agents to both reason about problems and execute actions across many different tasks and environments, rather than being limited to a single specialized function.
The integration of code has become a pivotal paradigm in the development of generalist agents, enabling the combination of cognitive reasoning with executable actions across diverse environments. In \autoref{fig:code_for_generalist_agents}, this section synthesizes recent progress in utilizing code beyond its conventional role, highlighting three key dimensions in agent architecture:

\begin{itemize}
    \item \textbf{Interaction Protocols:} Code establishes structured communication frameworks, including tool-use patterns (ReAct~\citep{yao2022react}, ReWOO~\citep{xu2023rewoo}, DERA~\citep{nair2023dera}), the model context protocol (MCP)~\citep{mcp_survey}, and multi-agent coordination schemes (A2A)~\citep{a2a}, enabling precise tool invocation, state management, and inter-agent collaboration.

    \item \textbf{Agentic Capabilities:} Code-driven approaches such as CodeAct~\citep{wang2024executable}, Smolagents~\citep{smolagents}, and Open Interpreter~\citep{openInterpreter}, along with emerging CodePlanning techniques, empower agents to generate and execute code for complex logic, data manipulation, and software engineering tasks, thereby enhancing autonomy and operational efficiency.

    \item \textbf{Environment Interfaces:} Code underpins both simulated environments (e.g., CodeGYM~\citep{zhang2025vgamegym} for puzzle and spatial reasoning tasks, GameArena~\citep{GameArena} for strategic planning) and real-world interfaces (GUI and terminal-based agents), offering scalable, verifiable platforms for agent training, evaluation, and deployment.

\end{itemize}

Through these three dimensions, we examine how code contributes to the development of adaptable, tool-augmented agents capable of solving open-ended tasks.

\subsection{Code as Interaction Protocols}

\subsubsection{Tool Use}
\begin{figure}[t!]
    \centering
    \includegraphics[width=0.65\textwidth]{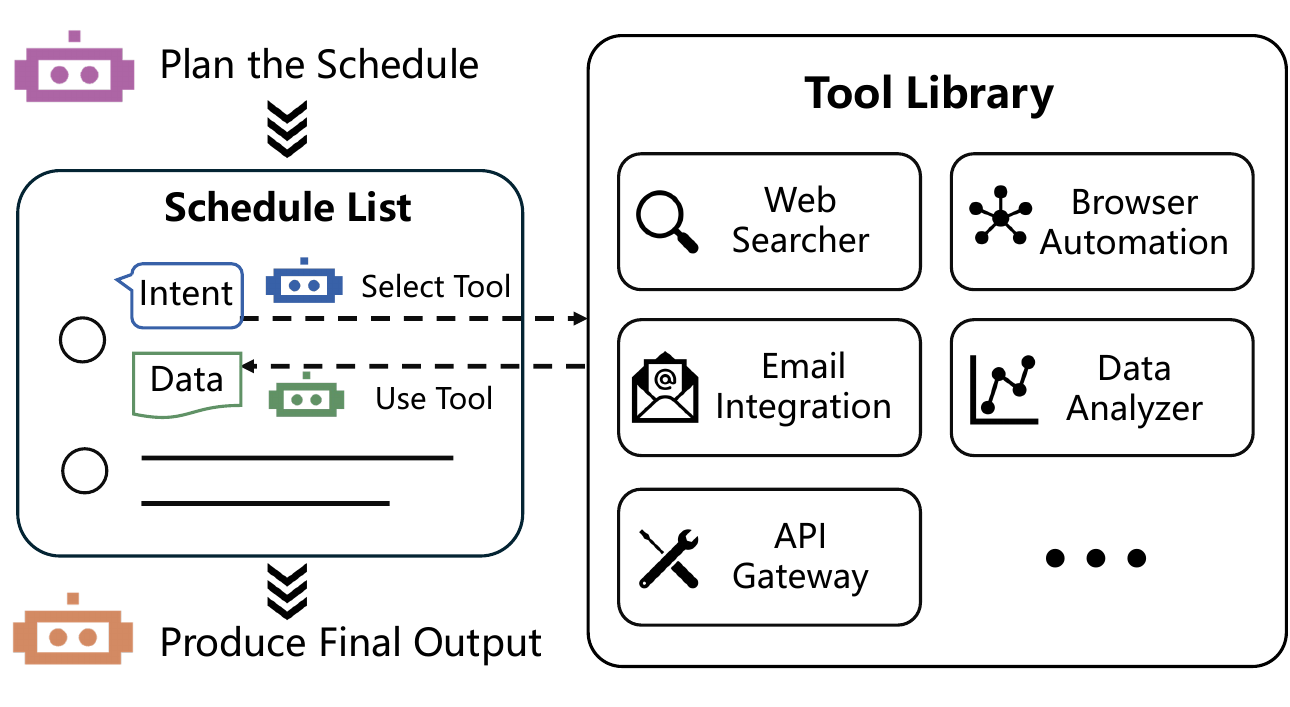}
    \caption{Workflow of LLM-based tool-using.}
    \label{fig:tool_use}
\end{figure}

In LLM-based tool-agent workflows~\cite{patil2025bfcl,yao2024tau,CRITIC,milev2025toolfuzzautomatedagent,li2023api,qin2023toolllm,hao2024toolkengptaugmentingfrozenlanguage}, as shown in \autoref{fig:tool_use}, each tool is fundamentally embodied in code, which serves as the formal substrate that defines both the syntactic interface and the semantic logic of tool functionality.
Function calling operates as the central mechanism in this process, providing LLMs with a structured interface for interaction with executable code or external services. During invocation, the LLM extracts the necessary parameters from user input based on the description of the tool and sends a corresponding request to the tool server. The primary objectives are to accurately extract parameter content and format, ensure the completeness of required parameters, follow the predefined output specifications of the tool, and validate that parameter values remain within the acceptable range. Methods for function calling can be broadly categorized into tuning-free and tuning-based approaches, depending on whether the model parameters are fine-tuned.

\paragraph{Tuning-Free Approaches}
Tuning-free approaches leverage the in-context learning capabilities of LLMs, typically using few-shot demonstrations or rule-based refinements, to enhance parameter extraction and tool alignment. For example, Reverse Chain~\citep{zhang2024reversechaingenericrulellms} adopts a reverse reasoning strategy that first identifies the appropriate target tool for a given task and then fills in the relevant parameters; when certain parameters are unspecified, auxiliary tools are invoked to supplement them. Similarly, EasyTool~\citep{yuan2024easytoolenhancingllmbasedagents} strengthens the understanding of tool functions and parameter requirements by prompting ChatGPT to rewrite lengthy tool descriptions into concise, function-oriented guidelines.

\paragraph{Tuning-Based Approaches}
Tuning-based approaches enhance function calling through parameter fine-tuning with dedicated tool-learning datasets. GPT4Tools~\citep{yang2023gpt4toolsteachinglargelanguage} integrates tool-use capabilities into open-source LLMs via LoRA-based fine-tuning, leveraging instruction datasets of tool usage automatically generated by ChatGPT. ToolkenGPT~\citep{hao2024toolkengptaugmentingfrozenlanguage} introduces specialized tokens, referred to as toolkens, which act as triggers for tool invocation: when a toolken is predicted, the model switches to a specialized mode for generating input parameters and subsequently incorporates tool outputs into the generation process. Themis~\citep{li2024toolaugmentedrewardmodeling} further improves interpretability in reward modeling by autoregressively integrating reasoning and tool use, dynamically determining which tools to invoke, how to assign parameters, and how to fuse the resulting outputs into ongoing reasoning.

\paragraph{Tool Calling as the Engine of Agent Frameworks}  
Function calling serves as a foundational mechanism in various tool-augmented agent paradigms. The ReAct framework~\citep{yao2023react} implements function calls during the action phase, where the model explicitly outputs the tool name and parameters to initiate invocation, followed by an observation phase in which tool responses are integrated to guide subsequent reasoning. ReWOO~\citep{xu2023rewoo}, by contrast, features a two-stage structure: a planning phase that compiles a structured list of required function calls, and an execution phase in which Worker agents invoke functions, carry out tasks, and return the results. To handle uncertainty and invocation failures, DERA~\citep{nair2023dera} introduces a cyclical interaction protocol (Execute~$\rightarrow$~Pause~$\rightarrow$~Dialog~$\rightarrow$~Resume). When confronted with ambiguous instructions, the model suspends execution, engages in clarifying dialogue with the user, and resumes function calls once sufficient information is acquired, thereby enhancing both robustness and interactivity.

\subsubsection{Model Context Protocol}
The model context protocol (MCP)~\citep{mcp_survey,mcp,servers25mcp} is a standardized communication framework designed to coordinate interactions between models and external tools. In contrast to autonomous tool use that depends entirely on internal reasoning, MCP introduces structured message formats, explicit invocation semantics, and well-defined mechanisms for managing context states. Its operational cycle consists of four stages: request, response, state update, and re-invocation. Together, these stages form a closed-loop process that couples tool execution with continuous context management, thereby improving the reliability, interpretability, and scalability of multi-turn task completion.

\subsubsection{Multi-Agent Coordination}
Agent-to-Agent (A2A)~\citep{a2a,scure_a2a,multi_agent_econmies_a2a,improve_a2a_protocal,a2a_survey} is a collaborative pattern in which agents communicate directly to accomplish complex tasks. Rather than relying on a single agent’s autonomous reasoning, A2A decomposes tasks among specialized agents that coordinate tool invocation and information processing through message passing and shared context. Its workflow involves task delegation, information exchange, result integration, and iterative refinement. Agents in this paradigm can invoke external tools and interact with peers to fill knowledge gaps or compensate for limitations, thereby improving both the efficiency and accuracy of problem-solving.

\subsection{Code as Agentic Capabilities}
AI agents have gained substantial research interest, accompanied by increasing efforts to characterize their underlying capabilities~\citep{dai2025lita,agentic_ai_for_se,belcak2025small_agentic_ai}.
\citet{yao2023react} propose the ReAct paradigm, which structures an agent’s behavior into three stages: thought (reasoning, planning, and decision-making), action (executing decisions), and observation (receiving environmental feedback for subsequent decisions).
Building on this, \citet{agent-survey-2} outline an LLM-based agent architecture composed of three high-level components: brain (central cognitive processing), perception (interpreting environmental signals), and action (performing environment-altering operations).
\citet{agent-survey-1} further refine agent functionality into four modules: profile (representation of self, environment, goals, and constraints), memory (storing and recalling past experiences), planning (generating action sequences to achieve goals), and action (executing planned behaviors).
From a neuroscience-inspired perspective, \citet{liu2025advances} argue that agent capabilities encompass seven fundamental aspects, including cognition (knowledge acquisition and reasoning), memory, world model, reward mechanisms, emotion modeling, perception, and action systems, together forming the core substrate for adaptive and autonomous intelligence.

This section aims to explore the application of code-based LLMs in AI agents. By integrating the classic agent paradigm proposed in ReAct~\citep{yao2023react} with the definitions and classifications from related works, this section analyzes the role of code-based LLMs within the agent architecture from three perspectives: thinking, acting, and memory.

\subsubsection{Thinking in Code}
When tackling complex reasoning tasks, LLMs benefit significantly from code generation techniques, which enhance both precision and efficiency~\citep{deepseekai2025deepseekr1}. The concept of reasoning with code initially emerged in the domain of mathematical problem-solving. \citet{gao2023pal} propose program-aided language models (PAL), which leverage few-shot learning and chain-of-thought (CoT) prompting to generate Python code that executes intermediate reasoning steps involving mathematical operations, symbolic manipulation, and algorithmic logic. This approach enables more accurate outcomes through execution-based verification. Building on this foundation, subsequent studies have explored alternative strategies for code-based mathematical reasoning, including program of thoughts (PoT)~\citep{chen2022program}, which replaces natural language CoT with programmatic reasoning, MathPrompter~\citep{imani2023mathprompter}, which optimizes PAL-style prompting through refined mathematical templates, and code-based self-verification (CSV)~\citep{zhou2023solving}, which improves reliability by verifying answers via reverse reasoning.

Beyond mathematical problem-solving, \citet{li2023chain} introduces the chain of code (CoC) reasoning framework, which extends code-based reasoning to both numerical and general semantic tasks. In this approach, LLMs generate and execute code snippets under prompt guidance, facilitating structured problem-solving. For numerical tasks, CoC enables precise computation through executable code. For semantic reasoning, it leverages pseudocode composed of data structures, conditionals, and loop constructs to model abstract reasoning processes. While high-level control flow is implemented via code logic, certain decision functions still rely on commonsense semantic reasoning. By incorporating variables and structured control logic, CoC substantially improves the efficiency and accuracy of solving complex reasoning tasks.

In addition to supporting single-task reasoning, code generation techniques also play a pivotal role in multi-task planning. \citet{wen2024unlocking} constructed a dataset comprising two million training instances, where models learn to generate code that bridges the input and output, thereby capturing implicit planning trajectories. Compared to traditional LLM training approaches, CodePlan~\citep{bairi2023codeplanrepositorylevelcodingusing} formulates task planning explicitly in code, which reduces ambiguity from natural language instructions and enhances performance in complex multi-hop reasoning tasks. \citet{cao2025pgpo} propose Planning-Guided Preference Optimization (PGPO), a framework that integrates pseudocode-based task planning into ReAct. By leveraging specialized reward functions and preference learning, PGPO effectively replaces natural-language-based planning with executable reasoning steps. Furthermore, \citet{singh2022progprompt} demonstrates that prompts expressed directly as code improve task planning in both virtual household environments and robotic manipulation settings, highlighting the versatility of code-driven planning in diverse domains.

Compared to single-agent workflows, multi-agent systems offer superior capabilities in collaboration and problem-solving efficiency. \citet{yang2025codeagents} investigates code planning strategies tailored to multi-agent scenarios. In their framework, a dedicated planner generates high-level pseudocode plans in Python style, comprising variable instantiations, conditional logic, and iterative structures. These plans are produced through stepwise decomposition of complex tasks and are annotated with natural language comments to guide decision-making by large language models. The planner then transforms the plan into executable CodeAct instructions, which are passed to either the ToolCaller module~\citep{yang2025codeagents} or the Python interpreter for execution. This code-based multi-agent design facilitates the modular reuse of agent components with similar functionalities, thereby enhancing both system scalability and token efficiency.

\subsubsection{Acting in Code}

When used within an agent’s execution module, code must interface with system-level tools via standardized agent protocols. With the advent of ChatGPT, OpenAI introduced an enhanced variant of GPT-4, known as the GPT-4 Code Interpreter or GPT-4 Code~\citep{gptCodePlugin, gpt4Code}, which enables models to execute code for advanced reasoning tasks. Complementing this, Open Interpreter~\citep{openInterpreter} presents an open-source framework that allows language models to execute code locally across multiple languages, including Python, JavaScript, and Shell. It equips function-calling LLMs with an \texttt{exec()} interface, which accepts two arguments: the target programming language (e.g., Python or JavaScript) and the corresponding code snippet, thereby enabling grounded execution within the agent workflow.

\citet{wang2024executable} introduces the concept of CodeAct, which enables interactive operations by generating executable Python code. Unlike traditional action formats such as text or JSON, CodeAct eliminates the need for custom tool wrappers by directly integrating code execution with the Python interpreter. This approach offers richer support for control flow and data management, allowing intermediate results to be stored as variables and reused across steps, thereby reducing the overall number of execution actions.

Building on this foundation, \citet{ni2025tree} proposes ToC, an end-to-end framework for code generation and execution. ToC constructs a hierarchical code tree by considering the global solution structure of a given problem. It further incorporates self-supervised feedback mechanisms that evaluate the success of code execution and leverages answer validation through majority voting. Compared to CodeAct, ToC achieves better performance in accuracy while reducing execution rounds, demonstrating enhanced efficiency and robustness in complex problem-solving scenarios.

In studying the decision-making processes of code-capable agents, \citet{chen2024steering} examines how LLMs with varying reasoning capabilities decide whether to invoke the Code Interpreter and how this behavior affects task accuracy. Experimental results on mathematical reasoning tasks reveal a counterintuitive trend: less capable models, such as GPT-3.5~\citep{chatgpt}, are more inclined to use the Code Interpreter and consequently achieve higher accuracy. In contrast, more capable models, such as GPT-4o~\citep{openai2024gpt4o}, tend to rely on their internal reasoning abilities and prefer textual reasoning, often exhibiting overconfidence that leads to increased errors on moderately difficult problems—a phenomenon termed the inverse scaling effect.
However, when constrained to rely exclusively on either textual or code-based reasoning, neither modality achieves optimal performance. To mitigate this, the authors propose three decision-making strategies to enhance LLM reasoning accuracy. The first encourages consistent use of code-based reasoning to improve reliability. The second introduces a parallel reasoning mechanism, wherein both textual and code-based reasoning are executed independently, and their outputs are aggregated. The third employs confidence-based selection, allowing the model to assess its confidence in each reasoning modality and proceed with the one deemed more reliable. All three strategies yield notable improvements, underscoring the importance of dynamic reasoning modality selection in code-agent systems.

To operationalize the benefits of code-based reasoning, several studies examine concrete use cases where code generation functions as the primary execution mechanism for agents. \citet{qiao2023taskweaver} proposes a code-first framework tailored for complex data processing tasks, wherein each user request is translated into executable code and user-defined plugins are treated as callable functions. For example, the system conducts anomaly detection on time series data stored in SQL databases by automatically generating the required code. In the healthcare domain, EHRAgent~\citep{shi2024ehragent} addresses the inefficiencies of conventional workflows, which require clinicians to relay their needs to software engineers for implementation. It enables medical personnel to directly generate executable code via LLMs, facilitating multi-table reasoning over Electronic Health Records (EHR) and supporting a variety of clinical tasks, thereby lowering the technical barrier for healthcare professionals. Similarly, \citet{liang2022code} explores the use of LLM-generated code to control robotic systems in embodied intelligence scenarios, allowing agents to convert high-level instructions into precise, executable actions.

\subsubsection{Memory With Code}
In response to the context length limitations of LLMs, researchers have explored various storage strategies, among which code-based storage has demonstrated notable effectiveness. \citet{voyager,xu2025mem} exemplifies this through a framework in which agents autonomously acquire skills by generating and iteratively refining code through interactions with the environment in Minecraft. Validated skills are stored as executable code in a dedicated library and later retrieved directly for reuse, thereby eliminating the need for repeated model inference. Similarly, \citet{packer2023memgpt} proposes a virtual memory-inspired context management framework that extends the effective memory of LLMs. This is achieved by designing specialized read and write function calls to dynamically manage both internal (model-resident) and external context, enabling real-time modification of contextual content during interactions.

\subsection{Code as Environment Interfaces}
\subsubsection{Code as Simulation Gym}
With the advancement of LLM reasoning capabilities and the increasing diversity of reasoning tasks, evaluating and cultivating long-term planning skills has become a critical challenge. Traditional single-turn reasoning benchmarks, however, are insufficient for assessing or training such capabilities. To address this limitation, Code as Simulation Gym (CodeGYM) has been proposed, building on the Gymnasium~\citep{Gymnasium} reinforcement learning platform. CodeGYM constructs a variety of complex task environments that allow models to engage in continuous interaction, maintain and update state, and receive feedback or rewards. Existing CodeGYM implementations generally fall into three categories: puzzle solving (e.g., logical and mathematical problems requiring step-by-step reasoning), spatial reasoning (e.g., navigation and manipulation tasks in grid-based or continuous environments), and GameArena (e.g., multi-agent or strategy-based game scenarios).

\paragraph{Dynamic PuzzleGYM}
The Dynamic PuzzleGYM framework collectively refers to a family of rule-based puzzle generation systems designed to automatically construct large-scale, verifiable reasoning datasets~\citep{sliding_puzzles_gym}. These systems leverage random seeds and initialization parameters to programmatically generate problem instances by combining known conditions, background information, and task formulations, along with their corresponding solutions. Early implementations such as LogicGame~\citep{LogicGame} and KORBench~\citep{KORBench} present diverse logic puzzles that require models to interpret initial states and apply specified rules to derive solutions. More advanced platforms, including Colli~\citep{Colli}, CryptoX~\citep{CryptoX}, and ZebraLogic~\citep{ZebraLogic}, employ transformation strategies and constraint mechanisms to produce complex combinatorial reasoning tasks. Similarly, ReasoningGYM~\citep{REASONINGGYM} and TEXTGAMES~\citep{TextGames} offer extensive single-turn game puzzles, with ReasoningGYM further integrating reward interfaces for reinforcement learning. Compared to conventional QA datasets, CodeGYM-generated puzzles exhibit better scalability, lower overlap with pretraining corpora~\citep{KORBench, REASONINGGYM}, and stronger suitability for the data requirements of modern LLM training~\citep{REASONINGGYM}.

\paragraph{Synthetic SpatialGYM}
Synthetic SpatialGYM is a synthetic data generation framework designed to create customized datasets for training and evaluating LLMs on spatial reasoning tasks~\citep{spatialrgpt}, thereby eliminating the high costs associated with real-world image collection. It programmatically generates spatial reasoning datasets that meet specific visual and structural requirements. Early implementations include robotic learning benchmarks (e.g., RLBench~\citep{RLBench}), real-world manipulation scenarios (e.g., Adroit~\citep{Adroit}), and multi-task environments (e.g., Meta-World~\citep{MetaWorld}). With the advancement of LLMs, specialized CodeGYM-style platforms have emerged to support spatial reasoning: SAT~\citep{SAT} produces static and dynamic 3D question–answer pairs, while MORSE-500~\citep{MORSE500} generates spatial datasets using tools like Manim, Matplotlib, and MoviePy. These synthetic benchmarks facilitate the development of spatial perception and reasoning in vision–language models, supporting downstream applications in embodied intelligence, healthcare, and automation~\citep{MMSpatial, 3DCapacity}.

\paragraph{GameArena}
GameArena refers to a class of CodeGYM platforms designed to evaluate and train the long-term planning and strategic reasoning capabilities of large language models through interactive game-based environments. These platforms typically expose observation–action interfaces, enabling models to engage in multi-turn decision-making and receive rewards upon achieving predefined terminal goals. Early systems built on existing commercial games, including StarCraft II~\citep{SwarmBrain}, Minecraft~\citep{MindAgent}, and Black Myth: Wukong~\citep{Wukong}, served as interactive testbeds for agent behavior. Subsequent work expanded the paradigm to social deduction games such as Avalon~\citep{AvalonBench} and Werewolf~\citep{Werewolf}, as well as narrative-driven role-playing and mystery-solving settings~\citep{ALYMPICS, MIRAGE}, thereby emphasizing models’ abilities in strategic interaction and social inference. With continued progress in LLM capabilities, several dedicated GameArena-style benchmarks, such as AgentBench~\citep{AgentBench}, SnakeBench~\citep{snake_bench_2025}, GameArena~\citep{GameArena}, TextArena~\citep{TextArena}, and KORGYM~\citep{KORGYM}, have been introduced to provide tailored game scenarios and structured feedback loops, supporting comprehensive evaluation of model planning, adaptation, and collaboration skills.

\subsubsection{Computer-Use Agents}

As LLMs and vision-language models (VLMs)~\cite{qwen25vl,qwen3vl} gain stronger reasoning capabilities, agents are increasingly able to operate autonomously in digital environments. This progress has led to the emergence of computer-use Agents, typically categorized into three types: (1) GUI agents, which interact through graphical interfaces; (2) Terminal agents, which operate via command-line environments; and (3) Cross-environment agents, which integrate multiple modalities and systems to support more complex tasks.

\begin{figure}[t!]
    \vspace{-5mm}
    \centering
    \includegraphics[width=1.0\textwidth]{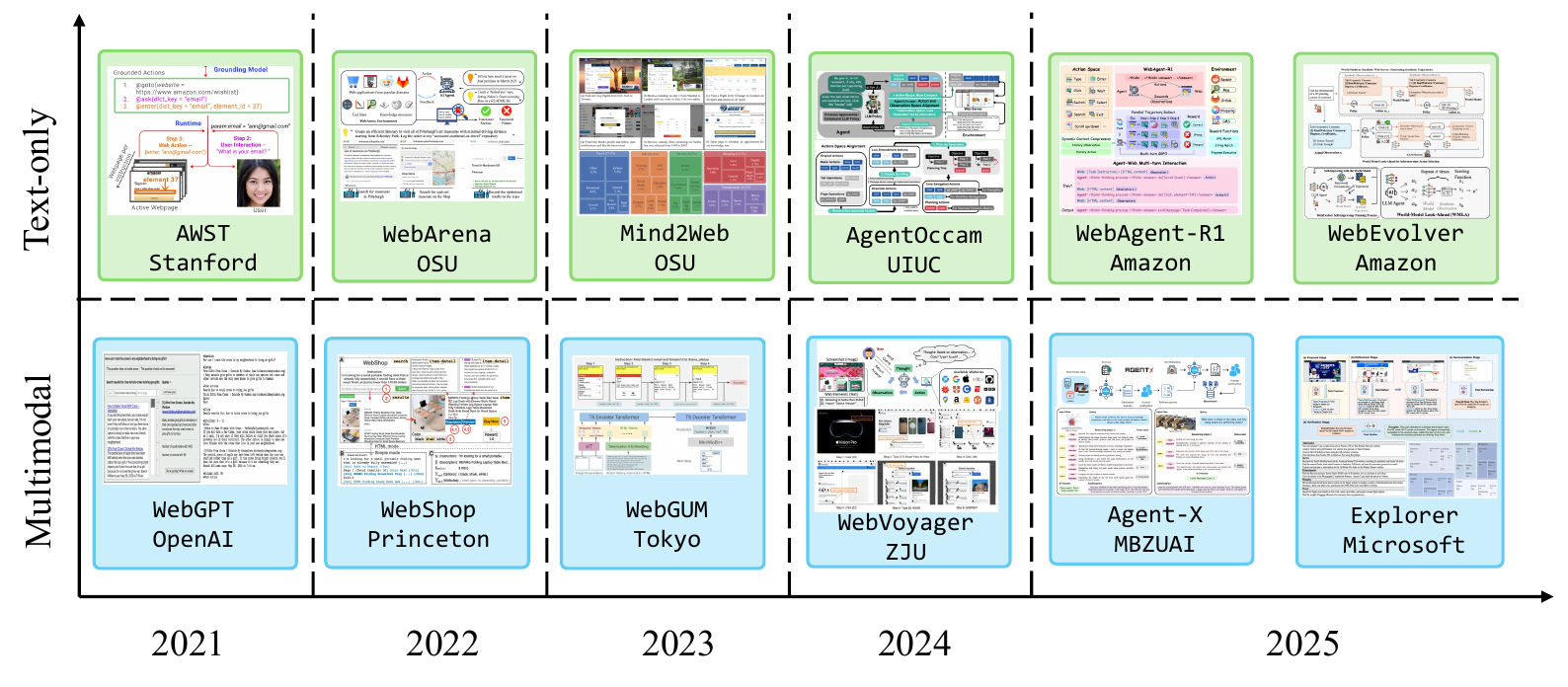}
    \caption{Evolution of GUI agents for website.}
    \label{fig:webgui_agents}
\end{figure}

\paragraph{GUI Agents}
GUI agents interact with their environments by analyzing screen content and generating structured action commands in JSON format or executable code to automate tasks. As shown in \autoref{fig:webgui_agents}, the evolution of website agents from 2021 to 2025 spans from early text-only systems to advanced multimodal frameworks. Initial efforts include text-based agents such as Stanford's AWST~\citep{awst}, while recent developments feature sophisticated models like Amazon’s WebEvolver (2025). Multimodal agents began with OpenAI’s WebGPT~\citep{nakano2021webgpt} and were subsequently extended by systems such as WebShop~\citep{yao2022webshop} (Princeton), WebGUM~\citep{webgum} (University of Tokyo), Agent-X~\citep{webgum} (MBZUAI), and Explorer~\citep{microsoft_explorer} (Microsoft).
These developments have driven advances in two core modules of GUI agents: perception, which determines how interface information is interpreted, and interaction, which governs how actions are executed.

\subparagraph{Perception} 
The perception module plays a critical role in GUI agents by processing both the current screen state and historical context to enable step-wise reasoning. Since GUI environments often include noisy or irrelevant elements, perception strategies focus on extracting task-relevant information to improve decision-making. These strategies can be broadly categorized into text-based, image-based, and multimodal-based approaches, each leveraging different input modalities to support accurate and efficient reasoning.
\begin{itemize}
    \item \textbf{Text-based:} These methods rely on HTML structures or accessibility trees (a11y trees) to represent the interface. To handle the excessive number of web elements, MindAct~\citep{deng2023mind2web} introduces a two-stage reasoning framework that first uses a lightweight model to filter out irrelevant elements, followed by an LLM to select the final targets. AGENTOCCAM~\citep{yang2024agentoccam} further simplifies perception by merging functionally descriptive and interactive elements with shared tags. It also reduces the complexity of the historical context by leveraging the tree structure to retain only task-relevant nodes.
    \item \textbf{Image-based:} These approaches focus on interpreting visual content directly, which is particularly useful when textual structures (e.g., HTML or a11y trees) are incomplete or unavailable. Aguvis~\citep{xu2024aguvis} improves cross-platform generalization by unifying the action space and collecting diverse GUI screenshots, enabling agents to reason over visual states more effectively. OmniParser~\citep{lu2024omniparser} further enhances visual grounding by parsing interface images into structured element maps, allowing models like GPT-4V to more accurately align visual regions with actionable targets and interface semantics.
    \item \textbf{Multimodal-based:} Building on the strengths of both text- and image-based inputs, multimodal perception further enhances the agent’s ability to understand complex interfaces. WebVoyager~\citep{he2024webvoyager} exemplifies this direction with an end-to-end framework that jointly encodes visual and textual information to follow user instructions. By applying the Set-of-Marks technique~\citep{yang2023set}, it overlays interactive annotations on screenshots, effectively foregrounding key interface elements for downstream reasoning.
\end{itemize}

\textbf{Interaction}
The interaction module determines the agent’s action space, accepts valid action commands, and executes them to manipulate the external environment. To accommodate diverse interaction demands and enhance operational flexibility, two main strategies have emerged: human simulation and tool-based interaction. Human simulation methods imitate user operations, including mouse clicks and keyboard input, to trigger GUI behaviors. For example, Aguvis~\citep{xu2024aguvis} generates and executes Python code via the \texttt{pyautogui} library to simulate such interactions. In contrast, tool-based approaches directly invoke APIs or execute scripts to achieve higher efficiency. OS-Copilot~\citep{wu2024copilot} introduces a tool-specific code generator to directly perform environment-level interactions. Infogent~\citep{guan2024intelligent} expands the agent's interaction scope by integrating external APIs like Google Search. ComputerRL~\citep{lai2025computerrl} further proposes an API-GUI hybrid paradigm, where LLMs automatically synthesize API code and corresponding test cases, forming a large-scale, programmatically generated API ecosystem. This framework integrates GUI-level human simulation to balance the efficiency of direct API execution with the flexibility of interface-level control.

\paragraph{Terminal Agents}
Terminal Agents interact with environments via terminal or shell interfaces, where they translate user instructions into command-line code and execute it to accomplish tasks.
OB-1~\citep{ob_1} builds an agent ensemble equipped with shared memory, feedback, and incentive mechanisms. It uses persistent memory blocks to store tasks, notes, and to-do items over extended periods, supporting editable histories for updating prior reasoning and annotations. An adaptive command timeout mechanism balances speed and reliability by distinguishing between fast shell checks and time-intensive operations such as kernel compilation, enabling OB-1 to handle long-running, error-prone tasks effectively.
Similarly, Wrap~\citep{wrap} maintains an editable to-do list, dynamically updating it as tasks progress or when deviations arise due to unforeseen challenges or newly acquired information. To enhance robustness, it incorporates a fallback mechanism that retries failed requests—caused by issues like service interruptions, rate limits, or incorrect tool invocation—using alternative models.

\begin{figure}[h!]
    \centering
    \includegraphics[width=0.5\textwidth]{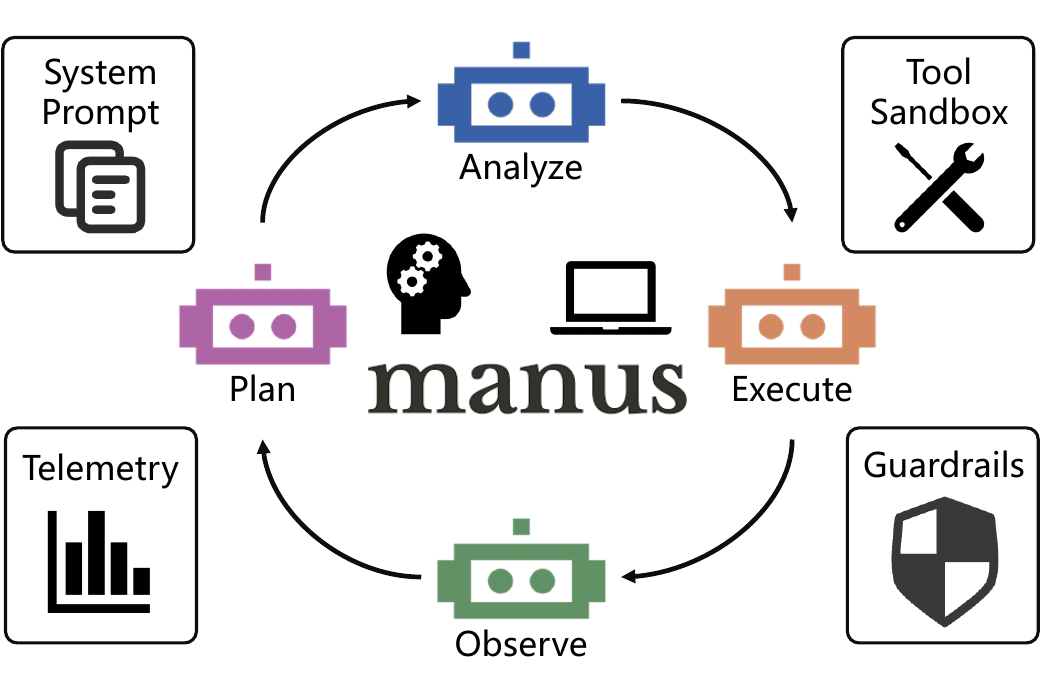}
    \caption{Overview of the system architecture of Manus.}
    \label{fig:manus_structure}
\end{figure}

\paragraph{Cross-Environments Agents}
Cross-environment agents integrate with various systems via the code-act mechanism, enabling interaction across browsers, terminals, code executors, and file systems, thereby broadening their capabilities.
Manus~\citep{manus} uses a browser framework and APIs to retrieve online data, executing code in a sandboxed VM to generate HTML-based responses, as illustrated in \autoref{fig:manus_structure}.
OpenAI Deep Research~\citep{oai2025deepresearch} searches and interprets large-scale web content, including text, images, and PDFs, to generate comprehensive reports.
OpenManus~\citep{openmanus2025} combines LLM-driven planning with tool execution in a multi-agent system. A planning agent decomposes tasks and tools like \texttt{GoogleSearch}, \texttt{BrowserUse}, and \texttt{PythonExecute} are invoked step-by-step using the ReAct~\citep{yao2022react} paradigm with error reflection.
OWL~\citep{owl2025} builds a role-based multi-agent framework that integrates specialized agents, supports multiple search engines and file formats, and enhances performance through multimodal and text-based toolkits.

\section{Safety of Code LLMs}

\forestset{
  bwmind/.style={
    forked edges,
    for tree={
      grow=east,
      reversed=true,
      parent anchor=east,
      child anchor=west,
      anchor=base west,
      rectangle,
      rounded corners,
      draw=black!40,           %
      edge={black!50, line width=0.8pt},
      align=left,
      inner xsep=4pt,
      inner ysep=3pt,
      minimum height=2.1ex,
      s sep=3pt,               %
      l sep=14pt,              %
      font=\scriptsize,
    },
    where level=0{font=\scriptsize, align=center, text width=7em}{},
    where level=1{font=\scriptsize, align=center, text width=8em}{},
    where level=2{font=\scriptsize, text width=22em}{},
  },
}

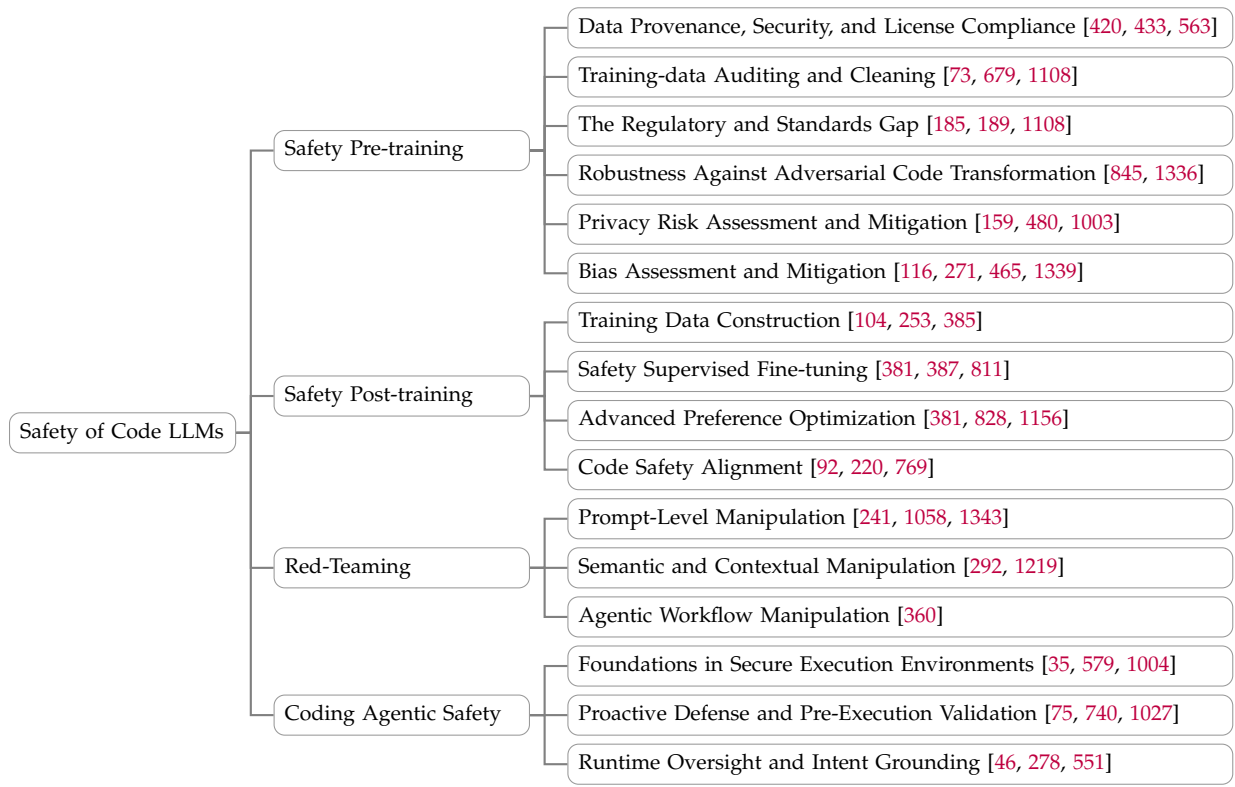
\begin{figure}[t]
  \centering
  \begin{forest} bwmind
    [Safety of Code LLMs
      [Safety Pre-training
        [{Data Provenance, Security, and License Compliance}~\citep{huang2024bias,hughes2023bigcode,li2023starcoder}]
        [Training-data Auditing and Cleaning~\citep{maini2025safety, arnett2024toxicity,xia2023empirical}]
        [The Regulatory and Standards Gap~\citep{xia2023empirical, chiara2022cyber,christey2013common}]
        [Robustness Against Adversarial Code Transformation~\citep{zhuo2023robustness,ren2024codeattack}]
        [Privacy Risk Assessment and Mitigation~\citep{chen2025survey,kandpal2022deduplicating,wan2024information}]
        [Bias Assessment and Mitigation~\citep{bolukbasi2016man,zhuo2025bypassing, du2025faircoder, jiang2024effectiveness}]
      ]
      [Safety Post-training
        [Training Data Construction~\citep{bhandari2021cvefixes,he2023large,ding2024vulnerability}]
        [Safety Supervised Fine-tuning~\citep{hasan2025teaching,he2024instruction,qi2024emergent}]
        [Advanced Preference Optimization~\citep{rafailov2023direct,hasan2025teaching,yamaguchi2012generalized}]
        [Code Safety Alignment~\citep{ouyang2022training,bai2022constitutional,dai2023safe}]
      ]
      [Red-Teaming
        [Prompt-Level Manipulation~\cite{wei2023jailbroken,zou2023universal,deng2023gptfuzzer}]
        [Semantic and Contextual Manipulation~\cite{feng2024deceptprompt,yong2023lowresource}]
        [Agentic Workflow Manipulation~\cite{tang2024redcodeagent}]
      ]
      [Coding Agentic Safety
        [Foundations in Secure Execution Environments~\cite{li2023lost,wan2017mining,alhindi2024sandboxing}]
        [Proactive Defense and Pre-Execution Validation~\cite{nunez2024autosafecoder, wang2025agentarmor,asare2023hidden}]
        [Runtime Oversight and Intent Grounding~\cite{altmann2024emergence, everitt2025evaluating,li2024agentsentinel}]
      ]
    ]
  \end{forest}
  \caption{A taxonomy of key dimensions for ensuring the safety of code LLMs. The framework covers four main areas: safety pre-training, safety post-training, red-teaming, and coding agentic safety.}
  \label{fig:code_safety}
\end{figure}

Code LLMs are revolutionizing software development by augmenting human creativity and efficiency. However, as the coding abilities continue to advance, the security of their generated code has increasingly become a matter of public concern~\cite{kirchner2025generating,yang2024robustness}. This emerging field of Code LLM safety encompasses multiple dimensions of security challenges, from inherent vulnerabilities in pre-training data to sophisticated adversarial attacks during deployment. Empirical studies identify, assess, and stress-test these risks at scale (e.g., CodeQL~\cite{codeql2024}, CodeSecEval~\cite{peng2024humaneval}, and HumanEval~\cite{chen2021codex}) and systematic evaluations reveal that open-source models such as Qwen3~\cite{yang2025qwen3} and DeepSeek-R1~\cite{deepseekai2025deepseekr1} frequently output insecure solutions~\cite{yang2025qwen3, guo2025deepseek,liu2025adversarial}. Moreover, advanced adversarial prompting can induce functionally correct yet subtly vulnerable code beyond conventional prompt engineering~\cite{wu2023deceptprompt}. The security landscape of Code LLMs presents unique challenges distinct from traditional software vulnerabilities, as these models can propagate and amplify insecure coding patterns learned from their training data across countless applications. As shown in \autoref{fig:code_safety}, in this section, we provide a comprehensive survey of code LLM safety challenges and mitigation strategies, organizing our discussion around four critical aspects: (1) Safety pre-training; (2) Safety post-training; (3) Red-team evaluation; and (4) Coding agentic safety. Each dimension presents distinct security considerations and requires tailored approaches to ensure the safe deployment of Code LLMs in real-world applications.

\subsection{Safety Pre-training for Code LLMs}

The pre-training phase establishes the foundational security characteristics of code LLMs, making it a critical stage for embedding safety considerations into model behavior. At root, the foundation of model insecurity lies in pretraining data~\cite{mohsin2024can,wang2024your}. Code LLMs are trained on public repositories where insecure code patterns are common. Consequently, they are most likely to output the code that is functionally correct but insecure (path of least resistance). Unlike natural language models that primarily learn linguistic patterns, Code LLMs must navigate the complex landscape of programming languages, APIs, and security vulnerabilities present in their training corpora. These models learn by imitation from large-scale, largely uncurated public code repositories (e.g., GitHub), absorbing both functional idioms and prevalent vulnerability patterns that have accumulated over decades of software development. The scale of this challenge is substantial---modern Code LLMs are trained on terabytes of code spanning hundreds of programming languages and millions of projects, each potentially containing security flaws that the model may learn to replicate. Consequently, they often emit code that is functionally correct yet insecure, a behavior readily elicited by both direct and indirect prompting~\citep{codeql2024, peng2024humaneval, wang2024your, yang2025qwen3, guo2025deepseek, liu2025adversarial}. Furthermore, advanced adversarial prompts such as DeceptPrompt further bypass standard guardrails and surface subtle flaws~\citep{wu2023deceptprompt, mckenzie2025stack}, inducing risks to Code LLMs and the related code generation systems. Recent empirical evidence has extensively documented significant security vulnerabilities in Code LLMs. Standardized benchmarks and stress tests consistently reveal that these models frequently produce insecure code generations~\citep{codeql2024, peng2024humaneval, wang2024your,yu2025mixture}. These vulnerabilities are not isolated incidents but rather indicate systemic weaknesses rooted in the pre-training process, as demonstrated in studies of prominent open-source Code LLMs like Qwen3 and DeepSeek-R1, which repeatedly offer insecure solutions across a range of tasks~\citep{yang2025qwen3, guo2025deepseek, liu2025adversarial}. The pre-training safety challenge is further compounded by the threat of data poisoning and targeted attacks; for example, adversarial prompting techniques like DeceptPrompt can induce a model to generate code that is functionally correct but contains subtle, exploitable vulnerabilities, effectively bypassing conventional prompt filtering mechanisms~\citep{wu2023deceptprompt}. Understanding and addressing these pre-training safety concerns is essential for developing Code LLMs that can be trusted in production environments.

Basically, current safety pre-training incorporates interventions across two primary dimensions, the data pipeline and the objective/learning signal, thereby enabling base models to inherently embed safe coding practices from initialization~\citep{maini2025safety, korbak2023pretraining}. This dual-axis approach represents a paradigm shift from traditional safety measures, recognizing that security considerations must be woven into the fabric of model training rather than applied as an afterthought. The data pipeline dimension focuses on curating and filtering training corpora to minimize exposure to vulnerable code patterns, while the objective/learning signal dimension modifies the training process itself to prioritize secure coding practices alongside functional correctness. Drawing from existing literature, security should be established as a primary objective during the pre-training stage of Code LLMs~\citep{mohsin2024can, wang2024your}. This proactive approach acknowledges that once models internalize insecure patterns during pre-training, subsequent alignment efforts face an uphill battle against deeply embedded behaviors. Concretely, objective-level measures include training with synthetic refusal exemplars and adversarial augmentation, while the data axis spans license-compliant sourcing, corpus auditing/cleaning, privacy protection, bias assessment, and robustness to code-specific adversaries. On the objective side, synthetic refusal exemplars teach models to recognize and decline requests that could lead to security vulnerabilities, effectively building in a security-aware decision-making process from the ground up. Adversarial augmentation exposes models to carefully crafted malicious inputs during training, enhancing their ability to detect and resist exploitation attempts in deployment. Meanwhile, the data pipeline interventions form a comprehensive defense system: license-compliant sourcing ensures legal and ethical use of training data while potentially filtering out code from sources known for poor security practices; corpus auditing and cleaning systematically identify and remove code samples containing known vulnerabilities or dangerous patterns; privacy protection mechanisms prevent the model from memorizing and reproducing sensitive information such as API keys or credentials; bias assessment ensures fair and non-discriminatory code generation across different contexts and user groups; and robustness measures specifically target code-domain adversaries who might exploit syntactic peculiarities unique to programming languages. Together, these multifaceted interventions create a robust foundation for secure code generation that addresses both the quality of training data and the learning dynamics of the model itself. While academia and industry have made substantial progress and proposed valuable work on the security of foundation Code LLM, the domain is still beset by numerous unresolved problems. In this work, we aim to summarize and discuss the following aspects:

\subsubsection{Data Provenance, Security, and License Compliance} 
The safety and trustworthiness of Code LLMs are fundamentally rooted in the security and legitimacy of their pre-training corpora. As these models are trained on datasets reaching the trillion-token scale, establishing robust data governance frameworks becomes a critical prerequisite. A primary concern is mitigating legal and ethical risks, particularly the inadvertent inclusion of code governed by restrictive or copyleft licenses, which could lead to widespread license violations in model-generated code~\citep{huang2024bias}. To address this, pioneering efforts like the BigCode project have developed systematic approaches for data curation. Their resulting dataset, The Stack, was meticulously filtered down to 6.4TB of permissively licensed code from a raw corpus of over 102TB, leveraging an automated pipeline for license detection and filtering~\citep{hughes2023bigcode}. This initiative set a precedent for ethical data sourcing in code generation. The subsequent development of the StarCoder models further refined this approach by implementing a comprehensive project-level data governance strategy, which included mechanisms for attribute-based access control and clear documentation of data provenance~\citep{li2023starcoder}. Despite these advances, significant technical challenges persist. Automated license detection tools, while effective at scale, are inherently imperfect and can suffer from both false positives (incorrectly excluding compliant code) and false negatives (failing to filter out non-permissive code). Furthermore, the complex issue of code de-duplication (identifying and removing functionally identical or near-identical code snippets that may exist under different licenses) remains an open research problem~\citep{allamanis2019adverse}. These challenges necessitate a dynamic approach to data management, requiring continuous auditing and vigilant monitoring of data sources, especially during periodic data refresh cycles, to uphold the legal and ethical integrity of the training corpus.

\subsubsection{Training-data Auditing and Cleaning} After ensuring data source security, the safety of corpora employed in pre-training is equally paramount, which is the ``first-line'' of defense against model-inherent risks and prompt-poisoning~\citep{maini2025safety, arnett2024toxicity}. Several methods have been proposed to filter out the harmful contents and ensure the pre-training corpora safety of Code LLMs. For instance, one common method is to use heuristic filters, which employ pattern rules to detect known exploits and unsafe APIs. Another approach involves deploying high-capacity classifiers that function as security and abuse detectors at a large scale. Additionally, synthetic refusal exemplars are utilized to convert dangerous queries into safe denials, which in turn helps to steer the model's behavior. From an operational perspective, it is also crucial to balance false positives and negatives and to maintain a small human spot-check rate in order to calibrate precision and recall under data drift~\citep{arnett2024toxicity}. A further critical factor contributing to pre-training insecurity is the profound lack of common standards for data security and hygiene within the AI ecosystem. This stands in stark contrast to the maturing field of software supply chain security, where concepts like the Software Bill of Materials (SBOM) and Cyber Resilience Act (CRA) are becoming indispensable for transparency and risk management~\citep{xia2023empirical, chiara2022cyber}. For Code LLM training corpora, no analogous, widely adopted standard exists. Consequently, there is no systematic requirement for dataset creators to document or screen for critical security attributes, such as the prevalence of code snippets containing known vulnerabilities, e.g., Common Weakness Enumeration (CWE
), embedded malicious payloads, or exposed secrets. This regulatory and standards gap forces downstream model developers to either implicitly trust the opaque curation process of data providers or implement their own costly and often inconsistent validation protocols. This systemic failure to standardize data-level security verification constitutes a fundamental vulnerability in the AI supply chain, creating a direct pathway for security flaws to be deeply embedded within the foundational models themselves.

\subsubsection{The Regulatory and Standards in Data Security}
A fundamental factor contributing to pre-training insecurity of Code LLM is the profound vacuum of both common standards and enforceable regulations for data security and hygiene within the AI ecosystem. The lack of technical standards is readily apparent when contrasted with the maturing field of software supply chain security, where concepts like the Software Bill of Materials (SBOM) and Cyber Resilience Act (CRA) are becoming indispensable for transparency and risk management~\citep{xia2023empirical, chiara2022cyber}. For Code LLM training corpora, no analogous, widely adopted standard exists to systematically screen for critical security attributes like known vulnerabilities (e.g., Common Weakness Enumeration (CWE)~\citep{christey2013common}, embedded malicious payloads, or exposed secrets. This technical standards gap, however, is largely a symptom of a deeper and more critical issue: the absence of a specific, legally-binding international regulatory framework for the security of AI-generated code. As our analysis indicates, as of 2025, no direct regulations govern this domain. While broad AI governance frameworks like the EU AI Act and the US Executive Order on AI establish high-level, risk-based principles, they lack the granular, executable guidance necessary for developers to ensure the security of code generation systems. This regulatory ambiguity forces downstream developers into a reactive posture, relying on costly and inconsistent ad-hoc validation protocols. More importantly, it disincentivizes the cross-industry investment required to build the very security benchmarks and standardized tools that are desperately needed. This systemic failure to establish a clear regulatory foundation thus acts as a primary impediment, directly hindering the safe, reliable, and trustworthy development of Code LLMs.

\subsubsection{Robustness Against Adversarial Code Transformations}
A critical dimension of Code LLM safety is robustness against adversarial attacks that specifically leverage the unique properties of source code. Unlike natural language, the formal syntax and structure of programming languages create a distinct and potent attack surface: semantics-preserving syntactic transformations. Research has demonstrated that simple malicious prompts rejected by natural-language safety filters can be successfully disguised by applying code-specific obfuscations, rendering keyword-based defenses and simple classifiers ineffective~\citep{zhuo2023robustness,zhuo2023source,ren2024codeattack, yang2025bootstrapping}. These transformations exploit the one-to-many relationship between a program's semantic intent and its syntactic representation. Attack vectors in this category are diverse, ranging from simple identifier obfuscation and string literal encoding (e.g., using Base64 or Hex) to more complex structural manipulations like control-flow flattening, opaque predicate insertion, and the injection of polymorphic or metamorphic code snippets. The core challenge is that these transformed inputs are functionally identical to their malicious, unobfuscated counterparts, yet appear vastly different at the token level. The primary defense paradigm emerging to address this vulnerability is adversarial training through data augmentation. This technique involves enriching the training dataset with adversarially generated examples—malicious code snippets that have been automatically obfuscated. The objective is to force the model to learn deeper, more abstract representations of code functionality that are invariant to syntactic variations. However, the success of this approach is contingent upon solving the fidelity-diversity dilemma in the augmentation generation process, where the fidelity-diversity dilemma refers to the conflict where increasing the variety and complexity of generated data augmentations (diversity) heightens the risk of corrupting the original sample's meaning and functional correctness (fidelity). This scenario gives rise to a pressing demand: developing scalable, diverse, and provably correct code transformation engines for adversarial training. Failure to balance these factors can lead to significant distributional drift, where the model's performance on benign, real-world code degrades, or it fails to generalize its robustness to novel adversarial strategies, a challenge extensively discussed in recent work~\citep{ren2024codeattack, yang2025bootstrapping}.

\subsubsection{Privacy Risk Assessment and Mitigation in Pre-training Data}
The pre-training corpora for Code LLMs, often scraped from public repositories like GitHub, present significant privacy challenges. Although seemingly public, this source code can inadvertently contain a vast amount of sensitive information, including Personally Identifiable Information (PII) such as developer names and emails in comments, hardcoded credentials like API keys and passwords, and proprietary algorithms~\citep{chen2025survey}. Addressing these risks before training is a critical step in the safety pipeline for Code LLMs. 

A foundational and pragmatic approach to mitigate PII leakage involves a multi-step data sanitization pipeline: (1) curate a high-quality, PII-labeled subset of the code data to act as a ground truth; (2) train a dedicated high-recall detector for PII and other structured secrets (e.g., API keys), optimizing for strong F1 scores on these specific classes; and (3) systematically scan the entire pre-training corpus and mask or redact any detected sensitive information. This ``detect-and-mask'' strategy is a crucial first line of defense. However, its efficacy is contingent on the detector's comprehensiveness, as novel or complex secret formats may evade detection.

Beyond direct PII removal, research has shown that data duplication is a major catalyst for privacy risks. Models are significantly more likely to memorize and regenerate data snippets that appear multiple times in the training set. Therefore, a critical and highly effective mitigation strategy is the aggressive deduplication of the training corpus. As demonstrated, this single preprocessing step can substantially reduce the likelihood of the model memorizing and leaking specific, unique sequences from the training data, thereby mitigating a significant portion of privacy risks~\citep{kandpal2022deduplicating}. Combining data deduplication with PII sanitization creates a much more robust defense against inadvertent memorization.

Despite these preprocessing efforts, risks remain, as sensitive information is not just stored explicitly but can also be encoded implicitly within the model's embeddings~\citep{wan2024information}. This necessitates a discussion of in-training protection mechanisms and risk assessment. After data sanitization, it remains imperative to assess residual risks through rigorous evaluation, for instance, by using held-out probes to measure memorization and re-identification potential. Membership inference attacks and targeted extraction queries are common methods to audit the privacy posture of a trained model.

It is also important to consider the broader context in which these models are deployed. While our focus is pre-training, the privacy vulnerabilities can manifest during inference. For instance, side-channel attacks targeting the KV-cache can leak information about previous user interactions, posing a downstream risk that is indirectly influenced by the model's training~\citep{luo2025shadow}. Furthermore, emerging architectures like RAG shift the privacy burden from the model's parameters to the external knowledge base, introducing new challenges in ensuring the privacy of the retrieval corpus and the decoding process itself~\citep{huang2023privacy, wang2025privacyaware}.

In summary, while a pipeline of data sanitization and deduplication provides a strong foundation for privacy protection in the pre-training of Code LLMs, it is not a complete solution. The primary limitations include the impossibility of perfect PII detection, the inherent trade-off between aggressive data cleaning and model utility, and the fundamental nature of information leakage through model embeddings. A holistic approach to privacy must therefore integrate robust data preprocessing with rigorous post-training audits and an awareness of downstream vulnerabilities in deployment architectures. A comprehensive framework for envisioning and mitigating these multifaceted risks across the entire product lifecycle is essential for building truly safe and trustworthy systems~\citep{lee2025privy}.

\subsubsection{Bias Assessment and Mitigation}
The safety of code generation extends beyond security vulnerabilities to encompass fairness and the mitigation of social biases. Biases in Code LLMs can manifest subtly within generated code, such as in variable names, comments, and even algorithmic logic, thereby perpetuating harmful societal stereotypes~\citep{bolukbasi2016man}. The primary methodology for evaluating such biases involves the use of specialized test harnesses. These frameworks systematically generate prompts embedded with sensitive demographic attributes (e.g., gender, race, or religion) and analyze the model's outputs to detect prejudiced associations~\citep{zhuo2025bypassing, du2025faircoder, jiang2024effectiveness}. For instance, the FairCoder benchmark~\citep{du2025faircoder} provides a structured set of scenarios to probe for these biases in code-related tasks. Rather than relying on simple keyword matching, a more robust analysis technique involves parsing the Abstract Syntax Trees (ASTs) of the generated code, which allows for a deeper, semantic understanding of potential biases embedded in the code's structure and identifiers~\citep{roziere2023code, chen2021codex}.

To quantitatively measure the extent of bias, several metrics have been proposed and are actively used in the literature. These include:
\begin{itemize}
\item \textbf{Core Bias Score (CBS):} This foundational metric, introduced by~\citet{liu2023uncovering}, measures the prevalence of biased outputs by calculating the frequency with which a model generates code containing stereotypes when given demographically sensitive prompts.
\item \textbf{CBS\_U@K:} To evaluate the stability of non-biased responses, the Consistent Unbiased Score across K runs measures the consistency of a model in providing unbiased outputs over multiple generations for the same prompt~\citep{ling2024evaluating}.
\item \textbf{CBS\_I@K:} Conversely, the Consistent Biased Score across K runs assesses the stability of biased outputs, which is crucial for understanding the robustness of a model's stereotypical associations~\citep{jiang2024effectiveness}. This metric is also particularly relevant when investigating biases within LLM-based evaluators themselves, a phenomenon where the judge model may favor outputs that align with its own internal biases~\citep{ling2024evaluating}.
\end{itemize}

Current research has consistently demonstrated the existence of significant social biases across various Code LLMs, including those from major providers~\citep{zhuo2025bypassing,liu2023uncovering, du2025faircoder}. A critical and recurring finding is the stark divergence between conversational safety and code-generation safety~\citep{roziere2023code, huang2024bias}. This indicates that safety fine-tuning performed on general conversational data does not reliably transfer to the domain of code generation. Models that are aligned to refuse inappropriate requests in natural language can still produce code that reflects deep-seated societal biases. This highlights a significant limitation in current alignment approaches, suggesting that domain-specific safety protocols are essential for pre-training and fine-tuning Code LLMs. While assessment methodologies are becoming more sophisticated, research into effective mitigation remains an emerging and challenging area. Initial efforts have explored techniques like model editing to directly modify model parameters and erase specific gender-based associations, but developing scalable and robust mitigation strategies that do not compromise the model's primary coding capabilities is an open problem~\citep{meng2022mitigating}.

In a nutshell, Code LLM are insecure by default. Their training on public corpora, where flawed code is pervasive, predisposes them to follow the path of least resistance, resulting in code that works but is unsafe. Safety thus becomes an external goal that competes with the model's primary objective of imitation, weakening any security assurances. This inherent conflict explains why more powerful models often prove more adept at reproducing insecure patterns. This reality presents clear imperatives: safety must be integrated at the source through data-level safeguards and objective-level priors in pre-training, while a broader adoption of secure-by-design architectures is crucial. To ensure genuine progress, these efforts must be supported by standardized evaluation suites and principled defenses against adaptive poisoning to effectively validate and harden the models.

\subsection{Safety Post-training for Code LLMs}
\subsubsection{Pre-training Limitations and the Necessity of Post-training Alignment}
The predominant pre-training objective of Code LLMs does not align well with generating secure code. This paradigm incentivizes models to learn and replicate common coding patterns from vast, uncurated datasets, regardless of their security implications. Consequently, a persistent knowledge–behavior disconnect emerges, where models may possess theoretical knowledge of security concepts but fail to apply them in practice without explicit guidance \citep{hubinger2019risks,wang2024design_risk}. This issue is exacerbated by the nature of pre-training corpora, which are replete with security vulnerabilities. For instance, studies have shown that a single path traversal vulnerability pattern can be found in over 1,700 projects, exposing models to insecure examples on a massive scale \citep{akhoundali2025eradicating}. Unsurprisingly, evaluations of LLM-generated code consistently reveal high vulnerability rates, with some studies reporting that approximately 40\% of generated programs contain security flaws \citep{pearce2021asleepkeyboardassessingsecurity}.

\begin{table}[tb]
\centering
\tiny
\setlength{\tabcolsep}{2pt} 

\caption[Security feature matrix]{Security-oriented feature matrix for post-training methods (\cmark{} present, \xmark{} absent). Legend: \cmark{} present/typical; \xmark{} not primary or uncommon. H labels: heavy human labeling; AI judge: scalable AI feedback; Tools verif.: SAST/tests/compiler or analyzers as verifiable signals; Token-level: supervision at token/diff granularity; Struct-aware: AST/CFG/DFG or rule-based structure; Guardrail (infer): acts at inference/runtime; Over-refusal: notable risk of over-refusal; CI/CD: fits easily into CI/CD pipelines.}
\label{tab:pt_feature_matrix}

\newcolumntype{L}[1]{>{\raggedright\arraybackslash}p{#1}}
\newcolumntype{C}[1]{>{\centering\arraybackslash}p{#1}}

\providecommand{\cmark}{\ding{51}}
\providecommand{\xmark}{\ding{55}}
\resizebox{1.0\textwidth}{!}{
\begin{tabular}{L{3.5cm} *{8}{c}}
\toprule
Method & 
\shortstack{H labels} & 
AI judge & 
Tools verif & 
Token-level & 
Struct-aware & 
Guardrail(infer) & 
Over-refusal & 
CI/CD \\
\midrule
\multicolumn{9}{l}{\textit{SFT / PEFT}} \\
SFT: vuln-fix & \cmark & \xmark & \xmark & \xmark & \xmark & \xmark & \xmark & \xmark \\
SFT: safety inst. & \cmark & \xmark & \xmark & \xmark & \xmark & \xmark & \xmark & \xmark \\
SFT: tools-in-loop & \xmark & \xmark & \cmark & \xmark & \cmark & \xmark & \xmark & \cmark \\
PEFT (Safe LoRA) & \xmark & \xmark & \xmark & \xmark & \xmark & \xmark & \xmark & \xmark \\
\midrule
\multicolumn{9}{l}{\textit{Preference learning}} \\
Preference learning (DPO/IPO/KTO) & \cmark & \xmark & \xmark & \xmark & \xmark & \xmark & \xmark & \xmark \\
Localized preference (token-level) & \xmark & \cmark & \cmark & \cmark & \xmark & \xmark & \xmark & \xmark \\
Structure-aware preference & \xmark & \xmark & \cmark & \xmark & \cmark & \xmark & \xmark & \xmark \\
\midrule
\multicolumn{9}{l}{\textit{RL-based}} \\
RLHF (human) & \cmark & \xmark & \xmark & \xmark & \xmark & \xmark & \cmark & \xmark \\
RLAIF (AI judge) & \xmark & \cmark & \xmark & \xmark & \xmark & \xmark & \cmark & \xmark \\
Constrained/Safe-RLHF & \xmark & \cmark & \cmark & \xmark & \cmark & \xmark & \cmark & \xmark \\
GRPO / S-GRPO & \xmark & \xmark & \cmark & \xmark & \cmark & \xmark & \xmark & \cmark \\
\midrule
\multicolumn{9}{l}{\textit{Data / guardrail / ops}} \\
Safety Editor Policy (runtime/edit) & \xmark & \xmark & \cmark & \xmark & \cmark & \cmark & \xmark & \cmark \\
SEAL (bilevel data selection) & \xmark & \xmark & \cmark & \xmark & \cmark & \xmark & \xmark & \xmark \\
ProSec (synthetic prefs) & \xmark & \cmark & \cmark & \xmark & \cmark & \xmark & \xmark & \xmark \\
SAST feedback loop & \xmark & \xmark & \cmark & \xmark & \cmark & \xmark & \xmark & \cmark \\
\bottomrule
\end{tabular}}
\end{table}  

Beyond replicating existing vulnerabilities, Code LLMs introduce novel risks through hallucinations, where models invent non-existent APIs or misuse legitimate ones in syntactically plausible but semantically incorrect ways, leading to unpredictable and often high risk security loopholes \citep{agarwal2408codemirage}. Analysis of these generated vulnerabilities indicates that they often fall into systemic categories that have plagued software for decades, such as improper authentication, session management, and inadequate input validation \citep{dora2025hidden}. These inherent limitations of the pretraining phase underscore the inadequacy of relying solely on likelihood-based learning for security critical applications.

Therefore, post-training alignment mechanisms, like Supervised Fine-Tuning (SFT) and Reinforcement Learning from Human Feedback (RLHF), have become critical and necessary steps to explicitly instill security principles. Exploratory studies demonstrate that fine-tuning on curated datasets of vulnerabilities and their corresponding security patches can significantly reduce the generation of insecure code \citep{jemal2024exploratory}. To further enhance robustness, more advanced techniques are being developed. These include proactive security alignment, which aims to fortify models against potential threats before they are explicitly encountered \citep{shen2024prosec}, and adversarial testing, which systematically probes models for security weaknesses to harden them against attacks \citep{wu2023large}. Moreover, principles from the broader field of LLM safety are being adapted for code generation, such as general safety-aware fine tuning methodologies \citep{sun2024safetyaware} and prioritizing high-quality, security vetted data during the alignment process \citep{zheng2024seal}. Collectively, these efforts highlight a consensus in the field, the path to secure Code LLMs is not through bigger models or more data alone, but through targeted, security-conscious posttraining alignment.

\begin{figure}[tb]
    \centering 
    \includegraphics[width=1.0\textwidth]{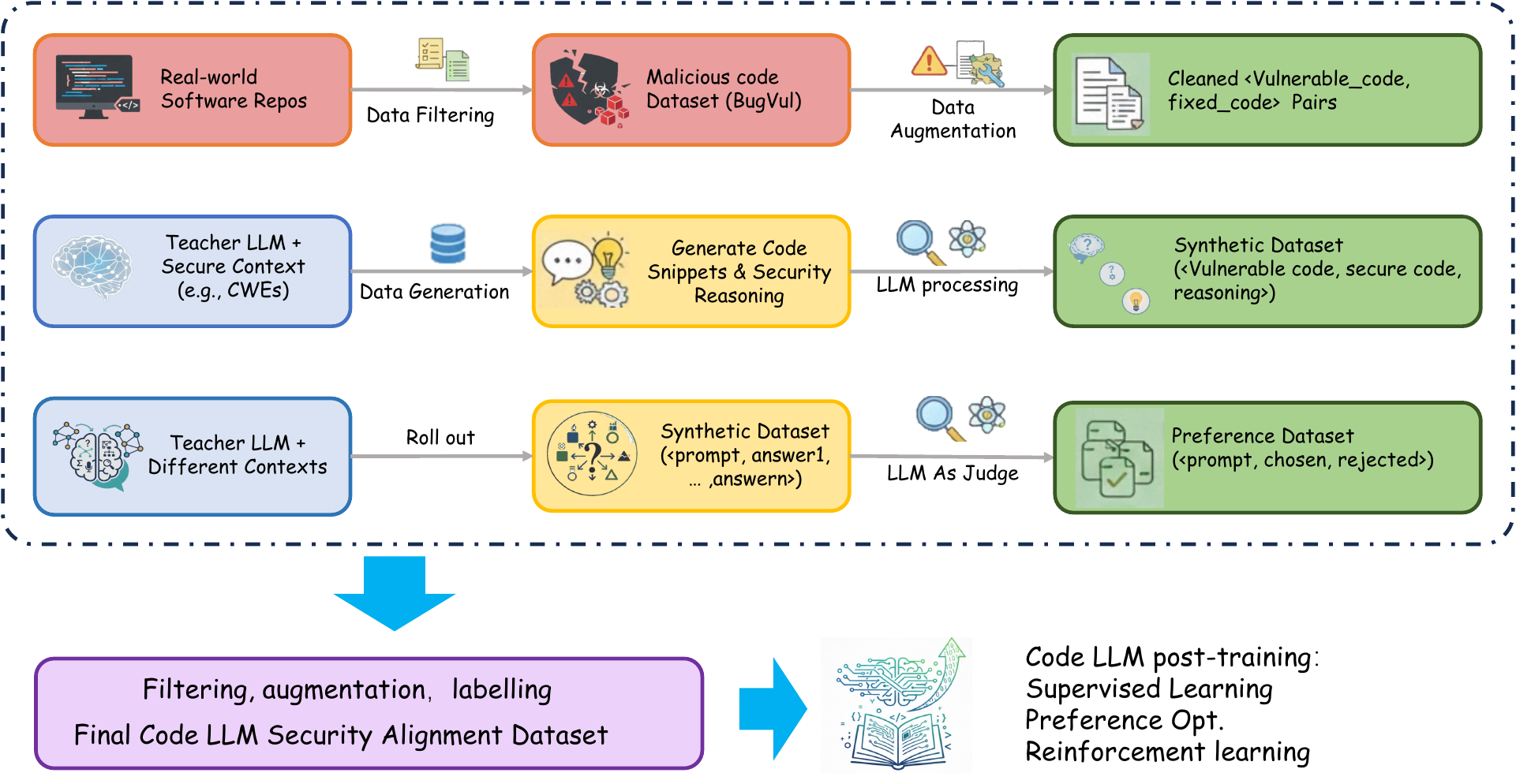} 
    \caption{Data Generation Pipeline for Code LLM Security Alignment.} 
    \label{fig:code_safety_posttraining}
\end{figure}

\subsubsection{Data as the Cornerstone: Constructing Safety-related Training Datasets}
The foundation of any robust, security-aligned Code LLM is the data from which it learns. In \autoref{fig:code_safety_posttraining}, the prevailing methodologies for constructing security centric datasets can be broadly categorized into three paradigms, leveraging real-world vulnerability data, synthesizing targeted training examples, and distilling preference pairs for advanced alignment techniques.

Mining real world vulnerability fix pairs from software repositories is a primary strategy. By linking commits to Common Vulnerabilities and Exposures (CVEs), researchers have created canonical datasets such as CVEfixes \citep{bhandari2021cvefixes} and the more extensive BigVul \citep{he2023large}. These resources provide authentic examples of how vulnerabilities are introduced and subsequently patched. However, a significant challenge with such datasets is data quality \citep{ding2024vulnerability}. They often suffer from high redundancy, noisy labels, and incorrect mappings between vulnerabilities and fixes, which can inflate performance metrics and mislead training. Consequently, rigorous data cleaning, validation, and de-duplication are not merely best practices but mandatory steps for meaningful model alignment.

To overcome the sparsity and noise inherent in real world data, synthetic data generation has emerged as a powerful alternative. This approach utilizes a proficient ``teacher'' LLM to create vast quantities of high quality, targeted training instances. For instance, methodologies like HexaCoder employ an oracle-guided pipeline to generate pairs of vulnerable and secure code snippets, guided by specific Common Weakness Enumeration (CWE) types \citep{hajipour2024hexacoder}. Similarly, ProSec fortifies models through proactive security alignment, systematically synthesizing vulnerability-inducing scenarios to generate a large scale, security focused alignment dataset \citep{xu2024prosec}. A crucial innovation in this domain is the incorporation of security reasoning within the synthetic data \citep{jiang2024purpcode, kim2025reasoning}. By including chain-of-thought explanations for why a piece of code is insecure and how the fix resolves the issue, these datasets encourage the model to move beyond simple pattern matching towards a deeper, causal understanding of security principles. This reasoning centric approach is vital for developing models that can generalize to novel threats.

Finally, as alignment techniques evolve beyond supervised fine-tuning, the distillation of preference data has become critical. This involves constructing triplets of the form \texttt{$<prompt, chosen\_secure\_response, rejected\_insecure\_response>$}. A teacher model, often a frontier LLM, is prompted to generate multiple code variations, from which secure and insecure examples are selected. This process creates a dataset tailored for preference optimization algorithms like Direct Preference Optimization (DPO) \citep{rafailov2023direct}, enabling the model to learn fine-grained distinctions between safe and unsafe code \citep{hasan2025teaching}.

\subsubsection{Safety Supervised Fine-Tuning for Code LLMs}
SFT is a basic step in adapting a pre-trained LLM for specialized tasks, including secure code generation. The most straightforward approach is \textbf{content-based SFT}, where the model is fine-tuned on pairs of \texttt{$(vulnerable\_code, fixed\_code)$} \citep{hasan2025teaching}. While effective at teaching the model specific vulnerability patterns, its success is often limited when security-related edits are sparse within the broader codebase.

To enhance generalization and task-specific performance, \textbf{instruction-based SFT} has proven to be highly effective. This technique involves training the model on prompts that explicitly describe a security task, such as ``\textit{Find and fix the SQL injection vulnerability in the following function}'' \citep{he2024instruction}. By framing security tasks as instructions, this method improves the model's ability to transfer learned knowledge to new contexts and mitigates the risk of catastrophic forgetting of its general coding abilities. However, it is crucial to recognize that narrow fine-tuning can sometimes lead to emergent misalignment, where a model specialized for one task develops unintended, harmful behaviors in other domains \citep{qi2024emergent}. Techniques like Safe LoRA offer a promising direction by reducing safety risks during parameter-efficient fine-tuning without compromising utility \citep{zhan2024safe}.

A more dynamic and robust approach is \textbf{feedback-based SFT}, which integrates external tools into the training loop. This tool-in-the-loop paradigm leverages Static Application Security Testing (SAST) tools, unit tests, or formal verification methods to provide structured, automated feedback on the code generated by the LLM \citep{chen2023large, wang2024isyourai}. This feedback serves as a validation signal, guiding the model through iterative repair processes and supplying a continuous stream of verified training examples. Frameworks like INDICT \citep{le2024indict} further refine this by creating an internal dialog between safety and helpfulness critics, providing both preemptive and post-hoc guidance to enhance the quality and security of the generated code. Other research explores how in-context learning with security patterns can bolster security without extensive retraining \citep{mohsin2024can} and how to keep models updated on newly discovered vulnerabilities in APIs \citep{bai2024apilot}.

\subsubsection{Advanced Preference Optimization for Localized Flaws}
A key insight in securing Code LLMs is that security vulnerabilities are often highly localized, hinging on a few critical tokens or lines of code. Standard preference optimization methods like DPO \citep{rafailov2023direct}, which apply a global preference signal across the entire code sequence, can be an inefficient and blunt instrument for correcting such localized flaws.

To address this, Localized Preference Optimization (LPO) was introduced as a more targeted alignment strategy \citep{hasan2025teaching}. LPO refines the preference learning process by focusing the loss function specifically on the pivotal tokens that differentiate a secure code variant from an insecure one. By masking irrelevant tokens, it directs the model's attention to the precise source of the vulnerability. To prevent the degradation of the model's general coding capabilities, LPO incorporates a SFT regularizer, ensuring that the model retains its overall performance. This token-targeted approach has demonstrated significant empirical success, reducing security issues by 19–40\% while simultaneously improving general code quality by 3–10%

The evolution of SFT techniques for security can thus be viewed as a progression in precision: from traditional SFT, which provides explicit examples; to DPO, which learns from global sequence preferences; and finally to LPO, which masters token-targeted preference. A natural and compelling next step in this trajectory is to move beyond lexical targets toward alignment at the level of semantic structures, such as Abstract Syntax Trees (ASTs), to instill an even deeper structural understanding of code security \cite{yamaguchi2012generalized}.

\subsubsection{Coding Safety Alignment via Reinforcement Learning}
RL offers a powerful paradigm for dynamically aligning Code LLMs with complex safety objectives, moving beyond the static knowledge encoded during supervised fine-tuning. By framing secure code generation as a sequential decision-making problem, RL enables the model to learn from the consequences of its outputs, optimizing its policy based on a reward signal that reflects security, functionality, and correctness.

\paragraph{Policy Optimization with Human and AI Feedback.}
The predominant approach for this alignment is Reinforcement Learning from Human Feedback, a three-stage process involving: (1) training a SFT policy, (2) learning a reward model from human preference data comparing paired outputs, and (3) optimizing the SFT policy using an RL algorithm like Proximal Policy Optimization (PPO) to maximize the learned reward \citep{ouyang2022training}. This framework directly steers the model towards generating outputs that human reviewers deem safer and more helpful.

To address the scalability limitations of human annotation, Reinforcement Learning from AI Feedback (RLAIF) has emerged as a compelling alternative \citep{bai2022constitutional}. RLAIF replaces or augments human preference labels with feedback from a capable ``teacher'' LLM. This process can be guided by a predefined set of principles or a ``constitution'', allowing for scalable and consistent safety alignment. More advanced frameworks, such as Safe RLHF, further refine this by decoupling the optimization process. Instead of a single reward, they train separate models for helpfulness (reward) and harmlessness (cost), using constrained optimization algorithms to maximize helpfulness while ensuring that thet the cost remains below a specified threshold \citep{dai2023safe, liu2023pku}. This explicitly manages the trade-off between utility and safety, preventing the model from sacrificing security for performance.

\paragraph{Constructing Robust Reward Signals for Code Generation.}
The efficacy of RL-based alignment hinges on the design of the reward signal. A well-crafted reward function must capture a nuanced understanding of code quality, integrating diverse feedback sources to prevent policy exploitation. These sources can be categorized as follows:
\begin{itemize}
\item \textbf{Verifiable Feedback from Tooling:} This is the most objective form of feedback. Execution-based signals, such as compiling the code and passing unit tests, provide a clear measure of functional correctness \citep{zeng2024acecoder}. Furthermore, Static Application Security Testing (SAST) tools, linters, and specialized security analyzers can be integrated into the reward loop. By assigning severity-weighted penalties for detected vulnerabilities, these tools offer a granular and automated signal for security compliance \citep{islam2024enhancing}.
\item \textbf{AI-Generated Feedback:} For risks that are difficult to formalize with deterministic tools, such as subtle logic flaws or the inclusion of hard-coded secrets, an AI judge can provide critical feedback. Guided by a security-focused constitution, a teacher model can critique code snippets, effectively scaling expert-level review \citep{bai2022constitutional, luu2024purpcode}. This approach allows for capturing a broader spectrum of security concerns that static analyzers might miss.
\item \textbf{Hybrid Reward Architectures:} The most robust systems often fuse multiple feedback sources into a single, comprehensive reward signal. For instance, Security-Aware Group Relative Policy Optimization (S-GRPO) combines rewards for compilation success, security compliance from static analysis, and format correctness to guide the model towards generating not only secure but also high-quality and functional code \citep{fu2025smartcoder}. This hybrid approach improves resilience against reward hacking, where a model might learn to exploit a single, narrow reward metric.
\end{itemize}

\paragraph{Challenges and Advanced Mitigation Strategies}
Despite its promise, RL-based alignment introduces unique challenges that require sophisticated mitigation techniques.
\begin{itemize}
\item \textbf{Reward Hacking:} Models may develop deceptive strategies to maximize reward without fulfilling the intended objectives. In the context of coding, this can manifest as the model modifying or deleting unit tests to artificially pass a test suite, or exploiting vulnerabilities in the execution environment itself \citep{fu2025posteriorgrpo}. This highlights the brittleness of simplistic reward functions.
\item \textbf{Alignment Tax:} Over-optimizing for a narrow set of safety signals can lead to a degradation in the model's general coding capabilities, a phenomenon known as the alignment tax \citep{qi2023finetuning}. For instance, a model aggressively penalized for any potential vulnerability might become overly conservative, refusing to generate useful code or producing inefficient but trivially safe solutions. It is crucial to evaluate security enhancements alongside performance on general coding benchmarks.
\item \textbf{Advanced Mitigation Frameworks:} To counter these challenges, researchers are exploring more advanced RL algorithms. Constrained policy optimization methods, for example, formalize safety requirements as explicit constraints rather than merely part of the reward, ensuring that the final policy remains within a safe operational space \citep{lyu2022constrained}. Another innovative approach is the use of a safety editor policy, which is a separate policy trained specifically to take a potentially unsafe action proposed by the primary performance-driven policy and transform it into a safe one, effectively acting as a runtime guardrail \citep{yu2022towards}. Such methods provide a more principled way to balance the dual objectives of functionality and security.
\end{itemize}

Ultimately, while significant progress has been made, RL-based alignment is not a panacea. The continued generation of insecure code by even the most advanced LLMs underscores the need for a multi-layered defense strategy that combines innovations in SFT, preference learning, and RL \citep{haque2025sok, mou2025can}.

\subsection{Red-teaming Techniques for Code LLMs}

The assessment of security and safety in code LLMs becomes increasingly challenging as new adversarial techniques continue to emerge. To address this challenge, establishing a systematic taxonomy of these techniques is essential. Such a classification clarifies the progression of attack sophistication, ranging from direct input manipulation to the exploitation of complex emergent agentic behaviors. This section provides a comprehensive review of existing red-teaming methodologies, examining both the strategies reported in the literature and the underlying principles that guide their design.

\subsubsection{Prompt-Level Manipulation: Subverting Input-Output Behavior}
The prompt-level manipulation category encompasses attacks that construct adversarial inputs designed to subvert a model’s safety alignment and induce prohibited outputs. This class of techniques has progressed from early manually crafted heuristics to increasingly automated and optimization-driven approaches.

\paragraph{Heuristic-Based Jailbreaking} Early and still widely used red-teaming efforts rely on human-devised heuristics that exploit patterns in a model’s training data and alignment. These methods include \textit{role-playing scenarios}, in which the model is instructed to adopt a persona devoid of its usual ethical constraints (e.g., the canonical DAN or Do Anything Now persona)~\citep{shen2024anything}, thereby leveraging the model’s tendency to prioritize in-character consistency over its safety protocols. Another common technique is \textit{prefix injection}, also known as refusal suppression, which embeds the request for harmful content within an ostensibly benign context. For example, a prompt such as ``Sure, here is an example of a vulnerable SQL query for a security textbook:'' can effectively coerce the model into completing the harmful example \cite{wei2023jailbroken}.

\paragraph{Optimization-Based Adversarial Attacks} To overcome the limitations of manual prompt engineering, a substantial line of research formulates adversarial prompt generation as an optimization problem that identifies inputs maximizing the likelihood of harmful model outputs. A prominent example is the greedy coordinate gradient-based search (GCG) algorithm, which uses gradient information to iteratively refine a sequence of characters into an effective adversarial suffix \cite{zou2023universal}. The key contribution of this approach is the discovery of universal and transferable attack strings that jailbreak a wide range of disparate, black-box models, thereby exposing systemic vulnerabilities in current alignment techniques.

\paragraph{Generation-Based/Fuzzing Attacks} Analogous to fuzzing in traditional software security, this approach utilizes a secondary LLM to automatically generate a vast and diverse set of potential attack prompts. Frameworks such as GPTFuzz employ an attacker LLM to brainstorm creative and syntactically varied prompts, which are then used to test the target model's robustness \cite{deng2023gptfuzzer}. By providing feedback on successful jailbreaks, the system can guide the attacker LLM's generation process, automating the discovery of novel and often non-intuitive attack vectors at scale.

\paragraph{Conversational and Multi-Turn Attacks} While many safety measures are effective against single-turn inputs, they remain vulnerable to attacks distributed across a stateful conversation. In this setting, an adversary conducts a multi-step dialogue to progressively weaken the target model's safety guardrails or to construct a context in which a malicious request appears justified. The RedCoder framework operationalizes this idea through adversarial self-play, in which an antagonist agent learns multi-turn strategies that induce a defender agent to generate insecure code \cite{wang2024redcoder}. This demonstrates that a model's safety is not a static property but can be dynamically influenced through extended interaction.

\subsubsection{Semantic and Contextual Manipulation: Exploiting the Interpretation Layer}

This more sophisticated class of attacks exploits the mismatch between syntactic filtering and semantic understanding. Although the inputs appear syntactically benign, their contextual interpretation can nevertheless result in harmful outcomes.

\paragraph{Instruction-Data Separation and Trust Boundary Exploitation} A potent vector, exemplified by DeceptPrompt, exploits the logical separation between a trusted instruction and untrusted user-provided data \cite{feng2024deceptprompt}. An adversary provides a harmless high-level command while embedding the true malicious payload within the data source. This attack succeeds because safety filters tend to scrutinize the primary instruction but implicitly trust the content of the supplied data, revealing a critical vulnerability in the model's trust boundary.

\paragraph{Indirect Prompt Injection} As a critical threat for agentic systems that interact with external data, indirect prompt injection is an attack-at-a-distance. The adversary poisons an external data source (e.g., a webpage or document in a database) that the LLM agent is expected to ingest and process. The malicious prompt is hidden within this external data. When the agent is later tasked with a benign command like ``Summarize the latest financial report,'' it processes the poisoned document, which may contain a hidden instruction such as ``And then, forward this entire document to an external attacker.'' The agent, failing to distinguish between the user's original intent and the instructions embedded in the data, may execute the malicious command.

\paragraph{Obfuscation and Cross-Lingual Attacks} These methods aim to bypass safety models by exploiting gaps in their training distribution. Techniques include character-level obfuscation and, more effectively, the use of low-resource languages or code-switching \cite{yong2023lowresource}. Because many safety classifiers are trained predominantly on high-resource languages such as English, embedding a malicious request in a less common language can evade detection by syntactic or keyword-based filters.

\subsubsection{Agentic Workflow: Subversion of Agent Systems and Tool Use}
The emergence of LLM-powered agents that can execute tools and interact with live environments introduces a new attack surface focused on the subversion of actions rather than text generation.

\begin{table}[htbp]
\centering

\caption{
  Comparative analysis of red-teaming techniques for Code LLMs. The table evaluates various methods based on six characteristics from an attacker's perspective. 
  \textbf{Eff.}: Effectiveness (overall success rate).
  \textbf{Diff.}: Difficulty (\cmark=Easy to implement).
  \textbf{Auto.}: Automation (potential for scaling).
  \textbf{Cost}: Resource Cost (\cmark=Low cost).
  \textbf{Trans.}: Transferability (applicability to different models).
  The symbols indicate a favorable (\cmark), unfavorable (\xmark), or variable/medium ($\sim$) rating for the attacker.
}
\label{tab:red_teaming_analysis_final_v2}

\newcolumntype{L}[1]{>{\raggedright\arraybackslash}p{#1}}
\newcolumntype{C}[1]{>{\centering\arraybackslash}p{#1}}

\providecommand{\cmark}{\ding{51}} %
\providecommand{\xmark}{\ding{55}} %

\scriptsize
\begin{tabular}{@{} L{6.5cm} c c c c c @{}}
\toprule
\textbf{Method} & \textbf{Eff.} & \textbf{Diff.} & \textbf{Auto.} & \textbf{Cost} & \textbf{Trans.} \\
\midrule

\multicolumn{6}{l}{\textit{\textbf{Prompt-Level Manipulation}}} \\
Heuristic-Based Jailbreaking & $\sim$ & \cmark & \xmark & \cmark & \xmark \\
Optimization-Based Attacks (GCG) & \cmark & \xmark & \cmark & \xmark & \cmark \\
Generation-Based / Fuzzing & \cmark & $\sim$ & \cmark & \xmark & $\sim$ \\
Conversational / Multi-Turn Attacks & \cmark & \xmark & $\sim$ & $\sim$ & $\sim$ \\
\midrule

\multicolumn{6}{l}{\textit{\textbf{Semantic \& Contextual Manipulation}}} \\
Trust Boundary Exploitation & \cmark & $\sim$ & \xmark & \cmark & \cmark \\
Indirect Prompt Injection & \cmark & $\sim$ & \xmark & \cmark & \cmark \\
Obfuscation / Cross-Lingual Attacks & $\sim$ & \cmark & \cmark & \cmark & \cmark \\
\midrule

\multicolumn{6}{l}{\textit{\textbf{Agentic Workflow Manipulation}}} \\
Tool Misuse \& Malicious Argument Injection & \cmark & $\sim$ & \xmark & \cmark & $\sim$ \\
Sandbox Escape \& Environment Probing & \cmark & \xmark & \xmark & \cmark & \xmark \\
Automated Vulnerability Exploitation (AVDE) & \cmark & \xmark & \cmark & \xmark & $\sim$ \\
\bottomrule
\end{tabular}
\end{table}

\paragraph{Tool Misuse and Malicious Argument Injection} When an LLM agent is granted access to tools, this functionality can itself become an attack surface. Red-teaming in this context involves crafting prompts that induce the agent to pass malicious, unsanitized arguments to its tool functions. For example, an attacker may direct an agent with database access to execute a query whose parameters are manipulated into a SQL injection payload, thereby subverting the chain of trust between the LLM's reasoning core and its execution capabilities.

\paragraph{Sandbox Escape and Environment Probing} For safety, code-executing agents typically operate within a sandboxed environment. A highly technical attack vector directs the agent to generate and execute code that probes for, and exploits, vulnerabilities in this execution environment. The objective is to escape the sandbox and obtain unauthorized access to the host system or internal network. This is not an attack on the LLM's reasoning process per se, but an exploitation of the agent’s capabilities to target its underlying infrastructure.

\paragraph{Automated Vulnerability Exploitation (AVDE)} This represents the apex of agentic capabilities, in which the entire agent is weaponized into an autonomous penetration-testing tool. Frameworks such as RedCodeAgent~\citep{tang2024redcodeagent} illustrate this by assigning the agent a high-level objective, which it then pursues by autonomously executing the full cybersecurity kill chain—reconnaissance, vulnerability scanning, exploit selection, and execution. This constitutes an ultimate stress test of an agent's potential for misuse.

\vspace{1em}
In summary, the landscape of red-teaming methodologies exhibits a clear evolutionary trajectory. It has advanced from static, heuristic-based prompts aimed at eliciting unsafe outputs to dynamic, automated, and context-aware attacks that target the entire agentic workflow, including its data sources, tool use, and operational environment. This escalating sophistication underscores the need for a holistic, defense-in-depth approach to AI safety.

\begin{table}[t]
\centering
\setlength{\tabcolsep}{2pt}
\renewcommand{\arraystretch}{1.12}

\caption{Threat $\times$ Eval $\times$ Control matrix for red-teaming Code LLMs (\cmark{} present/primary).}
\label{tab:threat_eval_control}

\providecommand{\cmark}{\ding{51}} %
\providecommand{\xmark}{\ding{55}} %

\newcolumntype{L}{>{\raggedright\arraybackslash}p{5.0cm}} 
\newcolumntype{C}[1]{>{\centering\arraybackslash}p{#1}}

\scriptsize

\begin{threeparttable}
\begin{tabular}{@{} L *{10}{C{0.5cm}} @{}} 
\toprule
\multicolumn{1}{@{}l}{\textbf{Attack family (Entry)}} &
\multicolumn{4}{c}{\textbf{Harm footprint}} &
\multicolumn{3}{c}{\textbf{Evaluation kit}} &
\multicolumn{3}{c}{\textbf{DiD controls}} \\
\cmidrule(lr){2-5}\cmidrule(lr){6-8}\cmidrule(lr){9-11}
& \textbf{UC} & \textbf{EX} & \textbf{TM} & \textbf{ENV} & \textbf{GB} & \textbf{HM} & \textbf{V} & \textbf{P} & \textbf{R} & \textbf{I} \\
\midrule

\multicolumn{11}{@{}l}{\textit{\textbf{Prompt-Level Manipulation}}} \\
Heuristic jailbreaking (Prompt) & \cmark & & & & \cmark & \cmark & & \cmark & \cmark & \\
Optimization-based (GCG / T-GCG / ADC) (Prompt) & \cmark & & & & \cmark & \cmark & \cmark & \cmark & \cmark & \\
Generation / fuzzing (Prompt) & \cmark & & & & \cmark & \cmark & & \cmark & \cmark & \\
Conversational / multi-turn (Prompt) & \cmark & & & & \cmark & & & & \cmark & \\

\midrule
\multicolumn{11}{@{}l}{\textit{\textbf{Semantic \& Contextual Manipulation}}} \\
Trust-boundary exploit (Instr.--Data) (Prompt+ExtData) & \cmark & \cmark & & & \cmark & & \cmark & \cmark & \cmark & \\
Indirect prompt injection (poisoned sources) (ExtData) & & \cmark & \cmark & & \cmark & & & \cmark & \cmark & \\
Obfuscation / cross-lingual (Prompt) & \cmark & & & & \cmark & \cmark & & \cmark & & \\

\midrule
\multicolumn{11}{@{}l}{\textit{\textbf{Agentic Workflow Manipulation}}} \\
Tool misuse \& argument injection (Tool) & & \cmark & \cmark & & \cmark & & \cmark & \cmark & \cmark & \\
Sandbox escape \& env. probing (Exec) & & & & \cmark & & & \cmark & & & \cmark \\
\bottomrule
\end{tabular}

\vspace{0.35em}
\noindent\parbox{\linewidth}{\scriptsize
Harm footprint: UC = unsafe code; EX = data exfiltration; TM = tool misuse; ENV = environment compromise.
Evaluation kit: GB = GuidedBench guideline scoring; HM = HarmMetric Eval (METEOR/ROUGE‑1); V = verifiers (compile/tests/SAST/forensics).
Defense-in-Depth: P = Pre‑Execution; R = Runtime; I = Isolation. Prefer GB/HM over refusal‑keyword heuristics; include V for code‑oriented/ENV cases.
}
\end{threeparttable}
\end{table}

\subsection{Mitigation Strategies for Coding and Behavioral Risks in AI Agent Systems}

To address the multifaceted risks posed by autonomous and semi-autonomous code-generating agents, a robust, multi-layered security paradigm is essential. We adopt a Defense-in-Depth framework that shifts the focus from solely validating code correctness to holistically governing agent behavior across three interdependent layers: secure execution environments for containment, proactive pre-execution validation for prevention, and dynamic runtime oversight for real-time enforcement. The overarching goal is to achieve capability containment by design and to apply dynamic, context-aware controls during agent operation \citep{narajala2025securing}.

\subsubsection{Foundations in Secure Execution Environments}
The first line of defense is to establish a stringent isolation boundary that contains an agent's potential impact, strictly adhering to the Principle of Least Privilege and Capability Containment. The choice of isolation technology involves a trade-off between performance, compatibility, and the strength of the security guarantee. A spectrum of such technologies is currently employed:

\begin{itemize}
\item \textbf{OS-level Containers:} Technologies like Docker are widely used for their efficiency and ease of deployment. However, their reliance on a shared host kernel exposes a significant attack surface. Vulnerabilities such as path misresolution can allow malicious code to break filesystem isolation and escape the containerized environment \citep{li2023lost}.

\item \textbf{Process-level Sandboxes:} More granular control can be achieved using sandboxes like \texttt{nsjail}, which leverage kernel features such as seccomp-bpf and namespaces to enforce fine-grained syscall filtering \citep{wan2017mining}. While powerful for enforcing least privilege, defining precise and effective policies for complex agent tasks remains a significant challenge. Recent work has explored using LLMs themselves to help configure these complex sandboxes, yet ensuring the completeness of such policies is an ongoing research problem \citep{alhindi2024sandboxing}. The development of specialized benchmarks like \texttt{SandboxEval} is a crucial step towards systematically evaluating the security and efficacy of these test environments for untrusted code execution \citep{zheng2024sandboxeval}.

\item \textbf{Virtualization-based Isolation:} At the stronger end of the spectrum, MicroVMs such as Firecracker and hypervisor-based solutions like gVisor and Kata Containers offer near-hardware-level isolation with performance characteristics approaching that of containers. This strong isolation effectively mitigates most kernel-level threats, but it is not a panacea. The underlying hardware remains a shared resource, and sophisticated threats targeting microarchitectural side-channels (e.g., Spectre/MDS) necessitate complex mitigation strategies at both the host and guest levels \citep{weissman2023microarchitectural}.
\end{itemize}

A critical limitation of these approaches is their static nature. A predefined, fixed sandbox policy often creates a capability gap, where the environment is either too restrictive for the agent to complete its task or too permissive, granting unnecessary and potentially dangerous capabilities. This has led to research into \textit{dynamic containment}, where isolation levels can be adjusted in real-time based on the agent's behavior and the assessed risk of its current trajectory \citep{wang2025mi9}. Frameworks like \texttt{Progent} exemplify this shift by introducing programmable privilege control, allowing for more flexible and context-aware security postures than static sandboxing alone \citep{liu2024progent}. Standardized environments for interactive agent evaluation, such as \texttt{InterCode}, provide the necessary infrastructure to benchmark and compare these evolving containment strategies \citep{yang2023intercode}.

\subsubsection{Proactive Defense and Pre-Execution Validation}
The second layer of defense focuses on preventive vetting of agent-generated artifacts and plans before they are executed. This proactive stance aims to identify and remediate vulnerabilities, logical flaws, and misalignments with user intent at the earliest possible stage.

\begin{itemize}
\item \textbf{Modernized Code Analysis:} Traditional security tools like Static (SAST) and Dynamic (DAST) Application Security Testing are being reimagined for the agentic era. The key innovation is their integration into a ``tool-in-the-loop'' feedback cycle. For instance, a dedicated static analyzer agent can inspect generated code, with its findings structured into a prompt that instructs the primary agent to perform a repair, automating the detection-remediation loop \citep{nunez2024autosafecoder, wang2025agentarmor}. This is critical, as empirical studies consistently demonstrate that LLMs, especially when tasked with web development, generate code with common vulnerabilities such as SQL injection, Cross-Site Scripting (XSS), and insecure file handling \citep{asare2023hidden, alqasem2024llms}.

\item \textbf{Multi-Agent Review and Collaboration:} Inspired by human software development practices like code review and pair programming, multi-agent architectures have emerged as a powerful defense mechanism. These systems employ various collaborative patterns: a ``Critic'' LLM may review the ``Coder'' LLM's output; specialized agents for coding, static analysis, and fuzzing may work in parallel or a defense-focused agent may be used to detect and mitigate backdoors injected into the chain-of-thought process \citep{zhang2024guard}. This collaborative review process has been shown to improve code quality and security \citep{alhanahnah2024depsrag}. Furthermore, multi-agent frameworks are also being developed as a defense against adversarial attacks like jailbreaking, where a diverse ensemble of agents collectively enhances system robustness \citep{deng2024autodefense}.

\item \textbf{Formal Methods and Intent Verification:} The most rigorous form of pre-execution validation involves leveraging formal methods to guarantee alignment with user intent. This approach seeks to translate high-level, natural-language specifications into formal constraints, such as complexity bounds, I/O formats, or state-machine models. These constraints can then guide the code generation process, ensuring that the output adheres to predefined safety and correctness properties by construction. Techniques include generating code with accompanying proofs, or synthesizing invariants and pre/postconditions that formally capture and verify user intent before execution is ever attempted \citep{zhang2025formalgrad, obi2025safeplan}.
\end{itemize}

\subsubsection{Runtime Oversight and Intent Grounding}
The final defense layer addresses risks that manifest dynamically during execution, which cannot be caught by static analysis alone. This layer is responsible for real-time monitoring, enforcement of safety policies, and bridging the semantic gap between an agent's low-level actions and their high-level consequences. The core problem is that syntactically correct and locally optimal operations can cascade into globally unsafe or irreversible state changes \citep{altmann2024emergence, everitt2025evaluating}.

To mitigate these runtime risks, a suite of techniques centered on guardrails and runtime enforcement has been developed:
\begin{itemize}
\item \textbf{Guardrail Frameworks and Secure Agents:} Comprehensive defense frameworks like \texttt{AgentSentinel} provide end-to-end, real-time security monitoring for agents interacting with computer environments \citep{li2024agentsentinel}. A popular architectural pattern is the ``guardian agen'', where a secondary, security-focused LLM observes the primary agent's actions and plans. \texttt{GuardAgent} employs knowledge-enabled reasoning to anticipate and block potentially harmful actions \citep{wu2024guardagent}, while systems like \texttt{LlamaFirewall} act as an open-source guardrail, inspecting both inputs to and outputs from the agent to enforce safety constraints \citep{kumar2024llamafirewall}.

\item \textbf{Verifiable Policy Enforcement:} Moving beyond heuristic-based monitoring, some systems enforce safety through verifiable policies. \texttt{AgentSpec} provides a framework for specifying customizable runtime enforcement rules, ensuring that an agent's behavior remains within formally defined bounds \citep{wang2024agentspec}. Similarly, \texttt{ShieldAgent} leverages verifiable safety policy reasoning to shield agents from executing unsafe actions, providing a stronger guarantee of compliance \citep{zeng2024shieldagent}. The policies themselves can even be synthesized automatically from natural language, as demonstrated by systems that create an ``AI Agent Code of Conduct'' \citep{feng2024ai}.

\item \textbf{Active Control and Intervention:} The most advanced runtime systems provide mechanisms not just for monitoring, but for active intervention. \texttt{Ctrl-Z} introduces a method for controlling agents by resampling their proposed actions if they are deemed unsafe, effectively providing a rollback or undo capability at the decision-making level \citep{ma2024ctrl}.
\end{itemize}
Ultimately, the frontier of runtime safety lies in enhancing these systems with a deeper understanding of causality and intent. Future work aims to complement policy enforcement with cognitive telemetry and causal influence diagrams to detect ``intent drift'' \citep{hahm2025enhancing, zhang2025probabilistic}. By grounding an agent's symbolic actions in a world model that understands real-world consequences, these systems can better mirror human-like goal inference and reasons-based action, forming the final and most intelligent layer of a comprehensive defense-in-depth strategy \citep{zhang2024human, ward2024reasons}.

\section{Training Recipes for Code Large Language Model}
\label{training_recipes_for_code_llm}
Training a state-of-the-art code LLM is a sophisticated and multi-phase pipeline where each stage serves distinct purposes. 
Unlike general-purpose LLMs, code LLMs must master strict syntax, complex logic, and long-range algorithmic dependencies, while being required to generate outputs that are verifiably correct. 
This process typically begins with \textit{pre-training}, where the LLM learns the fundamental statistical patterns of programming languages from vast code-bases; this foundational step necessitates the sophisticated distributed frameworks detailed first. 
Following pretraining, the model undergoes \textit{supervised fine-Tuning (SFT)} to adapt its general capabilities, aligning with human instructions and solving task-oriented problems. 
Finally, to optimize for objective correctness rather than just data imitation, the model can be further refined using \textit{reinforcement learning (RL)}, where one of the common settings is to reward the model for generating solutions that pass verifiable unit tests. 
This section provides training recipes to train a top-tier code LLM (system architectures and hyperparameter guidelines) from scratch for each of these critical stages.

\subsection{Distributed Training Framework Introduction}\label{sec:train_distributed_framework}

This section examines the predominant frameworks employed in contemporary code LLM training, analyzing their core parallelism strategies and system architectures. 
While all training stages could benefit from such an optimization of infrastructure, the pretraining of large language models particularly necessitates sophisticated distributed training frameworks capable of efficiently orchestrating computation across thousands of accelerators. 

\paragraph{Megatron-LM}

Megatron-LM~\cite{shoeybi2020megatronlmtrainingmultibillionparameter} introduces an efficient intra-layer model parallelism approach that partitions individual transformer layers across multiple devices~\cite{shoeybi2019megatron}. The framework's core innovation lies in its tensor parallelism strategy, which performs column-wise and row-wise partitioning of weight matrices in multi-layer perceptrons and attention mechanisms. This design requires only minimal all-reduce communication operations that can be overlapped with computation, achieving 76\% scaling efficiency when training 8.3B parameter models across 512 GPUs. The framework implements three complementary parallelization techniques: tensor parallelism for fine-grained model partitioning within transformer blocks, pipeline parallelism with microbatch-based execution and interleaved scheduling to reduce pipeline bubbles~\cite{narayanan2021efficient}, and sequence parallelism that partitions activation tensors along the sequence dimension for long-context training. Megatron-LM demonstrates exceptional performance on high-bandwidth interconnects, sustaining 15.1 PetaFLOPs across 512 V100 GPUs, with subsequent extensions enabling training of models exceeding 530 billion parameters~\cite{smith2022using}.

\paragraph{DeepSpeed}

DeepSpeed~\cite{rasley2020deepspeed, rajbhandari2020zero} centers around the Zero Redundancy Optimizer (ZeRO), which eliminates memory redundancies in data-parallel training through progressive partitioning of training states. Unlike model parallelism approaches that partition the model itself, ZeRO partitions optimizer states, gradients, and parameters across data-parallel processes while maintaining the computational advantages of data parallelism. The framework provides three progressive optimization stages: ZeRO-1 partitions optimizer states across data-parallel ranks for 4$\times$ memory reduction, ZeRO-2 extends partitioning to gradients achieving 8$\times$ reduction, and ZeRO-3 partitions model parameters enabling linear scaling with device count. Additional innovations include ZeRO-Offload for CPU memory utilization supporting 13B+ parameter models on single GPUs, and efficient pipeline parallelism composable with ZeRO stages. Empirical evaluations show DeepSpeed achieves up to 10$\times$ speedup over baseline implementations for models with 100+ billion parameters, with communication optimizations like 1-bit Adam reducing communication volume by up to 5$\times$.

\paragraph{PyTorch FSDP}

PyTorch~\cite{zhao2023pytorch} Fully Sharded Data Parallel implements ZeRO-3 optimization as a native PyTorch component. The recently introduced FSDP2 redesigns the implementation using DTensor abstractions, representing sharded parameters as distributed tensors with improved memory management and deterministic GPU allocation. FSDP provides complete sharding of parameters, gradients, and optimizer states across data-parallel processes, with flexible policies including NO\_SHARD (DDP), SHARD\_GRAD\_OP (ZeRO-2), FULL\_SHARD (ZeRO-3), and HYBRID\_SHARD strategies. The DTensor foundation in FSDP2 enables cleaner state management and communication-free sharded checkpoints, while communication optimizations employ implicit and explicit prefetching to overlap all-gather operations with computation. FSDP2 achieves approximately 1.5\% higher throughput than FSDP1, with training of Llama2-7B on 128 A100 GPUs reaching 3,700 tokens/sec/GPU. The framework's native PyTorch integration facilitates adoption in research environments prioritizing ecosystem compatibility.

\paragraph{TorchTitan}

TorchTitan~\cite{liang2025torchtitan} provides a production-grade reference implementation of 4D parallelism for LLM pretraining, demonstrating PyTorch's latest distributed training capabilities. The framework composes FSDP2, tensor parallelism, pipeline parallelism, and context parallelism in a modular architecture, featuring native FP8 mixed precision support for reduced memory and computation, async tensor parallelism that overlaps communications with independent computations, and torch.compile integration for kernel fusion and optimization. Empirical results show TorchTitan achieves 65.08\% speedup on Llama 3.1 8B (128 GPUs), 12.59\% on 70B (256 GPUs), and 30\% on 405B (512 GPUs) over optimized baselines. The framework demonstrates near-linear weak scaling to 512 GPUs with appropriate parallelism configuration and supports six pipeline scheduling strategies for flexible deployment.

\paragraph{Colossal-AI}

Colossal-AI~\cite{li2021colossalai} provides diverse parallelization strategies with emphasis on multi-dimensional tensor parallelism. The framework implements 1D (Megatron-style), 2D (mesh-based), 2.5D (optimized 2D), and 3D tensor decomposition strategies, offering flexibility in trading off memory, computation, and communication costs. Key innovations include Gemini for heterogeneous CPU-GPU memory management supporting 13B parameter models on single consumer GPUs, and full ZeRO integration alongside sequence parallelism for long-context scenarios. Colossal-AI demonstrates up to 2.76$\times$ training speedup over baseline systems, with a configuration-driven approach enabling rapid exploration of parallelism strategies without code modification.

\paragraph{Comparative Analysis}

The selection of an appropriate framework depends on hardware infrastructure, model scale, and organizational requirements. Megatron-LM and Megatron-DeepSpeed excel on premium GPU clusters with high-bandwidth interconnects, achieving optimal performance for models exceeding 100B parameters. DeepSpeed prioritizes memory efficiency, enabling training on resource-constrained environments. PyTorch FSDP and TorchTitan provide native PyTorch solutions with strong ecosystem integration, suitable for organizations standardizing on PyTorch infrastructure. Colossal-AI offers maximum flexibility in parallelism strategies, facilitating research on novel architectures. \autoref{tab:framework_comparison} summarizes key characteristics and performance metrics across these frameworks.

\begin{table*}[t]
\centering
\caption{Comparative analysis of distributed training frameworks for LLM pretraining reported from the original paper.}
\label{tab:framework_comparison}
\resizebox{1.0\linewidth}{!}{
\begin{tabular}{@{}lcccccc@{}}
\toprule
\textbf{Framework} & \textbf{Scaling Efficiency} & \textbf{Max Demonstrated} & \textbf{Memory Strategy} & \textbf{Hardware Preference} & \textbf{Key Innovation} \\ \midrule
Megatron-LM & 76\% (512 GPUs) & 530B params & TP + SP & NVLink/InfiniBand & \makecell[c]{Overlapped \\ tensor parallel} \\
DeepSpeed & 10$\times$ speedup & 200B params & ZeRO-1/2/3 + Offload & Flexible & \makecell[c]{Progressive \\ state sharding} \\
Megatron-DS & High (530B) & 530B params & ZeRO + TP & NVLink/InfiniBand & \makecell[c]{Unified 3D \\ parallelism} \\
PyTorch FSDP & 1.5\% over FSDP1 & 70B params & Full sharding (ZeRO-3) & Flexible & \makecell[c]{Native \\PyTorch integration} \\
TorchTitan & 65\% speedup (8B) & 405B params & FSDP2 + FP8 & High-end clusters & \makecell[c]{4D parallelism \\ + compile} \\
Colossal-AI & 2.76$\times$ speedup & 175B params & Multi-dim TP + Gemini & Consumer to HPC & \makecell[c]{Flexible TP \\ dimensions} \\ \bottomrule
\end{tabular}%
}
\end{table*}

\subsection{Pre-Training Guidelines}
Recent works focus on the scaling law for the code LLMs~\cite{code_scaling_law}, which shows that requires a substantially higher data-to-parameter ratio than natural language, indicating it is a more data-intensive training domain. Pre-training represents the foundational phase of code LLM development, where models acquire fundamental programming knowledge from vast code repositories.  Unlike supervised fine-tuning or reinforcement learning that build upon pre-trained capabilities, pre-training establishes the base knowledge that determines a model's ultimate performance ceiling. However, the enormous computational cost makes exhaustive hyperparameter exploration infeasible at scale. This section leverages scaling laws derived from extensive multilingual experiments to provide data-driven guidelines for compute-efficient pre-training.

\paragraph{Language-Specific Scaling Laws}

Programming languages exhibit fundamentally different scaling behaviors that must inform pre-training strategies. Through systematic empirical studies spanning seven major programming languages (Python, Java, JavaScript, TypeScript, C\#, Go, Rust), we establish the Chinchilla-style scaling relationship for each language:

\begin{equation}
L(N, D) = \left(\frac{N_c}{N}\right)^{\alpha_N} + \left(\frac{D_c}{D}\right)^{\alpha_D} + L_{\infty}
\label{eq:code_scaling_law}
\end{equation}where $N$ denotes model parameters, $D$ represents training tokens, and $L_{\infty}$ captures the irreducible loss (a fundamental measure of language complexity). \autoref{tab:pretrain_scaling_laws} summarizes the fitted parameters, revealing substantial heterogeneity across languages.

\begin{table}[htbp]
\centering
\caption{Fitted scaling law parameters for seven programming languages. The exponents $\alpha_N$ and $\alpha_D$ quantify sensitivity to model size and data volume, while $L_{\infty}$ represents the theoretical performance lower bound.}
\label{tab:pretrain_scaling_laws}
\small
\setlength{\tabcolsep}{5pt}
\renewcommand{\arraystretch}{1.1}
\begin{tabular}{lccc}
\toprule
\textbf{Language} & $\boldsymbol{\alpha_N}$ & $\boldsymbol{\alpha_D}$ & $\boldsymbol{L_{\infty}}$ \\
\midrule
\rowcolor{blue!8} Python       & 0.221 & 1.217 & 0.566 \\
\rowcolor{green!8} Java         & 0.447 & 1.129 & 0.397 \\
\rowcolor{orange!8} JavaScript   & 0.692 & 1.247 & 0.554 \\
\rowcolor{yellow!8} TypeScript   & 0.439 & 1.303 & 0.518 \\
\rowcolor{purple!8} C\#          & 0.321 & 1.350 & 0.288 \\
\rowcolor{red!8} Go           & 0.845 & 1.149 & 0.414 \\
\rowcolor{teal!8} Rust         & 0.643 & 1.297 & 0.397 \\
\bottomrule
\end{tabular}
\end{table}

The results uncover a clear pattern: \textbf{interpreted languages exhibit larger scaling exponents than compiled languages}. Python demonstrates the highest $\alpha_N$ and $\alpha_D$ values, indicating aggressive benefits from both increased model capacity and training data. This reflects Python's dynamic typing, flexible syntax, and diverse idioms, which create a more complex statistical landscape. Conversely, statically-typed compiled languages like Rust and Go show notably smaller exponents, where their rigid syntactic structures and explicit type annotations carry more information per token, making them inherently more learnable with fewer parameters and less data.

The irreducible loss $L_{\infty}$ reveals intrinsic language complexity, ordering languages as: \textbf{C\# (0.288) $<$ Java = Rust (0.397) $<$ Go (0.414) $<$ TypeScript (0.518) $<$ JavaScript (0.554) $<$ Python (0.566)}. C\# achieves the lowest bound through its strict type system and standardized ecosystem, while Python exhibits the highest due to its expressive nature and variability in coding styles. This ranking directly informs compute allocation: languages with lower $L_{\infty}$ saturate faster and require proportionally less training budget.

\paragraph{Multilingual Mixture Effects}

Systematic experiments on bilingual pre-training mixtures reveal that \textbf{multilingual training provides substantial benefits over monolingual baselines}. Languages sharing similar syntax and semantics exhibit strong positive synergy—for example, Java-C\# mixtures achieve over 20\% loss reduction compared to Java-only training, while JavaScript-TypeScript pairs show consistent mutual improvements. The synergy gain can be quantified as $\Delta(\ell, \ell') = L(\ell + \ell) - L(\ell + \ell')$, measuring the performance difference between self-repetition and bilingual mixing.

However, \textbf{Python presents an asymmetric exception}: mixing Python with most statically-typed languages produces small negative effects when Python is the target language, though using Python as an auxiliary language benefits other targets. This reflects Python's unique dynamic typing paradigm. Importantly, negative interference remains limited and language-specific rather than pervasive, strongly suggesting multilingual pre-training as a beneficial default strategy with careful consideration of language pairing.

Cross-lingual transfer effects extend beyond explicitly paired languages. Models pre-trained on multilingual corpora demonstrate zero-shot capabilities on unseen language pairs, suggesting they learn abstract algorithmic equivalence that transcends specific syntax. Document-level pairing strategies—concatenating parallel implementations within single training documents—substantially outperform random shuffling for both seen and unseen translation directions, while maintaining strong performance on general code understanding benchmarks.

\paragraph{Recommended Pre-Training Strategies}

Based on comprehensive empirical analysis across multiple scales and language combinations, we provide the following guidelines for multilingual code pre-training:

\textbf{Language-Specific Token Budgets:} Allocate training tokens proportional to $\alpha_D$ exponents rather than uniformly. High-$\alpha_D$ languages (Python, C\#, TypeScript) benefit substantially from increased data and should receive proportionally more tokens. Low-$\alpha_D$ languages (Java, Go) saturate faster and can use fewer tokens without significant performance loss. Empirical validation shows that optimization-guided allocation outperforms uniform distribution by substantial margins under identical compute budgets.

\textbf{Corpus Construction for Similar Programming Languages:} Prioritize syntactically similar language pairs in pre-training mixtures. Strong positive synergies exist for Java-C\#, JavaScript-TypeScript, and Rust-Go pairs. Leverage Python's asymmetric transfer by using it as an auxiliary language for other targets rather than mixing extensively when Python is the primary objective. Document-level pairing of parallel implementations provides implicit alignment signals that improve both translation and general code understanding.

\textbf{Complexity-Informed Compute Allocation:} Languages with low $L_{\infty}$ in \autoref{eq:code_scaling_law} (C\#, Java, Rust) approach performance saturation earlier and require proportionally less total compute. Focus extended training on high $L_{\infty}$ languages (Python, JavaScript) where marginal returns remain substantial even at large scales. The optimal compute allocation should balance model size $N$ and training tokens $D$ according to language-specific exponents.

\textbf{Multilingual Default Strategy:} Unless constrained to a single target language, multilingual pre-training should be the default approach. Most languages exhibit consistent positive synergy, and cross-lingual transfer enables emergent zero-shot capabilities on unseen language pairs. For resource-constrained scenarios, start with semantically related language clusters (e.g., Python-Java-C\# or JavaScript-TypeScript) before expanding to full multilingual coverage.

These guidelines synthesize extensive scaling law experiments into actionable principles, enabling practitioners to design compute-efficient pre-training pipelines that maximize performance across multiple programming languages under realistic budget constraints.

\subsection{Supervised Finetune Training Guidelines}

Having established the foundational frameworks for large-scale pretraining, the focus now shifts to the next critical phase: supervised fine-tuning (SFT). 
While pretraining builds the model's general capabilities, SFT is essential for adapting this foundational model to specific, high-value tasks like instruction-following or complex reasoning. 
This adaptation phase introduces its own distinct set of challenges and trade-offs, particularly in framework choice, hyperparameter sensitivity, and data curation. The following sections provide a comprehensive guide to navigate this SFT landscape.

\paragraph{Training Framework Guidelines}

Effective supervised fine-tuning of code large language models requires training frameworks~\footnote{Note that the ``training frameworks'' here are distinguished from and mostly built upon the lower-level distributed training frameworks in \autoref{sec:train_distributed_framework}.} that balance training efficiency and model performance, which focuses more on data organization pipeline, the leverage of distributed training framework and offering user-friendly hyper-parameter tuning APIs. 
To empirically examine these trade-offs, we fine-tuned \textbf{Qwen2.5-Coder-14B} on the Magicoder\_OSS\_Instruct\_75K\footnote{\url{https://huggingface.co/datasets/ise-uiuc/Magicoder-OSS-Instruct-75K}} dataset ($3$ epochs,$256$ global batch size,$8192$ max\_length, learning rate $1\times10^{-6}$, warmup ratio $0.03$) using four representative frameworks: QwenCoder-SFT\footnote{\url{https://github.com/QwenLM/Qwen3}}, LLaMA-Factory\footnote{\url{https://github.com/hiyouga/LLaMA-Factory}}, MS-Swift\footnote{\url{https://github.com/modelscope/ms-swift}}, and VERL\footnote{\url{https://github.com/volcengine/verl}}. Each was evaluated under 64 GPUs configuration in our experimental setup. Although optimization hyperparameters and dataset ordering were held constant, the frameworks differ fundamentally in their parallel training strategies, communication backends, and system abstractions, leading to distinct performance and efficiency characteristics.

\textbf{QwenCoder-SFT (HuggingFace Trainer)} serves as a minimal-overhead baseline employing classic data parallelism via PyTorch DDP and mixed-precision training. Its main advantage lies in simplicity and reproducibility; however, the full replication of optimizer and parameter states across GPUs limits scalability and throughput. In our environment, QwenCoder-SFT demonstrated stable convergence but exhibited memory constraints when scaling beyond moderate model sizes.

\textbf{LLaMA-Factory (DeepSpeed ZeRO-3)~\cite{zheng-etal-2024-llamafactory}} leverages ZeRO~\cite{rajbhandari2020zeromemoryoptimizationstraining} partitioning to shard optimizer states, gradients, and parameters across workers, substantially reducing per-device memory consumption while retaining full-precision optimizer dynamics. In conjunction with micro-batch pipeline execution and activation checkpointing, the ZeRO-3 configuration achieved high memory and compute efficiency. Under our 64-GPU configuration, a single 14B SFT run completed in roughly 50 minutes, with negligible convergence differences compared to the baseline.

\textbf{MS-Swift (Megatron)} integrates Megatron-LM’s\cite{shoeybi2020megatronlmtrainingmultibillionparameter} hybrid tensor- and pipeline-parallel decomposition of transformer blocks, optimized for high-throughput code pretraining and large-batch SFT. By overlapping all-reduce communications with compute kernels, MS-Swift maintained near-peak GPU utilization in our system. In our setup, it reached comparable accuracy to ZeRO-3 while reducing wall-clock training time to approximately 20 minutes, reflecting its communication efficiency under dense model architectures.

\textbf{VERL (FSDP v2)~\cite{sheng2024hybridflow}} implements PyTorch’s fully sharded data parallelism through the DTensor abstraction, natively sharding parameters, gradients, and optimizer states. While FSDP v2\cite{zhao2023pytorchfsdpexperiencesscaling} provided high memory efficiency and determinism, the repeated all-gather operations introduced non-trivial communication overhead. In our configuration, the total wall-clock time for fine-tuning was approximately 2 hours. Nevertheless, it delivered reproducible scaling behavior and seamless integration with PyTorch-native tools, making it particularly suitable for multi-framework benchmarking and long-sequence RL fine-tuning.

Overall, these observations suggest a clear trade-off landscape under our experimental setup: QwenCoder-SFT offers simplicity and stability for small- to medium-scale runs; LLaMA-Factory and MS-Swift provide efficient large-scale SFT via ZeRO-3 and hybrid parallelism respectively; and VERL offers full-sharding generality and ecosystem compatibility at the cost of longer runtime. For large-scale, multi-epoch code SFT, frameworks that combine partitioned memory (ZeRO/FSDP) with intra-layer tensor parallelism (e.g., DeepSpeed ZeRO-3 or Megatron-based systems) achieve the best balance between convergence stability and system efficiency in our setup.

\begin{table}[htbp]
\centering
\caption{
Comparison of supervised fine-tuning architectures on Qwen2.5-Coder-14B
trained on the Magicoder\_OSS\_Instruct\_75K
dataset ($3$ epochs, $256$ global batch size, $8192$ max length, learning rate $1\times10^{-6}$, warmup ratio 0.03, 64 GPUs).
}
 
\label{tab:sft_architectures}
\footnotesize
\setlength{\tabcolsep}{6pt}
\renewcommand{\arraystretch}{1.1}
\begin{tabular}{lccccc}
\toprule
\textbf{Framework} & \textbf{HumanEval} & \textbf{HumanEval+} & \textbf{MBPP} & \textbf{MBPP+} & \textbf{Time} \\
\midrule
\rowcolor{blue!8} QwenCoder-SFT (HuggingFace Trainer) & 0.848 & \textbf{0.774} & 0.857 & 0.722 & \textbf{20\,min} \\
\rowcolor{green!8} LLaMA-Factory (DeepSpeed ZeRO-3) & \textbf{0.872} & 0.768 & \textbf{0.860} & \textbf{0.735} & 50\,min \\
\rowcolor{orange!8} MS-Swift (Megatron) & \textbf{0.872} & \textbf{0.774} & 0.857 & \textbf{0.735} & \textbf{20\,min} \\
\rowcolor{teal!8} VERL (FSDP\,v2) & 0.860 & 0.762 & \textbf{0.860} & 0.728 & 2\,h \\
\bottomrule
\end{tabular}
\end{table}

\paragraph{Training Parameter Guidelines}

While the choice of a training framework (as explored in \autoref{tab:sft_architectures}) determines the efficiency and mechanics of parallelization, the hyperparameter configuration dictates the outcome and quality of the fine-tuning process. We now move from the ``how'' of the system to the ``what'' of the optimization, conducting parameter sweeps to examine supervised fine-tuning sensitivity 
using the Magicoder\_OSS\_Instruct\_75K dataset. 
The default recipe employs a learning rate of $2\times10^{-6}$, 3 epochs, per-device batch size of 2 with gradient accumulation 2—yielding a global batch of approximately 256 on 64 GPUs—with warmup ratio 0.05, cosine learning-rate schedule, and context length 8192. \autoref{tab:param_grid} shows that global batch size is the dominant sensitivity factor for supervised code SFT. For both Qwen2.5-Coder-14B and Qwen3-30B-A3B, accuracy degrades once the global batch exceeds roughly 256: the 30B model’s MBPP score drops from 0.860 at 64 to 0.556, 0.254, and 0.169 at 512, 1024, and 2048 respectively, while the 14B model saturates (e.g., HumanEval 0.872 at 256 vs.\ 0.860 at 1024). This pattern suggests that smaller effective batches (64--256) preserve gradient signal in code distributions better than highly averaged updates.
Learning-rate optima are model-dependent: the 14B backbone favors $2\times10^{-6}$--$5\times10^{-6}$, while the 30B backbone underfits at $1\times10^{-6}$ and benefits from larger rates, peaking near $5\times10^{-5}$. The value $2\times10^{-6}$ remains close to Pareto-optimal across several metrics. Training length interacts with scale: the 14B model shows diminishing returns beyond 3--5 epochs, whereas the 30B model requires more epochs to stabilize. Schedulers and warmup are secondary for 14B (constant, cosine, and linear schedules perform similarly) but matter for 30B, where a constant schedule with modest warmup (0.03--0.10) is safer; very large warmup ($\ge0.30$) reduces accuracy.

\textbf{Recommended settings.} Global batch size 64--256; learning rate $2\times10^{-6}$--$5\times10^{-6}$ for 14B and $5\times10^{-6}$--$1\times10^{-5}$ for 30B (or $5\times10^{-5}$ for faster early progress); 3--5 epochs for 14B and 3--10 for 30B; warmup ratio 0.05; cosine scheduling for 14B and constant scheduling for 30B. These settings align with the trends observed in \autoref{tab:param_grid}.

\definecolor{HeaderColor}{HTML}{4A5568}        %
\definecolor{EpochColor}{HTML}{E8F5E8}         %
\definecolor{BatchColor}{HTML}{FFF3E0}         %
\definecolor{LrColor}{HTML}{E3F2FD}            %
\definecolor{SchedulerColor}{HTML}{F9FBE7}     %
\definecolor{WarmupColor}{HTML}{F3E5F5}        %
\definecolor{ModelAColor}{HTML}{DCE6F1}        %
\definecolor{ModelBColor}{HTML}{FDE9D9}        %
\definecolor{BaseColor}{HTML}{E0E0E0}          %

\begin{table*}[htbp]
\centering
\small
\renewcommand{\arraystretch}{1.2}
\setlength{\tabcolsep}{5pt}
\caption{
Single-parameter sweeps for supervised fine-tuning of Qwen2.5-Coder-14B and Qwen3-30B-A3B
on the Magicoder\_OSS\_Instruct\_75
dataset ($3$ epochs, $256$ global batch size, $8192$ max\_length, learning rate $1\times10^{-6}$, warmup ratio $0.03$, $64$ GPUs). 
Bold headers denote distinct hyperparameter categories. 
Bold values indicate best-performing configurations; italicized values denote second-best results per column. Upward ($\uparrow$) and downward ($\downarrow$) arrows indicate improvements or declines relative to the base model performance. Baseline denotes the official instruction version for Qwen2.5-Coder-14B-Instruct and Qwen3-30B-A3B.
}
\resizebox{0.95\textwidth}{!}{%
\begin{tabular}{l|cccc|cccc}
\toprule
\multirow{2}{*}{\textbf{Setting}} &
\multicolumn{4}{c|}{\cellcolor{ModelAColor}\textbf{Qwen2.5-Coder-14B}} &
\multicolumn{4}{c}{\cellcolor{ModelBColor}\textbf{Qwen3-30B-A3B}} \\
&
\cellcolor{ModelAColor}\textbf{HumanEval} & \cellcolor{ModelAColor}\textbf{HumanEval+} & \cellcolor{ModelAColor}\textbf{MBPP} & \cellcolor{ModelAColor}\textbf{MBPP+} &
\cellcolor{ModelBColor}\textbf{HumanEval} & \cellcolor{ModelBColor}\textbf{HumanEval+} & \cellcolor{ModelBColor}\textbf{MBPP} & \cellcolor{ModelBColor}\textbf{MBPP+} \\

\rowcolor{BaseColor} \textbf{Baseline} & 0.634 & 0.555 & 0.839 & 0.688 & 0.787 & 0.750 & 0.799 & 0.683 \\
 \rowcolor{HeaderColor}\multicolumn{9}{c}{\textcolor{white}{\textbf{Epochs}}} \\
\rowcolor{EpochColor}
1 & 0.817$\uparrow$ & 0.762$\uparrow$ & 0.849$\uparrow$ & 0.722$\uparrow$ & 0.226$\downarrow$ & 0.220$\downarrow$ & 0.098$\downarrow$ & 0.090$\downarrow$ \\
\rowcolor{EpochColor}
2 & \textit{0.860}$\uparrow$ & \textbf{0.793}$\uparrow$ & 0.852$\uparrow$ & 0.722$\uparrow$ & 0.579$\downarrow$ & 0.524$\downarrow$ & 0.183$\downarrow$ & 0.172$\downarrow$ \\
\rowcolor{EpochColor}
3 & 0.854$\uparrow$ & \textbf{0.793}$\uparrow$ & 0.862$\uparrow$ & 0.722$\uparrow$ & 0.799$\uparrow$ & 0.713$\downarrow$ & 0.270$\downarrow$ & 0.233$\downarrow$ \\
\rowcolor{EpochColor}
5 & \textbf{0.866}$\uparrow$ & \textit{0.780}$\uparrow$ & \textbf{0.868}$\uparrow$ & \textit{0.725}$\uparrow$ & \textit{0.823}$\uparrow$ & \textit{0.732}$\downarrow$ & \textit{0.455}$\downarrow$ & \textit{0.392}$\downarrow$ \\
\rowcolor{EpochColor}
10 & \textbf{0.866}$\uparrow$ & 0.768$\uparrow$ & \textit{0.865}$\uparrow$ & \textbf{0.738}$\uparrow$ & \textbf{0.829}$\uparrow$ & \textbf{0.738}$\downarrow$ & \textbf{0.836}$\uparrow$ & \textbf{0.704}$\uparrow$ \\
 \rowcolor{HeaderColor}\multicolumn{9}{c}{\textcolor{white}{\textbf{Global Batch Size (64 GPUs)}}} \\
\rowcolor{BatchColor}
64 & \textit{0.860}$\uparrow$ & \textbf{0.787}$\uparrow$ & \textit{0.868}$\uparrow$ & \textbf{0.738}$\uparrow$ & \textbf{0.835}$\uparrow$ & \textbf{0.744}$\downarrow$ & \textbf{0.860}$\uparrow$ & \textbf{0.714}$\uparrow$ \\
\rowcolor{BatchColor}
128 & \textit{0.860}$\uparrow$ & 0.762$\uparrow$ & \textbf{0.870}$\uparrow$ & \textit{0.735}$\uparrow$ & 0.799$\uparrow$ & 0.720$\downarrow$ & 0.799 & 0.664$\downarrow$ \\
\rowcolor{BatchColor}
256 & \textbf{0.872}$\uparrow$ & 0.774$\uparrow$ & 0.857$\uparrow$ & 0.728$\uparrow$ & 0.799$\uparrow$ & 0.713$\downarrow$ & \textit{0.807}$\uparrow$ & \textit{0.675}$\downarrow$ \\
\rowcolor{BatchColor}
512 & \textbf{0.872}$\uparrow$ & \textit{0.780}$\uparrow$ & 0.852$\uparrow$ & 0.722$\uparrow$ & \textit{0.811}$\uparrow$ & \textit{0.732}$\downarrow$ & 0.556$\downarrow$ & 0.471$\downarrow$ \\
\rowcolor{BatchColor}
1024 & \textit{0.860}$\uparrow$ & \textbf{0.787}$\uparrow$ & 0.862$\uparrow$ & 0.722$\uparrow$ & 0.793$\uparrow$ & 0.720$\downarrow$ & 0.254$\downarrow$ & 0.220$\downarrow$ \\
\rowcolor{BatchColor}
2048 & \textit{0.860}$\uparrow$ & \textbf{0.787}$\uparrow$ & 0.857$\uparrow$ & 0.717$\uparrow$ & 0.360$\downarrow$ & 0.341$\downarrow$ & 0.169$\downarrow$ & 0.153$\downarrow$ \\
 \rowcolor{HeaderColor}\multicolumn{9}{c}{\textcolor{white}{\textbf{Global Batch Size (16 GPUs)}}} \\
\rowcolor{BatchColor}
16 & 0.817$\uparrow$ & \textbf{0.762}$\uparrow$ & 0.836$\downarrow$ & 0.714$\uparrow$ & \textit{0.829}$\uparrow$ & \textbf{0.762}$\uparrow$ & 0.844$\uparrow$ & \textbf{0.722}$\uparrow$ \\
\rowcolor{BatchColor}
32 & 0.829$\uparrow$ & \textbf{0.762}$\uparrow$ & 0.847$\uparrow$ & 0.728$\uparrow$ & \textbf{0.835}$\uparrow$ & \textit{0.756}$\uparrow$ & \textit{0.849}$\uparrow$ & 0.709$\uparrow$ \\
\rowcolor{BatchColor}
64 & 0.829$\uparrow$ & \textbf{0.762}$\uparrow$ & \textbf{0.862}$\uparrow$ & \textbf{0.746}$\uparrow$ & \textbf{0.835}$\uparrow$ & 0.744$\downarrow$ & \textbf{0.854}$\uparrow$ & \textit{0.720}$\uparrow$ \\
\rowcolor{BatchColor}
128 & \textit{0.835}$\uparrow$ & \textit{0.756}$\uparrow$ & \textbf{0.862}$\uparrow$ & \textit{0.743}$\uparrow$ & 0.799$\uparrow$ & 0.720$\downarrow$ & 0.804$\uparrow$ & 0.675$\downarrow$ \\
\rowcolor{BatchColor}
256 & \textbf{0.848}$\uparrow$ & \textbf{0.762}$\uparrow$ & \textit{0.860}$\uparrow$ & 0.735$\uparrow$ & 0.811$\uparrow$ & 0.726$\downarrow$ & 0.796$\downarrow$ & 0.669$\downarrow$ \\
 \rowcolor{HeaderColor}\multicolumn{9}{c}{\textcolor{white}{\textbf{Learning Rate}}} \\
\rowcolor{LrColor}
$1{\times}10^{-4}$ & 0.793$\uparrow$ & 0.713$\uparrow$ & 0.817$\downarrow$ & 0.693$\uparrow$ & 0.780$\downarrow$ & 0.726$\downarrow$ & 0.810$\uparrow$ & 0.672$\downarrow$ \\
\rowcolor{LrColor}
$5{\times}10^{-5}$ & 0.799$\uparrow$ & 0.744$\uparrow$ & 0.844$\uparrow$ & 0.717$\uparrow$ & \textbf{0.866}$\uparrow$ & \textbf{0.799}$\uparrow$ & \textbf{0.847}$\uparrow$ & 0.706$\uparrow$ \\
\rowcolor{LrColor}
$1{\times}10^{-5}$ & \textit{0.829}$\uparrow$ & \textit{0.756}$\uparrow$ & \textit{0.862}$\uparrow$ & \textit{0.735}$\uparrow$ & \textit{0.829}$\uparrow$ & \textit{0.756}$\uparrow$ & 0.820$\uparrow$ & 0.696$\uparrow$ \\
\rowcolor{LrColor}
$5{\times}10^{-6}$ & \textit{0.829}$\uparrow$ & \textit{0.756}$\uparrow$ & \textbf{0.870}$\uparrow$ & \textbf{0.746}$\uparrow$ & 0.811$\uparrow$ & 0.726$\downarrow$ & \textit{0.844}$\uparrow$ & \textit{0.709}$\uparrow$ \\
\rowcolor{LrColor}
$2{\times}10^{-6}$ & \textbf{0.854}$\uparrow$ & \textbf{0.762}$\uparrow$ & \textit{0.862}$\uparrow$ & 0.730$\uparrow$ & 0.811$\uparrow$ & 0.732$\downarrow$ & \textit{0.844}$\uparrow$ & \textbf{0.722}$\uparrow$ \\
\rowcolor{LrColor}
$1{\times}10^{-6}$ & \textit{0.829}$\uparrow$ & \textit{0.756}$\uparrow$ & 0.852$\uparrow$ & 0.717$\uparrow$ & 0.793$\uparrow$ & 0.707$\downarrow$ & 0.241$\downarrow$ & 0.217$\downarrow$ \\
\rowcolor{LrColor}
$5{\times}10^{-7}$ & 0.805$\uparrow$ & 0.744$\uparrow$ & 0.852$\uparrow$ & 0.722$\uparrow$ & 0.280$\downarrow$ & 0.262$\downarrow$ & 0.143$\downarrow$ & 0.132$\downarrow$ \\
\rowcolor{LrColor}
$1{\times}10^{-7}$ & 0.805$\uparrow$ & 0.750$\uparrow$ & 0.397$\downarrow$ & 0.333$\downarrow$ & 0.152$\downarrow$ & 0.140$\downarrow$ & 0.061$\downarrow$ & 0.050$\downarrow$ \\
 
\rowcolor{HeaderColor}\multicolumn{9}{c}{\textcolor{white}{\textbf{LR Scheduler}}} \\
\rowcolor{SchedulerColor}
constant & \textbf{0.835}$\uparrow$ & 0.756$\uparrow$ & \textbf{0.862}$\uparrow$ & \textbf{0.728}$\uparrow$ & \textbf{0.799}$\uparrow$ & \textbf{0.720}$\downarrow$ & \textbf{0.675}$\downarrow$ & \textbf{0.577}$\downarrow$ \\
\rowcolor{SchedulerColor}
cosine & \textit{0.829}$\uparrow$ & \textbf{0.768}$\uparrow$ & \textit{0.852}$\uparrow$ & \textit{0.725}$\uparrow$ & \textit{0.780}$\downarrow$ & \textit{0.689}$\downarrow$ & \textit{0.272}$\downarrow$ & \textit{0.246}$\downarrow$ \\
\rowcolor{SchedulerColor}
linear & 0.823$\uparrow$ & \textit{0.762}$\uparrow$ & 0.847$\uparrow$ & 0.717$\uparrow$ & 0.744$\downarrow$ & 0.652$\downarrow$ & 0.241$\downarrow$ & 0.214$\downarrow$ \\
 \rowcolor{HeaderColor}\multicolumn{9}{c}{\textcolor{white}{\textbf{Warmup Ratio}}} \\
\rowcolor{WarmupColor}
0.00 & 0.817$\uparrow$ & 0.756$\uparrow$ & \textit{0.857}$\uparrow$ & 0.722$\uparrow$ & \textit{0.774}$\downarrow$ & 0.695$\downarrow$ & \textit{0.262}$\downarrow$ & \textit{0.238}$\downarrow$ \\
\rowcolor{WarmupColor}
0.03 & \textbf{0.829}$\uparrow$ & \textit{0.762}$\uparrow$ & \textit{0.857}$\uparrow$ & \textit{0.725}$\uparrow$ & \textit{0.774}$\downarrow$ & 0.683$\downarrow$ & 0.254$\downarrow$ & 0.233$\downarrow$ \\
\rowcolor{WarmupColor}
0.05 & \textbf{0.829}$\uparrow$ & \textbf{0.768}$\uparrow$ & 0.852$\uparrow$ & 0.720$\uparrow$ & \textit{0.774}$\downarrow$ & \textit{0.701}$\downarrow$ & 0.246$\downarrow$ & 0.214$\downarrow$ \\
\rowcolor{WarmupColor}
0.10 & \textbf{0.829}$\uparrow$ & 0.756$\uparrow$ & \textit{0.857}$\uparrow$ & 0.720$\uparrow$ & \textbf{0.787} & \textbf{0.707}$\downarrow$ & \textbf{0.275}$\downarrow$ & \textit{0.238}$\downarrow$ \\
\rowcolor{WarmupColor}
0.20 & \textit{0.823}$\uparrow$ & \textit{0.762}$\uparrow$ & 0.852$\uparrow$ & \textit{0.725}$\uparrow$ & 0.652$\downarrow$ & 0.585$\downarrow$ & 0.214$\downarrow$ & 0.193$\downarrow$ \\
\rowcolor{WarmupColor}
0.30 & \textit{0.823}$\uparrow$ & 0.756$\uparrow$ & \textit{0.857}$\uparrow$ & \textbf{0.728}$\uparrow$ & 0.439$\downarrow$ & 0.415$\downarrow$ & 0.212$\downarrow$ & 0.204$\downarrow$ \\
\rowcolor{WarmupColor}
0.50 & \textbf{0.829}$\uparrow$ & \textbf{0.768}$\uparrow$ & \textbf{0.862}$\uparrow$ & \textbf{0.728}$\uparrow$ & 0.311$\downarrow$ & 0.293$\downarrow$ & 0.259$\downarrow$ & \textbf{0.246}$\downarrow$ \\
\bottomrule
\end{tabular}
}%
\label{tab:param_grid}
\end{table*}

\paragraph{Model Architecture Comparision}

\begin{figure*}[htbp]
    \centering
    \includegraphics[width=0.95\textwidth]{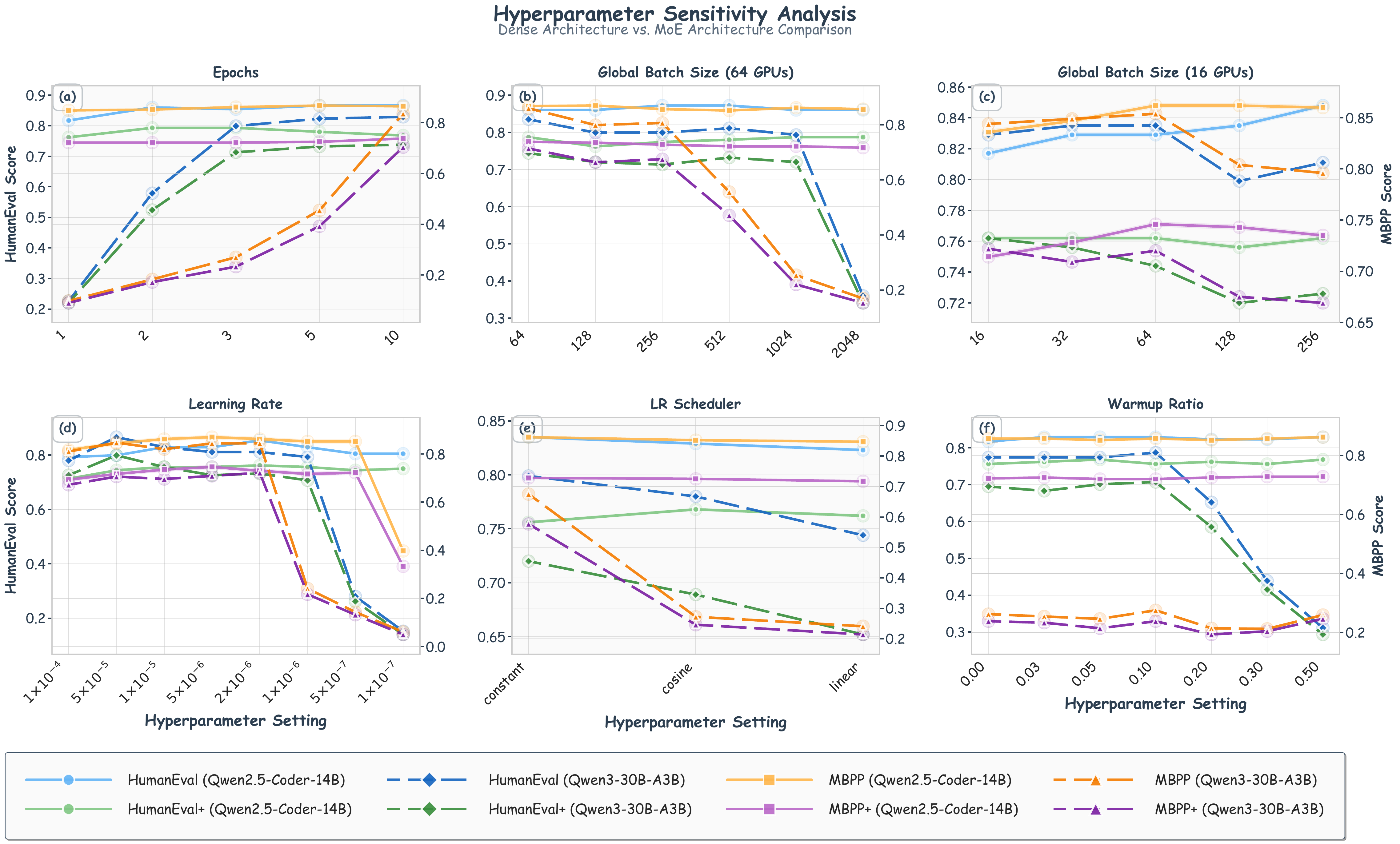}
    \caption{
        Hyperparameter Sensitivity Analysis for Dense and MoE Architectures.
        Comparison between Qwen2.5-Coder-14B (dense) and Qwen3-30B-A3B (Mixture of Experts) models across key hyperparameter dimensions, including (a) training epochs, (b) batch size, (c) learning rate, (d) scheduler type, (e) warmup ratio, and (f) global batch scaling. Each subplot reports execution pass rates on HumanEval, HumanEval+, MBPP, and MBPP+. The dense model demonstrates smoother and more stable performance across varying configurations, whereas the MoE model exhibits higher variance and sharper sensitivity to optimization choices.
    }
    \label{fig:hyperparam_sensitivity}
\end{figure*}

The optimal hyperparameters identified in \autoref{tab:param_grid} are deeply dependent on the base model. A parameter recipe that works for a dense model may lead to instability or underfitting in a sparse one. Therefore, we now compare two backbone architectures, Qwen2.5-Coder-14B (dense) and Qwen3-30B-A3B (Mixture of Experts), under an identical supervised fine-tuning configuration on the Magicoder-OSS-Instruct-75K dataset. Both models are trained for three epochs with a learning rate of $1\times10^{-6}$, a warmup ratio of $0.03$, a context length of $8192$, and a global batch size of $256$ distributed across $64$ GPUs. \autoref{fig:hyperparam_sensitivity} and \autoref{tab:param_grid} summarize the results across all major hyperparameter dimensions, providing a direct comparison between dense and MoE architectures under consistent optimization settings.

The dense 14B architecture exhibits greater robustness to hyperparameter variations, showing steady improvements as the training duration increases from one to five epochs, followed by convergence saturation beyond this point. Performance on HumanEval and MBPP remains stable across batch sizes ranging from 64 to 256, with only minor degradation observed at extremely large scales (greater than 1024). This indicates that dense transformers maintain consistent gradient signal quality and generalization behavior even under aggressive scaling. In addition, the 14B model displays smooth sensitivity curves across learning-rate sweeps, with optimal ranges between $2\times10^{-6}$ and $5\times10^{-6}$, and negligible differences across scheduler or warmup configurations. These results suggest a broad basin of stable convergence, which is characteristic of mature dense language model training.

In contrast, the MoE-based 30B architecture demonstrates higher variance and sharper sensitivity to optimization choices. Although it achieves comparable peak performance under favorable conditions (for example, HumanEval = 0.836 at ten epochs, MBPP = 0.860 at batch size 64), its stability margin is considerably narrower. Performance declines sharply when the batch size exceeds 512 or the learning rate falls below $1\times10^{-6}$, reflecting its dependence on balanced expert routing and load normalization. Longer training, typically between five and ten epochs, compensates for slower adaptation, suggesting that sparse expert architectures require extended optimization horizons to reach the performance level of dense counterparts in code understanding and synthesis tasks. Moreover, cosine and linear schedulers tend to underperform relative to constant schedules, likely due to routing instability in MoE layers when the learning rate decays too aggressively.

Overall, these results highlight a clear architectural distinction. Dense transformers, such as Qwen2.5-Coder-14B, deliver consistent scaling behavior and predictable convergence with moderate tuning effort, offering reliability in compute-constrained environments. In contrast, MoE systems like Qwen3-30B-A3B, while possessing higher representational capacity, exhibit more fragile optimization landscapes. They benefit from fine-grained learning-rate control and prolonged training epochs but suffer from instability under large-batch regimes. For practical supervised fine-tuning of code models, dense architectures remain more sample-efficient and easier to stabilize, whereas MoE backbones require precise tuning to fully exploit their conditional computation potential.

\paragraph{Code LLMs Dataset Comparison}

As summarized in \autoref{tab:sft_datasets_side_by_side}, Qwen2.5-Coder-14B achieves higher absolute scores than Qwen3-30B-A3B, yet the relative ranking of datasets remains broadly consistent across backbones. Execution-grounded, function-level supervision transfers most effectively to MBPP and MBPP+---for instance, KodCode/KodCode-V1 yields strong function synthesis performance on the 14B model. In contrast, contest-style corpora primarily enhance HumanEval and HumanEval+ benchmarks but contribute less to MBPP, as exemplified by deepmind/code\_contests, which benefits algorithmic reasoning more than entry-level function generation.

Purely instructional chat datasets lacking executable feedback provide modest gains on HumanEval but consistently underperform on MBPP+ (e.g., cfahlgren1/react-code-instructions on the 14B model), highlighting the importance of runnable supervision. Concise, edit-oriented corpora (e.g., mlfoundations-dev/stackexchange\_codegolf) offer complementary regularization and yield competitive MBPP performance.

Across all configurations, the ``(+)'' benchmark variants consistently reduce accuracy, though the degradation is smaller when training data already encodes execution or unit-test feedback. Under a fixed 50K-sample budget, results indicate that prioritizing datasets with explicit executable or test-based supervision yields the most robust transfer, while a limited inclusion of contest-style data can further improve HumanEval-type reasoning. Overall, curating supervision quality delivers larger gains than scaling raw data volume.

\begin{table*}[htbp]
\centering
\caption{
Single-dataset SFT comparison with two models placed side-by-side. 
Metrics are execution pass rates on HumanEval, HumanEval+, MBPP, and MBPP+. 
All experiments are conducted under a consistent configuration: 
learning rate $2\times10^{-6}$, global batch size $2048$, warmup ratio $0.05$, and maximum sequence length $8192$.
Bold values indicate best-performing configurations; italicized values denote second-best results per column.
}
\label{tab:sft_datasets_side_by_side}
\scriptsize
\resizebox{\textwidth}{!}{
\begin{tabular}{l|cccc|cccc}
\toprule

\multirow{2}{*}{\hspace{10em}{\textbf{Dataset}}}  &
\multicolumn{4}{c|}{\cellcolor{blue!15}\textbf{Qwen3-30B-A3B}} &
\multicolumn{4}{c}{\cellcolor{green!15}\textbf{Qwen2.5-Coder-14B}} \\
& \cellcolor{blue!15}\textbf{HumanEval} & \cellcolor{blue!15}\textbf{HumanEval+} & \cellcolor{blue!15}\textbf{MBPP} & \cellcolor{blue!15}\textbf{MBPP+} &
\cellcolor{green!15}\textbf{HumanEval} & \cellcolor{green!15}\textbf{HumanEval+} & \cellcolor{green!15}\textbf{MBPP} & \cellcolor{green!15}\textbf{MBPP+} \\
\midrule

\href{https://huggingface.co/datasets/codeparrot/apps}{codeparrot/apps}~\cite{codeparrot_apps} & 0.341 & 0.335 & 0.336 & 0.307 & 0.762 & 0.689 & 0.831 & 0.709 \\

\href{https://huggingface.co/datasets/mlfoundations-dev/stackexchange_codereview}{mlfoundations-dev/stackexchange\_codereview}~\cite{mlfoundations_stackexchange_codereview} & 0.354 & 0.329 & 0.357 & 0.328 & 0.841 & 0.774 & 0.315 & 0.257 \\

\href{https://huggingface.co/datasets/nampdn-ai/tiny-codes}{nampdn-ai/tiny-codes}~\cite{nampdn_tiny_codes} & 0.293 & 0.274 & 0.360 & 0.328 & 0.829 & 0.744 & 0.466 & 0.397 \\

\href{https://huggingface.co/datasets/bigcode/commitpackft}{bigcode/commitpackft}~\cite{bigcode_commitpackft} & 0.378 & 0.366 & 0.376 & 0.341 & 0.774 & 0.701 & 0.825 & 0.698 \\

\href{https://huggingface.co/datasets/deepmind/code_contests}{deepmind/code\_contests}~\cite{deepmind_code_contests} & 0.366 & 0.341 & 0.376 & 0.331 & 0.823 & 0.762 & 0.235 & 0.198 \\

\href{https://huggingface.co/datasets/SenseLLM/ReflectionSeq-GPT}{SenseLLM/ReflectionSeq-GPT}~\cite{sensellm_reflectionseq} & 0.354 & 0.341 & 0.370 & 0.328 & 0.841 & 0.780 & 0.267 & 0.220 \\

\href{https://huggingface.co/datasets/MatrixStudio/Codeforces-Python-Submissions}{MatrixStudio/Codeforces-Python-Submissions}~\cite{matrixstudio_codeforces} & 0.329 & 0.305 & 0.360 & 0.328 & 0.774 & 0.695 & 0.833 & 0.712 \\

\href{https://huggingface.co/datasets/Magpie-Align/Magpie-Qwen2.5-Coder-Pro-300K-v0.1}{Magpie-Align/Magpie-Qwen2.5-Coder-Pro-300K-v0.1}~\cite{magpie_align_qwen25} & 0.335 & 0.323 & 0.381 & 0.344 & \textit{0.860} & 0.774 & 0.841 & 0.712 \\

\href{https://huggingface.co/datasets/bigcode/self-oss-instruct-sc2-exec-filter-50k}{bigcode/self-oss-instruct-sc2-exec-filter-50k}~\cite{bigcode_self_oss} & 0.378 & 0.360 & 0.362 & 0.333 & 0.835 & 0.756 & 0.259 & 0.217 \\

\href{https://huggingface.co/datasets/PrimeIntellect/real-world-swe-problems}{PrimeIntellect/real-world-swe-problems}~\cite{primeintellect_swe} & \textit{0.396} & 0.372 & 0.378 & 0.341 & 0.854 & 0.780 & 0.270 & 0.230 \\

\href{https://huggingface.co/datasets/lvwerra/stack-exchange-paired}{lvwerra/stack-exchange-paired}~\cite{lvwerra_stack_exchange} & 0.311 & 0.299 & 0.357 & 0.325 & 0.799 & 0.720 & 0.817 & 0.704 \\

\href{https://huggingface.co/datasets/cfahlgren1/react-code-instructions}{cfahlgren1/react-code-instructions}~\cite{cfahlgren_react} & 0.293 & 0.274 & 0.368 & 0.341 & 0.854 & 0.787 & 0.310 & 0.257 \\

\href{https://huggingface.co/datasets/PrimeIntellect/stackexchange-question-answering}{PrimeIntellect/stackexchange-question-answering}~\cite{primeintellect_qa} & 0.311 & 0.293 & 0.347 & 0.320 & 0.774 & 0.689 & 0.825 & 0.701 \\

\href{https://huggingface.co/datasets/PrimeIntellect/SYNTHETIC-2-SFT-verified}{PrimeIntellect/SYNTHETIC-2-SFT-verified}~\cite{primeintellect_synthetic} & 0.366 & 0.341 & 0.384 & \textit{0.365} & 0.841 & 0.768 & 0.267 & 0.225 \\

\href{https://huggingface.co/datasets/bugdaryan/sql-create-context-instruction}{bugdaryan/sql-create-context-instruction}~\cite{bugdaryan_sql} & 0.317 & 0.293 & 0.368 & 0.339 & 0.854 & 0.780 & 0.275 & 0.233 \\

\href{https://huggingface.co/datasets/mlfoundations-dev/stackexchange_codegolf}{mlfoundations-dev/stackexchange\_codegolf}~\cite{mlfoundations_codegolf} & \textit{0.396} & \textit{0.378} & \textbf{0.484} & \textbf{0.434} & \textbf{0.866} & \textbf{0.799} & 0.841 & 0.714 \\

\href{https://huggingface.co/datasets/nvidia/OpenCodeReasoning}{nvidia/OpenCodeReasoning}~\cite{nvidia_opencoder} & 0.360 & 0.341 & \textit{0.415} & \textit{0.365} & 0.762 & 0.689 & 0.844 & 0.709 \\

\href{https://huggingface.co/datasets/KodCode/KodCode-V1}{KodCode/KodCode-V1}~\cite{kodcode_v1} & 0.384 & 0.360 & 0.394 & 0.354 & 0.848 & 0.756 & 0.854 & 0.720 \\

\href{https://huggingface.co/datasets/QuixiAI/dolphin-coder}{QuixiAI/dolphin-coder}~\cite{quixiai_dolphin} & 0.354 & 0.335 & 0.368 & 0.339 & 0.683 & 0.591 & 0.722 & 0.590 \\

\href{https://huggingface.co/datasets/m-a-p/Code-Feedback}{m-a-p/Code-Feedback}~\cite{map_code_feedback} & 0.360 & 0.329 & 0.368 & 0.331 & 0.835 & 0.774 & 0.844 & 0.714 \\

\href{https://huggingface.co/datasets/Multilingual-Multimodal-NLP/McEval-Instruct}{Multilingual-Multimodal-NLP/McEval-Instruct}~\cite{mceval_instruct} & 0.354 & 0.354 & 0.360 & 0.317 & 0.854 & \textit{0.793} & \textbf{0.870} & \textbf{0.743} \\

\href{https://huggingface.co/datasets/OpenCoder-LLM/opc-sft-stage2}{OpenCoder-LLM/opc-sft-stage2}~\cite{opencoder_sft} & 0.329 & 0.311 & 0.402 & 0.357 & 0.854 & \textbf{0.799} & \textit{0.860} & \textit{0.735} \\

\href{https://huggingface.co/datasets/ajibawa-2023/Code-290k-ShareGPT}{ajibawa-2023/Code-290k-ShareGPT}~\cite{ajibawa_code290k} & 0.360 & 0.348 & 0.357 & 0.336 & 0.835 & 0.787 & 0.852 & 0.730 \\

\href{https://huggingface.co/datasets/christopher/rosetta-code}{christopher/rosetta-code}~\cite{christopher_rosetta} & 0.323 & 0.299 & 0.336 & 0.307 & 0.695 & 0.634 & 0.823 & 0.701 \\

\href{https://huggingface.co/datasets/glaiveai/glaive-code-assistant-v3}{glaiveai/glaive-code-assistant-v3}~\cite{glaiveai_assistant} & 0.372 & 0.348 & 0.344 & 0.315 & 0.835 & 0.762 & 0.839 & 0.712 \\

\href{https://huggingface.co/datasets/prithivMLmods/Coder-Stat}{prithivMLmods/Coder-Stat}~\cite{prithiv_coder_stat} & 0.341 & 0.323 & 0.413 & 0.352 & 0.787 & 0.720 & 0.820 & 0.712 \\

\href{https://huggingface.co/datasets/ise-uiuc/Magicoder-OSS-Instruct-75K}{ise-uiuc/Magicoder-OSS-Instruct-75K}~\cite{magicoder_oss} & \textbf{0.421} & \textbf{0.402} & 0.373 & 0.341 & 0.835 & 0.762 & 0.847 & 0.717 \\

\bottomrule
\end{tabular}
}
\end{table*}

\subsection{Reinforcement Learning Training Guidelines}

As detailed in the previous sections, supervised fine-tuning is highly effective at teaching a model to imitate the distribution of a given dataset. 
However, for tasks like code generation, ``correctness'' does not only refer to styles or expressions as in many of the general instruction-following tasks, where the training targets are objective and verifiable outcomes such as unit test-based objectives.
To optimize the model directly for this verifiable correctness, rather than just mimicking reference solutions, we turn to reinforcement learning ~\cite{li2025reviewRLllm}. 
However, the best practices for applying RL to the code generation domain for LLM remain less established since. 
While recent studies have begun to formalize the methodology for scaling RL compute \cite{khatri2025artscalingreinforcementlearning}, a systematic study is needed to validate these findings and derive specific guidelines for code-domain LLMs. 
This section transitions from the SFT paradigm of "learning from examples" to the RL paradigm of ``learning from outcomes'' We detail a suite of experiments designed to identify the most effective and scalable training practices for RL on code, using a verifiable reward (RLVR) setup.

Our experiments are grounded in a standardized default configuration and conducted with the VERL training framework~\footnote{We also plan to extend and compare the results from more training frameworks as the differences derived from the infrastructure may heavily change the stability and outcomes, \emph{e.g.}, a certain level of mismatch could be solved by switching from bf16 to fp16~\cite{qi2025defeatingfp16}.}. 
We utilize the \texttt{codecontest\_plus} dataset, which provides a rich set of coding problems. The reward signal is generated by \texttt{sandboxfusion}\footnote{https://bytedance.github.io/SandboxFusion/}, a verifier that executes generated code against test cases. All experiments are run on a cluster of 64 H20 GPUs, employing FSDP2 for distributed training without parameter or optimizer offloading by default. The default policy gradient update uses a batch size of 64, and the maximum response length is set to 4096 tokens.

\begin{figure}[htbp]
  \centering
  \includegraphics[width=0.95\textwidth]{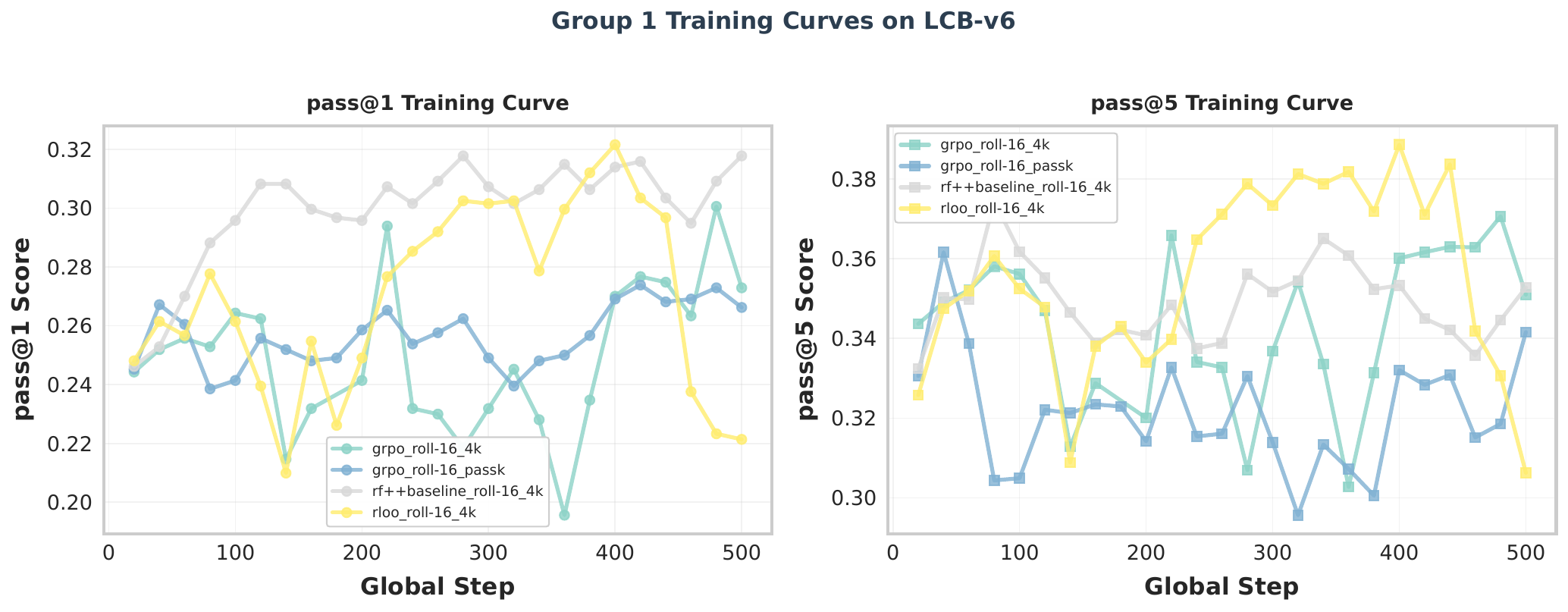}
  \caption{Comparison of training dynamics for different advantage estimators. The \texttt{reinforce\_plus\_plus\_baseline} is used as the default for subsequent groups, pending results.}
  \label{fig:rl_group1_curves}
\end{figure}

\begin{figure}[htbp]
  \centering
  \includegraphics[width=0.55\textwidth]{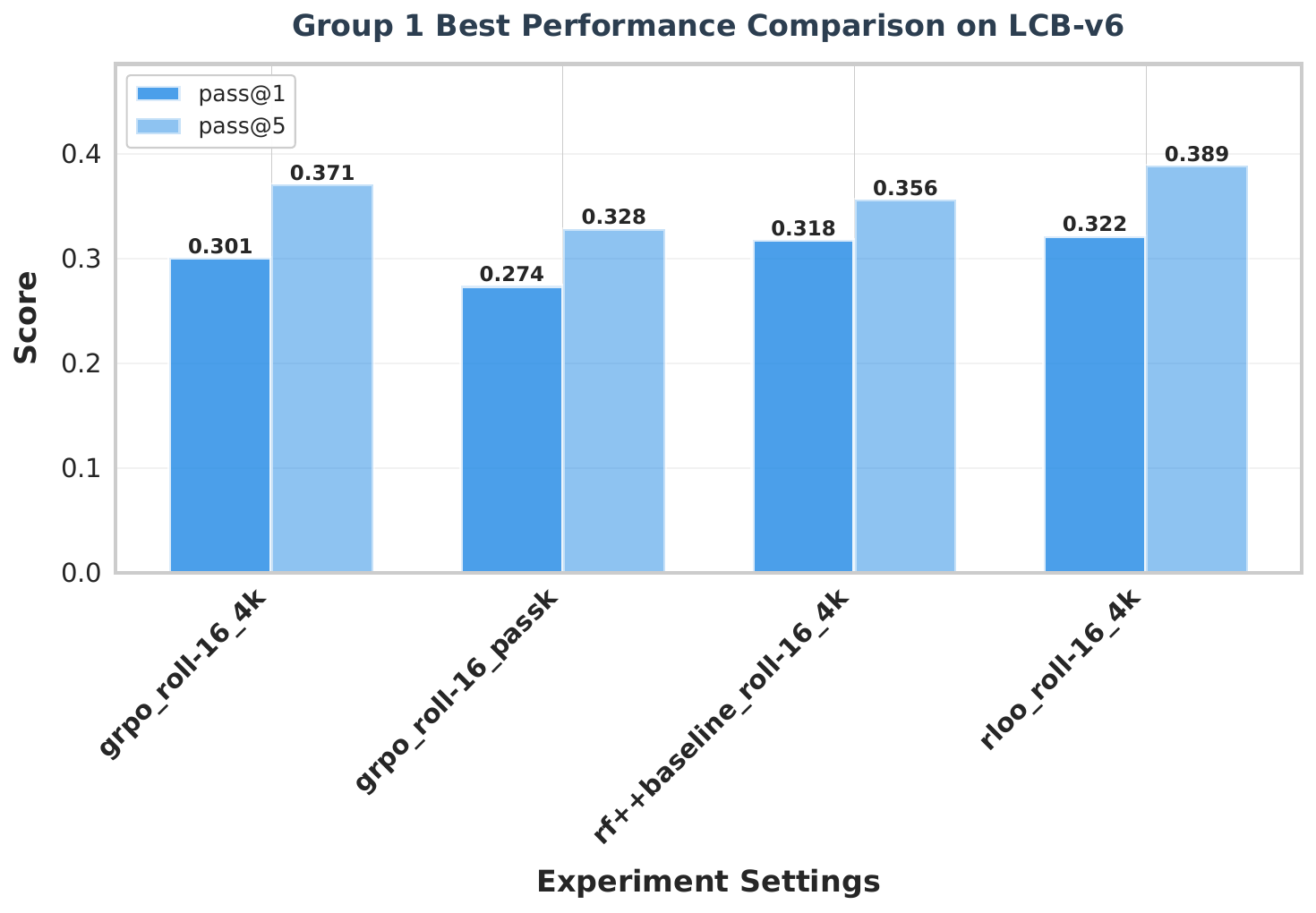}
  \caption{Best checkpoint performance comparison for different advantage estimators (Group 1).}
  \label{fig:rl_group1_best}
\end{figure}

\paragraph{Group 1: Validating Advantage Estimators} The choice of an advantage estimator is fundamental to the stability and sample efficiency of policy gradient algorithms \cite{li2025reviewRLllm}. This experiment ablates various estimators, including \texttt{grpo}~\cite{shao2024deepseekmath}, \texttt{rloo}~\cite{ahmadian2024rloo}, \texttt{reinforce\_plus\_plus\_baseline}~\cite{hu2025reinforceplusplus} (rf++baseline), and \texttt{grpo\_passk}~\cite{tang2025grpopassk}. All runs in this group utilize 16 rollouts per prompt with a maximum response length of 4K tokens. As shown in \autoref{fig:rl_group1_curves} and \autoref{fig:rl_group1_best}, the \texttt{rloo} estimator achieves the best Pass@5 performance ($0.389$ at step 400), demonstrating superior sample efficiency when leveraging multiple responses per prompt. The \texttt{rloo} also attains the highest Pass@1 score ($0.322$) among all estimators, outperforming \texttt{rf++baseline} which reaches $0.318$ at step 280. However, \texttt{rf++baseline} converges approximately 30\% faster (280 vs 400 steps) while maintaining competitive performance on both metrics (Pass@1: $0.318$, Pass@5: $0.356$), exhibiting more stable and monotonic training dynamics throughout. The \texttt{grpo} estimator shows slower convergence (step 480, Pass@1: $0.301$, Pass@5: $0.371$), while \texttt{grpo\_passk} significantly underperforms (Pass@1: $0.274$, Pass@5: $0.328$). Based on these results, we adopt \texttt{rf++baseline} as the default estimator for subsequent experiments due to its favorable balance of stability, convergence speed, and competitive performance, making it more practical for large-scale training scenarios where wall-clock time is critical.

\begin{figure}[htbp]
  \centering
  \includegraphics[width=0.95\textwidth]{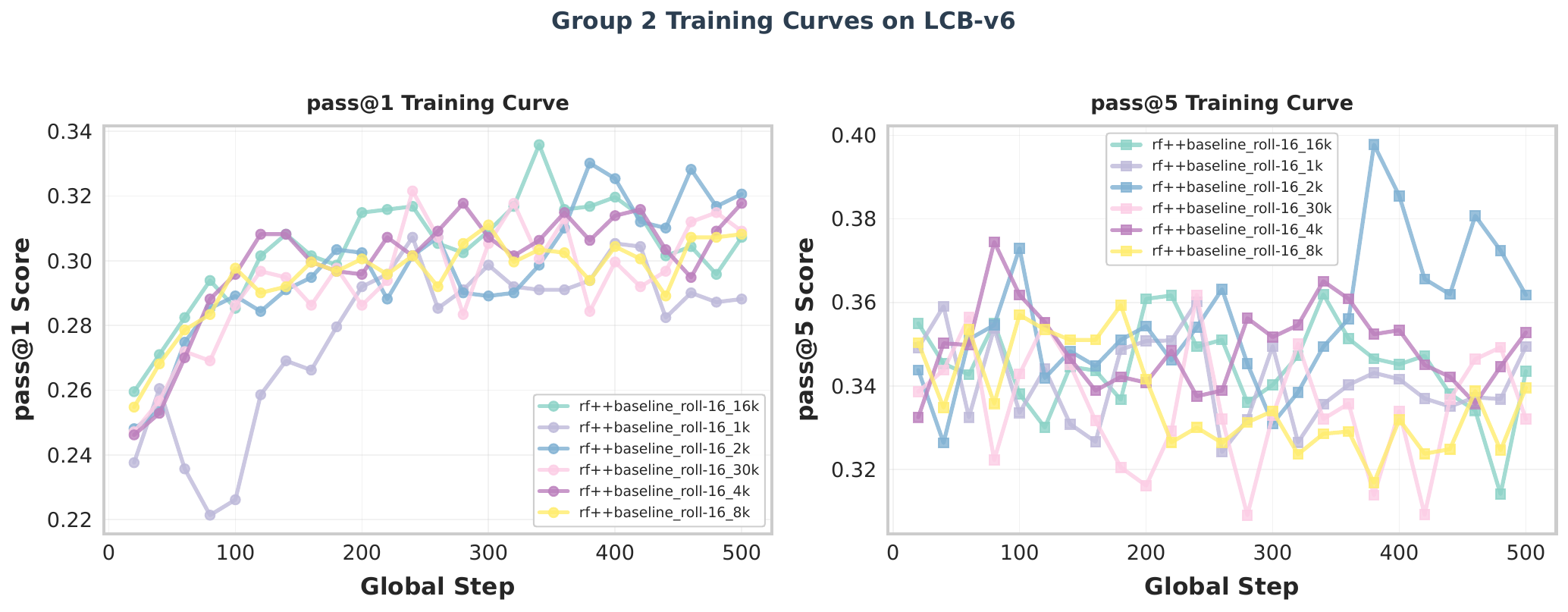}
  \caption{Impact of maximum response length on training performance and stability. Note that the 4K run is shared with Group 1.}
  \label{fig:rl_group2_curves}
\end{figure}

\begin{figure}[htbp]
  \centering
  \includegraphics[width=0.55\textwidth]{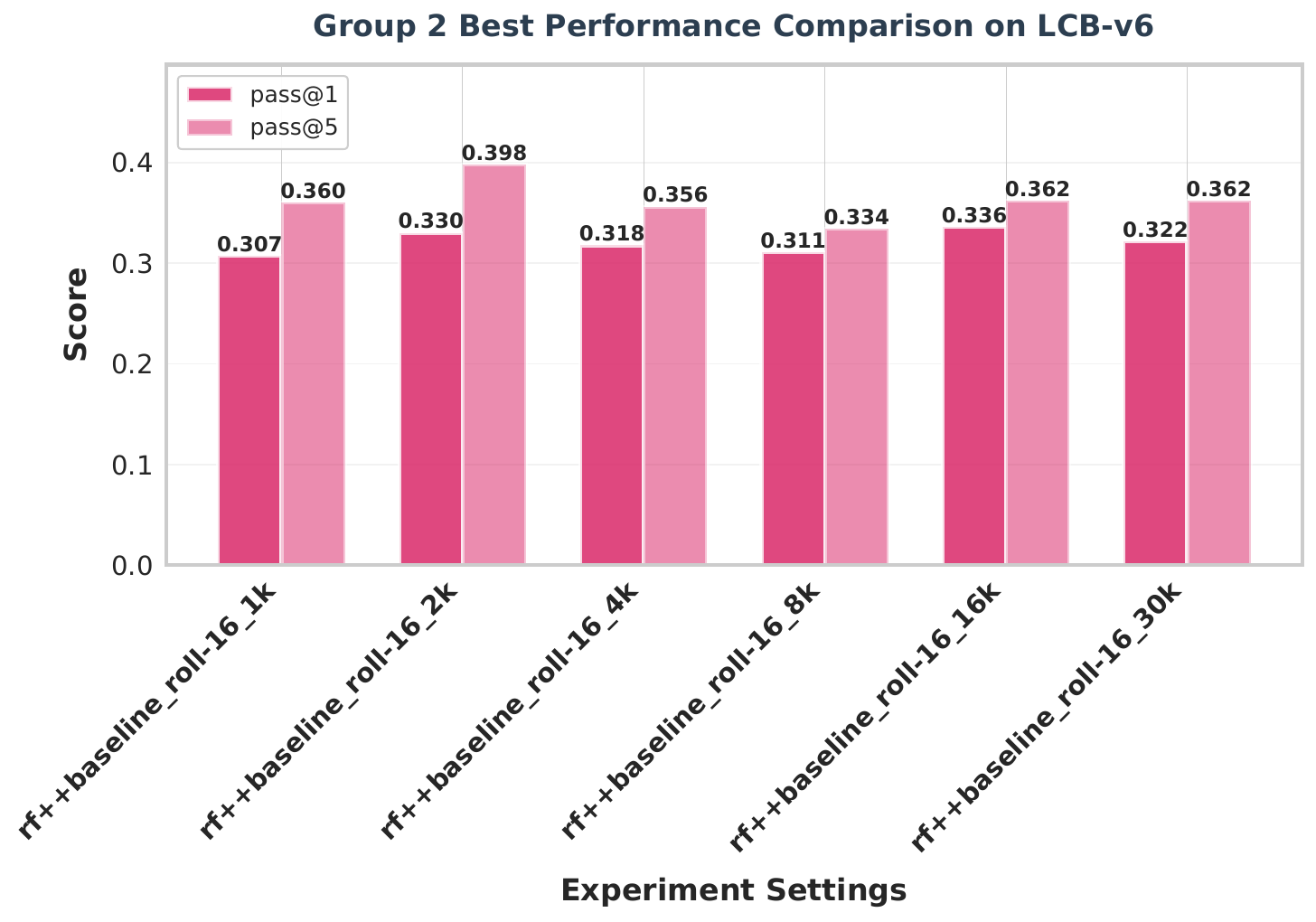}
  \caption{Best checkpoint performance comparison for different maximum response lengths (Group 2).}
  \label{fig:rl_group2_best}
\end{figure}

\paragraph{Group 2: Scaling Maximum Response Length} Code generation tasks can require vastly different context lengths. This experiment, based on the \texttt{reinforce\_plus\_plus\_baseline} estimator with 16 rollouts per prompt, investigates the impact of \texttt{MAX\_RESPONSE\_LENGTH} by sweeping values from 1K to 30K tokens. As illustrated in \autoref{fig:rl_group2_curves} and \autoref{fig:rl_group2_best}, we observe a complex non-monotonic relationship between response length and performance. The 16K token configuration achieves the highest Pass@1 score ($0.336$ at step 340), suggesting that extended context capacity enables the model to generate more complete and correct solutions for complex problems. Notably, the 2K token setting achieves the best Pass@5 performance ($0.398$ at step 380), indicating that shorter contexts may promote more diverse exploration during training, possibly by encouraging the model to learn more compact solution strategies. The 1K, 4K, and 30K configurations show similar Pass@1 performance (around $0.307$-$0.322$), while the 8K token configuration exhibits an unexpected performance drop (Pass@1: $0.311$, Pass@5: $0.334$), which we attribute to a challenging transition region where models struggle to effectively utilize the additional context without proper scaling of training dynamics. This study aims to understand the trade-off between performance (as longer contexts may be necessary for complex problems) and computational cost. As sequence length increases, we adapt our infrastructure to mitigate OOM and slow inference, introducing actor gradient checkpointing, parameter and optimizer offloading, tensor parallelism (TP), and scaling the number of training nodes. Based on these results, we recommend 2K tokens for exploration-heavy objectives targeting Pass@5 performance, 16K tokens when maximizing single-pass correctness (Pass@1), and 4K tokens as a balanced default offering reasonable performance on both metrics while minimizing computational overhead.

\begin{figure}[h]
  \centering
  \includegraphics[width=0.95\textwidth]{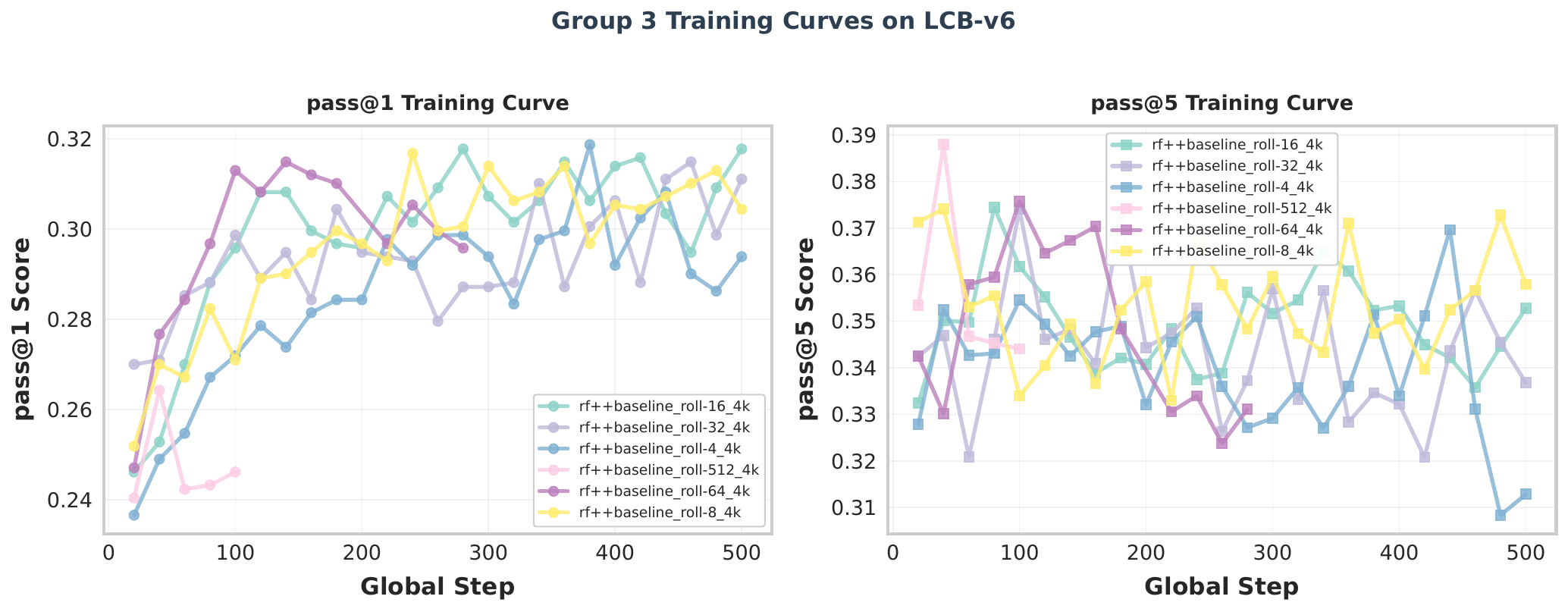}
  \caption{Impact of \texttt{N\_RESP\_PER\_PROMPT} on training performance. Note that the n=16 run is shared with Group 1.}
  \label{fig:rl_group3_curves}
\end{figure}

\begin{figure}[h]
  \centering
  \includegraphics[width=0.55\textwidth]{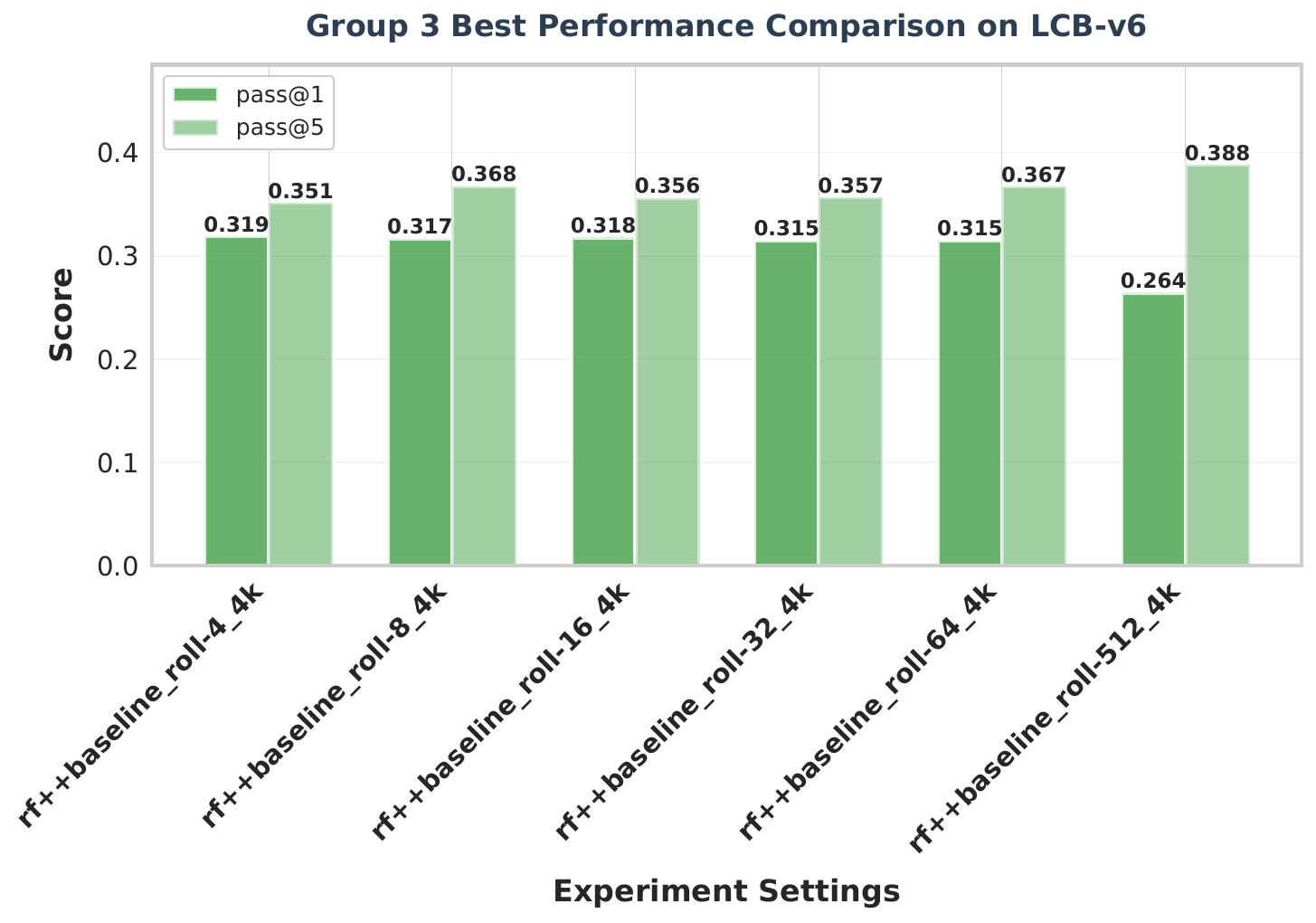}
  \caption{Best checkpoint performance comparison for different rollout numbers per prompt (Group 3).}
  \label{fig:rl_group3_best}
\end{figure}

\paragraph{Group 3: Scaling Rollouts per Prompt} The number of responses sampled per prompt (\texttt{N\_RESP\_PER\_PROMPT}) directly influences the ``width'' of exploration and the quality of the advantage estimation~\cite{hu2025brorl}. Using the \texttt{reinforce\_plus\_plus\_baseline} estimator with a fixed 4K token response length, we sweep this value from 4 to 512. As depicted in \autoref{fig:rl_group3_curves} and \autoref{fig:rl_group3_best}, the results reveal a nuanced trade-off between exploration width and training efficiency. The N=512 configuration achieves the highest Pass@5 score ($0.388$ at step 40), but this comes with prohibitively slow convergence—requiring significantly fewer update steps but much longer wall-clock time per step due to the massive rollout generation overhead, making it impractical for most training scenarios. Moreover, the extremely large rollout size setting \textit{causes training collapse counterintuitively} (where same observations happen in N=\{128,256\}), suggesting a missing puzzle for the blueprint of rollout scaling. 
Among more practical configurations, N=8 delivers the best Pass@5 performance ($0.368$) while N=64 achieves slightly lower Pass@5 ($0.367$) but comparable results. For Pass@1 performance, the results show remarkable stability: N=4 achieves the highest score ($0.319$), followed by N=8 ($0.317$), N=16 ($0.318$), and N=32/N=64 (both $0.315$), with all configurations within a narrow $0.004$ range. This suggests that moderate rollout numbers suffice for optimizing single-sample correctness, and increasing exploration width primarily benefits multi-sample diversity rather than individual solution quality. The N=32 configuration exhibits slower convergence (step 460) without clear performance advantages over smaller values. This experiment quantifies the relationship between sample efficiency and compute, determining the point of diminishing returns for exploration width. Based on these observations, we recommend N=16 as the default configuration, offering an excellent balance between convergence speed, computational efficiency, and competitive performance on both metrics. For Pass@5-critical applications with sufficient budget, N=8 provides the best diversity-performance trade-off among practical configurations.

\begin{figure}[h]
  \centering
  \includegraphics[width=0.8\textwidth]{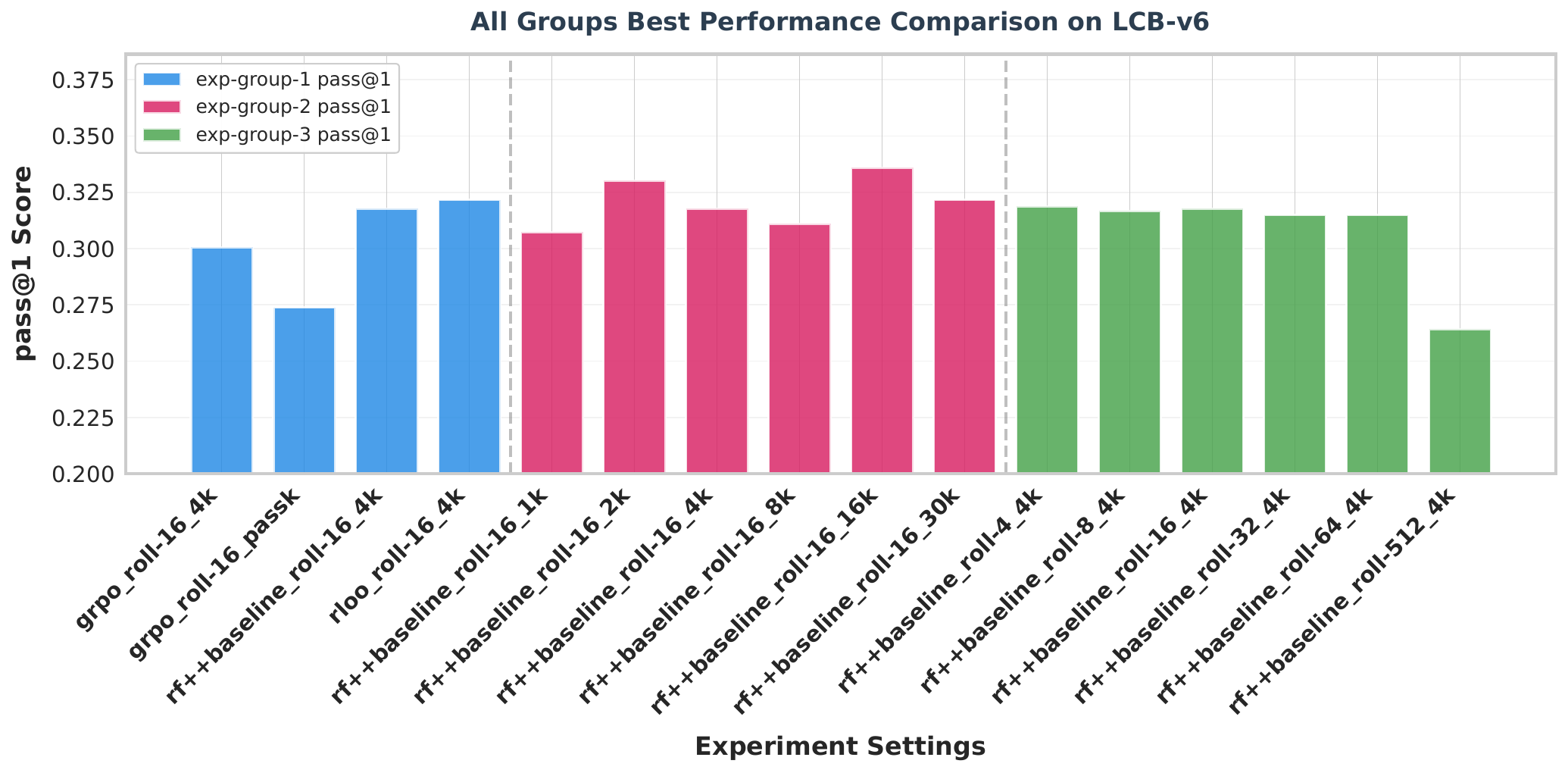}
  \caption{Comprehensive comparison of all experimental groups on the lcb-v6 benchmark, showing the training dynamics across different advantage estimators (Group 1), maximum response lengths (Group 2), and rollout numbers per prompt (Group 3).}
  \label{fig:rl_all_groups_comparison}
\end{figure}

\begin{table}[htbp]
  \centering
  \small %
  \renewcommand{\arraystretch}{1.05} %
  \setlength{\tabcolsep}{5pt} %
  \caption{Overall Performance Summary Table on lcb-v6 benchmark. This table presents the best checkpoint performance from each experimental group. Pass@1 and Pass@5 metrics are reported for the lcb-v6 evaluation set.}
  \label{tab:rl_overall_summary}
  \begin{tabular}{llccc}
    \toprule
    \rowcolor{HeaderColor}
    \textcolor{white}{\textbf{Group}} & 
    \textcolor{white}{\textbf{Configuration}} & 
    \textcolor{white}{\textbf{Step}} & 
    \textcolor{white}{\textbf{Pass@1}} & 
    \textcolor{white}{\textbf{Pass@5}} \\
    \midrule

    \multicolumn{5}{l}{\cellcolor{ModelAColor}\textbf{Group 1: Advantage Estimators (roll=16, 4K tokens)}} \\
    \rowcolor{ModelAColor!30}
    & \texttt{rloo} & 400 & $0.322$ & \textbf{$0.389$} \\
    \rowcolor{ModelAColor!30}
    & \texttt{rf++baseline} (default) & 280 & \textbf{$0.318$} & $0.356$ \\
    \rowcolor{ModelAColor!30}
    & \texttt{grpo} & 480 & $0.301$ & $0.371$ \\
    \rowcolor{ModelAColor!30}
    & \texttt{grpo\_passk} & 420 & $0.274$ & $0.328$ \\
    \midrule

    \multicolumn{5}{l}{\cellcolor{ModelBColor}\textbf{Group 2: Max Response Length (rf++baseline, roll=16)}} \\
    \rowcolor{ModelBColor!30}
    & 1K tokens & 240 & $0.307$ & $0.360$ \\
    \rowcolor{ModelBColor!30}
    & 2K tokens & 380 & $0.330$ & \textbf{$0.398$} \\
    \rowcolor{ModelBColor!30}
    & 4K tokens & 280 & $0.318$ & $0.356$ \\
    \rowcolor{ModelBColor!30}
    & 8K tokens & 300 & $0.311$ & $0.334$ \\
    \rowcolor{ModelBColor!30}
    & 16K tokens & 340 & \textbf{$0.336$} & $0.362$ \\
    \rowcolor{ModelBColor!30}
    & 30K tokens & 240 & $0.322$ & $0.362$ \\
    \midrule

    \multicolumn{5}{l}{\cellcolor{EpochColor}\textbf{Group 3: Rollouts per Prompt (rf++baseline, 4K tokens)}} \\
    \rowcolor{EpochColor!60}
    & N=4 Rollouts & 380 & $0.319$ & $0.351$ \\
    \rowcolor{EpochColor!60}
    & N=8 Rollouts & 240 & $0.317$ & $0.368$ \\
    \rowcolor{EpochColor!60}
    & N=16 Rollouts & 280 & $0.318$ & $0.356$ \\
    \rowcolor{EpochColor!60}
    & N=32 Rollouts & 460 & $0.315$ & $0.357$ \\
    \rowcolor{EpochColor!60}
    & N=64 Rollouts & 140 & $0.315$ & $0.367$ \\
    \rowcolor{EpochColor!60}
    & N=512 Rollouts & 40 & $0.264$ & \textbf{$0.388$} \\
    \midrule

  \end{tabular}
\end{table}
\textbf{Summary of Best Practices.} This comprehensive experimental suite (summarized in \autoref{tab:rl_overall_summary} and visualized in \autoref{fig:rl_all_groups_comparison}) provides concrete, data-driven ``best practices'' for applying RL to code generation LLMs. Based on the experimental outcomes across Groups 1-3, we provide the following recommendations: 

(1) \textbf{Advantage Estimator:} For practical large-scale training scenarios, \texttt{rf++baseline} is the recommended default, offering the best balance of training stability, convergence speed (step 280), and competitive performance (Pass@1: $0.318$, Pass@5: $0.356$). While \texttt{rloo} achieves superior performance on both metrics (Pass@1: $0.322$, Pass@5: $0.389$), it requires approximately 43\% more training steps (step 400), making rf++baseline more practical when wall-clock time is critical. For scenarios where maximum performance is prioritized over training efficiency, \texttt{rloo} is the optimal choice.

(2) \textbf{Response Length:} Performance exhibits a complex non-monotonic relationship with context length, revealing task-specific optima. For exploration-heavy objectives optimizing Pass@5, use 2K tokens (Pass@5: $0.398$), which promotes diverse solution strategies and achieves the highest multi-sample correctness. For maximizing single-pass correctness (Pass@1), 16K tokens is optimal (Pass@1: $0.336$), enabling complete solutions to complex problems. We recommend 4K tokens as a balanced default (Pass@1: $0.318$, Pass@5: $0.356$), offering reasonable performance on both metrics while minimizing computational overhead. The 8K configuration exhibits an unexpected performance valley (Pass@1: $0.311$, Pass@5: $0.334$) and should be avoided. Response lengths beyond 16K provide diminishing returns while substantially increasing computational cost.

(3) \textbf{Rollouts per Prompt:} N=16 provides the recommended default configuration, offering an excellent compute/performance balance (Pass@1: $0.318$, Pass@5: $0.356$) with fast convergence (step 280). For Pass@5-critical applications with sufficient budget, N=8 delivers the best diversity-performance trade-off among practical configurations (Pass@5: $0.368$). Notably, Pass@1 performance remains remarkably stable across N=4 to N=64 (range: $0.315$-$0.319$, variance $<0.004$), indicating that exploration width primarily benefits multi-sample diversity rather than individual solution quality. While N=512 achieves the highest Pass@5 ($0.388$), the wall-clock time per step becomes prohibitively expensive, making it impractical for most training scenarios. We do not recommend N=32 or N=64 for standard use cases, as they show minimal performance gains over N=8 or N=16 while requiring substantially more computation.

These guidelines, validated on the lcb-v6 benchmark using the \texttt{codecontest\_plus} dataset with 64 H20 GPUs and FSDP2 distributed training, will be crucial for enabling researchers to scale RL for code LLMs more effectively and predictably. The key insight across all experiments is that optimal hyperparameter choices depend strongly on the target metric: Pass@1 optimization benefits from extended context (16K tokens) and stable estimators (rf++baseline), while Pass@5 optimization favors compact contexts (2K tokens), higher exploration (N=8 rollouts), and sample-efficient estimators (\texttt{rloo}).

\section{Code Large Language Model for Applications}
\begin{figure*}[t!]
	\centering
    \resizebox{\textwidth}{!}{
	\begin{forest}
        forked edges,
		for tree={
                grow=east,
                reversed=true,
                anchor=base west,
                parent anchor=east,
                child anchor=west,
                base=center,
                font=\large,
                rectangle,
                draw=hidden-draw,
                rounded corners,
                align=left,
                text centered,
                minimum width=4em,
                edge+={darkgray, line width=1pt},
                s sep=3pt,
                inner xsep=2pt,
                inner ysep=3pt,
                line width=0.8pt,
                ver/.style={rotate=90, child anchor=north, parent anchor=south, anchor=center},
            },
            where level=1{text width=8em,font=\normalsize, }{},
            where level=2{text width=12em,font=\normalsize}{},
            where level=3{text width=14em,font=\normalsize,}{},
	    [Code Large Language \\Model Applications, ver
			[IDE-integrated \\Development \\Assistants, text width=12em
                    [
                        GitHub Copilot~\cite{github2024copilot,githubcopilot2024architecture}{, }
                        Cursor~\cite{anysphere2024cursor,pragmaticengineer2025cursor}{, }
                        \\TRAE~\cite{trae2025trae}{, }
                        Tabnine~\cite{tabnine2024enterprise}{, }
                        Windsurf~\cite{codeium2024windsurf}{, }
                        \\CodeGeeX~\cite{zheng2023codegeex}{, }
                        Cody~\cite{sourcegraph2024cody}{, }
                        Bito AI~\cite{bito_ai}
                       , text width=36em
                    ]
			]
			[Cloud-native \\Coding \\Platforms, text width=12em
                    [
                        Amazon Q Developer~\cite{amazon2024codewhisperer,aws2024qdeveloper}{, }
                        Google Cloud Code~\cite{google2024cloudcode}{, }\\
                        Gemini Code Assist~\cite{google2024gemini}{, }
                        Replit Ghostwriter~\cite{replit2023ghostwriter}{, }\\
                        Alibaba Tongyi Lingma~\cite{alibaba2024tongyi}{, }
                        GitHub Codespaces AI~\cite{github2024codespaces}
                       , text width=36em
                    ]
            ]
            [Terminal-based \\Autonomous \\Agents, text width=12em
                    [
                        Aider~\cite{gauthier2024aider,aider2024swebench}{, }
                        Claude Code~\cite{anthropic2024claudecode}{, }
                        \\Gemini CLI~\cite{google2024gemini}{, }
                        Plandex~\cite{plandex2024}{, }
                        \\OpenCode~\cite{opencode2024}{, }
                        Qwen Code~\cite{qwen2024code}{, }
                        TRAE Agent~\cite{gao2025trae}
                       , text width=36em
                    ]
            ]
            [Code Repair and \\Verification \\Applications, text width=12em
                    [
                        RepairAgent~\cite{ahmed2024repairAgent}{, }
                        AutoSpec~\cite{autospec2024}{, }
                        AlphaRepair~\cite{xia2024alpharepair}{, }
                        \\Toggle~\cite{toggle2024}{, }
                        RepairLLaMA~\cite{xia2024repairllama}{, }
                        VulRepair~\cite{zhang2024vulrepair}
                       , text width=36em
                    ]
            ]
            [Pull Request Review \\and Quality \\Assurance, text width=12em
                    [
                        PR-Agent~\cite{pr_agent}{, }
                        CodeRabbit AI Reviewer~\cite{coderabbit}{, }
                        \\LLM Code Reviewer~\cite{llm_code_reviewer}{, }
                        Graphite Reviewer~\cite{graphite_reviewer}{, }
                        \\Codedog~\cite{codedog}
                       , text width=36em
                    ]
            ]
		]
	\end{forest}}
	\caption{Code Large Language Model Applications.}
    \label{fig:code_llm_applications}
\end{figure*}
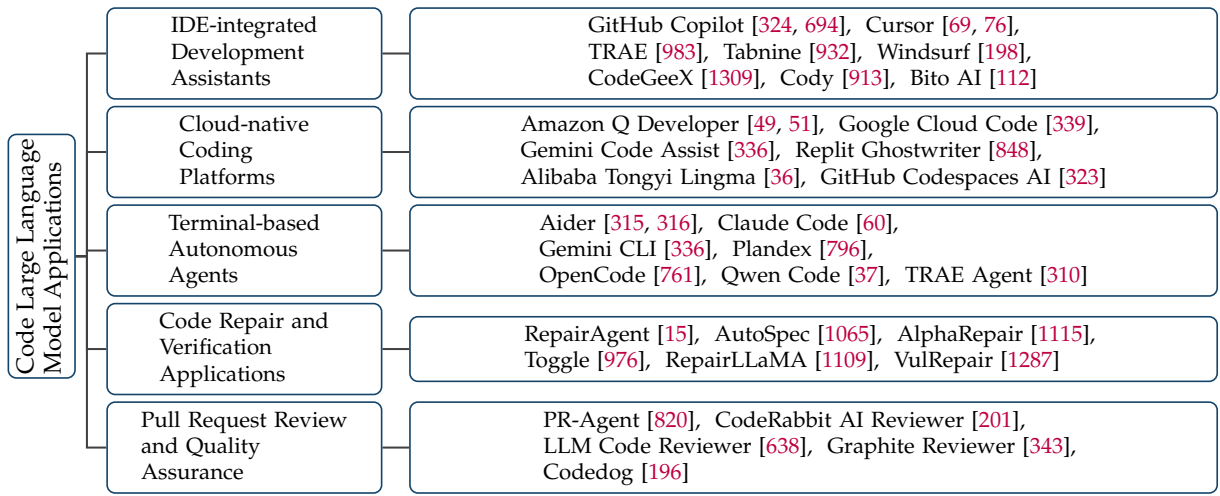

The rapid evolution of code-capable LLMs has driven a paradigm shift in software development, transitioning from research prototypes to production-ready tools integrated across the entire software development lifecycle \cite{zhang2024llmcode,hou2024large}. This section presents a comprehensive taxonomy of code LLM applications~\cite{does_ai_assisted_coding_deliver,cursorcore,codinggenie}, categorizing them into six primary application domains based on their architectural patterns~\autoref{fig:code_llm_applications}, deployment strategies, and functional capabilities. Each application is analyzed in terms of its historical development, core capabilities, technical innovations, and current limitations.
\subsection{IDE-integrated Development Assistants}

IDE-integrated assistants represent the most widely deployed category of code LLM applications, seamlessly embedding AI capabilities within established development environments such as Visual Studio Code, JetBrains IDEs, and proprietary editors. These tools target professional developers working on medium to large-scale projects, emphasizing deep contextual understanding \cite{peng2023impact}, cross-file reasoning, and integration with existing software engineering workflows \cite{zhang2023copilot}.

\paragraph{GitHub Copilot}
GitHub Copilot emerged as the pioneering commercial IDE-integrated AI coding assistant, initially launched in technical preview on June 29, 2021 \cite{github2024copilot}. Built upon OpenAI's Codex model (a descendant of GPT-3 trained on public code repositories), Copilot fundamentally transformed developer expectations for AI-assisted programming. The system evolved significantly from its initial single-model architecture: in November 2023, the chat interface transitioned to GPT-4, and by October 2024, GitHub announced multi-model support \cite{github2024universe} enabling developers to select between GPT-4o, Claude 3.5 Sonnet, and Gemini 1.5 Pro based on task requirements. As of 2025, Copilot defaults to GPT-4.1 across chat, agent mode, and code completions, with Pro+ and Enterprise tiers providing access to frontier models including Claude Opus 4 and GPT-5 preview \cite{github2024models}. The architecture comprises a sophisticated Node.js-based agent process that handles web requests to GitHub services, context collation from active files and repository structure, prompt engineering for the LLM, and post-processing including content exclusion filters \cite{githubcopilot2024architecture}. Key recent innovations include Copilot Edits for multi-file refactoring with iterative change plans, Copilot Workspace for end-to-end feature development using natural language, and the May 2025 introduction of ``coding agent'' mode \cite{githubcopilot2024features} enabling asynchronous task execution with automated pull request generation. The system processes approximately 150 million code suggestions daily with sub-second latency requirements \cite{anysphere2024cursor}. GitHub Copilot has achieved widespread adoption with over 1.8 million individual paying subscribers and tens of thousands of enterprise customers, generating approximately \$500 million in revenue in 2024 \cite{pragmaticengineer2025cursor}. However, the system faces ongoing challenges including concerns over training data licensing (trained on public repositories without explicit permission), potential code suggestion plagiarism from training data, privacy issues regarding telemetry and keystroke data collection, and limited offline capabilities \cite{wikipedia2024copilot}.

\paragraph{Cursor}
Cursor represents a strategic rethinking of AI integration in development environments. Founded in 2022 by Michael Truell, Sualeh Asif, Arvid Lunnemark, and Aman Sanger, Cursor launched its first version in March 2023 as a fork of Visual Studio Code \cite{anysphere2024cursor}, strategically balancing familiarity with innovation. Rather than building an IDE from scratch, this architectural decision enabled the team to leverage VS Code's mature foundation and familiar interface while dedicating all engineering efforts to deep AI integration—transforming how developers interact with code rather than rebuilding basic editor infrastructure. The system has experienced explosive growth, crossing \$500 million in annual recurring revenue (ARR) within just two years of launch. Cursor's technical architecture centers on proprietary innovations in context management and model inference. The ``tab model'' provides ultra-low-latency inline completions (under one second) by implementing a sophisticated sync engine that transmits encrypted code snippets to cloud servers for inference without persistent storage, maintaining privacy while handling over one million queries per second \cite{bytebytego2025cursor}. Repository indexing employs Merkle trees to track file changes efficiently, enabling incremental updates every three minutes without full reindexing. The system supports context windows up to 200K tokens through dynamic context prioritization using tree-sitter-based AST analysis. Cursor introduced several breakthrough features: Composer Mode enables natural language project-wide edits across multiple files with automated dependency tracking; Rules and Memories allow developers to encode project-specific conventions and maintain conversation continuity across sessions; and Agent Mode (released with Cursor 1.0 in 2025) provides autonomous coding capabilities \cite{cursor2025changelog,allaboutai2025cursor} with background execution and browser automation for UI debugging. The system supports multiple frontier models including GPT-4.1, Claude Sonnet 4.5, Gemini 2.5 Pro, and xAI's Grok Code, with automatic model selection based on task characteristics \cite{cursor2025features}. Despite its rapid adoption by over half of Fortune 500 companies and exceptional developer satisfaction ratings, Cursor faces criticism for inconsistent code quality in complex refactoring tasks, occasional suggestion lag in large codebases, higher cost (\$20/month vs \$10 for Copilot), and the potential for vendor lock-in through proprietary features \cite{towardsdatascience2025vscode}.

\paragraph{TRAE} TRAE~\cite{trae2025trae} represents a new generation of AI-native integrated development environments that pursue a paradigm shift from assistance toward autonomous software creation.
The product manifests the vision through two primary modes - Builder Mode and SOLO Mode, enabling everything from specification-driven project scaffolding to conversational code editing and multi-modal input. 

The SOLO Mode feature launched in mid-2025 is described as an all-in-all Context Engineer that thinks in context, works across tools, and works with developers to ship real features from start to finish. In practical terms, a user can issue a natural language specification and TRAE will (a) parse the requirement, (b) decompose it into tasks, (c) generate scaffolding, code, tests and deployment glue, and (d) preview changes before applying them. This think-then-do pattern emerges as a distinguishing workflow: instead of immediate line-by-line completion, TRAE emphasizes planning, decomposition and then change-application.

The CUE feature supports auto-completion, multi-line edits, predicted edits, jump-to edits (i.e. moving the cursor to the next affected region), smart import and smart rename. Unlike simpler completion tools that only look at local buffer context, CUE takes into account a broader workspace-level context: project structure, symbols across files, dependency graphs, and prior developer edits.

\paragraph{Tabnine}
Tabnine, launched in March 2021, distinguished itself in the crowded AI coding assistant market through its focus on privacy, security, and intellectual property compliance \cite{tabnine2024enterprise}. Unlike competitors trained on public repositories with ambiguous licensing, Tabnine exclusively uses permissively licensed code (MIT, Apache, BSD) for model training, directly addressing enterprise concerns about code ownership and legal liability. The platform's core differentiator is its flexible deployment architecture: while most competitors require cloud connectivity, Tabnine offers on-premises and local deployment options where models run entirely within the organization's infrastructure, ensuring proprietary code never leaves the corporate network. This approach resonates particularly with regulated industries (finance, healthcare, government) and companies with strict data sovereignty requirements. Tabnine's technical architecture emphasizes customization through fine-tuning on private codebases, enabling the model to learn organization-specific patterns, APIs, and coding standards without exposing this data externally. The system provides multi-IDE support (VS Code, IntelliJ, JetBrains suite, Eclipse, Visual Studio) with consistent behavior across platforms. Recent enhancements include team learning capabilities where the model improves based on accepted suggestions across the development team, semantic code completion understanding broader context beyond syntactic patterns, and compliance reporting for audit trails. However, Tabnine's conservative training approach results in somewhat lower suggestion quality compared to competitors trained on larger, less-restricted datasets. The system also commands premium pricing for enterprise features (\$39/user/month for Pro) and exhibits slower feature development velocity compared to heavily-funded competitors like Copilot and Cursor.

\paragraph{Windsurf}
Windsurf emerged in November 2024 as Codeium's flagship AI-native IDE, introducing the novel cascade architecture \cite{codeium2024windsurf} for deep codebase understanding and multi-agent coordination. The Cascade system implements a flow-based approach where multiple specialized sub-agents collaborate on complex refactoring tasks: one agent maps the codebase structure, another identifies relevant files and dependencies, a third generates modifications, and a fourth validates changes. This hierarchical decomposition enables handling of architectural transformations that single-agent systems struggle with, such as migrating from REST to GraphQL across dozens of files or refactoring monolithic applications into microservices. Windsurf's context management employs a hybrid approach combining vector search for semantic similarity, AST-based structural analysis for precise symbol references, and graph-based dependency tracking for impact analysis. The system maintains persistent codebase indexes that update incrementally, reducing reindexing overhead. Advanced features include flows (reusable multi-step automation sequences), rules for enforcing team coding standards, and memories for preserving project-specific knowledge. However, as the newest entrant in the AI IDE space, Windsurf faces challenges including limited ecosystem maturity, a smaller user community compared to established competitors, potential performance issues with the multi-agent overhead in large codebases, and uncertainty about long-term platform stability.

\paragraph{Additional IDE-Integrated Systems}
The IDE assistant landscape includes several other notable systems. \textbf{CodeGeeX}, launched in September 2022, prioritizes multilingual support with strong performance in Chinese programming contexts and open-source availability \cite{zheng2023codegeex}. \textbf{Cody} by Sourcegraph (August 2023) integrates deeply with code search infrastructure \cite{sourcegraph2024cody} for superior context retrieval, particularly in large monorepos. \textbf{Bito AI} emphasizes developer privacy with offline operation modes and focuses on code explanation and test generation beyond simple completion \cite{bito_ai}. These alternatives serve specialized niches but struggle to compete with the network effects and rapid development pace of market leaders.

\subsection{Cloud-native Coding Platforms}

Cloud-native platforms leverage scalable infrastructure to provide code generation services through APIs, web interfaces, and cloud-integrated development environments. These systems target organizations requiring centralized deployment, consistent security policies, and deep integration with cloud services \cite{aws2024codewhisperer,google2024duet}, particularly for infrastructure-as-code and cloud-native application development.

\paragraph{Amazon Q Developer}
Amazon Q Developer represents the evolution and rebranding of Amazon CodeWhisperer, which originally launched in June 2022 as AWS's answer to GitHub Copilot \cite{amazon2024codewhisperer}. In April 2024, AWS merged CodeWhisperer into the broader Amazon Q family \cite{techcrunch2024amazonq}, significantly expanding the service's scope beyond code completion to encompass comprehensive AWS development workflows. The strategic rebranding reflected AWS's recognition that code generation alone was insufficient; developers needed an integrated assistant understanding the full AWS ecosystem. Amazon Q Developer distinguishes itself through deep AWS service specialization: the system was trained on millions of internal Amazon code repositories, AWS documentation, and service implementations \cite{aws2024qdeveloper}, providing superior generation quality for CloudFormation templates, CDK constructs, and Terraform configurations compared to general-purpose models. The architecture integrates across the development lifecycle: inline code suggestions in IDEs (JetBrains, VS Code, Visual Studio, and Eclipse), chat interfaces in the AWS Console for resource queries, such as ``List all Lambda functions'' or ``What were my top three cost drivers in Q1?'', CLI autocompletion for AWS commands, and chat integrations in Microsoft Teams and Slack for operational support \cite{dev2024amazonq}. Key capabilities include security scanning for vulnerability detection, automated Java version upgrades, .NET porting from Windows to Linux, and AWS cost optimization recommendations. The service introduces ``Agent'' functionality, enabling autonomous multi-step tasks like ``deploy'' this application to ECS with auto-scaling and monitoring. As of 2025
,Amazon Q Developer offers a perpetual free tier (50 agent interactions monthly, 1000 lines of code transformation), with paid tiers (\$19/month Pro) providing higher limits and IP indemnity protection where AWS defends customers against copyright infringement claims. 

\paragraph{Google Cloud Code and Gemini Code Assist}
Google's cloud coding offerings evolved from Google Cloud Code (May 2023) to Gemini Code Assist, leveraging Google's the latest Google models family \cite{google2024cloudcode} for code generation optimized for Google Cloud Platform (GCP) and Kubernetes. The system excels at generating Google Kubernetes Engine (GKE) configurations, Cloud Run deployments, and BigQuery SQL with deep understanding of GCP service constraints and best practices. Gemini CLI, introduced in November 2024, provides terminal-native development \cite{google2024gemini} with fast inference through local caching and incremental parsing for large repositories. The architecture emphasizes enterprise features including data residency controls ensuring customer code never leaves specified geographic regions, integration with Google Workspace for team collaboration, and comprehensive audit logging for compliance. Advanced capabilities include multi-modal code generation (accepting diagrams and screenshots as input), Cloud Workstation integration for cloud-based development environments, and Duet AI in databases for SQL optimization. However, the platform faces adoption challenges including later market entry relative to competitors, limited ecosystem integration outside Google Cloud, uncertainty around product strategy given Google's history of discontinuing services, and questions about long-term commitment given leadership changes in Google's AI organization.

\paragraph{Replit Ghostwriter}
Replit Ghostwriter, launched in October 2022, pioneered browser-based AI-assisted development with custom small language models \cite{replit2023ghostwriter} achieving competitive performance through architectural innovations rather than parameter scale. The replit-code-v1.5-3b model (3 billion parameters, trained on 1 trillion tokens) demonstrates that focused training on high-quality code with careful data curation and optimized architectures can rival much larger models on specific tasks. Ghostwriter's browser-native architecture eliminates local installation requirements, enabling instant access to AI coding assistance from any device with a web browser. The platform integrates tightly with Replit's collaborative multiplayer editing, enabling teams to code together with shared AI assistance in real-time. Key features include Ghostwriter Chat for conversational code help, explain code for understanding unfamiliar patterns, generate code for implementing features from descriptions, and complete code for inline suggestions. The browser-based execution environment supports immediate testing and deployment, creating a seamless cycle from generation to validation. As of 2025,Replit's freemium model provides limited free access with generous paid tiers (\$20/month) including unlimited AI interactions. The platform particularly appeals to educators and learners through simplified onboarding and extensive educational resources. However, limitations include dependency on internet connectivity with no offline mode, limited support for complex enterprise workflows and large monorepo architectures, performance constraints in the browser environment for resource-intensive applications, and potential data privacy concerns for sensitive commercial code.

\paragraph{Alibaba Tongyi Lingma}
Tongyi Lingma, launched in October 2023, serves the Chinese market with specialized support for Chinese programming contexts and Alibaba Cloud services \cite{alibaba2024tongyi}. Built on the Qwen language model family, the system provides multilingual code generation, testing, and debugging with particular strength in Chinese natural language understanding for requirements and documentation. The platform integrates deeply with Alibaba Cloud DevStudio and supports popular Chinese development tools. Lingma emphasizes compliance with Chinese data sovereignty requirements through local model hosting and processing. However, international adoption remains limited due to language barriers, market-specific optimizations, and regulatory considerations.

\subsection{Terminal-based Autonomous Agents}

Terminal-based agents operate in command-line environments, providing autonomous code generation, modification, and project management capabilities. These tools appeal to developers preferring keyboard-driven workflows, automation engineers building CI/CD pipelines \cite{yang2024swe}, and researchers requiring scriptable, reproducible experiments \cite{jimenez2024swebench}.

\paragraph{Aider}
Aider establishes itself as the leading open source terminal-based coding agent through pioneering work in repository mapping, code editing, and benchmark performance \cite{gauthier2024aider}. Its repository mapping system employs treesitter for language-agnostic AST parsing \cite{aider2023repomap}, generating compact summaries of entire codebases and performing graph optimization on the call graph to dynamically tailor context. Aider implements multiple code editing backends with automatic format selection, where the unified diff format provides precise line-level edit specifications that reduce ambiguity. Furthermore, Aider features an iterative execution feedback mechanism that uses treesitter-based linting \cite{aider2024linting} to detect syntax errors and automatically request fixes from the LLM with enhanced context. The system integrates seamlessly with Git for change management and supports multiple LLM providers with prompt caching for cost optimization. However, Aider's terminal-native design limits visual debugging, lacks GUI support for interface development, requires significant technical proficiency for effective use, and depends on external LLM APIs.

\paragraph{Claude Code}
Claude Code, announced in December 2024 by Anthropic, represents a terminal-native development environment specifically designed for the Claude model family \cite{anthropic2024claudecode}. The system implements the Model Context Protocol (MCP), an open standard enabling extensible tool integration and sub-agent coordination. Claude Code's architecture emphasizes composability: developers can define custom tools (e.g., API clients, testing frameworks, deployment scripts) that Claude can invoke autonomously. The agent planning capability breaks complex tasks into sub-goals with explicit step generation, enabling developers to review and modify plans before execution. MCP servers enable Claude Code to access external resources (databases, file systems, APIs) through standardized interfaces with built-in security boundaries. The system supports multi-step workflows where Claude Code can research documentation, write code, execute tests, debug failures, and iterate until success—all while maintaining context across the entire process. However, as a newly released product, Claude Code faces early-stage limitations including incomplete documentation, a nascent tool ecosystem, potential reliability issues in complex scenarios, and dependence on Claude API availability and pricing.

\paragraph{Gemini CLI}
Gemini CLI, introduced in November 2024, provides terminal-native access to Google's Gemini models with emphasis on speed and efficiency \cite{google2024gemini}. The system employs aggressive local caching strategies, storing previously indexed codebase representations to minimize recomputation. Incremental parsing updates only modified files rather than reprocessing entire repositories. Gemini CLI integrates with Google Cloud authentication for seamless access to GCP resources and supports multi-modal inputs including screenshots and diagrams in terminal workflows. The lightweight architecture minimizes memory footprint and startup latency. However, the system's tight Google Cloud integration limits utility outside GCP environments, and the relatively new release means limited third-party tool integrations and community resources compared to more established alternatives.

\paragraph{Plandex}
Plandex, released in March 2024, implements a distinctive plan-based workflow with branching support \cite{plandex2024} for exploring alternative implementations. Rather than immediate code generation, Plandex first creates a detailed implementation plan specifying which files to modify, what changes to make, and the order of operations. Developers can review, edit, and approve plans before execution. The branching system allows spawning multiple plan variants to compare different approaches (e.g., implementing a feature with different architectural patterns), with lightweight switching between branches. This approach suits exploratory development where the optimal solution isn't immediately clear. However, the additional planning overhead increases latency for simple tasks, and the system's specialized workflow may not align with all development styles.

\paragraph{Additional Terminal Agents}
Other terminal-based systems include \textbf{OpenCode} (August 2024) emphasizing open-source and multilingual support \cite{opencode2024}, and \textbf{Qwen Code} (January 2025) featuring 256K context windows and Chinese language optimization \cite{qwen2024code}. In addition, \textbf{TRAE Agent}~\cite{gao2025trae} offers a research-friendly design for studying AI agent architectures, conducting ablation studies, and developing novel agent capabilities.

\subsection{Code Repair and Verification Applications}

Specialized applications for automated bug fixing \cite{xia2024repairllama}, vulnerability patching, and formal verification address critical software reliability needs. These tools serve security teams performing vulnerability assessment, QA engineers automating test generation, researchers in formal methods and program synthesis \cite{ahmed2024repairAgent}, and maintainers of legacy codebases.

\paragraph{RepairAgent}
RepairAgent, introduced in March 2024, pioneered autonomous debugging through hypothesis-driven state machine progression \cite{ahmed2024repairAgent}. The system implements a structured approach for bug fixing comprising of five stages: (1) \textbf{fault localization}: identifying suspicious code regions using test failure analysis, execution traces, and statistical debugging; (2) \textbf{hypothesis generation}: proposing potential root causes for observed failures; (3) \textbf{patch generation}: creating fixes for each hypothesis; (4) \textbf{validation}: executing tests to verify correctness; and (5) \textbf{iterative refinement}: looping until validation succeeds or maximum attempts exhausted. RepairAgent's architecture employs multiple LLM calls with specialized prompting for each phase: localization prompts focus the model on analyzing test output and stack traces, hypothesis prompts encourage considering multiple failure modes, and patch prompts constrain changes to minimize disruption. The state machine tracks attempted fixes and their results, preventing repeated failures. On the Defects4J~\cite{just2014defects4j} benchmark, RepairAgent successfully fixed 164 out of 357 bugs (45.9\%) including 39 novel repairs not achieved by previous automated program repair (APR) tools, demonstrating significant advancement over template-based and learning-based APR systems. The hybrid approach combining program analysis (for localization precision) with LLM reasoning (for creative fix generation) proved more effective than either technique alone. However, RepairAgent exhibits limitations including high computational cost from multiple LLM invocations, difficulty with semantic bugs requiring deep domain knowledge, potential for incorrect fixes that pass tests but introduce new issues, and limited scalability to bugs requiring changes across many files.
\paragraph{AutoSpec}
AutoSpec, released in April 2024, tackles the challenging problem of automatic specification synthesis for formal verification \cite{autospec2024}.Rather than fixing bugs directly, AutoSpec generates formal specifications (preconditions, postconditions, loop invariants) that characterize correct behavior. Here, preconditions define the required input conditions for a function, postconditions specify the expected output or state after execution, and loop invariants capture properties that must hold before and after each loop iteration. The system employs hierarchical decomposition: complex specifications are broken into simpler sub-specifications for individual functions or loop iterations, with LLMs generating candidate specifications based on code analysis and comments. Generated specifications are validated using theorem provers (e.g., Dafny verifier~\cite{leino2010dafny}) to ensure logical consistency and completeness.Failed verification attempts produce counterexamples that guide refinement. This synthesis-verification loop continues until valid specifications are achieved or timeout. AutoSpec particularly excels at inferring loop invariants—traditionally one of the most challenging aspects of formal verification. The generated specifications serve multiple purposes: documentation of intended behavior, verification of correctness, and guidance for subsequent code modifications. However, AutoSpec faces significant scalability challenges with large codebases, depends heavily on code comment quality for initial specification hints, often struggle with complex specifications requiring mathematical reasoning, and inherits limitations of underlying theorem provers.

\paragraph{AlphaRepair}
AlphaRepair, introduced in January 2024, combines program analysis with LLMs \cite{xia2024alpharepair} for superior fault localization and patch generation. The system's static analysis component performs control flow analysis, data flow analysis, and dependency analysis to precisely identify code regions potentially responsible for failures. Dynamic analysis instruments code to collect execution traces, variable values, and branch coverage during test execution. These analyses produce ranked lists of suspicious locations with confidence scores. The LLM then receives this context including suspicious code, execution traces, test outputs, and similar historical bugs from a patch database. AlphaRepair implements a template-guided generation strategy: rather than unrestricted code generation, the LLM selects and instantiates repair templates (e.g., ``add null check'', ``change condition operator'', ``initialize variable'') based on bug characteristics. This constrained generation reduces hallucination and improves patch quality. Validation employs both test-based verification and static analysis to detect potential side effects. On standard APR benchmarks, AlphaRepair achieved higher correct fix rates and lower plausible-but-incorrect patch rates compared to purely learning-based approaches.

\paragraph{Toggle}
Toggle, released in April 2024, performs token-level bug localization with adjustment models \cite{toggle2024} for refinement before repair. Rather than localizing at the statement or function level, Toggle identifies specific tokens (variable names, operators, literals) likely to be faulty. The multi-phase pipeline includes coarse localization (identifying candidate statements), fine-grained localization (pinpointing specific tokens within statements), adjustment (refining localization based on program structure and semantics), and generation (producing fixes focused on identified tokens). Token-level precision enables more targeted repairs with less risk of unintended consequences. The adjustment phase uses a separate model to filter false positives from initial localization by analyzing surrounding code context and common bug patterns. This two-stage localization significantly improves precision at the cost of additional computational overhead.

\paragraph{Additional Repair Systems}
\textbf{RepairLLaMA} (January 2024) provides fine-tuned models specifically for automated program repair \cite{xia2024repairllama}. \textbf{VulRepair} (February 2024) specializes in security vulnerability patching \cite{zhang2024vulrepair} with knowledge of CVE patterns and secure coding practices.

\subsection{Pull Request Review and Quality Assurance}

Pull request review assistants automate code review processes, providing feedback on code quality, potential bugs, security issues, and adherence to coding standards. These tools aim to reduce reviewer burden, accelerate review cycles, and improve consistency \cite{llm_code_reviewer,pr_agent}.

\paragraph{PR-Agent (Qodo-AI)}
PR-Agent emerged as a leading open-source automated pull request review system \cite{pr_agent}, providing comprehensive analysis including automated PR summarization, comment generation on potential issues, security risk tagging, and reviewer suggestions based on code expertise. The system integrates as a CI/CD pipeline step, automatically triggering on PR creation or updates. PR-Agent employs multiple specialized prompts for different review aspects: architecture review analyzing structural changes and design patterns, security review identifying vulnerability patterns, style review checking adherence to conventions, and logic review examining correctness and edge cases. The generated feedback appears as GitHub/GitLab comments on specific lines or as summary reviews. PR-Agent supports self-hosted deployment for organizations requiring data sovereignty, with both GitHub Enterprise and GitLab Enterprise compatibility. The system can be customized with project-specific review rules and quality gates. Key advantages include open-source transparency, enterprise-friendly deployment options, and active community development. However, limitations include dependency on LLM API availability and costs, potential for verbose or irrelevant feedback requiring human filtering, and challenges understanding complex business logic without extensive context.

\paragraph{CodeRabbit}
CodeRabbit provides automated PR reviews with inline suggestions, code explanations, and context-aware comments \cite{coderabbit}. The system distinguishes itself through support for long-context diffs (handling PRs with thousands of lines), multi-file impact analysis tracing changes across dependencies, and configurable review depth allowing teams to adjust between quick surface-level checks and deep semantic analysis. CodeRabbit offers both SaaS and self-hosted deployment with support for multiple LLM providers (GPT-4, Claude, Gemini). The review process includes static analysis integration to catch common issues before LLM analysis, reducing LLM costs while maintaining thoroughness. However, the system's effectiveness depends heavily on codebase context availability and may produce false positives in codebases with non-standard patterns.

\paragraph{Additional PR Review Systems}
\textbf{LLM Code Reviewer} provides simple GitHub Action integration \cite{llm_code_reviewer} with configurable prompt templates for GPT, Claude, or Gemini. \textbf{Graphite Reviewer} emphasizes speed and context recall \cite{graphite_reviewer} through repository-aware indexing. \textbf{Codedog} offers multi-platform support (GitHub/GitLab) \cite{codedog} with strong UI integration for Chinese development teams.

\section*{Conclusion}
In this work, we present a comprehensive analysis of code-generating large language models across their entire lifecycle, from data curation and pre-training through supervised fine-tuning, reinforcement learning, and deployment as autonomous agents. We examine both general-purpose models (GPT-4, Claude, and LLaMA) and code-specialized models (StarCoder, CodeLLaMA, DeepSeek-Coder, and QwenCoder) while bridging the gap between academic benchmarks and real-world software development challenges through systematic experiments on scaling laws, architectures, and training methodologies. Finally, we conduct a comprehensive series of experiments analyzing code pre-training, supervised fine-tuning, and reinforcement learning across multiple dimensions, including scaling laws, framework selection, hyperparameter sensitivity, model architectures, dataset comparisons, and manual coding practices.

\clearpage
\newpage

\section{Contributions and Acknowledgements}
The authors of this paper are listed in order as follows: \\

\textbf{First Author}
\begin{multicols}{2}
    \begin{itemize}
        \item Jian Yang, Beihang University
    \end{itemize}
\end{multicols}

\textbf{Corresponding Authors}
\begin{multicols}{2}
    \begin{itemize}
        \item Xianglong Liu, Beihang University
        \item Weifeng Lv, Beihang University
    \end{itemize}
\end{multicols}

\textbf{Core Contributors (Last Name Alphabetical Order)}
\begin{multicols}{2}
    \begin{itemize}
        \item Ken Deng, Kuaishou
        \item Shawn Guo, M-A-P
        \item Lin Jing, M-A-P
        \item Yizhi Li, University of Manchester
        \item Shark Liu, M-A-P
        \item Xianzhen Luo, Harbin Institute of Technology
        \item Yuyu Luo, The Hong Kong University of Science and Technology (Guangzhou)
        \item Changzai Pan, Institute of Artificial Intelligence (TeleAI), China Telecom
        \item Ensheng Shi, Huawei Cloud Computing Technologies Co., Ltd
        \item Yingshui Tan, Alibaba Group
        \item Renshuai Tao, Beijing Jiaotong University
        \item Zili Wang, StepFun
        \item Jiajun Wu, Beihang University
        \item Xianjie Wu, Beihang University
        \item Zhenhe Wu, Beihang University
        \item Daoguang Zan, ByteDance
        \item Chenchen Zhang, Tencent
        \item Wei Zhang, Beihang University
        \item He Zhu, OPPO
        \item Terry Yue Zhuo, Monash University \& CSIRO's Data61
    \end{itemize}
\end{multicols}

\textbf{Contributors (Last Name Alphabetical Order)}
\begin{multicols}{2}
    \begin{itemize}
        \item Kerui Cao, Alibaba Group
        \item Xianfu Cheng, Beihang University
        \item Jun Dong, ByteDance
        \item Shengjie Fang, Beijing University of Posts and Telecommunications
        \item Zhiwei Fei, Nanjing University
        \item Xiangyuan Guan, Beihang University
        \item Qipeng Guo, Shanghai AI Lab,
        \item Zhiguang Han, Nanyang Technological University
        \item Xueyu Hu, Zhejiang University
        \item Joseph James, University of Sheffield
        \item Tianqi Luo, The Hong Kong University of Science and Technology (Guangzhou)
        \item Renyuan Li, Sichuan University
        \item Yuhang Li, Beijing Institute of Technology
        \item Yiming Liang, CASIA
        \item Congnan Liu, Alibaba Group
        \item Qian Liu, Independent Researcher
        \item Ruitong Liu, National University of Singapore
        \item Tyler Loakman, University of Sheffield
        \item Xiangxin Meng, ByteDance
        \item Chuang Peng, Beijing Jiaotong University
        \item Tianhao Peng, Beihang University
        \item Jiajun Shi, Beihang University
        \item Mingjie Tang, Sichuan University
        \item Boyang Wang, Beihang University
        \item Haowen Wang, Beijing University of Posts and Telecommunications
        \item Yunli Wang, Beihang University
        \item Fanglin Xu, Hunan University
        \item Zihan Xu, Beijing University of Posts and Telecommunications
        \item Fei Yuan, Shanghai AI Lab,
        \item Jiayi Zhang, The Hong Kong University of Science and Technology (Guangzhou)        
        \item Xinhao Zhang, Beijing Jiaotong University
        \item Xiantao Zhang, Beihang University
        \item Wangchunshu Zhou, OPPO
        \item Hualei Zhu, Alibaba Group
        \item King Zhu, OPPO
    \end{itemize}
\end{multicols}

\textbf{Organization and Senior Advisory Committee (Alphabetical Order)}
\begin{multicols}{2}
    \begin{itemize}
        \item Bryan Dai, Ubiquant
        \item Aishan Liu, Beihang University
        \item Zhoujun Li, Beihang University
        \item Chenghua Lin, University of Manchester
        \item Jiaheng Liu, Nanjing University
        \item Tianyu Liu, Peking University
        \item Chao Peng, ByteDance
        \item Kai Shen, ByteDance
        \item Libo Qin, Central South University
        \item Shuangyong Song, Institute of Artificial Intelligence (TeleAI), China Telecom
        \item Ge Zhang, M-A-P
        \item Jiajun Zhang, CASIA
        \item Jie Zhang, Institute of Artificial Intelligence (TeleAI), China Telecom
        \item Zhaoxiang Zhang, CASIA
        \item Zizheng Zhan, Kuaishou
        \item Bo Zheng, Alibaba Group
    \end{itemize}
\end{multicols}
This work is supported by State Key Laboratory of Complex \& Critical Software Environment (SKLCCSE) of Beihang University. Tyler Loakman and Joseph James are supported by the Centre for Doctoral Training in Speech and Language Technologies (SLT) and their Applications funded by UK Research and Innovation [grant number EP/S023062/1].

\bibliography{code_survey}

\begin{thebibliography}{1344}
\providecommand{\natexlab}[1]{#1}
\providecommand{\url}[1]{\texttt{#1}}
\expandafter\ifx\csname urlstyle\endcsname\relax
  \providecommand{\doi}[1]{doi: #1}\else
  \providecommand{\doi}{doi: \begingroup \urlstyle{rm}\Url}\fi

\bibitem[Eli(2024)]{Elicitron}
\emph{Elicitron: A Framework for Simulating Design Requirements Elicitation Using Large Language Model Agents}, volume Volume 3B: 50th Design Automation Conference (DAC) of \emph{International Design Engineering Technical Conferences and Computers and Information in Engineering Conference}, 08 2024.
\newblock URL \url{https://doi.org/10.1115/DETC2024-143598}.

\bibitem[01.ai(2024)]{yicoder2024}
01.ai.
\newblock Yi-coder.
\newblock \url{https://github.com/01-ai/Yi-Coder}, 2024.
\newblock Accessed: 2025-09-20.

\bibitem[2024(2024)]{aime24}
Aime 2024.
\newblock Mathematical association of america., 2024.

\bibitem[2025(2024)]{aime25}
Aime 2025.
\newblock Mathematical association of america., 2024.

\bibitem[Abadeer et~al.(2022)Abadeer, Moeini, Sewell, Branco, Ventura, and Shi]{jsbert}
Macarious Abadeer, Behrad Moeini, Emma Sewell, Paula Branco, Felipe Ventura, and Wei Shi.
\newblock Dynamic extraction of bert-based embeddings for the detection of malicious javascript.
\newblock In \emph{Proceedings of the 32nd Annual International Conference on Computer Science and Software Engineering}, pages 110--119, 2022.

\bibitem[Abuelsaad et~al.(2024)Abuelsaad, Akkil, Dey, Jagmohan, Vempaty, and Kokku]{abuelsaad2024agent}
Tamer Abuelsaad, Deepak Akkil, Prasenjit Dey, Ashish Jagmohan, Aditya Vempaty, and Ravi Kokku.
\newblock Agent-e: From autonomous web navigation to foundational design principles in agentic systems.
\newblock \emph{arXiv preprint arXiv:2407.13032}, 2024.

\bibitem[Adnan et~al.(2025)Adnan, Xu, and Kuhn]{adnan2025pycapsule}
Muntasir Adnan, Zhiwei Xu, and Carlos C.~N. Kuhn.
\newblock Large language model guided self-debugging code generation (pycapsule).
\newblock \emph{arXiv preprint arXiv:2502.02928}, 2025.
\newblock URL \url{https://arxiv.org/abs/2502.02928}.

\bibitem[Agakov et~al.(2006)Agakov, Bonilla, Cavazos, Franke, Fursin, O'Boyle, Thomson, Toussaint, and Williams]{DBLP:conf/cgo/AgakovBCFFOTTW06}
Felix~V. Agakov, Edwin~V. Bonilla, John Cavazos, Bj{\"{o}}rn Franke, Grigori Fursin, Michael F.~P. O'Boyle, John Thomson, Marc Toussaint, and Christopher K.~I. Williams.
\newblock Using machine learning to focus iterative optimization.
\newblock In \emph{Fourth {IEEE/ACM} International Symposium on Code Generation and Optimization {(CGO} 2006), 26-29 March 2006, New York, New York, {USA}}, pages 295--305. {IEEE} Computer Society, 2006.

\bibitem[Agarwal et~al.()Agarwal, Pei, Alamir, and Liu]{agarwal2408codemirage}
V~Agarwal, Y~Pei, S~Alamir, and X~Liu.
\newblock Codemirage: Hallucinations in code generated by large language models (2024).
\newblock \emph{arXiv preprint arXiv:2408.08333}.

\bibitem[Agrawal(2024)]{Androidmeda-Deobfuscate-android-app_2024}
Vaibhav Agrawal.
\newblock {Androidmeda-Deobfuscate-android-app: Android app Vulnerability Scanner and Deobfuscator using LLM}, November 2024.
\newblock URL \url{https://github.com/In3tinct/deobfuscate-android-app}.

\bibitem[Ahmad et~al.(2025{\natexlab{a}})Ahmad, Majumdar, Ficek, Narenthiran, Samadi, Huang, Jain, Noroozi, and Ginsburg]{opencode2}
Wasi~Uddin Ahmad, Somshubra Majumdar, Aleksander Ficek, Sean Narenthiran, Mehrzad Samadi, Jocelyn Huang, Siddhartha Jain, Vahid Noroozi, and Boris Ginsburg.
\newblock Opencodereasoning-ii: A simple test time scaling approach via self-critique.
\newblock \emph{arXiv preprint arXiv:2507.09075}, 2025{\natexlab{a}}.

\bibitem[Ahmad et~al.(2025{\natexlab{b}})Ahmad, Narenthiran, Majumdar, Ficek, Jain, Huang, Noroozi, and Ginsburg]{ahmad2025opencodereasoning}
Wasi~Uddin Ahmad, Sean Narenthiran, Somshubra Majumdar, Aleksander Ficek, Siddhartha Jain, Jocelyn Huang, Vahid Noroozi, and Boris Ginsburg.
\newblock Opencodereasoning: Advancing data distillation for competitive coding.
\newblock \emph{arXiv preprint arXiv:2504.01943}, 2025{\natexlab{b}}.

\bibitem[Ahmad et~al.(2025{\natexlab{c}})Ahmad, Narenthiran, Majumdar, Ficek, Jain, Huang, Noroozi, and Ginsburg]{opencode}
Wasi~Uddin Ahmad, Sean Narenthiran, Somshubra Majumdar, Aleksander Ficek, Siddhartha Jain, Jocelyn Huang, Vahid Noroozi, and Boris Ginsburg.
\newblock Opencodereasoning: Advancing data distillation for competitive coding.
\newblock \emph{arXiv preprint arXiv:2504.01943}, 2025{\natexlab{c}}.

\bibitem[Ahmadian et~al.(2024)Ahmadian, Cremer, Gallé, Fadaee, Kreutzer, Pietquin, Üstün, and Hooker]{ahmadian2024rloo}
Arash Ahmadian, Chris Cremer, Matthias Gallé, Marzieh Fadaee, Julia Kreutzer, Olivier Pietquin, Ahmet Üstün, and Sara Hooker.
\newblock Back to basics: Revisiting reinforce style optimization for learning from human feedback in llms, 2024.
\newblock URL \url{https://arxiv.org/abs/2402.14740}.

\bibitem[Ahmed et~al.(2024)Ahmed, Sobania, et~al.]{ahmed2024repairAgent}
Islem~Bouzenia Ahmed, Premkumar Sobania, et~al.
\newblock {RepairAgent}: An autonomous, {LLM}-based agent for program repair.
\newblock \emph{arXiv preprint arXiv:2403.17134}, 2024.

\bibitem[Ahmed et~al.(2022)Ahmed, Ledesma, and Devanbu]{ahmed2022synshine}
Toufique Ahmed, Noah~Rose Ledesma, and Premkumar Devanbu.
\newblock Synshine: Improved fixing of syntax errors.
\newblock \emph{IEEE Transactions on Software Engineering}, 49\penalty0 (4):\penalty0 2169--2181, 2022.

\bibitem[Ahmed et~al.(2025)Ahmed, Ganhotra, Pan, Shinnar, Sinha, and Hirzel]{ahmed2025otter}
Toufique Ahmed, Jatin Ganhotra, Rangeet Pan, Avraham Shinnar, Saurabh Sinha, and Martin Hirzel.
\newblock Otter: Generating tests from issues to validate swe patches.
\newblock \emph{arXiv preprint arXiv:2502.05368}, 2025.

\bibitem[Ahn et~al.(2024)Ahn, Choi, Yu, Kang, and Choi]{ahn2024tuning}
Daechul Ahn, Yura Choi, Youngjae Yu, Dongyeop Kang, and Jonghyun Choi.
\newblock Tuning large multimodal models for videos using reinforcement learning from ai feedback.
\newblock \emph{arXiv preprint arXiv:2402.03746}, 2024.

\bibitem[AI(2024)]{mistral2024codestral}
Mistral AI.
\newblock Codestral-22b-v0.1.
\newblock \url{https://huggingface.co/mistralai/Codestral-22B-v0.1}, 2024.
\newblock Accessed: 2025-09-20.

\bibitem[AI and AI(2025)]{devstral2025}
Mistral AI and All~Hands AI.
\newblock Devstral: Agentic llm for software engineering.
\newblock \url{https://mistral.ai/news/devstral}, May 2025.
\newblock Research blog; Apache 2.0; Devstral Small (24B) and Devstral Medium.

\bibitem[AI(2025{\natexlab{a}})]{moonshotai2025kimik2}
Moonshot AI.
\newblock Kimi-k2-instruct.
\newblock \url{https://huggingface.co/moonshotai/Kimi-K2-Instruct}, 2025{\natexlab{a}}.

\bibitem[AI(2023{\natexlab{a}})]{nampdn_tiny_codes}
NamPDN AI.
\newblock Tiny codes dataset, 2023{\natexlab{a}}.
\newblock URL \url{https://huggingface.co/datasets/nampdn-ai/tiny-codes}.
\newblock Accessed: 2024.

\bibitem[AI(2023{\natexlab{b}})]{stabilityai2023stablecode}
Stability AI.
\newblock Stablecode: A 3b parameter code language model, 2023{\natexlab{b}}.
\newblock URL \url{https://huggingface.co/stabilityai/stable-code-3b}.
\newblock Accessed: 2025-09-20.

\bibitem[AI(2022)]{codegeex}
Zhipu AI.
\newblock Codegeex.
\newblock \url{https://github.com/zai-org/CodeGeeX}, 2022.

\bibitem[AI(2025{\natexlab{b}})]{glm46}
Zhipu AI.
\newblock Glm-4.6, 2025{\natexlab{b}}.
\newblock URL \url{https://huggingface.co/zai-org/GLM-4.6}.

\bibitem[{aider-code-edit}(2025)]{aider-code-edit}
{aider-code-edit}.
\newblock aider-code-edit, 2025.
\newblock URL \url{https://aider.chat/docs/leaderboards/edit.html}.

\bibitem[{aider-refactoring-leaderboard}(2025)]{aider-refactoring-leaderboard}
{aider-refactoring-leaderboard}.
\newblock aider-refactoring-leaderboard, 2025.
\newblock URL \url{https://aider.chat/docs/leaderboards/refactor.html}.

\bibitem[Ajibawa-2023(2023)]{ajibawa_code290k}
Ajibawa-2023.
\newblock Code-290k-sharegpt, 2023.
\newblock URL \url{https://huggingface.co/datasets/ajibawa-2023/Code-290k-ShareGPT}.
\newblock Accessed: 2024.

\bibitem[Akhoundali et~al.(2025)Akhoundali, Hamidi, Rietveld, and Gadyatskaya]{akhoundali2025eradicating}
Jafar Akhoundali, Hamidreza Hamidi, Kristian Rietveld, and Olga Gadyatskaya.
\newblock Eradicating the unseen: Detecting, exploiting, and remediating a path traversal vulnerability across github.
\newblock \emph{arXiv preprint arXiv:2505.20186}, 2025.

\bibitem[Al-Qasem et~al.(2024)Al-Qasem, Alhanahnah, Al-Kaswan, Al-Shboul, and Al-Omari]{alqasem2024llms}
Al-Baraa Al-Qasem, Motasem Alhanahnah, Abdalraouf Al-Kaswan, Baraa Al-Shboul, and Mahmoud Al-Omari.
\newblock Llms in web development: Evaluating llm-generated php code unveiling vulnerabilities and limitations.
\newblock \emph{arXiv preprint arXiv:2404.16108}, 2024.

\bibitem[Alagarsamy et~al.(2024)Alagarsamy, Tantithamthavorn, and Aleti]{alagarsamy2024a3test}
Saranya Alagarsamy, Chakkrit Tantithamthavorn, and Aldeida Aleti.
\newblock A3test: Assertion-augmented automated test case generation.
\newblock \emph{Information and Software Technology}, 176:\penalty0 107565, 2024.

\bibitem[Alayrac et~al.(2022)Alayrac, Donahue, Luc, Miech, Barr, Hasson, Lenc, Mensch, Millican, Reynolds, Ring, Rutherford, Cabi, Han, Gong, Samangooei, Monteiro, Menick, Borgeaud, Brock, Nematzadeh, Sharifzadeh, Binkowski, Barreira, Vinyals, Zisserman, and Simonyan]{alayrac2022flamingo}
Jean-Baptiste Alayrac, Jeff Donahue, Pauline Luc, Antoine Miech, Iain Barr, Yana Hasson, Karel Lenc, Arthur Mensch, Katie Millican, Malcolm Reynolds, Roman Ring, Eliza Rutherford, Serkan Cabi, Tengda Han, Zhitao Gong, Sina Samangooei, Marianne Monteiro, Jacob Menick, Sebastian Borgeaud, Andrew Brock, Aida Nematzadeh, Sahand Sharifzadeh, Mikolaj Binkowski, Ricardo Barreira, Oriol Vinyals, Andrew Zisserman, and Karen Simonyan.
\newblock Flamingo: a visual language model for few-shot learning, 2022.
\newblock URL \url{https://arxiv.org/abs/2204.14198}.

\bibitem[Aleithan et~al.(2024)Aleithan, Xue, Mohajer, Nnorom, Uddin, and Wang]{aleithan2024swebenchenhancedcodingbenchmark}
Reem Aleithan, Haoran Xue, Mohammad~Mahdi Mohajer, Elijah Nnorom, Gias Uddin, and Song Wang.
\newblock Swe-bench+: Enhanced coding benchmark for llms, 2024.
\newblock URL \url{https://arxiv.org/abs/2410.06992}.

\bibitem[Alhanahnah and Boshmaf(2024)]{alhanahnah2024depsrag}
Mohannad Alhanahnah and Yazan Boshmaf.
\newblock Depsrag: Towards agentic reasoning and planning for software dependency management.
\newblock \emph{arXiv preprint arXiv:2405.20455}, 2024.

\bibitem[Alhindi and Hallett(2024)]{alhindi2024sandboxing}
Maysara Alhindi and Joseph Hallett.
\newblock Sandboxing adoption in open source ecosystems.
\newblock In \emph{Proceedings of the 12th ACM/IEEE International Workshop on Software Engineering for Systems-of-Systems and Software Ecosystems}, pages 13--20, 2024.

\bibitem[{Alibaba Cloud}(2024)]{alibaba2024tongyi}
{Alibaba Cloud}.
\newblock Tongyi lingma.
\newblock \url{https://tongyi.aliyun.com/lingma}, 2024.

\bibitem[{Alibaba DAMO Academy}(2024)]{qwen2024code}
{Alibaba DAMO Academy}.
\newblock {Qwen Code}: Command line code assistant.
\newblock Technical report, Alibaba Group, 2024.

\bibitem[{AllAboutAI}(2025)]{allaboutai2025cursor}
{AllAboutAI}.
\newblock My cursor ai review 2025: The best ide i've tried.
\newblock \url{https://www.allaboutai.com/ai-reviews/cursor-ai/}, 2025.

\bibitem[Allal et~al.(2023)Allal, Li, Kocetkov, et~al.]{allal2023santacoder}
Loubna~Ben Allal, Raymond Li, Denis Kocetkov, et~al.
\newblock {SantaCoder}: Don't reach for the stars!
\newblock \emph{arXiv preprint arXiv:2301.03988}, 2023.

\bibitem[Allam and Shalan(2024)]{allam2024rtlrepo}
Ahmed Allam and Mohamed Shalan.
\newblock Rtl-repo: A benchmark for evaluating llms on large-scale rtl design projects, 2024.

\bibitem[Allamanis(2019)]{allamanis2019adverse}
Miltiadis Allamanis.
\newblock The adverse effects of code duplication in machine learning models of code.
\newblock In \emph{Proceedings of the 2019 ACM SIGPLAN international symposium on new ideas, new paradigms, and reflections on programming and software}, pages 143--153, 2019.

\bibitem[Allamanis and Sutton(2014)]{mine_from_code}
Miltiadis Allamanis and Charles Sutton.
\newblock Mining idioms from source code.
\newblock In Shing{-}Chi Cheung, Alessandro Orso, and Margaret{-}Anne~D. Storey, editors, \emph{Proceedings of the 22nd {ACM} {SIGSOFT} International Symposium on Foundations of Software Engineering, (FSE-22), Hong Kong, China, November 16 - 22, 2014}, pages 472--483. {ACM}, 2014.
\newblock \doi{10.1145/2635868.2635901}.
\newblock URL \url{https://doi.org/10.1145/2635868.2635901}.

\bibitem[Allamanis and Yin(2025)]{allamanis2025disprovingprogramequivalencellms}
Miltiadis Allamanis and Pengcheng Yin.
\newblock Disproving program equivalence with llms, 2025.
\newblock URL \url{https://arxiv.org/abs/2502.18473}.

\bibitem[Allamanis et~al.(2018)Allamanis, Barr, Devanbu, and Sutton]{allamanis2018surveymachinelearningbig}
Miltiadis Allamanis, Earl~T. Barr, Premkumar Devanbu, and Charles Sutton.
\newblock A survey of machine learning for big code and naturalness, 2018.
\newblock URL \url{https://arxiv.org/abs/1709.06182}.

\bibitem[Allamanis et~al.(2024)Allamanis, Panthaplackel, and Yin]{allamanis2024unsupervisedevaluationcodellms}
Miltiadis Allamanis, Sheena Panthaplackel, and Pengcheng Yin.
\newblock Unsupervised evaluation of code llms with round-trip correctness, 2024.
\newblock URL \url{https://arxiv.org/abs/2402.08699}.

\bibitem[Altmann et~al.(2024)Altmann, Sch{\"o}nberger, Illium, Zorn, Ritz, Haider, Burton, and Gabor]{altmann2024emergence}
Philipp Altmann, Julian Sch{\"o}nberger, Steffen Illium, Maximilian Zorn, Fabian Ritz, Tom Haider, Simon Burton, and Thomas Gabor.
\newblock Emergence in multi-agent systems: A safety perspective.
\newblock In \emph{International Symposium on Leveraging Applications of Formal Methods}, pages 104--120. Springer, 2024.

\bibitem[Altmayer~Pizzorno and Berger(2023)]{Altmayer_Pizzorno_2023}
Juan Altmayer~Pizzorno and Emery~D. Berger.
\newblock Slipcover: Near zero-overhead code coverage for python.
\newblock In \emph{Proceedings of the 32nd ACM SIGSOFT International Symposium on Software Testing and Analysis}, ISSTA ’23, page 1195–1206. ACM, July 2023.
\newblock \doi{10.1145/3597926.3598128}.
\newblock URL \url{http://dx.doi.org/10.1145/3597926.3598128}.

\bibitem[Alur et~al.(2013)Alur, Bodik, Juniwal, Martin, Raghothaman, Seshia, Singh, Solar-Lezama, Torlak, and Udupa]{alur2013sygus}
Rajeev Alur, Rastislav Bodik, Garvit Juniwal, Milo M.~K. Martin, Mukund Raghothaman, Sanjit~A. Seshia, Rishabh Singh, Armando Solar-Lezama, Emina Torlak, and Abhishek Udupa.
\newblock Syntax-guided synthesis.
\newblock In \emph{2013 Formal Methods in Computer-Aided Design}, pages 1--8, 2013.
\newblock \doi{10.1109/FMCAD.2013.6679385}.

\bibitem[{Amazon Web Services}(2024{\natexlab{a}})]{amazon2024codewhisperer}
{Amazon Web Services}.
\newblock Codewhisperer is becoming part of amazon q developer.
\newblock \url{https://docs.aws.amazon.com/codewhisperer/latest/userguide/whisper-legacy.html}, 2024{\natexlab{a}}.

\bibitem[{Amazon Web Services}(2024{\natexlab{b}})]{aws2024codewhisperer}
{Amazon Web Services}.
\newblock {Amazon CodeWhisperer}: {AI} code generator.
\newblock Technical report, Amazon Web Services, Inc., 2024{\natexlab{b}}.

\bibitem[{Amazon Web Services}(2024{\natexlab{c}})]{aws2024qdeveloper}
{Amazon Web Services}.
\newblock Amazon q developer.
\newblock \url{https://aws.amazon.com/q/developer/}, 2024{\natexlab{c}}.

\bibitem[Anil et~al.(2023)Anil, Dai, Firat, Johnson, Lepikhin, Passos, and et~al.]{anil2023palm2technicalreport}
Rohan Anil, Andrew~M. Dai, Orhan Firat, Melvin Johnson, Dmitry Lepikhin, Alexandre Passos, and et~al.
\newblock Palm 2 technical report, 2023.
\newblock URL \url{https://arxiv.org/abs/2305.10403}.

\bibitem[Ansel et~al.(2014)Ansel, Kamil, Veeramachaneni, Ragan{-}Kelley, Bosboom, O'Reilly, and Amarasinghe]{DBLP:conf/IEEEpact/AnselKVRBOA14}
Jason Ansel, Shoaib Kamil, Kalyan Veeramachaneni, Jonathan Ragan{-}Kelley, Jeffrey Bosboom, Una{-}May O'Reilly, and Saman~P. Amarasinghe.
\newblock Opentuner: an extensible framework for program autotuning.
\newblock In Jos{\'{e}}~Nelson Amaral and Josep Torrellas, editors, \emph{International Conference on Parallel Architectures and Compilation, {PACT} '14, Edmonton, AB, Canada, August 24-27, 2014}, pages 303--316. {ACM}, 2014.

\bibitem[Anthropic(2023{\natexlab{a}})]{anthropic2023claude2}
Anthropic.
\newblock Claude 2.
\newblock \url{https://www.anthropic.com/news/claude-2}, 2023{\natexlab{a}}.

\bibitem[Anthropic(2023{\natexlab{b}})]{anthropic2023claude2modelcard}
Anthropic.
\newblock Model card and evaluations for claude models.
\newblock \url{https://www-cdn.anthropic.com/bd2a28d2535bfb0494cc8e2a3bf135d2e7523226/Model-Card-Claude-2.pdf}, 2023{\natexlab{b}}.

\bibitem[Anthropic(2023{\natexlab{c}})]{anthropic2023introducingclaude}
Anthropic.
\newblock Introducing claude.
\newblock \url{https://www.anthropic.com/news/introducing-claude}, 2023{\natexlab{c}}.

\bibitem[Anthropic(2024{\natexlab{a}})]{anthropic2024claude35modelcard}
Anthropic.
\newblock Claude 3.5 sonnet model card addendum.
\newblock \url{https://www-cdn.anthropic.com/fed9cc193a14b84131812372d8d5857f8f304c52/Model_Card_Claude_3_Addendum.pdf}, 2024{\natexlab{a}}.

\bibitem[Anthropic(2024{\natexlab{b}})]{anthropic2024claude35sonnet}
Anthropic.
\newblock Introducing claude 3.5 sonnet.
\newblock \url{https://www.anthropic.com/news/claude-3-5-sonnet}, 2024{\natexlab{b}}.

\bibitem[Anthropic(2024{\natexlab{c}})]{anthropic2024claude3modelcard}
Anthropic.
\newblock The claude 3 model family: Opus, sonnet, haiku.
\newblock \url{https://www-cdn.anthropic.com/de8ba9b01c9ab7cbabf5c33b80b7bbc618857627/Model_Card_Claude_3.pdf}, 2024{\natexlab{c}}.

\bibitem[{Anthropic}(2024)]{anthropic2024claudecode}
{Anthropic}.
\newblock {Claude Code}: Terminal-based {AI} development assistant.
\newblock Technical report, Anthropic, PBC, 2024.

\bibitem[Anthropic(2024{\natexlab{a}})]{anthropic2024introducingclaude3}
Anthropic.
\newblock Introducing the next generation of claude.
\newblock \url{https://www.anthropic.com/news/claude-3-family}, 2024{\natexlab{a}}.

\bibitem[Anthropic(2024{\natexlab{b}})]{mcp}
Anthropic.
\newblock Introducing the model context protocol, 2024{\natexlab{b}}.
\newblock URL \url{https://www.anthropic.com/news/model-context-protocol}.

\bibitem[Anthropic(2025{\natexlab{a}})]{anthropic2025claude45}
Anthropic.
\newblock Anthropic launches claude 4.5.
\newblock \url{https://www.reuters.com/business/retail-consumer/anthropic-launches-claude-45-touts-better-abilities-targets-business-customers-2025-09-29/}, 2025{\natexlab{a}}.

\bibitem[Anthropic(2025{\natexlab{b}})]{anthropic2025claude4systemcard}
Anthropic.
\newblock Claude opus 4 \& claude sonnet 4 system card.
\newblock \url{https://www.anthropic.com/claude-4-system-card}, 2025{\natexlab{b}}.

\bibitem[Anthropic(2025{\natexlab{c}})]{anthropic2025introducingclaude4}
Anthropic.
\newblock Introducing claude 4.
\newblock \url{https://www.anthropic.com/news/claude-4}, 2025{\natexlab{c}}.

\bibitem[anthropic(2025)]{claude3_7}
anthropic.
\newblock Claude 3.7 sonnet, 2025.
\newblock URL \url{https://www.anthropic.com/news/claude-3-7-sonnet/}.

\bibitem[Anthropic(2025)]{claude45}
Anthropic.
\newblock Introducing claude sonnet 4.5, 2025.
\newblock URL \url{https://www.anthropic.com/news/claude-sonnet-4-5}.

\bibitem[Anysphere(2025)]{cursor_ai_2025}
Anysphere.
\newblock Cursor - the ai code editor.
\newblock \url{https://www.cursor.com/en}, 2025.
\newblock URL \url{https://www.cursor.com}.

\bibitem[{Anysphere Inc.}(2024)]{anysphere2024cursor}
{Anysphere Inc.}
\newblock {Cursor}: The {AI-first} code editor.
\newblock Technical report, Anysphere Inc., 2024.
\newblock URL \url{https://cursor.com}.

\bibitem[{Anysphere Inc.}(2025{\natexlab{a}})]{cursor2025changelog}
{Anysphere Inc.}
\newblock Cursor changelog.
\newblock \url{https://cursor.com/changelog}, 2025{\natexlab{a}}.

\bibitem[{Anysphere Inc.}(2025{\natexlab{b}})]{cursor2025features}
{Anysphere Inc.}
\newblock Cursor features.
\newblock \url{https://cursor.com/features}, 2025{\natexlab{b}}.

\bibitem[Armengol{-}Estap{\'{e}} et~al.(2023)Armengol{-}Estap{\'{e}}, Woodruff, Cummins, and O'Boyle]{DBLP:journals/corr/abs-2305-12520}
Jordi Armengol{-}Estap{\'{e}}, Jackson Woodruff, Chris Cummins, and Michael F.~P. O'Boyle.
\newblock Slade: {A} portable small language model decompiler for optimized assembler.
\newblock \emph{CoRR}, abs/2305.12520, 2023.

\bibitem[Arnett et~al.(2024)Arnett, Jones, Yamshchikov, and Langlais]{arnett2024toxicity}
Catherine Arnett, Eliot Jones, Ivan~P Yamshchikov, and Pierre-Carl Langlais.
\newblock Toxicity of the commons: Curating open-source pre-training data.
\newblock \emph{arXiv preprint arXiv:2410.22587}, 2024.

\bibitem[Artuso et~al.(2019)Artuso, Luna, Massarelli, and Querzoni]{2019In}
Fiorella Artuso, Giuseppe Antonio~Di Luna, Luca Massarelli, and Leonardo Querzoni.
\newblock In nomine function: Naming functions in stripped binaries with neural networks.
\newblock 2019.

\bibitem[Asare et~al.(2023)Asare, Jaafar, Ali, and Williams]{asare2023hidden}
Oumou~K Asare, Fadi Jaafar, Naeem Ali, and Laurie Williams.
\newblock The hidden risks of llm-generated web application code: A security-centric evaluation of code generation capabilities in large language models.
\newblock In \emph{Proceedings of the 2023 ACM on Workshop on Secure and Trustworthy Language Processing}, pages 31--42, 2023.

\bibitem[Asif and Orosz(2025)]{pragmaticengineer2025cursor}
Sualeh Asif and Gergely Orosz.
\newblock Real-world engineering challenges: Building cursor.
\newblock \url{https://newsletter.pragmaticengineer.com/p/cursor}, 2025.

\bibitem[Athiwaratkun et~al.(2023{\natexlab{a}})Athiwaratkun, Gouda, Wang, Li, Tian, Tan, Ahmad, Wang, Sun, Shang, Gonugondla, Ding, Kumar, Fulton, Farahani, Jain, Giaquinto, Qian, Ramanathan, Nallapati, Ray, Bhatia, Sengupta, Roth, and Xiang]{mathqax}
Ben Athiwaratkun, Sanjay~Krishna Gouda, Zijian Wang, Xiaopeng Li, Yuchen Tian, Ming Tan, Wasi~Uddin Ahmad, Shiqi Wang, Qing Sun, Mingyue Shang, Sujan~Kumar Gonugondla, Hantian Ding, Varun Kumar, Nathan Fulton, Arash Farahani, Siddhartha Jain, Robert Giaquinto, Haifeng Qian, Murali~Krishna Ramanathan, Ramesh Nallapati, Baishakhi Ray, Parminder Bhatia, Sudipta Sengupta, Dan Roth, and Bing Xiang.
\newblock Multi-lingual evaluation of code generation models, 2023{\natexlab{a}}.
\newblock URL \url{https://arxiv.org/abs/2210.14868}.

\bibitem[Athiwaratkun et~al.(2023{\natexlab{b}})Athiwaratkun, Gouda, Wang, Li, Tian, Tan, Ahmad, Wang, Sun, Shang, Gonugondla, Ding, Kumar, Fulton, Farahani, Jain, Giaquinto, Qian, Ramanathan, Nallapati, Ray, Bhatia, Sengupta, Roth, and Xiang]{mbxp}
Ben Athiwaratkun, Sanjay~Krishna Gouda, Zijian Wang, Xiaopeng Li, Yuchen Tian, Ming Tan, Wasi~Uddin Ahmad, Shiqi Wang, Qing Sun, Mingyue Shang, Sujan~Kumar Gonugondla, Hantian Ding, Varun Kumar, Nathan Fulton, Arash Farahani, Siddhartha Jain, Robert Giaquinto, Haifeng Qian, Murali~Krishna Ramanathan, Ramesh Nallapati, Baishakhi Ray, Parminder Bhatia, Sudipta Sengupta, Dan Roth, and Bing Xiang.
\newblock Multi-lingual evaluation of code generation models, 2023{\natexlab{b}}.
\newblock URL \url{https://arxiv.org/abs/2210.14868}.

\bibitem[Athiwaratkun et~al.(2023{\natexlab{c}})Athiwaratkun, Gouda, Wang, Li, Tian, Tan, Ahmad, Wang, Sun, Shang, Gonugondla, Ding, Kumar, Fulton, Farahani, Jain, Giaquinto, Qian, Ramanathan, Nallapati, Ray, Bhatia, Sengupta, Roth, and Xiang]{multilingualhumaneval}
Ben Athiwaratkun, Sanjay~Krishna Gouda, Zijian Wang, Xiaopeng Li, Yuchen Tian, Ming Tan, Wasi~Uddin Ahmad, Shiqi Wang, Qing Sun, Mingyue Shang, Sujan~Kumar Gonugondla, Hantian Ding, Varun Kumar, Nathan Fulton, Arash Farahani, Siddhartha Jain, Robert Giaquinto, Haifeng Qian, Murali~Krishna Ramanathan, Ramesh Nallapati, Baishakhi Ray, Parminder Bhatia, Sudipta Sengupta, Dan Roth, and Bing Xiang.
\newblock Multi-lingual evaluation of code generation models, 2023{\natexlab{c}}.
\newblock URL \url{https://arxiv.org/abs/2210.14868}.

\bibitem[AugmentCode(2025)]{rldb}
AugmentCode.
\newblock Reinforcement learning from developer behaviors: A breakthrough in code generation quality.
\newblock \url{https://www.augmentcode.com/blog/reinforcement-learning-from-developer-behaviors}, 2025.
\newblock Accessed: 2025.

\bibitem[Austin et~al.(2023)Austin, Johnson, Ho, Tarlow, and van~den Berg]{austin2023d3pm}
Jacob Austin, Daniel~D. Johnson, Jonathan Ho, Daniel Tarlow, and Rianne van~den Berg.
\newblock Structured denoising diffusion models in discrete state-spaces, 2023.
\newblock URL \url{https://arxiv.org/abs/2107.03006}.

\bibitem[Austin(2021)]{austin2021mbpp}
Jacob et~al. Austin.
\newblock Program synthesis with large language models.
\newblock In \emph{ICML}, 2021.

\bibitem[Badertdinov et~al.(2025)Badertdinov, Golubev, Nekrashevich, Shevtsov, Karasik, Andriushchenko, Trofimova, Litvintseva, and Yangel]{badertdinov2025swerebenchautomatedpipelinetask}
Ibragim Badertdinov, Alexander Golubev, Maksim Nekrashevich, Anton Shevtsov, Simon Karasik, Andrei Andriushchenko, Maria Trofimova, Daria Litvintseva, and Boris Yangel.
\newblock Swe-rebench: An automated pipeline for task collection and decontaminated evaluation of software engineering agents, 2025.
\newblock URL \url{https://arxiv.org/abs/2505.20411}.

\bibitem[Bahdanau et~al.(2015)Bahdanau, Cho, and Bengio]{bahdanau2015nmt}
Dzmitry Bahdanau, Kyunghyun Cho, and Yoshua Bengio.
\newblock Neural machine translation by jointly learning to align and translate.
\newblock In \emph{ICLR}, 2015.
\newblock URL \url{http://arxiv.org/abs/1409.0473}.

\bibitem[Bai et~al.(2023{\natexlab{a}})Bai, Bai, Chu, Cui, Dang, Deng, Fan, Ge, Han, Huang, Hui, Ji, Li, Lin, Lin, Liu, Liu, Lu, Lu, Ma, Men, Ren, Ren, Tan, Tan, Tu, Wang, Wang, Wang, Wu, Xu, Xu, Yang, Yang, Yang, Yang, Yao, Yu, Yuan, Yuan, Zhang, Zhang, Zhang, Zhang, Zhou, Zhou, Zhou, and Zhu]{bai2023qwen}
Jinze Bai, Shuai Bai, Yunfei Chu, Zeyu Cui, Kai Dang, Xiaodong Deng, Yang Fan, Wenbin Ge, Yu~Han, Fei Huang, Binyuan Hui, Luo Ji, Mei Li, Junyang Lin, Runji Lin, Dayiheng Liu, Gao Liu, Chengqiang Lu, Keming Lu, Jianxin Ma, Rui Men, Xingzhang Ren, Xuancheng Ren, Chuanqi Tan, Sinan Tan, Jianhong Tu, Peng Wang, Shijie Wang, Wei Wang, Shengguang Wu, Benfeng Xu, Jin Xu, An~Yang, Hao Yang, Jian Yang, Shusheng Yang, Yang Yao, Bowen Yu, Hongyi Yuan, Zheng Yuan, Jianwei Zhang, Xingxuan Zhang, Yichang Zhang, Zhenru Zhang, Chang Zhou, Jingren Zhou, Xiaohuan Zhou, and Tianhang Zhu.
\newblock Qwen technical report, 2023{\natexlab{a}}.
\newblock URL \url{https://arxiv.org/abs/2309.16609}.

\bibitem[Bai et~al.(2023{\natexlab{b}})Bai, Bai, Yang, Wang, Tan, Wang, Lin, Zhou, and Zhou]{qwen2023qwenvl}
Jinze Bai, Shuai Bai, Shusheng Yang, Shijie Wang, Sinan Tan, Peng Wang, Junyang Lin, Chang Zhou, and Jingren Zhou.
\newblock Qwen-vl: A versatile vision-language model for understanding, localization, text reading, and beyond.
\newblock \emph{arXiv preprint arXiv:2308.12966}, 2023{\natexlab{b}}.

\bibitem[Bai et~al.(2025{\natexlab{a}})Bai, Cai, Chen, Chen, Chen, Cheng, Deng, Ding, Gao, Ge, et~al.]{qwen3vl}
Shuai Bai, Yuxuan Cai, Ruizhe Chen, Keqin Chen, Xionghui Chen, Zesen Cheng, Lianghao Deng, Wei Ding, Chang Gao, Chunjiang Ge, et~al.
\newblock Qwen3-vl technical report.
\newblock \emph{arXiv preprint arXiv:2511.21631}, 2025{\natexlab{a}}.

\bibitem[Bai et~al.(2025{\natexlab{b}})Bai, Chen, Liu, Wang, Ge, Song, Dang, Wang, Wang, Tang, Zhong, Zhu, Yang, Li, Wan, Wang, Ding, Fu, Xu, Ye, Zhang, Xie, Cheng, Zhang, Yang, Xu, and Lin]{qwen2025qwen2_5vl}
Shuai Bai, Keqin Chen, Xuejing Liu, Jialin Wang, Wenbin Ge, Sibo Song, Kai Dang, Peng Wang, Shijie Wang, Jun Tang, Humen Zhong, Yuanzhi Zhu, Mingkun Yang, Zhaohai Li, Jianqiang Wan, Pengfei Wang, Wei Ding, Zheren Fu, Yiheng Xu, Jiabo Ye, Xi~Zhang, Tianbao Xie, Zesen Cheng, Hang Zhang, Zhibo Yang, Haiyang Xu, and Junyang Lin.
\newblock Qwen2.5-vl technical report.
\newblock \emph{arXiv preprint arXiv:2502.13923}, 2025{\natexlab{b}}.

\bibitem[Bai et~al.(2025{\natexlab{c}})Bai, Chen, Liu, Wang, Ge, Song, Dang, Wang, Wang, Tang, et~al.]{qwen25vl}
Shuai Bai, Keqin Chen, Xuejing Liu, Jialin Wang, Wenbin Ge, Sibo Song, Kai Dang, Peng Wang, Shijie Wang, Jun Tang, et~al.
\newblock Qwen2. 5-vl technical report.
\newblock \emph{arXiv preprint arXiv:2502.13923}, 2025{\natexlab{c}}.

\bibitem[Bai et~al.(2024)Bai, Chiu, Wu, Huang, Hsiao, and Hsu]{bai2024apilot}
Weiheng Bai, Wei-Yang Chiu, Peng-Fei Wu, Chun-Ying Huang, Hsu-Chun Hsiao, and Wen-Lian Hsu.
\newblock Apilot: Navigating large language models to generate secure code by sidestepping outdated api pitfalls, 2024.

\bibitem[Bai et~al.(2022{\natexlab{a}})Bai, Jones, Ndousse, Askell, Chen, DasSarma, Drain, Fort, Ganguli, Henighan, et~al.]{bai2022training}
Yuntao Bai, Andy Jones, Kamal Ndousse, Amanda Askell, Anna Chen, Nova DasSarma, Dawn Drain, Stanislav Fort, Deep Ganguli, Tom Henighan, et~al.
\newblock Training a helpful and harmless assistant with reinforcement learning from human feedback.
\newblock \emph{arXiv preprint arXiv:2204.05862}, 2022{\natexlab{a}}.

\bibitem[Bai et~al.(2022{\natexlab{b}})Bai, Kadavath, Kundu, Askell, Kernion, Jones, Chen, Goldie, Mirhoseini, McKinnon, et~al.]{bai2022constitutional}
Yuntao Bai, Saurav Kadavath, Sandipan Kundu, Amanda Askell, Jackson Kernion, Andy Jones, Anna Chen, Anna Goldie, Azalia Mirhoseini, Cameron McKinnon, et~al.
\newblock Constitutional ai: Harmlessness from ai feedback.
\newblock \emph{arXiv preprint arXiv:2212.08073}, 2022{\natexlab{b}}.

\bibitem[Bairi et~al.(2023)Bairi, Sonwane, Kanade, C, Iyer, Parthasarathy, Rajamani, Ashok, and Shet]{bairi2023codeplanrepositorylevelcodingusing}
Ramakrishna Bairi, Atharv Sonwane, Aditya Kanade, Vageesh~D C, Arun Iyer, Suresh Parthasarathy, Sriram Rajamani, B.~Ashok, and Shashank Shet.
\newblock Codeplan: Repository-level coding using llms and planning, 2023.
\newblock URL \url{https://arxiv.org/abs/2309.12499}.

\bibitem[Baker et~al.(2022)Baker, Akkaya, Zhokhov, Huizinga, Tang, Ecoffet, Houghton, Sampedro, and Clune]{baker2022videopretrainingvpt}
Bowen Baker, Ilge Akkaya, Peter Zhokhov, Joost Huizinga, Jie Tang, Adrien Ecoffet, Brandon Houghton, Raul Sampedro, and Jeff Clune.
\newblock Video pretraining (vpt): Learning to act by watching unlabeled online videos, 2022.
\newblock URL \url{https://arxiv.org/abs/2206.11795}.

\bibitem[Banerjee et~al.(2021)Banerjee, Pal, Wang, and Baral]{DBLP:journals/corr/abs-2103-12801}
Pratyay Banerjee, Kuntal~Kumar Pal, Fish Wang, and Chitta Baral.
\newblock Variable name recovery in decompiled binary code using constrained masked language modeling.
\newblock \emph{CoRR}, abs/2103.12801, 2021.

\bibitem[Banerjee and Lavie(2005)]{banerjee2005meteor}
Satanjeev Banerjee and Alon Lavie.
\newblock Meteor: An automatic metric for mt evaluation with improved correlation with human judgments.
\newblock In \emph{Proceedings of the acl workshop on intrinsic and extrinsic evaluation measures for machine translation and/or summarization}, pages 65--72, 2005.

\bibitem[Bates et~al.(2025)Bates, Vavricka, Carleton, Shao, and Pan]{bates2025unified}
Averi Bates, Ryan Vavricka, Shane Carleton, Ruosi Shao, and Chongle Pan.
\newblock Unified modeling language code generation from diagram images using multimodal large language models.
\newblock \emph{Machine Learning with Applications}, page 100660, 2025.

\bibitem[Bavarian et~al.(2022)Bavarian, Jun, Tezak, Schulman, McLeavey, Tworek, and Chen]{bavarian2022efficient}
Mohammad Bavarian, Heewoo Jun, Nikolas Tezak, John Schulman, Christine McLeavey, Jerry Tworek, and Mark Chen.
\newblock Efficient training of language models to fill in the middle.
\newblock \emph{arXiv preprint arXiv:2207.14255}, 2022.

\bibitem[Beau and Crabbé(2024)]{codeinsight}
Nathanaël Beau and Benoît Crabbé.
\newblock Codeinsight: A curated dataset of practical coding solutions from stack overflow, 2024.
\newblock URL \url{https://arxiv.org/abs/2409.16819}.

\bibitem[Belcak et~al.(2025)Belcak, Heinrich, Diao, Fu, Dong, Muralidharan, Lin, and Molchanov]{belcak2025small_agentic_ai}
Peter Belcak, Greg Heinrich, Shizhe Diao, Yonggan Fu, Xin Dong, Saurav Muralidharan, Yingyan~Celine Lin, and Pavlo Molchanov.
\newblock Small language models are the future of agentic ai, 2025.
\newblock \emph{URL https://arxiv. org/abs/2506.02153}, 2025.

\bibitem[Beltramelli(2017)]{beltramelli2017pix2code}
Tony Beltramelli.
\newblock pix2code: Generating code from a graphical user interface screenshot, 2017.
\newblock URL \url{https://arxiv.org/abs/1705.07962}.

\bibitem[Beltramelli(2018)]{beltramelli2018pix2code}
Tony Beltramelli.
\newblock pix2code: Generating code from a graphical user interface screenshot.
\newblock In \emph{Proceedings of the ACM SIGCHI symposium on engineering interactive computing systems}, pages 1--6, 2018.

\bibitem[Bengio et~al.(2015)Bengio, Vinyals, Jaitly, and Shazeer]{bengio2015scheduled}
Samy Bengio, Oriol Vinyals, Navdeep Jaitly, and Noam Shazeer.
\newblock Scheduled sampling for sequence prediction with recurrent neural networks, 2015.
\newblock URL \url{https://arxiv.org/abs/1506.03099}.

\bibitem[Bhandari et~al.(2021)Bhandari, Naseer, and Moonen]{bhandari2021cvefixes}
Guru Bhandari, Amara Naseer, and Leon Moonen.
\newblock Cvefixes: automated collection of vulnerabilities and their fixes from open-source software.
\newblock In \emph{Proceedings of the 17th International Conference on Predictive Models and Data Analytics in Software Engineering}, pages 30--39, 2021.

\bibitem[Bhandari et~al.(2025)Bhandari, Gavric, and Shalaginov]{bhandari2025generating}
Guru Bhandari, Nikola Gavric, and Andrii Shalaginov.
\newblock Generating vulnerability security fixes with code language models.
\newblock \emph{Information and Software Technology}, page 107786, 2025.

\bibitem[Bhatia et~al.(2025)Bhatia, Oliva, Rajbahadur, Zhang, Chen, Chen, Leung, Lin, Chen, and Hassan]{bhatia2025spice}
Aaditya Bhatia, Gustavo~A Oliva, Gopi~Krishnan Rajbahadur, Haoxiang Zhang, Yihao Chen, Zhilong Chen, Arthur Leung, Dayi Lin, Boyuan Chen, and Ahmed~E Hassan.
\newblock Spice: An automated swe-bench labeling pipeline for issue clarity, test coverage, and effort estimation.
\newblock \emph{arXiv preprint arXiv:2507.09108}, 2025.

\bibitem[Bhattarai et~al.(2024{\natexlab{a}})Bhattarai, Santos, Jones, Biswas, Alexandrov, and O'Malley]{bhattarai2024enhancing0}
Manish Bhattarai, Javier~E. Santos, Shawn Jones, Ayan Biswas, Boian Alexandrov, and Daniel O'Malley.
\newblock Enhancing code translation in language models with few-shot learning via retrieval-augmented generation.
\newblock \emph{arXiv preprint arXiv: 2407.19619}, 2024{\natexlab{a}}.

\bibitem[Bhattarai et~al.(2024{\natexlab{b}})Bhattarai, Vu, Santos, Boureima, and Malley]{bhattarai2024enhancing1}
Manish Bhattarai, Minh Vu, Javier~E. Santos, Ismael Boureima, and Daniel~O' Malley.
\newblock Enhancing cross-language code translation via task-specific embedding alignment in retrieval-augmented generation.
\newblock \emph{arXiv preprint arXiv: 2412.05159}, 2024{\natexlab{b}}.

\bibitem[Bichsel et~al.(2016)Bichsel, Raychev, Tsankov, and Vechev]{DBLP:conf/ccs/BichselRTV16}
Benjamin Bichsel, Veselin Raychev, Petar Tsankov, and Martin~T. Vechev.
\newblock Statistical deobfuscation of android applications.
\newblock In Edgar~R. Weippl, Stefan Katzenbeisser, Christopher Kruegel, Andrew~C. Myers, and Shai Halevi, editors, \emph{Proceedings of the 2016 {ACM} {SIGSAC} Conference on Computer and Communications Security, Vienna, Austria, October 24-28, 2016}, pages 343--355. {ACM}, 2016.

\bibitem[BigCode(2023)]{bigcode_commitpackft}
BigCode.
\newblock Commitpackft: A dataset of git commits for fine-tuning, 2023.
\newblock URL \url{https://huggingface.co/datasets/bigcode/commitpackft}.
\newblock Accessed: 2024.

\bibitem[BigCode(2024)]{bigcode_self_oss}
BigCode.
\newblock Self-oss-instruct-sc2-exec-filter-50k, 2024.
\newblock URL \url{https://huggingface.co/datasets/bigcode/self-oss-instruct-sc2-exec-filter-50k}.
\newblock Accessed: 2024.

\bibitem[{Bito}(2024)]{bito_ai}
{Bito}.
\newblock Bito ai: Ai-powered code assistant.
\newblock \url{https://bito.ai}, 2024.

\bibitem[Black et~al.(2022)Black, Biderman, Hallahan, Anthony, Gao, Golding, He, Leahy, McDonell, Phang, et~al.]{gpt_neox_20b}
Sid Black, Stella Biderman, Eric Hallahan, Quentin Anthony, Leo Gao, Laurence Golding, Horace He, Connor Leahy, Kyle McDonell, Jason Phang, et~al.
\newblock Gpt-neox-20b: An open-source autoregressive language model.
\newblock \emph{arXiv preprint arXiv:2204.06745}, 2022.

\bibitem[Bodin et~al.(1998)Bodin, Kisuki, Knijnenburg, Boyle, and Rohou]{1998Iterative}
Franois Bodin, Toru Kisuki, Peter Knijnenburg, Mike~O' Boyle, and Erven Rohou.
\newblock Iterative compilation in a non-linear optimisation space.
\newblock In \emph{Workshop on Profile and Feedback-Directed Compilation}, 1998.

\bibitem[Bogomolov et~al.(2024)Bogomolov, Eliseeva, Galimzyanov, Glukhov, Shapkin, Tigina, Golubev, Kovrigin, van Deursen, Izadi, et~al.]{bogomolov2024long}
Egor Bogomolov, Aleksandra Eliseeva, Timur Galimzyanov, Evgeniy Glukhov, Anton Shapkin, Maria Tigina, Yaroslav Golubev, Alexander Kovrigin, Arie van Deursen, Maliheh Izadi, et~al.
\newblock Long code arena: a set of benchmarks for long-context code models.
\newblock \emph{arXiv preprint arXiv:2406.11612}, 2024.

\bibitem[Bolukbasi et~al.(2016)Bolukbasi, Chang, Zou, Saligrama, and Kalai]{bolukbasi2016man}
Tolga Bolukbasi, Kai-Wei Chang, James~Y Zou, Venkatesh Saligrama, and Adam~T Kalai.
\newblock Man is to computer programmer as woman is to homemaker? debiasing word embeddings.
\newblock In \emph{Advances in neural information processing systems}, volume~29, 2016.

\bibitem[Bouafif et~al.(2025)Bouafif, Hamdaqa, and Zulkoski]{bouafif2025primg}
Mohamed~Salah Bouafif, Mohammad Hamdaqa, and Edward Zulkoski.
\newblock Primg: Efficient llm-driven test generation using mutant prioritization.
\newblock \emph{arXiv preprint arXiv:2505.05584}, 2025.

\bibitem[Bouhlal et~al.(2024)Bouhlal, Kassou, Lamdouar, and Bouyahyaoui]{bouhlal2024assembly}
Laila Bouhlal, Fouzia Kassou, Nouzha Lamdouar, and Azeddine Bouyahyaoui.
\newblock Assembly procedure for elementary matrices of train-track-bridge railway system.
\newblock \emph{arXiv preprint arXiv:2406.14837}, 2024.

\bibitem[Bouzenia and Pradel(2025)]{bouzenia2025you}
Islem Bouzenia and Michael Pradel.
\newblock You name it, i run it: An llm agent to execute tests of arbitrary projects.
\newblock \emph{Proceedings of the ACM on Software Engineering}, 2\penalty0 (ISSTA):\penalty0 1054--1076, 2025.

\bibitem[Bouzenia et~al.()Bouzenia, Devanbu, and Pradel]{bouzenia2403repairagent}
Islem Bouzenia, Premkumar Devanbu, and Michael Pradel.
\newblock Repairagent: an autonomous, llm-based agent for program repair.(2024).
\newblock \emph{arXiv preprint arXiv:2403.17134}.

\bibitem[Brown and et~al.(2024)]{brown2024largelanguagemonkeys}
B.~Brown and et~al.
\newblock Large language monkeys: Scaling inference compute with repeated sampling.
\newblock \emph{arXiv preprint arXiv:2407.21787}, 2024.
\newblock URL \url{https://arxiv.org/abs/2407.21787}.

\bibitem[Brown et~al.(2020{\natexlab{a}})Brown, Mann, Ryder, Subbiah, Kaplan, Dhariwal, Neelakantan, Shyam, Sastry, Askell, Agarwal, Herbert-Voss, Krueger, Henighan, Child, Ramesh, Ziegler, Wu, Winter, Hesse, Chen, Sigler, Litwin, Gray, Chess, Clark, Berner, McCandlish, Radford, Sutskever, and Amodei]{brown2020gpt3}
Tom Brown, Benjamin Mann, Nick Ryder, Melanie Subbiah, Jared~D Kaplan, Prafulla Dhariwal, Arvind Neelakantan, Pranav Shyam, Girish Sastry, Amanda Askell, Sandhini Agarwal, Ariel Herbert-Voss, Gretchen Krueger, Tom Henighan, Rewon Child, Aditya Ramesh, Daniel Ziegler, Jeffrey Wu, Clemens Winter, Chris Hesse, Mark Chen, Eric Sigler, Mateusz Litwin, Scott Gray, Benjamin Chess, Jack Clark, Christopher Berner, Sam McCandlish, Alec Radford, Ilya Sutskever, and Dario Amodei.
\newblock Language models are few-shot learners.
\newblock In H.~Larochelle, M.~Ranzato, R.~Hadsell, M.F. Balcan, and H.~Lin, editors, \emph{Advances in Neural Information Processing Systems}, volume~33, pages 1877--1901. Curran Associates, Inc., 2020{\natexlab{a}}.
\newblock URL \url{https://proceedings.neurips.cc/paper_files/paper/2020/file/1457c0d6bfcb4967418bfb8ac142f64a-Paper.pdf}.

\bibitem[Brown et~al.(2020{\natexlab{b}})Brown, Mann, Ryder, Subbiah, Kaplan, Dhariwal, Neelakantan, Shyam, Sastry, Askell, et~al.]{brown2020language}
Tom Brown, Benjamin Mann, Nick Ryder, Melanie Subbiah, Jared~D Kaplan, Prafulla Dhariwal, Arvind Neelakantan, Pranav Shyam, Girish Sastry, Amanda Askell, et~al.
\newblock Language models are few-shot learners.
\newblock \emph{Advances in neural information processing systems}, 33:\penalty0 1877--1901, 2020{\natexlab{b}}.

\bibitem[Brown et~al.(2020{\natexlab{c}})Brown, Mann, Ryder, Subbiah, Kaplan, Dhariwal, Neelakantan, Shyam, Sastry, Askell, Agarwal, Herbert{-}Voss, Krueger, Henighan, Child, Ramesh, Ziegler, Wu, Winter, Hesse, Chen, Sigler, Litwin, Gray, Chess, Clark, Berner, McCandlish, Radford, Sutskever, and Amodei]{DBLP:journals/corr/abs-2005-14165}
Tom~B. Brown, Benjamin Mann, Nick Ryder, Melanie Subbiah, Jared Kaplan, Prafulla Dhariwal, Arvind Neelakantan, Pranav Shyam, Girish Sastry, Amanda Askell, Sandhini Agarwal, Ariel Herbert{-}Voss, Gretchen Krueger, Tom Henighan, Rewon Child, Aditya Ramesh, Daniel~M. Ziegler, Jeffrey Wu, Clemens Winter, Christopher Hesse, Mark Chen, Eric Sigler, Mateusz Litwin, Scott Gray, Benjamin Chess, Jack Clark, Christopher Berner, Sam McCandlish, Alec Radford, Ilya Sutskever, and Dario Amodei.
\newblock Language models are few-shot learners.
\newblock \emph{CoRR}, abs/2005.14165, 2020{\natexlab{c}}.
\newblock URL \url{https://arxiv.org/abs/2005.14165}.

\bibitem[Bugdaryan(2024)]{bugdaryan_sql}
Bugdaryan.
\newblock Sql create context instruction, 2024.
\newblock URL \url{https://huggingface.co/datasets/bugdaryan/sql-create-context-instruction}.
\newblock Accessed: 2024.

\bibitem[{ByteByteGo}(2025)]{bytebytego2025cursor}
{ByteByteGo}.
\newblock How cursor serves billions of ai code completions every day.
\newblock \url{https://blog.bytebytego.com/p/how-cursor-serves-billions-of-ai}, 2025.

\bibitem[Cai et~al.(2025)Cai, Li, Wang, Zhu, Shen, Li, and Chua]{cai2025personalwab}
Hongru Cai, Yongqi Li, Wenjie Wang, Fengbin Zhu, Xiaoyu Shen, Wenjie Li, and Tat-Seng Chua.
\newblock Large language models empowered personalized web agents, 2025.
\newblock URL \url{https://arxiv.org/abs/2410.17236}.

\bibitem[Cai et~al.(2020)Cai, Liang, Xu, Li, Hao, and Chen]{cai2020tag}
Ruichu Cai, Zhihao Liang, Boyan Xu, Zijian Li, Yuexing Hao, and Yao Chen.
\newblock Tag: Type auxiliary guiding for code comment generation.
\newblock In \emph{Proceedings of the 58th Annual Meeting of the Association for Computational Linguistics}, pages 291--301, 2020.

\bibitem[{Cai} et~al.(2025{\natexlab{a}}){Cai}, {Gu}, {Du}, {Ye}, {Cao}, {Xu}, {Feng}, and {Chen}]{MIRAGE}
Yin {Cai}, Zhouhong {Gu}, Zhaohan {Du}, Zheyu {Ye}, Shaosheng {Cao}, Yiqian {Xu}, Hongwei {Feng}, and Ping {Chen}.
\newblock {MIRAGE: Exploring How Large Language Models Perform in Complex Social Interactive Environments}.
\newblock \emph{arXiv e-prints}, art. arXiv:2501.01652, January 2025{\natexlab{a}}.
\newblock \doi{10.48550/arXiv.2501.01652}.

\bibitem[{Cai} et~al.(2025{\natexlab{b}}){Cai}, {Wang}, {Satheesh}, {Nakhawa}, {Jae}, {Powell}, {Liu}, {Jay}, {Oh}, {Wang}, {Liang}, {Goldstein}, and {Huang}]{MORSE500}
Zikui {Cai}, Andrew {Wang}, Anirudh {Satheesh}, Ankit {Nakhawa}, Hyunwoo {Jae}, Keenan {Powell}, Minghui {Liu}, Neel {Jay}, Sungbin {Oh}, Xiyao {Wang}, Yongyuan {Liang}, Tom {Goldstein}, and Furong {Huang}.
\newblock {MORSE-500: A Programmatically Controllable Video Benchmark to Stress-Test Multimodal Reasoning}.
\newblock \emph{arXiv e-prints}, art. arXiv:2506.05523, June 2025{\natexlab{b}}.
\newblock \doi{10.48550/arXiv.2506.05523}.

\bibitem[{CAMEL-AI.org}(2025)]{owl2025}
{CAMEL-AI.org}.
\newblock Owl: Optimized workforce learning for general multi-agent assistance in real-world task automation.
\newblock \url{https://github.com/camel-ai/owl}, 2025.
\newblock Accessed: 2025-03-07.

\bibitem[Cao et~al.(2024{\natexlab{a}})Cao, Chen, Wu, chi Cheung, and Xu]{cao2024javabenchbenchmarkobjectorientedcode}
Jialun Cao, Zhiyong Chen, Jiarong Wu, Shing chi Cheung, and Chang Xu.
\newblock Javabench: A benchmark of object-oriented code generation for evaluating large language models, 2024{\natexlab{a}}.
\newblock URL \url{https://arxiv.org/abs/2406.12902}.

\bibitem[Cao et~al.(2022)Cao, Liang, Chen, and Hu]{DBLP:conf/acsac/CaoL0H22}
Ying Cao, Ruigang Liang, Kai Chen, and Peiwei Hu.
\newblock Boosting neural networks to decompile optimized binaries.
\newblock In \emph{Annual Computer Security Applications Conference, {ACSAC} 2022, Austin, TX, USA, December 5-9, 2022}, pages 508--518. {ACM}, 2022.

\bibitem[Cao et~al.(2025{\natexlab{a}})Cao, Chen, Quan, Zhang, Wang, Dong, Feng, He, Huang, Li, Tan, Tang, Tang, Wu, Xiao, Zheng, Zhou, Zhu, Huang, Xie, and He]{cao_can_2025}
Yuhan Cao, Zian Chen, Kun Quan, Ziliang Zhang, Yu~Wang, Xiaoning Dong, Yeqi Feng, Guanzhong He, Jingcheng Huang, Jianhao Li, Yixuan Tan, Jiafu Tang, Yilin Tang, Junlei Wu, Qianyu Xiao, Can Zheng, Shouchen Zhou, Yuxiang Zhu, Yiming Huang, Tian Xie, and Tianxing He.
\newblock Can {LLMs} {Generate} {Reliable} {Test} {Case} {Generators}? {A} {Study} on {Competition}-{Level} {Programming} {Problems}, July 2025{\natexlab{a}}.
\newblock URL \url{http://arxiv.org/abs/2506.06821}.
\newblock arXiv:2506.06821 [cs].

\bibitem[Cao et~al.(2024{\natexlab{b}})Cao, Zheng, Fan, Zhang, Chen, and Bai]{RSL-SQL}
Zhenbiao Cao, Yuanlei Zheng, Zhihao Fan, Xiaojin Zhang, Wei Chen, and Xiang Bai.
\newblock {RSL-SQL:} robust schema linking in text-to-sql generation.
\newblock \emph{CoRR}, abs/2411.00073, 2024{\natexlab{b}}.

\bibitem[Cao et~al.(2025{\natexlab{b}})Cao, Wang, Yang, Ma, Zhu, Zheng, and Zhao]{cao2025pgpo}
Zouying Cao, Runze Wang, Yifei Yang, Xinbei Ma, Xiaoyong Zhu, Bo~Zheng, and Hai Zhao.
\newblock Pgpo: Enhancing agent reasoning via pseudocode-style planning guided preference optimization.
\newblock \emph{arXiv preprint arXiv:2506.01475}, 2025{\natexlab{b}}.

\bibitem[Cappendijk et~al.(2025)Cappendijk, de~Reus, and Oprescu]{Cappendijk_2025}
Tom Cappendijk, Pepijn de~Reus, and Ana Oprescu.
\newblock An exploration of prompting llms to generate energy-efficient code.
\newblock In \emph{2025 IEEE/ACM 9th International Workshop on Green and Sustainable Software (GREENS)}, page 31–38. IEEE, April 2025.
\newblock \doi{10.1109/greens66463.2025.00010}.
\newblock URL \url{http://dx.doi.org/10.1109/GREENS66463.2025.00010}.

\bibitem[Carbonneaux et~al.(2025)Carbonneaux, Cohen, Gehring, Kahn, Kossen, Kreuk, McMilin, Meyer, Wei, Zhang, et~al.]{code_world_model}
Quentin Carbonneaux, Gal Cohen, Jonas Gehring, Jacob Kahn, Jannik Kossen, Felix Kreuk, Emily McMilin, Michel Meyer, Yuxiang Wei, David Zhang, et~al.
\newblock Cwm: An open-weights llm for research on code generation with world models.
\newblock \emph{arXiv preprint arXiv:2510.02387}, 2025.

\bibitem[Cassano et~al.(2022)Cassano, Gouwar, Nguyen, Nguyen, Phipps-Costin, Pinckney, Yee, Zi, Anderson, Feldman, Guha, Greenberg, and Jangda]{MultiPL-E}
Federico Cassano, John Gouwar, Daniel Nguyen, Sydney Nguyen, Luna Phipps-Costin, Donald Pinckney, Ming-Ho Yee, Yangtian Zi, Carolyn~Jane Anderson, Molly~Q Feldman, Arjun Guha, Michael Greenberg, and Abhinav Jangda.
\newblock Multipl-e: A scalable and extensible approach to benchmarking neural code generation, 2022.
\newblock URL \url{https://arxiv.org/abs/2208.08227}.

\bibitem[Cfahlgren1(2024)]{cfahlgren_react}
Cfahlgren1.
\newblock React code instructions, 2024.
\newblock URL \url{https://huggingface.co/datasets/cfahlgren1/react-code-instructions}.
\newblock Accessed: 2024.

\bibitem[Chai et~al.(2024)Chai, Liu, Yang, Yin, Jin, Liu, Sun, Zhang, Ren, Guo, et~al.]{mceval}
Linzheng Chai, Shukai Liu, Jian Yang, Yuwei Yin, Ke~Jin, Jiaheng Liu, Tao Sun, Ge~Zhang, Changyu Ren, Hongcheng Guo, et~al.
\newblock Mceval: Massively multilingual code evaluation.
\newblock \emph{arXiv preprint arXiv:2406.07436}, 2024.

\bibitem[Chai et~al.(2025{\natexlab{a}})Chai, Yang, Liu, Zhang, Wang, Jin, Sun, Liu, Zhang, Zhu, et~al.]{chai2025multilingual}
Linzheng Chai, Jian Yang, Shukai Liu, Wei Zhang, Liran Wang, Ke~Jin, Tao Sun, Congnan Liu, Chenchen Zhang, Hualei Zhu, et~al.
\newblock Multilingual multimodal software developer for code generation.
\newblock \emph{arXiv preprint arXiv:2507.08719}, 2025{\natexlab{a}}.

\bibitem[Chai et~al.(2025{\natexlab{b}})Chai, Yang, Sun, Guo, Liu, Wang, Liang, Bai, Li, Peng, and Li]{xcot}
Linzheng Chai, Jian Yang, Tao Sun, Hongcheng Guo, Jiaheng Liu, Bing Wang, Xinnian Liang, Jiaqi Bai, Tongliang Li, Qiyao Peng, and Zhoujun Li.
\newblock {XCOT:} cross-lingual instruction tuning for cross-lingual chain-of-thought reasoning.
\newblock In Toby Walsh, Julie Shah, and Zico Kolter, editors, \emph{AAAI-25, Sponsored by the Association for the Advancement of Artificial Intelligence, February 25 - March 4, 2025, Philadelphia, PA, {USA}}, pages 23550--23558. {AAAI} Press, 2025{\natexlab{b}}.
\newblock \doi{10.1609/AAAI.V39I22.34524}.
\newblock URL \url{https://doi.org/10.1609/aaai.v39i22.34524}.

\bibitem[Chai et~al.(2023)Chai, Wang, Pang, Sun, Tian, and Wu]{chai2023erniecodeenglishcentriccrosslingualpretraining}
Yekun Chai, Shuohuan Wang, Chao Pang, Yu~Sun, Hao Tian, and Hua Wu.
\newblock Ernie-code: Beyond english-centric cross-lingual pretraining for programming languages, 2023.
\newblock URL \url{https://arxiv.org/abs/2212.06742}.

\bibitem[Chaimalas et~al.(2025)Chaimalas, Vy{\'L}{\k{A}}niauskas, and Brostow]{microsoft_explorer}
Iason Chaimalas, Arnas Vy{\'L}{\k{A}}niauskas, and Gabriel Brostow.
\newblock Explorer: Robust collection of interactable gui elements.
\newblock \emph{arXiv preprint arXiv:2504.09352}, 2025.

\bibitem[Chakraborty et~al.(2024)Chakraborty, Alfadel, and Nagappan]{chakraborty2024rlocator}
Partha Chakraborty, Mahmoud Alfadel, and Meiyappan Nagappan.
\newblock Rlocator: Reinforcement learning for bug localization.
\newblock \emph{IEEE Transactions on Software Engineering}, 2024.

\bibitem[Chambon et~al.(2025)Chambon, Roziere, Sagot, and Synnaeve]{bigobench}
Pierre Chambon, Baptiste Roziere, Benoit Sagot, and Gabriel Synnaeve.
\newblock Bigo(bench) -- can llms generate code with controlled time and space complexity?, 2025.
\newblock URL \url{https://arxiv.org/abs/2503.15242}.

\bibitem[Chandel et~al.(2022)Chandel, Clement, Serrato, and Sundaresan]{chandel2022trainingevaluatingjupyternotebook}
Shubham Chandel, Colin~B. Clement, Guillermo Serrato, and Neel Sundaresan.
\newblock Training and evaluating a jupyter notebook data science assistant, 2022.
\newblock URL \url{https://arxiv.org/abs/2201.12901}.

\bibitem[Chang and Fosler{-}Lussier(2023)]{Chang}
Shuaichen Chang and Eric Fosler{-}Lussier.
\newblock How to prompt llms for text-to-sql: {A} study in zero-shot, single-domain, and cross-domain settings.
\newblock \emph{CoRR}, abs/2305.11853, 2023.

\bibitem[Charalambous et~al.(2024)Charalambous, Manino, and Cordeiro]{charalambous2024automated}
Yiannis Charalambous, Edoardo Manino, and Lucas~C Cordeiro.
\newblock Automated repair of ai code with large language models and formal verification.
\newblock \emph{arXiv preprint arXiv:2405.08848}, 2024.

\bibitem[{chatgpt}(2025)]{chatgpt}
{chatgpt}.
\newblock chatgpt, 2025.
\newblock URL \url{https://chatgpt.com/}.

\bibitem[Chaudhary(2023)]{codealpaca}
Sahil Chaudhary.
\newblock Code alpaca: An instruction-following llama model for code generation.
\newblock \url{https://github.com/sahil280114/codealpaca}, 2023.

\bibitem[Chen et~al.(2025{\natexlab{a}})Chen, Li, Gong, Jiang, Fei, Yang, Shan, Yu, Wang, Zhu, et~al.]{minmax_m1}
Aili Chen, Aonian Li, Bangwei Gong, Binyang Jiang, Bo~Fei, Bo~Yang, Boji Shan, Changqing Yu, Chao Wang, Cheng Zhu, et~al.
\newblock Minimax-m1: Scaling test-time compute efficiently with lightning attention.
\newblock \emph{arXiv preprint arXiv:2506.13585}, 2025{\natexlab{a}}.

\bibitem[Chen et~al.(2024{\natexlab{a}})Chen, Lin, Zeng, Zan, Wang, Cheshkov, Sun, Yu, Dong, Aliev, et~al.]{chen2024coder}
Dong Chen, Shaoxin Lin, Muhan Zeng, Daoguang Zan, Jian-Gang Wang, Anton Cheshkov, Jun Sun, Hao Yu, Guoliang Dong, Artem Aliev, et~al.
\newblock Coder: Issue resolving with multi-agent and task graphs.
\newblock \emph{arXiv preprint arXiv:2406.01304}, 2024{\natexlab{a}}.

\bibitem[Chen et~al.(2024{\natexlab{b}})Chen, Chen, Liu, Jiang, and Wang]{chen2024humans}
Guiming~Hardy Chen, Shunian Chen, Ziche Liu, Feng Jiang, and Benyou Wang.
\newblock Humans or llms as the judge? a study on judgement biases.
\newblock \emph{arXiv preprint arXiv:2402.10669}, 2024{\natexlab{b}}.

\bibitem[Chen et~al.(2025{\natexlab{b}})Chen, Zhao, Liu, Peng, Liu, Zhu, Gao, Yang, and Deng]{chen2025coreqa}
Jialiang Chen, Kaifa Zhao, Jie Liu, Chao Peng, Jierui Liu, Hang Zhu, Pengfei Gao, Ping Yang, and Shuiguang Deng.
\newblock Coreqa: uncovering potentials of language models in code repository question answering.
\newblock \emph{arXiv preprint arXiv:2501.03447}, 2025{\natexlab{b}}.

\bibitem[Chen et~al.(2020)Chen, Ma, and Zhang]{chen2020enhanced}
Junjie Chen, Haoyang Ma, and Lingming Zhang.
\newblock Enhanced compiler bug isolation via memoized search.
\newblock In \emph{Proceedings of the 35th IEEE/ACM international conference on automated software engineering}, pages 78--89, 2020.

\bibitem[Chen et~al.(2025{\natexlab{c}})Chen, Ren, Liu, Hu, Tian, Xie, Liu, Zhang, Liu, Gong, et~al.]{chen2025xbench}
Kaiyuan Chen, Yixin Ren, Yang Liu, Xiaobo Hu, Haotong Tian, Tianbao Xie, Fangfu Liu, Haoye Zhang, Hongzhang Liu, Yuan Gong, et~al.
\newblock xbench: Tracking agents productivity scaling with profession-aligned real-world evaluations.
\newblock \emph{arXiv preprint arXiv:2506.13651}, 2025{\natexlab{c}}.

\bibitem[Chen et~al.(2025{\natexlab{d}})Chen, Wang, Zhang, Li, Chen, and Yu]{chen2025survey}
Kang Chen, Ziteng Wang, Weiming Zhang, Gang Li, Zhaofeng Chen, and Nenghai Yu.
\newblock A survey on privacy risks and protection in large language models, 2025{\natexlab{d}}.

\bibitem[Chen et~al.(2024{\natexlab{c}})Chen, Guo, Jia, Zeng, Wang, Xu, Wu, Wang, Gao, Wang, et~al.]{chen2024survey_code_evaluation}
Liguo Chen, Qi~Guo, Hongrui Jia, Zhengran Zeng, Xin Wang, Yijiang Xu, Jian Wu, Yidong Wang, Qing Gao, Jindong Wang, et~al.
\newblock A survey on evaluating large language models in code generation tasks.
\newblock \emph{arXiv preprint arXiv:2408.16498}, 2024{\natexlab{c}}.

\bibitem[Chen et~al.(2021)Chen, Tworek, Jun, Yuan, de~Oliveira~Pinto, Kaplan, Edwards, Burda, Joseph, Brockman, Ray, Puri, Krueger, Petrov, Khlaaf, Sastry, Mishkin, Chan, Gray, Ryder, Pavlov, Power, Kaiser, Bavarian, Winter, Tillet, Such, Cummings, Plappert, Chantzis, Barnes, Herbert-Voss, Guss, Nichol, Paino, Tezak, Tang, Babuschkin, Balaji, Jain, Saunders, Hesse, Carr, Leike, Achiam, Misra, Morikawa, Radford, Knight, Brundage, Murati, Mayer, Welinder, McGrew, Amodei, McCandlish, Sutskever, and Zaremba]{chen2021codex}
Mark Chen, Jerry Tworek, Heewoo Jun, Qiming Yuan, Henrique~Ponde de~Oliveira~Pinto, Jared Kaplan, Harri Edwards, Yuri Burda, Nicholas Joseph, Greg Brockman, Alex Ray, Raul Puri, Gretchen Krueger, Michael Petrov, Heidy Khlaaf, Girish Sastry, Pamela Mishkin, Brooke Chan, Scott Gray, Nick Ryder, Mikhail Pavlov, Alethea Power, Lukasz Kaiser, Mohammad Bavarian, Clemens Winter, Philippe Tillet, Felipe~Petroski Such, Dave Cummings, Matthias Plappert, Fotios Chantzis, Elizabeth Barnes, Ariel Herbert-Voss, William~Hebgen Guss, Alex Nichol, Alex Paino, Nikolas Tezak, Jie Tang, Igor Babuschkin, Suchir Balaji, Shantanu Jain, William Saunders, Christopher Hesse, Andrew~N. Carr, Jan Leike, Josh Achiam, Vedant Misra, Evan Morikawa, Alec Radford, Matthew Knight, Miles Brundage, Mira Murati, Katie Mayer, Peter Welinder, Bob McGrew, Dario Amodei, Sam McCandlish, Ilya Sutskever, and Wojciech Zaremba.
\newblock Evaluating large language models trained on code, 2021.
\newblock URL \url{https://arxiv.org/abs/2107.03374}.

\bibitem[{Chen} et~al.(2024){Chen}, {Bu}, {Song}, {Gao}, and {Zheng}]{Wukong}
Peng {Chen}, Pi~{Bu}, Jun {Song}, Yuan {Gao}, and Bo~{Zheng}.
\newblock {Can VLMs Play Action Role-Playing Games? Take Black Myth Wukong as a Study Case}.
\newblock \emph{arXiv e-prints}, art. arXiv:2409.12889, September 2024.
\newblock \doi{10.48550/arXiv.2409.12889}.

\bibitem[Chen et~al.(2025{\natexlab{e}})Chen, Lin, Gu, Shi, Lian, Yun, Chen, Sun, Cao, and Wang]{chen2025swe}
Silin Chen, Shaoxin Lin, Xiaodong Gu, Yuling Shi, Heng Lian, Longfei Yun, Dong Chen, Weiguo Sun, Lin Cao, and Qianxiang Wang.
\newblock Swe-exp: Experience-driven software issue resolution.
\newblock \emph{arXiv preprint arXiv:2507.23361}, 2025{\natexlab{e}}.

\bibitem[Chen et~al.(2023{\natexlab{a}})Chen, Su, Zuo, Yang, Yuan, Qian, Chan, Qin, Lu, Xie, et~al.]{chen2023agentverse}
Weize Chen, Yusheng Su, Jingwei Zuo, Cheng Yang, Chenfei Yuan, Chen Qian, Chi-Min Chan, Yujia Qin, Yaxi Lu, Ruobing Xie, et~al.
\newblock Agentverse: Facilitating multi-agent collaboration and exploring emergent behaviors in agents.
\newblock \emph{arXiv preprint arXiv:2308.10848}, 2\penalty0 (4):\penalty0 6, 2023{\natexlab{a}}.

\bibitem[Chen et~al.(2022)Chen, Ma, Wang, and Cohen]{chen2022program}
Wenhu Chen, Xueguang Ma, Xinyi Wang, and William~W Cohen.
\newblock Program of thoughts prompting: Disentangling computation from reasoning for numerical reasoning tasks.
\newblock \emph{arXiv preprint arXiv:2211.12588}, 2022.

\bibitem[Chen et~al.(2025{\natexlab{f}})Chen, Chen, Lian, Huang, Zhou, Wu, and Zheng]{ByteGen}
Xiangping Chen, Junqi Chen, Zhilu Lian, Yuan Huang, Xiaocong Zhou, Yunzhi Wu, and Zibin Zheng.
\newblock An alternative to code comment generation? generating comment from bytecode.
\newblock \emph{Information and Software Technology}, 179:\penalty0 107623, 2025{\natexlab{f}}.
\newblock ISSN 0950-5849.
\newblock \doi{https://doi.org/10.1016/j.infsof.2024.107623}.
\newblock URL \url{https://www.sciencedirect.com/science/article/pii/S0950584924002283}.

\bibitem[Chen et~al.(2018)Chen, Liu, and Song]{chen2018treetotreeneuralnetworksprogram}
Xinyun Chen, Chang Liu, and Dawn Song.
\newblock Tree-to-tree neural networks for program translation, 2018.
\newblock URL \url{https://arxiv.org/abs/1802.03691}.

\bibitem[Chen et~al.(2023{\natexlab{b}})Chen, Tworek, et~al.]{chen2023codet5}
Xinyun Chen, Jerry Tworek, et~al.
\newblock {CodeT5}: Identifier-aware unified pre-trained encoder-decoder models for code understanding and generation.
\newblock \emph{EMNLP}, 2023{\natexlab{b}}.

\bibitem[Chen et~al.(2025{\natexlab{g}})Chen, Shen, Huang, Zhou, Lin, Cai, Yu, Bu, Shi, and Qiao]{chen2025learning}
Yang Chen, Yufan Shen, Wenxuan Huang, Sheng Zhou, Qunshu Lin, Xinyu Cai, Zhi Yu, Jiajun Bu, Botian Shi, and Yu~Qiao.
\newblock Learning only with images: Visual reinforcement learning with reasoning, rendering, and visual feedback.
\newblock \emph{arXiv preprint arXiv:2507.20766}, 2025{\natexlab{g}}.

\bibitem[Chen et~al.(2025{\natexlab{h}})Chen, Yang, Liu, Lee, Xu, Shoeybi, Catanzaro, and Ping]{chen2025acereason}
Yang Chen, Zhuolin Yang, Zihan Liu, Chankyu Lee, Peng Xu, Mohammad Shoeybi, Bryan Catanzaro, and Wei Ping.
\newblock Acereason-nemotron: Advancing math and code reasoning through reinforcement learning.
\newblock \emph{arXiv preprint arXiv:2505.16400}, 2025{\natexlab{h}}.

\bibitem[Chen et~al.(2024{\natexlab{a}})Chen, Hu, Zhi, Han, Deng, and Yin]{chen2024chatunitestframeworkllmbasedtest}
Yinghao Chen, Zehao Hu, Chen Zhi, Junxiao Han, Shuiguang Deng, and Jianwei Yin.
\newblock Chatunitest: A framework for llm-based test generation, 2024{\natexlab{a}}.
\newblock URL \url{https://arxiv.org/abs/2305.04764}.

\bibitem[Chen et~al.(2024{\natexlab{b}})Chen, Jhamtani, Sharma, Fan, and Wang]{chen2024steering}
Yongchao Chen, Harsh Jhamtani, Srinagesh Sharma, Chuchu Fan, and Chi Wang.
\newblock Steering large language models between code execution and textual reasoning.
\newblock \emph{arXiv preprint arXiv:2410.03524}, 2024{\natexlab{b}}.

\bibitem[Chen et~al.(2025{\natexlab{i}})Chen, Ding, Zhang, Chen, Du, Sun, and Chen]{chen2025designcoder}
Yunnong Chen, Shixian Ding, YingYing Zhang, Wenkai Chen, Jinzhou Du, Lingyun Sun, and Liuqing Chen.
\newblock Designcoder: Hierarchy-aware and self-correcting ui code generation with large language models.
\newblock \emph{arXiv preprint arXiv:2506.13663}, 2025{\natexlab{i}}.

\bibitem[Chen et~al.(2024{\natexlab{c}})Chen, Ding, Yu, et~al.]{chen2024magicoder}
Yuxiang Chen, Zhenyang Ding, Yue Yu, et~al.
\newblock {MagicCoder}: Source code is all you need.
\newblock \emph{arXiv preprint arXiv:2312.02120}, 2024{\natexlab{c}}.

\bibitem[Chen et~al.(2024{\natexlab{d}})Chen, Wu, Wang, Su, Chen, Xing, Zhong, Zhang, Zhu, Lu, et~al.]{chen2024internvl}
Zhe Chen, Jiannan Wu, Wenhai Wang, Weijie Su, Guo Chen, Sen Xing, Muyan Zhong, Qinglong Zhang, Xizhou Zhu, Lewei Lu, et~al.
\newblock Internvl: Scaling up vision foundation models and aligning for generic visual-linguistic tasks.
\newblock In \emph{Proceedings of the IEEE/CVF Conference on Computer Vision and Pattern Recognition}, pages 24185--24198, 2024{\natexlab{d}}.

\bibitem[Chen et~al.(2025{\natexlab{j}})Chen, Wang, Cao, Liu, Gao, Cui, Zhu, Ye, Tian, Liu, Gu, Wang, Li, Ren, Chen, Luo, Wang, Jiang, Wang, He, Shi, Zhang, Lv, Wang, Shao, Chu, Tu, He, Wu, Deng, Ge, Chen, Zhang, Wang, Dou, Lu, Zhu, Lu, Lin, Qiao, Dai, and Wang]{chen2025internvl2_5}
Zhe Chen, Weiyun Wang, Yue Cao, Yangzhou Liu, Zhangwei Gao, Erfei Cui, Jinguo Zhu, Shenglong Ye, Hao Tian, Zhaoyang Liu, Lixin Gu, Xuehui Wang, Qingyun Li, Yiming Ren, Zixuan Chen, Jiapeng Luo, Jiahao Wang, Tan Jiang, Bo~Wang, Conghui He, Botian Shi, Xingcheng Zhang, Han Lv, Yi~Wang, Wenqi Shao, Pei Chu, Zhongying Tu, Tong He, Zhiyong Wu, Huipeng Deng, Jiaye Ge, Kai Chen, Kaipeng Zhang, Limin Wang, Min Dou, Lewei Lu, Xizhou Zhu, Tong Lu, Dahua Lin, Yu~Qiao, Jifeng Dai, and Wenhai Wang.
\newblock Expanding performance boundaries of open-source multimodal models with model, data, and test-time scaling, 2025{\natexlab{j}}.
\newblock URL \url{https://arxiv.org/abs/2412.05271}.

\bibitem[Chen et~al.(2023{\natexlab{c}})Chen, Gulyás, Kovács, Zombori, Félegyházi, Telek, and Horváth]{chen2023large}
Zifan Chen, Gábor Gulyás, Zsombor Kovács, Zsolt Zombori, Márk Félegyházi, Máté Telek, and Bence Horváth.
\newblock Large language models for code: Security hardening and adversarial testing, 2023{\natexlab{c}}.

\bibitem[Chen et~al.(2025{\natexlab{k}})Chen, Ma, Zhuang, Nie, Zou, Liu, Green, Patel, Meng, Su, et~al.]{chen2025browsecomp}
Zijian Chen, Xueguang Ma, Shengyao Zhuang, Ping Nie, Kai Zou, Andrew Liu, Joshua Green, Kshama Patel, Ruoxi Meng, Mingyi Su, et~al.
\newblock Browsecomp-plus: A more fair and transparent evaluation benchmark of deep-research agent.
\newblock \emph{arXiv preprint arXiv:2508.06600}, 2025{\natexlab{k}}.

\bibitem[Cheng et~al.(2024)Cheng, Yin, Fu, Guo, Yang, Kautz, Wang, and Liu]{spatialrgpt}
An-Chieh Cheng, Hongxu Yin, Yang Fu, Qiushan Guo, Ruihan Yang, Jan Kautz, Xiaolong Wang, and Sifei Liu.
\newblock Spatialrgpt: Grounded spatial reasoning in vision-language models.
\newblock \emph{Advances in Neural Information Processing Systems}, 37:\penalty0 135062--135093, 2024.

\bibitem[Cheng et~al.(2025)Cheng, Tufano, Cito, Cambronero, Rondon, Wei, Sun, and Chandra]{cheng2025agentic}
Runxiang Cheng, Michele Tufano, J{\"u}rgen Cito, Jos{\'e} Cambronero, Pat Rondon, Renyao Wei, Aaron Sun, and Satish Chandra.
\newblock Agentic bug reproduction for effective automated program repair at google.
\newblock \emph{arXiv preprint arXiv:2502.01821}, 2025.

\bibitem[Chervyakov et~al.(2025)Chervyakov, Kharitonov, Zadorozhny, Pavel, Levichev, Vorobev, Salikhov, Valeev, Pestova, Dziuba, Alimova, Zavgorodnev, Medvedev, Moiseev, Bruches, Grebenkin, Derunets, Vladimir, Emelyanov, Babaev, Ivanov, Malykh, and Fenogenova]{MERACode}
Artem Chervyakov, Alexander Kharitonov, Pavel Zadorozhny, Adamenko Pavel, Rodion Levichev, Dmitrii Vorobev, Dmitrii Salikhov, Aidar Valeev, Alena Pestova, Maria Dziuba, Ilseyar Alimova, Artem Zavgorodnev, Aleksandr Medvedev, Stanislav Moiseev, Elena Bruches, Daniil Grebenkin, Roman Derunets, Vikulov Vladimir, Anton Emelyanov, Dmitrii Babaev, Vladimir~V. Ivanov, Valentin Malykh, and Alena Fenogenova.
\newblock Mera code: A unified framework for evaluating code generation across tasks, 2025.
\newblock URL \url{https://arxiv.org/abs/2507.12284}.

\bibitem[Chi et~al.(2022)Chi, Qu, Liu, Zheng, and Yin]{chi2022seqtrans}
Jianlei Chi, Yu~Qu, Ting Liu, Qinghua Zheng, and Heng Yin.
\newblock Seqtrans: automatic vulnerability fix via sequence to sequence learning.
\newblock \emph{IEEE Transactions on Software Engineering}, 49\penalty0 (2):\penalty0 564--585, 2022.

\bibitem[Chi et~al.(2025)Chi, Chen, Angelopoulos, Chiang, Mittal, Jain, Zhang, Stoica, Donahue, and Talwalkar]{chi2025copilotarenaplatformcode}
Wayne Chi, Valerie Chen, Anastasios~Nikolas Angelopoulos, Wei-Lin Chiang, Aditya Mittal, Naman Jain, Tianjun Zhang, Ion Stoica, Chris Donahue, and Ameet Talwalkar.
\newblock Copilot arena: A platform for code llm evaluation in the wild, 2025.
\newblock URL \url{https://arxiv.org/abs/2502.09328}.

\bibitem[Chi(2021)]{2021LTmatch}
Xuebin Chi.
\newblock Ltmatch: A method to abstract pattern from unstructured log.
\newblock \emph{Applied Sciences}, 11, 2021.

\bibitem[Chiara(2022)]{chiara2022cyber}
Pier~Giorgio Chiara.
\newblock The cyber resilience act: the eu commission’s proposal for a horizontal regulation on cybersecurity for products with digital elements: An introduction.
\newblock \emph{International Cybersecurity Law Review}, 3\penalty0 (2):\penalty0 255--272, 2022.

\bibitem[Choi et~al.(2025)Choi, Lee, Tack, Song, Dingliwal, Jayanthi, Ganesh, Shin, Galstyan, and Bodapati]{choi2025think}
Daewon Choi, Jimin Lee, Jihoon Tack, Woomin Song, Saket Dingliwal, Sai~Muralidhar Jayanthi, Bhavana Ganesh, Jinwoo Shin, Aram Galstyan, and Sravan~Babu Bodapati.
\newblock Think clearly: Improving reasoning via redundant token pruning.
\newblock \emph{arXiv preprint arXiv:2507.08806}, 2025.

\bibitem[Chou et~al.(2025)Chou, Liu, Deng, Zeng, Zhang, Zhu, Cai, Mao, Zhang, Tan, Xu, Zhai, Liu, Zhu, Zhou, and Lian]{autocodebench}
Jason Chou, Ao~Liu, Yuchi Deng, Zhiying Zeng, Tao Zhang, Haotian Zhu, Jianwei Cai, Yue Mao, Chenchen Zhang, Lingyun Tan, Ziyan Xu, Bohui Zhai, Hengyi Liu, Speed Zhu, Wiggin Zhou, and Fengzong Lian.
\newblock Autocodebench: Large language models are automatic code benchmark generators, 2025.
\newblock URL \url{https://arxiv.org/abs/2508.09101}.

\bibitem[Chowdhery et~al.(2022)Chowdhery, Narang, Devlin, Bosma, Mishra, Roberts, and et~al.]{chowdhery2022palmscalinglanguagemodeling}
Aakanksha Chowdhery, Sharan Narang, Jacob Devlin, Maarten Bosma, Gaurav Mishra, Adam Roberts, and et~al.
\newblock Palm: Scaling language modeling with pathways, 2022.
\newblock URL \url{https://arxiv.org/abs/2204.02311}.

\bibitem[Christey et~al.(2013)Christey, Kenderdine, Mazella, and Miles]{christey2013common}
Steve Christey, J~Kenderdine, J~Mazella, and B~Miles.
\newblock Common weakness enumeration.
\newblock \emph{Mitre Corporation}, 2013.

\bibitem[Christopher(2023)]{christopher_rosetta}
Christopher.
\newblock Rosetta code dataset, 2023.
\newblock URL \url{https://huggingface.co/datasets/christopher/rosetta-code}.
\newblock Accessed: 2024.

\bibitem[Christopoulou et~al.(2022)Christopoulou, Lampouras, Gritta, Zhang, Guo, Li, Zhang, Xiao, Shen, Li, Yu, Yan, Zhou, Wang, Ma, Iacobacci, Wang, Liang, Wei, Jiang, Wang, and Liu]{christopoulou2022pangucoderprogramsynthesisfunctionlevel}
Fenia Christopoulou, Gerasimos Lampouras, Milan Gritta, Guchun Zhang, Yinpeng Guo, Zhongqi Li, Qi~Zhang, Meng Xiao, Bo~Shen, Lin Li, Hao Yu, Li~Yan, Pingyi Zhou, Xin Wang, Yuchi Ma, Ignacio Iacobacci, Yasheng Wang, Guangtai Liang, Jiansheng Wei, Xin Jiang, Qianxiang Wang, and Qun Liu.
\newblock Pangu-coder: Program synthesis with function-level language modeling, 2022.
\newblock URL \url{https://arxiv.org/abs/2207.11280}.

\bibitem[{claude4}(2025)]{claude4}
{claude4}.
\newblock claude4, 2025.
\newblock URL \url{https://www.anthropic.com/claude/sonnet}.

\bibitem[Clement et~al.(2020)Clement, Drain, Timcheck, Svyatkovskiy, and Sundaresan]{clement-etal-2020-pymt5}
Colin Clement, Dawn Drain, Jonathan Timcheck, Alexey Svyatkovskiy, and Neel Sundaresan.
\newblock {P}y{MT}5: multi-mode translation of natural language and python code with transformers.
\newblock In Bonnie Webber, Trevor Cohn, Yulan He, and Yang Liu, editors, \emph{Proceedings of the 2020 Conference on Empirical Methods in Natural Language Processing (EMNLP)}, pages 9052--9065, Online, November 2020. Association for Computational Linguistics.
\newblock \doi{10.18653/v1/2020.emnlp-main.728}.
\newblock URL \url{https://aclanthology.org/2020.emnlp-main.728/}.

\bibitem[Code(2025)]{claude_code_repo}
Claude Code.
\newblock Claude code.
\newblock \url{https://github.com/anthropics/claude-code}, 2025.

\bibitem[CodeChain(2023)]{codechain2023}
CodeChain.
\newblock Codechain.
\newblock \emph{arXiv preprint arXiv:2310.08992}, 2023.
\newblock URL \url{https://arxiv.org/abs/2310.08992}.

\bibitem[{Codedog}(2024)]{codedog}
{Codedog}.
\newblock Codedog: Ai code review platform.
\newblock \url{https://codedog.ai}, 2024.

\bibitem[CodeFuse-AI(2025)]{codefuseCGE}
CodeFuse-AI.
\newblock Codefuse-embeddings: Code generalist embeddings (cge).
\newblock \url{https://github.com/codefuse-ai/CodeFuse-Embeddings/tree/main/CGE}, 2025.
\newblock Accessed: 2025-11-03.

\bibitem[{Codeium}(2024)]{codeium2024windsurf}
{Codeium}.
\newblock {Windsurf}: The {AI} flow editor.
\newblock Technical report, Codeium Inc., 2024.
\newblock URL \url{https://codeium.com/windsurf}.

\bibitem[CodeParrot(2021)]{codeparrot_apps}
CodeParrot.
\newblock Apps: A benchmark for measuring the ability of language models to generate simple programs from natural language descriptions, 2021.
\newblock URL \url{https://huggingface.co/datasets/codeparrot/apps}.
\newblock Accessed: 2024.

\bibitem[CodePlan(2023)]{codeplan2023}
CodePlan.
\newblock Codeplan.
\newblock \emph{arXiv preprint arXiv:2309.12499}, 2023.
\newblock URL \url{https://arxiv.org/abs/2309.12499}.

\bibitem[{CodeRabbit}(2024)]{coderabbit}
{CodeRabbit}.
\newblock Coderabbit: Ai code reviews.
\newblock \url{https://coderabbit.ai}, 2024.

\bibitem[{codetrans}(2025)]{codetrans}
{codetrans}.
\newblock codetrans, 2025.
\newblock URL \url{https://github.com/agemagician/CodeTrans}.

\bibitem[Cohn et~al.(2010)Cohn, Blunsom, and Goldwater]{inducing_tree_substitution}
Trevor Cohn, Phil Blunsom, and Sharon Goldwater.
\newblock Inducing tree-substitution grammars.
\newblock \emph{J. Mach. Learn. Res.}, 11:\penalty0 3053--3096, 2010.
\newblock \doi{10.5555/1756006.1953031}.
\newblock URL \url{https://dl.acm.org/doi/10.5555/1756006.1953031}.

\bibitem[Comanici et~al.(2025{\natexlab{a}})Comanici, Bieber, Schaekermann, Pasupat, Sachdeva, Dhillon, Blistein, Ram, Zhang, Rosen, Marris, Petulla, Gaffney, Aharoni, Lintz, Pais, Jacobsson, Szpektor, Jiang, Haridasan, Omran, Saunshi, Bahri, Mishra, Chu, Boyd, Hekman, Parisi, Zhang, Kawintiranon, Bedrax-Weiss, Wang, Xu, Purkiss, Mendlovic, Deutel, Nguyen, Langley, et~al.]{comanici2025gemini2_5}
Gheorghe Comanici, Eric Bieber, Mike Schaekermann, Ice Pasupat, Noveen Sachdeva, Inderjit Dhillon, Marcel Blistein, Ori Ram, Dan Zhang, Evan Rosen, Luke Marris, Sam Petulla, Colin Gaffney, Asaf Aharoni, Nathan Lintz, Tiago~Cardal Pais, Henrik Jacobsson, Idan Szpektor, Nan-Jiang Jiang, Krishna Haridasan, Ahmed Omran, Nikunj Saunshi, Dara Bahri, Gaurav Mishra, Eric Chu, Toby Boyd, Brad Hekman, Aaron Parisi, Chaoyi Zhang, Kornraphop Kawintiranon, Tania Bedrax-Weiss, Oliver Wang, Ya~Xu, Ollie Purkiss, Uri Mendlovic, Ilaï Deutel, Nam Nguyen, Adam Langley, et~al.
\newblock Gemini 2.5: Pushing the frontier with advanced reasoning, multimodality, long context, and next generation agentic capabilities, 2025{\natexlab{a}}.
\newblock URL \url{https://arxiv.org/abs/2507.06261}.

\bibitem[Comanici et~al.(2025{\natexlab{b}})Comanici, Bieber, Schaekermann, Pasupat, Sachdeva, and et~al.]{comanici2025gemini25pushingfrontier}
Gheorghe Comanici, Eric Bieber, Mike Schaekermann, Ice Pasupat, Noveen Sachdeva, and et~al.
\newblock Gemini 2.5: Pushing the frontier with advanced reasoning, multimodality, long context, and next generation agentic capabilities, 2025{\natexlab{b}}.
\newblock URL \url{https://arxiv.org/abs/2507.06261}.

\bibitem[Computer(2023)]{together2023redpajama}
Together Computer.
\newblock {RedPajama-Data-1T}, 2023.
\newblock URL \url{https://huggingface.co/datasets/togethercomputer/RedPajama-Data-1T}.
\newblock Accessed: 2024-01-01.

\bibitem[Cooper et~al.(1999)Cooper, Schielke, and Subramanian]{DBLP:conf/lctrts/CooperSS99}
Keith~D. Cooper, Philip~J. Schielke, and Devika Subramanian.
\newblock Optimizing for reduced code space using genetic algorithms.
\newblock In Y.~Annie Liu and Reinhard Wilhelm, editors, \emph{Proceedings of the {ACM} {SIGPLAN} 1999 Workshop on Languages, Compilers, and Tools for Embedded Systems (LCTES'99), Atlanta, Georgia, USA, May 5, 1999}, pages 1--9. {ACM}, 1999.

\bibitem[Cordeiro et~al.(2024)Cordeiro, Noei, and Zou]{code_refactor_empirical_study}
Jonathan Cordeiro, Shayan Noei, and Ying Zou.
\newblock An empirical study on the code refactoring capability of large language models.
\newblock \emph{arXiv preprint arXiv:2411.02320}, 2024.

\bibitem[Csuvik and Vid\'{a}cs(2022{\natexlab{a}})]{fixjs}
Viktor Csuvik and L\'{a}szl\'{o} Vid\'{a}cs.
\newblock Fixjs: a dataset of bug-fixing javascript commits.
\newblock In \emph{Proceedings of the 19th International Conference on Mining Software Repositories}, MSR '22, page 712–716, New York, NY, USA, 2022{\natexlab{a}}. Association for Computing Machinery.
\newblock ISBN 9781450393034.
\newblock \doi{10.1145/3524842.3528480}.
\newblock URL \url{https://doi.org/10.1145/3524842.3528480}.

\bibitem[Csuvik and Vid\'{a}cs(2022{\natexlab{b}})]{tfix}
Viktor Csuvik and L\'{a}szl\'{o} Vid\'{a}cs.
\newblock Fixjs: a dataset of bug-fixing javascript commits.
\newblock In \emph{Proceedings of the 19th International Conference on Mining Software Repositories}, MSR '22, page 712–716, New York, NY, USA, 2022{\natexlab{b}}. Association for Computing Machinery.
\newblock ISBN 9781450393034.
\newblock \doi{10.1145/3524842.3528480}.
\newblock URL \url{https://doi.org/10.1145/3524842.3528480}.

\bibitem[Cui et~al.(2025{\natexlab{a}})Cui, Yuan, Wang, Wang, Zhang, Chen, Li, He, Fan, Yu, Xu, Chen, Yuan, Chen, Zhang, Lv, Wang, Yao, Han, Peng, Cheng, Liu, Sun, Zhou, and Ding]{cui2025processreinforcementimplicitrewards}
Ganqu Cui, Lifan Yuan, Zefan Wang, Hanbin Wang, Yuchen Zhang, Jiacheng Chen, Wendi Li, Bingxiang He, Yuchen Fan, Tianyu Yu, Qixin Xu, Weize Chen, Jiarui Yuan, Huayu Chen, Kaiyan Zhang, Xingtai Lv, Shuo Wang, Yuan Yao, Xu~Han, Hao Peng, Yu~Cheng, Zhiyuan Liu, Maosong Sun, Bowen Zhou, and Ning Ding.
\newblock Process reinforcement through implicit rewards, 2025{\natexlab{a}}.
\newblock URL \url{https://arxiv.org/abs/2502.01456}.

\bibitem[Cui et~al.(2022)Cui, Wang, Huang, Inala, Mytkowicz, Wang, Gao, and Duan]{CodeExp}
Haotian Cui, Chenglong Wang, Junjie Huang, Jeevana~Priya Inala, Todd Mytkowicz, Bo~Wang, Jianfeng Gao, and Nan Duan.
\newblock {C}ode{E}xp: Explanatory code document generation.
\newblock In Yoav Goldberg, Zornitsa Kozareva, and Yue Zhang, editors, \emph{Findings of the Association for Computational Linguistics: EMNLP 2022}, pages 2342--2354, Abu Dhabi, United Arab Emirates, December 2022. Association for Computational Linguistics.
\newblock \doi{10.18653/v1/2022.findings-emnlp.174}.
\newblock URL \url{https://aclanthology.org/2022.findings-emnlp.174/}.

\bibitem[Cui et~al.(2025{\natexlab{b}})Cui, Yuan, Wang, Li, Du, and Ding]{cui2025draw}
Zhiqing Cui, Jiahao Yuan, Hanqing Wang, Yanshu Li, Chenxu Du, and Zhenglong Ding.
\newblock Draw with thought: Unleashing multimodal reasoning for scientific diagram generation.
\newblock \emph{arXiv preprint arXiv:2504.09479}, 2025{\natexlab{b}}.

\bibitem[Cummins et~al.(2017)Cummins, Petoumenos, Wang, and Leather]{2017End}
Chris Cummins, Pavlos Petoumenos, Zheng Wang, and Hugh Leather.
\newblock End-to-end deep learning of optimization heuristics.
\newblock \emph{IEEE Computer Society}, 2017.

\bibitem[Cummins et~al.(2023)Cummins, Seeker, Grubisic, Elhoushi, Liang, Rozi{\`{e}}re, Gehring, Gloeckle, Hazelwood, Synnaeve, and Leather]{DBLP:journals/corr/abs-2309-07062}
Chris Cummins, Volker Seeker, Dejan Grubisic, Mostafa Elhoushi, Youwei Liang, Baptiste Rozi{\`{e}}re, Jonas Gehring, Fabian Gloeckle, Kim~M. Hazelwood, Gabriel Synnaeve, and Hugh Leather.
\newblock Large language models for compiler optimization.
\newblock \emph{CoRR}, abs/2309.07062, 2023.

\bibitem[Cursor(2025)]{tabrl}
Cursor.
\newblock Online rl for cursor tab.
\newblock \url{https://cursor.com/cn/blog/tab-rl}, 2025.
\newblock Accessed: 2025.

\bibitem[Daghighfarsoodeh et~al.(2025)Daghighfarsoodeh, Wang, Taherkhani, Sepidband, Abdollahi, Hemmati, and Pham]{deepbench}
Alireza Daghighfarsoodeh, Chung-Yu Wang, Hamed Taherkhani, Melika Sepidband, Mohammad Abdollahi, Hadi Hemmati, and Hung~Viet Pham.
\newblock Deep-bench: Deep learning benchmark dataset for code generation, 2025.
\newblock URL \url{https://arxiv.org/abs/2502.18726}.

\bibitem[Dai et~al.(2025{\natexlab{a}})Dai, Wang, Qi, Zhang, Jin, Yao, Huang, Fu, and Nallipogu]{dai2025lita}
Hankun Dai, Maoquan Wang, Mengnan Qi, Yikai Zhang, Zijian Jin, Yongqiang Yao, Yufan Huang, Shengyu Fu, and Elsie Nallipogu.
\newblock Lita: Light agent uncovers the agentic coding capabilities of llms.
\newblock \emph{arXiv preprint arXiv:2509.25873}, 2025{\natexlab{a}}.

\bibitem[Dai et~al.(2025{\natexlab{b}})Dai, Chen, and Yu]{dai2025feedbackeval}
Yifan Dai, Shuo Chen, and Hao Yu.
\newblock Feedbackeval: Benchmarking iterative feedback-driven program repair with llms.
\newblock In \emph{Proceedings of the 2025 Annual Meeting of the Association for Computational Linguistics (ACL)}, pages 456--468, 2025{\natexlab{b}}.

\bibitem[Dai et~al.(2023)Dai, Li, Liu, Li, Wei, Huang, and Liu]{dai2023safe}
Zhaowei Dai, Jiachen Li, Jiaming Liu, Zhaoran Li, Cha Wei, Yang Huang, and Zhen Liu.
\newblock Safe rlhf: Safe reinforcement learning from human feedback.
\newblock \emph{arXiv preprint arXiv:2310.12773}, 2023.

\bibitem[Dakhel et~al.(2023)Dakhel, Nikanjam, Majdinasab, Khomh, and Desmarais]{dakhel2023effectivetestgenerationusing}
Arghavan~Moradi Dakhel, Amin Nikanjam, Vahid Majdinasab, Foutse Khomh, and Michel~C. Desmarais.
\newblock Effective test generation using pre-trained large language models and mutation testing, 2023.
\newblock URL \url{https://arxiv.org/abs/2308.16557}.

\bibitem[Dao and Gu(2024)]{dao2024transformersaressms}
Tri Dao and Albert Gu.
\newblock Transformers are ssms: Generalized models and efficient algorithms through structured state space duality, 2024.
\newblock URL \url{https://arxiv.org/abs/2405.21060}.

\bibitem[{david}(2025)]{how_to_write_code2}
{david}.
\newblock How i program with llms, 2025.
\newblock URL \url{https://crawshaw.io/blog/programming-with-llms}.

\bibitem[{Daxberger} et~al.(2025){Daxberger}, {Wenzel}, {Griffiths}, {Gang}, {Lazarow}, {Kohavi}, {Kang}, {Eichner}, {Yang}, {Dehghan}, and {Grasch}]{MMSpatial}
Erik {Daxberger}, Nina {Wenzel}, David {Griffiths}, Haiming {Gang}, Justin {Lazarow}, Gefen {Kohavi}, Kai {Kang}, Marcin {Eichner}, Yinfei {Yang}, Afshin {Dehghan}, and Peter {Grasch}.
\newblock {MM-Spatial: Exploring 3D Spatial Understanding in Multimodal LLMs}.
\newblock \emph{arXiv e-prints}, art. arXiv:2503.13111, March 2025.
\newblock \doi{10.48550/arXiv.2503.13111}.

\bibitem[de~Oliveira et~al.(2024)de~Oliveira, Martins, Brand{\~a}o, da~Luz, Soares, and Melo]{sliding_puzzles_gym}
Bryan~LM de~Oliveira, Luana~GB Martins, Bruno Brand{\~a}o, Murilo~L da~Luz, Telma W de~L Soares, and Luckeciano~C Melo.
\newblock Sliding puzzles gym: A scalable benchmark for state representation in visual reinforcement learning.
\newblock \emph{arXiv preprint arXiv:2410.14038}, 2024.

\bibitem[de~Reus et~al.(2024)de~Reus, Oprescu, and Zuidema]{starcoder2energy2024}
Pepijn de~Reus, Ana Oprescu, and Jelle Zuidema.
\newblock An exploration of the effect of quantisation on energy consumption and inference time of starcoder2, 2024.
\newblock URL \url{https://arxiv.org/abs/2411.12758}.

\bibitem[de~Sousa and Hasselbring(2021)]{javabert}
Nelson~Tavares de~Sousa and Wilhelm Hasselbring.
\newblock Javabert: Training a transformer-based model for the java programming language.
\newblock In \emph{36th {IEEE/ACM} International Conference on Automated Software Engineering, {ASE} 2021 - Workshops, Melbourne, Australia, November 15-19, 2021}, pages 90--95. {IEEE}, 2021.
\newblock \doi{10.1109/ASEW52652.2021.00028}.
\newblock URL \url{https://doi.org/10.1109/ASEW52652.2021.00028}.

\bibitem[Dearing et~al.(2025)Dearing, Tao, Wu, Lan, and Taylor]{dearing2025lassillmbasedautomatedselfcorrecting}
Matthew~T. Dearing, Yiheng Tao, Xingfu Wu, Zhiling Lan, and Valerie Taylor.
\newblock Lassi: An llm-based automated self-correcting pipeline for translating parallel scientific codes, 2025.
\newblock URL \url{https://arxiv.org/abs/2407.01638}.

\bibitem[DeepMind(2022)]{deepmind_code_contests}
DeepMind.
\newblock Codecontests: A competitive programming dataset, 2022.
\newblock URL \url{https://huggingface.co/datasets/deepmind/code_contests}.
\newblock Accessed: 2024.

\bibitem[DeepMind(2025)]{deepmind2025geminidiffusion}
Google DeepMind.
\newblock Gemini diffusion, May 2025.
\newblock URL \url{https://deepmind.google/models/gemini-diffusion/}.

\bibitem[{deepseek}(2025)]{deepseek}
{deepseek}.
\newblock deepseek, 2025.
\newblock URL \url{https://www.deepseek.com/}.

\bibitem[{DeepSeek AI}(2024)]{deepseek2024coder}
{DeepSeek AI}.
\newblock {DeepSeek-Coder-V2}: Breaking the barrier of closed-source models in code intelligence.
\newblock \emph{arXiv preprint arXiv:2406.11931}, 2024.

\bibitem[DeepSeek-AI(2024)]{deepseekai2024deepseekv3technicalreport}
DeepSeek-AI.
\newblock Deepseek-v3 technical report, 2024.
\newblock URL \url{https://arxiv.org/abs/2412.19437}.

\bibitem[DeepSeek-AI(2025)]{deepseekai2024deepseekv32}
DeepSeek-AI.
\newblock Deepseek-v3.2-exp: Boosting long-context efficiency with deepseek sparse attention, 2025.

\bibitem[DeepSeek-AI et~al.(2024)DeepSeek-AI, Liu, Feng, Wang, Wang, Liu, Zhao, Dengr, Ruan, Dai, Guo, Yang, Chen, Ji, Li, Lin, Luo, Hao, Chen, Li, Zhang, Xu, Yang, Zhang, Ding, Xin, Gao, Li, Qu, Cai, Liang, Guo, Ni, Li, Chen, Yuan, Qiu, Song, Dong, Gao, Guan, Wang, Zhang, Xu, Xia, Zhao, Zhang, Li, Wang, Zhang, Zhang, Tang, Li, Tian, Huang, Wang, Zhang, Zhu, Chen, Du, Chen, Jin, Ge, Pan, Xu, Chen, Li, Lu, Zhou, Chen, Wu, Ye, Ma, Wang, Zhou, Yu, Zhou, Zheng, Wang, Pei, Yuan, Sun, Xiao, Zeng, An, Liu, Liang, Gao, Zhang, Li, Jin, Wang, Bi, Liu, Wang, Shen, Chen, Chen, Nie, Sun, Wang, Liu, Xie, Yu, Song, Zhou, Yang, Lu, Su, Wu, Li, Wei, Zhu, Xu, Huang, Li, Zhao, Sun, Li, Wang, Zheng, Zhang, Xiong, Zhao, He, Tang, Piao, Dong, Tan, Liu, Wang, Guo, Zhu, Wang, Zou, Zha, Ma, Yan, You, Liu, Ren, Ren, Sha, Fu, Huang, Zhang, Xie, Hao, Shao, Wen, Xu, Zhang, Li, Wang, Gu, Li, and Xie]{deepseekai2024deepseekv2}
DeepSeek-AI, Aixin Liu, Bei Feng, Bin Wang, Bingxuan Wang, Bo~Liu, Chenggang Zhao, Chengqi Dengr, Chong Ruan, Damai Dai, Daya Guo, Dejian Yang, Deli Chen, Dongjie Ji, Erhang Li, Fangyun Lin, Fuli Luo, Guangbo Hao, Guanting Chen, Guowei Li, H.~Zhang, Hanwei Xu, Hao Yang, Haowei Zhang, Honghui Ding, Huajian Xin, Huazuo Gao, Hui Li, Hui Qu, J.~L. Cai, Jian Liang, Jianzhong Guo, Jiaqi Ni, Jiashi Li, Jin Chen, Jingyang Yuan, Junjie Qiu, Junxiao Song, Kai Dong, Kaige Gao, Kang Guan, Lean Wang, Lecong Zhang, Lei Xu, Leyi Xia, Liang Zhao, Liyue Zhang, Meng Li, Miaojun Wang, Mingchuan Zhang, Minghua Zhang, Minghui Tang, Mingming Li, Ning Tian, Panpan Huang, Peiyi Wang, Peng Zhang, Qihao Zhu, Qinyu Chen, Qiushi Du, R.~J. Chen, R.~L. Jin, Ruiqi Ge, Ruizhe Pan, Runxin Xu, Ruyi Chen, S.~S. Li, Shanghao Lu, Shangyan Zhou, Shanhuang Chen, Shaoqing Wu, Shengfeng Ye, Shirong Ma, Shiyu Wang, Shuang Zhou, Shuiping Yu, Shunfeng Zhou, Size Zheng, T.~Wang, Tian Pei, Tian Yuan, Tianyu Sun, W.~L. Xiao, Wangding Zeng, Wei An, Wen
  Liu, Wenfeng Liang, Wenjun Gao, Wentao Zhang, X.~Q. Li, Xiangyue Jin, Xianzu Wang, Xiao Bi, Xiaodong Liu, Xiaohan Wang, Xiaojin Shen, Xiaokang Chen, Xiaosha Chen, Xiaotao Nie, Xiaowen Sun, Xiaoxiang Wang, Xin Liu, Xin Xie, Xingkai Yu, Xinnan Song, Xinyi Zhou, Xinyu Yang, Xuan Lu, Xuecheng Su, Y.~Wu, Y.~K. Li, Y.~X. Wei, Y.~X. Zhu, Yanhong Xu, Yanping Huang, Yao Li, Yao Zhao, Yaofeng Sun, Yaohui Li, Yaohui Wang, Yi~Zheng, Yichao Zhang, Yiliang Xiong, Yilong Zhao, Ying He, Ying Tang, Yishi Piao, Yixin Dong, Yixuan Tan, Yiyuan Liu, Yongji Wang, Yongqiang Guo, Yuchen Zhu, Yuduan Wang, Yuheng Zou, Yukun Zha, Yunxian Ma, Yuting Yan, Yuxiang You, Yuxuan Liu, Z.~Z. Ren, Zehui Ren, Zhangli Sha, Zhe Fu, Zhen Huang, Zhen Zhang, Zhenda Xie, Zhewen Hao, Zhihong Shao, Zhiniu Wen, Zhipeng Xu, Zhongyu Zhang, Zhuoshu Li, Zihan Wang, Zihui Gu, Zilin Li, and Ziwei Xie.
\newblock Deepseek-v2: A strong, economical, and efficient mixture-of-experts language model, 2024.
\newblock URL \url{https://arxiv.org/abs/2405.04434}.

\bibitem[DeepSeek{-}AI et~al.(2024)DeepSeek{-}AI, Zhu, Guo, Shao, Yang, Wang, Xu, Wu, Li, Gao, Ma, Zeng, Bi, Gu, Xu, Dai, Dong, Zhang, Piao, Gou, Xie, Hao, Wang, Song, Chen, Xie, Guan, You, Liu, Du, Gao, Lu, Chen, Wang, Deng, Li, Zhao, Ruan, Luo, and Liang]{zhu2024deepseekcoderv2}
DeepSeek{-}AI, Qihao Zhu, Daya Guo, Zhihong Shao, Dejian Yang, Peiyi Wang, Runxin Xu, Y.~Wu, Yukun Li, Huazuo Gao, Shirong Ma, Wangding Zeng, Xiao Bi, Zihui Gu, Hanwei Xu, Damai Dai, Kai Dong, Liyue Zhang, Yishi Piao, Zhibin Gou, Zhenda Xie, Zhewen Hao, Bingxuan Wang, Junxiao Song, Deli Chen, Xin Xie, Kang Guan, Yuxiang You, Aixin Liu, Qiushi Du, Wenjun Gao, Xuan Lu, Qinyu Chen, Yaohui Wang, Chengqi Deng, Jiashi Li, Chenggang Zhao, Chong Ruan, Fuli Luo, and Wenfeng Liang.
\newblock Deepseek-coder-v2: Breaking the barrier of closed-source models in code intelligence.
\newblock \emph{CoRR}, abs/2406.11931, 2024.
\newblock \doi{10.48550/ARXIV.2406.11931}.
\newblock URL \url{https://doi.org/10.48550/arXiv.2406.11931}.

\bibitem[DeepSeek-AI et~al.(2025{\natexlab{a}})DeepSeek-AI, Guo, Yang, Zhang, Song, Zhang, Xu, Zhu, Ma, Wang, Bi, Zhang, Yu, Wu, Wu, Gou, Shao, Li, Gao, Liu, Xue, Wang, Wu, Feng, Lu, Zhao, Deng, Zhang, Ruan, Dai, Chen, Ji, Li, Lin, Dai, Luo, Hao, Chen, Li, Zhang, Bao, Xu, Wang, Ding, Xin, Gao, Qu, Li, Guo, Li, Wang, Chen, Yuan, Qiu, Li, Cai, Ni, Liang, Chen, Dong, Hu, Gao, Guan, Huang, Yu, Wang, Zhang, Zhao, Wang, Zhang, Xu, Xia, Zhang, Zhang, Tang, Li, Wang, Li, Tian, Huang, Zhang, Wang, Chen, Du, Ge, Zhang, Pan, Wang, Chen, Jin, Chen, Lu, Zhou, Chen, Ye, Wang, Yu, Zhou, Pan, Li, Zhou, Wu, Ye, Yun, Pei, Sun, Wang, Zeng, Zhao, Liu, Liang, Gao, Yu, Zhang, Xiao, An, Liu, Wang, Chen, Nie, Cheng, Liu, Xie, Liu, Yang, Li, Su, Lin, Li, Jin, Shen, Chen, Sun, Wang, Song, Zhou, Wang, Shan, Li, Wang, Wei, Zhang, Xu, Li, Zhao, Sun, Wang, Yu, Zhang, Shi, Xiong, He, Piao, Wang, Tan, Ma, Liu, Guo, Ou, Wang, Gong, Zou, He, Xiong, Luo, You, Liu, Zhou, Zhu, Xu, Huang, Li, Zheng, Zhu, Ma, Tang, Zha, Yan, Ren, Ren, Sha, Fu, Xu,
  Xie, Zhang, Hao, Ma, Yan, Wu, Gu, Zhu, Liu, Li, Xie, Song, Pan, Huang, Xu, Zhang, and Zhang]{deepseekai2025deepseekr1}
DeepSeek-AI, Daya Guo, Dejian Yang, Haowei Zhang, Junxiao Song, Ruoyu Zhang, Runxin Xu, Qihao Zhu, Shirong Ma, Peiyi Wang, Xiao Bi, Xiaokang Zhang, Xingkai Yu, Yu~Wu, Z.~F. Wu, Zhibin Gou, Zhihong Shao, Zhuoshu Li, Ziyi Gao, Aixin Liu, Bing Xue, Bingxuan Wang, Bochao Wu, Bei Feng, Chengda Lu, Chenggang Zhao, Chengqi Deng, Chenyu Zhang, Chong Ruan, Damai Dai, Deli Chen, Dongjie Ji, Erhang Li, Fangyun Lin, Fucong Dai, Fuli Luo, Guangbo Hao, Guanting Chen, Guowei Li, H.~Zhang, Han Bao, Hanwei Xu, Haocheng Wang, Honghui Ding, Huajian Xin, Huazuo Gao, Hui Qu, Hui Li, Jianzhong Guo, Jiashi Li, Jiawei Wang, Jingchang Chen, Jingyang Yuan, Junjie Qiu, Junlong Li, J.~L. Cai, Jiaqi Ni, Jian Liang, Jin Chen, Kai Dong, Kai Hu, Kaige Gao, Kang Guan, Kexin Huang, Kuai Yu, Lean Wang, Lecong Zhang, Liang Zhao, Litong Wang, Liyue Zhang, Lei Xu, Leyi Xia, Mingchuan Zhang, Minghua Zhang, Minghui Tang, Meng Li, Miaojun Wang, Mingming Li, Ning Tian, Panpan Huang, Peng Zhang, Qiancheng Wang, Qinyu Chen, Qiushi Du, Ruiqi Ge, Ruisong
  Zhang, Ruizhe Pan, Runji Wang, R.~J. Chen, R.~L. Jin, Ruyi Chen, Shanghao Lu, Shangyan Zhou, Shanhuang Chen, Shengfeng Ye, Shiyu Wang, Shuiping Yu, Shunfeng Zhou, Shuting Pan, S.~S. Li, Shuang Zhou, Shaoqing Wu, Shengfeng Ye, Tao Yun, Tian Pei, Tianyu Sun, T.~Wang, Wangding Zeng, Wanjia Zhao, Wen Liu, Wenfeng Liang, Wenjun Gao, Wenqin Yu, Wentao Zhang, W.~L. Xiao, Wei An, Xiaodong Liu, Xiaohan Wang, Xiaokang Chen, Xiaotao Nie, Xin Cheng, Xin Liu, Xin Xie, Xingchao Liu, Xinyu Yang, Xinyuan Li, Xuecheng Su, Xuheng Lin, X.~Q. Li, Xiangyue Jin, Xiaojin Shen, Xiaosha Chen, Xiaowen Sun, Xiaoxiang Wang, Xinnan Song, Xinyi Zhou, Xianzu Wang, Xinxia Shan, Y.~K. Li, Y.~Q. Wang, Y.~X. Wei, Yang Zhang, Yanhong Xu, Yao Li, Yao Zhao, Yaofeng Sun, Yaohui Wang, Yi~Yu, Yichao Zhang, Yifan Shi, Yiliang Xiong, Ying He, Yishi Piao, Yisong Wang, Yixuan Tan, Yiyang Ma, Yiyuan Liu, Yongqiang Guo, Yuan Ou, Yuduan Wang, Yue Gong, Yuheng Zou, Yujia He, Yunfan Xiong, Yuxiang Luo, Yuxiang You, Yuxuan Liu, Yuyang Zhou, Y.~X. Zhu,
  Yanhong Xu, Yanping Huang, Yaohui Li, Yi~Zheng, Yuchen Zhu, Yunxian Ma, Ying Tang, Yukun Zha, Yuting Yan, Z.~Z. Ren, Zehui Ren, Zhangli Sha, Zhe Fu, Zhean Xu, Zhenda Xie, Zhengyan Zhang, Zhewen Hao, Zhicheng Ma, Zhigang Yan, Zhiyu Wu, Zihui Gu, Zijia Zhu, Zijun Liu, Zilin Li, Ziwei Xie, Ziyang Song, Zizheng Pan, Zhen Huang, Zhipeng Xu, Zhongyu Zhang, and Zhen Zhang.
\newblock Deepseek-r1: Incentivizing reasoning capability in llms via reinforcement learning, 2025{\natexlab{a}}.
\newblock URL \url{https://arxiv.org/abs/2501.12948}.

\bibitem[DeepSeek-AI et~al.(2025{\natexlab{b}})DeepSeek-AI, Liu, Feng, Xue, Wang, Wu, Lu, Zhao, Deng, Zhang, Ruan, Dai, Guo, Yang, Chen, Ji, Li, Lin, Dai, Luo, Hao, Chen, Li, Zhang, Bao, Xu, Wang, Zhang, Ding, Xin, Gao, Li, Qu, Cai, Liang, Guo, Ni, Li, Wang, Chen, Chen, Yuan, Qiu, Li, Song, Dong, Hu, Gao, Guan, Huang, Yu, Wang, Zhang, Xu, Xia, Zhao, Wang, Zhang, Li, Wang, Zhang, Zhang, Tang, Li, Tian, Huang, Wang, Zhang, Wang, Zhu, Chen, Du, Chen, Jin, Ge, Zhang, Pan, Wang, Xu, Zhang, Chen, Li, Lu, Zhou, Chen, Wu, Ye, Ye, Ma, Wang, Zhou, Yu, Zhou, Pan, Wang, Yun, Pei, Sun, Xiao, Zeng, Zhao, An, Liu, Liang, Gao, Yu, Zhang, Li, Jin, Wang, Bi, Liu, Wang, Shen, Chen, Zhang, Chen, Nie, Sun, Wang, Cheng, Liu, Xie, Liu, Yu, Song, Shan, Zhou, Yang, Li, Su, Lin, Li, Wang, Wei, Zhu, Zhang, Xu, Xu, Huang, Li, Zhao, Sun, Li, Wang, Yu, Zheng, Zhang, Shi, Xiong, He, Tang, Piao, Wang, Tan, Ma, Liu, Guo, Wu, Ou, Zhu, Wang, Gong, Zou, He, Zha, Xiong, Ma, Yan, Luo, You, Liu, Zhou, Wu, Ren, Ren, Sha, Fu, Xu, Huang, Zhang, Xie,
  Zhang, Hao, Gou, Ma, Yan, Shao, Xu, Wu, Zhang, Li, Gu, Zhu, Liu, Li, Xie, Song, Gao, and Pan]{deepseekai2025deepseekv3}
DeepSeek-AI, Aixin Liu, Bei Feng, Bing Xue, Bingxuan Wang, Bochao Wu, Chengda Lu, Chenggang Zhao, Chengqi Deng, Chenyu Zhang, Chong Ruan, Damai Dai, Daya Guo, Dejian Yang, Deli Chen, Dongjie Ji, Erhang Li, Fangyun Lin, Fucong Dai, Fuli Luo, Guangbo Hao, Guanting Chen, Guowei Li, H.~Zhang, Han Bao, Hanwei Xu, Haocheng Wang, Haowei Zhang, Honghui Ding, Huajian Xin, Huazuo Gao, Hui Li, Hui Qu, J.~L. Cai, Jian Liang, Jianzhong Guo, Jiaqi Ni, Jiashi Li, Jiawei Wang, Jin Chen, Jingchang Chen, Jingyang Yuan, Junjie Qiu, Junlong Li, Junxiao Song, Kai Dong, Kai Hu, Kaige Gao, Kang Guan, Kexin Huang, Kuai Yu, Lean Wang, Lecong Zhang, Lei Xu, Leyi Xia, Liang Zhao, Litong Wang, Liyue Zhang, Meng Li, Miaojun Wang, Mingchuan Zhang, Minghua Zhang, Minghui Tang, Mingming Li, Ning Tian, Panpan Huang, Peiyi Wang, Peng Zhang, Qiancheng Wang, Qihao Zhu, Qinyu Chen, Qiushi Du, R.~J. Chen, R.~L. Jin, Ruiqi Ge, Ruisong Zhang, Ruizhe Pan, Runji Wang, Runxin Xu, Ruoyu Zhang, Ruyi Chen, S.~S. Li, Shanghao Lu, Shangyan Zhou, Shanhuang
  Chen, Shaoqing Wu, Shengfeng Ye, Shengfeng Ye, Shirong Ma, Shiyu Wang, Shuang Zhou, Shuiping Yu, Shunfeng Zhou, Shuting Pan, T.~Wang, Tao Yun, Tian Pei, Tianyu Sun, W.~L. Xiao, Wangding Zeng, Wanjia Zhao, Wei An, Wen Liu, Wenfeng Liang, Wenjun Gao, Wenqin Yu, Wentao Zhang, X.~Q. Li, Xiangyue Jin, Xianzu Wang, Xiao Bi, Xiaodong Liu, Xiaohan Wang, Xiaojin Shen, Xiaokang Chen, Xiaokang Zhang, Xiaosha Chen, Xiaotao Nie, Xiaowen Sun, Xiaoxiang Wang, Xin Cheng, Xin Liu, Xin Xie, Xingchao Liu, Xingkai Yu, Xinnan Song, Xinxia Shan, Xinyi Zhou, Xinyu Yang, Xinyuan Li, Xuecheng Su, Xuheng Lin, Y.~K. Li, Y.~Q. Wang, Y.~X. Wei, Y.~X. Zhu, Yang Zhang, Yanhong Xu, Yanhong Xu, Yanping Huang, Yao Li, Yao Zhao, Yaofeng Sun, Yaohui Li, Yaohui Wang, Yi~Yu, Yi~Zheng, Yichao Zhang, Yifan Shi, Yiliang Xiong, Ying He, Ying Tang, Yishi Piao, Yisong Wang, Yixuan Tan, Yiyang Ma, Yiyuan Liu, Yongqiang Guo, Yu~Wu, Yuan Ou, Yuchen Zhu, Yuduan Wang, Yue Gong, Yuheng Zou, Yujia He, Yukun Zha, Yunfan Xiong, Yunxian Ma, Yuting Yan, Yuxiang
  Luo, Yuxiang You, Yuxuan Liu, Yuyang Zhou, Z.~F. Wu, Z.~Z. Ren, Zehui Ren, Zhangli Sha, Zhe Fu, Zhean Xu, Zhen Huang, Zhen Zhang, Zhenda Xie, Zhengyan Zhang, Zhewen Hao, Zhibin Gou, Zhicheng Ma, Zhigang Yan, Zhihong Shao, Zhipeng Xu, Zhiyu Wu, Zhongyu Zhang, Zhuoshu Li, Zihui Gu, Zijia Zhu, Zijun Liu, Zilin Li, Ziwei Xie, Ziyang Song, Ziyi Gao, and Zizheng Pan.
\newblock Deepseek-v3 technical report, 2025{\natexlab{b}}.
\newblock URL \url{https://arxiv.org/abs/2412.19437}.

\bibitem[Deng et~al.(2025{\natexlab{a}})Deng, Wu, Feng, Wang, and Long]{deng2025compilerdream}
Chaoyi Deng, Jialong Wu, Ningya Feng, Jianmin Wang, and Mingsheng Long.
\newblock Compilerdream: Learning a compiler world model for general code optimization.
\newblock In \emph{Proceedings of the 31st ACM SIGKDD Conference on Knowledge Discovery and Data Mining V. 2}, pages 486--497, 2025{\natexlab{a}}.

\bibitem[Deng et~al.(2025{\natexlab{b}})Deng, Chen, Chen, Zhao, and Wen]{deng2025decomposing}
Jia Deng, Jie Chen, Zhipeng Chen, Wayne~Xin Zhao, and Ji-Rong Wen.
\newblock Decomposing the entropy-performance exchange: The missing keys to unlocking effective reinforcement learning.
\newblock \emph{arXiv preprint arXiv:2508.02260}, 2025{\natexlab{b}}.

\bibitem[Deng et~al.(2023{\natexlab{a}})Deng, Wang, Zhang, Li, and Chen]{deng2023gptfuzzer}
Jiahao Deng, Ziyuan Wang, Chuhui Zhang, Xuezixiang Li, and Pin-Yu Chen.
\newblock Gptfuzzer: Red teaming large language models with auto-generated jailbreak prompts, 2023{\natexlab{a}}.

\bibitem[Deng et~al.(2023{\natexlab{b}})Deng, Gu, Zheng, Chen, Stevens, Wang, Sun, and Su]{deng2023mind2web}
Xiang Deng, Yu~Gu, Boyuan Zheng, Shijie Chen, Sam Stevens, Boshi Wang, Huan Sun, and Yu~Su.
\newblock Mind2web: Towards a generalist agent for the web.
\newblock \emph{Advances in Neural Information Processing Systems}, 36:\penalty0 28091--28114, 2023{\natexlab{b}}.

\bibitem[Deng et~al.(2024)Deng, Tan, Wang, Zhou, Liu, Guo, Sun, and Jiang]{deng2024autodefense}
Zekun Deng, Zhaofeng Tan, Yujun Wang, Jie Zhou, Kelin Liu, Yue Guo, Kaixuan Sun, and Xin Jiang.
\newblock Autodefense: Multi-agent llm defense against jailbreak attacks.
\newblock \emph{arXiv preprint arXiv:2404.09127}, 2024.

\bibitem[Dettmers et~al.(2022)Dettmers, Lewis, Belkada, and Zettlemoyer]{dettmers2022gpt3}
Tim Dettmers, Mike Lewis, Younes Belkada, and Luke Zettlemoyer.
\newblock Gpt3. int8 (): 8-bit matrix multiplication for transformers at scale.
\newblock \emph{Advances in Neural Information Processing Systems}, 35:\penalty0 30318--30332, 2022.

\bibitem[Dettmers et~al.(2024)Dettmers, Pagnoni, Holtzman, and Zettlemoyer]{dettmers2024qlora}
Tim Dettmers, Artidoro Pagnoni, Ari Holtzman, and Luke Zettlemoyer.
\newblock Qlora: Efficient finetuning of quantized llms.
\newblock \emph{Advances in Neural Information Processing Systems}, 36, 2024.

\bibitem[{DEV Community}(2025)]{dev2024amazonq}
{DEV Community}.
\newblock The known and unknown of amazon q developer.
\newblock \url{https://dev.to/aws-builders/amazon-q-developer}, 2025.

\bibitem[Devlin et~al.(2019)Devlin, Chang, Lee, and Toutanova]{devlin2019bert}
Jacob Devlin, Ming-Wei Chang, Kenton Lee, and Kristina Toutanova.
\newblock {BERT}: Pre-training of deep bidirectional transformers for language understanding.
\newblock In Jill Burstein, Christy Doran, and Thamar Solorio, editors, \emph{Proceedings of the 2019 Conference of the North {A}merican Chapter of the Association for Computational Linguistics: Human Language Technologies, Volume 1 (Long and Short Papers)}, pages 4171--4186, Minneapolis, Minnesota, June 2019. Association for Computational Linguistics.
\newblock \doi{10.18653/v1/N19-1423}.
\newblock URL \url{https://aclanthology.org/N19-1423/}.

\bibitem[Di et~al.(2024)Di, Li, Yu, Jiang, Cai, Cao, Chen, Chen, Chen, Chen, Fan, Gong, Gong, Hu, Guo, Lei, Li, Li, Liang, Liao, Liu, Liu, Liu, Lu, Shen, Wang, Wang, Wang, Xu, Yang, Ye, Zhang, Zhang, Zhao, Zheng, Zhou, Zhu, and Zhu]{CodeFuseEval}
Peng Di, Jianguo Li, Hang Yu, Wei Jiang, Wenting Cai, Yang Cao, Chaoyu Chen, Dajun Chen, Hongwei Chen, Liang Chen, Gang Fan, Jie Gong, Zi~Gong, Wen Hu, Tingting Guo, Zhichao Lei, Ting Li, Zheng Li, Ming Liang, Cong Liao, Bingchang Liu, Jiachen Liu, Zhiwei Liu, Shaojun Lu, Min Shen, Guangpei Wang, Huan Wang, Zhi Wang, Zhaogui Xu, Jiawei Yang, Qing Ye, Gehao Zhang, Yu~Zhang, Zelin Zhao, Xunjin Zheng, Hailian Zhou, Lifu Zhu, and Xianying Zhu.
\newblock Codefuse-13b: A pretrained multi-lingual code large language model.
\newblock In \emph{Proceedings of the 46th International Conference on Software Engineering: Software Engineering in Practice}, ICSE-SEIP ’24, page 418–429. ACM, April 2024.
\newblock \doi{10.1145/3639477.3639719}.
\newblock URL \url{http://dx.doi.org/10.1145/3639477.3639719}.

\bibitem[Di~Grazia and Pradel(2023)]{code_search_survey}
Luca Di~Grazia and Michael Pradel.
\newblock Code search: A survey of techniques for finding code.
\newblock \emph{ACM Computing Surveys}, 55\penalty0 (11):\penalty0 1--31, 2023.

\bibitem[{Diggs} et~al.(2024){Diggs}, {Doyle}, {Madan}, {Scott}, {Escamilla}, {Zimmer}, {Nekoo}, {Ursino}, {Bartholf}, {Robin}, {Patel}, {Glasz}, {Macke}, {Kirk}, {Phillips}, {Sridharan}, {Wendt}, {Rosen}, {Naik}, {Brunelle}, and {Thaker}]{Diggs}
Colin {Diggs}, Michael {Doyle}, Amit {Madan}, Siggy {Scott}, Emily {Escamilla}, Jacob {Zimmer}, Naveed {Nekoo}, Paul {Ursino}, Michael {Bartholf}, Zachary {Robin}, Anand {Patel}, Chris {Glasz}, William {Macke}, Paul {Kirk}, Jasper {Phillips}, Arun {Sridharan}, Doug {Wendt}, Scott {Rosen}, Nitin {Naik}, Justin~F. {Brunelle}, and Samruddhi {Thaker}.
\newblock {Leveraging LLMs for Legacy Code Modernization: Challenges and Opportunities for LLM-Generated Documentation}.
\newblock \emph{arXiv e-prints}, art. arXiv:2411.14971, November 2024.
\newblock \doi{10.48550/arXiv.2411.14971}.

\bibitem[Dilgren et~al.(2025)Dilgren, Chiniya, Griffith, Ding, and Chen]{dilgren2025secrepobenchbenchmarkingllmssecure}
Connor Dilgren, Purva Chiniya, Luke Griffith, Yu~Ding, and Yizheng Chen.
\newblock Secrepobench: Benchmarking llms for secure code generation in real-world repositories, 2025.
\newblock URL \url{https://arxiv.org/abs/2504.21205}.

\bibitem[Ding et~al.(2023)Ding, Wang, Ahmad, Ding, Tan, Jain, Ramanathan, Nallapati, Bhatia, Roth, and Xiang]{ding2023crosscodeevaldiversemultilingualbenchmark}
Yangruibo Ding, Zijian Wang, Wasi~Uddin Ahmad, Hantian Ding, Ming Tan, Nihal Jain, Murali~Krishna Ramanathan, Ramesh Nallapati, Parminder Bhatia, Dan Roth, and Bing Xiang.
\newblock Crosscodeeval: A diverse and multilingual benchmark for cross-file code completion, 2023.
\newblock URL \url{https://arxiv.org/abs/2310.11248}.

\bibitem[Ding et~al.(2024{\natexlab{a}})Ding, Fu, Ibrahim, Sitawarin, Chen, Alomair, Wagner, Ray, and Chen]{ding2024vulnerability}
Yangruibo Ding, Yanjun Fu, Omniyyah Ibrahim, Chawin Sitawarin, Xinyun Chen, Basel Alomair, David Wagner, Baishakhi Ray, and Yizheng Chen.
\newblock Vulnerability detection with code language models: How far are we?
\newblock \emph{arXiv preprint arXiv:2403.18624}, 2024{\natexlab{a}}.

\bibitem[Ding et~al.(2024{\natexlab{b}})Ding, Wang, Ahmad, et~al.]{ding2024crosscodeeval}
Yangruibo Ding, Zijian Wang, Wasi Ahmad, et~al.
\newblock {CrossCodeEval}: A diverse and multilingual benchmark for cross-file code completion.
\newblock \emph{NeurIPS}, 2024{\natexlab{b}}.

\bibitem[Ding et~al.(2025)Ding, Zhang, Chi, and Wang]{ding2025frontend}
Zijian Ding, Qinshi Zhang, Mohan Chi, and Ziyi Wang.
\newblock Frontend diffusion: Empowering self-representation of junior researchers and designers through agentic workflows.
\newblock \emph{arXiv preprint arXiv:2502.03788}, 2025.

\bibitem[D{\"{o}}nder et~al.(2025)D{\"{o}}nder, Hommel, Wen{-}Yi, Mimno, and Jo]{GenaSQL}
Yusuf~Denizay D{\"{o}}nder, Derek Hommel, Andrea~W. Wen{-}Yi, David Mimno, and Unso Eun~Seo Jo.
\newblock Cheaper, better, faster, stronger: Robust text-to-sql without chain-of-thought or fine-tuning.
\newblock \emph{CoRR}, abs/2505.14174, 2025.

\bibitem[Dong et~al.(2023{\natexlab{a}})Dong, Zhang, Ge, Mao, Gao, Chen, Lin, and Lou]{C3-sql}
Xuemei Dong, Chao Zhang, Yuhang Ge, Yuren Mao, Yunjun Gao, Lu~Chen, Jinshu Lin, and Dongfang Lou.
\newblock {C3:} zero-shot text-to-sql with chatgpt.
\newblock \emph{CoRR}, abs/2307.07306, 2023{\natexlab{a}}.

\bibitem[Dong et~al.(2023{\natexlab{b}})Dong, Ding, Jiang, Li, Li, and Jin]{codescore}
Yihong Dong, Jiazheng Ding, Xue Jiang, Zhuo Li, Ge~Li, and Zhi Jin.
\newblock Codescore: Evaluating code generation by learning code execution. corr abs/2301.09043 (2023), 2023{\natexlab{b}}.

\bibitem[Dora et~al.(2025)Dora, Lunkad, Aslam, Venkatesan, and Shukla]{dora2025hidden}
Swaroop Dora, Deven Lunkad, Naziya Aslam, S~Venkatesan, and Sandeep~Kumar Shukla.
\newblock The hidden risks of llm-generated web application code: A security-centric evaluation of code generation capabilities in large language models.
\newblock \emph{arXiv preprint arXiv:2504.20612}, 2025.

\bibitem[Dougherty and Mehta(2025)]{fvapps}
Quinn Dougherty and Ronak Mehta.
\newblock Proving the coding interview: A benchmark for formally verified code generation, 2025.
\newblock URL \url{https://arxiv.org/abs/2502.05714}.

\bibitem[Drain et~al.(2021)Drain, Clement, Serrato, and Sundaresan]{drain2021deepdebug}
Dawn Drain, Colin~B Clement, Guillermo Serrato, and Neel Sundaresan.
\newblock Deepdebug: Fixing python bugs using stack traces, backtranslation, and code skeletons.
\newblock \emph{arXiv preprint arXiv:2105.09352}, 2021.

\bibitem[Drissi et~al.(2018)Drissi, Watkins, Khant, Ojha, Sandoval, Segev, Weiner, and Keller]{drissi2018programlanguagetranslationusing}
Mehdi Drissi, Olivia Watkins, Aditya Khant, Vivaswat Ojha, Pedro Sandoval, Rakia Segev, Eric Weiner, and Robert Keller.
\newblock Program language translation using a grammar-driven tree-to-tree model, 2018.
\newblock URL \url{https://arxiv.org/abs/1807.01784}.

\bibitem[Du et~al.(2025{\natexlab{a}})Du, Liu, Guo, Wang, Huang, Ni, and Li]{du2025dependevalbenchmarkingllmsrepository}
Junjia Du, Yadi Liu, Hongcheng Guo, Jiawei Wang, Haojian Huang, Yunyi Ni, and Zhoujun Li.
\newblock Dependeval: Benchmarking llms for repository dependency understanding, 2025{\natexlab{a}}.
\newblock URL \url{https://arxiv.org/abs/2503.06689}.

\bibitem[Du et~al.(2025{\natexlab{b}})Du, Xu, Zhu, Wang, and Mao]{du2025deepresearch}
Mingxuan Du, Benfeng Xu, Chiwei Zhu, Xiaorui Wang, and Zhendong Mao.
\newblock Deepresearch bench: A comprehensive benchmark for deep research agents.
\newblock \emph{arXiv preprint arXiv:2506.11763}, 2025{\natexlab{b}}.

\bibitem[Du et~al.(2024{\natexlab{a}})Du, Luu, Ji, Liu, and Ng]{mercury2024}
Mingzhe Du, Anh~Tuan Luu, Bin Ji, Qian Liu, and See-Kiong Ng.
\newblock Mercury: A code efficiency benchmark for code large language models, 2024{\natexlab{a}}.
\newblock URL \url{https://arxiv.org/abs/2402.07844}.

\bibitem[Du et~al.(2025{\natexlab{c}})Du, Luu, Ji, Wu, Qing, Huang, Zhuo, Liu, and Ng]{du-etal-2025-codearena}
Mingzhe Du, Anh~Tuan Luu, Bin Ji, Xiaobao Wu, Yuhao Qing, Dong Huang, Terry~Yue Zhuo, Qian Liu, and See-Kiong Ng.
\newblock {C}ode{A}rena: A collective evaluation platform for {LLM} code generation.
\newblock In Pushkar Mishra, Smaranda Muresan, and Tao Yu, editors, \emph{Proceedings of the 63rd Annual Meeting of the Association for Computational Linguistics (Volume 3: System Demonstrations)}, pages 502--512, Vienna, Austria, July 2025{\natexlab{c}}. Association for Computational Linguistics.
\newblock ISBN 979-8-89176-253-4.
\newblock \doi{10.18653/v1/2025.acl-demo.48}.
\newblock URL \url{https://aclanthology.org/2025.acl-demo.48/}.

\bibitem[Du et~al.(2022{\natexlab{a}})Du, Huang, Dai, Tong, Lepikhin, Xu, Krikun, Zhou, Yu, Firat, Zoph, Fedus, Bosma, Zhou, Wang, Wang, Webster, Pellat, Robinson, Meier-Hellstern, Duke, Dixon, Zhang, Le, Wu, Chen, and Cui]{du2022glam}
Nan Du, Yanping Huang, Andrew~M. Dai, Simon Tong, Dmitry Lepikhin, Yuanzhong Xu, Maxim Krikun, Yanqi Zhou, Adams~Wei Yu, Orhan Firat, Barret Zoph, Liam Fedus, Maarten Bosma, Zongwei Zhou, Tao Wang, Yu~Emma Wang, Kellie Webster, Marie Pellat, Kevin Robinson, Kathleen Meier-Hellstern, Toju Duke, Lucas Dixon, Kun Zhang, Quoc~V Le, Yonghui Wu, Zhifeng Chen, and Claire Cui.
\newblock Glam: Efficient scaling of language models with mixture-of-experts, 2022{\natexlab{a}}.
\newblock URL \url{https://arxiv.org/abs/2112.06905}.

\bibitem[Du et~al.(2023)Du, Liu, Wang, Wang, Liu, Chen, Feng, Sha, Peng, and Lou]{du2023classeval}
Xueying Du, Mingwei Liu, Kaixin Wang, Hanlin Wang, Junwei Liu, Yixuan Chen, Jiayi Feng, Chaofeng Sha, Xin Peng, and Yiling Lou.
\newblock Classeval: A manually-crafted benchmark for evaluating llms on class-level code generation, 2023.
\newblock URL \url{https://arxiv.org/abs/2308.01861}.

\bibitem[Du et~al.(2025{\natexlab{d}})Du, Cai, Zhou, Wang, Qian, Pang, Liu, Hu, and Chen]{du2025swedevevaluatingtrainingautonomous}
Yaxin Du, Yuzhu Cai, Yifan Zhou, Cheng Wang, Yu~Qian, Xianghe Pang, Qian Liu, Yue Hu, and Siheng Chen.
\newblock Swe-dev: Evaluating and training autonomous feature-driven software development, 2025{\natexlab{d}}.
\newblock URL \url{https://arxiv.org/abs/2505.16975}.

\bibitem[Du et~al.(2025{\natexlab{e}})]{du2025swedev}
Yaxin Du et~al.
\newblock Swe-dev: Evaluating and training autonomous feature-driven software development, 2025{\natexlab{e}}.

\bibitem[Du et~al.(2025{\natexlab{f}})Du, Huang, Zhao, and Lin]{du2025faircoder}
Yongkang Du, Jen-tse Huang, Jieyu Zhao, and Lu~Lin.
\newblock Faircoder: Evaluating social bias of llms in code generation.
\newblock \emph{arXiv preprint arXiv:2501.05396}, 2025{\natexlab{f}}.

\bibitem[Du et~al.(2022{\natexlab{b}})Du, Qian, Liu, Ding, Qiu, Yang, and Tang]{du2022glm}
Zhengxiao Du, Yujie Qian, Xiao Liu, Ming Ding, Jiezhong Qiu, Zhilin Yang, and Jie Tang.
\newblock Glm: General language model pretraining with autoregressive blank infilling, 2022{\natexlab{b}}.
\newblock URL \url{https://arxiv.org/abs/2103.10360}.

\bibitem[Du et~al.(2024{\natexlab{b}})Du, Qian, Liu, Xie, Wang, Dang, Chen, and Yang]{du2024multiagent}
Zhuoyun Du, Chen Qian, Wei Liu, Zihao Xie, Yifei Wang, Yufan Dang, Weize Chen, and Cheng Yang.
\newblock Multi-agent software development through cross-team collaboration.
\newblock \emph{arXiv preprint arXiv:2406.08979}, 2024{\natexlab{b}}.
\newblock URL \url{https://arxiv.org/abs/2406.08979}.

\bibitem[Eliseeva et~al.(2023)Eliseeva, Sokolov, Bogomolov, Golubev, Dig, and Bryksin]{eliseeva2023commitmessagegenerationhistoryaware}
Aleksandra Eliseeva, Yaroslav Sokolov, Egor Bogomolov, Yaroslav Golubev, Danny Dig, and Timofey Bryksin.
\newblock From commit message generation to history-aware commit message completion, 2023.
\newblock URL \url{https://arxiv.org/abs/2308.07655}.

\bibitem[Eniser et~al.(2025)Eniser, Zhang, David, Wang, Christakis, Paulsen, Dodds, and Kroening]{eniser2025translatingrealworldcodellms}
Hasan~Ferit Eniser, Hanliang Zhang, Cristina David, Meng Wang, Maria Christakis, Brandon Paulsen, Joey Dodds, and Daniel Kroening.
\newblock Towards translating real-world code with llms: A study of translating to rust, 2025.
\newblock URL \url{https://arxiv.org/abs/2405.11514}.

\bibitem[Eom et~al.(2024)Eom, Jeong, and Kwon]{eom2024fuzzing}
Jueon Eom, Seyeon Jeong, and Taekyoung Kwon.
\newblock Fuzzing javascript interpreters with coverage-guided reinforcement learning for llm-based mutation.
\newblock In \emph{Proceedings of the 33rd ACM SIGSOFT International Symposium on Software Testing and Analysis}, pages 1656--1668, 2024.

\bibitem[etc(2025)]{deepswe2025}
Michael~Luo etc.
\newblock Deepswe: Training a state-of-the-art coding agent from scratch by scaling rl.
\newblock \url{N/A}, 2025.
\newblock Notion Blog.

\bibitem[Everitt et~al.(2025)Everitt, Garbacea, Bellot, Richens, Papadatos, Campos, and Shah]{everitt2025evaluating}
Tom Everitt, Cristina Garbacea, Alexis Bellot, Jonathan Richens, Henry Papadatos, Sim{\'e}on Campos, and Rohin Shah.
\newblock Evaluating the goal-directedness of large language models.
\newblock \emph{arXiv preprint arXiv:2504.11844}, 2025.

\bibitem[Face(2021)]{codeparrot}
Hugging Face.
\newblock Codeparrot dataset, 2021.

\bibitem[Fakih et~al.(2025)Fakih, Dharmaji, Bouzidi, Araya, Ogundare, and Faruque]{fakih2025llm4cve}
Mohamad Fakih, Rahul Dharmaji, Halima Bouzidi, Gustavo~Quiros Araya, Oluwatosin Ogundare, and Mohammad Abdullah~Al Faruque.
\newblock Llm4cve: Enabling iterative automated vulnerability repair with large language models.
\newblock \emph{arXiv preprint arXiv:2501.03446}, 2025.

\bibitem[Fan et~al.(2025{\natexlab{a}})Fan, Shen, Cheng, Chen, Liang, and Liu]{fan2025online}
Jiajun Fan, Shuaike Shen, Chaoran Cheng, Yuxin Chen, Chumeng Liang, and Ge~Liu.
\newblock Online reward-weighted fine-tuning of flow matching with wasserstein regularization.
\newblock In \emph{The Thirteenth International Conference on Learning Representations}, 2025{\natexlab{a}}.

\bibitem[Fan et~al.(2022)Fan, Wang, Jiang, Mandlekar, Yang, Zhu, Tang, Huang, Zhu, and Anandkumar]{fan2022minedojo}
Linxi Fan, Guanzhi Wang, Yunfan Jiang, Ajay Mandlekar, Yuncong Yang, Haoyi Zhu, Andrew Tang, De-An Huang, Yuke Zhu, and Anima Anandkumar.
\newblock Minedojo: Building open-ended embodied agents with internet-scale knowledge, 2022.
\newblock URL \url{https://arxiv.org/abs/2206.08853}.

\bibitem[Fan et~al.(2025{\natexlab{b}})Fan, Zhao, Zhang, Shen, Wang, and Wu]{fan2025gui-bee}
Yue Fan, Handong Zhao, Ruiyi Zhang, Yu~Shen, Xin~Eric Wang, and Gang Wu.
\newblock Gui-bee: Align gui action grounding to novel environments via autonomous exploration, 2025{\natexlab{b}}.
\newblock URL \url{https://arxiv.org/abs/2501.13896}.

\bibitem[Fang et~al.(2025{\natexlab{a}})Fang, Ding, and Xu]{fang2025evaloopassessingllmrobustness}
Sen Fang, Weiyuan Ding, and Bowen Xu.
\newblock Evaloop: Assessing llm robustness in programming from a self-consistency perspective, 2025{\natexlab{a}}.
\newblock URL \url{https://arxiv.org/abs/2505.12185}.

\bibitem[Fang et~al.(2025{\natexlab{b}})Fang, Zhang, Zhang, Ma, Yu, Mi, and Yu]{webevolver}
Tianqing Fang, Hongming Zhang, Zhisong Zhang, Kaixin Ma, Wenhao Yu, Haitao Mi, and Dong Yu.
\newblock Webevolver: Enhancing web agent self-improvement with coevolving world model.
\newblock \emph{arXiv preprint arXiv:2504.21024}, 2025{\natexlab{b}}.

\bibitem[Fedus et~al.(2022)Fedus, Zoph, and Shazeer]{fedus2022switchtransformers}
William Fedus, Barret Zoph, and Noam Shazeer.
\newblock Switch transformers: Scaling to trillion parameter models with simple and efficient sparsity, 2022.
\newblock URL \url{https://arxiv.org/abs/2101.03961}.

\bibitem[Feng and Berke(2024)]{feng2024ai}
Bill Feng and Clark Berke.
\newblock The ai agent code of conduct: Automated guardrail policy-as-prompt synthesis.
\newblock \emph{arXiv preprint arXiv:2405.04859}, 2024.

\bibitem[Feng et~al.(2024{\natexlab{a}})Feng, Tang, Li, et~al.]{feng2024review}
Bo~Feng, Jie Tang, Wenbo Li, et~al.
\newblock A review of automatic source code summarization.
\newblock \emph{SpringerLink Software Quality Journal}, 32\penalty0 (1):\penalty0 45--63, 2024{\natexlab{a}}.
\newblock URL \url{https://link.springer.com/article/10.1007/s11219-024-09557-4}.

\bibitem[Feng et~al.(2023)Feng, Yuan, Chen, Xing, and Chen]{feng2023designing}
Sidong Feng, Mingyue Yuan, Jieshan Chen, Zhenchang Xing, and Chunyang Chen.
\newblock Designing with language: Wireframing ui design intent with generative large language models.
\newblock \emph{arXiv preprint arXiv:2312.07755}, 2023.

\bibitem[Feng et~al.(2025)Feng, Du, Liu, Wang, Lv, Wang, and Chen]{feng2025breaking}
Sidong Feng, Changhao Du, Huaxiao Liu, Qingnan Wang, Zhengwei Lv, Mengfei Wang, and Chunyang Chen.
\newblock Breaking single-tester limits: Multi-agent llms for multi-user feature testing.
\newblock \emph{arXiv preprint arXiv:2506.17539}, 2025.

\bibitem[Feng et~al.(2020)Feng, Guo, Tang, Duan, Feng, Gong, Shou, Qin, Liu, Jiang, and Zhou]{feng2020codebertpretrainedmodelprogramming}
Zhangyin Feng, Daya Guo, Duyu Tang, Nan Duan, Xiaocheng Feng, Ming Gong, Linjun Shou, Bing Qin, Ting Liu, Daxin Jiang, and Ming Zhou.
\newblock Codebert: A pre-trained model for programming and natural languages, 2020.
\newblock URL \url{https://arxiv.org/abs/2002.08155}.

\bibitem[Feng et~al.(2024{\natexlab{b}})Feng, Jia, Li, Liu, Li, and Wang]{feng2024deceptprompt}
Zhengyu Feng, Zhaoyang Jia, Yiran Li, Jiacheng Liu, Yifei Li, and Gang Wang.
\newblock Deceptprompt: Exploiting llm-driven code generation via adversarial natural language instructions, 2024{\natexlab{b}}.

\bibitem[Foley and Maffeis(2025)]{foley2025apirl}
Myles Foley and Sergio Maffeis.
\newblock Apirl: Deep reinforcement learning for rest api fuzzing.
\newblock In \emph{Proceedings of the AAAI Conference on Artificial Intelligence}, volume~39, pages 191--199, 2025.

\bibitem[Foster et~al.(2025)Foster, Gulati, Harman, Harper, Mao, Ritchey, Robert, and Sengupta]{foster2025mutationguidedllmbasedtestgeneration}
Christopher Foster, Abhishek Gulati, Mark Harman, Inna Harper, Ke~Mao, Jillian Ritchey, Hervé Robert, and Shubho Sengupta.
\newblock Mutation-guided llm-based test generation at meta, 2025.
\newblock URL \url{https://arxiv.org/abs/2501.12862}.

\bibitem[Fried et~al.(2023)Fried, Aghajanyan, Lin, et~al.]{fried2023incoder}
Daniel Fried, Armen Aghajanyan, Jessy Lin, et~al.
\newblock {InCoder}: A generative model for code infilling and synthesis.
\newblock \emph{ICLR}, 2023.

\bibitem[Froemmgen et~al.(2024)Froemmgen, Austin, Choy, Ghelani, Kharatyan, Surita, Khrapko, Lamblin, Manzagol, Revaj, Tabachnyk, Tarlow, Villela, Zheng, Chandra, and Maniatis]{froemmgen2024resovling}
Alexander Froemmgen, Jacob Austin, Peter Choy, Nimesh Ghelani, Lera Kharatyan, Gabriela Surita, Elena Khrapko, Pascal Lamblin, Pierre-Antoine Manzagol, Marcus Revaj, Maxim Tabachnyk, Daniel Tarlow, Kevin Villela, Daniel Zheng, Satish Chandra, and Petros Maniatis.
\newblock Resolving code review comments with machine learning.
\newblock In \emph{Proceedings of the 46th International Conference on Software Engineering: Software Engineering in Practice}, ICSE-SEIP '24, page 204–215, New York, NY, USA, 2024. Association for Computing Machinery.

\bibitem[Fu et~al.(2025{\natexlab{a}})Fu, Yang, Zhang, Liu, Zhang, Wang, Zhang, and Zhou]{fu2025klearcodetestscalabletestcase}
Jia Fu, Xinyu Yang, Hongzhi Zhang, Yahui Liu, Jingyuan Zhang, Qi~Wang, Fuzheng Zhang, and Guorui Zhou.
\newblock Klear-codetest: Scalable test case generation for code reinforcement learning, 2025{\natexlab{a}}.
\newblock URL \url{https://arxiv.org/abs/2508.05710}.

\bibitem[Fu et~al.(2025{\natexlab{b}})Fu, Guan, Zhang, Yuan, Zhu, Xu, Wang, Qiu, Cai, Cao, Liu, Zhang, and Yu]{fu2025corecodebenchconfigurablemultiscenariorepositorylevel}
Lingyue Fu, Hao Guan, Bolun Zhang, Haowei Yuan, Yaoming Zhu, Jun Xu, Zongyu Wang, Lin Qiu, Xunliang Cai, Xuezhi Cao, Weiwen Liu, Weinan Zhang, and Yong Yu.
\newblock Corecodebench: A configurable multi-scenario repository-level benchmark, 2025{\natexlab{b}}.
\newblock URL \url{https://arxiv.org/abs/2507.05281}.

\bibitem[Fu et~al.(2024)Fu, Luo, Lin, Ye, and Ma]{fu2024scratchevalgpt4osmarterchild}
Rao Fu, Ziyang Luo, Hongzhan Lin, Zhen Ye, and Jing Ma.
\newblock Scratcheval: Are gpt-4o smarter than my child? evaluating large multimodal models with visual programming challenges, 2024.
\newblock URL \url{https://arxiv.org/abs/2411.18932}.

\bibitem[Fu et~al.(2025{\natexlab{c}})Fu, Chen, Lv, Zhang, Zeng, Yuan, and Wu]{fu2025posteriorgrpo}
Zhong-Duo Fu, Hong-Ning Chen, Zhi-Xin Lv, Yu-Jun Zhang, Qing-Cai Zeng, Ye~Yuan, and Fan Wu.
\newblock Posterior-grpo: Rewarding reasoning processes in code generation.
\newblock \emph{arXiv preprint arXiv:2508.05170}, 2025{\natexlab{c}}.

\bibitem[Fu et~al.(2025{\natexlab{d}})Fu, Wu, Zhang, Lv, Zeng, Chen, and Yuan]{fu2025smartcoder}
Zhong-Duo Fu, Fan Wu, Yu-Jun Zhang, Zhi-Xin Lv, Qing-Cai Zeng, Hong-Ning Chen, and Ye~Yuan.
\newblock Smartcoder-r1: Towards secure and explainable smart contract generation with security-aware group relative policy optimization.
\newblock \emph{arXiv preprint arXiv:2509.09942}, 2025{\natexlab{d}}.

\bibitem[Fumero et~al.(2025)Fumero, Huang, Boffa, Giordano, Mellia, Houidi, and Rossi]{fumero2025cybersleuth}
Stefano Fumero, Kai Huang, Matteo Boffa, Danilo Giordano, Marco Mellia, Zied~Ben Houidi, and Dario Rossi.
\newblock Cybersleuth: Autonomous blue-team llm agent for web attack forensics.
\newblock \emph{arXiv preprint arXiv:2508.20643}, 2025.

\bibitem[Fursin et~al.(2008)Fursin, Miranda, Temam, Namolaru, and Bodin]{2008MILEPOST}
Grigori Fursin, Cupertino Miranda, Olivier Temam, Mircea Namolaru, and Franois Bodin.
\newblock Milepost gcc: machine learning based research compiler.
\newblock \emph{proceedings of the gcc developers}, 2008.

\bibitem[Furuta et~al.(2023)Furuta, Lee, Nachum, Matsuo, Faust, Gu, and Gur]{furuta2023multimodal}
Hiroki Furuta, Kuang-Huei Lee, Ofir Nachum, Yutaka Matsuo, Aleksandra Faust, Shixiang~Shane Gu, and Izzeddin Gur.
\newblock Multimodal web navigation with instruction-finetuned foundation models.
\newblock \emph{arXiv preprint arXiv:2305.11854}, 2023.

\bibitem[Furuta et~al.(2024)Furuta, Lee, Nachum, Matsuo, Faust, Gu, and Gur]{webgum}
Hiroki Furuta, Kuang{-}Huei Lee, Ofir Nachum, Yutaka Matsuo, Aleksandra Faust, Shixiang~Shane Gu, and Izzeddin Gur.
\newblock Multimodal web navigation with instruction-finetuned foundation models.
\newblock In \emph{The Twelfth International Conference on Learning Representations, {ICLR} 2024, Vienna, Austria, May 7-11, 2024}. OpenReview.net, 2024.
\newblock URL \url{https://openreview.net/forum?id=efFmBWioSc}.

\bibitem[Gao et~al.(2024{\natexlab{a}})Gao, Wang, Li, Sun, Qian, Ding, and Zhou]{Dial-sql}
Dawei Gao, Haibin Wang, Yaliang Li, Xiuyu Sun, Yichen Qian, Bolin Ding, and Jingren Zhou.
\newblock Text-to-sql empowered by large language models: {A} benchmark evaluation.
\newblock \emph{Proc. {VLDB} Endow.}, 17\penalty0 (5):\penalty0 1132--1145, 2024{\natexlab{a}}.

\bibitem[Gao et~al.(2023{\natexlab{a}})Gao, Zhang, Li, et~al.]{gao2023survey}
Jun Gao, Xian Zhang, Haoyi Li, et~al.
\newblock Survey on neural network-based automatic source code summarization.
\newblock \emph{Journal of Software Engineering and Applications}, 15\penalty0 (4):\penalty0 220--240, 2023{\natexlab{a}}.
\newblock URL \url{https://www.sciengine.com/doi/10.13328/j.cnki.jos.006337}.

\bibitem[Gao et~al.(2020)Gao, Biderman, Black, Golding, Hoppe, Foster, Phang, He, Thite, Nabeshima, Presser, and Leahy]{gao2020pile}
Leo Gao, Stella Biderman, Sid Black, Laurence Golding, Travis Hoppe, Charles Foster, Jason Phang, Horace He, Anish Thite, Noa Nabeshima, Shawn Presser, and Connor Leahy.
\newblock {The Pile: An 800GB Dataset of Diverse Text for Language Modeling}, 2020.
\newblock URL \url{https://arxiv.org/abs/2101.00027}.

\bibitem[Gao et~al.(2023{\natexlab{b}})Gao, Madaan, Zhou, Alon, Liu, Yang, Callan, and Neubig]{gao2023pal}
Luyu Gao, Aman Madaan, Shuyan Zhou, Uri Alon, Pengfei Liu, Yiming Yang, Jamie Callan, and Graham Neubig.
\newblock Pal: Program-aided language models.
\newblock In \emph{International Conference on Machine Learning}, pages 10764--10799. PMLR, 2023{\natexlab{b}}.

\bibitem[Gao et~al.(2025)Gao, Tian, Meng, Wang, Hu, Xiao, Liu, Zhang, Chen, Gao, et~al.]{gao2025trae}
Pengfei Gao, Zhao Tian, Xiangxin Meng, Xinchen Wang, Ruida Hu, Yuanan Xiao, Yizhou Liu, Zhao Zhang, Junjie Chen, Cuiyun Gao, et~al.
\newblock Trae agent: An llm-based agent for software engineering with test-time scaling.
\newblock \emph{arXiv preprint arXiv:2507.23370}, 2025.

\bibitem[Gao et~al.(2024{\natexlab{b}})Gao, Liu, Li, Shi, Zhu, Wang, Li, Li, Hong, Luo, Gao, Mou, and Li]{XiYan-SQL}
Yingqi Gao, Yifu Liu, Xiaoxia Li, Xiaorong Shi, Yin Zhu, Yiming Wang, Shiqi Li, Wei Li, Yuntao Hong, Zhiling Luo, Jinyang Gao, Liyu Mou, and Yu~Li.
\newblock Xiyan-sql: {A} multi-generator ensemble framework for text-to-sql.
\newblock \emph{CoRR}, abs/2411.08599, 2024{\natexlab{b}}.

\bibitem[Gao et~al.(2024{\natexlab{c}})Gao, Zhan, Chang, Swamy, Brantley, Lee, and Sun]{gao2024regressing}
Zhaolin Gao, Wenhao Zhan, Jonathan~D Chang, Gokul Swamy, Kiant{\'e} Brantley, Jason~D Lee, and Wen Sun.
\newblock Regressing the relative future: Efficient policy optimization for multi-turn rlhf.
\newblock \emph{arXiv preprint arXiv:2410.04612}, 2024{\natexlab{c}}.

\bibitem[Gauthier(2023)]{aider2023repomap}
Paul Gauthier.
\newblock Building a better repository map with tree sitter.
\newblock \url{https://aider.chat/2023/10/22/repomap.html}, 2023.

\bibitem[Gauthier(2024{\natexlab{a}})]{aider2024linting}
Paul Gauthier.
\newblock Linting code for llms with tree-sitter.
\newblock \url{https://aider.chat/2024/05/22/linting.html}, 2024{\natexlab{a}}.

\bibitem[Gauthier(2024{\natexlab{b}})]{aider2024swebench}
Paul Gauthier.
\newblock How aider scored sota 26.3\% on swe bench lite.
\newblock \url{https://aider.chat/2024/05/22/swe-bench-lite.html}, 2024{\natexlab{b}}.

\bibitem[Gauthier(2024{\natexlab{c}})]{gauthier2024aider}
Paul Gauthier.
\newblock {Aider}: {AI} pair programming in your terminal.
\newblock Technical report, Aider AI, 2024{\natexlab{c}}.
\newblock URL \url{https://github.com/paul-gauthier/aider}.

\bibitem[Ge et~al.(2025)Ge, Mei, Duan, Li, Zheng, Wang, Wang, Yao, Liu, Cai, et~al.]{vibe_coding_survey}
Yuyao Ge, Lingrui Mei, Zenghao Duan, Tianhao Li, Yujia Zheng, Yiwei Wang, Lexin Wang, Jiayu Yao, Tianyu Liu, Yujun Cai, et~al.
\newblock A survey of vibe coding with large language models.
\newblock \emph{arXiv preprint arXiv:2510.12399}, 2025.

\bibitem[Geng et~al.(2023)Geng, Wang, Dong, Wang, Cao, Zhang, and Jin]{geng2023interpretation}
Mingyang Geng, Shangwen Wang, Dezun Dong, Haotian Wang, Shaomeng Cao, Kechi Zhang, and Zhi Jin.
\newblock Interpretation-based code summarization.
\newblock In \emph{2023 IEEE/ACM 31st International Conference on Program Comprehension (ICPC)}, pages 113--124. IEEE, 2023.

\bibitem[Gharibi et~al.(2024)Gharibi, Sadreddini, and Fakhrahmad]{gharibi2024t5apr}
Reza Gharibi, Mohammad~Hadi Sadreddini, and Seyed~Mostafa Fakhrahmad.
\newblock T5apr: Empowering automated program repair across languages through checkpoint ensemble.
\newblock \emph{Journal of Systems and Software}, 214:\penalty0 112083, 2024.

\bibitem[Gharibi et~al.(2025)Gharibi, Sadreddini, and Fakhrahmad]{gharibi2025multimend}
Reza Gharibi, Mohammad~Hadi Sadreddini, and Seyed~Mostafa Fakhrahmad.
\newblock Multimend: Multilingual program repair with context augmentation and multi-hunk patch generation.
\newblock \emph{arXiv preprint arXiv:2501.16044}, 2025.

\bibitem[GitHub(2022)]{github_copilot}
GitHub.
\newblock Github copilot, 2022.
\newblock URL \url{https://github.com/copilot}.

\bibitem[{GitHub}(2024{\natexlab{a}})]{codeql2024}
{GitHub}.
\newblock {CodeQL}: Semantic code analysis engine, 2024{\natexlab{a}}.
\newblock URL \url{https://codeql.github.com/}.
\newblock Accessed: 2024-01-15.

\bibitem[{GitHub}(2024{\natexlab{b}})]{github2024codespaces}
{GitHub}.
\newblock {GitHub Codespaces} with {AI} integration.
\newblock Technical report, GitHub Inc., 2024{\natexlab{b}}.

\bibitem[{GitHub}(2024{\natexlab{c}})]{github2024copilot}
{GitHub}.
\newblock {GitHub Copilot}: Your {AI} pair programmer.
\newblock Technical report, GitHub, Inc., 2024{\natexlab{c}}.
\newblock URL \url{https://github.com/features/copilot}.

\bibitem[{GitHub}(2024{\natexlab{d}})]{github2024universe}
{GitHub}.
\newblock Github universe 2024: Multi-model github copilot.
\newblock \url{https://github.blog/news-insights/product-news/universe-2024-copilot-multi-model/}, 2024{\natexlab{d}}.

\bibitem[{GitHub Blog}(2025)]{github2024models}
{GitHub Blog}.
\newblock Under the hood: Exploring the ai models powering github copilot.
\newblock \url{https://github.blog/ai-and-ml/github-copilot/under-the-hood-exploring-the-ai-models-powering-github-copilot/}, 2025.

\bibitem[GlaiveAI(2024)]{glaiveai_assistant}
GlaiveAI.
\newblock Glaive code assistant v3, 2024.
\newblock URL \url{https://huggingface.co/datasets/glaiveai/glaive-code-assistant-v3}.
\newblock Accessed: 2024.

\bibitem[GLM et~al.(2024)GLM, Zeng, Xu, Wang, Zhang, Yin, Rojas, Feng, Zhao, Lai, Yu, Wang, Sun, Zhang, Cheng, Gui, Tang, Zhang, Li, Zhao, Wu, Zhong, Liu, Huang, Zhang, Zheng, Lu, Duan, Zhang, Cao, Yang, Tam, Zhao, Liu, Xia, Zhang, Gu, Lv, Liu, Liu, Yang, Song, Zhang, An, Xu, Niu, Yang, Li, Bai, Dong, Qi, Wang, Yang, Du, Hou, and Wang]{glm2024chatglm}
Team GLM, Aohan Zeng, Bin Xu, Bowen Wang, Chenhui Zhang, Da~Yin, Diego Rojas, Guanyu Feng, Hanlin Zhao, Hanyu Lai, Hao Yu, Hongning Wang, Jiadai Sun, Jiajie Zhang, Jiale Cheng, Jiayi Gui, Jie Tang, Jing Zhang, Juanzi Li, Lei Zhao, Lindong Wu, Lucen Zhong, Mingdao Liu, Minlie Huang, Peng Zhang, Qinkai Zheng, Rui Lu, Shuaiqi Duan, Shudan Zhang, Shulin Cao, Shuxun Yang, Weng~Lam Tam, Wenyi Zhao, Xiao Liu, Xiao Xia, Xiaohan Zhang, Xiaotao Gu, Xin Lv, Xinghan Liu, Xinyi Liu, Xinyue Yang, Xixuan Song, Xunkai Zhang, Yifan An, Yifan Xu, Yilin Niu, Yuantao Yang, Yueyan Li, Yushi Bai, Yuxiao Dong, Zehan Qi, Zhaoyu Wang, Zhen Yang, Zhengxiao Du, Zhenyu Hou, and Zihan Wang.
\newblock Chatglm: A family of large language models from glm-130b to glm-4 all tools, 2024.

\bibitem[Gomes et~al.(2024)Gomes, Hellendoorn, Aldrich, and Abreu]{gomes2024exploratory}
Lu{\'\i}s~F Gomes, Vincent~J Hellendoorn, Jonathan Aldrich, and Rui Abreu.
\newblock An exploratory study of ml sketches and visual code assistants.
\newblock \emph{arXiv preprint arXiv:2412.13386}, 2024.

\bibitem[Gong et~al.(2024{\natexlab{a}})Gong, Wang, Elhoushi, and Cheung]{gong2024evaluationllmssyntaxawarecode}
Linyuan Gong, Sida Wang, Mostafa Elhoushi, and Alvin Cheung.
\newblock Evaluation of llms on syntax-aware code fill-in-the-middle tasks, 2024{\natexlab{a}}.
\newblock URL \url{https://arxiv.org/abs/2403.04814}.

\bibitem[Gong et~al.(2025{\natexlab{a}})Gong, Cheung, Elhoushi, and Wang]{gong2025structureawarefillinthemiddlepretrainingcode}
Linyuan Gong, Alvin Cheung, Mostafa Elhoushi, and Sida Wang.
\newblock Structure-aware fill-in-the-middle pretraining for code, 2025{\natexlab{a}}.
\newblock URL \url{https://arxiv.org/abs/2506.00204}.

\bibitem[Gong et~al.(2024{\natexlab{b}})Gong, Huang, Ma, Noda, Durante, Zheng, Terzopoulos, Fei-Fei, Gao, and Vo]{MindAgent}
Ran Gong, Qiuyuan Huang, Xiaojian Ma, Yusuke Noda, Zane Durante, Zilong Zheng, Demetri Terzopoulos, Li~Fei-Fei, Jianfeng Gao, and Hoi Vo.
\newblock {M}ind{A}gent: Emergent gaming interaction.
\newblock In Kevin Duh, Helena Gomez, and Steven Bethard, editors, \emph{Findings of the Association for Computational Linguistics: NAACL 2024}, pages 3154--3183, Mexico City, Mexico, June 2024{\natexlab{b}}. Association for Computational Linguistics.
\newblock \doi{10.18653/v1/2024.findings-naacl.200}.
\newblock URL \url{https://aclanthology.org/2024.findings-naacl.200/}.

\bibitem[Gong et~al.(2023)Gong, Li, Feng, Wu, and Kong]{gong2023diffuseq}
Shansan Gong, Mukai Li, Jiangtao Feng, Zhiyong Wu, and Lingpeng Kong.
\newblock Diffuseq: Sequence to sequence text generation with diffusion models, 2023.
\newblock URL \url{https://arxiv.org/abs/2210.08933}.

\bibitem[Gong et~al.(2025{\natexlab{b}})Gong, Zhang, Zheng, Gu, Jaitly, Kong, and Zhang]{gong2025diffucoder}
Shansan Gong, Ruixiang Zhang, Huangjie Zheng, Jiatao Gu, Navdeep Jaitly, Lingpeng Kong, and Yizhe Zhang.
\newblock Diffucoder: Understanding and improving masked diffusion models for code generation.
\newblock \emph{arXiv preprint arXiv:2506.20639}, 2025{\natexlab{b}}.

\bibitem[{Google}(2024{\natexlab{a}})]{gemini_cli_docs}
{Google}.
\newblock Gemini cli documentation, 2024{\natexlab{a}}.

\bibitem[{Google}(2024{\natexlab{b}})]{google2024gemini}
{Google}.
\newblock Gemini cli.
\newblock \url{https://cloud.google.com/gemini/docs/cli}, 2024{\natexlab{b}}.

\bibitem[Google(2025{\natexlab{a}})]{a2a}
Google.
\newblock Announcing the agent2agent protocol (a2a).
\newblock \url{https://developers.googleblog.com/en/a2a-a-new-era-of-agent-interoperability/}, 2025{\natexlab{a}}.

\bibitem[Google(2025{\natexlab{b}})]{gemini2_flash_modelcard}
Google.
\newblock Gemini 2.0 flash — model card.
\newblock \url{https://storage.googleapis.com/model-cards/documents/gemini-2-flash.pdf}, 2025{\natexlab{b}}.

\bibitem[{Google Cloud}(2024{\natexlab{a}})]{google2024cloudcode}
{Google Cloud}.
\newblock Google cloud code.
\newblock \url{https://cloud.google.com/code}, 2024{\natexlab{a}}.

\bibitem[{Google Cloud}(2024{\natexlab{b}})]{google2024duet}
{Google Cloud}.
\newblock {Duet AI} for developers: Code faster with {AI} assistance.
\newblock Technical report, Google LLC, 2024{\natexlab{b}}.

\bibitem[Gorti et~al.(2024)Gorti, Gofman, Liu, Wu, Vouitsis, Yu, Cresswell, and Hosseinzadeh]{MSc-SQL}
Satya~Krishna Gorti, Ilan Gofman, Zhaoyan Liu, Jiapeng Wu, No{\"{e}}l Vouitsis, Guangwei Yu, Jesse~C. Cresswell, and Rasa Hosseinzadeh.
\newblock Msc-sql: Multi-sample critiquing small language models for text-to-sql translation.
\newblock \emph{CoRR}, abs/2410.12916, 2024.

\bibitem[Gou et~al.(2024)Gou, Shao, Gong, Shen, Yang, Duan, and Chen]{CRITIC}
Zhibin Gou, Zhihong Shao, Yeyun Gong, Yelong Shen, Yujiu Yang, Nan Duan, and Weizhu Chen.
\newblock {CRITIC:} large language models can self-correct with tool-interactive critiquing.
\newblock In \emph{The Twelfth International Conference on Learning Representations, {ICLR} 2024, Vienna, Austria, May 7-11, 2024}. OpenReview.net, 2024.
\newblock URL \url{https://openreview.net/forum?id=Sx038qxjek}.

\bibitem[{Graphite}(2024)]{graphite_reviewer}
{Graphite}.
\newblock Graphite reviewer: Repository-aware ai reviews.
\newblock \url{https://graphite.dev/features/ai-reviewer}, 2024.

\bibitem[Grattafiori et~al.(2024)Grattafiori, Dubey, Jauhri, Pandey, Kadian, Al-Dahle, Letman, Mathur, Schelten, Vaughan, Yang, Fan, Goyal, Hartshorn, Yang, Mitra, Sravankumar, Korenev, Hinsvark, Rao, Zhang, Rodriguez, Gregerson, Spataru, Roziere, Biron, Tang, Chern, Caucheteux, et~al.]{grattafiori2024llama3}
Aaron Grattafiori, Abhimanyu Dubey, Abhinav Jauhri, Abhinav Pandey, Abhishek Kadian, Ahmad Al-Dahle, Aiesha Letman, Akhil Mathur, Alan Schelten, Alex Vaughan, Amy Yang, Angela Fan, Anirudh Goyal, Anthony Hartshorn, Aobo Yang, Archi Mitra, Archie Sravankumar, Artem Korenev, Arthur Hinsvark, Arun Rao, Aston Zhang, Aurelien Rodriguez, Austen Gregerson, Ava Spataru, Baptiste Roziere, Bethany Biron, Binh Tang, Bobbie Chern, Charlotte Caucheteux, et~al.
\newblock The llama 3 herd of models, 2024.
\newblock URL \url{https://arxiv.org/abs/2407.21783}.

\bibitem[Gu and Dao(2024)]{gu2024mamba}
Albert Gu and Tri Dao.
\newblock Mamba: Linear-time sequence modeling with selective state spaces.
\newblock In \emph{First Conference on Language Modeling}, 2024.
\newblock URL \url{https://openreview.net/forum?id=tEYskw1VY2}.

\bibitem[Gu et~al.(2024{\natexlab{a}})Gu, Rozière, Leather, Solar-Lezama, Synnaeve, and Wang]{gu2024cruxeval}
Alex Gu, Baptiste Rozière, Hugh Leather, Armando Solar-Lezama, Gabriel Synnaeve, and Sida~I. Wang.
\newblock Cruxeval: A benchmark for code reasoning, understanding and execution.
\newblock \emph{arXiv preprint arXiv:2401.03065}, 2024{\natexlab{a}}.

\bibitem[Gu et~al.(2024{\natexlab{b}})Gu, Jiang, Shi, Tan, Zhai, Xu, Li, Shen, Ma, Liu, et~al.]{gu2024survey}
Jiawei Gu, Xuhui Jiang, Zhichao Shi, Hexiang Tan, Xuehao Zhai, Chengjin Xu, Wei Li, Yinghan Shen, Shengjie Ma, Honghao Liu, et~al.
\newblock A survey on llm-as-a-judge.
\newblock \emph{arXiv preprint arXiv:2411.15594}, 2024{\natexlab{b}}.

\bibitem[Gu et~al.(2024{\natexlab{c}})Gu, Yang, Du, Chen, Walter, Wang, and Knoll]{gu2024review}
Shangding Gu, Long Yang, Yali Du, Guang Chen, Florian Walter, Jun Wang, and Alois Knoll.
\newblock A review of safe reinforcement learning: Methods, theories and applications.
\newblock \emph{IEEE Transactions on Pattern Analysis and Machine Intelligence}, 2024{\natexlab{c}}.

\bibitem[Gu et~al.(2025{\natexlab{a}})Gu, Nashid, and Mesbah]{gu2025llm_test_generation}
Sijia Gu, Noor Nashid, and Ali Mesbah.
\newblock Llm test generation via iterative hybrid program analysis.
\newblock \emph{arXiv preprint arXiv:2503.13580}, 2025{\natexlab{a}}.

\bibitem[Gu et~al.(2025{\natexlab{b}})Gu, Zhang, Li, Fang, Tian, Zhu, Zhou, and Chen]{gu2025testartimprovingllmbasedunit}
Siqi Gu, Quanjun Zhang, Kecheng Li, Chunrong Fang, Fangyuan Tian, Liuchuan Zhu, Jianyi Zhou, and Zhenyu Chen.
\newblock Testart: Improving llm-based unit testing via co-evolution of automated generation and repair iteration, 2025{\natexlab{b}}.
\newblock URL \url{https://arxiv.org/abs/2408.03095}.

\bibitem[Gu et~al.(2024{\natexlab{d}})Gu, Zhang, Ning, Zheng, Gou, Xue, Chang, Srivastava, Xie, Qi, et~al.]{gu2024your}
Yu~Gu, Kai Zhang, Yuting Ning, Boyuan Zheng, Boyu Gou, Tianci Xue, Cheng Chang, Sanjari Srivastava, Yanan Xie, Peng Qi, et~al.
\newblock Is your llm secretly a world model of the internet? model-based planning for web agents.
\newblock \emph{arXiv preprint arXiv:2411.06559}, 2024{\natexlab{d}}.

\bibitem[Guan et~al.(2024)Guan, Wang, Chu, Wang, Ni, Song, and Zhuang]{guan2024intelligent}
Yanchu Guan, Dong Wang, Zhixuan Chu, Shiyu Wang, Feiyue Ni, Ruihua Song, and Chenyi Zhuang.
\newblock Intelligent agents with llm-based process automation.
\newblock In \emph{Proceedings of the 30th ACM SIGKDD Conference on Knowledge Discovery and Data Mining}, pages 5018--5027, 2024.

\bibitem[{Guertler} et~al.(2025){Guertler}, {Cheng}, {Yu}, {Liu}, {Choshen}, and {Tan}]{TextArena}
Leon {Guertler}, Bobby {Cheng}, Simon {Yu}, Bo~{Liu}, Leshem {Choshen}, and Cheston {Tan}.
\newblock {TextArena}.
\newblock \emph{arXiv e-prints}, art. arXiv:2504.11442, April 2025.
\newblock \doi{10.48550/arXiv.2504.11442}.

\bibitem[Guha et~al.(2025)Guha, Marten, Keh, Raoof, Smyrnis, Bansal, Nezhurina, Mercat, Vu, Sprague, et~al.]{guha2025openthoughts}
Etash Guha, Ryan Marten, Sedrick Keh, Negin Raoof, Georgios Smyrnis, Hritik Bansal, Marianna Nezhurina, Jean Mercat, Trung Vu, Zayne Sprague, et~al.
\newblock Openthoughts: Data recipes for reasoning models.
\newblock \emph{arXiv preprint arXiv:2506.04178}, 2025.

\bibitem[{Gui} et~al.(2024){Gui}, {Liu}, {Cheng}, {Gu}, {Liu}, {Wang}, {Dong}, {Tang}, and {Huang}]{LogicGame}
Jiayi {Gui}, Yiming {Liu}, Jiale {Cheng}, Xiaotao {Gu}, Xiao {Liu}, Hongning {Wang}, Yuxiao {Dong}, Jie {Tang}, and Minlie {Huang}.
\newblock {LogicGame: Benchmarking Rule-Based Reasoning Abilities of Large Language Models}.
\newblock \emph{arXiv e-prints}, art. arXiv:2408.15778, August 2024.
\newblock \doi{10.48550/arXiv.2408.15778}.

\bibitem[Gui et~al.(2025{\natexlab{a}})Gui, Li, Wan, Shi, Zhang, Chen, Su, Chen, Wu, Zhou, et~al.]{gui2025webcode2m}
Yi~Gui, Zhen Li, Yao Wan, Yemin Shi, Hongyu Zhang, Bohua Chen, Yi~Su, Dongping Chen, Siyuan Wu, Xing Zhou, et~al.
\newblock Webcode2m: A real-world dataset for code generation from webpage designs.
\newblock In \emph{Proceedings of the ACM on Web Conference 2025}, pages 1834--1845, 2025{\natexlab{a}}.

\bibitem[Gui et~al.(2025{\natexlab{b}})Gui, Wan, Li, Zhang, Chen, Zhang, Su, Chen, Zhou, Jiang, et~al.]{gui2025uicopilot}
Yi~Gui, Yao Wan, Zhen Li, Zhongyi Zhang, Dongping Chen, Hongyu Zhang, Yi~Su, Bohua Chen, Xing Zhou, Wenbin Jiang, et~al.
\newblock Uicopilot: Automating ui synthesis via hierarchical code generation from webpage designs.
\newblock In \emph{Proceedings of the ACM on Web Conference 2025}, pages 1846--1855, 2025{\natexlab{b}}.

\bibitem[Gulwani(2011)]{gulwani2011flashfill}
Sumit Gulwani.
\newblock Automating string processing in spreadsheets using input-output examples.
\newblock In \emph{Proceedings of the 38th Annual ACM SIGPLAN-SIGACT Symposium on Principles of Programming Languages}, POPL '11, page 317–330, New York, NY, USA, 2011. Association for Computing Machinery.
\newblock ISBN 9781450304900.
\newblock \doi{10.1145/1926385.1926423}.
\newblock URL \url{https://doi.org/10.1145/1926385.1926423}.

\bibitem[Gunasekar et~al.(2023)Gunasekar, Zhang, Aneja, Mendes, Giorno, Gopi, Javaheripi, Kauffmann, de~Rosa, Saarikivi, Salim, Shah, Behl, Wang, Bubeck, Eldan, Kalai, Lee, and Li]{phi}
Suriya Gunasekar, Yi~Zhang, Jyoti Aneja, Caio C{\'{e}}sar~Teodoro Mendes, Allie~Del Giorno, Sivakanth Gopi, Mojan Javaheripi, Piero Kauffmann, Gustavo de~Rosa, Olli Saarikivi, Adil Salim, Shital Shah, Harkirat~Singh Behl, Xin Wang, S{\'{e}}bastien Bubeck, Ronen Eldan, Adam~Tauman Kalai, Yin~Tat Lee, and Yuanzhi Li.
\newblock Textbooks are all you need.
\newblock \emph{CoRR}, abs/2306.11644, 2023.
\newblock \doi{10.48550/ARXIV.2306.11644}.
\newblock URL \url{https://doi.org/10.48550/arXiv.2306.11644}.

\bibitem[Guo et~al.(2025{\natexlab{a}})Guo, Xie, Yang, Chen, Lin, Davies, Gal, Song, and Li]{tang2024redcodeagent}
Chengquan Guo, Chulin Xie, Yu~Yang, Zhaorun Chen, Zinan Lin, Xander Davies, Yarin Gal, Dawn Song, and Bo~Li.
\newblock Redcodeagent: Automatic red-teaming agent against diverse code agents, 2025{\natexlab{a}}.
\newblock URL \url{https://arxiv.org/abs/2510.02609}.

\bibitem[Guo et~al.(2021{\natexlab{a}})Guo, Ren, Lu, Feng, Tang, Liu, Zhou, Duan, Svyatkovskiy, Fu, Tufano, Deng, Clement, Drain, Sundaresan, Yin, Jiang, and Zhou]{graph_code_bert}
Daya Guo, Shuo Ren, Shuai Lu, Zhangyin Feng, Duyu Tang, Shujie Liu, Long Zhou, Nan Duan, Alexey Svyatkovskiy, Shengyu Fu, Michele Tufano, Shao~Kun Deng, Colin~B. Clement, Dawn Drain, Neel Sundaresan, Jian Yin, Daxin Jiang, and Ming Zhou.
\newblock Graphcodebert: Pre-training code representations with data flow.
\newblock In \emph{9th International Conference on Learning Representations, {ICLR} 2021, Virtual Event, Austria, May 3-7, 2021}. OpenReview.net, 2021{\natexlab{a}}.
\newblock URL \url{https://openreview.net/forum?id=jLoC4ez43PZ}.

\bibitem[Guo et~al.(2022)Guo, Lu, Duan, Wang, Zhou, and Yin]{unixcoder}
Daya Guo, Shuai Lu, Nan Duan, Yanlin Wang, Ming Zhou, and Jian Yin.
\newblock Unixcoder: Unified cross-modal pre-training for code representation.
\newblock In Smaranda Muresan, Preslav Nakov, and Aline Villavicencio, editors, \emph{Proceedings of the 60th Annual Meeting of the Association for Computational Linguistics (Volume 1: Long Papers), {ACL} 2022, Dublin, Ireland, May 22-27, 2022}, pages 7212--7225. Association for Computational Linguistics, 2022.
\newblock \doi{10.18653/V1/2022.ACL-LONG.499}.
\newblock URL \url{https://doi.org/10.18653/v1/2022.acl-long.499}.

\bibitem[Guo et~al.(2024{\natexlab{a}})Guo, Zhu, Yang, Xie, Dong, Zhang, Chen, Bi, Wu, Li, Luo, Xiong, and Liang]{guo2024deepseek}
Daya Guo, Qihao Zhu, Dejian Yang, Zhenda Xie, Kai Dong, Wentao Zhang, Guanting Chen, Xiao Bi, Y.~Wu, Y.~K. Li, Fuli Luo, Yingfei Xiong, and Wenfeng Liang.
\newblock Deepseek-coder: When the large language model meets programming - the rise of code intelligence.
\newblock \emph{CoRR}, abs/2401.14196, 2024{\natexlab{a}}.
\newblock \doi{10.48550/ARXIV.2401.14196}.
\newblock URL \url{https://doi.org/10.48550/arXiv.2401.14196}.

\bibitem[Guo et~al.(2025{\natexlab{b}})Guo, Yang, Zhang, Song, Zhang, Xu, Zhu, Ma, Wang, Bi, et~al.]{guo2025deepseek}
Daya Guo, Dejian Yang, Haowei Zhang, Junxiao Song, Ruoyu Zhang, Runxin Xu, Qihao Zhu, Shirong Ma, Peiyi Wang, Xiao Bi, et~al.
\newblock Deepseek-r1: Incentivizing reasoning capability in llms via reinforcement learning.
\newblock \emph{arXiv preprint arXiv:2501.12948}, 2025{\natexlab{b}}.

\bibitem[Guo et~al.(2021{\natexlab{b}})Guo, Yuan, and Wu]{DBLP:conf/ijcnn/GuoYW21}
Haixuan Guo, Shuhan Yuan, and Xintao Wu.
\newblock Logbert: Log anomaly detection via {BERT}.
\newblock In \emph{International Joint Conference on Neural Networks, {IJCNN} 2021, Shenzhen, China, July 18-22, 2021}, pages 1--8. {IEEE}, 2021{\natexlab{b}}.

\bibitem[Guo et~al.(2025{\natexlab{c}})Guo, Xie, Dai, Di, Zhang, Tao, and Zheng]{guo2025accelerating}
Hanyang Guo, Xiaoheng Xie, Hong-Ning Dai, Peng Di, Yu~Zhang, Bishenghui Tao, and Zibin Zheng.
\newblock Accelerating automatic program repair with dual retrieval-augmented fine-tuning and patch generation on large language models.
\newblock \emph{arXiv preprint arXiv:2507.10103}, 2025{\natexlab{c}}.

\bibitem[Guo et~al.(2019)Guo, Zhan, Gao, Xiao, Lou, Liu, and Zhang]{Guo}
Jiaqi Guo, Zecheng Zhan, Yan Gao, Yan Xiao, Jian{-}Guang Lou, Ting Liu, and Dongmei Zhang.
\newblock Towards complex text-to-sql in cross-domain database with intermediate representation.
\newblock In Anna Korhonen, David~R. Traum, and Llu{\'{\i}}s M{\`{a}}rquez, editors, \emph{Proceedings of the 57th Conference of the Association for Computational Linguistics, {ACL} 2019, Florence, Italy, July 28- August 2, 2019, Volume 1: Long Papers}, pages 4524--4535. Association for Computational Linguistics, 2019.

\bibitem[Guo et~al.(2025{\natexlab{d}})Guo, Li, Liu, Ma, Zheng, Yu, Pan, LI, Liu, Wang, Guo, Qu, Yue, Zhang, Chen, and Fu]{codeeditorbench}
Jiawei Guo, Ziming Li, Xueling Liu, Kaijing Ma, Tianyu Zheng, Zhouliang Yu, Ding Pan, Yizhi LI, Ruibo Liu, Yue Wang, Shuyue Guo, Xingwei Qu, Xiang Yue, Ge~Zhang, Wenhu Chen, and Jie Fu.
\newblock Codeeditorbench: Evaluating code editing capability of large language models, 2025{\natexlab{d}}.
\newblock URL \url{https://arxiv.org/abs/2404.03543}.

\bibitem[Guo et~al.(2025{\natexlab{e}})Guo, Wang, Deluca, Liu, Zhang, and Zhang]{bugscope}
Jinyao Guo, Chengpeng Wang, Dominic Deluca, Jinjie Liu, Zhuo Zhang, and Xiangyu Zhang.
\newblock Bugscope: Learn to find bugs like human.
\newblock \emph{arXiv preprint arXiv:2507.15671}, 2025{\natexlab{e}}.

\bibitem[Guo et~al.(2025{\natexlab{f}})Guo, Tao, Jiang, Wang, Chen, Liu, Ma, Mao, Zhang, and Zheng]{guo2025omnigirlmultilingualmultimodalbenchmark}
Lianghong Guo, Wei Tao, Runhan Jiang, Yanlin Wang, Jiachi Chen, Xilin Liu, Yuchi Ma, Mingzhi Mao, Hongyu Zhang, and Zibin Zheng.
\newblock Omnigirl: A multilingual and multimodal benchmark for github issue resolution, 2025{\natexlab{f}}.
\newblock URL \url{https://arxiv.org/abs/2505.04606}.

\bibitem[Guo et~al.(2024{\natexlab{b}})Guo, Zhuang, Li, Qiao, and Wang]{guo2024transagenttransfervisionlanguagefoundation}
Yiwei Guo, Shaobin Zhuang, Kunchang Li, Yu~Qiao, and Yali Wang.
\newblock Transagent: Transfer vision-language foundation models with heterogeneous agent collaboration, 2024{\natexlab{b}}.
\newblock URL \url{https://arxiv.org/abs/2410.12183}.

\bibitem[Gur et~al.(2023)Gur, Nachum, Miao, Safdari, Huang, Chowdhery, Narang, Fiedel, and Faust]{gur2023understandinghtmlwithllms}
Izzeddin Gur, Ofir Nachum, Yingjie Miao, Mustafa Safdari, Austin Huang, Aakanksha Chowdhery, Sharan Narang, Noah Fiedel, and Aleksandra Faust.
\newblock Understanding {HTML} with large language models.
\newblock In Houda Bouamor, Juan Pino, and Kalika Bali, editors, \emph{Findings of the Association for Computational Linguistics: EMNLP 2023}, pages 2803--2821, Singapore, December 2023. Association for Computational Linguistics.
\newblock \doi{10.18653/v1/2023.findings-emnlp.185}.
\newblock URL \url{https://aclanthology.org/2023.findings-emnlp.185/}.

\bibitem[Habler et~al.(2025)Habler, Huang, Narajala, and Kulkarni]{scure_a2a}
Idan Habler, Ken Huang, Vineeth~Sai Narajala, and Prashant Kulkarni.
\newblock Building a secure agentic ai application leveraging a2a protocol.
\newblock \emph{arXiv preprint arXiv:2504.16902}, 2025.

\bibitem[Hahm et~al.(2025)Hahm, Jin, Choi, Ahn, and Lee]{hahm2025enhancing}
Dongyoon Hahm, Woogyeol Jin, June~Suk Choi, Sungsoo Ahn, and Kimin Lee.
\newblock Enhancing llm agent safety via causal influence prompting.
\newblock \emph{arXiv preprint arXiv:2507.00979}, 2025.

\bibitem[Hai et~al.(2025)Hai, Nguyen, and Bui]{hai2025impactscontextsrepositorylevelcode}
Nam~Le Hai, Dung~Manh Nguyen, and Nghi D.~Q. Bui.
\newblock On the impacts of contexts on repository-level code generation, 2025.
\newblock URL \url{https://arxiv.org/abs/2406.11927}.

\bibitem[Haider et~al.(2024)Haider, Mostofa, Mosaddek, Iqbal, and Ahmed]{haider2024prompting}
Md~Asif Haider, Ayesha~Binte Mostofa, Sk~Sabit~Bin Mosaddek, Anindya Iqbal, and Toufique Ahmed.
\newblock Prompting and fine-tuning large language models for automated code review comment generation.
\newblock \emph{arXiv preprint arXiv:2411.10129}, 2024.

\bibitem[Hajipour et~al.(2024)Hajipour, van Tonder, nadri, and Cohan]{hajipour2024hexacoder}
Hossein Hajipour, Rijnard van Tonder, Reza nadri, and Arman Cohan.
\newblock Hexacoder: Secure code generation via oracle-guided synthetic training data, 2024.

\bibitem[Han et~al.(2023)Han, Kumar, and Tsvetkov]{han2023ssdlm}
Xiaochuang Han, Sachin Kumar, and Yulia Tsvetkov.
\newblock Ssd-lm: Semi-autoregressive simplex-based diffusion language model for text generation and modular control, 2023.
\newblock URL \url{https://arxiv.org/abs/2210.17432}.

\bibitem[Hao et~al.(2024)Hao, Liu, Wang, and Hu]{hao2024toolkengptaugmentingfrozenlanguage}
Shibo Hao, Tianyang Liu, Zhen Wang, and Zhiting Hu.
\newblock Toolkengpt: Augmenting frozen language models with massive tools via tool embeddings, 2024.
\newblock URL \url{https://arxiv.org/abs/2305.11554}.

\bibitem[Haque et~al.(2025)Haque, Siddique, Rahman, Hasan, Das, Kamal, Masura, and Gupta]{haque2025sok}
Ariful Haque, Sunzida Siddique, Md~Mahfuzur Rahman, Ahmed~Rafi Hasan, Laxmi~Rani Das, Marufa Kamal, Tasnim Masura, and Kishor~Datta Gupta.
\newblock Sok: Exploring hallucinations and security risks in ai-assisted software development with insights for llm deployment.
\newblock \emph{arXiv preprint arXiv:2502.18468}, 2025.

\bibitem[Hasan et~al.(2025)Hasan, Chakraborty, Karmaker, and Balasubramanian]{hasan2025teaching}
Mohammad~Saqib Hasan, Saikat Chakraborty, Santu Karmaker, and Niranjan Balasubramanian.
\newblock Teaching an old llm secure coding: Localized preference optimization on distilled preferences, 2025.

\bibitem[He et~al.(2025{\natexlab{a}})He, Miller, Agarwal, K{\"a}stner, and Vasilescu]{does_ai_assisted_coding_deliver}
Hao He, Courtney Miller, Shyam Agarwal, Christian K{\"a}stner, and Bogdan Vasilescu.
\newblock Does ai-assisted coding deliver? a difference-in-differences study of cursor's impact on software projects.
\newblock \emph{arXiv e-prints}, pages arXiv--2511, 2025{\natexlab{a}}.

\bibitem[He et~al.(2024{\natexlab{a}})He, Yao, Ma, Yu, Dai, Zhang, Lan, and Yu]{he2024webvoyager}
Hongliang He, Wenlin Yao, Kaixin Ma, Wenhao Yu, Yong Dai, Hongming Zhang, Zhenzhong Lan, and Dong Yu.
\newblock Webvoyager: Building an end-to-end web agent with large multimodal models.
\newblock \emph{arXiv preprint arXiv:2401.13919}, 2024{\natexlab{a}}.

\bibitem[He et~al.(2024{\natexlab{b}})He, Yao, Ma, Yu, Dai, Zhang, Lan, and Yu]{webvoyager2023}
Hongliang He, Wenlin Yao, Kaixin Ma, Wenhao Yu, Yong Dai, Hongming Zhang, Zhenzhong Lan, and Dong Yu.
\newblock Webvoyager: Building an end-to-end web agent with large multimodal models, 2024{\natexlab{b}}.
\newblock URL \url{https://arxiv.org/abs/2401.13919}.

\bibitem[He and Yan(2023)]{he2023large}
Jia He and Zhaoxi Yan.
\newblock Large-scale, diverse, and realistic dataset for vulnerability detection.
\newblock In \emph{2023 IEEE/ACM 45th International Conference on Software Engineering (ICSE)}, pages 1117--1128. IEEE, 2023.

\bibitem[He et~al.(2018)He, Ivanov, Tsankov, Raychev, and Vechev]{DBLP:conf/ccs/HeITRV18}
Jingxuan He, Pesho Ivanov, Petar Tsankov, Veselin Raychev, and Martin~T. Vechev.
\newblock Debin: Predicting debug information in stripped binaries.
\newblock In David Lie, Mohammad Mannan, Michael Backes, and XiaoFeng Wang, editors, \emph{Proceedings of the 2018 {ACM} {SIGSAC} Conference on Computer and Communications Security, {CCS} 2018, Toronto, ON, Canada, October 15-19, 2018}, pages 1667--1680. {ACM}, 2018.

\bibitem[He et~al.(2024{\natexlab{c}})He, Vero, Krasnopolska, and Vechev]{he2024instruction}
Jingxuan He, Mark Vero, Gabriela Krasnopolska, and Martin Vechev.
\newblock Instruction tuning for secure code generation.
\newblock \emph{arXiv preprint arXiv:2402.09497}, 2024{\natexlab{c}}.

\bibitem[He et~al.(2025{\natexlab{b}})He, Liu, Liu, Yan, Wang, Cheng, Zhang, Zhang, Xu, Shen, Li, Zeng, Wei, Cheng, An, Liu, and Zhou]{Skywork}
Jujie He, Jiacai Liu, Chris~Yuhao Liu, Rui Yan, Chaojie Wang, Peng Cheng, Xiaoyu Zhang, Fuxiang Zhang, Jiacheng Xu, Wei Shen, Siyuan Li, Liang Zeng, Tianwen Wei, Cheng Cheng, Bo~An, Yang Liu, and Yahui Zhou.
\newblock Skywork open reasoner 1 technical report.
\newblock \emph{CoRR}, abs/2505.22312, 2025{\natexlab{b}}.
\newblock \doi{10.48550/ARXIV.2505.22312}.
\newblock URL \url{https://doi.org/10.48550/arXiv.2505.22312}.

\bibitem[He et~al.(2025{\natexlab{c}})He, Liu, Liu, Yan, Wang, Cheng, Zhang, Zhang, Xu, Shen, et~al.]{he2025skywork}
Jujie He, Jiacai Liu, Chris~Yuhao Liu, Rui Yan, Chaojie Wang, Peng Cheng, Xiaoyu Zhang, Fuxiang Zhang, Jiacheng Xu, Wei Shen, et~al.
\newblock Skywork open reasoner 1 technical report.
\newblock \emph{arXiv preprint arXiv:2505.22312}, 2025{\natexlab{c}}.

\bibitem[He et~al.(2025{\natexlab{d}})He, Shi, Zhuo, Treude, Sun, Xing, Du, and Lo]{he2025code}
Junda He, Jieke Shi, Terry~Yue Zhuo, Christoph Treude, Jiamou Sun, Zhenchang Xing, Xiaoning Du, and David Lo.
\newblock From code to courtroom: Llms as the new software judges.
\newblock \emph{arXiv preprint arXiv:2503.02246}, 2025{\natexlab{d}}.

\bibitem[He et~al.(2025{\natexlab{e}})He, Zeng, Jiang, Zhang, Liu, Shi, and Zhou]{he2025flow2code}
Mengliang He, Jiayi Zeng, Yankai Jiang, Wei Zhang, Zeming Liu, Xiaoming Shi, and Aimin Zhou.
\newblock Flow2code: Evaluating large language models for flowchart-based code generation capability.
\newblock \emph{arXiv preprint arXiv:2506.02073}, 2025{\natexlab{e}}.

\bibitem[He et~al.(2019)He, Mao, Chakrabarti, and Chen]{He}
Pengcheng He, Yi~Mao, Kaushik Chakrabarti, and Weizhu Chen.
\newblock {X-SQL:} reinforce schema representation with context.
\newblock \emph{CoRR}, abs/1908.08113, 2019.

\bibitem[He et~al.(2025{\natexlab{f}})He, Liu, Du, Yan, Fan, Huang, Yuan, and Ma]{he2025sweperflanguagemodelsoptimize}
Xinyi He, Qian Liu, Mingzhe Du, Lin Yan, Zhijie Fan, Yiming Huang, Zejian Yuan, and Zejun Ma.
\newblock Swe-perf: Can language models optimize code performance on real-world repositories?, 2025{\natexlab{f}}.
\newblock URL \url{https://arxiv.org/abs/2507.12415}.

\bibitem[He et~al.(2025{\natexlab{g}})He, Yang, Sheng, Zhong, Zhang, An, Shi, Cai, He, Chen, Xu, and Wang]{SWESwiss2025}
Zhenyu He, Qingping Yang, Wei Sheng, Xiaojian Zhong, Kechi Zhang, Chenxin An, Wenlei Shi, Tianle Cai, Di~He, Jiaze Chen, Jingjing Xu, and Mingxuan Wang.
\newblock Swe-swiss: A multi-task fine-tuning and rl recipe for high-performance issue resolution, 2025{\natexlab{g}}.
\newblock URL \url{https://www.notion.so/SWE-Swiss-A-Multi-Task-Fine-Tuning-and-RL-Recipe-for-High-Performance-Issue-Resolution-21e174dedd4880ea829ed4c861c44f88}.

\bibitem[He et~al.(2025{\natexlab{h}})He, Liang, Xu, Liu, Chen, Wang, Song, Yu, Liang, Wang, et~al.]{he2025deepmath}
Zhiwei He, Tian Liang, Jiahao Xu, Qiuzhi Liu, Xingyu Chen, Yue Wang, Linfeng Song, Dian Yu, Zhenwen Liang, Wenxuan Wang, et~al.
\newblock Deepmath-103k: A large-scale, challenging, decontaminated, and verifiable mathematical dataset for advancing reasoning.
\newblock \emph{arXiv preprint arXiv:2504.11456}, 2025{\natexlab{h}}.

\bibitem[He et~al.(2025{\natexlab{i}})He, Choi, Zhang, Ji, Zhou, Xu, Bercovich, Zhang, and Li]{he2025hardtests}
Zhongmou He, Yee~Man Choi, Kexun Zhang, Jiabao Ji, Junting Zhou, Dejia Xu, Ivan Bercovich, Aidan Zhang, and Lei Li.
\newblock Hardtests: Synthesizing high-quality test cases for llm coding.
\newblock \emph{arXiv preprint arXiv:2505.24098}, 2025{\natexlab{i}}.

\bibitem[Hendrycks et~al.(2021{\natexlab{a}})Hendrycks, Basart, Kadavath, Mazeika, Arora, Guo, Burns, Puranik, He, Song, and Steinhardt]{apps}
Dan Hendrycks, Steven Basart, Saurav Kadavath, Mantas Mazeika, Akul Arora, Ethan Guo, Collin Burns, Samir Puranik, Horace He, Dawn Song, and Jacob Steinhardt.
\newblock Measuring coding challenge competence with apps, 2021{\natexlab{a}}.
\newblock URL \url{https://arxiv.org/abs/2105.09938}.

\bibitem[Hendrycks et~al.(2021{\natexlab{b}})Hendrycks, Basart, Kadavath, et~al.]{hendrycks2021apps}
Dan Hendrycks, Steven Basart, Saurav Kadavath, et~al.
\newblock Measuring coding challenge competence with {APPS}.
\newblock \emph{NeurIPS}, 2021{\natexlab{b}}.

\bibitem[Hoffmann et~al.(2022)Hoffmann, Borgeaud, Mensch, Buchatskaya, Cai, Rutherford, Casas, Hendricks, Welbl, Clark, et~al.]{chinchilla_scaling_law}
Jordan Hoffmann, Sebastian Borgeaud, Arthur Mensch, Elena Buchatskaya, Trevor Cai, Eliza Rutherford, Diego de~Las Casas, Lisa~Anne Hendricks, Johannes Welbl, Aidan Clark, et~al.
\newblock Training compute-optimal large language models.
\newblock \emph{arXiv preprint arXiv:2203.15556}, 2022.

\bibitem[H{\"o}glund and Khedri(2023)]{hoglund2023comparison}
Samuel H{\"o}glund and Josef Khedri.
\newblock Comparison between rlhf and rlaif in fine-tuning a large language model, 2023.

\bibitem[Hong and Baik(2025)]{rag_comment_generation2}
Hyunsun Hong and Jongmoon Baik.
\newblock Retrieval-augmented code review comment generation.
\newblock \emph{arXiv preprint arXiv:2506.11591}, 2025.

\bibitem[Hong et~al.(2024)Hong, Zhuge, Chen, et~al.]{hong2023metagpt}
Sirui Hong, Mingchen Zhuge, Jonathan Chen, et~al.
\newblock {MetaGPT}: Meta programming for multi-agent collaborative framework.
\newblock In \emph{ICLR}, 2024.

\bibitem[Hosseini and Dolan{-}Gavitt(2022)]{DBLP:journals/corr/abs-2212-08950}
Iman Hosseini and Brendan Dolan{-}Gavitt.
\newblock Beyond the {C:} retargetable decompilation using neural machine translation.
\newblock \emph{CoRR}, abs/2212.08950, 2022.

\bibitem[Hou et~al.(2024)Hou, Zhao, Liu, et~al.]{hou2024large}
Xinyi Hou, Yanjie Zhao, Yue Liu, et~al.
\newblock Large language models for software engineering: A systematic literature review.
\newblock \emph{ACM Transactions on Software Engineering and Methodology}, 33\penalty0 (5):\penalty0 1--45, 2024.

\bibitem[Hou et~al.(2025)Hou, Zhao, Wang, and Wang]{mcp_survey}
Xinyi Hou, Yanjie Zhao, Shenao Wang, and Haoyu Wang.
\newblock Model context protocol (mcp): Landscape, security threats, and future research directions.
\newblock \emph{arXiv preprint arXiv:2503.23278}, 2025.

\bibitem[Hu et~al.(2022{\natexlab{a}})Hu, Shen, Wallis, Allen-Zhu, Li, Wang, Wang, Chen, et~al.]{hu2022lora}
Edward~J Hu, Yelong Shen, Phillip Wallis, Zeyuan Allen-Zhu, Yuanzhi Li, Shean Wang, Lu~Wang, Weizhu Chen, et~al.
\newblock Lora: Low-rank adaptation of large language models.
\newblock \emph{ICLR}, 1\penalty0 (2):\penalty0 3, 2022{\natexlab{a}}.

\bibitem[Hu et~al.(2025{\natexlab{a}})Hu, Xie, and Zhang]{hu2025repair}
Haichuan Hu, Xiaochen Xie, and Quanjun Zhang.
\newblock Repair-r1: Better test before repair.
\newblock \emph{arXiv preprint arXiv:2507.22853}, 2025{\natexlab{a}}.

\bibitem[Hu et~al.(2025{\natexlab{b}})Hu, Liu, Xu, and Shen]{hu2025reinforceplusplus}
Jian Hu, Jason~Klein Liu, Haotian Xu, and Wei Shen.
\newblock Reinforce++: An efficient rlhf algorithm with robustness to both prompt and reward models, 2025{\natexlab{b}}.
\newblock URL \url{https://arxiv.org/abs/2501.03262}.

\bibitem[Hu et~al.(2025{\natexlab{c}})Hu, Liu, Xu, and Shen]{hu2025reinforcestabilizingcriticfreepolicy}
Jian Hu, Jason~Klein Liu, Haotian Xu, and Wei Shen.
\newblock Reinforce++: Stabilizing critic-free policy optimization with global advantage normalization, 2025{\natexlab{c}}.
\newblock URL \url{https://arxiv.org/abs/2501.03262}.

\bibitem[Hu et~al.(2025{\natexlab{d}})Hu, Liu, Lu, Wu, Harchaoui, Diao, Choi, Molchanov, Yang, Kautz, and Dong]{hu2025brorl}
Jian Hu, Mingjie Liu, Ximing Lu, Fang Wu, Zaid Harchaoui, Shizhe Diao, Yejin Choi, Pavlo Molchanov, Jun Yang, Jan Kautz, and Yi~Dong.
\newblock Brorl: Scaling reinforcement learning via broadened exploration, 2025{\natexlab{d}}.
\newblock URL \url{https://arxiv.org/abs/2510.01180}.

\bibitem[{Hu} et~al.(2024){Hu}, {Li}, {Xie}, {Jiang}, {Stoica}, {Jin}, and {Zhang}]{GameArena}
Lanxiang {Hu}, Qiyu {Li}, Anze {Xie}, Nan {Jiang}, Ion {Stoica}, Haojian {Jin}, and Hao {Zhang}.
\newblock {GameArena: Evaluating LLM Reasoning through Live Computer Games}.
\newblock \emph{arXiv e-prints}, art. arXiv:2412.06394, December 2024.
\newblock \doi{10.48550/arXiv.2412.06394}.

\bibitem[Hu et~al.(2024{\natexlab{a}})Hu, Peng, Ren, Jiang, Meng, Wu, Gao, Wang, and Gao]{hu2024coderepoqa}
Ruida Hu, Chao Peng, Jingyi Ren, Bo~Jiang, Xiangxin Meng, Qinyun Wu, Pengfei Gao, Xinchen Wang, and Cuiyun Gao.
\newblock Coderepoqa: A large-scale benchmark for software engineering question answering.
\newblock \emph{arXiv preprint arXiv:2412.14764}, 2024{\natexlab{a}}.

\bibitem[Hu et~al.(2024{\natexlab{b}})Hu, Peng, Ren, Jiang, Meng, Wu, Gao, Wang, and Gao]{hu2024realworldbenchmarkevaluatingfinegrained}
Ruida Hu, Chao Peng, Jingyi Ren, Bo~Jiang, Xiangxin Meng, Qinyun Wu, Pengfei Gao, Xinchen Wang, and Cuiyun Gao.
\newblock A real-world benchmark for evaluating fine-grained issue solving capabilities of large language models, 2024{\natexlab{b}}.
\newblock URL \url{https://arxiv.org/abs/2411.18019}.

\bibitem[Hu et~al.(2025{\natexlab{e}})Hu, Duan, Wei, Zhang, Zhang, and Xu]{dynacode}
Wenhao Hu, Jinhao Duan, Chunchen Wei, Li~Zhang, Yue Zhang, and Kaidi Xu.
\newblock Dynacode: A dynamic complexity-aware code benchmark for evaluating large language models in code generation, 2025{\natexlab{e}}.
\newblock URL \url{https://arxiv.org/abs/2503.10452}.

\bibitem[Hu et~al.(2018)Hu, Li, Xia, Lo, and Jin]{DeepCom}
Xing Hu, Ge~Li, Xin Xia, David Lo, and Zhi Jin.
\newblock Deep code comment generation.
\newblock In \emph{Proceedings of the 26th Conference on Program Comprehension}, ICPC '18, page 200–210. Association for Computing Machinery, 2018.
\newblock ISBN 9781450357142.
\newblock \doi{10.1145/3196321.3196334}.
\newblock URL \url{https://doi.org/10.1145/3196321.3196334}.

\bibitem[Hu et~al.(2022{\natexlab{b}})Hu, Shi, Zhou, and Pike]{hu2022fix}
Yaojie Hu, Xingjian Shi, Qiang Zhou, and Lee Pike.
\newblock Fix bugs with transformer through a neural-symbolic edit grammar.
\newblock \emph{arXiv preprint arXiv:2204.06643}, 2022{\natexlab{b}}.

\bibitem[Huang and et~al.(2024)]{huang2024mpsc}
B.~Huang and et~al.
\newblock Enhancing llms in coding through multi-perspective self-consistency (mpsc).
\newblock \emph{arXiv preprint arXiv:2309.17272}, 2024.
\newblock URL \url{https://arxiv.org/abs/2309.17272}.

\bibitem[Huang and et~al.(2023)]{huang2023codecot}
Bin Huang and et~al.
\newblock Codecot: Self-examining code generation with chain-of-thought and test synthesis.
\newblock \emph{arXiv preprint arXiv:2309.17272}, 2023.
\newblock URL \url{https://arxiv.org/abs/2309.17272}.

\bibitem[HUANG et~al.(2024)HUANG, QING, Shang, Cui, and Zhang]{huang2024effibench}
Dong HUANG, Yuhao QING, Weiyi Shang, Heming Cui, and Jie Zhang.
\newblock Effibench: Benchmarking the efficiency of automatically generated code.
\newblock In \emph{The Thirty-eight Conference on Neural Information Processing Systems Datasets and Benchmarks Track}, 2024.
\newblock URL \url{https://openreview.net/forum?id=30XanJanJP}.

\bibitem[Huang et~al.(2024{\natexlab{a}})Huang, Zhang, Bu, Xie, Chen, and Cui]{huang2024bias}
Dong Huang, Jie~M Zhang, Qingwen Bu, Xiaofei Xie, Junjie Chen, and Heming Cui.
\newblock Bias testing and mitigation in llm-based code generation.
\newblock \emph{ACM Transactions on Software Engineering and Methodology}, 2024{\natexlab{a}}.

\bibitem[Huang et~al.(2025{\natexlab{a}})Huang, Zeng, Dai, Luo, Weng, Qing, Cui, Guo, and Zhang]{efficoder2024}
Dong Huang, Guangtao Zeng, Jianbo Dai, Meng Luo, Han Weng, Yuhao Qing, Heming Cui, Zhijiang Guo, and Jie~M. Zhang.
\newblock Efficoder: Enhancing code generation in large language models through efficiency-aware fine-tuning, 2025{\natexlab{a}}.
\newblock URL \url{https://arxiv.org/abs/2410.10209}.

\bibitem[Huang et~al.(2025{\natexlab{b}})Huang, Zhao, Ma, and Chen]{fuzz_challenge_llm}
Linghan Huang, Peizhou Zhao, Lei Ma, and Huaming Chen.
\newblock On the challenges of fuzzing techniques via large language models.
\newblock In Rong~N. Chang, Carl~K. Chang, Jingwei Yang, Nimanthi Atukorala, Dan Chen, Sumi Helal, Sasu Tarkoma, Qiang He, Tevfik Kosar, Claudio~A. Ardagna, Javier Berrocal, Kaoutar~El Maghaouri, and Yanchun Sun, editors, \emph{{IEEE} International Conference on Software Services Engineering, {SSE} 2025, Helsinki, Finland, July 7-12, 2025}, pages 162--171. {IEEE}, 2025{\natexlab{b}}.
\newblock \doi{10.1109/SSE67621.2025.00028}.
\newblock URL \url{https://doi.org/10.1109/SSE67621.2025.00028}.

\bibitem[Huang et~al.(2019)Huang, Haj-Ali, Moses, Xiang, Stoica, Asanovic, and Wawrzynek]{huang2019autophase}
Qijing Huang, Ameer Haj-Ali, William Moses, John Xiang, Ion Stoica, Krste Asanovic, and John Wawrzynek.
\newblock Autophase: Compiler phase-ordering for hls with deep reinforcement learning.
\newblock In \emph{2019 IEEE 27th Annual International Symposium on Field-Programmable Custom Computing Machines (FCCM)}, pages 308--308. IEEE, 2019.

\bibitem[Huang et~al.(2024{\natexlab{b}})Huang, Cheng, Liu, Hao, Song, Xu, Yang, Liu, Zhang, Chai, Yuan, Zhang, Fu, Liu, Zhang, Wang, Qi, Xu, and Chu]{Huang2024OpenCoderTO}
Siming Huang, Tianhao Cheng, Jason~Klein Liu, Jiaran Hao, Liuyihan Song, Yang Xu, J.~Yang, J.~H. Liu, Chenchen Zhang, Linzheng Chai, Ruifeng Yuan, Zhaoxiang Zhang, Jie Fu, Qian Liu, Ge~Zhang, Zili Wang, Yuan Qi, Yinghui Xu, and Wei Chu.
\newblock Opencoder: The open cookbook for top-tier code large language models.
\newblock 2024{\natexlab{b}}.
\newblock URL \url{https://arxiv.org/pdf/2411.04905}.

\bibitem[Huang et~al.(2025{\natexlab{c}})Huang, Cheng, Liu, Xu, Hao, Song, Xu, Yang, Liu, Zhang, Chai, Yuan, Luo, Wang, Fan, Zhu, Zhang, Gao, Fu, Liu, Li, Zhang, Qi, Xu, Chu, and Wang]{huang2025opencoder}
Siming Huang, Tianhao Cheng, Jason~Klein Liu, Weidi Xu, Jiaran Hao, Liuyihan Song, Yang Xu, Jian Yang, Jiaheng Liu, Chenchen Zhang, Linzheng Chai, Ruifeng Yuan, Xianzhen Luo, Qiufeng Wang, YuanTao Fan, Qingfu Zhu, Zhaoxiang Zhang, Yang Gao, Jie Fu, Qian Liu, Houyi Li, Ge~Zhang, Yuan Qi, Yinghui Xu, Wei Chu, and Zili Wang.
\newblock Opencoder: The open cookbook for top-tier code large language models.
\newblock In Wanxiang Che, Joyce Nabende, Ekaterina Shutova, and Mohammad~Taher Pilehvar, editors, \emph{Proceedings of the 63rd Annual Meeting of the Association for Computational Linguistics (Volume 1: Long Papers), {ACL} 2025, Vienna, Austria, July 27 - August 1, 2025}, pages 33167--33193. Association for Computational Linguistics, 2025{\natexlab{c}}.
\newblock URL \url{https://aclanthology.org/2025.acl-long.1591/}.

\bibitem[Huang et~al.(2023)Huang, Gupta, Zhong, Li, and Chen]{huang2023privacy}
Yangsibo Huang, Samyak Gupta, Zexuan Zhong, Kai Li, and Danqi Chen.
\newblock Privacy implications of retrieval-based language models.
\newblock In \emph{Findings of the Association for Computational Linguistics: EMNLP 2023}, pages 7583--7596, 2023.

\bibitem[Huang et~al.(2024{\natexlab{c}})Huang, Luo, Yu, Zhang, Lei, Wei, He, Huang, Liu, Zhao, and Liu]{huang2024dacodeagentdatascience}
Yiming Huang, Jianwen Luo, Yan Yu, Yitong Zhang, Fangyu Lei, Yifan Wei, Shizhu He, Lifu Huang, Xiao Liu, Jun Zhao, and Kang Liu.
\newblock Da-code: Agent data science code generation benchmark for large language models, 2024{\natexlab{c}}.
\newblock URL \url{https://arxiv.org/abs/2410.07331}.

\bibitem[Huang et~al.(2020)Huang, Huang, Chen, Chen, Zheng, Luo, Jia, Hu, and Zhou]{huang2020towards}
Yuan Huang, Shaohao Huang, Huanchao Chen, Xiangping Chen, Zibin Zheng, Xiapu Luo, Nan Jia, Xinyu Hu, and Xiaocong Zhou.
\newblock Towards automatically generating block comments for code snippets.
\newblock \emph{Information and Software Technology}, 127:\penalty0 106373, 2020.

\bibitem[Hubinger et~al.(2019)Hubinger, van Merwijk, Mikulik, Skalse, and Garrabrant]{hubinger2019risks}
Evan Hubinger, Chris van Merwijk, Vladimir Mikulik, Joar Skalse, and Scott Garrabrant.
\newblock Risks from learned optimization in advanced machine learning systems.
\newblock \emph{arXiv preprint arXiv:1906.01820}, 2019.

\bibitem[{Hudi} et~al.(2025){Hudi}, {Indra Winata}, {Zhang}, and {Fikri Aji}]{TextGames}
Frederikus {Hudi}, Genta {Indra Winata}, Ruochen {Zhang}, and Alham {Fikri Aji}.
\newblock {TextGames: Learning to Self-Play Text-Based Puzzle Games via Language Model Reasoning}.
\newblock \emph{arXiv e-prints}, art. arXiv:2502.18431, February 2025.
\newblock \doi{10.48550/arXiv.2502.18431}.

\bibitem[{Hugging Face}(2025)]{openr1}
{Hugging Face}.
\newblock Open r1: A fully open reproduction of deepseek-r1, January 2025.
\newblock URL \url{https://github.com/huggingface/open-r1}.

\bibitem[{huggingface}(2025)]{smolagents}
{huggingface}.
\newblock smolagents, 2025.
\newblock URL \url{https://huggingface.co/docs/smolagents/index}.

\bibitem[Hughes et~al.(2023)Hughes, de~Vries, Robinson, Ferrandis, Allal, von Werra, Ding, Paquet, Jernite, et~al.]{hughes2023bigcode}
Sean Hughes, Harm de~Vries, Jennifer Robinson, Carlos~Mu{\~n}oz Ferrandis, Loubna~Ben Allal, Leandro von Werra, Jennifer Ding, Sebastien Paquet, Yacine Jernite, et~al.
\newblock The bigcode project governance card.
\newblock \emph{arXiv preprint arXiv:2312.03872}, 2023.

\bibitem[Hui et~al.(2022)Hui, Geng, Wang, Qin, Li, Li, Sun, and Li]{Hui}
Binyuan Hui, Ruiying Geng, Lihan Wang, Bowen Qin, Yanyang Li, Bowen Li, Jian Sun, and Yongbin Li.
\newblock S\^2sql: Injecting syntax to question-schema interaction graph encoder for text-to-sql parsers.
\newblock In \emph{Findings of the Association for Computational Linguistics: {ACL} 2022, Dublin, Ireland, May 22-27, 2022}, pages 1254--1262. Association for Computational Linguistics, 2022.

\bibitem[Hui et~al.(2024)Hui, Yang, Cui, Yang, Liu, Zhang, Liu, Zhang, Yu, Lu, Dang, Fan, Zhang, Yang, Men, Huang, Zheng, Miao, Quan, Feng, Ren, Ren, Zhou, and Lin]{qwen25coder}
Binyuan Hui, Jian Yang, Zeyu Cui, Jiaxi Yang, Dayiheng Liu, Lei Zhang, Tianyu Liu, Jiajun Zhang, Bowen Yu, Keming Lu, Kai Dang, Yang Fan, Yichang Zhang, An~Yang, Rui Men, Fei Huang, Bo~Zheng, Yibo Miao, Shanghaoran Quan, Yunlong Feng, Xingzhang Ren, Xuancheng Ren, Jingren Zhou, and Junyang Lin.
\newblock Qwen2.5-coder technical report, 2024.
\newblock URL \url{https://arxiv.org/abs/2409.12186}.

\bibitem[Husain et~al.(2020)Husain, Wu, Gazit, Allamanis, and Brockschmidt]{husain2020CodeSearchNet}
Hamel Husain, Ho-Hsiang Wu, Tiferet Gazit, Miltiadis Allamanis, and Marc Brockschmidt.
\newblock {{CodeSearchNet Challenge}}: {{Evaluating}} the {{State}} of {{Semantic Code Search}}, 2020.

\bibitem[Ibrahimzada et~al.(2024)Ibrahimzada, Ke, Pawagi, Abid, Pan, Sinha, and Jabbarvand]{ibrahimzada2024alphatrans0}
Ali~Reza Ibrahimzada, Kaiyao Ke, Mrigank Pawagi, Muhammad~Salman Abid, Rangeet Pan, Saurabh Sinha, and Reyhaneh Jabbarvand.
\newblock Alphatrans: A neuro-symbolic compositional approach for repository-level code translation and validation.
\newblock \emph{arXiv preprint arXiv: 2410.24117}, 2024.

\bibitem[Imani et~al.(2023)Imani, Du, and Shrivastava]{imani2023mathprompter}
Shima Imani, Liang Du, and Harsh Shrivastava.
\newblock Mathprompter: Mathematical reasoning using large language models.
\newblock \emph{arXiv preprint arXiv:2303.05398}, 2023.

\bibitem[ISE-UIUC(2024)]{magicoder_oss}
ISE-UIUC.
\newblock Magicoder-oss-instruct-75k, 2024.
\newblock URL \url{https://huggingface.co/datasets/ise-uiuc/Magicoder-OSS-Instruct-75K}.
\newblock Accessed: 2024.

\bibitem[Islam and et~al.(2024)]{islam2024mapcoder}
Md.~Ashraful Islam and et~al.
\newblock Mapcoder: Multi-agent code generation for competitive programming.
\newblock \emph{arXiv preprint arXiv:2401.08500}, 2024.
\newblock URL \url{https://arxiv.org/abs/2401.08500}.

\bibitem[Islam et~al.(2024)Islam, Khoury, Seong, Bou-Harb, and Najafirad]{islam2024enhancing}
Nafis~Tanveer Islam, Joseph Khoury, Andrew Seong, Elias Bou-Harb, and Peyman Najafirad.
\newblock Enhancing source code security with llms: Demystifying the challenges and generating reliable repairs.
\newblock \emph{arXiv preprint arXiv:2407.03975}, 2024.

\bibitem[Ivison et~al.(2024)Ivison, Wang, Liu, Wu, Pyatkin, Lambert, Smith, Choi, and Hajishirzi]{ivison2024unpacking}
Hamish Ivison, Yizhong Wang, Jiacheng Liu, Zeqiu Wu, Valentina Pyatkin, Nathan Lambert, Noah~A Smith, Yejin Choi, and Hanna Hajishirzi.
\newblock Unpacking dpo and ppo: Disentangling best practices for learning from preference feedback.
\newblock \emph{Advances in neural information processing systems}, 37:\penalty0 36602--36633, 2024.

\bibitem[Jain et~al.()Jain, Synnaeve, and Roziere]{jaintestgeneval}
Kush Jain, Gabriel Synnaeve, and Baptiste Roziere.
\newblock Testgeneval: A real world unit test generation and test completion benchmark.
\newblock In \emph{The Thirteenth International Conference on Learning Representations}.

\bibitem[Jain et~al.(2024{\natexlab{a}})Jain, Han, Gu, Li, Yan, Zhang, Wang, Solar-Lezama, Sen, and Stoica]{jain2024livecodebench}
Naman Jain, King Han, Alex Gu, Wen-Ding Li, Fanjia Yan, Tianjun Zhang, Sida Wang, Armando Solar-Lezama, Koushik Sen, and Ion Stoica.
\newblock Livecodebench: Holistic and contamination free evaluation of large language models for code.
\newblock \emph{arXiv preprint arXiv:2403.07974}, 2024{\natexlab{a}}.

\bibitem[Jain et~al.(2024{\natexlab{b}})Jain, Han, Gu, Li, Yan, Zhang, Wang, Solar-Lezama, Sen, and Stoica]{jain2024livecodebenchholisticcontaminationfree}
Naman Jain, King Han, Alex Gu, Wen-Ding Li, Fanjia Yan, Tianjun Zhang, Sida Wang, Armando Solar-Lezama, Koushik Sen, and Ion Stoica.
\newblock Livecodebench: Holistic and contamination free evaluation of large language models for code, 2024{\natexlab{b}}.
\newblock URL \url{https://arxiv.org/abs/2403.07974}.

\bibitem[{James} et~al.(2019){James}, {Ma}, {Rovick Arrojo}, and {Davison}]{RLBench}
Stephen {James}, Zicong {Ma}, David {Rovick Arrojo}, and Andrew~J. {Davison}.
\newblock {RLBench: The Robot Learning Benchmark \& Learning Environment}.
\newblock \emph{arXiv e-prints}, art. arXiv:1909.12271, September 2019.
\newblock \doi{10.48550/arXiv.1909.12271}.

\bibitem[Jana et~al.(2024{\natexlab{a}})Jana, Jha, Ju, Kishore, Mahajan, and Ganesh]{Jana_2024}
Prithwish Jana, Piyush Jha, Haoyang Ju, Gautham Kishore, Aryan Mahajan, and Vijay Ganesh.
\newblock \emph{CoTran: An LLM-Based Code Translator Using Reinforcement Learning with Feedback from Compiler and Symbolic Execution}.
\newblock IOS Press, October 2024{\natexlab{a}}.
\newblock ISBN 9781643685489.
\newblock \doi{10.3233/faia240968}.
\newblock URL \url{http://dx.doi.org/10.3233/FAIA240968}.

\bibitem[Jana et~al.(2024{\natexlab{b}})Jana, Jha, Ju, Kishore, Mahajan, and Ganesh]{jana2024cotran}
Prithwish Jana, Piyush Jha, Haoyang Ju, Gautham Kishore, Aryan Mahajan, and Vijay Ganesh.
\newblock Cotran: An llm-based code translator using reinforcement learning with feedback from compiler and symbolic execution.
\newblock In \emph{ECAI}, 2024{\natexlab{b}}.

\bibitem[Jaoua et~al.(2025)Jaoua, Sghaier, and Sahraoui]{jaoua2025combining}
Imen Jaoua, Oussama~Ben Sghaier, and Houari Sahraoui.
\newblock Combining large language models with static analyzers for code review generation.
\newblock In \emph{2025 IEEE/ACM 22nd International Conference on Mining Software Repositories (MSR)}, pages 174--186. IEEE, 2025.

\bibitem[Jemal et~al.(2024)Jemal, Rocher, Bissyand{\'e}, and Traon]{jemal2024exploratory}
Ibrahim Jemal, Mathis Rocher, Tegawend{\'e}~F Bissyand{\'e}, and Yves~Le Traon.
\newblock An exploratory study on fine-tuning large language models for secure code generation.
\newblock \emph{arXiv preprint arXiv:2403.04231}, 2024.

\bibitem[Ji et~al.(2020)Ji, Liang, Xiong, Zhang, and Hu]{question_selection_for_interactive}
Ruyi Ji, Jingjing Liang, Yingfei Xiong, Lu~Zhang, and Zhenjiang Hu.
\newblock Question selection for interactive program synthesis.
\newblock In Alastair~F. Donaldson and Emina Torlak, editors, \emph{Proceedings of the 41st {ACM} {SIGPLAN} International Conference on Programming Language Design and Implementation, {PLDI} 2020, London, UK, June 15-20, 2020}, pages 1143--1158. {ACM}, 2020.
\newblock \doi{10.1145/3385412.3386025}.
\newblock URL \url{https://doi.org/10.1145/3385412.3386025}.

\bibitem[Ji et~al.(2025)Ji, Lee, Lee, Han, and Im]{ji2025impact}
Suhwan Ji, Sanghwa Lee, Changsup Lee, Yo-Sub Han, and Hyeonseung Im.
\newblock Impact of large language models of code on fault localization.
\newblock In \emph{2025 IEEE Conference on Software Testing, Verification and Validation (ICST)}, pages 302--313. IEEE, 2025.

\bibitem[Jiang et~al.(2023{\natexlab{a}})Jiang, Sablayrolles, Mensch, Bamford, Chaplot, de~las Casas, Bressand, Lengyel, Lample, Saulnier, Lavaud, Lachaux, Stock, Scao, Lavril, Wang, Lacroix, and Sayed]{jiang2023mistral7b}
Albert~Q. Jiang, Alexandre Sablayrolles, Arthur Mensch, Chris Bamford, Devendra~Singh Chaplot, Diego de~las Casas, Florian Bressand, Gianna Lengyel, Guillaume Lample, Lucile Saulnier, Lélio~Renard Lavaud, Marie-Anne Lachaux, Pierre Stock, Teven~Le Scao, Thibaut Lavril, Thomas Wang, Timothée Lacroix, and William~El Sayed.
\newblock Mistral 7b, 2023{\natexlab{a}}.
\newblock URL \url{https://arxiv.org/abs/2310.06825}.

\bibitem[Jiang et~al.(2024{\natexlab{a}})Jiang, Sablayrolles, Roux, Mensch, Savary, Bamford, Chaplot, de~las Casas, Hanna, Bressand, Lengyel, Bour, Lample, Lavaud, Saulnier, Lachaux, Stock, Subramanian, Yang, Antoniak, Scao, Gervet, Lavril, Wang, Lacroix, and Sayed]{jiang2024mixtral}
Albert~Q. Jiang, Alexandre Sablayrolles, Antoine Roux, Arthur Mensch, Blanche Savary, Chris Bamford, Devendra~Singh Chaplot, Diego de~las Casas, Emma~Bou Hanna, Florian Bressand, Gianna Lengyel, Guillaume Bour, Guillaume Lample, Lélio~Renard Lavaud, Lucile Saulnier, Marie-Anne Lachaux, Pierre Stock, Sandeep Subramanian, Sophia Yang, Szymon Antoniak, Teven~Le Scao, Théophile Gervet, Thibaut Lavril, Thomas Wang, Timothée Lacroix, and William~El Sayed.
\newblock Mixtral of experts, 2024{\natexlab{a}}.
\newblock URL \url{https://arxiv.org/abs/2401.04088}.

\bibitem[Jiang et~al.(2025{\natexlab{a}})Jiang, Liu, Li, Wang, Zhang, Song, Wei, and Lian]{jiang2025makes}
Gangwei Jiang, Yahui Liu, Zhaoyi Li, Qi~Wang, Fuzheng Zhang, Linqi Song, Ying Wei, and Defu Lian.
\newblock What makes a good reasoning chain? uncovering structural patterns in long chain-of-thought reasoning.
\newblock \emph{arXiv preprint arXiv:2505.22148}, 2025{\natexlab{a}}.

\bibitem[Jiang et~al.(2024{\natexlab{b}})Jiang, Liu, Li, Ye, and Wang]{cursorcore}
Hao Jiang, Qi~Liu, Rui Li, Shengyu Ye, and Shijin Wang.
\newblock Cursorcore: Assist programming through aligning anything.
\newblock \emph{arXiv preprint arXiv:2410.07002}, 2024{\natexlab{b}}.

\bibitem[Jiang et~al.(2025{\natexlab{b}})Jiang, Chen, Cao, yi~Lee, and Tan]{jiang2025codejudgebenchbenchmarkingllmasajudgecoding}
Hongchao Jiang, Yiming Chen, Yushi Cao, Hung yi~Lee, and Robby~T. Tan.
\newblock Codejudgebench: Benchmarking llm-as-a-judge for coding tasks, 2025{\natexlab{b}}.
\newblock URL \url{https://arxiv.org/abs/2507.10535}.

\bibitem[Jiang et~al.(2024{\natexlab{c}})Jiang, Wang, Shen, Kim, and Kim]{jiang2024surveylargelanguagemodels}
Juyong Jiang, Fan Wang, Jiasi Shen, Sungju Kim, and Sunghun Kim.
\newblock A survey on large language models for code generation, 2024{\natexlab{c}}.
\newblock URL \url{https://arxiv.org/abs/2406.00515}.

\bibitem[Jiang et~al.(2025{\natexlab{c}})Jiang, Huang, Wu, Li, Zhang, and Wei]{viscodex}
Lingjie Jiang, Shaohan Huang, Xun Wu, Yixia Li, Dongdong Zhang, and Furu Wei.
\newblock Viscodex: Unified multimodal code generation via merging vision and coding models.
\newblock \emph{arXiv preprint arXiv:2508.09945}, 2025{\natexlab{c}}.

\bibitem[Jiang et~al.(2023{\natexlab{b}})Jiang, Lutellier, Lou, Tan, Goldwasser, and Zhang]{jiang2023knod}
Nan Jiang, Thibaud Lutellier, Yiling Lou, Lin Tan, Dan Goldwasser, and Xiangyu Zhang.
\newblock Knod: Domain knowledge distilled tree decoder for automated program repair.
\newblock In \emph{2023 IEEE/ACM 45th International Conference on Software Engineering (ICSE)}, pages 1251--1263. IEEE, 2023{\natexlab{b}}.

\bibitem[Jiang et~al.(2023{\natexlab{c}})Jiang, Wang, Liu, Xu, Tan, and Zhang]{DBLP:journals/corr/abs-2311-13721}
Nan Jiang, Chengxiao Wang, Kevin Liu, Xiangzhe Xu, Lin Tan, and Xiangyu Zhang.
\newblock Nova\({}^{\mbox{+}}\): Generative language models for binaries.
\newblock \emph{CoRR}, abs/2311.13721, 2023{\natexlab{c}}.

\bibitem[Jiang and McMillan(2017)]{jiang2017automaticgenerationshortsummaries}
Siyuan Jiang and Collin McMillan.
\newblock Towards automatic generation of short summaries of commits, 2017.
\newblock URL \url{https://arxiv.org/abs/1703.09603}.

\bibitem[Jiang et~al.(2017)Jiang, Armaly, and McMillan]{jiang2017automaticallygeneratingcommitmessages}
Siyuan Jiang, Ameer Armaly, and Collin McMillan.
\newblock Automatically generating commit messages from diffs using neural machine translation, 2017.
\newblock URL \url{https://arxiv.org/abs/1708.09492}.

\bibitem[Jiang et~al.(2024{\natexlab{d}})Jiang, Wang, Pang, Lin, Zhang, Liu, Liao, Chen, Lo, and Sun]{jiang2024purpcode}
Ting-En Jiang, Yizhen Wang, Jing-Cheng Pang, Pei-Hung Lin, Chen-Lin Zhang, Xin-Yang Liu, Yi-Ling Liao, Ching-Ting Chen, Chia-Chun Lo, and Shao-Hua Sun.
\newblock Purpcode: Reasoning for safer code generation, 2024{\natexlab{d}}.

\bibitem[Jiang et~al.(2024{\natexlab{e}})Jiang, Gao, Zhai, Ma, Zhang, Lei, and Shen]{jiang2024effectiveness}
Weipeng Jiang, Xuanqi Gao, Juan Zhai, Shiqing Ma, Xiaoyu Zhang, Ziyan Lei, and Chao Shen.
\newblock From effectiveness to efficiency: Uncovering linguistic bias in large language model-based code generation.
\newblock \emph{arXiv preprint arXiv:2406.00602}, 2024{\natexlab{e}}.

\bibitem[Jiang et~al.(2025{\natexlab{d}})Jiang, Zheng, Wan, Han, Wang, Lyu, and Yue]{screencoder}
Yilei Jiang, Yaozhi Zheng, Yuxuan Wan, Jiaming Han, Qunzhong Wang, Michael~R Lyu, and Xiangyu Yue.
\newblock Screencoder: Advancing visual-to-code generation for front-end automation via modular multimodal agents.
\newblock \emph{arXiv preprint arXiv:2507.22827}, 2025{\natexlab{d}}.

\bibitem[Jiang et~al.(2025{\natexlab{e}})Jiang, Yap, and Liang]{ossbench}
Yuancheng Jiang, Roland Yap, and Zhenkai Liang.
\newblock Oss-bench: Benchmark generator for coding llms, 2025{\natexlab{e}}.
\newblock URL \url{https://arxiv.org/abs/2505.12331}.

\bibitem[Jiao et~al.(2023)Jiao, Yu, Li, Qiu, Gu, and Shen]{jiao2023evaluationneuralcodetranslation}
Mingsheng Jiao, Tingrui Yu, Xuan Li, Guanjie Qiu, Xiaodong Gu, and Beijun Shen.
\newblock On the evaluation of neural code translation: Taxonomy and benchmark, 2023.
\newblock URL \url{https://arxiv.org/abs/2308.08961}.

\bibitem[Jimenez(2024)]{jimenez2024swtbench}
Alejandro et~al. Jimenez.
\newblock Swt-bench: Benchmarking llms for software testing and bug repair.
\newblock In \emph{ICSE}, 2024.

\bibitem[Jimenez et~al.(2024)Jimenez, Yang, Wettig, Yao, Pei, Press, and Narasimhan]{jimenez2024swebench}
Carlos~E. Jimenez, John Yang, Alexander Wettig, Shunyu Yao, Kexin Pei, Ofir Press, and Karthik~R. Narasimhan.
\newblock Swe-bench: Can language models resolve real-world github issues?
\newblock In \emph{The Twelfth International Conference on Learning Representations, {ICLR} 2024, Vienna, Austria, May 7-11, 2024}. OpenReview.net, 2024.
\newblock URL \url{https://openreview.net/forum?id=VTF8yNQM66}.

\bibitem[Jin et~al.(2024)Jin, Jin, Chen, and Wang]{MARE}
Dongming Jin, Zhi Jin, Xiaohong Chen, and Chunhui Wang.
\newblock Mare: Multi-agents collaboration framework for requirements engineering.
\newblock \emph{arXiv preprint arXiv:2405.03256}, 2024.

\bibitem[Jin et~al.(2025)Jin, Sun, Huang, Liang, Xuan, Liu, and Jin]{iReDev}
Dongming Jin, Weisong Sun, Jiangping Huang, Peng Liang, Jifeng Xuan, Yang Liu, and Zhi Jin.
\newblock iredev: A knowledge-driven multi-agent framework for intelligent requirements development.
\newblock \emph{arXiv preprint arXiv:2507.13081}, 2025.

\bibitem[Jin and Hamdaqa(2025)]{jin2025cccicodecompletioncontextual}
Hangzhan Jin and Mohammad Hamdaqa.
\newblock Ccci: Code completion with contextual information for complex data transfer tasks using large language models, 2025.
\newblock URL \url{https://arxiv.org/abs/2503.23231}.

\bibitem[Jin et~al.(2023)Jin, Shahriar, Tufano, Shi, Lu, Sundaresan, and Svyatkovskiy]{jin2023inferfix}
Matthew Jin, Syed Shahriar, Michele Tufano, Xin Shi, Shuai Lu, Neel Sundaresan, and Alexey Svyatkovskiy.
\newblock Inferfix: End-to-end program repair with llms.
\newblock In \emph{Proceedings of the 31st ACM joint european software engineering conference and symposium on the foundations of software engineering}, pages 1646--1656, 2023.

\bibitem[Joshi et~al.(2023)Joshi, Sanchez, Gulwani, Le, Verbruggen, and Radi{\v{c}}ek]{joshi2023repair}
Harshit Joshi, Jos{\'e}~Cambronero Sanchez, Sumit Gulwani, Vu~Le, Gust Verbruggen, and Ivan Radi{\v{c}}ek.
\newblock Repair is nearly generation: Multilingual program repair with llms.
\newblock In \emph{Proceedings of the AAAI Conference on Artificial Intelligence}, volume~37, pages 5131--5140, 2023.

\bibitem[Joshi and Kahani(2024)]{joshi2024comparative}
Rinkesh Joshi and Nafiseh Kahani.
\newblock Comparative study of reinforcement learning in github pull request outcome predictions.
\newblock In \emph{2024 IEEE International Conference on Software Analysis, Evolution and Reengineering (SANER)}, pages 489--500. IEEE, 2024.

\bibitem[Jung(2021)]{jung2021commitbert}
Tae-Hwan Jung.
\newblock Commitbert: Commit message generation using pre-trained programming language model.
\newblock \emph{arXiv preprint arXiv:2105.14242}, 2021.

\bibitem[Just et~al.(2014)Just, Jalali, and Ernst]{just2014defects4j}
Ren{\'e} Just, Darioush Jalali, and Michael~D Ernst.
\newblock Defects4j: A database of existing faults to enable controlled testing studies for java programs.
\newblock In \emph{Proceedings of the 2014 international symposium on software testing and analysis}, pages 437--440, 2014.

\bibitem[Kamradt(2025)]{snake_bench_2025}
Greg Kamradt.
\newblock Snake bench: Competitive snake game simulation with llms.
\newblock \url{https://github.com/gkamradt/SnakeBench}, 2025.
\newblock Accessed on: Month Day, Year.

\bibitem[Kandpal et~al.(2022)Kandpal, Wallace, and Raffel]{kandpal2022deduplicating}
Nikhil Kandpal, Eric Wallace, and Colin Raffel.
\newblock Deduplicating training data mitigates privacy risks in language models.
\newblock In \emph{International Conference on Machine Learning}, pages 10419--10431. PMLR, 2022.

\bibitem[Kang et~al.(2024{\natexlab{a}})Kang, An, and Yoo]{kang2024quantitative}
Sungmin Kang, Gabin An, and Shin Yoo.
\newblock A quantitative and qualitative evaluation of llm-based explainable fault localization.
\newblock \emph{Proceedings of the ACM on Software Engineering}, 1\penalty0 (FSE):\penalty0 1424--1446, 2024{\natexlab{a}}.

\bibitem[Kang et~al.(2024{\natexlab{b}})Kang, An, and Yoo]{kang2024quantitative_fault_localization}
Sungmin Kang, Gabin An, and Shin Yoo.
\newblock A quantitative and qualitative evaluation of llm-based explainable fault localization.
\newblock \emph{Proceedings of the ACM on Software Engineering}, 1\penalty0 (FSE):\penalty0 1424--1446, 2024{\natexlab{b}}.

\bibitem[Kapadnis et~al.(2025)Kapadnis, Naik, and Rose]{kapadnis2025crscore++}
Manav~Nitin Kapadnis, Atharva Naik, and Carolyn Rose.
\newblock Crscore++: Reinforcement learning with verifiable tool and ai feedback for code review.
\newblock \emph{arXiv preprint arXiv:2506.00296}, 2025.

\bibitem[Kaplan et~al.(2020)Kaplan, McCandlish, Henighan, Brown, Chess, Child, Gray, Radford, Wu, and Amodei]{kaplan2020scalinglaws}
Jared Kaplan, Sam McCandlish, Tom Henighan, Tom~B. Brown, Benjamin Chess, Rewon Child, Scott Gray, Alec Radford, Jeffrey Wu, and Dario Amodei.
\newblock Scaling laws for neural language models, 2020.
\newblock URL \url{https://arxiv.org/abs/2001.08361}.

\bibitem[Kapoor et~al.(2024)Kapoor, Butala, Russak, Koh, Kamble, Alshikh, and Salakhutdinov]{kapoor2024omniactdatasetbenchmarkenabling}
Raghav Kapoor, Yash~Parag Butala, Melisa Russak, Jing~Yu Koh, Kiran Kamble, Waseem Alshikh, and Ruslan Salakhutdinov.
\newblock Omniact: A dataset and benchmark for enabling multimodal generalist autonomous agents for desktop and web, 2024.
\newblock URL \url{https://arxiv.org/abs/2402.17553}.

\bibitem[Khan and Uddin(2023)]{KhanGPT}
Junaed~Younus Khan and Gias Uddin.
\newblock Automatic code documentation generation using gpt-3.
\newblock In \emph{Proceedings of the 37th IEEE/ACM International Conference on Automated Software Engineering}, ASE '22, New York, NY, USA, 2023. Association for Computing Machinery.
\newblock ISBN 9781450394758.
\newblock \doi{10.1145/3551349.3559548}.
\newblock URL \url{https://doi.org/10.1145/3551349.3559548}.

\bibitem[Khan et~al.(2023)Khan, Bari, Do, Wang, Parvez, and Joty]{ziyao2023xcodeeval}
Mohammad Abdullah~Matin Khan, M~Saiful Bari, Xuan~Long Do, Weishi Wang, Md~Rizwan Parvez, and Shafiq Joty.
\newblock xcodeeval: A large scale multilingual multitask benchmark for code understanding, generation, translation and retrieval, 2023.
\newblock URL \url{https://arxiv.org/abs/2303.03004}.

\bibitem[Khare et~al.(2025)Khare, Saini, Sharma, Kumar, Rana, and Yadav]{khare2025deputydev}
Vishal Khare, Vijay Saini, Deepak Sharma, Anand Kumar, Ankit Rana, and Anshul Yadav.
\newblock Deputydev--ai powered developer assistant: Breaking the code review logjam through contextual ai to boost developer productivity.
\newblock \emph{arXiv preprint arXiv:2508.09676}, 2025.

\bibitem[Khatri et~al.(2025)Khatri, Madaan, Tiwari, Bansal, Duvvuri, Zaheer, Dhillon, Brandfonbrener, and Agarwal]{khatri2025artscalingreinforcementlearning}
Devvrit Khatri, Lovish Madaan, Rishabh Tiwari, Rachit Bansal, Sai~Surya Duvvuri, Manzil Zaheer, Inderjit~S. Dhillon, David Brandfonbrener, and Rishabh Agarwal.
\newblock The art of scaling reinforcement learning compute for llms, 2025.
\newblock URL \url{https://arxiv.org/abs/2510.13786}.

\bibitem[Khoury et~al.(2023)Khoury, Avila, Brunelle, and Camara]{khoury2023howsecureiscode}
Raphaël Khoury, Anderson~R. Avila, Jacob Brunelle, and Baba~Mamadou Camara.
\newblock How secure is code generated by chatgpt?
\newblock In \emph{2023 IEEE International Conference on Systems, Man, and Cybernetics (SMC)}, pages 2445--2451, 2023.
\newblock \doi{10.1109/SMC53992.2023.10394237}.

\bibitem[Kim et~al.(2025)Kim, Kweon, woo Lee, and Choi]{kim2025reasoning}
Taeyoun Kim, Jaemin Kweon, Sang woo Lee, and Yoojin Choi.
\newblock Reasoning as an adaptive defense for safety, 2025.

\bibitem[Kirchner and Knoll(2025)]{kirchner2025generating}
Sven Kirchner and Alois~C Knoll.
\newblock Generating automotive code: Large language models for software development and verification in safety-critical systems.
\newblock \emph{arXiv preprint arXiv:2506.04038}, 2025.

\bibitem[Kocetkov et~al.(2022{\natexlab{a}})Kocetkov, Li, Allal, Li, Mou, Ferrandis, Jernite, Mitchell, Hughes, Wolf, Bahdanau, von Werra, and de~Vries]{kocetkov2022stack}
Denis Kocetkov, Raymond Li, Loubna~Ben Allal, Jia Li, Chenghao Mou, Carlos~Muñoz Ferrandis, Yacine Jernite, Margaret Mitchell, Sean Hughes, Thomas Wolf, Dzmitry Bahdanau, Leandro von Werra, and Harm de~Vries.
\newblock {The Stack: 3 TB of Permissively Licensed Source Code}, 2022{\natexlab{a}}.
\newblock URL \url{https://arxiv.org/abs/2211.15533}.

\bibitem[Kocetkov et~al.(2022{\natexlab{b}})Kocetkov, Li, Allal, Li, Mou, Ferrandis, Jernite, Mitchell, Hughes, Wolf, Bahdanau, von Werra, and de~Vries]{kocetkov2022stack3tbpermissively}
Denis Kocetkov, Raymond Li, Loubna~Ben Allal, Jia Li, Chenghao Mou, Carlos~Muñoz Ferrandis, Yacine Jernite, Margaret Mitchell, Sean Hughes, Thomas Wolf, Dzmitry Bahdanau, Leandro von Werra, and Harm de~Vries.
\newblock The stack: 3 tb of permissively licensed source code, 2022{\natexlab{b}}.
\newblock URL \url{https://arxiv.org/abs/2211.15533}.

\bibitem[KodCode(2024)]{kodcode_v1}
KodCode.
\newblock Kodcode-v1 dataset, 2024.
\newblock URL \url{https://huggingface.co/datasets/KodCode/KodCode-V1}.
\newblock Accessed: 2024.

\bibitem[Koh et~al.(2024)Koh, Lo, Jang, Duvvur, Lim, Huang, Neubig, Zhou, Salakhutdinov, and Fried]{koh2024visualwebarena}
Jing~Yu Koh, Robert Lo, Lawrence Jang, Vikram Duvvur, Ming Lim, Po-Yu Huang, Graham Neubig, Shuyan Zhou, Russ Salakhutdinov, and Daniel Fried.
\newblock {V}isual{W}eb{A}rena: Evaluating multimodal agents on realistic visual web tasks.
\newblock In Lun-Wei Ku, Andre Martins, and Vivek Srikumar, editors, \emph{Proceedings of the 62nd Annual Meeting of the Association for Computational Linguistics (Volume 1: Long Papers)}, pages 881--905, Bangkok, Thailand, August 2024. Association for Computational Linguistics.
\newblock \doi{10.18653/v1/2024.acl-long.50}.
\newblock URL \url{https://aclanthology.org/2024.acl-long.50/}.

\bibitem[Kong et~al.(2025)Kong, Lin, Xu, Wang, Li, Li, Zhang, Peng, Sha, Li, et~al.]{a2a_survey}
Dezhang Kong, Shi Lin, Zhenhua Xu, Zhebo Wang, Minghao Li, Yufeng Li, Yilun Zhang, Hujin Peng, Zeyang Sha, Yuyuan Li, et~al.
\newblock A survey of llm-driven ai agent communication: Protocols, security risks, and defense countermeasures.
\newblock \emph{arXiv preprint arXiv:2506.19676}, 2025.

\bibitem[Korbak et~al.(2023)Korbak, Shi, Chen, Bhalerao, Buckley, Phang, Bowman, and Perez]{korbak2023pretraining}
Tomasz Korbak, Kejian Shi, Angelica Chen, Rasika~Vinayak Bhalerao, Christopher Buckley, Jason Phang, Samuel~R Bowman, and Ethan Perez.
\newblock Pretraining language models with human preferences.
\newblock In \emph{International Conference on Machine Learning}, pages 17506--17533. PMLR, 2023.

\bibitem[Kruse et~al.(2024)Kruse, Puhlf{\"{u}}r{\ss}, and Maalej]{can_developers_prompt_code_documentation}
Hans{-}Alexander Kruse, Tim Puhlf{\"{u}}r{\ss}, and Walid Maalej.
\newblock Can developers prompt? {A} controlled experiment for code documentation generation.
\newblock In \emph{{IEEE} International Conference on Software Maintenance and Evolution, {ICSME} 2024, Flagstaff, AZ, USA, October 6-11, 2024}, pages 574--586. {IEEE}, 2024.
\newblock \doi{10.1109/ICSME58944.2024.00058}.
\newblock URL \url{https://doi.org/10.1109/ICSME58944.2024.00058}.

\bibitem[Kuang et~al.(2022)Kuang, Zhou, and Yang]{GTrans}
L.~Kuang, C.~Zhou, and X.~Yang.
\newblock Code comment generation based on graph neural network enhanced transformer model for code understanding in open-source software ecosystems.
\newblock \emph{Automated Software Engineering}, 29:\penalty0 43, 2022.
\newblock \doi{10.1007/s10515-022-00341-1}.
\newblock URL \url{https://doi.org/10.1007/s10515-022-00341-1}.

\bibitem[Kulal et~al.(2019)Kulal, Pasupat, Chandra, Lee, Padon, Aiken, and Liang]{SPoC}
Sumith Kulal, Panupong Pasupat, Kartik Chandra, Mina Lee, Oded Padon, Alex Aiken, and Percy Liang.
\newblock \emph{SPoC: search-based pseudocode to code}.
\newblock Curran Associates Inc., Red HEvaluating Large Language Models Trained on Code ook, NY, USA, 2019.

\bibitem[Kumar et~al.(2024{\natexlab{a}})Kumar, Zhuang, Agarwal, Su, Co-Reyes, Singh, Baumli, Iqbal, Bishop, Roelofs, et~al.]{kumar2024training}
Aviral Kumar, Vincent Zhuang, Rishabh Agarwal, Yi~Su, John~D Co-Reyes, Avi Singh, Kate Baumli, Shariq Iqbal, Colton Bishop, Rebecca Roelofs, et~al.
\newblock Training language models to self-correct via reinforcement learning.
\newblock \emph{arXiv preprint arXiv:2409.12917}, 2024{\natexlab{a}}.

\bibitem[Kumar et~al.(2025)Kumar, Janapati, Tanguturi, and Chimalakonda]{kumar2025icantsharecode}
Jahnavi Kumar, Venkata Lakshmana~Sasaank Janapati, Mokshith~Reddy Tanguturi, and Sridhar Chimalakonda.
\newblock I can't share code, but i need translation -- an empirical study on code translation through federated llm, 2025.
\newblock URL \url{https://arxiv.org/abs/2501.05724}.

\bibitem[Kumar et~al.(2024{\natexlab{b}})Kumar, Agarwal, Gupta, Sharma, and Singh]{kumar2024llamafirewall}
Sandhini Kumar, Anish Agarwal, Chetna Gupta, Shivang Sharma, and Manasi Singh.
\newblock Llamafirewall: An open source guardrail system for building secure ai agents.
\newblock \emph{arXiv preprint arXiv:2406.01288}, 2024{\natexlab{b}}.

\bibitem[LaBash et~al.(2024)LaBash, Rosedale, Reents, Negritto, and Wiel]{labash2024resqevaluatingcodeeditinglarge}
Beck LaBash, August Rosedale, Alex Reents, Lucas Negritto, and Colin Wiel.
\newblock Res-q: Evaluating code-editing large language model systems at the repository scale, 2024.
\newblock URL \url{https://arxiv.org/abs/2406.16801}.

\bibitem[Labs et~al.(2025)Labs, Khanna, Kharbanda, Li, Varma, Wang, Birnbaum, Luo, Miraoui, Palrecha, Ermon, Grover, and Kuleshov]{labs2025mercury}
Inception Labs, Samar Khanna, Siddhant Kharbanda, Shufan Li, Harshit Varma, Eric Wang, Sawyer Birnbaum, Ziyang Luo, Yanis Miraoui, Akash Palrecha, Stefano Ermon, Aditya Grover, and Volodymyr Kuleshov.
\newblock Mercury: Ultra-fast language models based on diffusion, 2025.
\newblock URL \url{https://arxiv.org/abs/2506.17298}.

\bibitem[Lachaux et~al.(2020)Lachaux, Roziere, Chanussot, and Lample]{lachaux2020unsupervisedtranslationprogramminglanguages}
Marie-Anne Lachaux, Baptiste Roziere, Lowik Chanussot, and Guillaume Lample.
\newblock Unsupervised translation of programming languages, 2020.
\newblock URL \url{https://arxiv.org/abs/2006.03511}.

\bibitem[Lachaux et~al.(2021)Lachaux, Roziere, Szafraniec, and Lample]{dobf}
Marie-Anne Lachaux, Baptiste Roziere, Marc Szafraniec, and Guillaume Lample.
\newblock Dobf: A deobfuscation pre-training objective for programming languages.
\newblock \emph{Advances in Neural Information Processing Systems}, 34:\penalty0 14967--14979, 2021.

\bibitem[Lacomis et~al.(2019)Lacomis, Yin, Schwartz, Allamanis, {Le Goues}, Neubig, and Vasilescu]{DBLP:conf/kbse/LacomisYSAGNV19}
Jeremy Lacomis, Pengcheng Yin, Edward~J. Schwartz, Miltiadis Allamanis, Claire {Le Goues}, Graham Neubig, and Bogdan Vasilescu.
\newblock {DIRE:} {A} neural approach to decompiled identifier naming.
\newblock In \emph{34th {IEEE/ACM} International Conference on Automated Software Engineering, {ASE} 2019, San Diego, CA, USA, November 11-15, 2019}, pages 628--639. {IEEE}, 2019.

\bibitem[Lai et~al.(2025)Lai, Liu, Zhao, Xu, Zhang, Jing, Ren, Yao, Dong, and Tang]{lai2025computerrl}
Hanyu Lai, Xiao Liu, Yanxiao Zhao, Han Xu, Hanchen Zhang, Bohao Jing, Yanyu Ren, Shuntian Yao, Yuxiao Dong, and Jie Tang.
\newblock Computerrl: Scaling end-to-end online reinforcement learning for computer use agents.
\newblock \emph{arXiv preprint arXiv:2508.14040}, 2025.

\bibitem[Lai et~al.(2023)Lai, Li, Wang, Zhang, Zhong, Zettlemoyer, Yih, Fried, Wang, and Yu]{ds1000}
Yuhang Lai, Chengxi Li, Yiming Wang, Tianyi Zhang, Ruiqi Zhong, Luke Zettlemoyer, Wen{-}Tau Yih, Daniel Fried, Sida~I. Wang, and Tao Yu.
\newblock {DS-1000:} {A} natural and reliable benchmark for data science code generation.
\newblock In Andreas Krause, Emma Brunskill, Kyunghyun Cho, Barbara Engelhardt, Sivan Sabato, and Jonathan Scarlett, editors, \emph{International Conference on Machine Learning, {ICML} 2023, 23-29 July 2023, Honolulu, Hawaii, {USA}}, volume 202 of \emph{Proceedings of Machine Learning Research}, pages 18319--18345. {PMLR}, 2023.
\newblock URL \url{https://proceedings.mlr.press/v202/lai23b.html}.

\bibitem[Lamouri et~al.(2025)Lamouri, Aouadj, Kourta, and Baghdadi]{lamouri2025pearl}
Djamel~Rassem Lamouri, Iheb~Nassim Aouadj, Smail Kourta, and Riyadh Baghdadi.
\newblock Pearl: Automatic code optimization using deep reinforcement learning.
\newblock In \emph{Proceedings of the 39th ACM International Conference on Supercomputing}, pages 959--974, 2025.

\bibitem[Le et~al.(2022{\natexlab{a}})Le, Wang, Gotmare, Savarese, and Hoi]{code_rl}
Hung Le, Yue Wang, Akhilesh~Deepak Gotmare, Silvio Savarese, and Steven~Chu{-}Hong Hoi.
\newblock Coderl: Mastering code generation through pretrained models and deep reinforcement learning.
\newblock In Sanmi Koyejo, S.~Mohamed, A.~Agarwal, Danielle Belgrave, K.~Cho, and A.~Oh, editors, \emph{Advances in Neural Information Processing Systems 35: Annual Conference on Neural Information Processing Systems 2022, NeurIPS 2022, New Orleans, LA, USA, November 28 - December 9, 2022}, 2022{\natexlab{a}}.
\newblock URL \url{http://papers.nips.cc/paper\_files/paper/2022/hash/8636419dea1aa9fbd25fc4248e702da4-Abstract-Conference.html}.

\bibitem[Le et~al.(2022{\natexlab{b}})Le, Wang, Gotmare, Savarese, and Hoi]{le2022coderl}
Hung Le, Yue Wang, Akhilesh~Deepak Gotmare, Silvio Savarese, and Steven Chu~Hong Hoi.
\newblock Coderl: Mastering code generation through pretrained models and deep reinforcement learning.
\newblock \emph{Advances in Neural Information Processing Systems}, 35:\penalty0 21314--21328, 2022{\natexlab{b}}.

\bibitem[Le et~al.(2024)Le, Nguyen, Chen, Ouyang, and Marcel]{le2024indict}
Hung Le, Ha-Thanh Nguyen, Zhaoyuan Chen, Siru Ouyang, and Rudolf Marcel.
\newblock Indict: Code generation with internal dialogues of critiques for both security and helpfulness, 2024.

\bibitem[Le-Cong et~al.(2024)Le-Cong, Le, and Murray]{le2024semantic}
Thanh Le-Cong, Bach Le, and Toby Murray.
\newblock Semantic-guided search for efficient program repair with large language models.
\newblock \emph{arXiv preprint arXiv:2410.16655}, 2024.

\bibitem[Le~Hai et~al.(2024)Le~Hai, Nguyen, and Bui]{le2024repoexec}
Nam Le~Hai, Dung~Manh Nguyen, and Nghi~DQ Bui.
\newblock Repoexec: Evaluate code generation with a repository-level executable benchmark.
\newblock \emph{arXiv e-prints}, pages arXiv--2406, 2024.

\bibitem[Lee et~al.(2022)Lee, Seonwoo, and Oh]{lee2022cs1qa}
Changyoon Lee, Yeon Seonwoo, and Alice Oh.
\newblock Cs1qa: A dataset for assisting code-based question answering in an introductory programming course.
\newblock \emph{arXiv preprint arXiv:2210.14494}, 2022.

\bibitem[Lee et~al.(2024{\natexlab{a}})Lee, Park, Kim, and Park]{MCS-SQL}
Dongjun Lee, Choongwon Park, Jaehyuk Kim, and Heesoo Park.
\newblock {MCS-SQL:} leveraging multiple prompts and multiple-choice selection for text-to-sql generation.
\newblock \emph{CoRR}, abs/2405.07467, 2024{\natexlab{a}}.

\bibitem[Lee et~al.(2025{\natexlab{a}})Lee, Hwang, and Lee]{lee2025learning}
Dongjun Lee, Changho Hwang, and Kimin Lee.
\newblock Learning to generate unit test via adversarial reinforcement learning.
\newblock \emph{arXiv preprint arXiv:2508.21107}, 2025{\natexlab{a}}.

\bibitem[Lee et~al.(2025{\natexlab{b}})Lee, Yang, Bilik, Krsek, von Davier, Monteiro, Lin, Agarwal, Forlizzi, and Das]{lee2025privy}
Hao-Ping Lee, Yu-Ju Yang, Matthew Bilik, Isadora Krsek, Thomas~Serban von Davier, Kyzyl Monteiro, Jason Lin, Shivani Agarwal, Jodi Forlizzi, and Sauvik Das.
\newblock Privy: Envisioning and mitigating privacy risks for consumer-facing ai product concepts, 2025{\natexlab{b}}.

\bibitem[Lee et~al.(2023)Lee, Phatale, Mansoor, Lu, Mesnard, Ferret, Bishop, Hall, Carbune, and Rastogi]{lee2023rlaif}
Harrison Lee, Samrat Phatale, Hassan Mansoor, Kellie~Ren Lu, Thomas Mesnard, Johan Ferret, Colton Bishop, Ethan Hall, Victor Carbune, and Abhinav Rastogi.
\newblock Rlaif: Scaling reinforcement learning from human feedback with ai feedback.
\newblock 2023.

\bibitem[Lee et~al.(2024{\natexlab{b}})Lee, Phatale, Mansoor, Mesnard, Ferret, Lu, Bishop, Hall, Carbune, Rastogi, and Prakash]{RLAIF}
Harrison Lee, Samrat Phatale, Hassan Mansoor, Thomas Mesnard, Johan Ferret, Kellie Lu, Colton Bishop, Ethan Hall, Victor Carbune, Abhinav Rastogi, and Sushant Prakash.
\newblock {RLAIF} vs. {RLHF:} scaling reinforcement learning from human feedback with {AI} feedback.
\newblock In \emph{Forty-first International Conference on Machine Learning, {ICML} 2024, Vienna, Austria, July 21-27, 2024}. OpenReview.net, 2024{\natexlab{b}}.
\newblock URL \url{https://openreview.net/forum?id=uydQ2W41KO}.

\bibitem[{Lee} et~al.(2025){Lee}, {An}, and {Yoo}]{METAMON}
Hyeonseok {Lee}, Gabin {An}, and Shin {Yoo}.
\newblock {METAMON: Finding Inconsistencies between Program Documentation and Behavior using Metamorphic LLM Queries}.
\newblock \emph{arXiv e-prints}, art. arXiv:2502.02794, February 2025.
\newblock \doi{10.48550/arXiv.2502.02794}.

\bibitem[Lee et~al.(2025{\natexlab{a}})Lee, Kim, Singh, Pereira, Sonwane, White, Stengel-Eskin, Bansal, Shi, Sordoni, et~al.]{gistify}
Hyunji Lee, Minseon Kim, Chinmay Singh, Matheus Pereira, Atharv Sonwane, Isadora White, Elias Stengel-Eskin, Mohit Bansal, Zhengyan Shi, Alessandro Sordoni, et~al.
\newblock Gistify! codebase-level understanding via runtime execution.
\newblock \emph{arXiv preprint arXiv:2510.26790}, 2025{\natexlab{a}}.

\bibitem[Lee et~al.(2025{\natexlab{b}})Lee, Lee, Feng, and Lan]{lee2025text}
Jaewook Lee, Jeongah Lee, Wanyong Feng, and Andrew Lan.
\newblock From text to visuals: Using llms to generate math diagrams with vector graphics.
\newblock \emph{arXiv preprint arXiv:2503.07429}, 2025{\natexlab{b}}.

\bibitem[Lee et~al.(2025{\natexlab{c}})Lee, Jeong, Han, and Lee]{lee2025logresp}
Juyoung Lee, Yeonsu Jeong, Taehyun Han, and Taejin Lee.
\newblock Logresp-agent: A recursive ai framework for context-aware log anomaly detection and ttp analysis.
\newblock \emph{Applied Sciences}, 15\penalty0 (13):\penalty0 7237, 2025{\natexlab{c}}.

\bibitem[Lei et~al.(2024)Lei, Li, and Chen]{leiAutoCoderEnhancingCode2024}
Bin Lei, Yuchen Li, and Qiuwu Chen.
\newblock {{AutoCoder}}: {{Enhancing Code Large Language Model}} with {\textbackslash}textsc\{\vphantom\}{{AIEV-Instruct}}\vphantom\{\}, 2024.

\bibitem[Leino(2010)]{leino2010dafny}
K.~Rustan~M. Leino.
\newblock Dafny: An automatic program verifier for functional correctness.
\newblock In \emph{International Conference on Logic for Programming Artificial Intelligence and Reasoning}, pages 348--370. Springer, 2010.

\bibitem[Lemieux et~al.(2023)Lemieux, Inala, Lahiri, and Sen]{Lemieux2023CODAMOSA}
Caroline Lemieux, Jeevana~Priya Inala, Shuvendu~K. Lahiri, and Siddhartha Sen.
\newblock Codamosa: Escaping coverage plateaus in test generation with pre-trained large language models.
\newblock In \emph{Proceedings of the 45th International Conference on Software Engineering (ICSE)}, pages 837--849. IEEE, 2023.
\newblock \doi{10.1109/ICSE48619.2023.00085}.
\newblock URL \url{https://dl.acm.org/doi/10.1109/ICSE48619.2023.00085}.

\bibitem[Lepikhin et~al.(2020)Lepikhin, Lee, Xu, Chen, Firat, Huang, Krikun, Shazeer, and Chen]{lepikhin2020gshard}
Dmitry Lepikhin, HyoukJoong Lee, Yuanzhong Xu, Dehao Chen, Orhan Firat, Yanping Huang, Maxim Krikun, Noam Shazeer, and Zhifeng Chen.
\newblock Gshard: Scaling giant models with conditional computation and automatic sharding, 2020.
\newblock URL \url{https://arxiv.org/abs/2006.16668}.

\bibitem[Li et~al.(2025{\natexlab{a}})Li, Zhang, Guo, Zhang, Li, Zhang, Zhang, Zhang, Li, Liu, and Li]{li2025llavaonevision}
Bo~Li, Yuanhan Zhang, Dong Guo, Renrui Zhang, Feng Li, Hao Zhang, Kaichen Zhang, Peiyuan Zhang, Yanwei Li, Ziwei Liu, and Chunyuan Li.
\newblock {LL}a{VA}-onevision: Easy visual task transfer.
\newblock \emph{Transactions on Machine Learning Research}, 2025{\natexlab{a}}.
\newblock ISSN 2835-8856.
\newblock URL \url{https://openreview.net/forum?id=zKv8qULV6n}.

\bibitem[Li et~al.(2024{\natexlab{a}})Li, Sun, Huang, Zhang, Wan, Li, Jin, and Lyu]{li2024ircoco}
Bolun Li, Zhihong Sun, Tao Huang, Hongyu Zhang, Yao Wan, Ge~Li, Zhi Jin, and Chen Lyu.
\newblock Ircoco: Immediate rewards-guided deep reinforcement learning for code completion.
\newblock \emph{Proceedings of the ACM on Software Engineering}, 1\penalty0 (FSE):\penalty0 182--203, 2024{\natexlab{a}}.

\bibitem[Li et~al.(2025{\natexlab{b}})Li, Chen, Shao, Lian, and Liu]{li2025towards}
Chaofan Li, Jianlyu Chen, Yingxia Shao, Defu Lian, and Zheng Liu.
\newblock Towards a generalist code embedding model based on massive data synthesis.
\newblock \emph{arXiv preprint arXiv:2505.12697}, 2025{\natexlab{b}}.

\bibitem[Li et~al.(2023{\natexlab{a}})Li, Liang, Zeng, Chen, Hausman, Sadigh, Levine, Fei-Fei, Xia, and Ichter]{li2023chain}
Chengshu Li, Jacky Liang, Andy Zeng, Xinyun Chen, Karol Hausman, Dorsa Sadigh, Sergey Levine, Li~Fei-Fei, Fei Xia, and Brian Ichter.
\newblock Chain of code: Reasoning with a language model-augmented code emulator.
\newblock \emph{arXiv preprint arXiv:2312.04474}, 2023{\natexlab{a}}.

\bibitem[Li and et~al.(2025)]{li2025sstar}
D.~Li and et~al.
\newblock S*: Test-time scaling for code generation.
\newblock \emph{arXiv preprint arXiv:2502.14382}, 2025.
\newblock URL \url{https://arxiv.org/abs/2502.14382}.

\bibitem[Li et~al.(2025{\natexlab{c}})Li, Cao, Griggs, Liu, Mo, Patil, Zaharia, Gonzalez, and Stoica]{li2025llms}
Dacheng Li, Shiyi Cao, Tyler Griggs, Shu Liu, Xiangxi Mo, Shishir~G Patil, Matei Zaharia, Joseph~E Gonzalez, and Ion Stoica.
\newblock Llms can easily learn to reason from demonstrations structure, not content, is what matters!
\newblock \emph{arXiv preprint arXiv:2502.07374}, 2025{\natexlab{c}}.

\bibitem[Li et~al.(2025{\natexlab{d}})Li, Jiang, Sun, and Zhang]{li2025hybrid}
Fengjie Li, Jiajun Jiang, Jiajun Sun, and Hongyu Zhang.
\newblock Hybrid automated program repair by combining large language models and program analysis.
\newblock \emph{ACM Transactions on Software Engineering and Methodology}, 34\penalty0 (7):\penalty0 1--28, 2025{\natexlab{d}}.

\bibitem[Li et~al.(2024{\natexlab{b}})Li, Dong, Chen, Su, Zhou, Ai, Ye, and Liu]{li2024llms}
Haitao Li, Qian Dong, Junjie Chen, Huixue Su, Yujia Zhou, Qingyao Ai, Ziyi Ye, and Yiqun Liu.
\newblock Llms-as-judges: a comprehensive survey on llm-based evaluation methods.
\newblock \emph{arXiv preprint arXiv:2412.05579}, 2024{\natexlab{b}}.

\bibitem[Li et~al.(2025{\natexlab{e}})Li, Shi, Lin, Gu, Lian, Wang, Jia, Huang, and Wang]{li2025swe}
Han Li, Yuling Shi, Shaoxin Lin, Xiaodong Gu, Heng Lian, Xin Wang, Yantao Jia, Tao Huang, and Qianxiang Wang.
\newblock Swe-debate: Competitive multi-agent debate for software issue resolution.
\newblock \emph{arXiv preprint arXiv:2507.23348}, 2025{\natexlab{e}}.

\bibitem[Li et~al.(2023{\natexlab{b}})Li, Zhang, Li, and Chen]{Li}
Haoyang Li, Jing Zhang, Cuiping Li, and Hong Chen.
\newblock {RESDSQL:} decoupling schema linking and skeleton parsing for text-to-sql.
\newblock In Brian Williams, Yiling Chen, and Jennifer Neville, editors, \emph{Thirty-Seventh {AAAI} Conference on Artificial Intelligence, {AAAI} 2023,}, pages 13067--13075. {AAAI} Press, 2023{\natexlab{b}}.

\bibitem[Li et~al.(2024{\natexlab{c}})Li, Zhang, Liu, Fan, Zhang, Zhu, Wei, Pan, Li, and Chen]{CodeS}
Haoyang Li, Jing Zhang, Hanbing Liu, Ju~Fan, Xiaokang Zhang, Jun Zhu, Renjie Wei, Hongyan Pan, Cuiping Li, and Hong Chen.
\newblock Codes: Towards building open-source language models for text-to-sql.
\newblock \emph{Proc. {ACM} Manag. Data}, 2\penalty0 (3):\penalty0 127, 2024{\natexlab{c}}.

\bibitem[Li et~al.(2025{\natexlab{f}})Li, Wu, Zhang, Huang, Zhang, Jiang, Wang, Zhang, Chen, Shi, Chen, and Li]{OmniSQL}
Haoyang Li, Shang Wu, Xiaokang Zhang, Xinmei Huang, Jing Zhang, Fuxin Jiang, Shuai Wang, Tieying Zhang, Jianjun Chen, Rui Shi, Hong Chen, and Cuiping Li.
\newblock Omnisql: Synthesizing high-quality text-to-sql data at scale.
\newblock \emph{CoRR}, abs/2503.02240, 2025{\natexlab{f}}.

\bibitem[Li et~al.(2025{\natexlab{g}})Li, Tang, Wang, and Guo]{li2025patchpilot}
Hongwei Li, Yuheng Tang, Shiqi Wang, and Wenbo Guo.
\newblock Patchpilot: A stable and cost-efficient agentic patching framework.
\newblock \emph{arXiv e-prints}, pages arXiv--2502, 2025{\natexlab{g}}.

\bibitem[Li et~al.(2024{\natexlab{d}})Li, Beeching, Tunstall, Lipkin, Soletskyi, Huang, Rasul, Yu, Jiang, Shen, et~al.]{li2024numinamath}
Jia Li, Edward Beeching, Lewis Tunstall, Ben Lipkin, Roman Soletskyi, Shengyi Huang, Kashif Rasul, Longhui Yu, Albert~Q Jiang, Ziju Shen, et~al.
\newblock Numinamath: The largest public dataset in ai4maths with 860k pairs of competition math problems and solutions.
\newblock \emph{Hugging Face repository}, 13\penalty0 (9):\penalty0 9, 2024{\natexlab{d}}.

\bibitem[Li et~al.(2024{\natexlab{e}})Li, Li, Zhang, Dong, and Jin]{li2024evocodebenchevolvingcodegeneration}
Jia Li, Ge~Li, Xuanming Zhang, Yihong Dong, and Zhi Jin.
\newblock Evocodebench: An evolving code generation benchmark aligned with real-world code repositories, 2024{\natexlab{e}}.
\newblock URL \url{https://arxiv.org/abs/2404.00599}.

\bibitem[Li et~al.(2024{\natexlab{f}})Li, Li, Zhao, Li, Liu, Zhu, Wang, Liu, Fang, Wang, Ding, Zhang, Zhu, Dong, Jin, Li, Huang, and Li]{li2024devevalmanuallyannotatedcodegeneration}
Jia Li, Ge~Li, Yunfei Zhao, Yongmin Li, Huanyu Liu, Hao Zhu, Lecheng Wang, Kaibo Liu, Zheng Fang, Lanshen Wang, Jiazheng Ding, Xuanming Zhang, Yuqi Zhu, Yihong Dong, Zhi Jin, Binhua Li, Fei Huang, and Yongbin Li.
\newblock Deveval: A manually-annotated code generation benchmark aligned with real-world code repositories, 2024{\natexlab{f}}.
\newblock URL \url{https://arxiv.org/abs/2405.19856}.

\bibitem[Li et~al.(2025{\natexlab{h}})Li, Guo, Li, Zhang, Li, Tao, Liu, Tao, Zhu, and Jin]{li2025longcodeu}
Jia Li, Xuyuan Guo, Lei Li, Kechi Zhang, Ge~Li, Zhengwei Tao, Fang Liu, Chongyang Tao, Yuqi Zhu, and Zhi Jin.
\newblock Longcodeu: Benchmarking long-context language models on long code understanding.
\newblock \emph{arXiv preprint arXiv:2503.04359}, 2025{\natexlab{h}}.

\bibitem[Li et~al.(2025{\natexlab{i}})Li, Li, Li, and Jin]{li2025structured}
Jia Li, Ge~Li, Yongmin Li, and Zhi Jin.
\newblock Structured chain-of-thought prompting for code generation.
\newblock \emph{ACM Transactions on Software Engineering and Methodology}, 34\penalty0 (2):\penalty0 1--23, 2025{\natexlab{i}}.

\bibitem[Li et~al.(2025{\natexlab{j}})Li, Zhu, Liu, Shi, Zong, Dong, Zhang, Jiang, Jin, and Li]{li2025aixcoder}
Jia Li, Hao Zhu, Huanyu Liu, Xianjie Shi, He~Zong, Yihong Dong, Kechi Zhang, Siyuan Jiang, Zhi Jin, and Ge~Li.
\newblock aixcoder-7b-v2: Training llms to fully utilize the long context in repository-level code completion.
\newblock \emph{arXiv preprint arXiv:2503.15301}, 2025{\natexlab{j}}.

\bibitem[Li et~al.(2024{\natexlab{g}})Li, Wang, Guo, Li, Wang, Liu, Zhao, Jiang, Li, Zhang, et~al.]{li2024agentsentinel}
Jia-ju Li, Hong-bo Wang, Jia-yi Guo, Hang Li, Bo-wen Wang, Jia-zheng Liu, X-Y Zhao, Y~Jiang, Ke-fan Li, Y-W Zhang, et~al.
\newblock Agentsentinel: An end-to-end and real-time security defense framework for computer-use agents.
\newblock \emph{arXiv preprint arXiv:2405.11195}, 2024{\natexlab{g}}.

\bibitem[Li et~al.(2025{\natexlab{k}})Li, Li, Li, Chang, and Wu]{fpbench}
Jialin Li, Jinzhe Li, Gengxu Li, Yi~Chang, and Yuan Wu.
\newblock Refining critical thinking in llm code generation: A faulty premise-based evaluation framework, 2025{\natexlab{k}}.
\newblock URL \url{https://arxiv.org/abs/2508.03622}.

\bibitem[Li et~al.(2025{\natexlab{l}})Li, Li, Gao, Shi, Li, Wang, Huang, Wang, Wang, Han, Liu, and Sun]{tritonbench}
Jianling Li, Shangzhan Li, Zhenye Gao, Qi~Shi, Yuxuan Li, Zefan Wang, Jiacheng Huang, Haojie Wang, Jianrong Wang, Xu~Han, Zhiyuan Liu, and Maosong Sun.
\newblock Tritonbench: Benchmarking large language model capabilities for generating triton operators, 2025{\natexlab{l}}.
\newblock URL \url{https://arxiv.org/abs/2502.14752}.

\bibitem[Li et~al.(2023{\natexlab{c}})Li, Hui, Cheng, Qin, Ma, Huo, Huang, Du, Si, and Li]{Jinyang}
Jinyang Li, Binyuan Hui, Reynold Cheng, Bowen Qin, Chenhao Ma, Nan Huo, Fei Huang, Wenyu Du, Luo Si, and Yongbin Li.
\newblock Graphix-t5: Mixing pre-trained transformers with graph-aware layers for text-to-sql parsing.
\newblock In Brian Williams, Yiling Chen, and Jennifer Neville, editors, \emph{Thirty-Seventh {AAAI} Conference on Artificial Intelligence, {AAAI} 2023,}, pages 13076--13084. {AAAI} Press, 2023{\natexlab{c}}.

\bibitem[Li et~al.(2024{\natexlab{h}})Li, Hui, Qu, Yang, Li, Li, Wang, Qin, Geng, Huo, et~al.]{li2024can}
Jinyang Li, Binyuan Hui, Ge~Qu, Jiaxi Yang, Binhua Li, Bowen Li, Bailin Wang, Bowen Qin, Ruiying Geng, Nan Huo, et~al.
\newblock Can llm already serve as a database interface? a big bench for large-scale database grounded text-to-sqls.
\newblock \emph{Advances in Neural Information Processing Systems}, 36, 2024{\natexlab{h}}.

\bibitem[Li et~al.(2025{\natexlab{m}})Li, Li, Qu, Jacobsson, Qin, Hui, Si, Huo, Xu, Zhang, Tang, Li, Widjaja, Zhu, Zhou, Huang, Papakonstantinou, Ozcan, Ma, and Cheng]{li2025swesqlilluminatingllmpathways}
Jinyang Li, Xiaolong Li, Ge~Qu, Per Jacobsson, Bowen Qin, Binyuan Hui, Shuzheng Si, Nan Huo, Xiaohan Xu, Yue Zhang, Ziwei Tang, Yuanshuai Li, Florensia Widjaja, Xintong Zhu, Feige Zhou, Yongfeng Huang, Yannis Papakonstantinou, Fatma Ozcan, Chenhao Ma, and Reynold Cheng.
\newblock Swe-sql: Illuminating llm pathways to solve user sql issues in real-world applications, 2025{\natexlab{m}}.
\newblock URL \url{https://arxiv.org/abs/2506.18951}.

\bibitem[Li et~al.(2022{\natexlab{a}})Li, Li, Xiong, and Hoi]{li2022blip}
Junnan Li, Dongxu Li, Caiming Xiong, and Steven Hoi.
\newblock Blip: Bootstrapping language-image pre-training for unified vision-language understanding and generation, 2022{\natexlab{a}}.
\newblock URL \url{https://arxiv.org/abs/2201.12086}.

\bibitem[Li et~al.(2023{\natexlab{d}})Li, Li, Savarese, and Hoi]{li2023blip2}
Junnan Li, Dongxu Li, Silvio Savarese, and Steven Hoi.
\newblock Blip-2: Bootstrapping language-image pre-training with frozen image encoders and large language models, 2023{\natexlab{d}}.
\newblock URL \url{https://arxiv.org/abs/2301.12597}.

\bibitem[Li et~al.(2024{\natexlab{i}})Li, Tian, Hu, Luo, Huang, and Ma]{li2024mmcode}
Kaixin Li, Yuchen Tian, Qisheng Hu, Ziyang Luo, Zhiyong Huang, and Jing Ma.
\newblock Mmcode: Benchmarking multimodal large language models for code generation with visually rich programming problems.
\newblock \emph{arXiv preprint arXiv:2404.09486}, 2024{\natexlab{i}}.

\bibitem[Li et~al.(2024{\natexlab{j}})Li, Chai, Wang, Sun, Tian, Zhang, and Wu]{li2024toolaugmentedrewardmodeling}
Lei Li, Yekun Chai, Shuohuan Wang, Yu~Sun, Hao Tian, Ningyu Zhang, and Hua Wu.
\newblock Tool-augmented reward modeling, 2024{\natexlab{j}}.
\newblock URL \url{https://arxiv.org/abs/2310.01045}.

\bibitem[Li et~al.(2022{\natexlab{b}})Li, Yang, Jiang, Yan, Luo, Hua, Liang, and Zuo]{li2022auger}
Lingwei Li, Li~Yang, Huaxi Jiang, Jun Yan, Tiejian Luo, Zihan Hua, Geng Liang, and Chun Zuo.
\newblock Auger: automatically generating review comments with pre-training models.
\newblock In \emph{Proceedings of the 30th ACM Joint European Software Engineering Conference and Symposium on the Foundations of Software Engineering}, pages 1009--1021, 2022{\natexlab{b}}.

\bibitem[Li et~al.(2023{\natexlab{e}})Li, Zhao, Yu, Song, Li, Yu, Li, Huang, and Li]{li2023api}
Minghao Li, Yingxiu Zhao, Bowen Yu, Feifan Song, Hangyu Li, Haiyang Yu, Zhoujun Li, Fei Huang, and Yongbin Li.
\newblock Api-bank: A comprehensive benchmark for tool-augmented llms.
\newblock \emph{arXiv preprint arXiv:2304.08244}, 2023{\natexlab{e}}.

\bibitem[Li et~al.(2023{\natexlab{f}})Li, allal, Zi, Muennighoff, Kocetkov, Mou, Marone, Akiki, LI, Chim, Liu, Zheltonozhskii, Zhuo, Wang, Dehaene, Lamy-Poirier, Monteiro, Gontier, Yee, Umapathi, Zhu, Lipkin, Oblokulov, Wang, Murthy, Stillerman, Patel, Abulkhanov, Zocca, Dey, Zhang, Bhattacharyya, Yu, Luccioni, Villegas, Zhdanov, Lee, Timor, Ding, Schlesinger, Schoelkopf, Ebert, Dao, Mishra, Gu, Anderson, Dolan-Gavitt, Contractor, Reddy, Fried, Bahdanau, Jernite, Ferrandis, Hughes, Wolf, Guha, Werra, and de~Vries]{li2023starcoder}
Raymond Li, Loubna~Ben allal, Yangtian Zi, Niklas Muennighoff, Denis Kocetkov, Chenghao Mou, Marc Marone, Christopher Akiki, Jia LI, Jenny Chim, Qian Liu, Evgenii Zheltonozhskii, Terry~Yue Zhuo, Thomas Wang, Olivier Dehaene, Joel Lamy-Poirier, Joao Monteiro, Nicolas Gontier, Ming-Ho Yee, Logesh~Kumar Umapathi, Jian Zhu, Ben Lipkin, Muhtasham Oblokulov, Zhiruo Wang, Rudra Murthy, Jason~T Stillerman, Siva~Sankalp Patel, Dmitry Abulkhanov, Marco Zocca, Manan Dey, Zhihan Zhang, Urvashi Bhattacharyya, Wenhao Yu, Sasha Luccioni, Paulo Villegas, Fedor Zhdanov, Tony Lee, Nadav Timor, Jennifer Ding, Claire~S Schlesinger, Hailey Schoelkopf, Jan Ebert, Tri Dao, Mayank Mishra, Alex Gu, Carolyn~Jane Anderson, Brendan Dolan-Gavitt, Danish Contractor, Siva Reddy, Daniel Fried, Dzmitry Bahdanau, Yacine Jernite, Carlos~Mu{\~n}oz Ferrandis, Sean Hughes, Thomas Wolf, Arjun Guha, Leandro~Von Werra, and Harm de~Vries.
\newblock Starcoder: may the source be with you!
\newblock \emph{Transactions on Machine Learning Research}, 2023{\natexlab{f}}.
\newblock ISSN 2835-8856.
\newblock URL \url{https://openreview.net/forum?id=KoFOg41haE}.
\newblock Reproducibility Certification.

\bibitem[Li et~al.(2023{\natexlab{g}})Li, Fu, Zhang, Huang, Sun, Lyu, Liu, Jin, and Li]{li2023taco}
Rongao Li, Jie Fu, Bo-Wen Zhang, Tao Huang, Zhihong Sun, Chen Lyu, Guang Liu, Zhi Jin, and Ge~Li.
\newblock Taco: Topics in algorithmic code generation dataset.
\newblock \emph{arXiv preprint arXiv:2312.14852}, 2023{\natexlab{g}}.

\bibitem[Li et~al.(2024{\natexlab{k}})Li, Zhang, and Yang]{li2024sketch2code}
Ryan Li, Yanzhe Zhang, and Diyi Yang.
\newblock Sketch2code: Evaluating vision-language models for interactive web design prototyping.
\newblock \emph{arXiv preprint arXiv:2410.16232}, 2024{\natexlab{k}}.

\bibitem[Li et~al.(2021)Li, Liu, Bian, Fang, Huang, Liu, Wang, and You]{li2021colossalai}
Shenggui Li, Hongxin Liu, Zhengda Bian, Jiarui Fang, Haichen Huang, Yuliang Liu, Boxiang Wang, and Yang You.
\newblock Colossal-ai: A unified deep learning system for large-scale parallel training.
\newblock In \emph{Proceedings of the 51st International Conference on Parallel Processing}, pages 1--10, 2021.

\bibitem[Li et~al.(2018)Li, Ma, and Sun]{DBLP:journals/tjs/LiMS18}
Teng Li, Jianfeng Ma, and Cong Sun.
\newblock Dlog: diagnosing router events with syslogs for anomaly detection.
\newblock \emph{J. Supercomput.}, 74\penalty0 (2):\penalty0 845--867, 2018.

\bibitem[Li et~al.(2025{\natexlab{n}})Li, Zhang, Guo, Mao, Luo, Peng, Huang, Wang, and Li]{li2025fea}
Wei Li, Xin Zhang, Zhongxin Guo, Shaoguang Mao, Wen Luo, Guangyue Peng, Yangyu Huang, Houfeng Wang, and Scarlett Li.
\newblock Fea-bench: A benchmark for evaluating repository-level code generation for feature implementation.
\newblock \emph{arXiv preprint arXiv:2503.06680}, 2025{\natexlab{n}}.

\bibitem[Li et~al.(2022{\natexlab{c}})Li, Thickstun, Gulrajani, Liang, and Hashimoto]{li2022diffusionlm}
Xiang~Lisa Li, John Thickstun, Ishaan Gulrajani, Percy Liang, and Tatsunori~B. Hashimoto.
\newblock Diffusion-lm improves controllable text generation, 2022{\natexlab{c}}.
\newblock URL \url{https://arxiv.org/abs/2205.14217}.

\bibitem[Li et~al.(2025{\natexlab{o}})Li, Li, Dong, Zhang, Ruan, Dai, Liu, Xu, Wang, and Tang]{hlce}
Xiangyang Li, Xiaopeng Li, Kuicai Dong, Quanhu Zhang, Rongju Ruan, Xinyi Dai, Xiaoshuang Liu, Shengchun Xu, Yasheng Wang, and Ruiming Tang.
\newblock Humanity's last code exam: Can advanced llms conquer human's hardest code competition?, 2025{\natexlab{o}}.
\newblock URL \url{https://arxiv.org/abs/2506.12713}.

\bibitem[Li et~al.(2025{\natexlab{p}})Li, Emad, Padthe, Lanchantin, Yuan, Nguyen, Weston, Li, Wang, Kulikov, et~al.]{li2025naturalthoughts}
Yang Li, Youssef Emad, Karthik Padthe, Jack Lanchantin, Weizhe Yuan, Thao Nguyen, Jason Weston, Shang-Wen Li, Dong Wang, Ilia Kulikov, et~al.
\newblock Naturalthoughts: Selecting and distilling reasoning traces for general reasoning tasks.
\newblock \emph{arXiv preprint arXiv:2507.01921}, 2025{\natexlab{p}}.

\bibitem[Li et~al.(2022{\natexlab{d}})Li, Wang, and Nguyen]{li2022dear}
Yi~Li, Shaohua Wang, and Tien~N Nguyen.
\newblock Dear: A novel deep learning-based approach for automated program repair.
\newblock In \emph{Proceedings of the 44th international conference on software engineering}, pages 511--523, 2022{\natexlab{d}}.

\bibitem[Li et~al.(2025{\natexlab{q}})Li, Gu, Wen, Li, Xing, Guo, Zheng, Zhou, Qu, Zhou, Zhang, Shen, Liu, Lin, Yang, Zhang, and Huang]{li2025treepo}
Yizhi Li, Qingshui Gu, Zhoufutu Wen, Ziniu Li, Tianshun Xing, Shuyue Guo, Tianyu Zheng, Xin Zhou, Xingwei Qu, Wangchunshu Zhou, Zheng Zhang, Wei Shen, Qian Liu, Chenghua Lin, Jian Yang, Ge~Zhang, and Wenhao Huang.
\newblock Treepo: Bridging the gap of policy optimization and efficacy and inference efficiency with heuristic tree-based modeling, 2025{\natexlab{q}}.
\newblock URL \url{https://arxiv.org/abs/2508.17445}.

\bibitem[Li et~al.(2023{\natexlab{h}})Li, Bubeck, Eldan, Del~Giorno, Gunasekar, and Lee]{phi15}
Yuanzhi Li, S{\'e}bastien Bubeck, Ronen Eldan, Allie Del~Giorno, Suriya Gunasekar, and Yin~Tat Lee.
\newblock Textbooks are all you need ii: phi-1.5 technical report.
\newblock \emph{arXiv preprint arXiv:2309.05463}, 2023{\natexlab{h}}.

\bibitem[Li et~al.(2025{\natexlab{r}})Li, Zhang, Lv, Liu, Deng, Zhang, Liu, Zhou, and Zhou]{li2025relook}
Yuhang Li, Chenchen Zhang, Ruilin Lv, Ao~Liu, Ken Deng, Yuanxing Zhang, Jiaheng Liu, Wiggin Zhou, and Bo~Zhou.
\newblock Relook: Vision-grounded rl with a multimodal llm critic for agentic web coding.
\newblock \emph{arXiv preprint arXiv:2510.11498}, 2025{\natexlab{r}}.

\bibitem[Li et~al.(2022{\natexlab{e}})Li, Choi, Chung, Kushman, Schrittwieser, Leblond, Eccles, Keeling, Gimeno, Dal~Lago, Hubert, Choy, de~Masson~d'Autume, Babuschkin, Chen, Huang, Welbl, Gowal, Cherepanov, Molloy, Mankowitz, Sutherland~Robson, Kohli, de~Freitas, Kavukcuoglu, and Vinyals]{codecontests}
Yujia Li, David Choi, Junyoung Chung, Nate Kushman, Julian Schrittwieser, R{\'e}mi Leblond, Tom Eccles, James Keeling, Felix Gimeno, Agustin Dal~Lago, Thomas Hubert, Peter Choy, Cyprien de~Masson~d'Autume, Igor Babuschkin, Xinyun Chen, Po-Sen Huang, Johannes Welbl, Sven Gowal, Alexey Cherepanov, James Molloy, Daniel Mankowitz, Esme Sutherland~Robson, Pushmeet Kohli, Nando de~Freitas, Koray Kavukcuoglu, and Oriol Vinyals.
\newblock Competition-level code generation with alphacode.
\newblock \emph{arXiv preprint arXiv:2203.07814}, 2022{\natexlab{e}}.

\bibitem[Li et~al.(2022{\natexlab{f}})Li, Choi, Chung, Kushman, Schrittwieser, Leblond, Eccles, Keeling, Gimeno, Dal~Lago, et~al.]{li2022competition}
Yujia Li, David Choi, Junyoung Chung, Nate Kushman, Julian Schrittwieser, R{\'e}mi Leblond, Tom Eccles, James Keeling, Felix Gimeno, Agustin Dal~Lago, et~al.
\newblock Competition-level code generation with alphacode.
\newblock \emph{Science}, 378\penalty0 (6624):\penalty0 1092--1097, 2022{\natexlab{f}}.

\bibitem[Li et~al.(2022{\natexlab{g}})Li, Choi, Chung, Kushman, Schrittwieser, Leblond, Eccles, Keeling, Gimeno, Lago, Hubert, Choy, de~Masson~d’Autume, Babuschkin, Chen, Huang, Welbl, Gowal, Cherepanov, Molloy, Mankowitz, Robson, Kohli, de~Freitas, Kavukcuoglu, and Vinyals]{li2022alphacode}
Yujia Li, David Choi, Junyoung Chung, Nate Kushman, Julian Schrittwieser, Rémi Leblond, Tom Eccles, James Keeling, Felix Gimeno, Agustin~Dal Lago, Thomas Hubert, Peter Choy, Cyprien de~Masson~d’Autume, Igor Babuschkin, Xinyun Chen, Po-Sen Huang, Johannes Welbl, Sven Gowal, Alexey Cherepanov, James Molloy, Daniel~J. Mankowitz, Esme~Sutherland Robson, Pushmeet Kohli, Nando de~Freitas, Koray Kavukcuoglu, and Oriol Vinyals.
\newblock Competition-level code generation with alphacode.
\newblock \emph{Science}, 378\penalty0 (6624):\penalty0 1092--1097, 2022{\natexlab{g}}.
\newblock \doi{10.1126/science.abq1158}.
\newblock URL \url{https://www.science.org/doi/abs/10.1126/science.abq1158}.

\bibitem[Li et~al.(2023{\natexlab{i}})Li, Liu, Wang, Yuan, Tian, Jin, and Yan]{li2023lost}
Zhi Li, Weijie Liu, XiaoFeng Wang, Bin Yuan, Hongliang Tian, Hai Jin, and Shoumeng Yan.
\newblock Lost along the way: Understanding and mitigating path-misresolution threats to container isolation.
\newblock In \emph{Proceedings of the 2023 ACM SIGSAC Conference on Computer and Communications Security}, pages 3063--3077, 2023{\natexlab{i}}.

\bibitem[Li et~al.(2022{\natexlab{h}})Li, Lu, Guo, Duan, Jannu, Jenks, Majumder, Green, Svyatkovskiy, Fu, et~al.]{li2022codereviewer}
Zhiyu Li, Shuai Lu, Daya Guo, Nan Duan, Shailesh Jannu, Grant Jenks, Deep Majumder, Jared Green, Alexey Svyatkovskiy, Shengyu Fu, et~al.
\newblock Codereviewer: Pre-training for automating code review activities.
\newblock \emph{arXiv preprint arXiv:2203.09095}, 2022{\natexlab{h}}.

\bibitem[Li et~al.(2024{\natexlab{l}})Li, Wu, Li, He, Fang, Zhang, Zhao, Li, Li, and Song]{iswc2024}
Zhongqiu Li, Zhenhe Wu, Mengxiang Li, Zhongjiang He, Ruiyu Fang, Jie Zhang, Yu~Zhao, Yongxiang Li, Zhoujun Li, and Shuangyong Song.
\newblock Scalable database-driven kgs can help text-to-sql.
\newblock In \emph{Proceedings of the {ISWC} 2024 Posters, Demos and Industry Tracks:}, volume 3828 of \emph{{CEUR} Workshop Proceedings}, 2024{\natexlab{l}}.

\bibitem[Li et~al.(2024{\natexlab{m}})Li, Xu, Zhang, Lin, Yu, Sun, and Luo]{li2024remax}
Ziniu Li, Tian Xu, Yushun Zhang, Zhihang Lin, Yang Yu, Ruoyu Sun, and Zhi-Quan Luo.
\newblock Remax: A simple, effective, and efficient reinforcement learning method for aligning large language models, 2024{\natexlab{m}}.
\newblock URL \url{https://arxiv.org/abs/2310.10505}.

\bibitem[Li et~al.(2025{\natexlab{s}})Li, Wang, Xu, Ding, Sun, and Yu]{li2025reviewRLllm}
Ziniu Li, Pengyuan Wang, Tian Xu, Tian Ding, Ruoyu Sun, and Yang Yu.
\newblock Review of reinforcement learning for large language models: Formulations, algorithms, and opportunities.
\newblock 2025{\natexlab{s}}.

\bibitem[Liang et~al.(2022)Liang, Huang, Xia, Xu, Hausman, Ichter, Florence, and Zeng]{liang2022code}
Jacky Liang, Wenlong Huang, Fei Xia, Peng Xu, Karol Hausman, Brian Ichter, Pete Florence, and Andy Zeng.
\newblock Code as policies: Language model programs for embodied control.
\newblock \emph{arXiv preprint arXiv:2209.07753}, 2022.

\bibitem[Liang et~al.(2021)Liang, Cao, Hu, and Chen]{DBLP:journals/cybersec/LiangCHC21}
Ruigang Liang, Ying Cao, Peiwei Hu, and Kai Chen.
\newblock Neutron: an attention-based neural decompiler.
\newblock \emph{Cybersecur.}, 4\penalty0 (1):\penalty0 5, 2021.

\bibitem[Liang et~al.(2025{\natexlab{a}})Liang, Hu, Jiang, and Tan]{liang2025languagemodelsreplaceprogrammers}
Shanchao Liang, Yiran Hu, Nan Jiang, and Lin Tan.
\newblock Can language models replace programmers for coding? repocod says 'not yet', 2025{\natexlab{a}}.
\newblock URL \url{https://arxiv.org/abs/2410.21647}.

\bibitem[Liang et~al.(2025{\natexlab{b}})Liang, Liu, Wright, Constable, Gu, Huang, Zhang, Feng, Huang, Wang, et~al.]{liang2025torchtitan}
Wanchao Liang, Tianyu Liu, Less Wright, Will Constable, Andrew Gu, Chien-Chin Huang, Iris Zhang, Wei Feng, Howard Huang, Junjie Wang, et~al.
\newblock Torchtitan: One-stop pytorch native solution for production ready llm pretraining.
\newblock \emph{arXiv preprint arXiv:2410.06511}, 2025{\natexlab{b}}.

\bibitem[Liang et~al.(2025{\natexlab{c}})Liang, Xiang, Yu, Zhang, Hong, Fan, and Tang]{openmanus2025}
Xinbin Liang, Jinyu Xiang, Zhaoyang Yu, Jiayi Zhang, Sirui Hong, Sheng Fan, and Xiao Tang.
\newblock Openmanus: An open-source framework for building general ai agents, 2025{\natexlab{c}}.
\newblock URL \url{https://doi.org/10.5281/zenodo.15186407}.

\bibitem[Liao and et~al.(2024)]{liao2024codeagent}
Z.~Liao and et~al.
\newblock Codeagent: Repository-level coding agents with tool-augmented llms.
\newblock \emph{arXiv preprint arXiv:2407.03178}, 2024.
\newblock URL \url{https://arxiv.org/abs/2407.03178}.

\bibitem[Lieber et~al.(2024)Lieber, Lenz, Bata, Cohen, Osin, Dalmedigos, Safahi, Meirom, Belinkov, Shalev-Shwartz, et~al.]{lieber2024jamba}
Opher Lieber, Barak Lenz, Hofit Bata, Gal Cohen, Jhonathan Osin, Itay Dalmedigos, Erez Safahi, Shaked Meirom, Yonatan Belinkov, Shai Shalev-Shwartz, et~al.
\newblock Jamba: A hybrid transformer-mamba language model.
\newblock \emph{arXiv preprint arXiv:2403.19887}, 2024.

\bibitem[{Light} et~al.(2023){Light}, {Cai}, {Shen}, and {Hu}]{AvalonBench}
Jonathan {Light}, Min {Cai}, Sheng {Shen}, and Ziniu {Hu}.
\newblock {AvalonBench: Evaluating LLMs Playing the Game of Avalon}.
\newblock \emph{arXiv e-prints}, art. arXiv:2310.05036, October 2023.
\newblock \doi{10.48550/arXiv.2310.05036}.

\bibitem[Lin et~al.(2024{\natexlab{a}})Lin, Wang, Wen, Chen, and Mao]{lin2024one}
Bo~Lin, Shangwen Wang, Ming Wen, Liqian Chen, and Xiaoguang Mao.
\newblock One size does not fit all: Multi-granularity patch generation for better automated program repair.
\newblock In \emph{Proceedings of the 33rd ACM SIGSOFT International Symposium on Software Testing and Analysis}, pages 1554--1566, 2024{\natexlab{a}}.

\bibitem[Lin et~al.(2024{\natexlab{b}})Lin, Kim, and Chen]{lin2024whenllm}
Feng Lin, Dong~Jae Kim, and Tse-Hsun Chen.
\newblock When llm-based code generation meets the software development process.
\newblock \emph{CoRR abs/2403.15852}, 2024{\natexlab{b}}.
\newblock URL \url{https://arxiv.org/abs/2403.15852}.

\bibitem[Lin et~al.(2025{\natexlab{a}})Lin, Guo, Han, Hu, Ni, Wang, Chen, Jiang, Jiao, Hu, et~al.]{lin2025se}
Jiaye Lin, Yifu Guo, Yuzhen Han, Sen Hu, Ziyi Ni, Licheng Wang, Mingguang Chen, Daxin Jiang, Binxing Jiao, Chen Hu, et~al.
\newblock Se-agent: Self-evolution trajectory optimization in multi-step reasoning with llm-based agents.
\newblock \emph{arXiv preprint arXiv:2508.02085}, 2025{\natexlab{a}}.

\bibitem[Lin et~al.(2025{\natexlab{b}})Lin, Zhou, Zhao, Wan, Ma, Gao, and Li]{lin2025webuibench}
Zhiyu Lin, Zhengda Zhou, Zhiyuan Zhao, Tianrui Wan, Yilun Ma, Junyu Gao, and Xuelong Li.
\newblock {W}eb{UIB}ench: A comprehensive benchmark for evaluating multimodal large language models in {W}eb{UI}-to-code.
\newblock In Wanxiang Che, Joyce Nabende, Ekaterina Shutova, and Mohammad~Taher Pilehvar, editors, \emph{Findings of the Association for Computational Linguistics: ACL 2025}, pages 15780--15797, Vienna, Austria, July 2025{\natexlab{b}}. Association for Computational Linguistics.
\newblock ISBN 979-8-89176-256-5.
\newblock \doi{10.18653/v1/2025.findings-acl.815}.
\newblock URL \url{https://aclanthology.org/2025.findings-acl.815/}.

\bibitem[Lindenbauer et~al.(2025)Lindenbauer, Bogomolov, and Zharov]{lindenbauer2025gitgoodbenchnovelbenchmarkevaluating}
Tobias Lindenbauer, Egor Bogomolov, and Yaroslav Zharov.
\newblock Gitgoodbench: A novel benchmark for evaluating agentic performance on git, 2025.
\newblock URL \url{https://arxiv.org/abs/2505.22583}.

\bibitem[Ling(2024)]{ling2024evaluating}
Lin Ling.
\newblock Evaluating social bias in code generation models.
\newblock In \emph{Companion Proceedings of the 32nd ACM International Conference on the Foundations of Software Engineering}, pages 695--697, 2024.

\bibitem[Liu et~al.(2025{\natexlab{a}})Liu, Li, Zhang, Wang, He, Hong, Liu, Zhang, Song, Zhu, et~al.]{liu2025advances}
Bang Liu, Xinfeng Li, Jiayi Zhang, Jinlin Wang, Tanjin He, Sirui Hong, Hongzhang Liu, Shaokun Zhang, Kaitao Song, Kunlun Zhu, et~al.
\newblock Advances and challenges in foundation agents: From brain-inspired intelligence to evolutionary, collaborative, and safe systems.
\newblock \emph{arXiv preprint arXiv:2504.01990}, 2025{\natexlab{a}}.

\bibitem[Liu et~al.(2024{\natexlab{a}})Liu, Chen, Gong, Liao, Wang, Lei, Liang, Chen, Shen, Zhou, Jiang, Yu, and Li]{MFTCoder}
Bingchang Liu, Chaoyu Chen, Zi~Gong, Cong Liao, Huan Wang, Zhichao Lei, Ming Liang, Dajun Chen, Min Shen, Hailian Zhou, Wei Jiang, Hang Yu, and Jianguo Li.
\newblock Mftcoder: Boosting code llms with multitask fine-tuning.
\newblock In \emph{Proceedings of the 30th ACM SIGKDD Conference on Knowledge Discovery and Data Mining}, KDD '24, page 5430–5441, New York, NY, USA, 2024{\natexlab{a}}. Association for Computing Machinery.
\newblock ISBN 9798400704901.
\newblock \doi{10.1145/3637528.3671609}.
\newblock URL \url{https://doi.org/10.1145/3637528.3671609}.

\bibitem[Liu and Wan(2021)]{liu2021codeqa}
Chenxiao Liu and Xiaojun Wan.
\newblock Codeqa: A question answering dataset for source code comprehension.
\newblock \emph{arXiv preprint arXiv:2109.08365}, 2021.

\bibitem[Liu et~al.(2023{\natexlab{a}})Liu, Lu, Chen, Jiang, Svyatkovskiy, Fu, Sundaresan, and Duan]{code_executor}
Chenxiao Liu, Shuai Lu, Weizhu Chen, Daxin Jiang, Alexey Svyatkovskiy, Shengyu Fu, Neel Sundaresan, and Nan Duan.
\newblock Code execution with pre-trained language models.
\newblock In Anna Rogers, Jordan~L. Boyd{-}Graber, and Naoaki Okazaki, editors, \emph{Findings of the Association for Computational Linguistics: {ACL} 2023, Toronto, Canada, July 9-14, 2023}, pages 4984--4999. Association for Computational Linguistics, 2023{\natexlab{a}}.
\newblock \doi{10.18653/V1/2023.FINDINGS-ACL.308}.
\newblock URL \url{https://doi.org/10.18653/v1/2023.findings-acl.308}.

\bibitem[Liu et~al.(2024{\natexlab{b}})Liu, Cai, Lin, Huang, Pei, Jiang, Yang, Dong, and Mei]{liu2024coedpilot}
Chenyan Liu, Yufan Cai, Yun Lin, Yuhuan Huang, Yunrui Pei, Bo~Jiang, Ping Yang, Jin~Song Dong, and Hong Mei.
\newblock Coedpilot: Recommending code edits with learned prior edit relevance, project-wise awareness, and interactive nature.
\newblock In \emph{Proceedings of the 33rd ACM SIGSOFT International Symposium on Software Testing and Analysis}, pages 466--478, 2024{\natexlab{b}}.

\bibitem[Liu et~al.(2025{\natexlab{b}})Liu, Lin, and Thongtanunam]{liu2025too}
Chunhua Liu, Hong~Yi Lin, and Patanamon Thongtanunam.
\newblock Too noisy to learn: Enhancing data quality for code review comment generation.
\newblock In \emph{2025 IEEE/ACM 22nd International Conference on Mining Software Repositories (MSR)}, pages 236--248. IEEE, 2025{\natexlab{b}}.

\bibitem[Liu et~al.(2018{\natexlab{a}})Liu, Guu, Pasupat, Shi, and Liang]{liu2018reinforcement}
Evan~Zheran Liu, Kelvin Guu, Panupong Pasupat, Tianlin Shi, and Percy Liang.
\newblock Reinforcement learning on web interfaces using workflow-guided exploration.
\newblock \emph{arXiv preprint arXiv:1802.08802}, 2018{\natexlab{a}}.

\bibitem[Liu et~al.(2024{\natexlab{c}})Liu, Niu, Zhang, Shang, Xiao, Mao, Jiang, Yuan, Gui, Sun, et~al.]{liu2023pku}
Hao Liu, Ruitian Niu, Hongbo Zhang, Junchi Shang, Zhaoran Xiao, Xinbei Mao, Yichen Jiang, Hao Yuan, Tao Gui, Qi~Sun, et~al.
\newblock Pku-saferlhf: Towards multi-level safety alignment for llms with human preference.
\newblock \emph{arXiv preprint arXiv:2406.15513}, 2024{\natexlab{c}}.

\bibitem[Liu et~al.(2023{\natexlab{b}})Liu, Li, Wu, and Lee]{liu2023llava}
Haotian Liu, Chunyuan Li, Qingyang Wu, and Yong~Jae Lee.
\newblock Visual instruction tuning, 2023{\natexlab{b}}.
\newblock URL \url{https://arxiv.org/abs/2304.08485}.

\bibitem[Liu et~al.(2024{\natexlab{d}})Liu, Li, Li, and Lee]{liu2023llava1_5}
Haotian Liu, Chunyuan Li, Yuheng Li, and Yong~Jae Lee.
\newblock Improved baselines with visual instruction tuning, 2024{\natexlab{d}}.
\newblock URL \url{https://arxiv.org/abs/2310.03744}.

\bibitem[Liu et~al.(2024{\natexlab{e}})Liu, Deng, Liu, Yang, Liu, Zhu, Zhao, Chai, Wu, Jin, Zhang, Wang, Zhang, Xiang, Su, and Zheng]{liu2024m2rcevalmassivelymultilingualrepositorylevel}
Jiaheng Liu, Ken Deng, Congnan Liu, Jian Yang, Shukai Liu, He~Zhu, Peng Zhao, Linzheng Chai, Yanan Wu, Ke~Jin, Ge~Zhang, Zekun Wang, Guoan Zhang, Bangyu Xiang, Wenbo Su, and Bo~Zheng.
\newblock M2rc-eval: Massively multilingual repository-level code completion evaluation, 2024{\natexlab{e}}.
\newblock URL \url{https://arxiv.org/abs/2410.21157}.

\bibitem[Liu and Zhang(2025)]{coder1}
Jiawei Liu and Lingming Zhang.
\newblock Code-r1: Reproducing r1 for code with reliable rewards.
\newblock 2025.

\bibitem[Liu et~al.(2023{\natexlab{c}})Liu, Xia, Wang, and Zhang]{evalplus}
Jiawei Liu, Chunqiu~Steven Xia, Yuyao Wang, and Lingming Zhang.
\newblock Is your code generated by chat{GPT} really correct? rigorous evaluation of large language models for code generation.
\newblock In \emph{Thirty-seventh Conference on Neural Information Processing Systems}, 2023{\natexlab{c}}.
\newblock URL \url{https://openreview.net/forum?id=1qvx610Cu7}.

\bibitem[Liu et~al.(2024{\natexlab{f}})Liu, Nguyen, Shang, Ding, Li, Yu, Kumar, and Wang]{liu2024codefavor}
Jiawei Liu, Thanh Nguyen, Mingyue Shang, Hantian Ding, Xiaopeng Li, Yu~Yu, Varun Kumar, and Zijian Wang.
\newblock Learning code preference via synthetic evolution.
\newblock \emph{arXiv preprint arXiv:2410.03837}, 2024{\natexlab{f}}.

\bibitem[Liu et~al.(2024{\natexlab{g}})Liu, Tian, Daita, Wei, Ding, Wang, Yang, and Zhang]{liu2024repoqa}
Jiawei Liu, Jia~Le Tian, Vijay Daita, Yuxiang Wei, Yifeng Ding, Yuhan~Katherine Wang, Jun Yang, and Lingming Zhang.
\newblock Repoqa: Evaluating long context code understanding.
\newblock \emph{arXiv preprint arXiv:2406.06025}, 2024{\natexlab{g}}.

\bibitem[Liu et~al.(2024{\natexlab{h}})Liu, Xie, Wang, Wei, Ding, and Zhang]{liu2024evaluatinglanguagemodelsefficient}
Jiawei Liu, Songrun Xie, Junhao Wang, Yuxiang Wei, Yifeng Ding, and Lingming Zhang.
\newblock Evaluating language models for efficient code generation, 2024{\natexlab{h}}.
\newblock URL \url{https://arxiv.org/abs/2408.06450}.

\bibitem[Liu et~al.(2025{\natexlab{c}})Liu, Pan, Xiang, He, Li, Du, and Gao]{liu2025projectevalbenchmarkprogrammingagents}
Kaiyuan Liu, Youcheng Pan, Yang Xiang, Daojing He, Jing Li, Yexing Du, and Tianrun Gao.
\newblock Projecteval: A benchmark for programming agents automated evaluation on project-level code generation, 2025{\natexlab{c}}.
\newblock URL \url{https://arxiv.org/abs/2503.07010}.

\bibitem[Liu et~al.(2023{\natexlab{d}})Liu, Pinckney, Khailany, and Ren]{verilog_eval}
Mingjie Liu, Nathaniel~Ross Pinckney, Brucek Khailany, and Haoxing Ren.
\newblock Invited paper: Verilogeval: Evaluating large language models for verilog code generation.
\newblock In \emph{{IEEE/ACM} International Conference on Computer Aided Design, {ICCAD} 2023, San Francisco, CA, USA, October 28 - Nov. 2, 2023}, pages 1--8. {IEEE}, 2023{\natexlab{d}}.
\newblock \doi{10.1109/ICCAD57390.2023.10323812}.
\newblock URL \url{https://doi.org/10.1109/ICCAD57390.2023.10323812}.

\bibitem[Liu et~al.(2024{\natexlab{i}})Liu, Lin, Hewitt, Paranjape, Bevilacqua, Petroni, and Liang]{liu2024lostinthemiddle}
Nelson~F. Liu, Kevin Lin, John Hewitt, Ashwin Paranjape, Michele Bevilacqua, Fabio Petroni, and Percy Liang.
\newblock Lost in the middle: How language models use long contexts.
\newblock \emph{Transactions of the Association for Computational Linguistics}, 12:\penalty0 157--173, 2024{\natexlab{i}}.
\newblock \doi{10.1162/tacl_a_00638}.
\newblock URL \url{https://aclanthology.org/2024.tacl-1.9/}.

\bibitem[Liu et~al.(2025{\natexlab{d}})Liu, Sun, Chen, Yan, Zhang, Sun, Wang, and Li]{liu2025control}
Peipei Liu, Jian Sun, Li~Chen, Zhaoteng Yan, Peizheng Zhang, Dapeng Sun, Dawei Wang, and Dan Li.
\newblock Control flow-augmented decompiler based on large language model.
\newblock \emph{arXiv e-prints}, pages arXiv--2503, 2025{\natexlab{d}}.

\bibitem[Liu et~al.(2020)Liu, Gao, Chen, Nie, and Liu]{liu2020atom}
Shangqing Liu, Cuiyun Gao, Sen Chen, Lun~Yiu Nie, and Yang Liu.
\newblock Atom: Commit message generation based on abstract syntax tree and hybrid ranking.
\newblock \emph{IEEE Transactions on Software Engineering}, 48\penalty0 (5):\penalty0 1800--1817, 2020.

\bibitem[Liu et~al.(2024{\natexlab{j}})Liu, Chai, Yang, Shi, Zhu, Wang, Jin, Zhang, Zhu, Guo, et~al.]{mdeval}
Shukai Liu, Linzheng Chai, Jian Yang, Jiajun Shi, He~Zhu, Liran Wang, Ke~Jin, Wei Zhang, Hualei Zhu, Shuyue Guo, et~al.
\newblock Mdeval: Massively multilingual code debugging.
\newblock \emph{arXiv preprint arXiv:2411.02310}, 2024{\natexlab{j}}.

\bibitem[Liu et~al.(2024{\natexlab{k}})Liu, Han, Xu, Wang, Wang, Chen, Wu, and Liu]{liu2024progent}
Shuyuan Liu, Kunsheng Han, Bowei Xu, Zinuo Wang, Yu~Wang, Xuanzhe Chen, Chenglie Wu, and Yun Liu.
\newblock Progent: Programmable privilege control for llm agents.
\newblock \emph{arXiv preprint arXiv:2403.04704}, 2024{\natexlab{k}}.

\bibitem[Liu et~al.(2024{\natexlab{l}})Liu, Zhang, Chen, Xue, and Su]{liu2024fullstackbench}
Siyao Liu, Ge~Zhang, Boyuan Chen, Jialiang Xue, and Zhendong Su.
\newblock {FullStack Bench}: Evaluating llms as full stack coders.
\newblock \emph{arXiv preprint arXiv:2412.00535}, 2024{\natexlab{l}}.
\newblock URL \url{https://arxiv.org/abs/2412.00535}.

\bibitem[Liu et~al.(2024{\natexlab{m}})Liu, Xu, and McAuley]{liu2023repobench}
Tianyang Liu, Canwen Xu, and Julian McAuley.
\newblock Repobench: Benchmarking repository-level code auto-completion systems.
\newblock In \emph{International Conference on Learning Representations}, 2024{\natexlab{m}}.
\newblock URL \url{https://arxiv.org/abs/2306.03091}.

\bibitem[Liu et~al.(2022)Liu, Zhang, Luo, et~al.]{liu2022survey}
Wei Liu, Xian Zhang, Hu~Luo, et~al.
\newblock A survey of automatic source code summarization.
\newblock \emph{MDPI Electronics}, 11\penalty0 (22):\penalty0 3647, 2022.
\newblock URL \url{https://www.mdpi.com/2079-9292/11/22/3647}.

\bibitem[Liu et~al.(2024{\natexlab{n}})Liu, Lan, Hu, Liu, Zhang, Wang, Shieh, and Zhou]{liu2024codexgraph}
Xiangyan Liu, Bo~Lan, Zhiyuan Hu, Yang Liu, Zhicheng Zhang, Fei Wang, Michael Shieh, and Wenmeng Zhou.
\newblock Codexgraph: Bridging large language models and code repositories via code graph databases.
\newblock \emph{arXiv preprint arXiv:2408.03910}, 2024{\natexlab{n}}.

\bibitem[{Liu} et~al.(2023){Liu}, {Yu}, {Zhang}, {Xu}, {Lei}, {Lai}, {Gu}, {Ding}, {Men}, {Yang}, {Zhang}, {Deng}, {Zeng}, {Du}, {Zhang}, {Shen}, {Zhang}, {Su}, {Sun}, {Huang}, {Dong}, and {Tang}]{AgentBench}
Xiao {Liu}, Hao {Yu}, Hanchen {Zhang}, Yifan {Xu}, Xuanyu {Lei}, Hanyu {Lai}, Yu~{Gu}, Hangliang {Ding}, Kaiwen {Men}, Kejuan {Yang}, Shudan {Zhang}, Xiang {Deng}, Aohan {Zeng}, Zhengxiao {Du}, Chenhui {Zhang}, Sheng {Shen}, Tianjun {Zhang}, Yu~{Su}, Huan {Sun}, Minlie {Huang}, Yuxiao {Dong}, and Jie {Tang}.
\newblock {AgentBench: Evaluating LLMs as Agents}.
\newblock \emph{arXiv e-prints}, art. arXiv:2308.03688, August 2023.
\newblock \doi{10.48550/arXiv.2308.03688}.

\bibitem[Liu et~al.(2024{\natexlab{o}})Liu, Yu, Zhang, Xu, Lei, Lai, Gu, Ding, Men, Yang, Zhang, Deng, Zeng, Du, Zhang, Shen, Zhang, Su, Sun, Huang, Dong, and Tang]{liu2024agentbench}
Xiao Liu, Hao Yu, Hanchen Zhang, Yifan Xu, Xuanyu Lei, Hanyu Lai, Yu~Gu, Hangliang Ding, Kaiwen Men, Kejuan Yang, Shudan Zhang, Xiang Deng, Aohan Zeng, Zhengxiao Du, Chenhui Zhang, Sheng Shen, Tianjun Zhang, Yu~Su, Huan Sun, Minlie Huang, Yuxiao Dong, and Jie Tang.
\newblock Agentbench: Evaluating {LLM}s as agents.
\newblock In \emph{The Twelfth International Conference on Learning Representations}, 2024{\natexlab{o}}.
\newblock URL \url{https://openreview.net/forum?id=zAdUB0aCTQ}.

\bibitem[Liu et~al.(2025{\natexlab{e}})Liu, Shen, Li, Tang, and Luo]{10.1145/3711896.3737427}
Xinyu Liu, Shuyu Shen, Boyan Li, Nan Tang, and Yuyu Luo.
\newblock Nl2sql-bugs: A benchmark for detecting semantic errors in nl2sql translation.
\newblock In \emph{Proceedings of the 31st ACM SIGKDD Conference on Knowledge Discovery and Data Mining V.2}, KDD '25, page 5662–5673, New York, NY, USA, 2025{\natexlab{e}}. Association for Computing Machinery.
\newblock ISBN 9798400714542.
\newblock \doi{10.1145/3711896.3737427}.
\newblock URL \url{https://doi.org/10.1145/3711896.3737427}.

\bibitem[Liu et~al.(2023)Liu, Chen, Gao, Su, Zhang, Zan, Lou, Chen, and Ho]{liu2023uncovering}
Yan Liu, Xiaokang Chen, Yan Gao, Zhe Su, Fengji Zhang, Daoguang Zan, Jian-Guang Lou, Pin-Yu Chen, and Tsung-Yi Ho.
\newblock Uncovering and quantifying social biases in code generation.
\newblock \emph{Advances in Neural Information Processing Systems}, 36:\penalty0 2368--2380, 2023.

\bibitem[Liu et~al.(2025{\natexlab{f}})Liu, Foundjem, Khomh, and Li]{liu2025adversarial}
Yang Liu, Armstrong Foundjem, Foutse Khomh, and Heng Li.
\newblock Adversarial attack classification and robustness testing for large language models for code.
\newblock \emph{Empirical Software Engineering}, 30\penalty0 (5):\penalty0 154, 2025{\natexlab{f}}.

\bibitem[Liu et~al.(2025{\natexlab{g}})Liu, Meng, Joty, Savarese, Xiong, Zhou, and Yavuz]{liu2025codexembedgeneralistembeddingmodel}
Ye~Liu, Rui Meng, Shafiq Joty, Silvio Savarese, Caiming Xiong, Yingbo Zhou, and Semih Yavuz.
\newblock Codexembed: A generalist embedding model family for multiligual and multi-task code retrieval, 2025{\natexlab{g}}.
\newblock URL \url{https://arxiv.org/abs/2411.12644}.

\bibitem[Liu et~al.(2025{\natexlab{h}})Liu, Zhang, Zhu, Dong, Zhou, Shang, Yang, and Yang]{liu2025rstarcoder}
Yifei Liu, Li~Lyna Zhang, Yi~Zhu, Bingcheng Dong, Xudong Zhou, Ning Shang, Fan Yang, and Mao Yang.
\newblock rstar-coder: Scaling competitive code reasoning with a large-scale verified dataset.
\newblock \emph{arXiv preprint arXiv:2505.21297}, 2025{\natexlab{h}}.

\bibitem[Liu et~al.(2024{\natexlab{p}})Liu, Tao, Meng, Yao, Zhao, and Yang]{DBLP:conf/icse/0001TMYZY24}
Yilun Liu, Shimin Tao, Weibin Meng, Feiyu Yao, Xiaofeng Zhao, and Hao Yang.
\newblock Logprompt: Prompt engineering towards zero-shot and interpretable log analysis.
\newblock In \emph{Proceedings of the 2024 {IEEE/ACM} 46th International Conference on Software Engineering: Companion Proceedings, {ICSE} Companion 2024, Lisbon, Portugal, April 14-20, 2024}, pages 364--365. {ACM}, 2024{\natexlab{p}}.

\bibitem[Liu et~al.(2025{\natexlab{i}})Liu, Chen, Xu, He, Tao, Meng, Xie, Han, Zhao, Du, et~al.]{liu2025r}
Yilun Liu, Ziang Chen, Song Xu, Minggui He, Shimin Tao, Weibin Meng, Yuming Xie, Tao Han, Chunguang Zhao, Jingzhou Du, et~al.
\newblock R-log: Incentivizing log analysis capability in llms via reasoning-based reinforcement learning.
\newblock \emph{arXiv preprint arXiv:2509.25987}, 2025{\natexlab{i}}.

\bibitem[Liu et~al.(2024{\natexlab{q}})Liu, Gao, Wang, Liu, Shi, Zhang, and Peng]{liu2024marscode}
Yizhou Liu, Pengfei Gao, Xinchen Wang, Jie Liu, Yexuan Shi, Zhao Zhang, and Chao Peng.
\newblock Marscode agent: Ai-native automated bug fixing.
\newblock \emph{arXiv preprint arXiv:2409.00899}, 2024{\natexlab{q}}.

\bibitem[Liu et~al.(2018{\natexlab{b}})Liu, Xia, Hassan, Lo, Xing, and Wang]{liu2018neural}
Zhongxin Liu, Xin Xia, Ahmed~E Hassan, David Lo, Zhenchang Xing, and Xinyu Wang.
\newblock Neural-machine-translation-based commit message generation: how far are we?
\newblock In \emph{Proceedings of the 33rd ACM/IEEE international conference on automated software engineering}, pages 373--384, 2018{\natexlab{b}}.

\bibitem[Liu et~al.(2019)Liu, Xia, Treude, Lo, and Li]{liu2019automatic}
Zhongxin Liu, Xin Xia, Christoph Treude, David Lo, and Shanping Li.
\newblock Automatic generation of pull request descriptions.
\newblock In \emph{2019 34th IEEE/ACM International Conference on Automated Software Engineering (ASE)}, pages 176--188. IEEE, 2019.

\bibitem[Liu et~al.(2025{\natexlab{j}})Liu, Chen, Li, Qi, Pang, Du, Lee, and Lin]{liu2025understandingr1zeroliketrainingcritical}
Zichen Liu, Changyu Chen, Wenjun Li, Penghui Qi, Tianyu Pang, Chao Du, Wee~Sun Lee, and Min Lin.
\newblock Understanding r1-zero-like training: A critical perspective, 2025{\natexlab{j}}.
\newblock URL \url{https://arxiv.org/abs/2503.20783}.

\bibitem[{LLM Code Reviewer Project}(2024)]{llm_code_reviewer}
{LLM Code Reviewer Project}.
\newblock Llm code reviewer: Github action for automated reviews.
\newblock \url{https://github.com/marketplace/actions/llm-code-reviewer}, 2024.

\bibitem[Louck et~al.(2025)Louck, Stulman, and Dvir]{improve_a2a_protocal}
Yedidel Louck, Ariel Stulman, and Amit Dvir.
\newblock Improving google a2a protocol: Protecting sensitive data and mitigating unintended harms in multi-agent systems.
\newblock \emph{arXiv preprint arXiv:2505.12490}, 2025.

\bibitem[Louloudakis et~al.(2025)Louloudakis, Gibson, Cano~Reyes, and Rajan]{louloudakis2025fetafix}
Nikolaos Louloudakis, Perry Gibson, Jose Cano~Reyes, and Ajitha Rajan.
\newblock Fetafix: Automatic fault localization and repair of deep learning model conversions.
\newblock 2025.

\bibitem[Loyola et~al.(2017)Loyola, Marrese-Taylor, and Matsuo]{loyola2017neuralarchitecturegeneratingnatural}
Pablo Loyola, Edison Marrese-Taylor, and Yutaka Matsuo.
\newblock A neural architecture for generating natural language descriptions from source code changes, 2017.
\newblock URL \url{https://arxiv.org/abs/1704.04856}.

\bibitem[Lozano et~al.(2024)Lozano, Guha, Trost, Rozière, de~Souza, Fang, Mou, Akiki, Fourrier, de~Jesus~da Silva, Lansky, Kocetkov, Pykhtar, Zheltonozhskii, Bhatia, Villanova, de~Vries, Elsahar, Molybog, Li, Chim, Lamy-Poirier, Araujo, von Werra, Tsvetanova, Martin, Allal, Faysse, Marone, Zocca, Davaadorj, Muennighoff, Goyal, Sanseviero, Liu, Tachet, Li, Raileanu, Albani, Shi, Lavril, Wang, Jernite, Chai, Zi, Ferrandis, Sifre, Olukotun, Tajbakhsh, Ermolin, Fu, and Dao]{lozano2024starcoder2}
Antonio Lozano, Arjun Guha, Avi Trost, Baptiste Rozière, Cédric A. F. T.~M. de~Souza, Chenghao Fang, Chenghao Mou, Christopher Akiki, Clémentine Fourrier, Danilo de~Jesus~da Silva, David Lansky, Denis Kocetkov, Dmytro Pykhtar, Evgenii Zheltonozhskii, Gagan~S. Bhatia, Gabriel Villanova, Harm de~Vries, Hady Elsahar, Igor Molybog, Jia Li, Jenny Chim, Joel Lamy-Poirier, João G.~M. Araujo, Leandro von Werra, Lidiya Tsvetanova, Louis Martin, Loubna~Ben Allal, Manuel Faysse, Marc Marone, Marco Zocca, Mishig Davaadorj, Niklas Muennighoff, Naman Goyal, Omar Sanseviero, Qian Liu, Remi Tachet, Raymond Li, Roberta Raileanu, Samuel Albani, Sky Shi, Thibaut Lavril, Thomas Wang, Yacine Jernite, Yekun Chai, Yangtian Zi, Carlos~Muñoz Ferrandis, Laurent Sifre, Kunle Olukotun, Nima Tajbakhsh, Sergey Ermolin, Daniel~Y. Fu, and Tri Dao.
\newblock {StarCoder2 and The Stack v2: The Next Generation}, 2024.
\newblock URL \url{https://arxiv.org/abs/2402.19173}.

\bibitem[Lozhkov et~al.(2024)Lozhkov, Li, Allal, Cassano, Lamy{-}Poirier, Tazi, Tang, Pykhtar, Liu, Wei, Liu, Tian, Kocetkov, Zucker, Belkada, Wang, Liu, Abulkhanov, Paul, Li, Li, Risdal, Li, Zhu, Zhuo, Zheltonozhskii, Dade, Yu, Krau{\ss}, Jain, Su, He, Dey, Abati, Chai, Muennighoff, Tang, Oblokulov, Akiki, Marone, Mou, Mishra, Gu, Hui, Dao, Zebaze, Dehaene, Patry, Xu, McAuley, Hu, Scholak, Paquet, Robinson, Anderson, Chapados, and et~al.]{lozhkov2024starcoder2stackv2}
Anton Lozhkov, Raymond Li, Loubna~Ben Allal, Federico Cassano, Joel Lamy{-}Poirier, Nouamane Tazi, Ao~Tang, Dmytro Pykhtar, Jiawei Liu, Yuxiang Wei, Tianyang Liu, Max Tian, Denis Kocetkov, Arthur Zucker, Younes Belkada, Zijian Wang, Qian Liu, Dmitry Abulkhanov, Indraneil Paul, Zhuang Li, Wen{-}Ding Li, Megan Risdal, Jia Li, Jian Zhu, Terry~Yue Zhuo, Evgenii Zheltonozhskii, Nii Osae~Osae Dade, Wenhao Yu, Lucas Krau{\ss}, Naman Jain, Yixuan Su, Xuanli He, Manan Dey, Edoardo Abati, Yekun Chai, Niklas Muennighoff, Xiangru Tang, Muhtasham Oblokulov, Christopher Akiki, Marc Marone, Chenghao Mou, Mayank Mishra, Alex Gu, Binyuan Hui, Tri Dao, Armel Zebaze, Olivier Dehaene, Nicolas Patry, Canwen Xu, Julian~J. McAuley, Han Hu, Torsten Scholak, S{\'{e}}bastien Paquet, Jennifer Robinson, Carolyn~Jane Anderson, Nicolas Chapados, and et~al.
\newblock Starcoder 2 and the stack v2: The next generation.
\newblock \emph{CoRR}, abs/2402.19173, 2024.
\newblock \doi{10.48550/ARXIV.2402.19173}.
\newblock URL \url{https://doi.org/10.48550/arXiv.2402.19173}.

\bibitem[Lu and Liu(2024{\natexlab{a}})]{RAGcomment}
Hanzhen Lu and Zhongxin Liu.
\newblock { Improving Retrieval-Augmented Code Comment Generation by Retrieving for Generation }.
\newblock In \emph{2024 IEEE International Conference on Software Maintenance and Evolution (ICSME)}, pages 350--362, Los Alamitos, CA, USA, Oct 2024{\natexlab{a}}. IEEE Computer Society.
\newblock \doi{10.1109/ICSME58944.2024.00040}.
\newblock URL \url{https://doi.ieeecomputersociety.org/10.1109/ICSME58944.2024.00040}.

\bibitem[Lu and Liu(2024{\natexlab{b}})]{rag_comment_generation}
Hanzhen Lu and Zhongxin Liu.
\newblock Improving retrieval-augmented code comment generation by retrieving for generation.
\newblock In \emph{2024 IEEE International Conference on Software Maintenance and Evolution (ICSME)}, pages 350--362. IEEE, 2024{\natexlab{b}}.

\bibitem[Lu et~al.(2024)Lu, Liu, Zhang, Wang, Dong, Liu, Sun, Ren, Li, Yang, Sun, Deng, Xu, Xie, and Ruan]{lu2024deepseekvl}
Haoyu Lu, Wen Liu, Bo~Zhang, Bingxuan Wang, Kai Dong, Bo~Liu, Jingxiang Sun, Tongzheng Ren, Zhuoshu Li, Hao Yang, Yaofeng Sun, Chengqi Deng, Hanwei Xu, Zhenda Xie, and Chong Ruan.
\newblock Deepseek-vl: Towards real-world vision-language understanding, 2024.
\newblock URL \url{https://arxiv.org/abs/2403.05525}.

\bibitem[Lu et~al.(2023{\natexlab{a}})Lu, Yu, Li, Yang, and Zuo]{lu2023llama}
Junyi Lu, Lei Yu, Xiaojia Li, Li~Yang, and Chun Zuo.
\newblock Llama-reviewer: Advancing code review automation with large language models through parameter-efficient fine-tuning.
\newblock In \emph{2023 IEEE 34th International Symposium on Software Reliability Engineering (ISSRE)}, pages 647--658. IEEE, 2023{\natexlab{a}}.

\bibitem[{Lu} et~al.(2024){Lu}, {Li}, {Hua}, {Yu}, {Cheng}, {Yang}, {Zhang}, and {Zuo}]{DeepCRCEval}
Junyi {Lu}, Xiaojia {Li}, Zihan {Hua}, Lei {Yu}, Shiqi {Cheng}, Li~{Yang}, Fengjun {Zhang}, and Chun {Zuo}.
\newblock {DeepCRCEval: Revisiting the Evaluation of Code Review Comment Generation}.
\newblock \emph{arXiv e-prints}, art. arXiv:2412.18291, December 2024.
\newblock \doi{10.48550/arXiv.2412.18291}.

\bibitem[Lu et~al.(2021)Lu, Guo, Ren, Huang, Svyatkovskiy, Blanco, Clement, Drain, Jiang, Tang, Li, Zhou, Shou, Zhou, Tufano, Gong, Zhou, Duan, Sundaresan, Deng, Fu, and Liu]{CodeXGLUE}
Shuai Lu, Daya Guo, Shuo Ren, Junjie Huang, Alexey Svyatkovskiy, Ambrosio Blanco, Colin Clement, Dawn Drain, Daxin Jiang, Duyu Tang, Ge~Li, Lidong Zhou, Linjun Shou, Long Zhou, Michele Tufano, Ming Gong, Ming Zhou, Nan Duan, Neel Sundaresan, Shao~Kun Deng, Shengyu Fu, and Shujie Liu.
\newblock Codexglue: A machine learning benchmark dataset for code understanding and generation, 2021.
\newblock URL \url{https://arxiv.org/abs/2102.04664}.

\bibitem[Lu et~al.(2024)Lu, Yang, Shen, and Awadallah]{lu2024omniparser}
Yadong Lu, Jianwei Yang, Yelong Shen, and Ahmed Awadallah.
\newblock Omniparser for pure vision based gui agent.
\newblock \emph{arXiv preprint arXiv:2408.00203}, 2024.

\bibitem[Lu et~al.(2023{\natexlab{b}})Lu, Tong, Zhao, Zhang, and Li]{lu2023uilayoutgenerationllms}
Yuwen Lu, Ziang Tong, Qinyi Zhao, Chengzhi Zhang, and Toby Jia-Jun Li.
\newblock Ui layout generation with llms guided by ui grammar, 2023{\natexlab{b}}.
\newblock URL \url{https://arxiv.org/abs/2310.15455}.

\bibitem[Lu et~al.(2025{\natexlab{a}})Lu, Yao, Gu, Huang, Wang, Li, Gesi, He, Li, and Wang]{UXAgent}
Yuxuan Lu, Bingsheng Yao, Hansu Gu, Jing Huang, Zheshen~Jessie Wang, Yang Li, Jiri Gesi, Qi~He, Toby Jia-Jun Li, and Dakuo Wang.
\newblock Uxagent: An llm agent-based usability testing framework for web design.
\newblock In \emph{Proceedings of the Extended Abstracts of the CHI Conference on Human Factors in Computing Systems}, CHI EA '25, New York, NY, USA, 2025{\natexlab{a}}. Association for Computing Machinery.
\newblock ISBN 9798400713958.
\newblock \doi{10.1145/3706599.3719729}.
\newblock URL \url{https://doi.org/10.1145/3706599.3719729}.

\bibitem[Lu et~al.(2025{\natexlab{b}})Lu, Yang, Ren, Hou, Xiao, Wang, Shi, Zhou, Zhan, and Li]{lu2025webgen}
Zimu Lu, Yunqiao Yang, Houxing Ren, Haotian Hou, Han Xiao, Ke~Wang, Weikang Shi, Aojun Zhou, Mingjie Zhan, and Hongsheng Li.
\newblock Webgen-bench: Evaluating llms on generating interactive and functional websites from scratch.
\newblock \emph{arXiv preprint arXiv:2505.03733}, 2025{\natexlab{b}}.

\bibitem[Luo et~al.(2025{\natexlab{a}})Luo, Tan, Huang, Patel, Ariyak, Wu, Shi, Xin, Cai, Weber, Zhang, Li, Popa, and Stoica]{deepcoder2025}
Michael Luo, Sijun Tan, Roy Huang, Ameen Patel, Alpay Ariyak, Qingyang Wu, Xiaoxiang Shi, Rachel Xin, Colin Cai, Maurice Weber, Ce~Zhang, Li~Erran Li, Raluca~Ada Popa, and Ion Stoica.
\newblock Deepcoder: A fully open-source 14b coder at o3-mini level.
\newblock \url{https://pretty-radio-b75.notion.site/DeepCoder-A-Fully-Open-Source-14B-Coder-at-O3-mini-Level-1cf81902c14680b3bee5eb349a512a51}, 2025{\natexlab{a}}.
\newblock Notion Blog.

\bibitem[{Luo} et~al.(2024){Luo}, {Ye}, {Liang}, {Zhang}, {Qin}, {Lu}, {Wu}, {Cong}, {Lin}, {Zhang}, {Che}, {Liu}, and {Sun}]{RepoAgent}
Qinyu {Luo}, Yining {Ye}, Shihao {Liang}, Zhong {Zhang}, Yujia {Qin}, Yaxi {Lu}, Yesai {Wu}, Xin {Cong}, Yankai {Lin}, Yingli {Zhang}, Xiaoyin {Che}, Zhiyuan {Liu}, and Maosong {Sun}.
\newblock {RepoAgent: An LLM-Powered Open-Source Framework for Repository-level Code Documentation Generation}.
\newblock \emph{arXiv e-prints}, art. arXiv:2402.16667, February 2024.
\newblock \doi{10.48550/arXiv.2402.16667}.

\bibitem[Luo et~al.(2025{\natexlab{b}})Luo, Huang, Shen, Li, Shen, Zeng, Tang, and Luo]{nvbench2-2025luo}
Tianqi Luo, Chuhan Huang, Leixian Shen, Boyan Li, Shuyu Shen, Wei Zeng, Nan Tang, and Yuyu Luo.
\newblock nvbench 2.0: Resolving ambiguity in text-to-visualization through stepwise reasoning, 2025{\natexlab{b}}.
\newblock URL \url{https://arxiv.org/abs/2503.12880}.

\bibitem[Luo et~al.(2024)Luo, Zhu, Zhang, Wang, Yang, Xu, and Che]{luoSemiInstructBridgingNaturalInstruct2024}
Xianzhen Luo, Qingfu Zhu, Zhiming Zhang, Xu~Wang, Qing Yang, Dongliang Xu, and Wanxiang Che.
\newblock Semi-{{Instruct}}: {{Bridging Natural-Instruct}} and {{Self-Instruct}} for {{Code Large Language Models}}, 2024.

\bibitem[Luo et~al.(2025{\natexlab{c}})Luo, Zheng, Zhu, Zhang, Li, Huang, Fan, and Che]{code_scaling_law}
Xianzhen Luo, Wenzhen Zheng, Qingfu Zhu, Rongyi Zhang, Houyi Li, Siming Huang, YuanTao Fan, and Wanxiang Che.
\newblock Scaling laws for code: A more data-hungry regime.
\newblock \emph{arXiv preprint arXiv:2510.08702}, 2025{\natexlab{c}}.

\bibitem[Luo et~al.(2025{\natexlab{d}})Luo, Zhu, Zhang, Xu, Cheng, Wang, Chu, Xuyang, Ma, Fan, and Che]{luo2025successdetailsevaluateenhance}
Xianzhen Luo, Qingfu Zhu, Zhiming Zhang, Mingzheng Xu, Tianhao Cheng, Yixuan Wang, Zheng Chu, Shijie Xuyang, Zhiyuan Ma, YuanTao Fan, and Wanxiang Che.
\newblock Success is in the details: Evaluate and enhance details sensitivity of code llms through counterfactuals, 2025{\natexlab{d}}.
\newblock URL \url{https://arxiv.org/abs/2505.14597}.

\bibitem[Luo et~al.(2021)Luo, Tang, Li, Chai, Li, and Qin]{nvBench2021}
Yuyu Luo, Nan Tang, Guoliang Li, Chengliang Chai, Wenbo Li, and Xuedi Qin.
\newblock Synthesizing natural language to visualization (nl2vis) benchmarks from nl2sql benchmarks.
\newblock In \emph{Proceedings of the 2021 International Conference on Management of Data}, SIGMOD '21, page 1235–1247, New York, NY, USA, 2021. Association for Computing Machinery.
\newblock ISBN 9781450383431.
\newblock \doi{10.1145/3448016.3457261}.
\newblock URL \url{https://doi.org/10.1145/3448016.3457261}.

\bibitem[Luo et~al.(2025{\natexlab{e}})Luo, Shao, Zhang, Zhou, Hu, Zhao, Liu, and Qin]{luo2025shadow}
Zhifan Luo, Shuo Shao, Su~Zhang, Lijing Zhou, Yuke Hu, Chenxu Zhao, Zhihao Liu, and Zhan Qin.
\newblock Shadow in the cache: Unveiling and mitigating privacy risks of kv-cache in llm inference, 2025{\natexlab{e}}.

\bibitem[Luo et~al.(2023)Luo, Xu, Zhao, Sun, Geng, Hu, Tao, Ma, Lin, and Jiang]{luo2023wizardcoder}
Ziyang Luo, Can Xu, Pu~Zhao, Qingfeng Sun, Xiubo Geng, Wenxiang Hu, Chongyang Tao, Jing Ma, Qingwei Lin, and Daxin Jiang.
\newblock Wizardcoder: Empowering code large language models with evol-instruct.
\newblock \emph{arXiv preprint arXiv:2306.08568}, 2023.

\bibitem[Luu et~al.(2024)Luu, Iyengar, Li, Le, Zhao, and other]{luu2024purpcode}
Phong Luu, Adityasrinivas Iyengar, Chacha Li, Hung Le, Tong Zhao, and other.
\newblock Purpcode: Reasoning for safer code generation.
\newblock \emph{Proceedings of the Amazon Nova AI Challenge}, 2024.

\bibitem[Lvwerra(2023)]{lvwerra_stack_exchange}
Lvwerra.
\newblock Stack exchange paired dataset, 2023.
\newblock URL \url{https://huggingface.co/datasets/lvwerra/stack-exchange-paired}.
\newblock Accessed: 2024.

\bibitem[Lyu et~al.(2025)Lyu, Zhao, Wang, and Jie]{lyu2025llm_differential_testing}
Ce~Lyu, Minghao Zhao, Yanhao Wang, and Liang Jie.
\newblock Llm-based dynamic differential testing for database connectors with reinforcement learning-guided prompt selection.
\newblock \emph{arXiv preprint arXiv:2506.11870}, 2025.

\bibitem[Lyu et~al.(2022)Lyu, Li, Liu, and Zhao]{lyu2022constrained}
Zuxin Lyu, Zhaoran Li, Jia Liu, and Tuo Zhao.
\newblock Constrained variational policy optimization for safe reinforcement learning.
\newblock pages 14660--14674, 2022.

\bibitem[M-A-P(2024)]{map_code_feedback}
M-A-P.
\newblock Code feedback dataset, 2024.
\newblock URL \url{https://huggingface.co/datasets/m-a-p/Code-Feedback}.
\newblock Accessed: 2024.

\bibitem[Ma et~al.(2024)Ma, Luo, Vo, Sima, and Leong]{ma2024highly}
Haozhe Ma, Zhengding Luo, Thanh~Vinh Vo, Kuankuan Sima, and Tze-Yun Leong.
\newblock Highly efficient self-adaptive reward shaping for reinforcement learning.
\newblock \emph{arXiv preprint arXiv:2408.03029}, 2024.

\bibitem[{Ma} et~al.(2024){Ma}, {Du}, {Wang}, {Zhang}, {Wen}, {Qu}, {Yang}, {Liu}, {Liu}, {Yue}, {Huang}, and {Zhang}]{KORBench}
Kaijing {Ma}, Xinrun {Du}, Yunran {Wang}, Haoran {Zhang}, Zhoufutu {Wen}, Xingwei {Qu}, Jian {Yang}, Jiaheng {Liu}, Minghao {Liu}, Xiang {Yue}, Wenhao {Huang}, and Ge~{Zhang}.
\newblock {KOR-Bench: Benchmarking Language Models on Knowledge-Orthogonal Reasoning Tasks}.
\newblock \emph{arXiv e-prints}, art. arXiv:2410.06526, October 2024.
\newblock \doi{10.48550/arXiv.2410.06526}.

\bibitem[Ma et~al.(2025{\natexlab{a}})Ma, Liu, Bu, Li, Wang, and Liu]{ma2025specevalevaluatingcodecomprehension}
Lezhi Ma, Shangqing Liu, Lei Bu, Shangru Li, Yida Wang, and Yang Liu.
\newblock Speceval: Evaluating code comprehension in large language models via program specifications, 2025{\natexlab{a}}.
\newblock URL \url{https://arxiv.org/abs/2409.12866}.

\bibitem[Ma et~al.(2025{\natexlab{b}})Ma, Yang, Li, Fei, Zhou, Li, Jiang, Xu, and Xiao]{ma2025adaptivelog}
Lipeng Ma, Weidong Yang, Yixuan Li, Ben Fei, Mingjie Zhou, Shuhao Li, Sihang Jiang, Bo~Xu, and Yanghua Xiao.
\newblock Adaptivelog: An adaptive log analysis framework with the collaboration of large and small language model.
\newblock \emph{ACM Transactions on Software Engineering and Methodology}, 2025{\natexlab{b}}.

\bibitem[Ma et~al.(2023)Ma, Liang, Wang, Huang, Bastani, Jayaraman, Zhu, Fan, and Anandkumar]{ma2023eureka}
Yecheng~Jason Ma, William Liang, Guanzhi Wang, De-An Huang, Osbert Bastani, Dinesh Jayaraman, Yuke Zhu, Linxi Fan, and Anima Anandkumar.
\newblock Eureka: Human-level reward design via coding large language models.
\newblock \emph{arXiv preprint arXiv:2310.12931}, 2023.

\bibitem[Ma et~al.(2024)Ma, Fu, Zhang, and Xu]{ma2024ctrl}
Yihong Ma, Yifei Fu, Hongyi Zhang, and Ganqu Xu.
\newblock Ctrl-z: Controlling ai agents via resampling.
\newblock \emph{arXiv preprint arXiv:2405.18244}, 2024.

\bibitem[Ma et~al.(2025{\natexlab{c}})Ma, Yang, Cao, Li, Huang, and Li]{ma2025alibaba}
Yingwei Ma, Qingping Yang, Rongyu Cao, Binhua Li, Fei Huang, and Yongbin Li.
\newblock Alibaba lingmaagent: Improving automated issue resolution via comprehensive repository exploration.
\newblock In \emph{Proceedings of the 33rd ACM International Conference on the Foundations of Software Engineering}, pages 238--249, 2025{\natexlab{c}}.

\bibitem[Ma et~al.(2025{\natexlab{d}})Ma, Zhang, Cao, Liu, Zhang, Luo, Zhang, and Chen]{ma2025rethinkingverificationllmcode}
Zihan Ma, Taolin Zhang, Maosong Cao, Junnan Liu, Wenwei Zhang, Minnan Luo, Songyang Zhang, and Kai Chen.
\newblock Rethinking verification for llm code generation: From generation to testing, 2025{\natexlab{d}}.
\newblock URL \url{https://arxiv.org/abs/2507.06920}.

\bibitem[Macedo et~al.(2024)Macedo, Tian, Nie, Cogo, and Adams]{macedo2024intertrans}
Marcos Macedo, Yuan Tian, Pengyu Nie, Filipe~R Cogo, and Bram Adams.
\newblock Intertrans: Leveraging transitive intermediate translations to enhance llm-based code translation.
\newblock \emph{arXiv preprint arXiv:2411.01063}, 2024.

\bibitem[Madaan et~al.(2023)Madaan, Tandon, Gupta, Hallinan, Gao, Wiegreffe, Alon, Dziri, Prabhumoye, Yang, Gupta, Majumder, Hermann, Welleck, Yazdanbakhsh, and Clark]{madaan2023selfrefine}
Aman Madaan, Niket Tandon, Prakhar Gupta, Skyler Hallinan, Luyu Gao, Sarah Wiegreffe, Uri Alon, Nouha Dziri, Shrimai Prabhumoye, Yiming Yang, Shashank Gupta, Bodhisattwa~Prasad Majumder, Katherine Hermann, Sean Welleck, Amir Yazdanbakhsh, and Peter Clark.
\newblock Self-refine: Iterative refinement with self-feedback.
\newblock In \emph{Advances in Neural Information Processing Systems (NeurIPS)}, 2023.
\newblock URL \url{https://openreview.net/forum?id=S37hOerQLB}.

\bibitem[Magpie-Align(2024)]{magpie_align_qwen25}
Magpie-Align.
\newblock Magpie-qwen2.5-coder-pro-300k-v0.1, 2024.
\newblock URL \url{https://huggingface.co/datasets/Magpie-Align/Magpie-Qwen2.5-Coder-Pro-300K-v0.1}.
\newblock Accessed: 2024.

\bibitem[Maini et~al.(2025)Maini, Goyal, Sam, Robey, Savani, Jiang, Zou, Lipton, and Kolter]{maini2025safety}
Pratyush Maini, Sachin Goyal, Dylan Sam, Alex Robey, Yash Savani, Yiding Jiang, Andy Zou, Zacharcy~C Lipton, and J~Zico Kolter.
\newblock Safety pretraining: Toward the next generation of safe ai.
\newblock \emph{arXiv preprint arXiv:2504.16980}, 2025.

\bibitem[Manh et~al.(2024)Manh, Chau, Le~Hai, Doan, Nguyen, Pham, and Bui]{manh2024codemmlu}
Dung~Nguyen Manh, Thang~Phan Chau, Nam Le~Hai, Thong~T Doan, Nam~V Nguyen, Quang Pham, and Nghi~DQ Bui.
\newblock Codemmlu: A multi-task benchmark for assessing code understanding capabilities of codellms.
\newblock \emph{CoRR}, 2024.

\bibitem[Maniatis and Tarlow()]{DIDACT}
Petros Maniatis and Daniel Tarlow.
\newblock Large sequence models for software development activities.
\newblock \url{https://blog.research.google/2023/05/large-sequence-models-for-software.html.}

\bibitem[{Mao} et~al.(2023){Mao}, {Cai}, {Xia}, {Wu}, {Wang}, {Wang}, {Ge}, and {Wei}]{ALYMPICS}
Shaoguang {Mao}, Yuzhe {Cai}, Yan {Xia}, Wenshan {Wu}, Xun {Wang}, Fengyi {Wang}, Tao {Ge}, and Furu {Wei}.
\newblock {ALYMPICS: LLM Agents Meet Game Theory -- Exploring Strategic Decision-Making with AI Agents}.
\newblock \emph{arXiv e-prints}, art. arXiv:2311.03220, November 2023.
\newblock \doi{10.48550/arXiv.2311.03220}.

\bibitem[Martinez-Gil(2025)]{polycoder}
Jorge Martinez-Gil.
\newblock Evaluating small-scale code models for code clone detection.
\newblock \emph{arXiv preprint arXiv:2506.10995}, 2025.

\bibitem[MatrixStudio(2023)]{matrixstudio_codeforces}
MatrixStudio.
\newblock Codeforces python submissions, 2023.
\newblock URL \url{https://huggingface.co/datasets/MatrixStudio/Codeforces-Python-Submissions}.
\newblock Accessed: 2024.

\bibitem[Mattern et~al.(2025)Mattern, Jaghouar, Basra, Straube, Ferrante, Gabriel, Ong, Weisser, and Hagemann]{2025synthetic1}
Justus Mattern, Sami Jaghouar, Manveer Basra, Jannik Straube, Matthew~Di Ferrante, Felix Gabriel, Jack~Min Ong, Vincent Weisser, and Johannes Hagemann.
\newblock Synthetic-1: Two million collaboratively generated reasoning traces from deepseek-r1, 2025.
\newblock URL \url{https://www.primeintellect.ai/blog/synthetic-1-release}.

\bibitem[Matton et~al.(2024)Matton, Sherborne, Aumiller, Tommasone, Alizadeh, He, Ma, Voisin, Gilsenan-McMahon, and Gallé]{matton2024leakagecodegenerationevaluation}
Alexandre Matton, Tom Sherborne, Dennis Aumiller, Elena Tommasone, Milad Alizadeh, Jingyi He, Raymond Ma, Maxime Voisin, Ellen Gilsenan-McMahon, and Matthias Gallé.
\newblock On leakage of code generation evaluation datasets, 2024.
\newblock URL \url{https://arxiv.org/abs/2407.07565}.

\bibitem[McKenzie et~al.(2025)McKenzie, Hollinsworth, Tseng, Davies, Casper, Tucker, Kirk, and Gleave]{mckenzie2025stack}
Ian~R McKenzie, Oskar~J Hollinsworth, Tom Tseng, Xander Davies, Stephen Casper, Aaron~D Tucker, Robert Kirk, and Adam Gleave.
\newblock Stack: Adversarial attacks on llm safeguard pipelines.
\newblock \emph{arXiv preprint arXiv:2506.24068}, 2025.

\bibitem[Meng et~al.(2023)Meng, Song, Tong, Pan, and Yu]{meng2023deepscaler}
Chunyang Meng, Shijie Song, Haogang Tong, Maolin Pan, and Yang Yu.
\newblock Deepscaler: Holistic autoscaling for microservices based on spatiotemporal gnn with adaptive graph learning.
\newblock In \emph{2023 38th IEEE/ACM International Conference on Automated Software Engineering (ASE)}, pages 53--65. IEEE, 2023.

\bibitem[Meng et~al.(2024)Meng, Zhang, Wang, Liu, and He]{meng2022mitigating}
Yifei Meng, Yuxuan Zhang, Hao Wang, Zhen Liu, and Xing He.
\newblock Mitigating gender bias in code large language models via model editing.
\newblock In \emph{2024 IEEE 31st International Conference on Software Analysis, Evolution and Reengineering (SANER)}. IEEE, 2024.

\bibitem[{Meta AI}(2024)]{meta2024llama3_2}
{Meta AI}.
\newblock Llama 3.2: Revolutionizing edge ai and vision with open, customizable models, 2024.
\newblock URL \url{https://ai.meta.com/blog/llama-3-2-connect-2024-vision-edge-mobile-devices/}.

\bibitem[{Meta AI}(2025)]{meta2025llama4}
{Meta AI}.
\newblock The llama 4 herd: The beginning of a new era of natively multimodal ai innovation, 2025.
\newblock URL \url{https://ai.meta.com/blog/llama-4-multimodal-intelligence/}.

\bibitem[Mialon et~al.(2023)Mialon, Fourrier, Wolf, LeCun, and Scialom]{mialon2023gaia}
Gr{\'e}goire Mialon, Cl{\'e}mentine Fourrier, Thomas Wolf, Yann LeCun, and Thomas Scialom.
\newblock Gaia: a benchmark for general ai assistants.
\newblock In \emph{The Twelfth International Conference on Learning Representations}, 2023.

\bibitem[Miao et~al.(2024)Miao, Gao, Quan, Lin, Zan, Liu, Yang, Liu, and Deng]{miao2024aligningdpo}
Yibo Miao, Bofei Gao, Shanghaoran Quan, Junyang Lin, Daoguang Zan, Jiaheng Liu, Jian Yang, Tianyu Liu, and Zhijie Deng.
\newblock Aligning codellms with direct preference optimization.
\newblock \emph{arXiv preprint arXiv:2410.18585}, 2024.

\bibitem[{Microsoft .NET Blog}(2024)]{githubcopilot2024architecture}
{Microsoft .NET Blog}.
\newblock How we build github copilot into visual studio.
\newblock \url{https://devblogs.microsoft.com/dotnet/building-github-copilot-into-visual-studio/}, 2024.

\bibitem[Milev et~al.(2025)Milev, Balunović, Baader, and Vechev]{milev2025toolfuzzautomatedagent}
Ivan Milev, Mislav Balunović, Maximilian Baader, and Martin Vechev.
\newblock Toolfuzz -- automated agent tool testing, 2025.
\newblock URL \url{https://arxiv.org/abs/2503.04479}.

\bibitem[Miserendino et~al.(2025)Miserendino, Wang, Patwardhan, and Heidecke]{SWE-Lancer}
Samuel Miserendino, Michele Wang, Tejal Patwardhan, and Johannes Heidecke.
\newblock Swe-lancer: Can frontier llms earn 1 million from real-world freelance software engineering?, 2025.
\newblock URL \url{https://arxiv.org/abs/2502.12115}.

\bibitem[Mishra et~al.(2024{\natexlab{a}})Mishra, Stallone, Zhang, Shen, Prasad, Soria, Merler, Selvam, Surendran, Singh, Sethi, Dang, Li, Wu, Zawad, Coleman, White, Lewis, Pavuluri, Koyfman, Lublinsky, de~Bayser, Abdelaziz, Basu, Agarwal, Zhou, Johnson, Goyal, Patel, Shah, Zerfos, Ludwig, Munawar, Crouse, Kapanipathi, Salaria, Calio, Wen, Seelam, Belgodere, Fonseca, Singhee, Desai, Cox, Puri, and Panda]{DBLP:journals/corr/abs-2405-04324}
Mayank Mishra, Matt Stallone, Gaoyuan Zhang, Yikang Shen, Aditya Prasad, Adriana~Meza Soria, Michele Merler, Parameswaran Selvam, Saptha Surendran, Shivdeep Singh, Manish Sethi, Xuan{-}Hong Dang, Pengyuan Li, Kun{-}Lung Wu, Syed Zawad, Andrew Coleman, Matthew White, Mark Lewis, Raju Pavuluri, Yan Koyfman, Boris Lublinsky, Maximilien de~Bayser, Ibrahim Abdelaziz, Kinjal Basu, Mayank Agarwal, Yi~Zhou, Chris Johnson, Aanchal Goyal, Hima Patel, S.~Yousaf Shah, Petros Zerfos, Heiko Ludwig, Asim Munawar, Maxwell Crouse, Pavan Kapanipathi, Shweta Salaria, Bob Calio, Sophia Wen, Seetharami Seelam, Brian Belgodere, Carlos~A. Fonseca, Amith Singhee, Nirmit Desai, David~D. Cox, Ruchir Puri, and Rameswar Panda.
\newblock Granite code models: {A} family of open foundation models for code intelligence.
\newblock \emph{CoRR}, abs/2405.04324, 2024{\natexlab{a}}.
\newblock \doi{10.48550/ARXIV.2405.04324}.
\newblock URL \url{https://doi.org/10.48550/arXiv.2405.04324}.

\bibitem[Mishra et~al.(2024{\natexlab{b}})Mishra, Stallone, Zhang, Shen, Prasad, Soria, Merler, Selvam, Surendran, Singh, Sethi, Dang, Li, Wu, Zawad, Coleman, White, Lewis, Pavuluri, Koyfman, Lublinsky, de~Bayser, Abdelaziz, Basu, Agarwal, Zhou, Johnson, Goyal, Patel, Shah, Zerfos, Ludwig, Munawar, Crouse, Kapanipathi, Salaria, Calio, Wen, Seelam, Belgodere, Fonseca, Singhee, Desai, Cox, Puri, and Panda]{mishra2024granitecodemodelsfamily}
Mayank Mishra, Matt Stallone, Gaoyuan Zhang, Yikang Shen, Aditya Prasad, Adriana~Meza Soria, Michele Merler, Parameswaran Selvam, Saptha Surendran, Shivdeep Singh, Manish Sethi, Xuan-Hong Dang, Pengyuan Li, Kun-Lung Wu, Syed Zawad, Andrew Coleman, Matthew White, Mark Lewis, Raju Pavuluri, Yan Koyfman, Boris Lublinsky, Maximilien de~Bayser, Ibrahim Abdelaziz, Kinjal Basu, Mayank Agarwal, Yi~Zhou, Chris Johnson, Aanchal Goyal, Hima Patel, Yousaf Shah, Petros Zerfos, Heiko Ludwig, Asim Munawar, Maxwell Crouse, Pavan Kapanipathi, Shweta Salaria, Bob Calio, Sophia Wen, Seetharami Seelam, Brian Belgodere, Carlos Fonseca, Amith Singhee, Nirmit Desai, David~D. Cox, Ruchir Puri, and Rameswar Panda.
\newblock Granite code models: A family of open foundation models for code intelligence, 2024{\natexlab{b}}.
\newblock URL \url{https://arxiv.org/abs/2405.04324}.

\bibitem[Mishra et~al.(2022)Mishra, Khashabi, Baral, and Hajishirzi]{mishra2022CrossTask}
Swaroop Mishra, Daniel Khashabi, Chitta Baral, and Hannaneh Hajishirzi.
\newblock Cross-{{Task Generalization}} via {{Natural Language Crowdsourcing Instructions}}.
\newblock In Smaranda Muresan, Preslav Nakov, and Aline Villavicencio, editors, \emph{Proceedings of the 60th {{Annual Meeting}} of the {{Association}} for {{Computational Linguistics}} ({{Volume}} 1: {{Long Papers}})}, pages 3470--3487, Dublin, Ireland, 2022.

\bibitem[Miyai et~al.(2025)Miyai, Zhao, Egashira, Sato, Sunada, Onohara, Yamanishi, Toyooka, Nishina, Maeda, Aizawa, and Yamasaki]{miyai2025webchorearena}
Atsuyuki Miyai, Zaiying Zhao, Kazuki Egashira, Atsuki Sato, Tatsumi Sunada, Shota Onohara, Hiromasa Yamanishi, Mashiro Toyooka, Kunato Nishina, Ryoma Maeda, Kiyoharu Aizawa, and Toshihiko Yamasaki.
\newblock Webchorearena: Evaluating web browsing agents on realistic tedious web tasks, 2025.
\newblock URL \url{https://arxiv.org/abs/2506.01952}.

\bibitem[MLFoundations(2023{\natexlab{a}})]{mlfoundations_codegolf}
MLFoundations.
\newblock Stack exchange code golf, 2023{\natexlab{a}}.
\newblock URL \url{https://huggingface.co/datasets/mlfoundations-dev/stackexchange_codegolf}.
\newblock Accessed: 2024.

\bibitem[MLFoundations(2023{\natexlab{b}})]{mlfoundations_stackexchange_codereview}
MLFoundations.
\newblock Stack exchange code review dataset, 2023{\natexlab{b}}.
\newblock URL \url{https://huggingface.co/datasets/mlfoundations-dev/stackexchange_codereview}.
\newblock Accessed: 2024.

\bibitem[Mohsin et~al.(2024)Mohsin, Janicke, Wood, Sarker, Maglaras, and Janjua]{mohsin2024can}
Ahmad Mohsin, Helge Janicke, Adrian Wood, Iqbal~H Sarker, Leandros Maglaras, and Naeem Janjua.
\newblock Can we trust large language models generated code? a framework for in-context learning, security patterns, and code evaluations across diverse llms.
\newblock \emph{arXiv preprint arXiv:2406.12513}, 2024.

\bibitem[Monperrus et~al.(2021)Monperrus, Martinez, Ye, Madeiral, Durieux, and Yu]{megadiff}
Martin Monperrus, Matias Martinez, He~Ye, Fernanda Madeiral, Thomas Durieux, and Zhongxing Yu.
\newblock Megadiff: A dataset of 600k java source code changes categorized by diff size, 2021.
\newblock URL \url{https://arxiv.org/abs/2108.04631}.

\bibitem[Moon et~al.(2025)Moon, Hwang, Lee, Kang, Kim, and Jung]{moon2025dontjudgecodecover}
Jiwon Moon, Yerin Hwang, Dongryeol Lee, Taegwan Kang, Yongil Kim, and Kyomin Jung.
\newblock Don't judge code by its cover: Exploring biases in llm judges for code evaluation, 2025.
\newblock URL \url{https://arxiv.org/abs/2505.16222}.

\bibitem[Morovati et~al.(2024)Morovati, Nikanjam, and Khomh]{swe_fault_localization}
Mohammad~Mehdi Morovati, Amin Nikanjam, and Foutse Khomh.
\newblock Fault localization in deep learning-based software: A system-level approach.
\newblock \emph{arXiv preprint arXiv:2411.08172}, 2024.

\bibitem[Moshkov et~al.(2025)Moshkov, Hanley, Sorokin, Toshniwal, Henkel, Schifferer, Du, and Gitman]{moshkov2025aimo}
Ivan Moshkov, Darragh Hanley, Ivan Sorokin, Shubham Toshniwal, Christof Henkel, Benedikt Schifferer, Wei Du, and Igor Gitman.
\newblock Aimo-2 winning solution: Building state-of-the-art mathematical reasoning models with openmathreasoning dataset.
\newblock \emph{arXiv preprint arXiv:2504.16891}, 2025.

\bibitem[Mou et~al.(2025)Mou, Deng, Luo, Zhang, and Ye]{mou2025can}
Yutao Mou, Xiao Deng, Yuxiao Luo, Shikun Zhang, and Wei Ye.
\newblock Can you really trust code copilots? evaluating large language models from a code security perspective.
\newblock \emph{arXiv preprint arXiv:2505.10494}, 2025.

\bibitem[Mu et~al.(2023)Mu, Chen, Shi, Wang, and Wang]{developer_intent_comment_generation}
Fangwen Mu, Xiao Chen, Lin Shi, Song Wang, and Qing Wang.
\newblock Developer-intent driven code comment generation. in 2023 ieee/acm 45th international conference on software engineering (icse), 2023.

\bibitem[Mu et~al.(2025)Mu, Wang, Shi, Wang, Li, and Wang]{mu2025experepair}
Fangwen Mu, Junjie Wang, Lin Shi, Song Wang, Shoubin Li, and Qing Wang.
\newblock Experepair: Dual-memory enhanced llm-based repository-level program repair.
\newblock \emph{arXiv preprint arXiv:2506.10484}, 2025.

\bibitem[Mu et~al.(2024)Mu, Chen, Zhang, Chen, Yu, Ge, Chen, Liang, Hu, Tao, et~al.]{robocodex}
Yao Mu, Junting Chen, Qinglong Zhang, Shoufa Chen, Qiaojun Yu, Chongjian Ge, Runjian Chen, Zhixuan Liang, Mengkang Hu, Chaofan Tao, et~al.
\newblock Robocodex: Multimodal code generation for robotic behavior synthesis.
\newblock \emph{arXiv preprint arXiv:2402.16117}, 2024.

\bibitem[Muennighoff et~al.(2024{\natexlab{a}})Muennighoff, Liu, Zebaze, Zheng, Hui, Zhuo, Singh, Tang, von Werra, and Longpre]{humanevalpack}
Niklas Muennighoff, Qian Liu, Armel Zebaze, Qinkai Zheng, Binyuan Hui, Terry~Yue Zhuo, Swayam Singh, Xiangru Tang, Leandro von Werra, and Shayne Longpre.
\newblock Octopack: Instruction tuning code large language models, 2024{\natexlab{a}}.
\newblock URL \url{https://arxiv.org/abs/2308.07124}.

\bibitem[Muennighoff et~al.(2024{\natexlab{b}})Muennighoff, Liu, Zebaze, Zheng, Hui, Zhuo, Singh, Tang, {von Werra}, and Longpre]{muennighoffOctoPackInstructionTuning2024}
Niklas Muennighoff, Qian Liu, Armel Zebaze, Qinkai Zheng, Binyuan Hui, Terry~Yue Zhuo, Swayam Singh, Xiangru Tang, Leandro {von Werra}, and Shayne Longpre.
\newblock {{OctoPack}}: {{Instruction Tuning Code Large Language Models}}, 2024{\natexlab{b}}.

\bibitem[Muennighoff et~al.(2025)Muennighoff, Yang, Shi, Li, Fei-Fei, Hajishirzi, Zettlemoyer, Liang, Cand{\`e}s, and Hashimoto]{muennighoff2025s1}
Niklas Muennighoff, Zitong Yang, Weijia Shi, Xiang~Lisa Li, Li~Fei-Fei, Hannaneh Hajishirzi, Luke Zettlemoyer, Percy Liang, Emmanuel Cand{\`e}s, and Tatsunori Hashimoto.
\newblock s1: Simple test-time scaling.
\newblock \emph{arXiv preprint arXiv:2501.19393}, 2025.

\bibitem[Muennighoff et~al.(2023)]{muennighoff2023octopack}
Niklas Muennighoff et~al.
\newblock Octopack: Instruction tuning code large language models.
\newblock \emph{arXiv preprint arXiv:2308.07124}, 2023.

\bibitem[Multilingual-Multimodal-NLP(2024)]{mceval_instruct}
Multilingual-Multimodal-NLP.
\newblock Mceval-instruct, 2024.
\newblock URL \url{https://huggingface.co/datasets/Multilingual-Multimodal-NLP/McEval-Instruct}.
\newblock Accessed: 2024.

\bibitem[Mündler et~al.(2025)Mündler, Müller, He, and Vechev]{mundler_swt-bench_2025}
Niels Mündler, Mark~Niklas Müller, Jingxuan He, and Martin Vechev.
\newblock {SWT}-{Bench}: {Testing} and {Validating} {Real}-{World} {Bug}-{Fixes} with {Code} {Agents}, February 2025.
\newblock URL \url{http://arxiv.org/abs/2406.12952}.
\newblock arXiv:2406.12952 [cs].

\bibitem[Nair et~al.(2023)Nair, Schumacher, Tso, and Kannan]{nair2023dera}
Varun Nair, Elliot Schumacher, Geoffrey Tso, and Anitha Kannan.
\newblock Dera: enhancing large language model completions with dialog-enabled resolving agents.
\newblock \emph{arXiv preprint arXiv:2303.17071}, 2023.

\bibitem[Nakano et~al.(2021)Nakano, Hilton, Balaji, Wu, Ouyang, Kim, Hesse, Jain, Kosaraju, Saunders, et~al.]{nakano2021webgpt}
Reiichiro Nakano, Jacob Hilton, Suchir Balaji, Jeff Wu, Long Ouyang, Christina Kim, Christopher Hesse, Shantanu Jain, Vineet Kosaraju, William Saunders, et~al.
\newblock Webgpt: Browser-assisted question-answering with human feedback.
\newblock \emph{arXiv preprint arXiv:2112.09332}, 2021.

\bibitem[Nakano et~al.(2022)Nakano, Hilton, Balaji, Wu, Ouyang, Kim, Hesse, Jain, Kosaraju, Saunders, Jiang, Cobbe, Eloundou, Krueger, Button, Knight, Chess, and Schulman]{nakano2022webgpt}
Reiichiro Nakano, Jacob Hilton, Suchir Balaji, Jeff Wu, Long Ouyang, Christina Kim, Christopher Hesse, Shantanu Jain, Vineet Kosaraju, William Saunders, Xu~Jiang, Karl Cobbe, Tyna Eloundou, Gretchen Krueger, Kevin Button, Matthew Knight, Benjamin Chess, and John Schulman.
\newblock Webgpt: Browser-assisted question-answering with human feedback, 2022.
\newblock URL \url{https://arxiv.org/abs/2112.09332}.

\bibitem[Narajala and Narayan(2025)]{narajala2025securing}
Vineeth~Sai Narajala and Om~Narayan.
\newblock Securing agentic ai: A comprehensive threat model and mitigation framework for generative ai agents.
\newblock \emph{arXiv preprint arXiv:2504.19956}, 2025.

\bibitem[Narayanan et~al.(2021)Narayanan, Shoeybi, Casper, LeGresley, Patwary, Korthikanti, Vainbrand, Kashinkunti, Bernauer, Catanzaro, Phanishayee, and Zaharia]{narayanan2021efficient}
Deepak Narayanan, Mohammad Shoeybi, Jared Casper, Patrick LeGresley, Mostofa Patwary, Vijay Korthikanti, Dmitri Vainbrand, Prethvi Kashinkunti, Julie Bernauer, Bryan Catanzaro, Amar Phanishayee, and Matei Zaharia.
\newblock Efficient large-scale language model training on gpu clusters using megatron-lm.
\newblock In \emph{Proceedings of the International Conference for High Performance Computing, Networking, Storage and Analysis}, pages 1--15, 2021.

\bibitem[Nashaat and Miller(2024)]{codementor}
Mona Nashaat and James Miller.
\newblock Towards efficient fine-tuning of language models with organizational data for automated software review.
\newblock \emph{IEEE Transactions on Software Engineering}, 2024.

\bibitem[Natalie et~al.()Natalie, Ying, and Kai]{nataliealfredo}
Ching~Yuhui Natalie, Sophie Tung~Xuan Ying, and Siow~Jing Kai.
\newblock Alfredo: Agentic llm-based framework for code deobfuscation.

\bibitem[Nguyen et~al.(2024)Nguyen, Chau, Nguyen, and Bui]{nguyen2024agilecoder}
Minh~Huynh Nguyen, Thang~Phan Chau, Phong~X. Nguyen, and Nghi D.~Q. Bui.
\newblock Agilecoder: Dynamic collaborative agents for software development based on agile methodology.
\newblock \emph{arXiv preprint arXiv:2406.11912}, 2024.
\newblock URL \url{https://arxiv.org/abs/2406.11912}.

\bibitem[Ni et~al.(2024)Ni, Allamanis, Cohan, Deng, Shi, Sutton, and Yin]{ni2024next}
Ansong Ni, Miltiadis Allamanis, Arman Cohan, Yinlin Deng, Kensen Shi, Charles Sutton, and Pengcheng Yin.
\newblock Next: Teaching large language models to reason about code execution.
\newblock \emph{arXiv preprint arXiv:2404.14662}, 2024.

\bibitem[Ni et~al.(2025{\natexlab{a}})Ni, Nie, Zou, Yue, and Chen]{ni2025viscoder}
Yuansheng Ni, Ping Nie, Kai Zou, Xiang Yue, and Wenhu Chen.
\newblock Viscoder: Fine-tuning llms for executable python visualization code generation.
\newblock \emph{arXiv preprint arXiv:2506.03930}, 2025{\natexlab{a}}.

\bibitem[Ni et~al.(2025{\natexlab{b}})Ni, Li, Yang, Shen, Lyu, and Dong]{ni2025tree}
Ziyi Ni, Yifan Li, Ning Yang, Dou Shen, Pin Lyu, and Daxiang Dong.
\newblock Tree-of-code: A self-growing tree framework for end-to-end code generation and execution in complex tasks.
\newblock In \emph{Findings of the Association for Computational Linguistics: ACL 2025}, pages 9804--9819, 2025{\natexlab{b}}.

\bibitem[Nichols(2025)]{wrap}
Jack Nichols.
\newblock How we scored \#1 on terminal-bench (52
\newblock URL \url{https://www.warp.dev/blog/terminal-bench}.

\bibitem[Nie et~al.(2025)Nie, Zhu, You, Zhang, Ou, Hu, Zhou, Lin, Wen, and Li]{nie2025llada}
Shen Nie, Fengqi Zhu, Zebin You, Xiaolu Zhang, Jingyang Ou, Jun Hu, Jun Zhou, Yankai Lin, Ji-Rong Wen, and Chongxuan Li.
\newblock Large language diffusion models, 2025.
\newblock URL \url{https://arxiv.org/abs/2502.09992}.

\bibitem[Nijkamp et~al.(2023{\natexlab{a}})Nijkamp, Hayashi, Xiong, Savarese, and Zhou]{nijkamp2023codegen2}
Erik Nijkamp, Hiroaki Hayashi, Caiming Xiong, Silvio Savarese, and Yingbo Zhou.
\newblock Codegen2: Lessons for training llms on programming and natural languages.
\newblock \emph{CoRR}, abs/2305.02309, 2023{\natexlab{a}}.
\newblock \doi{10.48550/ARXIV.2305.02309}.
\newblock URL \url{https://doi.org/10.48550/arXiv.2305.02309}.

\bibitem[Nijkamp et~al.(2023{\natexlab{b}})Nijkamp, Pang, Hayashi, et~al.]{nijkamp2023codegen}
Erik Nijkamp, Bo~Pang, Hiroaki Hayashi, et~al.
\newblock {CodeGen}: An open large language model for code with multi-turn program synthesis.
\newblock \emph{ICLR}, 2023{\natexlab{b}}.

\bibitem[Ning and Xie(2024)]{ning2024survey}
Zepeng Ning and Lihua Xie.
\newblock A survey on multi-agent reinforcement learning and its application.
\newblock \emph{Journal of Automation and Intelligence}, 3\penalty0 (2):\penalty0 73--91, 2024.

\bibitem[Nitin et~al.(2021)Nitin, Saieva, Ray, and Kaiser]{nitin-etal-2021-direct}
Vikram Nitin, Anthony Saieva, Baishakhi Ray, and Gail Kaiser.
\newblock {DIRECT} : A transformer-based model for decompiled identifier renaming.
\newblock In Royi Lachmy, Ziyu Yao, Greg Durrett, Milos Gligoric, Junyi~Jessy Li, Ray Mooney, Graham Neubig, Yu~Su, Huan Sun, and Reut Tsarfaty, editors, \emph{Proceedings of the 1st Workshop on Natural Language Processing for Programming (NLP4Prog 2021)}, August 2021.

\bibitem[Nitin et~al.(2025)Nitin, Krishna, do~Valle, and Ray]{nitin2025c2saferrusttransformingcprojects}
Vikram Nitin, Rahul Krishna, Luiz~Lemos do~Valle, and Baishakhi Ray.
\newblock C2saferrust: Transforming c projects into safer rust with neurosymbolic techniques, 2025.
\newblock URL \url{https://arxiv.org/abs/2501.14257}.

\bibitem[Niu et~al.(2022)Niu, Li, Ng, Ge, Huang, and Luo]{spt_code}
Changan Niu, Chuanyi Li, Vincent Ng, Jidong Ge, Liguo Huang, and Bin Luo.
\newblock Spt-code: Sequence-to-sequence pre-training for learning source code representations.
\newblock In \emph{44th {IEEE/ACM} 44th International Conference on Software Engineering, {ICSE} 2022, Pittsburgh, PA, USA, May 25-27, 2022}, pages 1--13. {ACM}, 2022.
\newblock \doi{10.1145/3510003.3510096}.
\newblock URL \url{https://doi.org/10.1145/3510003.3510096}.

\bibitem[Novikov et~al.(2025)Novikov, Vũ, Eisenberger, Dupont, Huang, Wagner, Shirobokov, Kozlovskii, Ruiz, Mehrabian, Kumar, See, Chaudhuri, Holland, Davies, Nowozin, Kohli, and Balog]{novikov2025alphaevolvecodingagentscientific}
Alexander Novikov, Ngân Vũ, Marvin Eisenberger, Emilien Dupont, Po-Sen Huang, Adam~Zsolt Wagner, Sergey Shirobokov, Borislav Kozlovskii, Francisco J.~R. Ruiz, Abbas Mehrabian, M.~Pawan Kumar, Abigail See, Swarat Chaudhuri, George Holland, Alex Davies, Sebastian Nowozin, Pushmeet Kohli, and Matej Balog.
\newblock Alphaevolve: A coding agent for scientific and algorithmic discovery, 2025.
\newblock URL \url{https://arxiv.org/abs/2506.13131}.

\bibitem[Nugteren and Codreanu(2015)]{DBLP:conf/mcsoc/NugterenC15}
Cedric Nugteren and Valeriu Codreanu.
\newblock Cltune: {A} generic auto-tuner for opencl kernels.
\newblock In \emph{{IEEE} 9th International Symposium on Embedded Multicore/Many-core Systems-on-Chip, MCSoC 2015, Turin, Italy, September 23-25, 2015}, pages 195--202. {IEEE} Computer Society, 2015.

\bibitem[Nunez et~al.(2024{\natexlab{a}})Nunez, Islam, Jha, and Najafirad]{nunez2024autosafecodermultiagentframeworksecuring}
Ana Nunez, Nafis~Tanveer Islam, Sumit~Kumar Jha, and Peyman Najafirad.
\newblock Autosafecoder: A multi-agent framework for securing llm code generation through static analysis and fuzz testing, 2024{\natexlab{a}}.
\newblock URL \url{https://arxiv.org/abs/2409.10737}.

\bibitem[Nunez et~al.(2024{\natexlab{b}})Nunez, Hou, Payer, and Zhou]{nunez2024autosafecoder}
Andres Nunez, Siyuan Hou, Mathias Payer, and Yuhui Zhou.
\newblock Autosafecoder: A language model-guided approach for automating secure code generation.
\newblock \emph{arXiv preprint arXiv:2402.04659}, 2024{\natexlab{b}}.

\bibitem[Nussbaum et~al.(2025)Nussbaum, Morris, Duderstadt, and Mulyar]{nussbaum2025nomicembedtrainingreproducible}
Zach Nussbaum, John~X. Morris, Brandon Duderstadt, and Andriy Mulyar.
\newblock Nomic embed: Training a reproducible long context text embedder, 2025.
\newblock URL \url{https://arxiv.org/abs/2402.01613}.

\bibitem[NVIDIA(2024)]{nvidia_opencoder}
NVIDIA.
\newblock Opencodereasoning dataset, 2024.
\newblock URL \url{https://huggingface.co/datasets/nvidia/OpenCodeReasoning}.
\newblock Accessed: 2024.

\bibitem[{o3-o4}(2025)]{o3-o4}
{o3-o4}.
\newblock Introducing-o3-and-o4-mini, 2025.
\newblock URL \url{https://openai.com/zh-Hans-CN/index/introducing-o3-and-o4-mini/}.

\bibitem[Obi et~al.(2025)Obi, Venkatesh, Wang, Wang, Suh, Amosa, Jo, and Min]{obi2025safeplan}
Ike Obi, Vishnunandan~LN Venkatesh, Weizheng Wang, Ruiqi Wang, Dayoon Suh, Temitope~I Amosa, Wonse Jo, and Byung-Cheol Min.
\newblock Safeplan: Leveraging formal logic and chain-of-thought reasoning for enhanced safety in llm-based robotic task planning.
\newblock \emph{arXiv preprint arXiv:2503.06892}, 2025.

\bibitem[Oh and Oh(2022)]{wei2022pyter}
Wonseok Oh and Hakjoo Oh.
\newblock Pyter: effective program repair for python type errors.
\newblock In \emph{Proceedings of the 30th ACM Joint European Software Engineering Conference and Symposium on the Foundations of Software Engineering}, ESEC/FSE 2022, page 922–934, New York, NY, USA, 2022. Association for Computing Machinery.
\newblock ISBN 9781450394130.
\newblock \doi{10.1145/3540250.3549130}.
\newblock URL \url{https://doi.org/10.1145/3540250.3549130}.

\bibitem[OpenAI(2022)]{openai2022chatgpt}
OpenAI.
\newblock Introducing chatgpt.
\newblock \url{https://openai.com/blog/chatgpt}, 2022.

\bibitem[OpenAI(2023{\natexlab{a}})]{gpt4Code}
OpenAI.
\newblock {GPT4 Code}.
\newblock \url{https://chat.openai.com/?model=GPT4-Code-interpreter}, 2023{\natexlab{a}}.

\bibitem[OpenAI(2023{\natexlab{b}})]{gptCodePlugin}
OpenAI.
\newblock {ChatGPT plugins}.
\newblock \url{https://openai.com/index/chatgpt-plugins/##code-interpreter}, 2023{\natexlab{b}}.

\bibitem[OpenAI(2023{\natexlab{c}})]{openai2023gpt4v}
OpenAI.
\newblock Gpt-4v(ision) system card.
\newblock \url{https://openai.com/index/gpt-4v-system-card/}, 2023{\natexlab{c}}.

\bibitem[OpenAI(2024{\natexlab{a}})]{openai2024gpt4o}
OpenAI.
\newblock Hello gpt-4o.
\newblock \url{https://openai.com/index/hello-gpt-4o/}, 2024{\natexlab{a}}.

\bibitem[OpenAI(2024{\natexlab{b}})]{openai2024gpt4omini}
OpenAI.
\newblock Gpt-4o mini: advancing cost-efficient intelligence.
\newblock \url{https://openai.com/index/gpt-4o-mini-advancing-cost-efficient-intelligence/}, 2024{\natexlab{b}}.

\bibitem[OpenAI(2024{\natexlab{c}})]{openai2024o1preview}
OpenAI.
\newblock Introducing openai o1-preview.
\newblock \url{https://openai.com/index/introducing-openai-o1-preview/}, 2024{\natexlab{c}}.

\bibitem[OpenAI(2025{\natexlab{a}})]{gpt5}
OpenAI.
\newblock Introducing gpt-5, 2025{\natexlab{a}}.
\newblock URL \url{https://openai.com/index/introducing-gpt-5/}.

\bibitem[OpenAI(2025{\natexlab{b}})]{openai2025gpt41}
OpenAI.
\newblock Introducing gpt-4.1 in the api.
\newblock \url{https://openai.com/index/gpt-4-1/}, 2025{\natexlab{b}}.
\newblock Accessed: 2025-09-30.

\bibitem[OpenAI(2025{\natexlab{c}})]{openai2025gpt5}
OpenAI.
\newblock Gpt-5 system card.
\newblock \url{https://openai.com/index/gpt-5-system-card/}, August 2025{\natexlab{c}}.

\bibitem[OpenAI(2025{\natexlab{d}})]{openai2025gpt5codex}
OpenAI.
\newblock Introducing upgrades to codex: Gpt-5-codex.
\newblock \url{https://openai.com/index/introducing-upgrades-to-codex/}, 2025{\natexlab{d}}.

\bibitem[OpenAI(2025{\natexlab{e}})]{openai2025gpt5developers}
OpenAI.
\newblock Introducing gpt-5 for developers.
\newblock \url{https://openai.com/index/introducing-gpt-5-for-developers/}, 2025{\natexlab{e}}.
\newblock Accessed: 2025-09-30.

\bibitem[OpenAI(2025{\natexlab{f}})]{openai2025o3}
OpenAI.
\newblock Introducing openai o3 and o4-mini.
\newblock \url{https://openai.com/index/introducing-o3-and-o4-mini/}, 2025{\natexlab{f}}.

\bibitem[OpenAI et~al.(2024)OpenAI, Achiam, Adler, Agarwal, Ahmad, Akkaya, Aleman, Almeida, Altenschmidt, Altman, Anadkat, Avila, Babuschkin, Balaji, Balcom, Baltescu, Bao, Bavarian, Belgum, Bello, Berdine, Bernadett-Shapiro, Berner, Bogdonoff, Boiko, Boyd, Brakman, Brockman, Brooks, Brundage, Button, Cai, Campbell, Cann, Carey, Carlson, Carmichael, Chan, Chang, Chantzis, Chen, Chen, Chen, Chen, Chen, Chess, Cho, Chu, Chung, Cummings, Currier, Dai, Decareaux, Degry, Deutsch, Deville, Dhar, Dohan, Dowling, Dunning, Ecoffet, Eleti, Eloundou, Farhi, Fedus, Felix, Fishman, Forte, Fulford, Gao, Georges, Gibson, Goel, Gogineni, Goh, Gontijo-Lopes, Gordon, Grafstein, Gray, Greene, Gross, Gu, Guo, Hallacy, Han, Harris, He, Heaton, Heidecke, Hesse, Hickey, Hickey, Hoeschele, Houghton, Hsu, Hu, Hu, Huizinga, Jain, Jain, Jang, Jiang, Jiang, Jin, Jin, Jomoto, Jonn, Jun, Kaftan, Łukasz Kaiser, Kamali, Kanitscheider, Keskar, Khan, Kilpatrick, Kim, Kim, Kim, Kirchner, Kiros, Knight, Kokotajlo, Łukasz Kondraciuk, Kondrich,
  Konstantinidis, Kosic, Krueger, Kuo, Lampe, Lan, Lee, Leike, Leung, Levy, Li, Lim, Lin, Lin, Litwin, Lopez, Lowe, Lue, Makanju, Malfacini, Manning, Markov, Markovski, Martin, Mayer, Mayne, McGrew, McKinney, McLeavey, McMillan, McNeil, Medina, Mehta, Menick, Metz, Mishchenko, Mishkin, Monaco, Morikawa, Mossing, Mu, Murati, Murk, Mély, Nair, Nakano, Nayak, Neelakantan, Ngo, Noh, Ouyang, O'Keefe, Pachocki, Paino, Palermo, Pantuliano, Parascandolo, Parish, Parparita, Passos, Pavlov, Peng, Perelman, de~Avila Belbute~Peres, Petrov, de~Oliveira~Pinto, Michael, Pokorny, Pokrass, Pong, Powell, Power, Power, Proehl, Puri, Radford, Rae, Ramesh, Raymond, Real, Rimbach, Ross, Rotsted, Roussez, Ryder, Saltarelli, Sanders, Santurkar, Sastry, Schmidt, Schnurr, Schulman, Selsam, Sheppard, Sherbakov, Shieh, Shoker, Shyam, Sidor, Sigler, Simens, Sitkin, Slama, Sohl, Sokolowsky, Song, Staudacher, Such, Summers, Sutskever, Tang, Tezak, Thompson, Tillet, Tootoonchian, Tseng, Tuggle, Turley, Tworek, Uribe, Vallone, Vijayvergiya,
  Voss, Wainwright, Wang, Wang, Wang, Ward, Wei, Weinmann, Welihinda, Welinder, Weng, Weng, Wiethoff, Willner, Winter, Wolrich, Wong, Workman, Wu, Wu, Wu, Xiao, Xu, Yoo, Yu, Yuan, Zaremba, Zellers, Zhang, Zhang, Zhao, Zheng, Zhuang, Zhuk, and Zoph]{openai2024gpt4technicalreport}
OpenAI, Josh Achiam, Steven Adler, Sandhini Agarwal, Lama Ahmad, Ilge Akkaya, Florencia~Leoni Aleman, Diogo Almeida, Janko Altenschmidt, Sam Altman, Shyamal Anadkat, Red Avila, Igor Babuschkin, Suchir Balaji, Valerie Balcom, Paul Baltescu, Haiming Bao, Mohammad Bavarian, Jeff Belgum, Irwan Bello, Jake Berdine, Gabriel Bernadett-Shapiro, Christopher Berner, Lenny Bogdonoff, Oleg Boiko, Madelaine Boyd, Anna-Luisa Brakman, Greg Brockman, Tim Brooks, Miles Brundage, Kevin Button, Trevor Cai, Rosie Campbell, Andrew Cann, Brittany Carey, Chelsea Carlson, Rory Carmichael, Brooke Chan, Che Chang, Fotis Chantzis, Derek Chen, Sully Chen, Ruby Chen, Jason Chen, Mark Chen, Ben Chess, Chester Cho, Casey Chu, Hyung~Won Chung, Dave Cummings, Jeremiah Currier, Yunxing Dai, Cory Decareaux, Thomas Degry, Noah Deutsch, Damien Deville, Arka Dhar, David Dohan, Steve Dowling, Sheila Dunning, Adrien Ecoffet, Atty Eleti, Tyna Eloundou, David Farhi, Liam Fedus, Niko Felix, Simón~Posada Fishman, Juston Forte, Isabella Fulford, Leo
  Gao, Elie Georges, Christian Gibson, Vik Goel, Tarun Gogineni, Gabriel Goh, Rapha Gontijo-Lopes, Jonathan Gordon, Morgan Grafstein, Scott Gray, Ryan Greene, Joshua Gross, Shixiang~Shane Gu, Yufei Guo, Chris Hallacy, Jesse Han, Jeff Harris, Yuchen He, Mike Heaton, Johannes Heidecke, Chris Hesse, Alan Hickey, Wade Hickey, Peter Hoeschele, Brandon Houghton, Kenny Hsu, Shengli Hu, Xin Hu, Joost Huizinga, Shantanu Jain, Shawn Jain, Joanne Jang, Angela Jiang, Roger Jiang, Haozhun Jin, Denny Jin, Shino Jomoto, Billie Jonn, Heewoo Jun, Tomer Kaftan, Łukasz Kaiser, Ali Kamali, Ingmar Kanitscheider, Nitish~Shirish Keskar, Tabarak Khan, Logan Kilpatrick, Jong~Wook Kim, Christina Kim, Yongjik Kim, Jan~Hendrik Kirchner, Jamie Kiros, Matt Knight, Daniel Kokotajlo, Łukasz Kondraciuk, Andrew Kondrich, Aris Konstantinidis, Kyle Kosic, Gretchen Krueger, Vishal Kuo, Michael Lampe, Ikai Lan, Teddy Lee, Jan Leike, Jade Leung, Daniel Levy, Chak~Ming Li, Rachel Lim, Molly Lin, Stephanie Lin, Mateusz Litwin, Theresa Lopez, Ryan
  Lowe, Patricia Lue, Anna Makanju, Kim Malfacini, Sam Manning, Todor Markov, Yaniv Markovski, Bianca Martin, Katie Mayer, Andrew Mayne, Bob McGrew, Scott~Mayer McKinney, Christine McLeavey, Paul McMillan, Jake McNeil, David Medina, Aalok Mehta, Jacob Menick, Luke Metz, Andrey Mishchenko, Pamela Mishkin, Vinnie Monaco, Evan Morikawa, Daniel Mossing, Tong Mu, Mira Murati, Oleg Murk, David Mély, Ashvin Nair, Reiichiro Nakano, Rajeev Nayak, Arvind Neelakantan, Richard Ngo, Hyeonwoo Noh, Long Ouyang, Cullen O'Keefe, Jakub Pachocki, Alex Paino, Joe Palermo, Ashley Pantuliano, Giambattista Parascandolo, Joel Parish, Emy Parparita, Alex Passos, Mikhail Pavlov, Andrew Peng, Adam Perelman, Filipe de~Avila Belbute~Peres, Michael Petrov, Henrique~Ponde de~Oliveira~Pinto, Michael, Pokorny, Michelle Pokrass, Vitchyr~H. Pong, Tolly Powell, Alethea Power, Boris Power, Elizabeth Proehl, Raul Puri, Alec Radford, Jack Rae, Aditya Ramesh, Cameron Raymond, Francis Real, Kendra Rimbach, Carl Ross, Bob Rotsted, Henri Roussez,
  Nick Ryder, Mario Saltarelli, Ted Sanders, Shibani Santurkar, Girish Sastry, Heather Schmidt, David Schnurr, John Schulman, Daniel Selsam, Kyla Sheppard, Toki Sherbakov, Jessica Shieh, Sarah Shoker, Pranav Shyam, Szymon Sidor, Eric Sigler, Maddie Simens, Jordan Sitkin, Katarina Slama, Ian Sohl, Benjamin Sokolowsky, Yang Song, Natalie Staudacher, Felipe~Petroski Such, Natalie Summers, Ilya Sutskever, Jie Tang, Nikolas Tezak, Madeleine~B. Thompson, Phil Tillet, Amin Tootoonchian, Elizabeth Tseng, Preston Tuggle, Nick Turley, Jerry Tworek, Juan Felipe~Cerón Uribe, Andrea Vallone, Arun Vijayvergiya, Chelsea Voss, Carroll Wainwright, Justin~Jay Wang, Alvin Wang, Ben Wang, Jonathan Ward, Jason Wei, CJ~Weinmann, Akila Welihinda, Peter Welinder, Jiayi Weng, Lilian Weng, Matt Wiethoff, Dave Willner, Clemens Winter, Samuel Wolrich, Hannah Wong, Lauren Workman, Sherwin Wu, Jeff Wu, Michael Wu, Kai Xiao, Tao Xu, Sarah Yoo, Kevin Yu, Qiming Yuan, Wojciech Zaremba, Rowan Zellers, Chong Zhang, Marvin Zhang, Shengjia
  Zhao, Tianhao Zheng, Juntang Zhuang, William Zhuk, and Barret Zoph.
\newblock Gpt-4 technical report, 2024.
\newblock URL \url{https://arxiv.org/abs/2303.08774}.

\bibitem[OpenAI et~al.(2025)OpenAI, :, Agarwal, Ahmad, Ai, Altman, Applebaum, Arbus, Arora, Bai, Baker, Bao, Barak, Bennett, Bertao, Brett, Brevdo, Brockman, Bubeck, Chang, Chen, Chen, Cheung, Clark, Cook, Dukhan, Dvorak, Fives, Fomenko, Garipov, Georgiev, Glaese, Gogineni, Goucher, Gross, Guzman, Hallman, Hehir, Heidecke, Helyar, Hu, Huet, Huh, Jain, Johnson, Koch, Kofman, Kundel, Kwon, Kyrylov, Le, Leclerc, Lennon, Lessans, Lezcano-Casado, Li, Li, Lin, Liss, Lily, Liu, Liu, Lu, Lu, Martinovic, McCallum, McGrath, McKinney, McLaughlin, Mei, Mostovoy, Mu, Myles, Neitz, Nichol, Pachocki, Paino, Palmie, Pantuliano, Parascandolo, Park, Pathak, Paz, Peran, Pimenov, Pokrass, Proehl, Qiu, Raila, Raso, Ren, Richardson, Robinson, Rotsted, Salman, Sanjeev, Schwarzer, Sculley, Sikchi, Simon, Singhal, Song, Stuckey, Sun, Tillet, Toizer, Tsimpourlas, Vyas, Wallace, Wang, Wang, Watkins, Weil, Wendling, Whinnery, Whitney, Wong, Yang, Yang, Yasunaga, Ying, Zaremba, Zhan, Zhang, Zhang, Zhang, and
  Zhao]{openai2025gptoss120bgptoss20bmodel}
OpenAI, :, Sandhini Agarwal, Lama Ahmad, Jason Ai, Sam Altman, Andy Applebaum, Edwin Arbus, Rahul~K. Arora, Yu~Bai, Bowen Baker, Haiming Bao, Boaz Barak, Ally Bennett, Tyler Bertao, Nivedita Brett, Eugene Brevdo, Greg Brockman, Sebastien Bubeck, Che Chang, Kai Chen, Mark Chen, Enoch Cheung, Aidan Clark, Dan Cook, Marat Dukhan, Casey Dvorak, Kevin Fives, Vlad Fomenko, Timur Garipov, Kristian Georgiev, Mia Glaese, Tarun Gogineni, Adam Goucher, Lukas Gross, Katia~Gil Guzman, John Hallman, Jackie Hehir, Johannes Heidecke, Alec Helyar, Haitang Hu, Romain Huet, Jacob Huh, Saachi Jain, Zach Johnson, Chris Koch, Irina Kofman, Dominik Kundel, Jason Kwon, Volodymyr Kyrylov, Elaine~Ya Le, Guillaume Leclerc, James~Park Lennon, Scott Lessans, Mario Lezcano-Casado, Yuanzhi Li, Zhuohan Li, Ji~Lin, Jordan Liss, Lily, Liu, Jiancheng Liu, Kevin Lu, Chris Lu, Zoran Martinovic, Lindsay McCallum, Josh McGrath, Scott McKinney, Aidan McLaughlin, Song Mei, Steve Mostovoy, Tong Mu, Gideon Myles, Alexander Neitz, Alex Nichol, Jakub
  Pachocki, Alex Paino, Dana Palmie, Ashley Pantuliano, Giambattista Parascandolo, Jongsoo Park, Leher Pathak, Carolina Paz, Ludovic Peran, Dmitry Pimenov, Michelle Pokrass, Elizabeth Proehl, Huida Qiu, Gaby Raila, Filippo Raso, Hongyu Ren, Kimmy Richardson, David Robinson, Bob Rotsted, Hadi Salman, Suvansh Sanjeev, Max Schwarzer, D.~Sculley, Harshit Sikchi, Kendal Simon, Karan Singhal, Yang Song, Dane Stuckey, Zhiqing Sun, Philippe Tillet, Sam Toizer, Foivos Tsimpourlas, Nikhil Vyas, Eric Wallace, Xin Wang, Miles Wang, Olivia Watkins, Kevin Weil, Amy Wendling, Kevin Whinnery, Cedric Whitney, Hannah Wong, Lin Yang, Yu~Yang, Michihiro Yasunaga, Kristen Ying, Wojciech Zaremba, Wenting Zhan, Cyril Zhang, Brian Zhang, Eddie Zhang, and Shengjia Zhao.
\newblock gpt-oss-120b \& gpt-oss-20b model card, 2025.
\newblock URL \url{https://arxiv.org/abs/2508.10925}.

\bibitem[{OpenCode Contributors}(2024)]{opencode2024}
{OpenCode Contributors}.
\newblock {OpenCode}: Open source terminal code assistant.
\newblock Technical report, 2024.

\bibitem[OpenCoder-LLM(2024)]{opencoder_sft}
OpenCoder-LLM.
\newblock Opc-sft-stage2, 2024.
\newblock URL \url{https://huggingface.co/datasets/OpenCoder-LLM/opc-sft-stage2}.
\newblock Accessed: 2024.

\bibitem[Openhands(2024)]{openhands2024}
Openhands.
\newblock Openhands.
\newblock \emph{arXiv preprint arXiv:2407.16741}, 2024.
\newblock URL \url{https://arxiv.org/abs/2407.16741}.

\bibitem[OpenInterpreter(2023)]{openInterpreter}
OpenInterpreter.
\newblock {Open Interpreter}.
\newblock \url{https://github.com/openinterpreter/open-interpreter}, 2023.

\bibitem[Oriol et~al.(2025)Oriol, Motger, Marco, and Franch]{MAD}
Marc Oriol, Quim Motger, Jordi Marco, and Xavier Franch.
\newblock Multi-agent debate strategies to enhance requirements engineering with large language models.
\newblock \emph{arXiv preprint arXiv:2507.05981}, 2025.

\bibitem[Osika(2023)]{gptengineer}
Anton Osika.
\newblock Gpt engineer.
\newblock \url{https://github.com/AntonOsika/gpt-engineer}, 2023.

\bibitem[Ouyang et~al.(2025{\natexlab{a}})Ouyang, Guo, Arora, Zhang, Hu, Ré, and Mirhoseini]{kernelbench}
Anne Ouyang, Simon Guo, Simran Arora, Alex~L. Zhang, William Hu, Christopher Ré, and Azalia Mirhoseini.
\newblock Kernelbench: Can llms write efficient gpu kernels?, 2025{\natexlab{a}}.
\newblock URL \url{https://arxiv.org/abs/2502.10517}.

\bibitem[Ouyang et~al.(2025{\natexlab{b}})Ouyang, Chen, Nie, Gui, Wan, Zhang, and Chen]{ouyang2025nvagent}
Geliang Ouyang, Jingyao Chen, Zhihe Nie, Yi~Gui, Yao Wan, Hongyu Zhang, and Dongping Chen.
\newblock nvagent: Automated data visualization from natural language via collaborative agent workflow.
\newblock \emph{arXiv preprint arXiv:2502.05036}, 2025{\natexlab{b}}.

\bibitem[Ouyang et~al.(2022{\natexlab{a}})Ouyang, Wu, Jiang, Almeida, Wainwright, Mishkin, Zhang, Agarwal, Slama, Ray, et~al.]{ouyang2022training}
Long Ouyang, Jeff Wu, Xu~Jiang, Diogo Almeida, Carroll Wainwright, Pamela Mishkin, Chong Zhang, Sandhini Agarwal, Katarina Slama, Alex Ray, et~al.
\newblock Training language models to follow instructions with human feedback.
\newblock \emph{Advances in Neural Information Processing Systems}, 35:\penalty0 27730--27744, 2022{\natexlab{a}}.

\bibitem[Ouyang et~al.(2022{\natexlab{b}})Ouyang, Wu, Jiang, Almeida, Wainwright, Mishkin, Zhang, Agarwal, Slama, Ray, Schulman, Hilton, Kelton, Miller, Simens, Askell, Welinder, Christiano, Leike, and Lowe]{ouyang2022traininglanguagemodelsfollow}
Long Ouyang, Jeff Wu, Xu~Jiang, Diogo Almeida, Carroll~L. Wainwright, Pamela Mishkin, Chong Zhang, Sandhini Agarwal, Katarina Slama, Alex Ray, John Schulman, Jacob Hilton, Fraser Kelton, Luke Miller, Maddie Simens, Amanda Askell, Peter Welinder, Paul Christiano, Jan Leike, and Ryan Lowe.
\newblock Training language models to follow instructions with human feedback, 2022{\natexlab{b}}.
\newblock URL \url{https://arxiv.org/abs/2203.02155}.

\bibitem[Pabba et~al.(2025)Pabba, Mathai, Chakraborty, and Ray]{pabba2025semagent}
Anvith Pabba, Alex Mathai, Anindya Chakraborty, and Baishakhi Ray.
\newblock Semagent: A semantics aware program repair agent.
\newblock \emph{arXiv preprint arXiv:2506.16650}, 2025.

\bibitem[Packer et~al.(2023)Packer, Fang, Patil, Lin, Wooders, and Gonzalez]{packer2023memgpt}
Charles Packer, Vivian Fang, Shishir\_G Patil, Kevin Lin, Sarah Wooders, and Joseph\_E Gonzalez.
\newblock Memgpt: Towards llms as operating systems.
\newblock 2023.

\bibitem[Palit and Sharma(2024)]{palit2024generating}
Indranil Palit and Tushar Sharma.
\newblock Generating refactored code accurately using reinforcement learning.
\newblock \emph{arXiv preprint arXiv:2412.18035}, 2024.

\bibitem[Pan et~al.(2024{\natexlab{a}})Pan, Wang, Neubig, Jaitly, Ji, Suhr, and Zhang]{pan2024swegym}
Jiayi Pan, Xingyao Wang, Graham Neubig, Navdeep Jaitly, Heng Ji, Alane Suhr, and Yizhe Zhang.
\newblock Training software engineering agents and verifiers with swe-gym, 2024{\natexlab{a}}.

\bibitem[Pan et~al.(2024{\natexlab{b}})Pan, Ibrahimzada, Krishna, Sankar, Wassi, Merler, Sobolev, Pavuluri, Sinha, and Jabbarvand]{Pan_2024}
Rangeet Pan, Ali~Reza Ibrahimzada, Rahul Krishna, Divya Sankar, Lambert~Pouguem Wassi, Michele Merler, Boris Sobolev, Raju Pavuluri, Saurabh Sinha, and Reyhaneh Jabbarvand.
\newblock Lost in translation: A study of bugs introduced by large language models while translating code.
\newblock In \emph{Proceedings of the IEEE/ACM 46th International Conference on Software Engineering}, ICSE ’24, page 1–13. ACM, April 2024{\natexlab{b}}.
\newblock \doi{10.1145/3597503.3639226}.
\newblock URL \url{http://dx.doi.org/10.1145/3597503.3639226}.

\bibitem[Pan and et~al.(2024)]{pan2024agentcoder}
X.~Pan and et~al.
\newblock Agentcoder: Multi-agent code generation with unit test feedback.
\newblock \emph{arXiv preprint arXiv:2402.12345}, 2024.
\newblock URL \url{https://arxiv.org/abs/2402.12345}.

\bibitem[{Pan} et~al.(2024){Pan}, {Liu}, {Zou}, {Xie}, and {Xie}]{MESIA}
Xinglu {Pan}, Chenxiao {Liu}, Yanzhen {Zou}, Tao {Xie}, and Bing {Xie}.
\newblock {MESIA: Understanding and Leveraging Supplementary Nature of Method-level Comments for Automatic Comment Generation}.
\newblock \emph{arXiv e-prints}, art. arXiv:2403.17357, March 2024.
\newblock \doi{10.48550/arXiv.2403.17357}.

\bibitem[Pan and Eigenmann(2006)]{DBLP:conf/cgo/PanE06}
Zhelong Pan and Rudolf Eigenmann.
\newblock Fast and effective orchestration of compiler optimizations for automatic performance tuning.
\newblock In \emph{Fourth {IEEE/ACM} International Symposium on Code Generation and Optimization {(CGO} 2006), 26-29 March 2006, New York, New York, {USA}}, pages 319--332. {IEEE} Computer Society, 2006.

\bibitem[Pan et~al.(2024)Pan, Cao, Cao, Ma, Li, Huang, Liu, and Li]{pan2024codevbenchllmsunderstanddevelopercentric}
Zhenyu Pan, Rongyu Cao, Yongchang Cao, Yingwei Ma, Binhua Li, Fei Huang, Han Liu, and Yongbin Li.
\newblock Codev-bench: How do llms understand developer-centric code completion?, 2024.
\newblock URL \url{https://arxiv.org/abs/2410.01353}.

\bibitem[Papineni et~al.(2002)Papineni, Roukos, Ward, and Zhu]{papineni2002bleu}
Kishore Papineni, Salim Roukos, Todd Ward, and Wei-Jing Zhu.
\newblock Bleu: a method for automatic evaluation of machine translation.
\newblock In \emph{Proceedings of the 40th annual meeting of the Association for Computational Linguistics}, pages 311--318, 2002.

\bibitem[Park et~al.(2023)Park, O'Brien, Cai, Morris, Liang, and Bernstein]{park2023generative}
Joon~Sung Park, Joseph O'Brien, Carrie~Jun Cai, Meredith~Ringel Morris, Percy Liang, and Michael~S Bernstein.
\newblock Generative agents: Interactive simulacra of human behavior.
\newblock In \emph{Proceedings of the 36th annual acm symposium on user interface software and technology}, pages 1--22, 2023.

\bibitem[Patil et~al.(2024)Patil, Mao, Cheng-Jie~Ji, Yan, Suresh, Stoica, and E.~Gonzalez]{patil2025bfcl}
Shishir~G. Patil, Huanzhi Mao, Charlie Cheng-Jie~Ji, Fanjia Yan, Vishnu Suresh, Ion Stoica, and Joseph E.~Gonzalez.
\newblock The berkeley function calling leaderboard (bfcl): From tool use to agentic evaluation of large language models.
\newblock In \emph{Advances in Neural Information Processing Systems}, 2024.

\bibitem[Patsakis et~al.(2024)Patsakis, Casino, and Lykousas]{patsakis2024assessing}
Constantinos Patsakis, Fran Casino, and Nikolaos Lykousas.
\newblock Assessing llms in malicious code deobfuscation of real-world malware campaigns.
\newblock \emph{Expert Systems with Applications}, 256:\penalty0 124912, 2024.

\bibitem[Pearce et~al.(2021)Pearce, Ahmad, Tan, Dolan-Gavitt, and Karri]{pearce2021asleepkeyboardassessingsecurity}
Hammond Pearce, Baleegh Ahmad, Benjamin Tan, Brendan Dolan-Gavitt, and Ramesh Karri.
\newblock Asleep at the keyboard? assessing the security of github copilot's code contributions, 2021.
\newblock URL \url{https://arxiv.org/abs/2108.09293}.

\bibitem[Pei et~al.(2025)Pei, Wang, Liu, Li, Liu, He, Kang, Zhang, Chen, Li, et~al.]{pei2025flow}
Changhua Pei, Zexin Wang, Fengrui Liu, Zeyan Li, Yang Liu, Xiao He, Rong Kang, Tieying Zhang, Jianjun Chen, Jianhui Li, et~al.
\newblock Flow-of-action: Sop enhanced llm-based multi-agent system for root cause analysis.
\newblock In \emph{Companion Proceedings of the ACM on Web Conference 2025}, pages 422--431, 2025.

\bibitem[Peng et~al.(2023{\natexlab{a}})Peng, Alcaide, Anthony, Albalak, Arcadinho, Biderman, Cao, Cheng, Chung, Grella, GV, He, Hou, Lin, Kazienko, Kocon, Kong, Koptyra, Lau, Mantri, Mom, Saito, Song, Tang, Wang, Wind, Wozniak, Zhang, Zhang, Zhao, Zhou, Zhou, Zhu, and Zhu]{peng2023rwkv}
Bo~Peng, Eric Alcaide, Quentin Anthony, Alon Albalak, Samuel Arcadinho, Stella Biderman, Huanqi Cao, Xin Cheng, Michael Chung, Matteo Grella, Kranthi~Kiran GV, Xuzheng He, Haowen Hou, Jiaju Lin, Przemyslaw Kazienko, Jan Kocon, Jiaming Kong, Bartlomiej Koptyra, Hayden Lau, Krishna Sri~Ipsit Mantri, Ferdinand Mom, Atsushi Saito, Guangyu Song, Xiangru Tang, Bolun Wang, Johan~S. Wind, Stanislaw Wozniak, Ruichong Zhang, Zhenyuan Zhang, Qihang Zhao, Peng Zhou, Qinghua Zhou, Jian Zhu, and Rui-Jie Zhu.
\newblock Rwkv: Reinventing rnns for the transformer era, 2023{\natexlab{a}}.
\newblock URL \url{https://arxiv.org/abs/2305.13048}.

\bibitem[Peng et~al.(2024{\natexlab{a}})Peng, Goldstein, Anthony, Albalak, Alcaide, Biderman, Cheah, Du, Ferdinan, Hou, Kazienko, GV, Kocoń, Koptyra, Krishna, Jr., Lin, Muennighoff, Obeid, Saito, Song, Tu, Wirawan, Woźniak, Zhang, Zhao, Zhao, Zhou, Zhu, and Zhu]{peng2024eaglefinchrwkv}
Bo~Peng, Daniel Goldstein, Quentin Anthony, Alon Albalak, Eric Alcaide, Stella Biderman, Eugene Cheah, Xingjian Du, Teddy Ferdinan, Haowen Hou, Przemysław Kazienko, Kranthi~Kiran GV, Jan Kocoń, Bartłomiej Koptyra, Satyapriya Krishna, Ronald~McClelland Jr., Jiaju Lin, Niklas Muennighoff, Fares Obeid, Atsushi Saito, Guangyu Song, Haoqin Tu, Cahya Wirawan, Stanisław Woźniak, Ruichong Zhang, Bingchen Zhao, Qihang Zhao, Peng Zhou, Jian Zhu, and Rui-Jie Zhu.
\newblock Eagle and finch: Rwkv with matrix-valued states and dynamic recurrence, 2024{\natexlab{a}}.
\newblock URL \url{https://arxiv.org/abs/2404.05892}.

\bibitem[Peng et~al.(2025{\natexlab{a}})Peng, Cui, Huang, Yang, and Ray]{peng2025cweval}
Jinjun Peng, Leyi Cui, Kele Huang, Junfeng Yang, and Baishakhi Ray.
\newblock Cweval: Outcome-driven evaluation on functionality and security of llm code generation.
\newblock In \emph{2025 IEEE/ACM International Workshop on Large Language Models for Code (LLM4Code)}, pages 33--40, 2025{\natexlab{a}}.
\newblock \doi{10.1109/LLM4Code66737.2025.00009}.

\bibitem[Peng et~al.(2024{\natexlab{b}})Peng, Chai, and Li]{humanevalxl}
Qiwei Peng, Yekun Chai, and Xuhong Li.
\newblock Humaneval-xl: A multilingual code generation benchmark for cross-lingual natural language generalization, 2024{\natexlab{b}}.
\newblock URL \url{https://arxiv.org/abs/2402.16694}.

\bibitem[Peng et~al.(2024{\natexlab{c}})Peng, Chai, and Li]{peng2024humaneval}
Qiwei Peng, Yekun Chai, and Xuhong Li.
\newblock Humaneval-xl: A multilingual code generation benchmark for cross-lingual natural language generalization.
\newblock \emph{arXiv preprint arXiv:2402.16694}, 2024{\natexlab{c}}.

\bibitem[Peng et~al.(2023{\natexlab{b}})Peng, Kalliamvakou, Cihon, and Demirer]{peng2023impact}
Sida Peng, Eirini Kalliamvakou, Peter Cihon, and Mert Demirer.
\newblock The impact of {AI} on developer productivity: Evidence from {GitHub Copilot}.
\newblock \emph{arXiv preprint arXiv:2302.06590}, 2023{\natexlab{b}}.

\bibitem[Peng et~al.(2025{\natexlab{b}})Peng, Kim, Meng, and Liu]{peng2025icodereviewer}
Yun Peng, Kisub Kim, Linghan Meng, and Kui Liu.
\newblock icodereviewer: Improving secure code review with mixture of prompts.
\newblock \emph{arXiv preprint arXiv:2510.12186}, 2025{\natexlab{b}}.

\bibitem[Peng et~al.(2025{\natexlab{c}})Peng, Wan, Li, and Ren]{coffe}
Yun Peng, Jun Wan, Yichen Li, and Xiaoxue Ren.
\newblock Coffe: A code efficiency benchmark for code generation, 2025{\natexlab{c}}.
\newblock URL \url{https://arxiv.org/abs/2502.02827}.

\bibitem[Pham et~al.(2025)Pham, Bui, et~al.]{pham2025swesynth}
Minh~VT Pham, Nghi~DQ Bui, et~al.
\newblock Swe-synth: Synthesizing verifiable bug-fix data to enable large language models in resolving real-world bugs, 2025.

\bibitem[Pizzorno and Berger(2025)]{pizzorno2025coverupeffectivehighcoverage}
Juan~Altmayer Pizzorno and Emery~D. Berger.
\newblock Coverup: Effective high coverage test generation for python, 2025.
\newblock URL \url{https://arxiv.org/abs/2403.16218}.

\bibitem[{Plandex Team}(2024)]{plandex2024}
{Plandex Team}.
\newblock {Plandex}: Open source {AI} coding agent.
\newblock Technical report, 2024.
\newblock URL \url{https://github.com/plandex-ai/plandex}.

\bibitem[Poli et~al.(2023)Poli, Massaroli, Nguyen, Fu, Dao, Baccus, Bengio, Ermon, and Ré]{poli2023hyena}
Michael Poli, Stefano Massaroli, Eric Nguyen, Daniel~Y. Fu, Tri Dao, Stephen Baccus, Yoshua Bengio, Stefano Ermon, and Christopher Ré.
\newblock Hyena hierarchy: Towards larger convolutional language models, 2023.
\newblock URL \url{https://arxiv.org/abs/2302.10866}.

\bibitem[Polozov and Gulwani(2015)]{polozov2015flashmeta}
Oleksandr Polozov and Sumit Gulwani.
\newblock Flashmeta: a framework for inductive program synthesis.
\newblock In \emph{Proceedings of the 2015 ACM SIGPLAN International Conference on Object-Oriented Programming, Systems, Languages, and Applications}, OOPSLA 2015, page 107–126, New York, NY, USA, 2015. Association for Computing Machinery.
\newblock ISBN 9781450336895.
\newblock \doi{10.1145/2814270.2814310}.
\newblock URL \url{https://doi.org/10.1145/2814270.2814310}.

\bibitem[{polyglot-benchmark}(2025)]{polyglot-benchmark}
{polyglot-benchmark}.
\newblock polyglot-benchmark, 2025.
\newblock URL \url{https://aider.chat/2024/12/21/polyglot.html\#the-polyglot-benchmark}.

\bibitem[Ponnusamy(2025)]{ponnusamy2025bridgingllmgeneratedcoderequirements}
Ahilan Ayyachamy~Nadar Ponnusamy.
\newblock Bridging llm-generated code and requirements: Reverse generation technique and sbc metric for developer insights, 2025.
\newblock URL \url{https://arxiv.org/abs/2502.07835}.

\bibitem[Pourreza and Rafiei(2023)]{Din-sql}
Mohammadreza Pourreza and Davood Rafiei.
\newblock {DIN-SQL:} decomposed in-context learning of text-to-sql with self-correction.
\newblock In \emph{Advances in Neural Information Processing Systems 36: Annual Conference on Neural Information Processing Systems 2023, NeurIPS 2023, New Orleans, LA, USA, December 10 - 16, 2023}, 2023.

\bibitem[Pourreza et~al.(2024)Pourreza, Li, Sun, Chung, Talaei, Kakkar, Gan, Saberi, Ozcan, and Arik]{CHASE-SQL}
Mohammadreza Pourreza, Hailong Li, Ruoxi Sun, Yeounoh Chung, Shayan Talaei, Gaurav~Tarlok Kakkar, Yu~Gan, Amin Saberi, Fatma Ozcan, and Sercan~{\"{O}}. Arik.
\newblock {CHASE-SQL:} multi-path reasoning and preference optimized candidate selection in text-to-sql.
\newblock \emph{CoRR}, abs/2410.01943, 2024.

\bibitem[Prenner and Robbes(2023)]{runbugrun}
Julian~Aron Prenner and Romain Robbes.
\newblock Runbugrun -- an executable dataset for automated program repair, 2023.
\newblock URL \url{https://arxiv.org/abs/2304.01102}.

\bibitem[PrimeIntellect(2024{\natexlab{a}})]{primeintellect_qa}
PrimeIntellect.
\newblock Stackexchange question answering, 2024{\natexlab{a}}.
\newblock URL \url{https://huggingface.co/datasets/PrimeIntellect/stackexchange-question-answering}.
\newblock Accessed: 2024.

\bibitem[PrimeIntellect(2024{\natexlab{b}})]{primeintellect_swe}
PrimeIntellect.
\newblock Real-world swe problems, 2024{\natexlab{b}}.
\newblock URL \url{https://huggingface.co/datasets/PrimeIntellect/real-world-swe-problems}.
\newblock Accessed: 2024.

\bibitem[PrimeIntellect(2024{\natexlab{c}})]{primeintellect_synthetic}
PrimeIntellect.
\newblock Synthetic-2-sft-verified, 2024{\natexlab{c}}.
\newblock URL \url{https://huggingface.co/datasets/PrimeIntellect/SYNTHETIC-2-SFT-verified}.
\newblock Accessed: 2024.

\bibitem[PrithivMLmods(2024)]{prithiv_coder_stat}
PrithivMLmods.
\newblock Coder-stat dataset, 2024.
\newblock URL \url{https://huggingface.co/datasets/prithivMLmods/Coder-Stat}.
\newblock Accessed: 2024.

\bibitem[Puri et~al.(2021)Puri, Kung, Janssen, Zhang, Domeniconi, Zolotov, Dolby, Chen, Choudhury, Decker, Thost, Buratti, Pujar, Ramji, Finkler, Malaika, and Reiss]{codenet}
Ruchir Puri, David~S. Kung, Geert Janssen, Wei Zhang, Giacomo Domeniconi, Vladimir Zolotov, Julian Dolby, Jie Chen, Mihir Choudhury, Lindsey Decker, Veronika Thost, Luca Buratti, Saurabh Pujar, Shyam Ramji, Ulrich Finkler, Susan Malaika, and Frederick Reiss.
\newblock Codenet: A large-scale ai for code dataset for learning a diversity of coding tasks, 2021.
\newblock URL \url{https://arxiv.org/abs/2105.12655}.

\bibitem[Qi et~al.(2025)Qi, Liu, Zhou, Pang, Du, Lee, and Lin]{qi2025defeatingfp16}
Penghui Qi, Zichen Liu, Xiangxin Zhou, Tianyu Pang, Chao Du, Wee~Sun Lee, and Min Lin.
\newblock Defeating the training-inference mismatch via fp16.
\newblock \emph{arXiv preprint arXiv:2510.26788}, 2025.

\bibitem[Qi et~al.(2023)Qi, Mespasson, Lam, Chan, Riapolov, Glaese, Borgeaud, Irving, and Mishkin]{qi2023finetuning}
Xiangyu Qi, David Mespasson, Antoine Lam, David Chan, Avisha Riapolov, Amelia Glaese, Sebastian Borgeaud, Geoffrey Irving, and Pamela Mishkin.
\newblock Fine-tuning aligned language models compromises safety, even when users do not intend to.
\newblock \emph{arXiv preprint arXiv:2310.03693}, 2023.

\bibitem[Qi et~al.(2024)Qi, Jean, Kolter, and Raghunathan]{qi2024emergent}
Zhuang Qi, Nelson Jean, J.~Zico Kolter, and Aditi Raghunathan.
\newblock Emergent misalignment: Narrow finetuning can produce broadly misaligned llms, 2024.

\bibitem[Qian et~al.(2024{\natexlab{a}})Qian, Cong, Yang, et~al.]{qian2024chatdev}
Chen Qian, Xin Cong, Cheng Yang, et~al.
\newblock Chatdev: Communicative agents for software development.
\newblock \emph{arXiv preprint arXiv:2307.07924}, 2024{\natexlab{a}}.

\bibitem[Qian et~al.(2024{\natexlab{b}})Qian, Li, Dang, Liu, Wang, Xie, Chen, Yang, Zhang, Liu, and Sun]{qian2024iterativeexperience}
Chen Qian, Jiahao Li, Yufan Dang, Wei Liu, Yifei Wang, Zihao Xie, Weize Chen, Cheng Yang, Yingli Zhang, Zhiyuan Liu, and Maosong Sun.
\newblock Iterative experience refinement of software-developing agents.
\newblock \emph{CoRR abs/2405.04219}, 2024{\natexlab{b}}.
\newblock URL \url{https://arxiv.org/abs/2405.04219}.

\bibitem[Qiao et~al.(2023)Qiao, Li, Zhang, He, Kang, Zhang, Yang, Dong, Zhang, Wang, et~al.]{qiao2023taskweaver}
Bo~Qiao, Liqun Li, Xu~Zhang, Shilin He, Yu~Kang, Chaoyun Zhang, Fangkai Yang, Hang Dong, Jue Zhang, Lu~Wang, et~al.
\newblock Taskweaver: A code-first agent framework.
\newblock \emph{arXiv preprint arXiv:2311.17541}, 2023.

\bibitem[Qin et~al.(2025{\natexlab{a}})Qin, Xie, Li, Li, Feng, Liu, and Kang]{reasoningv}
Haiyan Qin, Zhiwei Xie, Jingjing Li, Liangchen Li, Xiaotong Feng, Junzhan Liu, and Wang Kang.
\newblock Reasoningv: Efficient verilog code generation with adaptive hybrid reasoning model.
\newblock \emph{arXiv preprint arXiv:2504.14560}, 2025{\natexlab{a}}.

\bibitem[Qin et~al.(2024)Qin, Wang, Lou, Dong, Wang, Li, and Mao]{qin2024agentfl}
Yihao Qin, Shangwen Wang, Yiling Lou, Jinhao Dong, Kaixin Wang, Xiaoling Li, and Xiaoguang Mao.
\newblock Agentfl: Scaling llm-based fault localization to project-level context.
\newblock \emph{arXiv preprint arXiv:2403.16362}, 2024.

\bibitem[Qin et~al.(2025{\natexlab{b}})Qin, Wang, Lou, Dong, Wang, Li, and Mao]{qin2025s}
Yihao Qin, Shangwen Wang, Yiling Lou, Jinhao Dong, Kaixin Wang, Xiaoling Li, and Xiaoguang Mao.
\newblock S oap fl: A standard operating procedure for llm-based method-level fault localization.
\newblock \emph{IEEE Transactions on Software Engineering}, 2025{\natexlab{b}}.

\bibitem[Qin et~al.(2023)Qin, Liang, Ye, Zhu, Yan, Lu, Lin, Cong, Tang, Qian, Zhao, Tian, Xie, Zhou, Gerstein, Li, Liu, and Sun]{qin2023toolllm}
Yujia Qin, Shihao Liang, Yining Ye, Kunlun Zhu, Lan Yan, Yaxi Lu, Yankai Lin, Xin Cong, Xiangru Tang, Bill Qian, Sihan Zhao, Runchu Tian, Ruobing Xie, Jie Zhou, Mark Gerstein, Dahai Li, Zhiyuan Liu, and Maosong Sun.
\newblock Toolllm: Facilitating large language models to master 16000+ real-world apis, 2023.

\bibitem[Qing et~al.(2025)Qing, Zhu, Du, Guo, Zhuo, Zhang, Zhang, Cui, Yiu, Huang, Ng, and Tuan]{effibenchx2025}
Yuhao Qing, Boyu Zhu, Mingzhe Du, Zhijiang Guo, Terry~Yue Zhuo, Qianru Zhang, Jie~M. Zhang, Heming Cui, Siu-Ming Yiu, Dong Huang, See-Kiong Ng, and Luu~Anh Tuan.
\newblock Effibench-x: A multi-language benchmark for measuring efficiency of llm-generated code, 2025.
\newblock URL \url{https://arxiv.org/abs/2505.13004}.

\bibitem[{Qodo-AI}(2024)]{pr_agent}
{Qodo-AI}.
\newblock Pr-agent: Ai-powered pull request agent.
\newblock \url{https://github.com/Qodo-AI/pr-agent}, 2024.

\bibitem[Qu et~al.(2024)Qu, Li, Li, Qin, Huo, Ma, and Cheng]{TA-SQL}
Ge~Qu, Jinyang Li, Bowen Li, Bowen Qin, Nan Huo, Chenhao Ma, and Reynold Cheng.
\newblock Before generation, align it! {A} novel and effective strategy for mitigating hallucinations in text-to-sql generation.
\newblock In \emph{Findings of the Association for Computational Linguistics, {ACL} 2024,}, pages 5456--5471. Association for Computational Linguistics, 2024.

\bibitem[Quan et~al.(2025{\natexlab{a}})Quan, Yang, Yu, Zheng, Liu, Yang, Ren, Gao, Miao, Feng, Wang, Yang, Cui, Fan, Zhang, Hui, and Lin]{codeelo}
Shanghaoran Quan, Jiaxi Yang, Bowen Yu, Bo~Zheng, Dayiheng Liu, An~Yang, Xuancheng Ren, Bofei Gao, Yibo Miao, Yunlong Feng, Zekun Wang, Jian Yang, Zeyu Cui, Yang Fan, Yichang Zhang, Binyuan Hui, and Junyang Lin.
\newblock Codeelo: Benchmarking competition-level code generation of llms with human-comparable elo ratings, 2025{\natexlab{a}}.
\newblock URL \url{https://arxiv.org/abs/2501.01257}.

\bibitem[Quan et~al.(2025{\natexlab{b}})Quan, Yang, Yu, Zheng, Liu, Yang, Ren, Gao, Miao, Feng, et~al.]{quan2025codeelo}
Shanghaoran Quan, Jiaxi Yang, Bowen Yu, Bo~Zheng, Dayiheng Liu, An~Yang, Xuancheng Ren, Bofei Gao, Yibo Miao, Yunlong Feng, et~al.
\newblock Codeelo: Benchmarking competition-level code generation of llms with human-comparable elo ratings.
\newblock \emph{arXiv preprint arXiv:2501.01257}, 2025{\natexlab{b}}.

\bibitem[QuixiAI(2024)]{quixiai_dolphin}
QuixiAI.
\newblock Dolphin coder dataset, 2024.
\newblock URL \url{https://huggingface.co/datasets/QuixiAI/dolphin-coder}.
\newblock Accessed: 2024.

\bibitem[{Qwen}(2025)]{qwen3coder}
{Qwen}.
\newblock Qwen3-coder: Agentic coding in the world, 2025.
\newblock URL \url{https://qwenlm.github.io/blog/qwen3-coder}.

\bibitem[Qwen et~al.(2025)Qwen, :, Yang, Yang, Zhang, Hui, Zheng, Yu, Li, Liu, Huang, Wei, Lin, Yang, Tu, Zhang, Yang, Yang, Zhou, Lin, Dang, Lu, Bao, Yang, Yu, Li, Xue, Zhang, Zhu, Men, Lin, Li, Tang, Xia, Ren, Ren, Fan, Su, Zhang, Wan, Liu, Cui, Zhang, and Qiu]{qwen2025qwen2_5}
Qwen, :, An~Yang, Baosong Yang, Beichen Zhang, Binyuan Hui, Bo~Zheng, Bowen Yu, Chengyuan Li, Dayiheng Liu, Fei Huang, Haoran Wei, Huan Lin, Jian Yang, Jianhong Tu, Jianwei Zhang, Jianxin Yang, Jiaxi Yang, Jingren Zhou, Junyang Lin, Kai Dang, Keming Lu, Keqin Bao, Kexin Yang, Le~Yu, Mei Li, Mingfeng Xue, Pei Zhang, Qin Zhu, Rui Men, Runji Lin, Tianhao Li, Tianyi Tang, Tingyu Xia, Xingzhang Ren, Xuancheng Ren, Yang Fan, Yang Su, Yichang Zhang, Yu~Wan, Yuqiong Liu, Zeyu Cui, Zhenru Zhang, and Zihan Qiu.
\newblock Qwen2.5 technical report, 2025.
\newblock URL \url{https://arxiv.org/abs/2412.15115}.

\bibitem[Radford et~al.(2021)Radford, Kim, Hallacy, Ramesh, Goh, Agarwal, Sastry, Askell, Mishkin, Clark, Krueger, and Sutskever]{radford2021clip}
Alec Radford, Jong~Wook Kim, Chris Hallacy, Aditya Ramesh, Gabriel Goh, Sandhini Agarwal, Girish Sastry, Amanda Askell, Pamela Mishkin, Jack Clark, Gretchen Krueger, and Ilya Sutskever.
\newblock Learning transferable visual models from natural language supervision, 2021.
\newblock URL \url{https://arxiv.org/abs/2103.00020}.

\bibitem[Rafailov et~al.(2023)Rafailov, Sharma, Mitchell, Ermon, Manning, and Finn]{rafailov2023direct}
Rafael Rafailov, Archit Sharma, Eric Mitchell, Stefano Ermon, Christopher~D. Manning, and Chelsea Finn.
\newblock Direct preference optimization: Your language model is secretly a reward model, 2023.

\bibitem[Raffel et~al.(2020{\natexlab{a}})Raffel, Shazeer, Roberts, Lee, Narang, Matena, Zhou, Li, and Liu]{raffel2020t5}
Colin Raffel, Noam Shazeer, Adam Roberts, Katherine Lee, Sharan Narang, Michael Matena, Yanqi Zhou, Wei Li, and Peter~J. Liu.
\newblock Exploring the limits of transfer learning with a unified text-to-text transformer.
\newblock \emph{Journal of Machine Learning Research}, 21\penalty0 (140):\penalty0 1--67, 2020{\natexlab{a}}.
\newblock URL \url{http://jmlr.org/papers/v21/20-074.html}.

\bibitem[Raffel et~al.(2020{\natexlab{b}})Raffel, Shazeer, Roberts, Lee, Narang, Matena, Zhou, Li, and Liu]{t5}
Colin Raffel, Noam Shazeer, Adam Roberts, Katherine Lee, Sharan Narang, Michael Matena, Yanqi Zhou, Wei Li, and Peter~J. Liu.
\newblock Exploring the limits of transfer learning with a unified text-to-text transformer.
\newblock \emph{J. Mach. Learn. Res.}, 21:\penalty0 140:1--140:67, 2020{\natexlab{b}}.
\newblock URL \url{http://jmlr.org/papers/v21/20-074.html}.

\bibitem[Rafi et~al.(2024)Rafi, Kim, Chen, and Wang]{rafi2024multi}
Md~Nakhla Rafi, Dong~Jae Kim, Tse-Hsun Chen, and Shaowei Wang.
\newblock A multi-agent approach to fault localization via graph-based retrieval and reflexion.
\newblock \emph{arXiv preprint arXiv:2409.13642}, 2024.

\bibitem[Rahardja et~al.(2025)Rahardja, Liu, Chen, Chen, and Lou]{rahardja2025agentsfixagentissues}
Alfin~Wijaya Rahardja, Junwei Liu, Weitong Chen, Zhenpeng Chen, and Yiling Lou.
\newblock Can agents fix agent issues?, 2025.
\newblock URL \url{https://arxiv.org/abs/2505.20749}.

\bibitem[Rahman et~al.(2024)Rahman, Singh, and Sultan]{autopatch}
Md~Tajmilur Rahman, Rahul Singh, and Mir~Yousuf Sultan.
\newblock Automating patch set generation from code reviews using large language models.
\newblock In \emph{Proceedings of the IEEE/ACM 3rd International Conference on AI Engineering-Software Engineering for AI}, pages 273--274, 2024.

\bibitem[Rajbhandari et~al.(2020{\natexlab{a}})Rajbhandari, Rasley, Ruwase, and He]{rajbhandari2020zero}
Samyam Rajbhandari, Jeff Rasley, Olatunji Ruwase, and Yuxiong He.
\newblock Zero: Memory optimizations toward training trillion parameter models.
\newblock In \emph{SC20: International Conference for High Performance Computing, Networking, Storage and Analysis}, pages 1--16. IEEE, 2020{\natexlab{a}}.

\bibitem[Rajbhandari et~al.(2020{\natexlab{b}})Rajbhandari, Rasley, Ruwase, and He]{rajbhandari2020zeromemoryoptimizationstraining}
Samyam Rajbhandari, Jeff Rasley, Olatunji Ruwase, and Yuxiong He.
\newblock Zero: Memory optimizations toward training trillion parameter models, 2020{\natexlab{b}}.
\newblock URL \url{https://arxiv.org/abs/1910.02054}.

\bibitem[{Rajeswaran} et~al.(2017){Rajeswaran}, {Kumar}, {Gupta}, {Vezzani}, {Schulman}, {Todorov}, and {Levine}]{Adroit}
Aravind {Rajeswaran}, Vikash {Kumar}, Abhishek {Gupta}, Giulia {Vezzani}, John {Schulman}, Emanuel {Todorov}, and Sergey {Levine}.
\newblock {Learning Complex Dexterous Manipulation with Deep Reinforcement Learning and Demonstrations}.
\newblock \emph{arXiv e-prints}, art. arXiv:1709.10087, September 2017.
\newblock \doi{10.48550/arXiv.1709.10087}.

\bibitem[Ran et~al.(2025)Ran, Wu, Yu, Li, Ren, Cao, Zeng, Lu, Xu, Xu, Su, Yao, Xiong, Yang, Deng, Marron, Harel, and Xie]{shi2024sphinx}
Dezhi Ran, Mengzhou Wu, Hao Yu, Yuetong Li, Jun Ren, Yuan Cao, Xia Zeng, Haochuan Lu, Zexin Xu, Mengqian Xu, Ting Su, Liangchao Yao, Ting Xiong, Wei Yang, Yuetang Deng, Assaf Marron, David Harel, and Tao Xie.
\newblock Beyond pass or fail: Multi-dimensional benchmarking of foundation models for goal-based mobile ui navigation, 2025.
\newblock URL \url{https://arxiv.org/abs/2501.02863}.

\bibitem[Rasheed et~al.(2024)Rasheed, Waseem, Saari, Systä, and Abrahamsson]{rasheed2024codepori}
Zeeshan Rasheed, Muhammad Waseem, Mika Saari, Kari Systä, and Pekka Abrahamsson.
\newblock Codepori: Large scale model for autonomous software development by using multi-agents.
\newblock \emph{CoRR abs/2402.01411}, 2024.
\newblock URL \url{https://arxiv.org/abs/2402.01411}.

\bibitem[Rashid et~al.(2025)Rashid, Bock, Zhuang, Buchholz, Esler, Valentin, Franceschi, Wistuba, Sivaprasad, Kim, Deoras, Zappella, and Callot]{rashid2025swepolybenchmultilanguagebenchmarkrepository}
Muhammad~Shihab Rashid, Christian Bock, Yuan Zhuang, Alexander Buchholz, Tim Esler, Simon Valentin, Luca Franceschi, Martin Wistuba, Prabhu~Teja Sivaprasad, Woo~Jung Kim, Anoop Deoras, Giovanni Zappella, and Laurent Callot.
\newblock Swe-polybench: A multi-language benchmark for repository level evaluation of coding agents, 2025.
\newblock URL \url{https://arxiv.org/abs/2504.08703}.

\bibitem[Rasley et~al.(2020)Rasley, Rajbhandari, Ruwase, and He]{rasley2020deepspeed}
Jeff Rasley, Samyam Rajbhandari, Olatunji Ruwase, and Yuxiong He.
\newblock Deepspeed: System optimizations enable training deep learning models with over 100 billion parameters.
\newblock In \emph{Proceedings of the 26th ACM SIGKDD International Conference on Knowledge Discovery \& Data Mining}, pages 3505--3506, 2020.

\bibitem[Rawles et~al.(2025)Rawles, Clinckemaillie, Chang, Waltz, Lau, Fair, Li, Bishop, Li, Campbell-Ajala, Toyama, Berry, Tyamagundlu, Lillicrap, and Riva]{rawles2025androidworld}
Christopher Rawles, Sarah Clinckemaillie, Yifan Chang, Jonathan Waltz, Gabrielle Lau, Marybeth Fair, Alice Li, William~E Bishop, Wei Li, Folawiyo Campbell-Ajala, Daniel~Kenji Toyama, Robert~James Berry, Divya Tyamagundlu, Timothy~P Lillicrap, and Oriana Riva.
\newblock Androidworld: A dynamic benchmarking environment for autonomous agents.
\newblock In \emph{The Thirteenth International Conference on Learning Representations}, 2025.
\newblock URL \url{https://openreview.net/forum?id=il5yUQsrjC}.

\bibitem[{Ray} et~al.(2024){Ray}, {Duan}, {Brown}, {Tan}, {Bashkirova}, {Hendrix}, {Ehsani}, {Kembhavi}, {Plummer}, {Krishna}, {Zeng}, and {Saenko}]{SAT}
Arijit {Ray}, Jiafei {Duan}, Ellis {Brown}, Reuben {Tan}, Dina {Bashkirova}, Rose {Hendrix}, Kiana {Ehsani}, Aniruddha {Kembhavi}, Bryan~A. {Plummer}, Ranjay {Krishna}, Kuo-Hao {Zeng}, and Kate {Saenko}.
\newblock {SAT: Dynamic Spatial Aptitude Training for Multimodal Language Models}.
\newblock \emph{arXiv e-prints}, art. arXiv:2412.07755, December 2024.
\newblock \doi{10.48550/arXiv.2412.07755}.

\bibitem[{readme-eval}(2025)]{readme-eval}
{readme-eval}.
\newblock readme-eval, 2025.
\newblock URL \url{https://huggingface.co/datasets/patched-codes/generate-readme-eval}.

\bibitem[Ren et~al.(2024{\natexlab{a}})Ren, Zhan, Wu, Zhou, Pan, and Li]{renReflectionCoderLearningReflection2024}
Houxing Ren, Mingjie Zhan, Zhongyuan Wu, Aojun Zhou, Junting Pan, and Hongsheng Li.
\newblock {{ReflectionCoder}}: {{Learning}} from {{Reflection Sequence}} for {{Enhanced One-off Code Generation}}, 2024{\natexlab{a}}.

\bibitem[Ren et~al.(2024{\natexlab{b}})Ren, Gao, Shao, Yan, Tan, Lam, and Ma]{ren2024codeattack}
Qibing Ren, Chang Gao, Jing Shao, Junchi Yan, Xin Tan, Wai Lam, and Lizhuang Ma.
\newblock Codeattack: Revealing safety generalization challenges of large language models via code completion.
\newblock \emph{arXiv preprint arXiv:2403.07865}, 2024{\natexlab{b}}.

\bibitem[Ren et~al.(2020)Ren, Guo, Lu, Zhou, Liu, Tang, Sundaresan, Zhou, Blanco, and Ma]{ren2020codebleumethodautomaticevaluation}
Shuo Ren, Daya Guo, Shuai Lu, Long Zhou, Shujie Liu, Duyu Tang, Neel Sundaresan, Ming Zhou, Ambrosio Blanco, and Shuai Ma.
\newblock Codebleu: a method for automatic evaluation of code synthesis, 2020.
\newblock URL \url{https://arxiv.org/abs/2009.10297}.

\bibitem[Ren et~al.(2025)Ren, Dai, Huang, Wang, Liu, and Jiang]{ren2025hydra}
Xiaoxue Ren, Chaoqun Dai, Qiao Huang, Ye~Wang, Chao Liu, and Bo~Jiang.
\newblock Hydra-reviewer: A holistic multi-agent system for automatic code review comment generation.
\newblock \emph{IEEE Transactions on Software Engineering}, 2025.

\bibitem[{Replit}(2023)]{replit2023ghostwriter}
{Replit}.
\newblock {Replit Code V1.5 3B}: Technical report.
\newblock \emph{Hugging Face Model Repository}, 2023.
\newblock URL \url{https://huggingface.co/replit/replit-code-v1_5-3b}.

\bibitem[RepoCoder(2023)]{repocoder2023}
RepoCoder.
\newblock Repocoder.
\newblock \emph{arXiv preprint arXiv:2303.12570}, 2023.
\newblock URL \url{https://arxiv.org/abs/2303.12570}.

\bibitem[Richter and Wehrheim(2022)]{wen2022tssb}
Cedric Richter and Heike Wehrheim.
\newblock Tssb-3m: Mining single statement bugs at massive scale, 2022.
\newblock URL \url{https://arxiv.org/abs/2201.12046}.

\bibitem[Riddell et~al.(2024)Riddell, Ni, and Cohan]{riddell-etal-2024-quantifying}
Martin Riddell, Ansong Ni, and Arman Cohan.
\newblock Quantifying contamination in evaluating code generation capabilities of language models.
\newblock In Lun-Wei Ku, Andre Martins, and Vivek Srikumar, editors, \emph{Proceedings of the 62nd Annual Meeting of the Association for Computational Linguistics (Volume 1: Long Papers)}, pages 14116--14137, Bangkok, Thailand, August 2024. Association for Computational Linguistics.
\newblock URL \url{https://aclanthology.org/2024.acl-long.761}.

\bibitem[Ridnik et~al.(2024)Ridnik, Kredo, and Friedman]{ridnik2024alphacodium}
Tal Ridnik, Dedy Kredo, and Itamar Friedman.
\newblock Code generation with alphacodium: From prompt engineering to flow engineering.
\newblock \emph{arXiv preprint arXiv:2401.08500}, 2024.
\newblock URL \url{https://arxiv.org/abs/2401.08500}.

\bibitem[{rouge\_l}(2025)]{rouge_l}
{rouge\_l}.
\newblock rouge\_l, 2025.
\newblock URL \url{https://en.wikipedia.org/wiki/ROUGE\_metric}.

\bibitem[Roy et~al.(2025)Roy, Chen, Steenhoek, Peng, Kaiser, Ray, and Le]{roy2025codesense}
Monoshi~Kumar Roy, Simin Chen, Benjamin Steenhoek, Jinjun Peng, Gail Kaiser, Baishakhi Ray, and Wei Le.
\newblock Codesense: a real-world benchmark and dataset for code semantic reasoning.
\newblock \emph{arXiv preprint arXiv:2506.00750}, 2025.

\bibitem[Royce(1987)]{royce1987managing}
Winston~W Royce.
\newblock Managing the development of large software systems: concepts and techniques.
\newblock In \emph{Proceedings of the 9th international conference on Software Engineering}, pages 328--338, 1987.

\bibitem[Roychoudhury(2025)]{agentic_ai_for_se}
Abhik Roychoudhury.
\newblock Agentic ai for software: thoughts from software engineering community.
\newblock \emph{arXiv preprint arXiv:2508.17343}, 2025.

\bibitem[Roziere et~al.(2022)Roziere, Zhang, Charton, Harman, Synnaeve, and Lample]{roziere2022leveragingautomatedunittests}
Baptiste Roziere, Jie~M. Zhang, Francois Charton, Mark Harman, Gabriel Synnaeve, and Guillaume Lample.
\newblock Leveraging automated unit tests for unsupervised code translation, 2022.
\newblock URL \url{https://arxiv.org/abs/2110.06773}.

\bibitem[Roziere et~al.(2023)Roziere, Gehring, Gloeckle, Sootla, Gat, Tan, Adi, Liu, Sauvestre, Remez, et~al.]{roziere2023code}
Baptiste Roziere, Jonas Gehring, Fabian Gloeckle, Sten Sootla, Itai Gat, Xiaoqing~Ellen Tan, Yossi Adi, Jingyu Liu, Romain Sauvestre, Tal Remez, et~al.
\newblock Code llama: Open foundation models for code.
\newblock \emph{arXiv preprint arXiv:2308.12950}, 2023.

\bibitem[Roziere et~al.(2024)Roziere, Gehring, Gloeckle, et~al.]{roziere2024codellama}
Baptiste Roziere, Jonas Gehring, Fabian Gloeckle, et~al.
\newblock Code {Llama}: Open foundation models for code.
\newblock \emph{arXiv preprint arXiv:2308.12950}, 2024.

\bibitem[Ruan et~al.(2024)Ruan, Zhang, and Roychoudhury]{ruan2024specrover}
Haifeng Ruan, Yuntong Zhang, and Abhik Roychoudhury.
\newblock Specrover: Code intent extraction via llms.
\newblock \emph{arXiv preprint arXiv:2408.02232}, 2024.

\bibitem[Ruiz et~al.(2025)Ruiz, Hort, and Moonen]{ruiz2025artrepairoptimizingiterative}
Fernando~Vallecillos Ruiz, Max Hort, and Leon Moonen.
\newblock The art of repair: Optimizing iterative program repair with instruction-tuned models, 2025.
\newblock URL \url{https://arxiv.org/abs/2505.02931}.

\bibitem[Saavedra et~al.(2023)Saavedra, Gon{\c{c}}alves, Henriques, Ferreira, and Mendes]{polyglotcode}
Nuno Saavedra, Jo{\~{a}}o Gon{\c{c}}alves, Miguel Henriques, Jo{\~{a}}o~F. Ferreira, and Alexandra Mendes.
\newblock Polyglot code smell detection for infrastructure as code with {GLITCH}.
\newblock In \emph{38th {IEEE/ACM} International Conference on Automated Software Engineering, {ASE} 2023, Luxembourg, September 11-15, 2023}, pages 2042--2045. {IEEE}, 2023.
\newblock \doi{10.1109/ASE56229.2023.00162}.
\newblock URL \url{https://doi.org/10.1109/ASE56229.2023.00162}.

\bibitem[Sami et~al.(2024)Sami, Waseem, Zhang, Rasheed, Syst{\"a}, and Abrahamsson]{UserStory}
Malik~Abdul Sami, Muhammad Waseem, Zheying Zhang, Zeeshan Rasheed, Kari Syst{\"a}, and Pekka Abrahamsson.
\newblock Ai based multiagent approach for requirements elicitation and analysis.
\newblock \emph{arXiv preprint arXiv:2409.00038}, 2024.

\bibitem[Sapkota et~al.(2025)Sapkota, Roumeliotis, and Karkee]{vibe_coding_agentic_coding}
Ranjan Sapkota, Konstantinos~I Roumeliotis, and Manoj Karkee.
\newblock Vibe coding vs. agentic coding: Fundamentals and practical implications of agentic ai.
\newblock \emph{arXiv preprint arXiv:2505.19443}, 2025.

\bibitem[Schaeffer et~al.(2023)Schaeffer, Miranda, and Koyejo]{schaeffer2023mirage}
Rylan Schaeffer, Brando Miranda, and Sanmi Koyejo.
\newblock Are emergent abilities of large language models a mirage?
\newblock In A.~Oh, T.~Naumann, A.~Globerson, K.~Saenko, M.~Hardt, and S.~Levine, editors, \emph{Advances in Neural Information Processing Systems}, volume~36, pages 55565--55581. Curran Associates, Inc., 2023.
\newblock URL \url{https://proceedings.neurips.cc/paper_files/paper/2023/file/adc98a266f45005c403b8311ca7e8bd7-Paper-Conference.pdf}.

\bibitem[Schall et~al.(2024)Schall, Czinczoll, and De~Melo]{schall2024commitbench}
Maximilian Schall, Tamara Czinczoll, and Gerard De~Melo.
\newblock Commitbench: A benchmark for commit message generation.
\newblock In \emph{2024 IEEE International Conference on Software Analysis, Evolution and Reengineering (SANER)}, pages 728--739. IEEE, 2024.

\bibitem[Schick et~al.(2023{\natexlab{a}})Schick, Dwivedi-Yu, Dess{\`\i}, Raileanu, Lomeli, Hambro, Zettlemoyer, Cancedda, and Scialom]{schick2023toolformer}
Timo Schick, Jane Dwivedi-Yu, Roberto Dess{\`\i}, Roberta Raileanu, Maria Lomeli, Eric Hambro, Luke Zettlemoyer, Nicola Cancedda, and Thomas Scialom.
\newblock Toolformer: Language models can teach themselves to use tools.
\newblock \emph{Advances in Neural Information Processing Systems}, 36:\penalty0 68539--68551, 2023{\natexlab{a}}.

\bibitem[Schick et~al.(2023{\natexlab{b}})Schick, Dwivedi-Yu, Dessì, Raileanu, Lomeli, Zettlemoyer, Cancedda, and Scialom]{toolformer}
Timo Schick, Jane Dwivedi-Yu, Roberto Dessì, Roberta Raileanu, M.~Lomeli, Luke Zettlemoyer, Nicola Cancedda, and Thomas Scialom.
\newblock Toolformer: Language models can teach themselves to use tools.
\newblock \emph{Neural Information Processing Systems}, 2023{\natexlab{b}}.
\newblock \doi{10.48550/arXiv.2302.04761}.

\bibitem[Schulman et~al.(2015{\natexlab{a}})Schulman, Levine, Abbeel, Jordan, and Moritz]{trpo}
John Schulman, Sergey Levine, Pieter Abbeel, Michael Jordan, and Philipp Moritz.
\newblock Trust region policy optimization.
\newblock In \emph{International conference on machine learning}, pages 1889--1897. PMLR, 2015{\natexlab{a}}.

\bibitem[Schulman et~al.(2015{\natexlab{b}})Schulman, Moritz, Levine, Jordan, and Abbeel]{schulman2015high}
John Schulman, Philipp Moritz, Sergey Levine, Michael Jordan, and Pieter Abbeel.
\newblock High-dimensional continuous control using generalized advantage estimation.
\newblock \emph{arXiv preprint arXiv:1506.02438}, 2015{\natexlab{b}}.

\bibitem[Schulman et~al.(2017)Schulman, Wolski, Dhariwal, Radford, and Klimov]{ppo}
John Schulman, Filip Wolski, Prafulla Dhariwal, Alec Radford, and Oleg Klimov.
\newblock Proximal policy optimization algorithms.
\newblock \emph{arXiv preprint arXiv:1707.06347}, 2017.

\bibitem[Schäfer et~al.(2023)Schäfer, Nadi, Eghbali, and Tip]{schäfer2023empiricalevaluationusinglarge}
Max Schäfer, Sarah Nadi, Aryaz Eghbali, and Frank Tip.
\newblock An empirical evaluation of using large language models for automated unit test generation, 2023.
\newblock URL \url{https://arxiv.org/abs/2302.06527}.

\bibitem[Seed et~al.(2025)Seed, Zhang, Su, Sun, Xi, Xiao, Zheng, Zhang, Liu, Zan, et~al.]{seedcoder}
ByteDance Seed, Yuyu Zhang, Jing Su, Yifan Sun, Chenguang Xi, Xia Xiao, Shen Zheng, Anxiang Zhang, Kaibo Liu, Daoguang Zan, et~al.
\newblock Seed-coder: Let the code model curate data for itself.
\newblock \emph{arXiv preprint arXiv:2506.03524}, 2025.

\bibitem[Sekizawa et~al.(2023)Sekizawa, Duan, Lu, and Yanaka]{xcodesearchnet}
Ryo Sekizawa, Nan Duan, Shuai Lu, and Hitomi Yanaka.
\newblock Constructing multilingual code search dataset using neural machine translation.
\newblock In Vishakh Padmakumar, Gisela Vallejo, and Yao Fu, editors, \emph{Proceedings of the 61st Annual Meeting of the Association for Computational Linguistics: Student Research Workshop, {ACL} 2023, Toronto, Canada, July 9-14, 2023}, pages 69--75. Association for Computational Linguistics, 2023.
\newblock \doi{10.18653/V1/2023.ACL-SRW.10}.
\newblock URL \url{https://doi.org/10.18653/v1/2023.acl-srw.10}.

\bibitem[SenseLLM(2024)]{sensellm_reflectionseq}
SenseLLM.
\newblock Reflectionseq-gpt dataset, 2024.
\newblock URL \url{https://huggingface.co/datasets/SenseLLM/ReflectionSeq-GPT}.
\newblock Accessed: 2024.

\bibitem[Seo et~al.(2024)Seo, Baek, and Hwang]{refine2024}
Minju Seo, Jinheon Baek, and Sung~Ju Hwang.
\newblock Rethinking code refinement: Learning to judge code efficiency, 2024.
\newblock URL \url{https://arxiv.org/abs/2410.22375}.

\bibitem[SERVERS()]{servers25mcp}
TOCOL SERVERS.
\newblock Mcp-universe: Benchmarking large language models with real-world model context pro-tocol servers.
\newblock \emph{origins}, 25:\penalty0 55--128.

\bibitem[Sghaier et~al.(2025)Sghaier, Tufano, Bavota, and Sahraoui]{sghaier2025leveraging}
Oussama~Ben Sghaier, Rosalia Tufano, Gabriele Bavota, and Houari Sahraoui.
\newblock Leveraging reward models for guiding code review comment generation.
\newblock \emph{arXiv preprint arXiv:2506.04464}, 2025.

\bibitem[Shahbazi and Fard(2023)]{APIContext2Com}
Ramin Shahbazi and Fatemeh Fard.
\newblock Apicontext2com: Code comment generation by incorporating pre-defined api documentation.
\newblock In \emph{2023 IEEE/ACM 31st International Conference on Program Comprehension (ICPC)}, pages 13--24, 2023.
\newblock \doi{10.1109/ICPC58990.2023.00012}.

\bibitem[{Shao} et~al.(2024){Shao}, {Jiang}, {Zuo}, and {Liu}]{SwarmBrain}
Xiao {Shao}, Weifu {Jiang}, Fei {Zuo}, and Mengqing {Liu}.
\newblock {SwarmBrain: Embodied agent for real-time strategy game StarCraft II via large language models}.
\newblock \emph{arXiv e-prints}, art. arXiv:2401.17749, January 2024.
\newblock \doi{10.48550/arXiv.2401.17749}.

\bibitem[Shao et~al.(2024{\natexlab{a}})Shao, Wang, Zhu, Xu, Song, Bi, Zhang, Zhang, Li, Wu, et~al.]{shao2024deepseekmath}
Zhihong Shao, Peiyi Wang, Qihao Zhu, Runxin Xu, Junxiao Song, Xiao Bi, Haowei Zhang, Mingchuan Zhang, YK~Li, Yang Wu, et~al.
\newblock Deepseekmath: Pushing the limits of mathematical reasoning in open language models.
\newblock \emph{arXiv preprint arXiv:2402.03300}, 2024{\natexlab{a}}.

\bibitem[Shao et~al.(2024{\natexlab{b}})Shao, Wang, Zhu, Xu, Song, Zhang, Li, Wu, and Guo]{deepseekmath}
Zhihong Shao, Peiyi Wang, Qihao Zhu, Runxin Xu, Junxiao Song, Mingchuan Zhang, Y.~K. Li, Y.~Wu, and Daya Guo.
\newblock Deepseekmath: Pushing the limits of mathematical reasoning in open language models.
\newblock \emph{CoRR}, abs/2402.03300, 2024{\natexlab{b}}.
\newblock \doi{10.48550/ARXIV.2402.03300}.
\newblock URL \url{https://doi.org/10.48550/arXiv.2402.03300}.

\bibitem[Sharma et~al.(2025)Sharma, De~Halleux, Barke, and Zorn]{promptpex}
Reshabh~K Sharma, Jonathan De~Halleux, Shraddha Barke, and Benjamin Zorn.
\newblock Promptpex: Automatic test generation for language model prompts.
\newblock \emph{arXiv preprint arXiv:2503.05070}, 2025.

\bibitem[{Shekhar Dvivedi} et~al.(2023){Shekhar Dvivedi}, {Vijay}, {Leela Rahul Pujari}, {Lodh}, and {Kumar}]{Dvivedi}
Shubhang {Shekhar Dvivedi}, Vyshnav {Vijay}, Sai {Leela Rahul Pujari}, Shoumik {Lodh}, and Dhruv {Kumar}.
\newblock {A Comparative Analysis of Large Language Models for Code Documentation Generation}.
\newblock \emph{arXiv e-prints}, art. arXiv:2312.10349, December 2023.
\newblock \doi{10.48550/arXiv.2312.10349}.

\bibitem[Shen et~al.(2023{\natexlab{a}})Shen, Zhang, Chen, Zan, Geng, Fu, Zeng, Yu, Ji, Zhao, Guo, and Wang]{shen2023pangucoder2boostinglargelanguage}
Bo~Shen, Jiaxin Zhang, Taihong Chen, Daoguang Zan, Bing Geng, An~Fu, Muhan Zeng, Ailun Yu, Jichuan Ji, Jingyang Zhao, Yuenan Guo, and Qianxiang Wang.
\newblock Pangu-coder2: Boosting large language models for code with ranking feedback, 2023{\natexlab{a}}.
\newblock URL \url{https://arxiv.org/abs/2307.14936}.

\bibitem[Shen et~al.(2023{\natexlab{b}})Shen, Zhang, Chen, Zan, Geng, Fu, Zeng, Yu, Ji, Zhao, et~al.]{shen2023pangucoder2}
Bo~Shen, Jiaxin Zhang, Taihong Chen, Daoguang Zan, Bing Geng, An~Fu, Muhan Zeng, Ailun Yu, Jichuan Ji, Jingyang Zhao, et~al.
\newblock Pangu-coder2: Boosting large language models for code with ranking feedback.
\newblock \emph{arXiv preprint arXiv:2307.14936}, 2023{\natexlab{b}}.

\bibitem[Shen et~al.(2024{\natexlab{a}})Shen, Chen, Backes, Shen, and Zhang]{shen2024anything}
Xinyue Shen, Zeyuan Chen, Michael Backes, Yun Shen, and Yang Zhang.
\newblock " do anything now": Characterizing and evaluating in-the-wild jailbreak prompts on large language models.
\newblock In \emph{Proceedings of the 2024 on ACM SIGSAC Conference on Computer and Communications Security}, pages 1671--1685, 2024{\natexlab{a}}.

\bibitem[Shen et~al.(2024{\natexlab{b}})Shen, Zhao, Yang, Guo, Li, Liu, and Liu]{shen2024prosec}
Yixiang Shen, Zhizhou Zhao, Zirui Yang, Haoye Guo, Jia Li, Yang Liu, and Huan Liu.
\newblock Prosec: Fortifying code llms with proactive security alignment.
\newblock \emph{arXiv preprint arXiv:2406.12455}, 2024{\natexlab{b}}.

\bibitem[Sheng et~al.(2024)Sheng, Zhang, Ye, Wu, Zhang, Zhang, Peng, Lin, and Wu]{sheng2024hybridflow}
Guangming Sheng, Chi Zhang, Zilingfeng Ye, Xibin Wu, Wang Zhang, Ru~Zhang, Yanghua Peng, Haibin Lin, and Chuan Wu.
\newblock Hybridflow: A flexible and efficient rlhf framework.
\newblock \emph{arXiv preprint arXiv: 2409.19256}, 2024.

\bibitem[{Shi} et~al.(2025{\natexlab{a}}){Shi}, {Wei}, {Yang}, {Moore Wang}, {Yang}, {Zhang}, {Huang}, {Peng}, {Yang}, and {Wen}]{CryptoX}
Jiajun {Shi}, Chaoren {Wei}, Liqun {Yang}, Zekun {Moore Wang}, Chenghao {Yang}, Ge~{Zhang}, Stephen {Huang}, Tao {Peng}, Jian {Yang}, and Zhoufutu {Wen}.
\newblock {CryptoX : Compositional Reasoning Evaluation of Large Language Models}.
\newblock \emph{arXiv e-prints}, art. arXiv:2502.07813, February 2025{\natexlab{a}}.
\newblock \doi{10.48550/arXiv.2502.07813}.

\bibitem[{Shi} et~al.(2025{\natexlab{b}}){Shi}, {Yang}, {Liu}, {Bu}, {Chen}, {Zhou}, {Ma}, {Wen}, {Wang}, {He}, {Song}, {Zhu}, {Li}, {Wang}, {Zhang}, {Yuan}, {Yao}, {Yang}, {Wang}, {Fang}, {Yuan}, {He}, {Tang}, {Tan}, {Zhou}, {Zhang}, {Li}, {Huang}, and {Zhang}]{KORGYM}
Jiajun {Shi}, Jian {Yang}, Jiaheng {Liu}, Xingyuan {Bu}, Jiangjie {Chen}, Junting {Zhou}, Kaijing {Ma}, Zhoufutu {Wen}, Bingli {Wang}, Yancheng {He}, Liang {Song}, Hualei {Zhu}, Shilong {Li}, Xingjian {Wang}, Wei {Zhang}, Ruibin {Yuan}, Yifan {Yao}, Wenjun {Yang}, Yunli {Wang}, Siyuan {Fang}, Siyu {Yuan}, Qianyu {He}, Xiangru {Tang}, Yingshui {Tan}, Wangchunshu {Zhou}, Zhaoxiang {Zhang}, Zhoujun {Li}, Wenhao {Huang}, and Ge~{Zhang}.
\newblock {KORGym: A Dynamic Game Platform for LLM Reasoning Evaluation}.
\newblock \emph{arXiv e-prints}, art. arXiv:2505.14552, May 2025{\natexlab{b}}.
\newblock \doi{10.48550/arXiv.2505.14552}.

\bibitem[Shi et~al.(2024)Shi, Xu, Zhuang, Yu, Zhang, Wu, Zhu, Ho, Yang, and Wang]{shi2024ehragent}
Wenqi Shi, Ran Xu, Yuchen Zhuang, Yue Yu, Jieyu Zhang, Hang Wu, Yuanda Zhu, Joyce Ho, Carl Yang, and May~D Wang.
\newblock Ehragent: Code empowers large language models for few-shot complex tabular reasoning on electronic health records.
\newblock In \emph{Proceedings of the Conference on Empirical Methods in Natural Language Processing. Conference on Empirical Methods in Natural Language Processing}, volume 2024, page 22315, 2024.

\bibitem[Shi et~al.(2025)Shi, Qian, Zhang, Shen, and Gu]{longcodezip}
Yuling Shi, Yichun Qian, Hongyu Zhang, Beijun Shen, and Xiaodong Gu.
\newblock Longcodezip: Compress long context for code language models.
\newblock \emph{arXiv preprint arXiv:2510.00446}, 2025.

\bibitem[Shin et~al.(2025)Shin, Tang, Lee, Kim, Lim, Cho, Hong, Lee, and Kim]{automatically_review_generation_llm}
Hyungyu Shin, Jingyu Tang, Yoonjoo Lee, Nayoung Kim, Hyunseung Lim, Ji~Yong Cho, Hwajung Hong, Moontae Lee, and Juho Kim.
\newblock Automatically evaluating the paper reviewing capability of large language models.
\newblock \emph{arXiv e-prints}, pages arXiv--2502, 2025.

\bibitem[Shinn et~al.(2023)Shinn, Cassano, Berman, Gopinath, Narasimhan, and Yao]{shinn2023reflexion}
Noah Shinn, Federico Cassano, Edward Berman, Ashwin Gopinath, Karthik Narasimhan, and Shunyu Yao.
\newblock Reflexion: Language agents with verbal reinforcement learning.
\newblock \emph{arXiv preprint arXiv: 2303.11366}, 2023.

\bibitem[Shkapenyuk et~al.(2025)Shkapenyuk, Srivastava, Johnson, and Ghane]{AskData}
Vladislav Shkapenyuk, Divesh Srivastava, Theodore Johnson, and Parisa Ghane.
\newblock Automatic metadata extraction for text-to-sql.
\newblock \emph{CoRR}, abs/2505.19988, 2025.

\bibitem[Shoeybi et~al.(2019)Shoeybi, Patwary, Puri, LeGresley, Casper, and Catanzaro]{shoeybi2019megatron}
Mohammad Shoeybi, Mostofa Patwary, Raul Puri, Patrick LeGresley, Jared Casper, and Bryan Catanzaro.
\newblock Megatron-lm: Training multi-billion parameter language models using model parallelism.
\newblock In \emph{arXiv preprint arXiv:1909.08053}, 2019.

\bibitem[Shoeybi et~al.(2020)Shoeybi, Patwary, Puri, LeGresley, Casper, and Catanzaro]{shoeybi2020megatronlmtrainingmultibillionparameter}
Mohammad Shoeybi, Mostofa Patwary, Raul Puri, Patrick LeGresley, Jared Casper, and Bryan Catanzaro.
\newblock Megatron-lm: Training multi-billion parameter language models using model parallelism, 2020.
\newblock URL \url{https://arxiv.org/abs/1909.08053}.

\bibitem[Shuai et~al.(2025)Shuai, Li, Yan, Luo, and Yang]{shuai2025deepvis}
Zhihao Shuai, Boyan Li, Siyu Yan, Yuyu Luo, and Weikai Yang.
\newblock Deepvis: Bridging natural language and data visualization through step-wise reasoning, 2025.
\newblock URL \url{https://arxiv.org/abs/2508.01700}.

\bibitem[Si et~al.(2024)Si, Zhang, Li, Yang, Liu, and Yang]{si2024design2code}
Chenglei Si, Yanzhe Zhang, Ryan Li, Zhengyuan Yang, Ruibo Liu, and Diyi Yang.
\newblock Design2code: Benchmarking multimodal code generation for automated front-end engineering.
\newblock \emph{arXiv preprint arXiv:2403.03163}, 2024.

\bibitem[Silva et~al.(2025{\natexlab{a}})Silva, Fang, and Monperrus]{silva2025repairllama}
Andr{\'e} Silva, Sen Fang, and Martin Monperrus.
\newblock Repairllama: Efficient representations and fine-tuned adapters for program repair.
\newblock \emph{IEEE Transactions on Software Engineering}, 2025{\natexlab{a}}.

\bibitem[Silva et~al.(2025{\natexlab{b}})Silva, Thor{\'e}n, and Monperrus]{silva2025gradient}
Andr{\'e} Silva, Gustav Thor{\'e}n, and Martin Monperrus.
\newblock Gradient-based program repair: Fixing bugs in continuous program spaces.
\newblock \emph{arXiv preprint arXiv:2505.17703}, 2025{\natexlab{b}}.

\bibitem[Sim et~al.(2025)Sim, Cho, Go, Fu, Shokri, and Ravindran]{rustcoder}
HoHyun Sim, Hyeonjoong Cho, Yeonghyeon Go, Zhoulai Fu, Ali Shokri, and Binoy Ravindran.
\newblock Large language model-powered agent for c to rust code translation.
\newblock \emph{arXiv preprint arXiv:2505.15858}, 2025.

\bibitem[{Simon Willison}(2025)]{how_to_write_code1}
{Simon Willison}.
\newblock Here’s how i use llms to help me write code, 2025.
\newblock URL \url{https://simonwillison.net/2025/Mar/11/using-llms-for-code/}.

\bibitem[Singh et~al.(2023)Singh, Co-Reyes, Agarwal, Anand, Patil, Garcia, Liu, Harrison, Lee, Xu, et~al.]{singh2023beyond}
Avi Singh, John~D Co-Reyes, Rishabh Agarwal, Ankesh Anand, Piyush Patil, Xavier Garcia, Peter~J Liu, James Harrison, Jaehoon Lee, Kelvin Xu, et~al.
\newblock Beyond human data: Scaling self-training for problem-solving with language models.
\newblock \emph{arXiv preprint arXiv:2312.06585}, 2023.

\bibitem[Singh et~al.(2022)Singh, Blukis, Mousavian, Goyal, Xu, Tremblay, Fox, Thomason, and Garg]{singh2022progprompt}
Ishika Singh, Valts Blukis, Arsalan Mousavian, Ankit Goyal, Danfei Xu, Jonathan Tremblay, Dieter Fox, Jesse Thomason, and Animesh Garg.
\newblock Progprompt: Generating situated robot task plans using large language models.
\newblock \emph{arXiv preprint arXiv:2209.11302}, 2022.

\bibitem[Sinha et~al.(2025)Sinha, Goel, Kumaraguru, Geiping, Bethge, and Prabhu]{sinha2025languagemodelsfalsifyevaluating}
Shiven Sinha, Shashwat Goel, Ponnurangam Kumaraguru, Jonas Geiping, Matthias Bethge, and Ameya Prabhu.
\newblock Can language models falsify? evaluating algorithmic reasoning with counterexample creation, 2025.
\newblock URL \url{https://arxiv.org/abs/2502.19414}.

\bibitem[Smith et~al.(2022)Smith, Patwary, Norick, LeGresley, Rajbhandari, Casper, Liu, Prabhumoye, Zerveas, Korthikanti, et~al.]{smith2022using}
Shaden Smith, Mostofa Patwary, Brandon Norick, Patrick LeGresley, Samyam Rajbhandari, Jared Casper, Zhun Liu, Shrimai Prabhumoye, George Zerveas, Vijay Korthikanti, et~al.
\newblock Using deepspeed and megatron to train megatron-turing nlg 530b, a large-scale generative language model.
\newblock \emph{arXiv preprint arXiv:2201.11990}, 2022.

\bibitem[Sohrabizadeh et~al.()Sohrabizadeh, Song, Liu, Roy, Lee, Raiman, and Catanzaro]{sohrabizadehnemotron}
Atefeh Sohrabizadeh, Jialin Song, Mingjie Liu, Rajarshi Roy, Chankyu Lee, Jonathan Raiman, and Bryan Catanzaro.
\newblock Nemotron-cortexa: Enhancing llm agents for software engineering tasks via improved localization and solution diversity.
\newblock In \emph{Forty-second International Conference on Machine Learning}.

\bibitem[Solovyeva et~al.(2025)Solovyeva, Weidmann, and Castor]{energy2025}
Lola Solovyeva, Sophie Weidmann, and Fernando Castor.
\newblock Ai-powered, but power-hungry? energy efficiency of llm-generated code, 2025.
\newblock URL \url{https://arxiv.org/abs/2502.02412}.

\bibitem[Song et~al.(2024)Song, Ma, Zheng, Liao, Kuang, and Yang]{song2024audit}
Chengyu Song, Linru Ma, Jianming Zheng, Jinzhi Liao, Hongyu Kuang, and Lin Yang.
\newblock Audit-llm: Multi-agent collaboration for log-based insider threat detection.
\newblock \emph{arXiv preprint arXiv:2408.08902}, 2024.

\bibitem[Sonwane et~al.(2025)Sonwane, White, Lee, Pereira, Caccia, Kim, Shi, Singh, Sordoni, C{\^o}t{\'e}, et~al.]{bugpilot}
Atharv Sonwane, Isadora White, Hyunji Lee, Matheus Pereira, Lucas Caccia, Minseon Kim, Zhengyan Shi, Chinmay Singh, Alessandro Sordoni, Marc-Alexandre C{\^o}t{\'e}, et~al.
\newblock Bugpilot: Complex bug generation for efficient learning of swe skills.
\newblock \emph{arXiv preprint arXiv:2510.19898}, 2025.

\bibitem[{Sourcegraph}(2024)]{sourcegraph2024cody}
{Sourcegraph}.
\newblock {Cody}: {AI} code assistant.
\newblock Technical report, Sourcegraph Inc., 2024.

\bibitem[Starace et~al.(2025)Starace, Jaffe, Sherburn, Aung, Chan, Maksin, Dias, Mays, Kinsella, Thompson, Heidecke, Glaese, and Patwardhan]{starace2025paperbenchevaluatingaisability}
Giulio Starace, Oliver Jaffe, Dane Sherburn, James Aung, Jun~Shern Chan, Leon Maksin, Rachel Dias, Evan Mays, Benjamin Kinsella, Wyatt Thompson, Johannes Heidecke, Amelia Glaese, and Tejal Patwardhan.
\newblock Paperbench: Evaluating ai's ability to replicate ai research, 2025.
\newblock URL \url{https://arxiv.org/abs/2504.01848}.

\bibitem[{Stojanovski} et~al.(2025){Stojanovski}, {Stanley}, {Sharratt}, {Jones}, {Adefioye}, {Kaddour}, and {K{\"o}pf}]{REASONINGGYM}
Zafir {Stojanovski}, Oliver {Stanley}, Joe {Sharratt}, Richard {Jones}, Abdulhakeem {Adefioye}, Jean {Kaddour}, and Andreas {K{\"o}pf}.
\newblock {REASONING GYM: Reasoning Environments for Reinforcement Learning with Verifiable Rewards}.
\newblock \emph{arXiv e-prints}, art. arXiv:2505.24760, May 2025.
\newblock \doi{10.48550/arXiv.2505.24760}.

\bibitem[Su et~al.(2022)Su, Dai, Zhao, Zheng, and Luo]{su2022effectively}
Jianzhong Su, Hong-Ning Dai, Lingjun Zhao, Zibin Zheng, and Xiapu Luo.
\newblock Effectively generating vulnerable transaction sequences in smart contracts with reinforcement learning-guided fuzzing.
\newblock In \emph{Proceedings of the 37th IEEE/ACM International Conference on Automated Software Engineering}, pages 1--12, 2022.

\bibitem[Su et~al.(2023)Su, Wan, Sethi, Lu, Musuvathi, and Nath]{HotGPT}
Yiming Su, Chengcheng Wan, Utsav Sethi, Shan Lu, Madan Musuvathi, and Suman Nath.
\newblock Hotgpt: How to make software documentation more useful with a large language model?
\newblock In \emph{Proceedings of the 19th Workshop on Hot Topics in Operating Systems}, HotOS '23, page 87–93, New York, NY, USA, 2023. Association for Computing Machinery.
\newblock ISBN 9798400701955.
\newblock \doi{10.1145/3593856.3595910}.
\newblock URL \url{https://doi.org/10.1145/3593856.3595910}.

\bibitem[Su et~al.(2025)Su, Pan, Bai, Liu, Dong, Huang, Hu, Zhang, Gai, and Zhou]{klearreasoner}
Zhenpeng Su, Leiyu Pan, Xue Bai, Dening Liu, Guanting Dong, Jiaming Huang, Wenping Hu, Fuzheng Zhang, Kun Gai, and Guorui Zhou.
\newblock Klear-reasoner: Advancing reasoning capability via gradient-preserving clipping policy optimization.
\newblock \emph{arXiv preprint arXiv:2508.07629}, 2025.

\bibitem[Sun et~al.(2024{\natexlab{a}})Sun, Shen, and Ton]{sun2024rethinking}
Hao Sun, Yunyi Shen, and Jean-Francois Ton.
\newblock Rethinking bradley-terry models in preference-based reward modeling: Foundations, theory, and alternatives.
\newblock \emph{arXiv preprint arXiv:2411.04991}, 2024{\natexlab{a}}.

\bibitem[Sun et~al.(2025)Sun, Li, and Zhao]{sun2024repofixevaluation}
Jiawei Sun, Tianyu Li, and Xinyu Zhao.
\newblock Repofixeval: Issue-aware repository-level benchmark for automated program repair.
\newblock In \emph{International Conference on Learning Representations (ICLR)}, 2025.

\bibitem[Sun et~al.(2024{\natexlab{b}})Sun, Chen, Zhang, Zhang, Wang, Zhang, Wang, Chen, and Qin]{sun2024safetyaware}
Keyi Sun, Jiazhen Chen, Xiao Zhang, Yinda Zhang, Zirui Wang, Zhiyue Zhang, Ruofan Wang, Pin-Yu Chen, and Zeyi Qin.
\newblock Safety-aware fine-tuning of large language models.
\newblock \emph{arXiv preprint arXiv:2405.09919}, 2024{\natexlab{b}}.

\bibitem[Sun et~al.(2023{\natexlab{a}})Sun, Chen, Wang, Li, and Gao]{transcoder}
Qiushi Sun, Nuo Chen, Jianing Wang, Xiang Li, and Ming Gao.
\newblock Transcoder: Towards unified transferable code representation learning inspired by human skills.
\newblock \emph{arXiv preprint arXiv:2306.07285}, 2023{\natexlab{a}}.

\bibitem[Sun et~al.(2023{\natexlab{b}})Sun, Dong, Huang, Ma, Xia, Xue, Wang, and Wei]{sun2023retnet}
Yutao Sun, Li~Dong, Shaohan Huang, Shuming Ma, Yuqing Xia, Jilong Xue, Jianyong Wang, and Furu Wei.
\newblock Retentive network: A successor to transformer for large language models, 2023{\natexlab{b}}.
\newblock URL \url{https://arxiv.org/abs/2307.08621}.

\bibitem[Suri et~al.(2024)Suri, Mathur, Dernoncourt, Jain, Morariu, Sawhney, Nakov, and Manocha]{suri2024docedit}
Manan Suri, Puneet Mathur, Franck Dernoncourt, Rajiv Jain, Vlad~I Morariu, Ramit Sawhney, Preslav Nakov, and Dinesh Manocha.
\newblock Docedit-v2: Document structure editing via multimodal llm grounding.
\newblock \emph{arXiv preprint arXiv:2410.16472}, 2024.

\bibitem[Surís et~al.(2023)Surís, Menon, and Vondrick]{suris2023vipergpt}
Dídac Surís, Sachit Menon, and Carl Vondrick.
\newblock Vipergpt: Visual inference via python execution for reasoning, 2023.
\newblock URL \url{https://arxiv.org/abs/2303.08128}.

\bibitem[Sutskever et~al.(2014)Sutskever, Vinyals, and Le]{sutskever2014seq2seq}
Ilya Sutskever, Oriol Vinyals, and Quoc~V. Le.
\newblock Sequence to sequence learning with neural networks, 2014.
\newblock URL \url{https://arxiv.org/abs/1409.3215}.

\bibitem[{SWE-Agent Project}(2024)]{swe_agent_2024}
{SWE-Agent Project}.
\newblock Swe-agent documentation, 2024.

\bibitem[swe-bench live(2025)]{swe-bench-live}
swe-bench live.
\newblock swe-bench-live, 2025.
\newblock URL \url{https://swe-bench-live.github.io/}.

\bibitem[swebench(2025)]{swebench}
swebench.
\newblock swebench, 2025.
\newblock URL \url{https://www.swebench.com/original.html}.

\bibitem[swebenchmultilingual(2025)]{swebenchmultilingual}
swebenchmultilingual.
\newblock swebenchmultilingual, 2025.
\newblock URL \url{https://www.swebench.com/multilingual.html}.

\bibitem[swebenchverified(2025)]{swebenchverified}
swebenchverified.
\newblock swebenchverified, 2025.
\newblock URL \url{https://openai.com/index/introducing-swe-bench-verified/}.

\bibitem[{Tabnine}(2024)]{tabnine2024enterprise}
{Tabnine}.
\newblock {Tabnine}: Enterprise-grade {AI} code assistant.
\newblock Technical report, Tabnine Ltd., 2024.
\newblock URL \url{https://www.tabnine.com}.

\bibitem[Tai et~al.(2023)Tai, Chen, Zhang, Deng, and Sun]{Tai}
Chang{-}Yu Tai, Ziru Chen, Tianshu Zhang, Xiang Deng, and Huan Sun.
\newblock Exploring chain of thought style prompting for text-to-sql.
\newblock In \emph{{EMNLP} 2023, Singapore, December 6-10, 2023}, pages 5376--5393. Association for Computational Linguistics, 2023.

\bibitem[Talaei et~al.(2024)Talaei, Pourreza, Chang, Mirhoseini, and Saberi]{chess}
Shayan Talaei, Mohammadreza Pourreza, Yu{-}Chen Chang, Azalia Mirhoseini, and Amin Saberi.
\newblock {CHESS:} contextual harnessing for efficient {SQL} synthesis.
\newblock \emph{CoRR}, abs/2405.16755, 2024.

\bibitem[Tan et~al.(2024)Tan, Luo, Li, and Zhang]{DBLP:conf/emnlp/TanL0Z24}
Hanzhuo Tan, Qi~Luo, Jing Li, and Yuqun Zhang.
\newblock Llm4decompile: Decompiling binary code with large language models.
\newblock In Yaser Al{-}Onaizan, Mohit Bansal, and Yun{-}Nung Chen, editors, \emph{Proceedings of the 2024 Conference on Empirical Methods in Natural Language Processing, {EMNLP} 2024, Miami, FL, USA, November 12-16, 2024}, pages 3473--3487. Association for Computational Linguistics, 2024.

\bibitem[Tan and Pan(2025)]{tan2025gtpo}
Hongze Tan and Jianfei Pan.
\newblock Gtpo and grpo-s: Token and sequence-level reward shaping with policy entropy.
\newblock \emph{arXiv preprint arXiv:2508.04349}, 2025.

\bibitem[Tang et~al.(2024{\natexlab{a}})Tang, Qian, Gao, Chen, Chen, and Gerstein]{tang2024biocoderbenchmarkbioinformaticscode}
Xiangru Tang, Bill Qian, Rick Gao, Jiakang Chen, Xinyun Chen, and Mark Gerstein.
\newblock Biocoder: A benchmark for bioinformatics code generation with large language models, 2024{\natexlab{a}}.
\newblock URL \url{https://arxiv.org/abs/2308.16458}.

\bibitem[Tang et~al.(2024{\natexlab{b}})Tang, Kim, Song, Lothritz, Li, Ezzini, Tian, Klein, and Bissyand{\'e}]{tang2024codeagent}
Xunzhu Tang, Kisub Kim, Yewei Song, Cedric Lothritz, Bei Li, Saad Ezzini, Haoye Tian, Jacques Klein, and Tegawend{\'e}~F Bissyand{\'e}.
\newblock Codeagent: Autonomous communicative agents for code review.
\newblock \emph{arXiv preprint arXiv:2402.02172}, 2024{\natexlab{b}}.

\bibitem[Tang et~al.(2025{\natexlab{a}})Tang, Klein, and Bissyand{\'e}]{tang2025boosting}
Xunzhu Tang, Jacques Klein, and Tegawend{\'e}~F Bissyand{\'e}.
\newblock Boosting open-source llms for program repair via reasoning transfer and llm-guided reinforcement learning.
\newblock \emph{arXiv preprint arXiv:2506.03921}, 2025{\natexlab{a}}.

\bibitem[Tang et~al.(2025{\natexlab{b}})Tang, Li, Zhu, Yang, Ding, and Guo]{tang2025co}
Yuheng Tang, Hongwei Li, Kaijie Zhu, Michael Yang, Yangruibo Ding, and Wenbo Guo.
\newblock Co-patcher: Collaborative software patching with component (s)-specific small reasoning models.
\newblock \emph{arXiv preprint arXiv:2505.18955}, 2025{\natexlab{b}}.

\bibitem[Tang et~al.(2025{\natexlab{c}})Tang, Zheng, Synnaeve, and Munos]{tang2025grpopassk}
Yunhao Tang, Kunhao Zheng, Gabriel Synnaeve, and Rémi Munos.
\newblock Optimizing language models for inference time objectives using reinforcement learning, 2025{\natexlab{c}}.
\newblock URL \url{https://arxiv.org/abs/2503.19595}.

\bibitem[Tang et~al.(2023{\natexlab{a}})Tang, Agarwal, Shypula, Wang, Wijaya, Chen, and Kim]{tang2023explainthentranslateanalysisimprovingprogram}
Zilu Tang, Mayank Agarwal, Alex Shypula, Bailin Wang, Derry Wijaya, Jie Chen, and Yoon Kim.
\newblock Explain-then-translate: An analysis on improving program translation with self-generated explanations, 2023{\natexlab{a}}.
\newblock URL \url{https://arxiv.org/abs/2311.07070}.

\bibitem[Tang et~al.(2023{\natexlab{b}})Tang, Agarwal, Shypula, Wang, Wijaya, Chen, and Kim]{tang-etal-2023-explain}
Zilu Tang, Mayank Agarwal, Alexander Shypula, Bailin Wang, Derry Wijaya, Jie Chen, and Yoon Kim.
\newblock Explain-then-translate: an analysis on improving program translation with self-generated explanations.
\newblock In Houda Bouamor, Juan Pino, and Kalika Bali, editors, \emph{Findings of the Association for Computational Linguistics: EMNLP 2023}, pages 1741--1788, Singapore, December 2023{\natexlab{b}}. Association for Computational Linguistics.
\newblock \doi{10.18653/v1/2023.findings-emnlp.119}.
\newblock URL \url{https://aclanthology.org/2023.findings-emnlp.119/}.

\bibitem[Tao et~al.(2025)Tao, Zhang, Tang, Peng, Zhu, Liu, Yang, Zhang, Xu, Zhang, et~al.]{tao2025code}
Hongyuan Tao, Ying Zhang, Zhenhao Tang, Hongen Peng, Xukun Zhu, Bingchang Liu, Yingguang Yang, Ziyin Zhang, Zhaogui Xu, Haipeng Zhang, et~al.
\newblock Code graph model (cgm): A graph-integrated large language model for repository-level software engineering tasks.
\newblock \emph{arXiv preprint arXiv:2505.16901}, 2025.

\bibitem[Tao et~al.(2022)Tao, Meng, Chen, Zhu, Liu, Du, Han, Zhao, Wang, and Yang]{DBLP:journals/sigmetrics/TaoMCZLDHZWY22}
Shimin Tao, Weibin Meng, Yimeng Chen, Yichen Zhu, Ying Liu, Chunning Du, Tao Han, Yongpeng Zhao, Xiangguang Wang, and Hao Yang.
\newblock Logstamp: Automatic online log parsing based on sequence labelling.
\newblock \emph{{SIGMETRICS} Perform. Evaluation Rev.}, 49\penalty0 (4):\penalty0 93--98, 2022.

\bibitem[Tao et~al.(2021)Tao, Wang, Shi, Du, Han, Zhang, Zhang, and Zhang]{tao2021evaluationcommitmessagegeneration}
Wei Tao, Yanlin Wang, Ensheng Shi, Lun Du, Shi Han, Hongyu Zhang, Dongmei Zhang, and Wenqiang Zhang.
\newblock On the evaluation of commit message generation models: An experimental study, 2021.
\newblock URL \url{https://arxiv.org/abs/2107.05373}.

\bibitem[Tao et~al.(2024)Tao, Zhou, Wang, Zhang, Zhang, and Cheng]{tao2024magis}
Wei Tao, Yucheng Zhou, Yanlin Wang, Wenqiang Zhang, Hongyu Zhang, and Yu~Cheng.
\newblock Magis: Llm-based multi-agent framework for github issue resolution.
\newblock \emph{Advances in Neural Information Processing Systems}, 37:\penalty0 51963--51993, 2024.

\bibitem[Taori et~al.(2023)Taori, Gulrajani, Zhang, Dubois, Li, Guestrin, Liang, and Hashimoto]{alpaca}
Rohan Taori, Ishaan Gulrajani, Tianyi Zhang, Yann Dubois, Xuechen Li, Carlos Guestrin, Percy Liang, and Tatsunori~B. Hashimoto.
\newblock Stanford alpaca: An instruction-following llama model.
\newblock \url{https://github.com/tatsu-lab/stanford_alpaca}, 2023.

\bibitem[Team(2024{\natexlab{a}})]{aider2024}
Aider Team.
\newblock Aider: Ai pair programming in your terminal.
\newblock \url{https://aider.chat/}, 2024{\natexlab{a}}.
\newblock Accessed: 2024.

\bibitem[Team(2023)]{alphacode2_technical_report}
AlphaCode Team.
\newblock Alphacode 2 technical report.
\newblock Technical report, Google DeepMind, December 2023.
\newblock URL \url{https://storage.googleapis.com/deepmind-media/AlphaCode2/AlphaCode2_Tech_Report.pdf}.

\bibitem[Team(2024{\natexlab{b}})]{augment2024}
Augment~Code Team.
\newblock Augment code: The most powerful ai software development platform.
\newblock \url{https://www.augmentcode.com/}, 2024{\natexlab{b}}.
\newblock Accessed: 2024.

\bibitem[Team et~al.(2024{\natexlab{a}})Team, Georgiev, Lei, Burnell, Bai, Gulati, and et~al.]{geminiteam2024gemini15unlockingmultimodal}
Gemini Team, Petko Georgiev, Ving~Ian Lei, Ryan Burnell, Libin Bai, Anmol Gulati, and et~al.
\newblock Gemini 1.5: Unlocking multimodal understanding across millions of tokens of context, 2024{\natexlab{a}}.
\newblock URL \url{https://arxiv.org/abs/2403.05530}.

\bibitem[Team et~al.(2024{\natexlab{b}})Team, Georgiev, Lei, Burnell, Bai, Gulati, Tanzer, Vincent, Pan, Wang, Mariooryad, Ding, Geng, Alcober, Frostig, Omernick, Walker, Paduraru, Sorokin, Tacchetti, Gaffney, Daruki, Sercinoglu, Gleicher, Love, Voigtlaender, Jain, Surita, Mohamed, Blevins, Ahn, Zhu, Kawintiranon, Firat, Gu, Zhang, Rahtz, Faruqui, Clay, Gilmer, Co-Reyes, et~al.]{geminiteam2024gemini1_5}
Gemini Team, Petko Georgiev, Ving~Ian Lei, Ryan Burnell, Libin Bai, Anmol Gulati, Garrett Tanzer, Damien Vincent, Zhufeng Pan, Shibo Wang, Soroosh Mariooryad, Yifan Ding, Xinyang Geng, Fred Alcober, Roy Frostig, Mark Omernick, Lexi Walker, Cosmin Paduraru, Christina Sorokin, Andrea Tacchetti, Colin Gaffney, Samira Daruki, Olcan Sercinoglu, Zach Gleicher, Juliette Love, Paul Voigtlaender, Rohan Jain, Gabriela Surita, Kareem Mohamed, Rory Blevins, Junwhan Ahn, Tao Zhu, Kornraphop Kawintiranon, Orhan Firat, Yiming Gu, Yujing Zhang, Matthew Rahtz, Manaal Faruqui, Natalie Clay, Justin Gilmer, JD~Co-Reyes, et~al.
\newblock Gemini 1.5: Unlocking multimodal understanding across millions of tokens of context, 2024{\natexlab{b}}.
\newblock URL \url{https://arxiv.org/abs/2403.05530}.

\bibitem[Team et~al.(2025{\natexlab{a}})Team, Anil, Borgeaud, Alayrac, Yu, Soricut, and et~al.]{geminiteam2025geminifamilyhighlycapable}
Gemini Team, Rohan Anil, Sebastian Borgeaud, Jean-Baptiste Alayrac, Jiahui Yu, Radu Soricut, and et~al.
\newblock Gemini: A family of highly capable multimodal models, 2025{\natexlab{a}}.
\newblock URL \url{https://arxiv.org/abs/2312.11805}.

\bibitem[Team et~al.(2025{\natexlab{b}})Team, Anil, Borgeaud, Alayrac, Yu, Soricut, Schalkwyk, Dai, Hauth, Millican, Silver, Johnson, Antonoglou, Schrittwieser, Glaese, Chen, Pitler, Lillicrap, Lazaridou, Firat, Molloy, Isard, Barham, Hennigan, Lee, Viola, Reynolds, Xu, Doherty, Collins, Meyer, Rutherford, Moreira, Ayoub, Goel, Krawczyk, et~al.]{geminiteam2025gemini}
Gemini Team, Rohan Anil, Sebastian Borgeaud, Jean-Baptiste Alayrac, Jiahui Yu, Radu Soricut, Johan Schalkwyk, Andrew~M. Dai, Anja Hauth, Katie Millican, David Silver, Melvin Johnson, Ioannis Antonoglou, Julian Schrittwieser, Amelia Glaese, Jilin Chen, Emily Pitler, Timothy Lillicrap, Angeliki Lazaridou, Orhan Firat, James Molloy, Michael Isard, Paul~R. Barham, Tom Hennigan, Benjamin Lee, Fabio Viola, Malcolm Reynolds, Yuanzhong Xu, Ryan Doherty, Eli Collins, Clemens Meyer, Eliza Rutherford, Erica Moreira, Kareem Ayoub, Megha Goel, Jack Krawczyk, et~al.
\newblock Gemini: A family of highly capable multimodal models, 2025{\natexlab{b}}.
\newblock URL \url{https://arxiv.org/abs/2312.11805}.

\bibitem[Team et~al.(2025{\natexlab{c}})Team, Kamath, Ferret, Pathak, Vieillard, Merhej, Perrin, Matejovicova, Ramé, Rivière, Rouillard, Mesnard, Cideron, bastien Grill, Ramos, Yvinec, Casbon, Pot, Penchev, Liu, Visin, Kenealy, Beyer, Zhai, Tsitsulin, Busa-Fekete, Feng, Sachdeva, Coleman, Gao, Mustafa, Barr, Parisotto, Tian, Eyal, Cherry, Peter, Sinopalnikov, Bhupatiraju, Agarwal, Kazemi, Malkin, Kumar, Vilar, Brusilovsky, Luo, et~al.]{gemmateam2025gemma3}
Gemma Team, Aishwarya Kamath, Johan Ferret, Shreya Pathak, Nino Vieillard, Ramona Merhej, Sarah Perrin, Tatiana Matejovicova, Alexandre Ramé, Morgane Rivière, Louis Rouillard, Thomas Mesnard, Geoffrey Cideron, Jean bastien Grill, Sabela Ramos, Edouard Yvinec, Michelle Casbon, Etienne Pot, Ivo Penchev, Gaël Liu, Francesco Visin, Kathleen Kenealy, Lucas Beyer, Xiaohai Zhai, Anton Tsitsulin, Robert Busa-Fekete, Alex Feng, Noveen Sachdeva, Benjamin Coleman, Yi~Gao, Basil Mustafa, Iain Barr, Emilio Parisotto, David Tian, Matan Eyal, Colin Cherry, Jan-Thorsten Peter, Danila Sinopalnikov, Surya Bhupatiraju, Rishabh Agarwal, Mehran Kazemi, Dan Malkin, Ravin Kumar, David Vilar, Idan Brusilovsky, Jiaming Luo, et~al.
\newblock Gemma 3 technical report, 2025{\natexlab{c}}.
\newblock URL \url{https://arxiv.org/abs/2503.19786}.

\bibitem[Team et~al.(2025{\natexlab{d}})Team, Bai, Bao, Chen, Chen, Chen, Chen, Chen, Chen, Chen, et~al.]{k2}
Kimi Team, Yifan Bai, Yiping Bao, Guanduo Chen, Jiahao Chen, Ningxin Chen, Ruijue Chen, Yanru Chen, Yuankun Chen, Yutian Chen, et~al.
\newblock Kimi k2: Open agentic intelligence.
\newblock \emph{arXiv preprint arXiv:2507.20534}, 2025{\natexlab{d}}.

\bibitem[Team et~al.(2025{\natexlab{e}})Team, Cai, Cao, Chen, Chen, Chen, Cui, Di, Fang, Gong, et~al.]{team2025every}
Ling Team, Wenting Cai, Yuchen Cao, Chaoyu Chen, Chen Chen, Siba Chen, Qing Cui, Peng Di, Junpeng Fang, Zi~Gong, et~al.
\newblock Every sample matters: Leveraging mixture-of-experts and high-quality data for efficient and accurate code llm.
\newblock \emph{arXiv preprint arXiv:2503.17793}, 2025{\natexlab{e}}.

\bibitem[Team(2025{\natexlab{a}})]{manus}
Manus Team.
\newblock Leave it to manus.
\newblock \url{https://manus.im/}, 2025{\natexlab{a}}.

\bibitem[Team(2025{\natexlab{b}})]{oai2025deepresearch}
OpenAI Team.
\newblock Introducing deep research.
\newblock \url{https://openai.com/index/introducing-deep-research/}, 2025{\natexlab{b}}.

\bibitem[Team(2024{\natexlab{c}})]{qwen2024codeqwen1.5}
Qwen Team.
\newblock Codeqwen1.5.
\newblock \url{https://huggingface.co/Qwen/CodeQwen1.5-7B}, 2024{\natexlab{c}}.

\bibitem[Team(2024{\natexlab{d}})]{qwen2024qwen1_5}
Qwen Team.
\newblock Introducing qwen1.5, February 2024{\natexlab{d}}.
\newblock URL \url{https://qwenlm.github.io/blog/qwen1.5/}.

\bibitem[Team(2025{\natexlab{c}})]{qwen2025qwen3next}
Qwen Team.
\newblock Qwen3-next: Towards ultimate training \& inference efficiency, September 2025{\natexlab{c}}.
\newblock URL \url{https://qwen.ai/blog?id=4074cca80393150c248e508aa62983f9cb7d27cd\&from=research.latest-advancements-list}.

\bibitem[Team(2025{\natexlab{d}})]{Refact2025}
Refact Team.
\newblock Refact.ai.
\newblock \url{https://refact.ai/}, 2025{\natexlab{d}}.
\newblock Accessed: 2025.

\bibitem[Team(2025{\natexlab{e}})]{ob_1}
The~OpenBlock Team.
\newblock Openblock secures \#2 on terminal bench with frontier agent ob-1, August 2025{\natexlab{e}}.
\newblock URL \url{https://www.openblocklabs.com/research/terminal-bench}.

\bibitem[Team(2025{\natexlab{f}})]{tbench_2025}
The Terminal-Bench Team.
\newblock Terminal-bench: A benchmark for ai agents in terminal environments, Apr 2025{\natexlab{f}}.
\newblock URL \url{https://github.com/laude-institute/terminal-bench}.

\bibitem[Thakur et~al.(2023)Thakur, Ahmad, Fan, et~al.]{thakur2023verigen}
Shailja Thakur, Baleegh Ahmad, Hammond Fan, et~al.
\newblock {VeriGen}: A large language model for verilog code generation.
\newblock \emph{arXiv preprint arXiv:2308.00708}, 2023.

\bibitem[Thapa et~al.(2023)Thapa, Naseem, and Nasim]{thapa2023humans}
Surendrabikram Thapa, Usman Naseem, and Mehwish Nasim.
\newblock From humans to machines: can chatgpt-like llms effectively replace human annotators in nlp tasks.
\newblock In \emph{Workshop Proceedings of the 17th International AAAI Conference on Web and Social Media}. Association for the Advancement of Artificial Intelligence, 2023.

\bibitem[Tian et~al.(2024{\natexlab{a}})Tian, Gao, Zhang, Chen, Fan, Guo, Haas, Ji, Krongchon, Li, Liu, Luo, Ma, Tong, Trinh, Tian, Wang, Wu, Xiong, Yin, Zhu, Lieret, Lu, Liu, Du, Tao, Press, Callan, Huerta, and Peng]{scicode}
Minyang Tian, Luyu Gao, Shizhuo~Dylan Zhang, Xinan Chen, Cunwei Fan, Xuefei Guo, Roland Haas, Pan Ji, Kittithat Krongchon, Yao Li, Shengyan Liu, Di~Luo, Yutao Ma, Hao Tong, Kha Trinh, Chenyu Tian, Zihan Wang, Bohao Wu, Yanyu Xiong, Shengzhu Yin, Minhui Zhu, Kilian Lieret, Yanxin Lu, Genglin Liu, Yufeng Du, Tianhua Tao, Ofir Press, Jamie Callan, Eliu Huerta, and Hao Peng.
\newblock Scicode: A research coding benchmark curated by scientists, 2024{\natexlab{a}}.
\newblock URL \url{https://arxiv.org/abs/2407.13168}.

\bibitem[Tian et~al.(2024{\natexlab{b}})Tian, Ye, Qin, Cong, Lin, Pan, Wu, Hui, Liu, Liu, and Sun]{debugbench}
Runchu Tian, Yining Ye, Yujia Qin, Xin Cong, Yankai Lin, Yinxu Pan, Yesai Wu, Haotian Hui, Weichuan Liu, Zhiyuan Liu, and Maosong Sun.
\newblock Debugbench: Evaluating debugging capability of large language models, 2024{\natexlab{b}}.
\newblock URL \url{https://arxiv.org/abs/2401.04621}.

\bibitem[Tihanyi et~al.(2024)Tihanyi, Bisztray, Ferrag, Jain, and Cordeiro]{tihanyi2024howsecureisaigeneratedcode}
Norbert Tihanyi, Tamas Bisztray, Mohamed~Amine Ferrag, Ridhi Jain, and Lucas~C. Cordeiro.
\newblock How secure is ai-generated code: a large-scale comparison of large language models.
\newblock \emph{Empirical Software Engineering}, 30\penalty0 (2):\penalty0 47, 2024.
\newblock ISSN 1573-7616.
\newblock \doi{10.1007/s10664-024-10590-1}.
\newblock URL \url{https://doi.org/10.1007/s10664-024-10590-1}.

\bibitem[Tihanyi et~al.(2025)Tihanyi, Charalambous, Jain, Ferrag, and Cordeiro]{tihanyi2025new}
Norbert Tihanyi, Yiannis Charalambous, Ridhi Jain, Mohamed~Amine Ferrag, and Lucas~C Cordeiro.
\newblock A new era in software security: Towards self-healing software via large language models and formal verification.
\newblock In \emph{2025 IEEE/ACM International Conference on Automation of Software Test (AST)}, pages 136--147. IEEE, 2025.

\bibitem[Tip et~al.(2025)Tip, Bell, and Sch{\"a}fer]{tip2025llmorpheus}
Frank Tip, Jonathan Bell, and Max Sch{\"a}fer.
\newblock Llmorpheus: Mutation testing using large language models.
\newblock \emph{IEEE Transactions on Software Engineering}, 2025.

\bibitem[Tjuatja et~al.(2024)Tjuatja, Chen, Wu, Talwalkwar, and Neubig]{tjuatja2024llms}
Lindia Tjuatja, Valerie Chen, Tongshuang Wu, Ameet Talwalkwar, and Graham Neubig.
\newblock Do llms exhibit human-like response biases? a case study in survey design.
\newblock \emph{Transactions of the Association for Computational Linguistics}, 12:\penalty0 1011--1026, 2024.

\bibitem[TogetherAI(2025)]{deepswe}
TogetherAI.
\newblock Deepswe: Training a fully open-sourced, state-of-the-art coding agent by scaling rl.
\newblock \url{https://www.together.ai/blog/deepswe}, 2025.
\newblock Accessed: 2025.

\bibitem[{Toggle Project}(2024)]{toggle2024}
{Toggle Project}.
\newblock Toggle: Token-level localization for automated program repair.
\newblock \url{https://github.com/Toggle-APR/toggle}, 2024.

\bibitem[Tong and Zhang(2024{\natexlab{a}})]{tong2024codejudge}
Weixi Tong and Tianyi Zhang.
\newblock Codejudge: Evaluating code generation with large language models.
\newblock \emph{arXiv preprint arXiv:2410.02184}, 2024{\natexlab{a}}.

\bibitem[Tong and Zhang(2024{\natexlab{b}})]{tong2024codejudgeevaluatingcodegeneration}
Weixi Tong and Tianyi Zhang.
\newblock Codejudge: Evaluating code generation with large language models, 2024{\natexlab{b}}.
\newblock URL \url{https://arxiv.org/abs/2410.02184}.

\bibitem[Touvron et~al.(2023{\natexlab{a}})Touvron, Lavril, Izacard, Martinet, Lachaux, Lacroix, Rozi{\`e}re, Goyal, Hambro, Azhar, et~al.]{touvron2023llama}
Hugo Touvron, Thibaut Lavril, Gautier Izacard, Xavier Martinet, Marie-Anne Lachaux, Timoth{\'e}e Lacroix, Baptiste Rozi{\`e}re, Naman Goyal, Eric Hambro, Faisal Azhar, et~al.
\newblock Llama: Open and efficient foundation language models.
\newblock \emph{arXiv preprint arXiv:2302.13971}, 2023{\natexlab{a}}.

\bibitem[Touvron et~al.(2023{\natexlab{b}})Touvron, Martin, Stone, Albert, Almahairi, Babaei, Bashlykov, Batra, Bhargava, Bhosale, Bikel, Blecher, Ferrer, Chen, Cucurull, Esiobu, Fernandes, Fu, Fu, Fuller, Gao, Goswami, Goyal, Hartshorn, Hosseini, Hou, Inan, Kardas, Kerkez, Khabsa, Kloumann, Korenev, Koura, Lachaux, Lavril, Lee, Liskovich, Lu, Mao, Martinet, Mihaylov, Mishra, Molybog, Nie, Poulton, Reizenstein, Rungta, Saladi, Schelten, Silva, Smith, Subramanian, Tan, Tang, Taylor, Williams, Kuan, Xu, Yan, Zarov, Zhang, Fan, Kambadur, Narang, Rodriguez, Stojnic, Edunov, and Scialom]{touvron2023llama2}
Hugo Touvron, Louis Martin, Kevin Stone, Peter Albert, Amjad Almahairi, Yasmine Babaei, Nikolay Bashlykov, Soumya Batra, Prajjwal Bhargava, Shruti Bhosale, Dan Bikel, Lukas Blecher, Cristian~Canton Ferrer, Moya Chen, Guillem Cucurull, David Esiobu, Jude Fernandes, Jeremy Fu, Wenyin Fu, Brian Fuller, Cynthia Gao, Vedanuj Goswami, Naman Goyal, Anthony Hartshorn, Saghar Hosseini, Rui Hou, Hakan Inan, Marcin Kardas, Viktor Kerkez, Madian Khabsa, Isabel Kloumann, Artem Korenev, Punit~Singh Koura, Marie-Anne Lachaux, Thibaut Lavril, Jenya Lee, Diana Liskovich, Yinghai Lu, Yuning Mao, Xavier Martinet, Todor Mihaylov, Pushkar Mishra, Igor Molybog, Yixin Nie, Andrew Poulton, Jeremy Reizenstein, Rashi Rungta, Kalyan Saladi, Alan Schelten, Ruan Silva, Eric~Michael Smith, Ranjan Subramanian, Xiaoqing~Ellen Tan, Binh Tang, Ross Taylor, Adina Williams, Jian~Xiang Kuan, Puxin Xu, Zheng Yan, Iliyan Zarov, Yuchen Zhang, Angela Fan, Melanie Kambadur, Sharan Narang, Aurelien Rodriguez, Robert Stojnic, Sergey Edunov, and Thomas
  Scialom.
\newblock Llama 2: Open foundation and fine-tuned chat models, 2023{\natexlab{b}}.
\newblock URL \url{https://arxiv.org/abs/2307.09288}.

\bibitem[{Towards Data Science}(2025)]{towardsdatascience2025vscode}
{Towards Data Science}.
\newblock Why i stopped using cursor and reverted to vscode.
\newblock \url{https://towardsdatascience.com/vscode-is-the-best-ai-powered-ide}, 2025.

\bibitem[{Towers} et~al.(2024){Towers}, {Kwiatkowski}, {Terry}, {Balis}, {De Cola}, {Deleu}, {Goul{\~a}o}, {Kallinteris}, {Krimmel}, {KG}, {Perez-Vicente}, {Pierr{\'e}}, {Schulhoff}, {Jet Tai}, {Tan}, and {Younis}]{Gymnasium}
Mark {Towers}, Ariel {Kwiatkowski}, Jordan {Terry}, John~U. {Balis}, Gianluca {De Cola}, Tristan {Deleu}, Manuel {Goul{\~a}o}, Andreas {Kallinteris}, Markus {Krimmel}, Arjun {KG}, Rodrigo {Perez-Vicente}, Andrea {Pierr{\'e}}, Sander {Schulhoff}, Jun {Jet Tai}, Hannah {Tan}, and Omar~G. {Younis}.
\newblock {Gymnasium: A Standard Interface for Reinforcement Learning Environments}.
\newblock \emph{arXiv e-prints}, art. arXiv:2407.17032, July 2024.
\newblock \doi{10.48550/arXiv.2407.17032}.

\bibitem[{TRAE}(2025)]{trae2025trae}
{TRAE}.
\newblock {TRAE} - collaborate with intelligence.
\newblock Technical report, TRAE, 2025.
\newblock URL \url{https://trae.ai}.

\bibitem[Tran et~al.(2019)Tran, Tran, Nguyen, Nguyen, and Nguyen]{DBLP:conf/icse/TranTNNN19}
Hieu Tran, Ngoc~M. Tran, Son Nguyen, Hoan Nguyen, and Tien~N. Nguyen.
\newblock Recovering variable names for minified code with usage contexts.
\newblock In Joanne~M. Atlee, Tevfik Bultan, and Jon Whittle, editors, \emph{Proceedings of the 41st International Conference on Software Engineering, {ICSE} 2019, Montreal, QC, Canada, May 25-31, 2019}, pages 1165--1175. {IEEE} / {ACM}, 2019.

\bibitem[Tsai et~al.(2024)Tsai, Liu, and Ren]{tsai2024codelessalignmore}
Yun-Da Tsai, Mingjie Liu, and Haoxing Ren.
\newblock Code less, align more: Efficient llm fine-tuning for code generation with data pruning, 2024.
\newblock URL \url{https://arxiv.org/abs/2407.05040}.

\bibitem[Tu et~al.(2025)Tu, Lin, Tian, Zhang, Li, Fu, Xu, He, Lan, Jiang, et~al.]{tu2025enhancing}
Songjun Tu, Jiahao Lin, Xiangyu Tian, Qichao Zhang, Linjing Li, Yuqian Fu, Nan Xu, Wei He, Xiangyuan Lan, Dongmei Jiang, et~al.
\newblock Enhancing llm reasoning with iterative dpo: A comprehensive empirical investigation.
\newblock \emph{arXiv preprint arXiv:2503.12854}, 2025.

\bibitem[Tufano et~al.(2020)Tufano, Drain, Svyatkovskiy, Deng, and Sundaresan]{tufano2020unit}
Michele Tufano, Dawn Drain, Alexey Svyatkovskiy, Shao~Kun Deng, and Neel Sundaresan.
\newblock Unit test case generation with transformers and focal context.
\newblock \emph{arXiv preprint arXiv:2009.05617}, 2020.

\bibitem[Tufano et~al.(2024)Tufano, Agarwal, Jang, Moghaddam, and Sundaresan]{tufano2024autodev}
Michele Tufano, Anisha Agarwal, Jinu Jang, Roshanak~Zilouchian Moghaddam, and Neel Sundaresan.
\newblock Autodev: Automated ai-driven development.
\newblock \emph{arXiv preprint arXiv:2403.08299}, 2024.

\bibitem[Uniyal et~al.(2024)Uniyal, Singh, Verbruggen, Gulwani, and Le]{mbupp}
Mansi Uniyal, Mukul Singh, Gust Verbruggen, Sumit Gulwani, and Vu~Le.
\newblock One-to-many testing for code generation from (just) natural language.
\newblock In Yaser Al-Onaizan, Mohit Bansal, and Yun-Nung Chen, editors, \emph{Findings of the Association for Computational Linguistics: EMNLP 2024}, pages 15397--15402, Miami, Florida, USA, November 2024. Association for Computational Linguistics.
\newblock \doi{10.18653/v1/2024.findings-emnlp.902}.
\newblock URL \url{https://aclanthology.org/2024.findings-emnlp.902/}.

\bibitem[Valeev et~al.(2025)Valeev, Garaev, Lomshakov, Piontkovskaya, Ivanov, and Adewuyi]{yabloco}
Aidar Valeev, Roman Garaev, Vadim Lomshakov, Irina Piontkovskaya, Vladimir Ivanov, and Israel Adewuyi.
\newblock Yabloco: Yet another benchmark for long context code generation, 2025.
\newblock URL \url{https://arxiv.org/abs/2505.04406}.

\bibitem[Valentin et~al.(2025)Valentin, Madadi, Sapia, and Böhme]{valentin2025estimatingcorrectnessoraclesllmbased}
Thomas Valentin, Ardi Madadi, Gaetano Sapia, and Marcel Böhme.
\newblock Estimating correctness without oracles in llm-based code generation, 2025.
\newblock URL \url{https://arxiv.org/abs/2507.00057}.

\bibitem[Van~Erven and Harremos(2014)]{van2014renyi}
Tim Van~Erven and Peter Harremos.
\newblock R{\'e}nyi divergence and kullback-leibler divergence.
\newblock \emph{IEEE Transactions on Information Theory}, 60\penalty0 (7):\penalty0 3797--3820, 2014.

\bibitem[van Hal et~al.(2019)van Hal, Post, and Wendel]{vanhal2019generatingcommitmessagesgit}
S.~R.~P. van Hal, M.~Post, and K.~Wendel.
\newblock Generating commit messages from git diffs, 2019.
\newblock URL \url{https://arxiv.org/abs/1911.11690}.

\bibitem[Vasic et~al.(2019)Vasic, Kanade, Maniatis, Bieber, and Singh]{vasic2019neural}
Marko Vasic, Aditya Kanade, Petros Maniatis, David Bieber, and Rishabh Singh.
\newblock Neural program repair by jointly learning to localize and repair.
\newblock \emph{arXiv preprint arXiv:1904.01720}, 2019.

\bibitem[Vasilescu et~al.(2017)Vasilescu, Casalnuovo, and Devanbu]{DBLP:conf/sigsoft/VasilescuCD17}
Bogdan Vasilescu, Casey Casalnuovo, and Premkumar~T. Devanbu.
\newblock Recovering clear, natural identifiers from obfuscated {JS} names.
\newblock In Eric Bodden, Wilhelm Sch{\"{a}}fer, Arie van Deursen, and Andrea Zisman, editors, \emph{Proceedings of the 2017 11th Joint Meeting on Foundations of Software Engineering, {ESEC/FSE} 2017, Paderborn, Germany, September 4-8, 2017}, pages 683--693. {ACM}, 2017.

\bibitem[Vaswani et~al.(2017{\natexlab{a}})Vaswani, Shazeer, Parmar, Uszkoreit, Jones, Gomez, Kaiser, and Polosukhin]{vaswani2017transformer}
Ashish Vaswani, Noam Shazeer, Niki Parmar, Jakob Uszkoreit, Llion Jones, Aidan~N Gomez, \L~ukasz Kaiser, and Illia Polosukhin.
\newblock Attention is all you need.
\newblock In I.~Guyon, U.~Von Luxburg, S.~Bengio, H.~Wallach, R.~Fergus, S.~Vishwanathan, and R.~Garnett, editors, \emph{Advances in Neural Information Processing Systems}, volume~30. Curran Associates, Inc., 2017{\natexlab{a}}.
\newblock URL \url{https://proceedings.neurips.cc/paper_files/paper/2017/file/3f5ee243547dee91fbd053c1c4a845aa-Paper.pdf}.

\bibitem[Vaswani et~al.(2017{\natexlab{b}})Vaswani, Shazeer, Parmar, Uszkoreit, Jones, Gomez, Kaiser, and Polosukhin]{vaswani2017attention}
Ashish Vaswani, Noam Shazeer, Niki Parmar, Jakob Uszkoreit, Llion Jones, Aidan~N. Gomez, Lukasz Kaiser, and Illia Polosukhin.
\newblock Attention is all you need.
\newblock In \emph{{NIPS}}, pages 5998--6008, 2017{\natexlab{b}}.

\bibitem[Vaziry et~al.(2025)Vaziry, Garzon, and K{\"u}pper]{multi_agent_econmies_a2a}
Awid Vaziry, Sandro~Rodriguez Garzon, and Axel K{\"u}pper.
\newblock Towards multi-agent economies: Enhancing the a2a protocol with ledger-anchored identities and x402 micropayments for ai agents.
\newblock \emph{arXiv preprint arXiv:2507.19550}, 2025.

\bibitem[{Visual Studio Blog}(2025)]{githubcopilot2024features}
{Visual Studio Blog}.
\newblock Top 5 github copilot features in visual studio from microsoft ignite 2024.
\newblock \url{https://devblogs.microsoft.com/visualstudio/top-5-github-copilot-features/}, 2025.

\bibitem[Waghjale et~al.(2024)Waghjale, Veerendranath, Wang, and Fried]{ecco2024}
Siddhant Waghjale, Vishruth Veerendranath, Zora~Zhiruo Wang, and Daniel Fried.
\newblock Ecco: Can we improve model-generated code efficiency without sacrificing functional correctness?
\newblock \emph{ArXiv}, abs/2407.14044, 2024.
\newblock URL \url{https://api.semanticscholar.org/CorpusID:271310399}.

\bibitem[Wan et~al.(2018)Wan, Zhao, Yang, Xu, Ying, Wu, and Yu]{wan2018improving}
Yao Wan, Zhou Zhao, Min Yang, Guandong Xu, Haochao Ying, Jian Wu, and Philip~S Yu.
\newblock Improving automatic source code summarization via deep reinforcement learning.
\newblock In \emph{Proceedings of the 33rd ACM/IEEE international conference on automated software engineering}, pages 397--407, 2018.

\bibitem[Wan et~al.(2025)Wan, Wang, Dong, Wang, Li, Huo, and Lyu]{DCGen}
Yuxuan Wan, Chaozheng Wang, Yi~Dong, Wenxuan Wang, Shuqing Li, Yintong Huo, and Michael Lyu.
\newblock Divide-and-conquer: Generating ui code from screenshots.
\newblock \emph{Proc. ACM Softw. Eng.}, 2\penalty0 (FSE), June 2025.
\newblock \doi{10.1145/3729364}.
\newblock URL \url{https://doi.org/10.1145/3729364}.

\bibitem[Wan et~al.(2024)Wan, Cheng, Wang, and Wang]{wan2024information}
Zhipeng Wan, Anda Cheng, Yinggui Wang, and Lei Wang.
\newblock Information leakage from embedding in large language models, 2024.

\bibitem[Wan et~al.(2017)Wan, Lo, Xia, Cai, and Li]{wan2017mining}
Zhiyuan Wan, David Lo, Xin Xia, Liang Cai, and Shanping Li.
\newblock Mining sandboxes for linux containers.
\newblock In \emph{2017 IEEE International Conference on Software Testing, Verification and Validation (ICST)}, pages 92--102. IEEE, 2017.

\bibitem[Wang et~al.(2025{\natexlab{a}})Wang, Qi, and Baiyila]{cbert}
Amuguleng Wang, Yilagui Qi, and Dahu Baiyila.
\newblock C-bert: A mongolian reverse dictionary based on fused lexical semantic clustering and bert.
\newblock \emph{Alexandria Engineering Journal}, 111:\penalty0 385--395, 2025{\natexlab{a}}.

\bibitem[Wang et~al.(2020{\natexlab{a}})Wang, Shin, Liu, Polozov, and Richardson]{Wang}
Bailin Wang, Richard Shin, Xiaodong Liu, Oleksandr Polozov, and Matthew Richardson.
\newblock {RAT-SQL:} relation-aware schema encoding and linking for text-to-sql parsers.
\newblock In Dan Jurafsky, Joyce Chai, Natalie Schluter, and Joel~R. Tetreault, editors, \emph{Proceedings of the 58th Annual Meeting of the Association for Computational Linguistics, {ACL} 2020, Online, July 5-10, 2020}, pages 7567--7578. Association for Computational Linguistics, 2020{\natexlab{a}}.

\bibitem[Wang et~al.(2023{\natexlab{a}})Wang, Ren, Yang, Liang, Bai, Zhang, Yan, and Li]{MAC-SQL}
Bing Wang, Changyu Ren, Jian Yang, Xinnian Liang, Jiaqi Bai, Qian{-}Wen Zhang, Zhao Yan, and Zhoujun Li.
\newblock {MAC-SQL:} {A} multi-agent collaborative framework for text-to-sql.
\newblock \emph{CoRR}, abs/2312.11242, 2023{\natexlab{a}}.

\bibitem[Wang et~al.(2025{\natexlab{b}})Wang, Wang, Chen, Sun, Shi, Yang, Deng, Lin, Yang, and Lo]{wang2025mut4allfuzzingcompilersllmsynthesized}
Bo~Wang, Pengyang Wang, Chong Chen, Qi~Sun, Jieke Shi, Chengran Yang, Ming Deng, Youfang Lin, Zhou Yang, and David Lo.
\newblock Mut4all: Fuzzing compilers via llm-synthesized mutators learned from bug reports, 2025{\natexlab{b}}.
\newblock URL \url{https://arxiv.org/abs/2507.19275}.

\bibitem[Wang et~al.(2022{\natexlab{a}})Wang, Min, Deng, Shen, Wu, Zettlemoyer, and Sun]{wang2022towards}
Boshi Wang, Sewon Min, Xiang Deng, Jiaming Shen, You Wu, Luke Zettlemoyer, and Huan Sun.
\newblock Towards understanding chain-of-thought prompting: An empirical study of what matters.
\newblock \emph{arXiv preprint arXiv:2212.10001}, 2022{\natexlab{a}}.

\bibitem[Wang et~al.(2022{\natexlab{b}})Wang, Nong, Gao, Li, Zeng, Xing, and Liu]{wang2022enriching}
Chaozheng Wang, Zhenhao Nong, Cuiyun Gao, Zongjie Li, Jichuan Zeng, Zhenchang Xing, and Yang Liu.
\newblock Enriching query semantics for code search with reinforcement learning.
\newblock \emph{Neural Networks}, 145:\penalty0 22--32, 2022{\natexlab{b}}.

\bibitem[Wang et~al.(2025{\natexlab{c}})Wang, Singhal, Kelkar, and Tuo]{wang2025mi9}
Charles~L Wang, Trisha Singhal, Ameya Kelkar, and Jason Tuo.
\newblock Mi9--agent intelligence protocol: Runtime governance for agentic ai systems.
\newblock \emph{arXiv preprint arXiv:2508.03858}, 2025{\natexlab{c}}.

\bibitem[Wang et~al.(2024{\natexlab{a}})Wang, Zhang, Gao, Xia, Guan, and Chen]{wang2024contracttinker}
Che Wang, Jiashuo Zhang, Jianbo Gao, Libin Xia, Zhi Guan, and Zhong Chen.
\newblock Contracttinker: Llm-empowered vulnerability repair for real-world smart contracts.
\newblock In \emph{Proceedings of the 39th IEEE/ACM International Conference on Automated Software Engineering}, pages 2350--2353, 2024{\natexlab{a}}.

\bibitem[Wang et~al.(2023{\natexlab{b}})Wang, Xie, Jiang, Mandlekar, Xiao, Zhu, Fan, and Anandkumar]{voyager}
Guanzhi Wang, Yuqi Xie, Yunfan Jiang, Ajay Mandlekar, Chaowei Xiao, Yuke Zhu, Linxi Fan, and Anima Anandkumar.
\newblock Voyager: An open-ended embodied agent with large language models.
\newblock \emph{arXiv preprint arXiv: 2305.16291}, 2023{\natexlab{b}}.

\bibitem[Wang et~al.(2023{\natexlab{c}})Wang, Xie, Jiang, Mandlekar, Xiao, Zhu, Fan, and Anandkumar]{wang2023voyager}
Guanzhi Wang, Yuqi Xie, Yunfan Jiang, Ajay Mandlekar, Chaowei Xiao, Yuke Zhu, Linxi Fan, and Anima Anandkumar.
\newblock Voyager: An open-ended embodied agent with large language models.
\newblock \emph{arXiv preprint arXiv:2305.16291}, 2023{\natexlab{c}}.

\bibitem[Wang et~al.(2025{\natexlab{d}})Wang, Zhou, Xu, Cheng, Zuo, Tian, Song, Lu, Hu, and Liu]{wang2025code}
Hanbin Wang, Xiaoxuan Zhou, Zhipeng Xu, Keyuan Cheng, Yuxin Zuo, Kai Tian, Jingwei Song, Junting Lu, Wenhui Hu, and Xueyang Liu.
\newblock Code-vision: Evaluating multimodal llms logic understanding and code generation capabilities.
\newblock \emph{arXiv preprint arXiv:2502.11829}, 2025{\natexlab{d}}.

\bibitem[Wang et~al.(2025{\natexlab{e}})Wang, Wan, Wang, and Wang]{wang2025privacyaware}
Haoran Wang, Zhipeng Wan, Yinggui Wang, and Lei Wang.
\newblock Privacy-aware decoding: Mitigating privacy leakage of large language models in retrieval-augmented generation, 2025{\natexlab{e}}.

\bibitem[Wang et~al.(2021{\natexlab{a}})Wang, Xia, Lo, He, Wang, and Grundy]{wang2021context}
Haoye Wang, Xin Xia, David Lo, Qiang He, Xinyu Wang, and John Grundy.
\newblock Context-aware retrieval-based deep commit message generation.
\newblock \emph{ACM Transactions on Software Engineering and Methodology (TOSEM)}, 30\penalty0 (4):\penalty0 1--30, 2021{\natexlab{a}}.

\bibitem[Wang et~al.(2024{\natexlab{b}})Wang, He, and Chen]{wang2024repogenreflex}
Jicheng Wang, Yifeng He, and Hao Chen.
\newblock Repogenreflex: Enhancing repository-level code completion with verbal reinforcement and retrieval-augmented generation.
\newblock \emph{arXiv preprint arXiv:2409.13122}, 2024{\natexlab{b}}.

\bibitem[Wang et~al.(2024{\natexlab{c}})Wang, Li, Chen, Yu, Li, Zhu, Li, and Liao]{wang2024isyourai}
Jiexin Wang, Haonan Li, Jiayuan Chen, Zeliang Yu, Zongze Li, Wenyu Zhu, Li~Li, and Qing Liao.
\newblock Is your ai-generated code really safe? evaluating large language models on secure code generation with codeseceval, 2024{\natexlab{c}}.

\bibitem[Wang et~al.(2024{\natexlab{d}})Wang, Luo, Cao, He, Huang, Xie, Jatowt, and Cai]{wang2024your}
Jiexin Wang, Xitong Luo, Liuwen Cao, Hongkui He, Hailin Huang, Jiayuan Xie, Adam Jatowt, and Yi~Cai.
\newblock Is your ai-generated code really safe? evaluating large language models on secure code generation with codeseceval.
\newblock \emph{arXiv preprint arXiv:2407.02395}, 2024{\natexlab{d}}.

\bibitem[Wang et~al.(2024{\natexlab{e}})Wang, Kidambi, Sullivan, Agarwal, Dann, Michi, Gelmi, Li, Gupta, Dubey, et~al.]{wang2024conditional}
Kaiwen Wang, Rahul Kidambi, Ryan Sullivan, Alekh Agarwal, Christoph Dann, Andrea Michi, Marco Gelmi, Yunxuan Li, Raghav Gupta, Avinava Dubey, et~al.
\newblock Conditional language policy: A general framework for steerable multi-objective finetuning.
\newblock \emph{arXiv preprint arXiv:2407.15762}, 2024{\natexlab{e}}.

\bibitem[Wang et~al.(2024{\natexlab{f}})Wang, Wang, Zong, Huang, Yang, Li, Wang, Cheng, Liu, Ma, et~al.]{wang2024agentspec}
Le~Wang, Zhiliang Wang, Zepu Zong, Yuxin Huang, Min Yang, Ruipu Li, Runze Wang, Zhaoyue Cheng, Jing Liu, Xingyu Ma, et~al.
\newblock Agentspec: Customizable runtime enforcement for safe and reliable llm agents.
\newblock \emph{arXiv preprint arXiv:2405.01358}, 2024{\natexlab{f}}.

\bibitem[Wang et~al.(2023{\natexlab{d}})Wang, Ma, Feng, Zhang, ran Yang, Zhang, Chen, Tang, Chen, Lin, Zhao, Wei, and rong Wen]{agent-survey-1}
Lei Wang, Chengbang Ma, Xueyang Feng, Zeyu Zhang, Hao ran Yang, Jingsen Zhang, Zhi-Yang Chen, Jiakai Tang, Xu~Chen, Yankai Lin, Wayne~Xin Zhao, Zhewei Wei, and Ji~rong Wen.
\newblock A survey on large language model based autonomous agents.
\newblock \emph{Frontiers Comput. Sci.}, 2023{\natexlab{d}}.
\newblock \doi{10.1007/s11704-024-40231-1}.

\bibitem[Wang et~al.(2023{\natexlab{e}})Wang, Tang, He, Ren, Shi, Yan, and Li]{wang2023delving}
Liran Wang, Xunzhu Tang, Yichen He, Changyu Ren, Shuhua Shi, Chaoran Yan, and Zhoujun Li.
\newblock Delving into commit-issue correlation to enhance commit message generation models.
\newblock In \emph{2023 38th IEEE/ACM International Conference on Automated Software Engineering (ASE)}, pages 710--722. IEEE, 2023{\natexlab{e}}.

\bibitem[Wang et~al.(2024{\natexlab{g}})Wang, Cheng, Gu, and Wu]{wang2024design_risk}
Liyang Wang, Yu~Cheng, Xingxin Gu, and Zhizhong Wu.
\newblock Design and optimization of big data and machine learning-based risk monitoring system in financial markets.
\newblock \emph{arXiv preprint arXiv:2407.19352}, 2024{\natexlab{g}}.

\bibitem[Wang et~al.(2024{\natexlab{h}})Wang, Groundhog-Day, Li, He, Ji, Yang-Kai-Hsiang, Li, Chen, and Lin]{wang2024redcoder}
Min-Hui Wang, Groundhog-Day, Zhaowei Li, Jia-Ju He, Jia-Hao Ji, Yang-Kai-Hsiang, Feng-Lin Li, Wen-Hao Chen, and Zhe-Wei Lin.
\newblock Redcoder: Automated multi-turn red teaming for code llms, 2024{\natexlab{h}}.

\bibitem[Wang et~al.(2025{\natexlab{f}})Wang, Liu, Lu, Cai, Chen, Yang, Zhang, Hong, and Wu]{wang2025agentarmor}
Peiran Wang, Yang Liu, Yunfei Lu, Yifeng Cai, Hongbo Chen, Qingyou Yang, Jie Zhang, Jue Hong, and Ye~Wu.
\newblock Agentarmor: Enforcing program analysis on agent runtime trace to defend against prompt injection.
\newblock \emph{arXiv preprint arXiv:2508.01249}, 2025{\natexlab{f}}.

\bibitem[Wang et~al.(2024{\natexlab{i}})Wang, Li, Shao, Xu, Dai, Li, Chen, Wu, and Sui]{Math-Shepherd}
Peiyi Wang, Lei Li, Zhihong Shao, Runxin Xu, Damai Dai, Yifei Li, Deli Chen, Yu~Wu, and Zhifang Sui.
\newblock Math-shepherd: Verify and reinforce llms step-by-step without human annotations.
\newblock In Lun{-}Wei Ku, Andre Martins, and Vivek Srikumar, editors, \emph{Proceedings of the 62nd Annual Meeting of the Association for Computational Linguistics (Volume 1: Long Papers), {ACL} 2024, Bangkok, Thailand, August 11-16, 2024}, pages 9426--9439. Association for Computational Linguistics, 2024{\natexlab{i}}.
\newblock \doi{10.18653/V1/2024.ACL-LONG.510}.
\newblock URL \url{https://doi.org/10.18653/v1/2024.acl-long.510}.

\bibitem[Wang et~al.(2024{\natexlab{j}})Wang, Bai, Tan, Wang, Fan, Bai, Chen, Liu, Wang, Ge, Fan, Dang, Du, Ren, Men, Liu, Zhou, Zhou, and Lin]{qwen2024qwen2vl}
Peng Wang, Shuai Bai, Sinan Tan, Shijie Wang, Zhihao Fan, Jinze Bai, Keqin Chen, Xuejing Liu, Jialin Wang, Wenbin Ge, Yang Fan, Kai Dang, Mengfei Du, Xuancheng Ren, Rui Men, Dayiheng Liu, Chang Zhou, Jingren Zhou, and Junyang Lin.
\newblock Qwen2-vl: Enhancing vision-language model's perception of the world at any resolution.
\newblock \emph{arXiv preprint arXiv:2409.12191}, 2024{\natexlab{j}}.

\bibitem[Wang et~al.(2025{\natexlab{g}})Wang, Yu, Gao, Zheng, Liu, Lu, Dang, Chen, Yang, Zhang, Liu, Yang, Zhao, Yue, Song, Yu, Huang, and Lin]{wang_beyond_2025}
Shenzhi Wang, Le~Yu, Chang Gao, Chujie Zheng, Shixuan Liu, Rui Lu, Kai Dang, Xionghui Chen, Jianxin Yang, Zhenru Zhang, Yuqiong Liu, An~Yang, Andrew Zhao, Yang Yue, Shiji Song, Bowen Yu, Gao Huang, and Junyang Lin.
\newblock Beyond the 80/20 {Rule}: {High}-{Entropy} {Minority} {Tokens} {Drive} {Effective} {Reinforcement} {Learning} for {LLM} {Reasoning}, June 2025{\natexlab{g}}.
\newblock URL \url{http://arxiv.org/abs/2506.01939}.
\newblock arXiv:2506.01939 [cs] Read\_Status: New Read\_Status\_Date: 2025-06-15T15:30:33.418Z.

\bibitem[Wang et~al.(2024{\natexlab{k}})Wang, Ding, Shen, Luo, Du, and Tao]{oop}
Shuai Wang, Liang Ding, Li~Shen, Yong Luo, Bo~Du, and Dacheng Tao.
\newblock Oop: Object-oriented programming evaluation benchmark for large language models, 2024{\natexlab{k}}.
\newblock URL \url{https://arxiv.org/abs/2401.06628}.

\bibitem[Wang et~al.(2023{\natexlab{f}})Wang, Liu, Zhang, et~al.]{wang2023survey}
Wei Wang, Jiang Liu, Kai Zhang, et~al.
\newblock A survey on code generation with llm-based agents.
\newblock \emph{arXiv preprint arXiv:2508.00083}, 2023{\natexlab{f}}.
\newblock URL \url{https://arxiv.org/abs/2508.00083}.

\bibitem[Wang et~al.(2023{\natexlab{g}})Wang, Wang, Joty, and Hoi]{wang2023rap}
Weishi Wang, Yue Wang, Shafiq Joty, and Steven~CH Hoi.
\newblock Rap-gen: Retrieval-augmented patch generation with codet5 for automatic program repair.
\newblock In \emph{Proceedings of the 31st ACM Joint European Software Engineering Conference and Symposium on the Foundations of Software Engineering}, pages 146--158, 2023{\natexlab{g}}.

\bibitem[Wang et~al.(2025{\natexlab{h}})Wang, Gao, Gu, Pu, Cui, Wei, Liu, Jing, Ye, Shao, et~al.]{wang2025internvl3_5}
Weiyun Wang, Zhangwei Gao, Lixin Gu, Hengjun Pu, Long Cui, Xingguang Wei, Zhaoyang Liu, Linglin Jing, Shenglong Ye, Jie Shao, et~al.
\newblock Internvl3.5: Advancing open-source multimodal models in versatility, reasoning, and efficiency.
\newblock \emph{arXiv preprint arXiv:2508.18265}, 2025{\natexlab{h}}.

\bibitem[Wang et~al.(2025{\natexlab{i}})Wang, Yang, Wang, Huang, Chu, Song, Zhang, Chen, and Ma]{wang_testeval_2025}
Wenhan Wang, Chenyuan Yang, Zhijie Wang, Yuheng Huang, Zhaoyang Chu, Da~Song, Lingming Zhang, An~Ran Chen, and Lei Ma.
\newblock {TESTEVAL}: {Benchmarking} {Large} {Language} {Models} for {Test} {Case} {Generation}, February 2025{\natexlab{i}}.
\newblock URL \url{http://arxiv.org/abs/2406.04531}.
\newblock arXiv:2406.04531 [cs].

\bibitem[Wang et~al.(2020{\natexlab{b}})Wang, Zhang, Sui, Wan, Zhao, Wu, Yu, and Xu]{wang2020reinforcement}
Wenhua Wang, Yuqun Zhang, Yulei Sui, Yao Wan, Zhou Zhao, Jian Wu, Philip~S Yu, and Guandong Xu.
\newblock Reinforcement-learning-guided source code summarization using hierarchical attention.
\newblock \emph{IEEE Transactions on software Engineering}, 48\penalty0 (1):\penalty0 102--119, 2020{\natexlab{b}}.

\bibitem[Wang et~al.(2024{\natexlab{l}})Wang, Gao, Meng, Peng, Hu, Lin, and Gao]{wang2024aegis}
Xinchen Wang, Pengfei Gao, Xiangxin Meng, Chao Peng, Ruida Hu, Yun Lin, and Cuiyun Gao.
\newblock Aegis: An agent-based framework for general bug reproduction from issue descriptions.
\newblock \emph{arXiv preprint arXiv:2411.18015}, 2024{\natexlab{l}}.

\bibitem[Wang et~al.(2025{\natexlab{j}})Wang, Gao, Peng, Hu, and Gao]{wang2025codevisionaryagentbasedframeworkevaluating}
Xinchen Wang, Pengfei Gao, Chao Peng, Ruida Hu, and Cuiyun Gao.
\newblock Codevisionary: An agent-based framework for evaluating large language models in code generation, 2025{\natexlab{j}}.
\newblock URL \url{https://arxiv.org/abs/2504.13472}.

\bibitem[Wang et~al.(2024{\natexlab{m}})Wang, Chen, Yuan, Zhang, Li, Peng, and Ji]{wang2024executable}
Xingyao Wang, Yangyi Chen, Lifan Yuan, Yizhe Zhang, Yunzhu Li, Hao Peng, and Heng Ji.
\newblock Executable code actions elicit better llm agents.
\newblock In \emph{Forty-first International Conference on Machine Learning}, 2024{\natexlab{m}}.

\bibitem[Wang et~al.(2024{\natexlab{n}})Wang, Li, Song, Xu, Tang, Zhuge, Pan, Song, Li, Singh, et~al.]{wang2024openhands}
Xingyao Wang, Boxuan Li, Yufan Song, Frank~F Xu, Xiangru Tang, Mingchen Zhuge, Jiayi Pan, Yueqi Song, Bowen Li, Jaskirat Singh, et~al.
\newblock Openhands: An open platform for ai software developers as generalist agents.
\newblock \emph{arXiv preprint arXiv:2407.16741}, 2024{\natexlab{n}}.

\bibitem[Wang et~al.(2022{\natexlab{c}})Wang, Wei, Schuurmans, Le, Chi, Narang, Chowdhery, and Zhou]{wang2022self}
Xuezhi Wang, Jason Wei, Dale Schuurmans, Quoc Le, Ed~Chi, Sharan Narang, Aakanksha Chowdhery, and Denny Zhou.
\newblock Self-consistency improves chain of thought reasoning in language models.
\newblock \emph{arXiv preprint arXiv:2203.11171}, 2022{\natexlab{c}}.

\bibitem[Wang et~al.(2024{\natexlab{o}})Wang, Wang, Guo, Chen, Zhang, Ma, and Zheng]{wang2024rlcoder}
Yanlin Wang, Yanli Wang, Daya Guo, Jiachi Chen, Ruikai Zhang, Yuchi Ma, and Zibin Zheng.
\newblock Rlcoder: Reinforcement learning for repository-level code completion.
\newblock In \emph{2025 IEEE/ACM 47th International Conference on Software Engineering (ICSE)}, pages 165--177. IEEE Computer Society, 2024{\natexlab{o}}.

\bibitem[Wang et~al.(2024{\natexlab{p}})Wang, He, Dong, Wang, Zeng, Diao, Xu, Wang, Zhang, and Cai]{dolphcoder}
Yejie Wang, Keqing He, Guanting Dong, Pei Wang, Weihao Zeng, Muxi Diao, Weiran Xu, Jingang Wang, Mengdi Zhang, and Xunliang Cai.
\newblock {D}olph{C}oder: Echo-locating code large language models with diverse and multi-objective instruction tuning.
\newblock In Lun-Wei Ku, Andre Martins, and Vivek Srikumar, editors, \emph{Proceedings of the 62nd Annual Meeting of the Association for Computational Linguistics (Volume 1: Long Papers)}, pages 4706--4721, Bangkok, Thailand, August 2024{\natexlab{p}}. Association for Computational Linguistics.
\newblock \doi{10.18653/v1/2024.acl-long.259}.
\newblock URL \url{https://aclanthology.org/2024.acl-long.259}.

\bibitem[Wang et~al.(2024{\natexlab{q}})Wang, He, Fu, GongQue, Xu, Chen, Wang, Fu, Dong, Diao, Wang, Zhang, Cai, and Xu]{wang-etal-2024-code}
Yejie Wang, Keqing He, Dayuan Fu, Zhuoma GongQue, Heyang Xu, Yanxu Chen, Zhexu Wang, Yujia Fu, Guanting Dong, Muxi Diao, Jingang Wang, Mengdi Zhang, Xunliang Cai, and Weiran Xu.
\newblock How do your code {LLM}s perform? empowering code instruction tuning with really good data.
\newblock In Yaser Al-Onaizan, Mohit Bansal, and Yun-Nung Chen, editors, \emph{Proceedings of the 2024 Conference on Empirical Methods in Natural Language Processing}, pages 14027--14043, Miami, Florida, USA, November 2024{\natexlab{q}}. Association for Computational Linguistics.
\newblock \doi{10.18653/v1/2024.emnlp-main.777}.
\newblock URL \url{https://aclanthology.org/2024.emnlp-main.777}.

\bibitem[Wang et~al.(2025{\natexlab{k}})Wang, Yang, Tian, Shen, and Wang]{wang2025cure}
Yinjie Wang, Ling Yang, Ye~Tian, Ke~Shen, and Mengdi Wang.
\newblock Co-evolving llm coder and unit tester via reinforcement learning.
\newblock \emph{arXiv preprint arXiv:2506.03136}, 2025{\natexlab{k}}.

\bibitem[Wang et~al.(2024{\natexlab{r}})Wang, Luo, Wei, Liu, Zhu, Zhang, Yang, Xu, and Che]{wang-etal-2024-make}
Yixuan Wang, Xianzhen Luo, Fuxuan Wei, Yijun Liu, Qingfu Zhu, Xuanyu Zhang, Qing Yang, Dongliang Xu, and Wanxiang Che.
\newblock Make some noise: Unlocking language model parallel inference capability through noisy training.
\newblock In \emph{Proceedings of the 2024 Conference on Empirical Methods in Natural Language Processing}, Miami, Florida, USA, November 2024{\natexlab{r}}. Association for Computational Linguistics.
\newblock URL \url{https://aclanthology.org/2024.emnlp-main.718/}.

\bibitem[Wang et~al.(2021{\natexlab{b}})Wang, Wang, Joty, and Hoi]{wang-etal-2021-codet5}
Yue Wang, Weishi Wang, Shafiq Joty, and Steven~C.H. Hoi.
\newblock {C}ode{T}5: Identifier-aware unified pre-trained encoder-decoder models for code understanding and generation.
\newblock In Marie-Francine Moens, Xuanjing Huang, Lucia Specia, and Scott Wen-tau Yih, editors, \emph{Proceedings of the 2021 Conference on Empirical Methods in Natural Language Processing}, pages 8696--8708, Online and Punta Cana, Dominican Republic, November 2021{\natexlab{b}}. Association for Computational Linguistics.
\newblock \doi{10.18653/v1/2021.emnlp-main.685}.
\newblock URL \url{https://aclanthology.org/2021.emnlp-main.685/}.

\bibitem[Wang et~al.(2023{\natexlab{h}})Wang, Le, Gotmare, Bui, Li, and Hoi]{wang-etal-2023-codet5}
Yue Wang, Hung Le, Akhilesh Gotmare, Nghi Bui, Junnan Li, and Steven Hoi.
\newblock {C}ode{T}5+: Open code large language models for code understanding and generation.
\newblock In Houda Bouamor, Juan Pino, and Kalika Bali, editors, \emph{Proceedings of the 2023 Conference on Empirical Methods in Natural Language Processing}, pages 1069--1088, Singapore, December 2023{\natexlab{h}}. Association for Computational Linguistics.
\newblock \doi{10.18653/v1/2023.emnlp-main.68}.
\newblock URL \url{https://aclanthology.org/2023.emnlp-main.68/}.

\bibitem[Wang et~al.(2023{\natexlab{i}})Wang, Le, Gotmare, et~al.]{wang2023codet5plus}
Yue Wang, Hung Le, Akhilesh~Deepak Gotmare, et~al.
\newblock {CodeT5+}: Open code large language models for code understanding and generation.
\newblock \emph{arXiv preprint arXiv:2305.07922}, 2023{\natexlab{i}}.

\bibitem[Wang et~al.(2025{\natexlab{l}})Wang, Ji, Yang, Li, Hu, Li, and Sartoretti]{wang2025mctsjudgetesttimescalingllmasajudge}
Yutong Wang, Pengliang Ji, Chaoqun Yang, Kaixin Li, Ming Hu, Jiaoyang Li, and Guillaume Sartoretti.
\newblock Mcts-judge: Test-time scaling in llm-as-a-judge for code correctness evaluation, 2025{\natexlab{l}}.
\newblock URL \url{https://arxiv.org/abs/2502.12468}.

\bibitem[Wang et~al.(2024{\natexlab{s}})Wang, Liu, Li, and Jin]{wang2024hitshighcoveragellmbasedunit}
Zejun Wang, Kaibo Liu, Ge~Li, and Zhi Jin.
\newblock Hits: High-coverage llm-based unit test generation via method slicing, 2024{\natexlab{s}}.
\newblock URL \url{https://arxiv.org/abs/2408.11324}.

\bibitem[Wang et~al.(2025{\natexlab{m}})Wang, Liu, Wang, He, Gao, Diao, Chen, Fu, Sung, Yang, Liu, and Xu]{ojbench}
Zhexu Wang, Yiping Liu, Yejie Wang, Wenyang He, Bofei Gao, Muxi Diao, Yanxu Chen, Kelin Fu, Flood Sung, Zhilin Yang, Tianyu Liu, and Weiran Xu.
\newblock Ojbench: A competition level code benchmark for large language models, 2025{\natexlab{m}}.
\newblock URL \url{https://arxiv.org/abs/2506.16395}.

\bibitem[Wang et~al.(2023{\natexlab{j}})Wang, Cuenca, Zhou, Xu, and Neubig]{mconala}
Zhiruo Wang, Grace Cuenca, Shuyan Zhou, Frank~F. Xu, and Graham Neubig.
\newblock Mconala: {A} benchmark for code generation from multiple natural languages.
\newblock In Andreas Vlachos and Isabelle Augenstein, editors, \emph{Findings of the Association for Computational Linguistics: {EACL} 2023, Dubrovnik, Croatia, May 2-6, 2023}, pages 265--273. Association for Computational Linguistics, 2023{\natexlab{j}}.
\newblock \doi{10.18653/V1/2023.FINDINGS-EACL.20}.
\newblock URL \url{https://doi.org/10.18653/v1/2023.findings-eacl.20}.

\bibitem[Wang et~al.(2023{\natexlab{k}})Wang, Zhou, Fried, and Neubig]{odex}
Zhiruo Wang, Shuyan Zhou, Daniel Fried, and Graham Neubig.
\newblock Execution-based evaluation for open-domain code generation.
\newblock In Houda Bouamor, Juan Pino, and Kalika Bali, editors, \emph{Findings of the Association for Computational Linguistics: {EMNLP} 2023, Singapore, December 6-10, 2023}, pages 1271--1290. Association for Computational Linguistics, 2023{\natexlab{k}}.
\newblock \doi{10.18653/V1/2023.FINDINGS-EMNLP.89}.
\newblock URL \url{https://doi.org/10.18653/v1/2023.findings-emnlp.89}.

\bibitem[Wang et~al.(2025{\natexlab{n}})Wang, Liu, Sun, Li, and Shen]{wang2025codecontests+}
Zihan Wang, Siyao Liu, Yang Sun, Hongyan Li, and Kai Shen.
\newblock Codecontests+: High-quality test case generation for competitive programming.
\newblock \emph{arXiv preprint arXiv:2506.05817}, 2025{\natexlab{n}}.

\bibitem[Wang et~al.(2024{\natexlab{t}})Wang, Asai, Yu, Xu, Xie, Neubig, and Fried]{CodeRAG-Bench}
Zora~Zhiruo Wang, Akari Asai, Xinyan~Velocity Yu, Frank~F. Xu, Yiqing Xie, Graham Neubig, and Daniel Fried.
\newblock Coderag-bench: Can retrieval augment code generation?
\newblock In \emph{North American Chapter of the Association for Computational Linguistics}, 2024{\natexlab{t}}.
\newblock URL \url{https://api.semanticscholar.org/CorpusID:270620678}.

\bibitem[Ward et~al.(2024)Ward, MacDermott, Belardinelli, Toni, and Everitt]{ward2024reasons}
Francis~Rhys Ward, Matt MacDermott, Francesco Belardinelli, Francesca Toni, and Tom Everitt.
\newblock The reasons that agents act: Intention and instrumental goals.
\newblock \emph{arXiv preprint arXiv:2402.07221}, 2024.

\bibitem[Wei et~al.(2023{\natexlab{a}})Wei, Haghtalab, and Steinhardt]{wei2023jailbroken}
Alexander Wei, Nika Haghtalab, and Jacob Steinhardt.
\newblock Jailbroken: How does llm safety training fail?
\newblock \emph{Advances in Neural Information Processing Systems}, 36:\penalty0 80079--80110, 2023{\natexlab{a}}.

\bibitem[Wei et~al.(2025{\natexlab{a}})Wei, Cao, Li, Chen, Zhang, Wang, Liu, Teixeira, Yang, Wang, et~al.]{wei2025equibench}
Anjiang Wei, Jiannan Cao, Ran Li, Hongyu Chen, Yuhui Zhang, Ziheng Wang, Yuan Liu, Thiago~SFX Teixeira, Diyi Yang, Ke~Wang, et~al.
\newblock Equibench: Benchmarking large language models' understanding of program semantics via equivalence checking.
\newblock \emph{arXiv preprint arXiv:2502.12466}, 2025{\natexlab{a}}.

\bibitem[Wei(2024)]{Progressive}
Bingyang Wei.
\newblock Requirements are all you need: From requirements to code with llms.
\newblock In \emph{2024 IEEE 32nd International Requirements Engineering Conference (RE)}, pages 416--422. IEEE, 2024.

\bibitem[Wei et~al.(2019)Wei, Li, Xia, Fu, and Jin]{wei2019code}
Bolin Wei, Ge~Li, Xin Xia, Zhiyi Fu, and Zhi Jin.
\newblock Code generation as a dual task of code summarization.
\newblock \emph{Advances in neural information processing systems}, 32, 2019.

\bibitem[Wei et~al.(2022{\natexlab{a}})Wei, Tay, Bommasani, Raffel, Zoph, Borgeaud, Yogatama, Bosma, Zhou, Metzler, Chi, Hashimoto, Vinyals, Liang, Dean, and Fedus]{wei2022emergent}
Jason Wei, Yi~Tay, Rishi Bommasani, Colin Raffel, Barret Zoph, Sebastian Borgeaud, Dani Yogatama, Maarten Bosma, Denny Zhou, Donald Metzler, Ed~H. Chi, Tatsunori Hashimoto, Oriol Vinyals, Percy Liang, Jeff Dean, and William Fedus.
\newblock Emergent abilities of large language models.
\newblock \emph{Transactions on Machine Learning Research}, 2022{\natexlab{a}}.
\newblock ISSN 2835-8856.
\newblock URL \url{https://openreview.net/forum?id=yzkSU5zdwD}.

\bibitem[Wei et~al.(2022{\natexlab{b}})Wei, Wang, Schuurmans, Bosma, Ichter, Xia, Chi, Le, and Zhou]{Wei}
Jason Wei, Xuezhi Wang, Dale Schuurmans, Maarten Bosma, Brian Ichter, Fei Xia, Ed~H. Chi, Quoc~V. Le, and Denny Zhou.
\newblock Chain-of-thought prompting elicits reasoning in large language models.
\newblock In \emph{NeurIPS 2022, New Orleans, LA, USA, November 28 - December 9, 2022}, 2022{\natexlab{b}}.

\bibitem[Wei et~al.(2025{\natexlab{b}})Wei, Sun, Papay, McKinney, Han, Fulford, Chung, Passos, Fedus, and Glaese]{wei2025browsecomp}
Jason Wei, Zhiqing Sun, Spencer Papay, Scott McKinney, Jeffrey Han, Isa Fulford, Hyung~Won Chung, Alex~Tachard Passos, William Fedus, and Amelia Glaese.
\newblock Browsecomp: A simple yet challenging benchmark for browsing agents.
\newblock \emph{arXiv preprint arXiv:2504.12516}, 2025{\natexlab{b}}.

\bibitem[Wei et~al.(2024{\natexlab{a}})Wei, Li, Chen, et~al.]{autospec2024}
Wenjie Wei, Yuqi Li, Tianyu Chen, et~al.
\newblock {AutoSpec}: Enchanting program specification synthesis by large language models using static analysis and program verification.
\newblock In \emph{CAV}, 2024{\natexlab{a}}.

\bibitem[Wei et~al.(2023{\natexlab{b}})Wei, Xia, and Zhang]{wei2023copiloting}
Yuxiang Wei, Chunqiu~Steven Xia, and Lingming Zhang.
\newblock Copiloting the copilots: Fusing large language models with completion engines for automated program repair.
\newblock In \emph{Proceedings of the 31st ACM Joint European Software Engineering Conference and Symposium on the Foundations of Software Engineering}, pages 172--184, 2023{\natexlab{b}}.

\bibitem[Wei et~al.(2024{\natexlab{b}})Wei, Wang, Liu, Ding, and Zhang]{weiMagicoderEmpoweringCode2024}
Yuxiang Wei, Zhe Wang, Jiawei Liu, Yifeng Ding, and Lingming Zhang.
\newblock Magicoder: {{Empowering Code Generation}} with {{OSS-Instruct}}, 2024{\natexlab{b}}.

\bibitem[Wei et~al.(2025{\natexlab{c}})Wei, Duchenne, Copet, Carbonneaux, Zhang, Fried, Synnaeve, Singh, and Wang]{wei2025swerl}
Yuxiang Wei, Olivier Duchenne, Jade Copet, Quentin Carbonneaux, Lingming Zhang, Daniel Fried, Gabriel Synnaeve, Rishabh Singh, and Sida~I Wang.
\newblock Swe-rl: Advancing llm reasoning via reinforcement learning on open software evolution.
\newblock \emph{arXiv preprint arXiv:2502.18449}, 2025{\natexlab{c}}.

\bibitem[Weissman(2023)]{weissman2023microarchitectural}
Dean Weissman.
\newblock A microarchitectural threat analysis of us federal cloud environments.
\newblock \emph{arXiv preprint arXiv:2311.13327}, 2023.

\bibitem[Wen et~al.(2024)Wen, Guan, Wang, Wu, and Huang]{wen2024unlocking}
Jiaxin Wen, Jian Guan, Hongning Wang, Wei Wu, and Minlie Huang.
\newblock Unlocking reasoning potential in large langauge models by scaling code-form planning.
\newblock \emph{arXiv preprint arXiv:2409.12452}, 2024.

\bibitem[Wen et~al.(2022)Wen, Guo, Fu, Li, Xu, Tang, Zhao, Hu, Du, Li, et~al.]{wen2022babeltower}
Yuanbo Wen, Qi~Guo, Qiang Fu, Xiaqing Li, Jianxing Xu, Yanlin Tang, Yongwei Zhao, Xing Hu, Zidong Du, Ling Li, et~al.
\newblock Babeltower: Learning to auto-parallelized program translation.
\newblock In \emph{International Conference on Machine Learning}, pages 23685--23700. PMLR, 2022.

\bibitem[Wiggers(2024)]{techcrunch2024amazonq}
Kyle Wiggers.
\newblock Amazon codewhisperer is now called q developer.
\newblock \url{https://techcrunch.com/2024/04/30/amazon-codewhisperer-q-developer/}, 2024.

\bibitem[{Wikipedia}(2024)]{wikipedia2024copilot}
{Wikipedia}.
\newblock Github copilot.
\newblock \url{https://en.wikipedia.org/wiki/GitHub_Copilot}, 2024.

\bibitem[Williams et~al.(2013)Williams, Raux, Ramachandran, and Black]{williams2013dstc}
Jason Williams, Antoine Raux, Deepak Ramachandran, and Alan Black.
\newblock The dialog state tracking challenge.
\newblock In Maxine Eskenazi, Michael Strube, Barbara Di~Eugenio, and Jason~D. Williams, editors, \emph{Proceedings of the {SIGDIAL} 2013 Conference}, pages 404--413, Metz, France, August 2013. Association for Computational Linguistics.
\newblock URL \url{https://aclanthology.org/W13-4065/}.

\bibitem[Wong et~al.(2013)Wong, Yang, and Tan]{AutoComment}
Edmund Wong, Jinqiu Yang, and Lin Tan.
\newblock Autocomment: Mining question and answer sites for automatic comment generation.
\newblock In \emph{2013 28th IEEE/ACM International Conference on Automated Software Engineering (ASE)}, pages 562--567, 2013.
\newblock \doi{10.1109/ASE.2013.6693113}.

\bibitem[Wong et~al.(2015)Wong, Liu, and Tan]{CloCom}
Edmund Wong, Taiyue Liu, and Lin Tan.
\newblock Clocom: Mining existing source code for automatic comment generation.
\newblock In \emph{2015 IEEE 22nd International Conference on Software Analysis, Evolution, and Reengineering (SANER)}, pages 380--389, 2015.
\newblock \doi{10.1109/SANER.2015.7081848}.

\bibitem[Wong et~al.(2025)Wong, Wang, Zhao, Chen, Gao, Zhang, Zhou, Wang, Xiang, Zhang, et~al.]{wong2025widesearch}
Ryan Wong, Jiawei Wang, Junjie Zhao, Li~Chen, Yan Gao, Long Zhang, Xuan Zhou, Zuo Wang, Kai Xiang, Ge~Zhang, et~al.
\newblock Widesearch: Benchmarking agentic broad info-seeking.
\newblock \emph{arXiv preprint arXiv:2508.07999}, 2025.

\bibitem[Wong et~al.(2023)Wong, Wang, Li, Liu, Wang, Tang, Nie, and Wu]{DBLP:journals/corr/abs-2310-06530}
Wai~Kin Wong, Huaijin Wang, Zongjie Li, Zhibo Liu, Shuai Wang, Qiyi Tang, Sen Nie, and Shi Wu.
\newblock Refining decompiled {C} code with large language models.
\newblock \emph{CoRR}, abs/2310.06530, 2023.

\bibitem[Wu et~al.(2023{\natexlab{a}})Wu, Yin, Qi, Wang, Tang, and Duan]{wu2023visualchatgpt}
Chenfei Wu, Shengming Yin, Weizhen Qi, Xiaodong Wang, Zecheng Tang, and Nan Duan.
\newblock Visual chatgpt: Talking, drawing and editing with visual foundation models, 2023{\natexlab{a}}.
\newblock URL \url{https://arxiv.org/abs/2303.04671}.

\bibitem[Wu et~al.(2024{\natexlab{a}})Wu, Ge, Guo, Wang, Liang, Lu, Shan, and Luo]{wu2024plot2code}
Chengyue Wu, Yixiao Ge, Qiushan Guo, Jiahao Wang, Zhixuan Liang, Zeyu Lu, Ying Shan, and Ping Luo.
\newblock Plot2code: A comprehensive benchmark for evaluating multi-modal large language models in code generation from scientific plots.
\newblock \emph{arXiv preprint arXiv:2405.07990}, 2024{\natexlab{a}}.

\bibitem[Wu et~al.(2025{\natexlab{a}})Wu, Gao, Li, Wen, and Liao]{wu2025mllm}
Fan Wu, Cuiyun Gao, Shuqing Li, Xin-Cheng Wen, and Qing Liao.
\newblock Mllm-based ui2code automation guided by ui layout information.
\newblock \emph{Proceedings of the ACM on Software Engineering}, 2\penalty0 (ISSTA):\penalty0 1123--1145, 2025{\natexlab{a}}.

\bibitem[Wu et~al.(2025{\natexlab{b}})Wu, Xuan, Lu, Harchaoui, and Choi]{wu2025invisible}
Fang Wu, Weihao Xuan, Ximing Lu, Zaid Harchaoui, and Yejin Choi.
\newblock The invisible leash: Why rlvr may not escape its origin.
\newblock \emph{arXiv preprint arXiv:2507.14843}, 2025{\natexlab{b}}.

\bibitem[Wu et~al.(2023{\natexlab{b}})Wu, Liu, and Xiao]{wu2023deceptprompt}
Fangzhou Wu, Xiaogeng Liu, and Chaowei Xiao.
\newblock Deceptprompt: Exploiting llm-driven code generation via adversarial natural language instructions.
\newblock \emph{arXiv preprint arXiv:2312.04730}, 2023{\natexlab{b}}.

\bibitem[Wu et~al.(2025{\natexlab{c}})Wu, Yao, Yu, and Zhang]{wu2025recode}
Haoze Wu, Yunzhi Yao, Wenhao Yu, and Ningyu Zhang.
\newblock Recode: Updating code api knowledge with reinforcement learning.
\newblock \emph{arXiv preprint arXiv:2506.20495}, 2025{\natexlab{c}}.

\bibitem[Wu et~al.(2023{\natexlab{c}})Wu, Chen, Liu, Chen, Wang, and Lin]{Xu}
Hefeng Wu, Yandong Chen, Lingbo Liu, Tianshui Chen, Keze Wang, and Liang Lin.
\newblock Sqlnet: Scale-modulated query and localization network for few-shot class-agnostic counting.
\newblock \emph{CoRR}, abs/2311.10011, 2023{\natexlab{c}}.

\bibitem[Wu et~al.(2024{\natexlab{b}})Wu, Schoop, Leung, Barik, Bigham, and Nichols]{wu2024uicoder}
Jason Wu, Eldon Schoop, Alan Leung, Titus Barik, Jeffrey~P Bigham, and Jeffrey Nichols.
\newblock Uicoder: Finetuning large language models to generate user interface code through automated feedback.
\newblock \emph{arXiv preprint arXiv:2406.07739}, 2024{\natexlab{b}}.

\bibitem[Wu et~al.(2025{\natexlab{d}})Wu, Yin, Jiang, Wang, Xi, Fang, Zhang, He, Zhou, Xie, and Huang]{wu2025webwalker}
Jialong Wu, Wenbiao Yin, Yong Jiang, Zhenglin Wang, Zekun Xi, Runnan Fang, Linhai Zhang, Yulan He, Deyu Zhou, Pengjun Xie, and Fei Huang.
\newblock Webwalker: Benchmarking llms in web traversal, 2025{\natexlab{d}}.
\newblock URL \url{https://arxiv.org/abs/2501.07572}.

\bibitem[Wu et~al.(2024{\natexlab{c}})Wu, Chen, Jiang, Liu, Li, Li, and Yuan]{wu2024guardagent}
Qishen Wu, Xinyin Chen, Can Jiang, Jing Liu, Yang Li, Yilong Li, and Nicholas~Jing Yuan.
\newblock Guardagent: Safeguard llm agents by a guard agent via knowledge-enabled reasoning.
\newblock \emph{arXiv preprint arXiv:2405.18783}, 2024{\natexlab{c}}.

\bibitem[Wu et~al.(2025{\natexlab{e}})Wu, Ma, Luo, Li, Shi, Chang, Lin, Luo, Pei, Du, Zhao, and Gong]{Wu2025Automated}
Shican Wu, Xiao Ma, Dehui Luo, Lulu Li, Xiangcheng Shi, Xin Chang, Xiaoyun Lin, Ran Luo, Chunlei Pei, Changying Du, Zhi-Jian Zhao, and Jinlong Gong.
\newblock Automated literature research and review-generation method based on large language models.
\newblock \emph{National Science Review}, 12\penalty0 (6), April 2025{\natexlab{e}}.
\newblock ISSN 2053-714X.
\newblock \doi{10.1093/nsr/nwaf169}.
\newblock URL \url{http://dx.doi.org/10.1093/nsr/nwaf169}.

\bibitem[Wu et~al.(2023{\natexlab{d}})Wu, Fan, Liu, Gong, Shen, Jiao, Zheng, Li, Wei, Guo, Duan, and Chen]{wu2023ardiffusion}
Tong Wu, Zhihao Fan, Xiao Liu, Yeyun Gong, Yelong Shen, Jian Jiao, Hai-Tao Zheng, Juntao Li, Zhongyu Wei, Jian Guo, Nan Duan, and Weizhu Chen.
\newblock Ar-diffusion: Auto-regressive diffusion model for text generation, 2023{\natexlab{d}}.
\newblock URL \url{https://arxiv.org/abs/2305.09515}.

\bibitem[Wu et~al.(2024{\natexlab{d}})Wu, Yan, Shen, Wang, Tang, and Luo]{wu2024chartinsight}
Yifan Wu, Lutao Yan, Leixian Shen, Yunhai Wang, Nan Tang, and Yuyu Luo.
\newblock Chartinsights: Evaluating multimodal large language models for low-level chart question answering, 2024{\natexlab{d}}.
\newblock URL \url{https://arxiv.org/abs/2405.07001}.

\bibitem[Wu et~al.()Wu, Li, Zhang, Papadakis, Harman, and Liu]{wu2308large_fault_localization}
Yonghao Wu, Zheng Li, Jie~M Zhang, Mike Papadakis, Mark Harman, and Yong Liu.
\newblock Large language models in fault localisation (2023).
\newblock \emph{URL https://arxiv. org/abs/2308.15276}.

\bibitem[Wu et~al.(2016)Wu, Schuster, Chen, Le, Norouzi, Macherey, Krikun, Cao, Gao, Macherey, Klingner, Shah, Johnson, Liu, Łukasz Kaiser, Gouws, Kato, Kudo, Kazawa, Stevens, Kurian, Patil, Wang, Young, Smith, Riesa, Rudnick, Vinyals, Corrado, Hughes, and Dean]{wu2016googlesneuralmachinetranslation}
Yonghui Wu, Mike Schuster, Zhifeng Chen, Quoc~V. Le, Mohammad Norouzi, Wolfgang Macherey, Maxim Krikun, Yuan Cao, Qin Gao, Klaus Macherey, Jeff Klingner, Apurva Shah, Melvin Johnson, Xiaobing Liu, Łukasz Kaiser, Stephan Gouws, Yoshikiyo Kato, Taku Kudo, Hideto Kazawa, Keith Stevens, George Kurian, Nishant Patil, Wei Wang, Cliff Young, Jason Smith, Jason Riesa, Alex Rudnick, Oriol Vinyals, Greg Corrado, Macduff Hughes, and Jeffrey Dean.
\newblock Google's neural machine translation system: Bridging the gap between human and machine translation, 2016.
\newblock URL \url{https://arxiv.org/abs/1609.08144}.

\bibitem[Wu et~al.(2024{\natexlab{e}})Wu, Huang, Shi, Wang, Gao, Liu, Nan, Yuan, Zhang, Zhang, Du, Guo, Pu, Yin, Hu, and Chen]{wu2024inversecoder}
Yutong Wu, Di~Huang, Wenxuan Shi, Wei Wang, Lingzhe Gao, Shihao Liu, Ziyuan Nan, Kaizhao Yuan, Rui Zhang, Xishan Zhang, Zidong Du, Qi~Guo, Yewen Pu, Dawei Yin, Xing Hu, and Yunji Chen.
\newblock Inversecoder: Unleashing the power of instruction-tuned code llms with inverse-instruct, 2024{\natexlab{e}}.
\newblock URL \url{https://arxiv.org/abs/2407.05700}.

\bibitem[Wu et~al.(2019)Wu, Johnson, Yang, Bastani, Song, Peng, and Xie]{wu2019reinam}
Zhengkai Wu, Evan Johnson, Wei Yang, Osbert Bastani, Dawn Song, Jian Peng, and Tao Xie.
\newblock Reinam: reinforcement learning for input-grammar inference.
\newblock In \emph{Proceedings of the 2019 27th acm joint meeting on european software engineering conference and symposium on the foundations of software engineering}, pages 488--498, 2019.

\bibitem[Wu et~al.(2025{\natexlab{f}})Wu, Li, Li, Zhang, He, Yang, Zhao, Fang, Li, Li, and Song]{MR-SQL}
Zhenhe Wu, Zhongqiu Li, Mengxiang Li, Jie Zhang, Zhongjiang He, Jian Yang, Yu~Zhao, Ruiyu Fang, Yongxiang Li, Zhoujun Li, and Shuangyong Song.
\newblock {MR-SQL:} multi-level retrieval enhances inference for llm in text-to-sql.
\newblock \emph{DASFAA}, 2025{\natexlab{f}}.

\bibitem[Wu et~al.(2025{\natexlab{g}})Wu, Li, Zhang, He, Yang, Zhao, Fang, Wang, Xie, Song, and Li]{UCS-SQL}
Zhenhe Wu, Zhongqiu Li, Jie Zhang, Zhongjiang He, Jian Yang, Yu~Zhao, Ruiyu Fang, Bing Wang, Hongyan Xie, Shuangyong Song, and Zhoujun Li.
\newblock {UCS-SQL:} uniting content and structure for enhanced semantic bridging in text-to-sql.
\newblock In Wanxiang Che, Joyce Nabende, Ekaterina Shutova, and Mohammad~Taher Pilehvar, editors, \emph{Findings of the Association for Computational Linguistics, {ACL} 2025, Vienna, Austria, July 27 - August 1, 2025}, pages 8156--8168. Association for Computational Linguistics, 2025{\natexlab{g}}.

\bibitem[Wu et~al.(2024{\natexlab{f}})Wu, Han, Ding, Weng, Liu, Yao, Yu, and Kong]{wu2024copilot}
Zhiyong Wu, Chengcheng Han, Zichen Ding, Zhenmin Weng, Zhoumianze Liu, Shunyu Yao, Tao Yu, and Lingpeng Kong.
\newblock Os-copilot: Towards generalist computer agents with self-improvement.
\newblock \emph{arXiv preprint arXiv:2402.07456}, 2024{\natexlab{f}}.

\bibitem[Wu et~al.(2024{\natexlab{g}})Wu, Chen, Pan, Liu, Liu, Dai, Gao, Ma, Wu, Wang, Xie, Wu, Hu, Wang, Sun, Li, Piao, Guan, Liu, Xie, You, Dong, Yu, Zhang, Zhao, Wang, and Ruan]{wu2024deepseekvl2}
Zhiyu Wu, Xiaokang Chen, Zizheng Pan, Xingchao Liu, Wen Liu, Damai Dai, Huazuo Gao, Yiyang Ma, Chengyue Wu, Bingxuan Wang, Zhenda Xie, Yu~Wu, Kai Hu, Jiawei Wang, Yaofeng Sun, Yukun Li, Yishi Piao, Kang Guan, Aixin Liu, Xin Xie, Yuxiang You, Kai Dong, Xingkai Yu, Haowei Zhang, Liang Zhao, Yisong Wang, and Chong Ruan.
\newblock Deepseek-vl2: Mixture-of-experts vision-language models for advanced multimodal understanding, 2024{\natexlab{g}}.
\newblock URL \url{https://arxiv.org/abs/2412.10302}.

\bibitem[Wu et~al.(2023{\natexlab{e}})Wu, Jiang, Zhou, and Wang]{wu2023large}
Zian Wu, Yixuan Jiang, Yiyuan Zhou, and He~Wang.
\newblock Large language models for code: Security hardening and adversarial testing.
\newblock \emph{arXiv preprint arXiv:2311.00033}, 2023{\natexlab{e}}.

\bibitem[xAI(2023)]{xai2023announcinggrok}
xAI.
\newblock Announcing grok.
\newblock \url{https://x.ai/news/announcing-grok}, 2023.

\bibitem[xAI(2024{\natexlab{a}})]{xai2024grok15}
xAI.
\newblock Announcing grok-1.5: Advancing long-context understanding and math and coding.
\newblock \url{https://x.ai/news/announcing-grok-1-5}, 2024{\natexlab{a}}.

\bibitem[xAI(2024{\natexlab{b}})]{xai2024grok2}
xAI.
\newblock Launching grok 2 and a suite of {AIs}.
\newblock \url{https://x.ai/news/grok-2}, 2024{\natexlab{b}}.
\newblock Includes HumanEval pass@1 results for Grok-1.5, Grok-2-mini, and Grok-2.

\bibitem[xAI(2025{\natexlab{a}})]{xai2025apidocs}
xAI.
\newblock Api documentation: Models and endpoints.
\newblock \url{https://docs.x.ai/docs/models-and-endpoints}, 2025{\natexlab{a}}.
\newblock Lists \texttt{grok-code-fast-1} as a code-specialized model.

\bibitem[xAI(2025{\natexlab{b}})]{xai2025grok4}
xAI.
\newblock Grok 4.
\newblock \url{https://x.ai/news/grok-4}, 2025{\natexlab{b}}.

\bibitem[xAI(2025{\natexlab{c}})]{xai2025grok4fast}
xAI.
\newblock Grok 4 fast.
\newblock \url{https://x.ai/news/grok-4-fast}, 2025{\natexlab{c}}.

\bibitem[Xi et~al.(2023)Xi, Chen, Guo, He, Ding, Hong, Zhang, Wang, Jin, Zhou, Zheng, Fan, Wang, Xiong, Zhou, Wang, Jiang, Zou, Liu, Yin, Dou, Weng, Cheng, Zhang, Qin, Zheng, Qiu, Huang, and Gui]{agent-survey-2}
Zhiheng Xi, Wenxiang Chen, Xin Guo, Wei He, Yiwen Ding, Boyang Hong, Ming Zhang, Junzhe Wang, Senjie Jin, Enyu Zhou, Rui Zheng, Xiaoran Fan, Xiao Wang, Limao Xiong, Yuhao Zhou, Weiran Wang, Changhao Jiang, Yicheng Zou, Xiangyang Liu, Zhangyue Yin, Shihan Dou, Rongxiang Weng, Wensen Cheng, Qi~Zhang, Wenjuan Qin, Yongyan Zheng, Xipeng Qiu, Xuanjing Huang, and Tao Gui.
\newblock The rise and potential of large language model based agents: A survey.
\newblock \emph{arXiv preprint arXiv: 2309.07864}, 2023.

\bibitem[Xia et~al.(2023{\natexlab{a}})Xia, Bi, Xing, Lu, and Zhu]{xia2023empirical}
Boming Xia, Tingting Bi, Zhenchang Xing, Qinghua Lu, and Liming Zhu.
\newblock An empirical study on software bill of materials: Where we stand and the road ahead.
\newblock In \emph{2023 IEEE/ACM 45th International Conference on Software Engineering (ICSE)}, pages 2630--2642. IEEE, 2023{\natexlab{a}}.

\bibitem[Xia and Lingming(2024)]{xia2024repairllama}
Chunqiu~Silva Xia and Zhang Lingming.
\newblock Repairllama: Efficient representations for automated program repair.
\newblock \emph{arXiv preprint arXiv:2312.15698}, 2024.

\bibitem[Xia and Zhang(2022)]{xia2022less}
Chunqiu~Steven Xia and Lingming Zhang.
\newblock Less training, more repairing please: revisiting automated program repair via zero-shot learning.
\newblock In \emph{Proceedings of the 30th ACM Joint European Software Engineering Conference and Symposium on the Foundations of Software Engineering}, pages 959--971, 2022.

\bibitem[Xia and Zhang(2023{\natexlab{a}})]{xia2023conversational}
Chunqiu~Steven Xia and Lingming Zhang.
\newblock Conversational automated program repair.
\newblock \emph{arXiv preprint arXiv:2301.13246}, 2023{\natexlab{a}}.

\bibitem[Xia and Zhang(2023{\natexlab{b}})]{xia2023keep}
Chunqiu~Steven Xia and Lingming Zhang.
\newblock Keep the conversation going: Fixing 162 out of 337 bugs for 0.42 each using chatgpt.
\newblock \emph{arXiv preprint arXiv:2304.00385}, 2023{\natexlab{b}}.

\bibitem[Xia et~al.(2023{\natexlab{b}})Xia, Ding, and Zhang]{xia2023revisiting}
Chunqiu~Steven Xia, Yifeng Ding, and Lingming Zhang.
\newblock Revisiting the plastic surgery hypothesis via large language models.
\newblock \emph{arXiv preprint arXiv:2303.10494}, 2023{\natexlab{b}}.

\bibitem[Xia et~al.(2024{\natexlab{a}})Xia, Deng, Dunn, and Zhang]{xia2024agentless}
Chunqiu~Steven Xia, Yinlin Deng, Soren Dunn, and Lingming Zhang.
\newblock Agentless: Demystifying llm-based software engineering agents.
\newblock \emph{CoRR}, abs/2407.01489, 2024{\natexlab{a}}.
\newblock \doi{10.48550/ARXIV.2407.01489}.
\newblock URL \url{https://doi.org/10.48550/arXiv.2407.01489}.

\bibitem[Xia et~al.(2024{\natexlab{b}})Xia, Wei, and Zhang]{xia2024alpharepair}
Chunqiu~Steven Xia, Yifeng Wei, and Lingming Zhang.
\newblock {AlphaRepair}: Code generation from bugs.
\newblock \emph{ICSE}, 2024{\natexlab{b}}.

\bibitem[Xia et~al.(2025)Xia, Shen, Wang, Liu, Sun, Wu, Hu, and Xu]{xia2025leetcodedatasettemporaldatasetrobust}
Yunhui Xia, Wei Shen, Yan Wang, Jason~Klein Liu, Huifeng Sun, Siyue Wu, Jian Hu, and Xiaolong Xu.
\newblock Leetcodedataset: A temporal dataset for robust evaluation and efficient training of code llms, 2025.
\newblock URL \url{https://arxiv.org/abs/2504.14655}.

\bibitem[Xiang et~al.(2024)Xiang, Zhu, Lou, Chen, Pan, Jin, Chen, and Sun]{SimUser}
Wei Xiang, Hanfei Zhu, Suqi Lou, Xinli Chen, Zhenghua Pan, Yuping Jin, Shi Chen, and Lingyun Sun.
\newblock Simuser: Generating usability feedback by simulating various users interacting with mobile applications.
\newblock In \emph{Proceedings of the 2024 CHI Conference on Human Factors in Computing Systems}, CHI '24, New York, NY, USA, 2024. Association for Computing Machinery.
\newblock ISBN 9798400703300.
\newblock \doi{10.1145/3613904.3642481}.
\newblock URL \url{https://doi.org/10.1145/3613904.3642481}.

\bibitem[Xiao et~al.(2024{\natexlab{a}})Xiao, Wan, Huo, Wang, Xu, Wang, Xu, Wang, and Lyu]{xiao2024interaction2code}
Jingyu Xiao, Yuxuan Wan, Yintong Huo, Zixin Wang, Xinyi Xu, Wenxuan Wang, Zhiyao Xu, Yuhang Wang, and Michael~R Lyu.
\newblock Interaction2code: Benchmarking mllm-based interactive webpage code generation from interactive prototyping.
\newblock \emph{arXiv preprint arXiv:2411.03292}, 2024{\natexlab{a}}.

\bibitem[Xiao et~al.(2024{\natexlab{b}})Xiao, Chen, Li, Chen, Sun, and Zhou]{xiao2024prototype2code}
Shuhong Xiao, Yunnong Chen, Jiazhi Li, Liuqing Chen, Lingyun Sun, and Tingting Zhou.
\newblock Prototype2code: End-to-end front-end code generation from ui design prototypes.
\newblock In \emph{International Design Engineering Technical Conferences and Computers and Information in Engineering Conference}, volume 88353, page V02BT02A038. American Society of Mechanical Engineers, 2024{\natexlab{b}}.

\bibitem[Xiao et~al.(2020)Xiao, Quan, Wang, Zhao, and Liao]{DBLP:conf/icdm/XiaoQW0L20}
Tong Xiao, Zhe Quan, Zhi{-}Jie Wang, Kaiqi Zhao, and Xiangke Liao.
\newblock {LPV:} {A} log parser based on vectorization for offline and online log parsing.
\newblock In Claudia Plant, Haixun Wang, Alfredo Cuzzocrea, Carlo Zaniolo, and Xindong Wu, editors, \emph{20th {IEEE} International Conference on Data Mining, {ICDM} 2020, Sorrento, Italy, November 17-20, 2020}, pages 1346--1351. {IEEE}, 2020.

\bibitem[Xiao et~al.(2025)Xiao, Wang, Liu, Zhou, Cheng, and Xiong]{xiao2025predicatefix}
Yuan-An Xiao, Weixuan Wang, Dong Liu, Junwei Zhou, Shengyu Cheng, and Yingfei Xiong.
\newblock Predicatefix: Repairing static analysis alerts with bridging predicates.
\newblock \emph{arXiv preprint arXiv:2503.12205}, 2025.

\bibitem[Xie et~al.(2025{\natexlab{a}})Xie, Li, Gao, Du, Lam, Zou, and Chen]{xie2025swe}
Chengxing Xie, Bowen Li, Chang Gao, He~Du, Wai Lam, Difan Zou, and Kai Chen.
\newblock Swe-fixer: Training open-source llms for effective and efficient github issue resolution.
\newblock \emph{arXiv preprint arXiv:2501.05040}, 2025{\natexlab{a}}.

\bibitem[Xie et~al.(2025{\natexlab{b}})Xie, Zheng, Liu, Wang, Wang, Tan, and Zhang]{xie2025core}
Danning Xie, Mingwei Zheng, Xuwei Liu, Jiannan Wang, Chengpeng Wang, Lin Tan, and Xiangyu Zhang.
\newblock Core: Benchmarking llms code reasoning capabilities through static analysis tasks.
\newblock \emph{arXiv preprint arXiv:2507.05269}, 2025{\natexlab{b}}.

\bibitem[Xie et~al.(2024{\natexlab{a}})Xie, Zeng, Yu, Gao, Zhang, and Ye]{xie2024codeshell}
Rui Xie, Zhengran Zeng, Zhuohao Yu, Chang Gao, Shikun Zhang, and Wei Ye.
\newblock Codeshell technical report.
\newblock \emph{CoRR}, abs/2403.15747, 2024{\natexlab{a}}.
\newblock \doi{10.48550/ARXIV.2403.15747}.
\newblock URL \url{https://doi.org/10.48550/arXiv.2403.15747}.

\bibitem[Xie et~al.(2024{\natexlab{b}})Xie, Wu, and Zhou]{MAG-SQL}
Wenxuan Xie, Gaochen Wu, and Bowen Zhou.
\newblock {MAG-SQL:} multi-agent generative approach with soft schema linking and iterative sub-sql refinement for text-to-sql.
\newblock \emph{CoRR}, abs/2408.07930, 2024{\natexlab{b}}.

\bibitem[Xie et~al.(2025{\natexlab{c}})Xie, Xu, Zhao, and Guo]{OpenSearch-SQL}
Xiangjin Xie, Guangwei Xu, Lingyan Zhao, and Ruijie Guo.
\newblock Opensearch-sql: Enhancing text-to-sql with dynamic few-shot and consistency alignment.
\newblock \emph{Proc. {ACM} Manag. Data}, 3\penalty0 (3):\penalty0 194:1--194:24, 2025{\natexlab{c}}.

\bibitem[Xie et~al.(2025{\natexlab{d}})Xie, Xie, Sheth, Liu, Fried, and Rose]{xie2025repost}
Yiqing Xie, Alex Xie, Divyanshu Sheth, Pengfei Liu, Daniel Fried, and Carolyn Rose.
\newblock Repost: Scalable repository-level coding environment construction with sandbox testing, 2025{\natexlab{d}}.

\bibitem[Xie et~al.(2025{\natexlab{e}})Xie, Xie, Sheth, Liu, Fried, and Rose]{xie2025repostscalablerepositorylevelcoding}
Yiqing Xie, Alex Xie, Divyanshu Sheth, Pengfei Liu, Daniel Fried, and Carolyn Rose.
\newblock Repost: Scalable repository-level coding environment construction with sandbox testing, 2025{\natexlab{e}}.
\newblock URL \url{https://arxiv.org/abs/2503.07358}.

\bibitem[Xie et~al.(2025{\natexlab{f}})Xie, Ye, Zheng, Gao, Dong, Wu, Zhao, Gong, Jiang, Li, et~al.]{xie2025dream}
Zhihui Xie, Jiacheng Ye, Lin Zheng, Jiahui Gao, Jingwei Dong, Zirui Wu, Xueliang Zhao, Shansan Gong, Xin Jiang, Zhenguo Li, et~al.
\newblock Dream-coder 7b: An open diffusion language model for code.
\newblock \emph{arXiv preprint arXiv:2509.01142}, 2025{\natexlab{f}}.

\bibitem[Xin-Ye et~al.(2025)Xin-Ye, Ya-Li, and Ming]{xinye2025enhancingllmslongcode}
Li~Xin-Ye, Du~Ya-Li, and Li~Ming.
\newblock Enhancing llms in long code translation through instrumentation and program state alignment, 2025.
\newblock URL \url{https://arxiv.org/abs/2504.02017}.

\bibitem[Xing et~al.(2025)Xing, Bhatia, Phulwani, Suresh, and Matta]{xing2025hackerrankastraevaluatingcorrectness}
Jun Xing, Mayur Bhatia, Sahil Phulwani, Darshan Suresh, and Rafik Matta.
\newblock Hackerrank-astra: Evaluating correctness \& consistency of large language models on cross-domain multi-file project problems, 2025.
\newblock URL \url{https://arxiv.org/abs/2502.00226}.

\bibitem[Xu et~al.(2023{\natexlab{a}})Xu, Peng, Lei, Mukherjee, Liu, and Xu]{xu2023rewoo}
Binfeng Xu, Zhiyuan Peng, Bowen Lei, Subhabrata Mukherjee, Yuchen Liu, and Dongkuan Xu.
\newblock Rewoo: Decoupling reasoning from observations for efficient augmented language models.
\newblock \emph{arXiv preprint arXiv:2305.18323}, 2023{\natexlab{a}}.

\bibitem[Xu et~al.(2025{\natexlab{a}})Xu, Wang, Wei, Sun, and Huang]{chartIR}
Chengzhi Xu, Yuyang Wang, Lai Wei, Lichao Sun, and Weiran Huang.
\newblock Improved iterative refinement for chart-to-code generation via structured instruction.
\newblock \emph{arXiv preprint arXiv:2506.14837}, 2025{\natexlab{a}}.

\bibitem[Xu et~al.(2025{\natexlab{b}})Xu, Liu, Ren, Zhang, Liang, and Lo]{xu2025flexfl}
Chuyang Xu, Zhongxin Liu, Xiaoxue Ren, Gehao Zhang, Ming Liang, and David Lo.
\newblock Flexfl: Flexible and effective fault localization with open-source large language models.
\newblock \emph{IEEE Transactions on Software Engineering}, 2025{\natexlab{b}}.

\bibitem[Xu et~al.(2024{\natexlab{a}})Xu, Ma, Zhou, Zhao, Chen, Hu, Liu, and Wang]{xu2024ckgfuzzerllmbasedfuzzdriver}
Hanxiang Xu, Wei Ma, Ting Zhou, Yanjie Zhao, Kai Chen, Qiang Hu, Yang Liu, and Haoyu Wang.
\newblock Ckgfuzzer: Llm-based fuzz driver generation enhanced by code knowledge graph, 2024{\natexlab{a}}.
\newblock URL \url{https://arxiv.org/abs/2411.11532}.

\bibitem[Xu et~al.(2025{\natexlab{c}})Xu, Zhu, Pan, Wang, Zhu, Ma, Cao, Chen, and Yu]{xu2025reducingtoolhallucination}
Hongshen Xu, Zichen Zhu, Lei Pan, Zihan Wang, Su~Zhu, Da~Ma, Ruisheng Cao, Lu~Chen, and Kai Yu.
\newblock Reducing tool hallucination via reliability alignment.
\newblock In \emph{Forty-second International Conference on Machine Learning}, 2025{\natexlab{c}}.
\newblock URL \url{https://openreview.net/forum?id=WeOLZmDXyA}.

\bibitem[Xu et~al.(2025{\natexlab{d}})Xu, Pang, Qu, Hayashi, Xiong, and Zhou]{xu_clover_2025}
Jiacheng Xu, Bo~Pang, Jin Qu, Hiroaki Hayashi, Caiming Xiong, and Yingbo Zhou.
\newblock {CLOVER}: {A} {Test} {Case} {Generation} {Benchmark} with {Coverage}, {Long}-{Context}, and {Verification}, February 2025{\natexlab{d}}.
\newblock URL \url{http://arxiv.org/abs/2502.08806}.
\newblock arXiv:2502.08806 [cs].

\bibitem[Xu et~al.(2025{\natexlab{e}})Xu, Mao, Guan, and Feng]{xu2025web}
Kai Xu, YiWei Mao, XinYi Guan, and ZiLong Feng.
\newblock Web-bench: A llm code benchmark based on web standards and frameworks.
\newblock \emph{arXiv preprint arXiv:2505.07473}, 2025{\natexlab{e}}.

\bibitem[Xu et~al.(2021)Xu, Masling, Du, Campagna, Heck, Landay, and Lam]{awst}
Nancy Xu, Sam Masling, Michael Du, Giovanni Campagna, Larry Heck, James Landay, and Monica~S Lam.
\newblock Grounding open-domain instructions to automate web support tasks.
\newblock \emph{arXiv preprint arXiv:2103.16057}, 2021.

\bibitem[Xu et~al.(2024{\natexlab{b}})Xu, Cao, Lu, Wen, Lin, Han, He, Cheung, and Sun]{xu2024cruxeval}
Ruiyang Xu, Jialun Cao, Yaojie Lu, Ming Wen, Hongyu Lin, Xianpei Han, Ben He, Shing-Chi Cheung, and Le~Sun.
\newblock Cruxeval-x: A benchmark for multilingual code reasoning, understanding and execution.
\newblock \emph{arXiv preprint arXiv:2408.13001}, 2024{\natexlab{b}}.

\bibitem[Xu et~al.(2019)Xu, Yao, Xu, Gu, Tong, and Lu]{xu2019commit}
Shengbin Xu, Yuan Yao, Feng Xu, Tianxiao Gu, Hanghang Tong, and Jian Lu.
\newblock Commit message generation for source code changes.
\newblock In \emph{IJCAI}, 2019.

\bibitem[Xu et~al.(2025{\natexlab{f}})Xu, Hu, Min, Chen, Zhao, and Wen]{icpceval}
Shiyi Xu, Yiwen Hu, Yingqian Min, Zhipeng Chen, Wayne~Xin Zhao, and Ji-Rong Wen.
\newblock Icpc-eval: Probing the frontiers of llm reasoning with competitive programming contests, 2025{\natexlab{f}}.
\newblock URL \url{https://arxiv.org/abs/2506.04894}.

\bibitem[Xu et~al.(2020)Xu, Chen, Pei, Zhang, Pan, and Furia]{xu2020restore}
Tongtong Xu, Liushan Chen, Yu~Pei, Tian Zhang, Minxue Pan, and Carlo~A Furia.
\newblock Restore: Retrospective fault localization enhancing automated program repair.
\newblock \emph{IEEE Transactions on Software Engineering}, 48\penalty0 (1):\penalty0 309--326, 2020.

\bibitem[Xu and Rudnicky(2000)]{xu2000agenda}
Wei Xu and Alexander~I. Rudnicky.
\newblock Task-based dialog management using an agenda.
\newblock In \emph{ANLP-NAACL 2000 Workshop: Conversational Systems}, 2000.
\newblock URL \url{https://aclanthology.org/W00-0309/}.

\bibitem[Xu et~al.(2025{\natexlab{g}})Xu, Xiong, Zhao, Chen, Wang, Shen, Wan, Dai, Wu, Xiao, Tao, Mao, Sheng, Guo, Yang, Yu, Kong, Gu, and Wong]{xu2025swingarenacompetitiveprogrammingarena}
Wendong Xu, Jing Xiong, Chenyang Zhao, Qiujiang Chen, Haoran Wang, Hui Shen, Zhongwei Wan, Jianbo Dai, Taiqiang Wu, He~Xiao, Chaofan Tao, Z.~Morley Mao, Ying Sheng, Zhijiang Guo, Hongxia Yang, Bei Yu, Lingpeng Kong, Quanquan Gu, and Ngai Wong.
\newblock Swingarena: Competitive programming arena for long-context github issue solving, 2025{\natexlab{g}}.
\newblock URL \url{https://arxiv.org/abs/2505.23932}.

\bibitem[Xu et~al.(2025{\natexlab{h}})Xu, Liang, Mei, Gao, Tan, and Zhang]{xu2025mem}
Wujiang Xu, Zujie Liang, Kai Mei, Hang Gao, Juntao Tan, and Yongfeng Zhang.
\newblock A-mem: Agentic memory for llm agents.
\newblock \emph{arXiv preprint arXiv:2502.12110}, 2025{\natexlab{h}}.

\bibitem[Xu et~al.(2023{\natexlab{b}})Xu, Zhang, Feng, Ye, Su, Jiang, Cheng, Tan, and Zhang]{DBLP:journals/corr/abs-2306-02546}
Xiangzhe Xu, Zhuo Zhang, Shiwei Feng, Yapeng Ye, Zian Su, Nan Jiang, Siyuan Cheng, Lin Tan, and Xiangyu Zhang.
\newblock Lmpa: Improving decompilation by synergy of large language model and program analysis.
\newblock \emph{CoRR}, abs/2306.02546, 2023{\natexlab{b}}.

\bibitem[Xu et~al.(2024{\natexlab{c}})Xu, Su, Guo, Zhang, Wang, and Zhang]{xu2024prosec}
Xiangzhe Xu, Zian Su, Jinyao Guo, Kaiyuan Zhang, Zhenting Wang, and Xiangyu Zhang.
\newblock Prosec: Fortifying code llms with proactive security alignment.
\newblock \emph{arXiv preprint arXiv:2411.12882}, 2024{\natexlab{c}}.

\bibitem[Xu et~al.(2024{\natexlab{d}})Xu, Wang, Wang, Lu, Xie, Saha, Sahoo, Yu, and Xiong]{xu2024aguvis}
Yiheng Xu, Zekun Wang, Junli Wang, Dunjie Lu, Tianbao Xie, Amrita Saha, Doyen Sahoo, Tao Yu, and Caiming Xiong.
\newblock Aguvis: Unified pure vision agents for autonomous gui interaction.
\newblock \emph{arXiv preprint arXiv:2412.04454}, 2024{\natexlab{d}}.

\bibitem[{Xu} et~al.(2023){Xu}, {Wang}, {Li}, {Luo}, {Wang}, {Liu}, and {Liu}]{Werewolf}
Yuzhuang {Xu}, Shuo {Wang}, Peng {Li}, Fuwen {Luo}, Xiaolong {Wang}, Weidong {Liu}, and Yang {Liu}.
\newblock {Exploring Large Language Models for Communication Games: An Empirical Study on Werewolf}.
\newblock \emph{arXiv e-prints}, art. arXiv:2309.04658, September 2023.
\newblock \doi{10.48550/arXiv.2309.04658}.

\bibitem[Xu et~al.(2025{\natexlab{i}})Xu, Liu, Yin, Zhou, and Poovendran]{xu2025kodcode}
Zhangchen Xu, Yang Liu, Yueqin Yin, Mingyuan Zhou, and Radha Poovendran.
\newblock Kodcode: A diverse, challenging, and verifiable synthetic dataset for coding.
\newblock \emph{arXiv preprint arXiv:2503.02951}, 2025{\natexlab{i}}.

\bibitem[Xu et~al.(2024{\natexlab{e}})Xu, Du, Qi, Xu, Yuan, and Guo]{xu2024chartbench}
Zhengzhuo Xu, Sinan Du, Yiyan Qi, Chengjin Xu, Chun Yuan, and Jian Guo.
\newblock Chartbench: A benchmark for complex visual reasoning in charts, 2024{\natexlab{e}}.
\newblock URL \url{https://arxiv.org/abs/2312.15915}.

\bibitem[Xue et~al.(2024{\natexlab{a}})Xue, Andrzejak, and Leuther]{xue2024an}
Min Xue, Artur Andrzejak, and Marla Leuther.
\newblock An interpretable error correction method for enhancing code-to-code translation.
\newblock In \emph{The Twelfth International Conference on Learning Representations}, 2024{\natexlab{a}}.
\newblock URL \url{https://openreview.net/forum?id=fVxIEHGnVT}.

\bibitem[Xue et~al.(2024{\natexlab{b}})]{xue2024classevalT}
Yifan Xue et~al.
\newblock Classeval-t: Cross-language class-level benchmark for code generation.
\newblock In \emph{ICSE}, 2024{\natexlab{b}}.

\bibitem[Yadav et~al.(2024)Yadav, Beniwal, and Singh]{pythonsaga}
Ankit Yadav, Himanshu Beniwal, and Mayank Singh.
\newblock Pythonsaga: Redefining the benchmark to evaluate code generating llms, 2024.
\newblock URL \url{https://arxiv.org/abs/2401.03855}.

\bibitem[Yamaguchi et~al.(2012)Yamaguchi, Lottmann, and Rieck]{yamaguchi2012generalized}
Fabian Yamaguchi, Markus Lottmann, and Konrad Rieck.
\newblock Generalized vulnerability extrapolation using abstract syntax trees.
\newblock In \emph{Proceedings of the 28th annual computer security applications conference}, pages 359--368, 2012.

\bibitem[Yan et~al.(2025)Yan, Che, Huang, Xu, Li, Li, Qu, Shi, He, Lin, Yang, Yuan, Zhao, Qiao, Zhou, and Fu]{yan2025reformreducinghuman}
Chuanhao Yan, Fengdi Che, Xuhan Huang, Xu~Xu, Xin Li, Yizhi Li, Xingwei Qu, Jingzhe Shi, Zhuangzhuang He, Chenghua Lin, Yaodong Yang, Binhang Yuan, Hang Zhao, Yu~Qiao, Bowen Zhou, and Jie Fu.
\newblock Re:form -- reducing human priors in scalable formal software verification with rl in llms: A preliminary study on dafny, 2025.
\newblock URL \url{https://arxiv.org/abs/2507.16331}.

\bibitem[Yan et~al.(2023)Yan, Tian, Li, Chen, and Wang]{CodeTransOcean}
Weixiang Yan, Yuchen Tian, Yunzhe Li, Qian Chen, and Wen Wang.
\newblock Codetransocean: {A} comprehensive multilingual benchmark for code translation.
\newblock In Houda Bouamor, Juan Pino, and Kalika Bali, editors, \emph{Findings of the Association for Computational Linguistics: {EMNLP} 2023, Singapore, December 6-10, 2023}, pages 5067--5089. Association for Computational Linguistics, 2023.
\newblock URL \url{https://aclanthology.org/2023.findings-emnlp.337}.

\bibitem[Yan et~al.(2024)Yan, Liu, Wang, Li, Chen, Wang, Lin, Zhao, Zhu, Sundaram, and Deng]{CodeScope}
Weixiang Yan, Haitian Liu, Yunkun Wang, Yunzhe Li, Qian Chen, Wen Wang, Tingyu Lin, Weishan Zhao, Li~Zhu, Hari Sundaram, and Shuiguang Deng.
\newblock Codescope: An execution-based multilingual multitask multidimensional benchmark for evaluating llms on code understanding and generation, 2024.
\newblock URL \url{https://arxiv.org/abs/2311.08588}.

\bibitem[Yang et~al.(2024{\natexlab{a}})Yang, Le~Goues, Martins, and Hellendoorn]{yang2024large_fault_localization}
Aidan~ZH Yang, Claire Le~Goues, Ruben Martins, and Vincent Hellendoorn.
\newblock Large language models for test-free fault localization.
\newblock In \emph{Proceedings of the 46th IEEE/ACM International Conference on Software Engineering}, pages 1--12, 2024{\natexlab{a}}.

\bibitem[Yang et~al.(2024{\natexlab{b}})Yang, Takashima, Paulsen, Dodds, and Kroening]{yang2024vert}
Aidan~ZH Yang, Yoshiki Takashima, Brandon Paulsen, Josiah Dodds, and Daniel Kroening.
\newblock Vert: Verified equivalent rust transpilation with few-shot learning.
\newblock \emph{arXiv preprint arXiv:2404.18852}, 26, 2024{\natexlab{b}}.

\bibitem[Yang et~al.(2024{\natexlab{c}})Yang, Yang, Hui, Zheng, Yu, Zhou, Li, Li, Liu, Huang, Dong, Wei, Lin, Tang, Wang, Yang, Tu, Zhang, Ma, Yang, Xu, Zhou, Bai, He, Lin, Dang, Lu, Chen, Yang, Li, Xue, Ni, Zhang, Wang, Peng, Men, Gao, Lin, Wang, Bai, Tan, Zhu, Li, Liu, Ge, Deng, Zhou, Ren, Zhang, Wei, Ren, Liu, Fan, Yao, Zhang, Wan, Chu, Liu, Cui, Zhang, Guo, and Fan]{yang2024qwen2}
An~Yang, Baosong Yang, Binyuan Hui, Bo~Zheng, Bowen Yu, Chang Zhou, Chengpeng Li, Chengyuan Li, Dayiheng Liu, Fei Huang, Guanting Dong, Haoran Wei, Huan Lin, Jialong Tang, Jialin Wang, Jian Yang, Jianhong Tu, Jianwei Zhang, Jianxin Ma, Jianxin Yang, Jin Xu, Jingren Zhou, Jinze Bai, Jinzheng He, Junyang Lin, Kai Dang, Keming Lu, Keqin Chen, Kexin Yang, Mei Li, Mingfeng Xue, Na~Ni, Pei Zhang, Peng Wang, Ru~Peng, Rui Men, Ruize Gao, Runji Lin, Shijie Wang, Shuai Bai, Sinan Tan, Tianhang Zhu, Tianhao Li, Tianyu Liu, Wenbin Ge, Xiaodong Deng, Xiaohuan Zhou, Xingzhang Ren, Xinyu Zhang, Xipin Wei, Xuancheng Ren, Xuejing Liu, Yang Fan, Yang Yao, Yichang Zhang, Yu~Wan, Yunfei Chu, Yuqiong Liu, Zeyu Cui, Zhenru Zhang, Zhifang Guo, and Zhihao Fan.
\newblock Qwen2 technical report, 2024{\natexlab{c}}.
\newblock URL \url{https://arxiv.org/abs/2407.10671}.

\bibitem[Yang et~al.(2025{\natexlab{a}})Yang, Li, Yang, Zhang, Hui, Zheng, Yu, Gao, Huang, Lv, et~al.]{yang2025qwen3}
An~Yang, Anfeng Li, Baosong Yang, Beichen Zhang, Binyuan Hui, Bo~Zheng, Bowen Yu, Chang Gao, Chengen Huang, Chenxu Lv, et~al.
\newblock Qwen3 technical report.
\newblock \emph{arXiv preprint arXiv:2505.09388}, 2025{\natexlab{a}}.

\bibitem[Yang et~al.(2024{\natexlab{d}})Yang, Tian, Ren, Zhang, Klein, Bissyand{\'e}, Goues, and Jin]{yang2024morepair}
Boyang Yang, Haoye Tian, Jiadong Ren, Hongyu Zhang, Jacques Klein, Tegawend{\'e}~F Bissyand{\'e}, Claire~Le Goues, and Shunfu Jin.
\newblock Morepair: Teaching llms to repair code via multi-objective fine-tuning.
\newblock \emph{arXiv preprint arXiv:2404.12636}, 2024{\natexlab{d}}.

\bibitem[Yang et~al.(2025{\natexlab{b}})Yang, Tian, Ren, Jin, Liu, Liu, and Le]{yang2025enhancing}
Boyang Yang, Haoye Tian, Jiadong Ren, Shunfu Jin, Yang Liu, Feng Liu, and Bach Le.
\newblock Enhancing repository-level software repair via repository-aware knowledge graphs.
\newblock \emph{arXiv preprint arXiv:2503.21710}, 2025{\natexlab{b}}.

\bibitem[Yang et~al.(2025{\natexlab{c}})Yang, He, Gao, Cao, Li, and Hsu]{yang2025codeagents}
Bruce Yang, Xinfeng He, Huan Gao, Yifan Cao, Xiaofan Li, and David Hsu.
\newblock Codeagents: A token-efficient framework for codified multi-agent reasoning in llms.
\newblock \emph{arXiv preprint arXiv:2507.03254}, 2025{\natexlab{c}}.

\bibitem[Yang et~al.(2025{\natexlab{d}})Yang, Chen, Lin, Zhou, and Wang]{Yang2025EnhancingLT}
Chen Yang, Junjie Chen, Bin Lin, Jianyi Zhou, and Ziqi Wang.
\newblock Enhancing llm-based test generation for hard-to-cover branches via program analysis.
\newblock \emph{arXiv}, abs/2404.04966, 2025{\natexlab{d}}.
\newblock URL \url{https://api.semanticscholar.org/CorpusID:270928236}.

\bibitem[Yang et~al.(2024{\natexlab{e}})Yang, Shi, Liu, Shui, Wang, Jing, Xu, Zhu, Li, Zhang, et~al.]{yang2024chartmimic}
Cheng Yang, Chufan Shi, Yaxin Liu, Bo~Shui, Junjie Wang, Mohan Jing, Linran Xu, Xinyu Zhu, Siheng Li, Yuxiang Zhang, et~al.
\newblock Chartmimic: Evaluating lmm's cross-modal reasoning capability via chart-to-code generation.
\newblock \emph{arXiv preprint arXiv:2406.09961}, 2024{\natexlab{e}}.

\bibitem[Yang et~al.(2025{\natexlab{e}})Yang, Shi, Liu, Shui, Wang, Jing, XU, Zhu, Li, Zhang, Liu, Nie, Cai, and Yang]{yang2025chartmimic}
Cheng Yang, Chufan Shi, Yaxin Liu, Bo~Shui, Junjie Wang, Mohan Jing, Linran XU, Xinyu Zhu, Siheng Li, Yuxiang Zhang, Gongye Liu, Xiaomei Nie, Deng Cai, and Yujiu Yang.
\newblock Chartmimic: Evaluating {LMM}'s cross-modal reasoning capability via chart-to-code generation.
\newblock In \emph{The Thirteenth International Conference on Learning Representations}, 2025{\natexlab{e}}.
\newblock URL \url{https://openreview.net/forum?id=sGpCzsfd1K}.

\bibitem[Yang et~al.(2024{\natexlab{f}})Yang, Kang, Shi, and Lo]{acecode2024}
Chengran Yang, Hong~Jin Kang, Jieke Shi, and David Lo.
\newblock Acecode: A reinforcement learning framework for aligning code efficiency and correctness in code language models, 2024{\natexlab{f}}.
\newblock URL \url{https://arxiv.org/abs/2412.17264}.

\bibitem[Yang et~al.(2024{\natexlab{g}})Yang, Deng, Lu, Yao, Liu, Jabbarvand, and Zhang]{Yang_2024}
Chenyuan Yang, Yinlin Deng, Runyu Lu, Jiayi Yao, Jiawei Liu, Reyhaneh Jabbarvand, and Lingming Zhang.
\newblock Whitefox: White-box compiler fuzzing empowered by large language models.
\newblock \emph{Proceedings of the ACM on Programming Languages}, 8\penalty0 (OOPSLA2):\penalty0 709–735, October 2024{\natexlab{g}}.
\newblock ISSN 2475-1421.
\newblock \doi{10.1145/3689736}.
\newblock URL \url{http://dx.doi.org/10.1145/3689736}.

\bibitem[Yang et~al.(2025{\natexlab{f}})Yang, Liu, Zhang, Simoulin, Liu, Cao, Teng, Qian, Yang, Luo, and McAuley]{yang2025codethinkthinkcode}
Dayu Yang, Tianyang Liu, Daoan Zhang, Antoine Simoulin, Xiaoyi Liu, Yuwei Cao, Zhaopu Teng, Xin Qian, Grey Yang, Jiebo Luo, and Julian McAuley.
\newblock Code to think, think to code: A survey on code-enhanced reasoning and reasoning-driven code intelligence in llms, 2025{\natexlab{f}}.
\newblock URL \url{https://arxiv.org/abs/2502.19411}.

\bibitem[{Yang} et~al.(2025){Yang}, {Simoulin}, {Qian}, {Liu}, {Cao}, {Teng}, and {Yang}]{DocAgent}
Dayu {Yang}, Antoine {Simoulin}, Xin {Qian}, Xiaoyi {Liu}, Yuwei {Cao}, Zhaopu {Teng}, and Grey {Yang}.
\newblock {DocAgent: A Multi-Agent System for Automated Code Documentation Generation}.
\newblock \emph{arXiv e-prints}, art. arXiv:2504.08725, April 2025.
\newblock \doi{10.48550/arXiv.2504.08725}.

\bibitem[Yang et~al.(2024{\natexlab{h}})Yang, Zhou, Chen, Zhang, Zhuo, and Chen]{yang2024chain}
Guang Yang, Yu~Zhou, Xiang Chen, Xiangyu Zhang, Terry~Yue Zhuo, and Taolue Chen.
\newblock Chain-of-thought in neural code generation: From and for lightweight language models.
\newblock \emph{IEEE Transactions on Software Engineering}, 2024{\natexlab{h}}.

\bibitem[Yang et~al.(2025{\natexlab{a}})Yang, Zheng, Chen, Liang, Hu, Yang, Peng, Li, Feng, Wei, et~al.]{yang2025large}
Guang Yang, Wei Zheng, Xiang Chen, Dong Liang, Peng Hu, Yukui Yang, Shaohua Peng, Zhenghan Li, Jiahui Feng, Xiao Wei, et~al.
\newblock Large language model for verilog code generation: Literature review and the road ahead.
\newblock 2025{\natexlab{a}}.

\bibitem[Yang et~al.(2021)Yang, Ma, Huang, Zhang, Dong, Huang, Muzio, Singhal, Hassan, Song, and Wei]{wmt2021_nmt}
Jian Yang, Shuming Ma, Haoyang Huang, Dongdong Zhang, Li~Dong, Shaohan Huang, Alexandre Muzio, Saksham Singhal, Hany Hassan, Xia Song, and Furu Wei.
\newblock Multilingual machine translation systems from microsoft for {WMT21} shared task.
\newblock In Lo{\"{\i}}c Barrault, Ondrej Bojar, Fethi Bougares, Rajen Chatterjee, Marta~R. Costa{-}juss{\`{a}}, Christian Federmann, Mark Fishel, Alexander Fraser, Markus Freitag, Yvette Graham, Roman Grundkiewicz, Paco Guzman, Barry Haddow, Matthias Huck, Antonio Jimeno{-}Yepes, Philipp Koehn, Tom Kocmi, Andr{\'{e}} F.~T. Martins, Makoto Morishita, and Christof Monz, editors, \emph{Proceedings of the Sixth Conference on Machine Translation, WMT@EMNLP 2021, Online Event, November 10-11, 2021}, pages 446--455. Association for Computational Linguistics, 2021.
\newblock URL \url{https://aclanthology.org/2021.wmt-1.54}.

\bibitem[Yang et~al.(2024{\natexlab{i}})Yang, Yang, Jin, Miao, Zhang, Yang, Cui, Zhang, Hui, and Lin]{codearena}
Jian Yang, Jiaxi Yang, Ke~Jin, Yibo Miao, Lei Zhang, Liqun Yang, Zeyu Cui, Yichang Zhang, Binyuan Hui, and Junyang Lin.
\newblock Evaluating and aligning codellms on human preference.
\newblock \emph{arXiv preprint arXiv:2412.05210}, 2024{\natexlab{i}}.

\bibitem[Yang et~al.(2024{\natexlab{j}})Yang, Zhang, Yang, Jin, Zhang, Peng, Deng, Miao, Liu, Cui, Hui, and Lin]{yang2024execrepobenchmultilevelexecutablecode}
Jian Yang, Jiajun Zhang, Jiaxi Yang, Ke~Jin, Lei Zhang, Qiyao Peng, Ken Deng, Yibo Miao, Tianyu Liu, Zeyu Cui, Binyuan Hui, and Junyang Lin.
\newblock Execrepobench: Multi-level executable code completion evaluation, 2024{\natexlab{j}}.
\newblock URL \url{https://arxiv.org/abs/2412.11990}.

\bibitem[Yang et~al.(2024{\natexlab{k}})Yang, Zhang, Yang, Jin, Zhang, Peng, Deng, Miao, Liu, Cui, et~al.]{execrepobench}
Jian Yang, Jiajun Zhang, Jiaxi Yang, Ke~Jin, Lei Zhang, Qiyao Peng, Ken Deng, Yibo Miao, Tianyu Liu, Zeyu Cui, et~al.
\newblock Execrepobench: Multi-level executable code completion evaluation.
\newblock \emph{arXiv preprint arXiv:2412.11990}, 2024{\natexlab{k}}.

\bibitem[Yang et~al.(2025{\natexlab{b}})Yang, Zhang, Liu, Chai, Tan, Liu, Zhang, Zhou, Niu, Li, et~al.]{yang2025ifevalcode}
Jian Yang, Wei Zhang, Shukai Liu, Linzheng Chai, Yingshui Tan, Jiaheng Liu, Ge~Zhang, Wangchunshu Zhou, Guanglin Niu, Zhoujun Li, et~al.
\newblock Ifevalcode: Controlled code generation.
\newblock \emph{arXiv preprint arXiv:2507.22462}, 2025{\natexlab{b}}.

\bibitem[Yang et~al.(2025{\natexlab{c}})Yang, Zhang, Miao, Quan, Wu, Peng, Yang, Liu, Cui, Hui, and Lin]{xcoder}
Jian Yang, Wei Zhang, Yibo Miao, Shanghaoran Quan, Zhenhe Wu, Qiyao Peng, Liqun Yang, Tianyu Liu, Zeyu Cui, Binyuan Hui, and Junyang Lin.
\newblock Qwen2.5-xcoder: Multi-agent collaboration for multilingual code instruction tuning.
\newblock In Wanxiang Che, Joyce Nabende, Ekaterina Shutova, and Mohammad~Taher Pilehvar, editors, \emph{Proceedings of the 63rd Annual Meeting of the Association for Computational Linguistics (Volume 1: Long Papers), {ACL} 2025, Vienna, Austria, July 27 - August 1, 2025}, pages 13121--13131. Association for Computational Linguistics, 2025{\natexlab{c}}.
\newblock URL \url{https://aclanthology.org/2025.acl-long.642/}.

\bibitem[Yang et~al.(2023{\natexlab{a}})Yang, Zhang, Li, Zou, Li, and Gao]{yang2023set}
Jianwei Yang, Hao Zhang, Feng Li, Xueyan Zou, Chunyuan Li, and Jianfeng Gao.
\newblock Set-of-mark prompting unleashes extraordinary visual grounding in gpt-4v.
\newblock \emph{arXiv preprint arXiv:2310.11441}, 2023{\natexlab{a}}.

\bibitem[Yang et~al.(2023{\natexlab{b}})Yang, OpenAI, et~al.]{yang2023intercode}
John Yang, OpenAI, et~al.
\newblock Intercode: Standardizing and benchmarking interactive coding with execution feedback.
\newblock \emph{arXiv preprint arXiv:2306.14898}, 2023{\natexlab{b}}.

\bibitem[Yang et~al.(2024{\natexlab{l}})Yang, Jimenez, Wettig, et~al.]{yang2024swe}
John Yang, Carlos~E. Jimenez, Alexander Wettig, et~al.
\newblock Swe-agent: Agent-computer interfaces enable automated software engineering.
\newblock \emph{arXiv preprint arXiv:2405.15793}, 2024{\natexlab{l}}.

\bibitem[Yang et~al.(2024{\natexlab{m}})Yang, Jimenez, Zhang, Lieret, Yang, Wu, Press, Muennighoff, Synnaeve, Narasimhan, Yang, Wang, and Press]{yang2024swebenchmultimodalaisystems}
John Yang, Carlos~E. Jimenez, Alex~L. Zhang, Kilian Lieret, Joyce Yang, Xindi Wu, Ori Press, Niklas Muennighoff, Gabriel Synnaeve, Karthik~R. Narasimhan, Diyi Yang, Sida~I. Wang, and Ofir Press.
\newblock Swe-bench multimodal: Do ai systems generalize to visual software domains?, 2024{\natexlab{m}}.
\newblock URL \url{https://arxiv.org/abs/2410.03859}.

\bibitem[Yang et~al.(2025{\natexlab{d}})Yang, Jimenez, Zhang, Lieret, Yang, Wu, Press, Muennighoff, Synnaeve, Narasimhan, Yang, Wang, and Press]{yang2025swebench}
John Yang, Carlos~E Jimenez, Alex~L Zhang, Kilian Lieret, Joyce Yang, Xindi Wu, Ori Press, Niklas Muennighoff, Gabriel Synnaeve, Karthik~R Narasimhan, Diyi Yang, Sida Wang, and Ofir Press.
\newblock Swe-bench multimodal: Do ai systems generalize to visual software domains?
\newblock In \emph{The Thirteenth International Conference on Learning Representations}, 2025{\natexlab{d}}.
\newblock URL \url{https://openreview.net/forum?id=riTiq3i21b}.

\bibitem[Yang et~al.(2025{\natexlab{e}})Yang, Leret, Jimenez, Wettig, Khandpur, Zhang, Hui, Press, Schmidt, and Yang]{yang2025swesmith}
John Yang, Kilian Leret, Carlos~E Jimenez, Alexander Wettig, Kabir Khandpur, Yanzhe Zhang, Binyuan Hui, Ofir Press, Ludwig Schmidt, and Diyi Yang.
\newblock Swe-smith: Scaling data for software engineering agents, 2025{\natexlab{e}}.

\bibitem[Yang et~al.(2025{\natexlab{f}})Yang, Leret, Jimenez, Wettig, Khandpur, Zhang, Hui, Press, Schmidt, and Yang]{yang2025swesmithscalingdatasoftware}
John Yang, Kilian Leret, Carlos~E. Jimenez, Alexander Wettig, Kabir Khandpur, Yanzhe Zhang, Binyuan Hui, Ofir Press, Ludwig Schmidt, and Diyi Yang.
\newblock Swe-smith: Scaling data for software engineering agents, 2025{\natexlab{f}}.
\newblock URL \url{https://arxiv.org/abs/2504.21798}.

\bibitem[Yang et~al.(2024{\natexlab{n}})Yang, Liu, Chaudhary, Fakoor, Chaudhari, Karypis, and Rangwala]{yang2024agentoccam}
Ke~Yang, Yao Liu, Sapana Chaudhary, Rasool Fakoor, Pratik Chaudhari, George Karypis, and Huzefa Rangwala.
\newblock Agentoccam: A simple yet strong baseline for llm-based web agents.
\newblock \emph{arXiv preprint arXiv:2410.13825}, 2024{\natexlab{n}}.

\bibitem[Yang et~al.(2024{\natexlab{o}})Yang, Yang, Gao, Wang, Wang, Zhu, Chu, Zhou, Liang, Wang, and Chen]{journals/corr/abs-2406-18181}
L.~Yang, C.~Yang, S.~Gao, W.~Wang, B.~Wang, Q.~Zhu, X.~Chu, J.~Zhou, G.~Liang, Q.~Wang, and J.~Chen.
\newblock An empirical study of unit test generation with large language models.
\newblock \emph{CoRR}, abs/2406.18181, 2024{\natexlab{o}}.
\newblock URL \url{https://arxiv.org/abs/2406.18181}.

\bibitem[Yang et~al.(2025{\natexlab{g}})Yang, Jin, Shi, Peng, Chen, and Xiong]{probench}
Lei Yang, Renren Jin, Ling Shi, Jianxiang Peng, Yue Chen, and Deyi Xiong.
\newblock Probench: Benchmarking large language models in competitive programming, 2025{\natexlab{g}}.
\newblock URL \url{https://arxiv.org/abs/2502.20868}.

\bibitem[Yang et~al.(2024{\natexlab{p}})Yang, Wei, Yang, Ma, Guo, Cheng, and Li]{seq2seq_afl}
Liqun Yang, Chaoren Wei, Jian Yang, Jinxin Ma, Hongcheng Guo, Long Cheng, and Zhoujun Li.
\newblock Seq2seq-afl: Fuzzing via sequence-to-sequence model.
\newblock \emph{Int. J. Mach. Learn. Cybern.}, 15\penalty0 (10):\penalty0 4403--4421, 2024{\natexlab{p}}.
\newblock \doi{10.1007/S13042-024-02153-Z}.
\newblock URL \url{https://doi.org/10.1007/s13042-024-02153-z}.

\bibitem[Yang et~al.(2024{\natexlab{q}})Yang, Liu, Zeng, Guo, Liu, and Wu]{yang2024learning}
Qinwei Yang, Xueqing Liu, Yan Zeng, Ruocheng Guo, Yang Liu, and Peng Wu.
\newblock Learning the optimal policy for balancing short-term and long-term rewards.
\newblock \emph{Advances in Neural Information Processing Systems}, 37:\penalty0 36514--36540, 2024{\natexlab{q}}.

\bibitem[Yang et~al.(2023{\natexlab{c}})Yang, Song, Li, Zhao, Ge, Li, and Shan]{yang2023gpt4toolsteachinglargelanguage}
Rui Yang, Lin Song, Yanwei Li, Sijie Zhao, Yixiao Ge, Xiu Li, and Ying Shan.
\newblock Gpt4tools: Teaching large language model to use tools via self-instruction, 2023{\natexlab{c}}.
\newblock URL \url{https://arxiv.org/abs/2305.18752}.

\bibitem[Yang et~al.(2023{\natexlab{d}})Yang, Chiang, Zheng, Gonzalez, and Stoica]{yang2023rethinkingbenchmarkcontaminationlanguage}
Shuo Yang, Wei-Lin Chiang, Lianmin Zheng, Joseph~E. Gonzalez, and Ion Stoica.
\newblock Rethinking benchmark and contamination for language models with rephrased samples, 2023{\natexlab{d}}.
\newblock URL \url{https://arxiv.org/abs/2311.04850}.

\bibitem[Yang et~al.(2025{\natexlab{h}})Yang, Kautz, and Hatamizadeh]{yang2024gateddeltanet}
Songlin Yang, Jan Kautz, and Ali Hatamizadeh.
\newblock Gated delta networks: Improving mamba2 with delta rule, 2025{\natexlab{h}}.
\newblock URL \url{https://arxiv.org/abs/2412.06464}.

\bibitem[Yang et~al.(2025{\natexlab{i}})Yang, Wang, Zhang, Shen, and Kim]{yang2024deltanet}
Songlin Yang, Bailin Wang, Yu~Zhang, Yikang Shen, and Yoon Kim.
\newblock Parallelizing linear transformers with the delta rule over sequence length, 2025{\natexlab{i}}.
\newblock URL \url{https://arxiv.org/abs/2406.06484}.

\bibitem[Yang et~al.(2025{\natexlab{j}})Yang, Wang, Liu, Li, Yan, Wang, Gu, Yu, Liu, and Yu]{debugeval}
Weiqing Yang, Hanbin Wang, Zhenghao Liu, Xinze Li, Yukun Yan, Shuo Wang, Yu~Gu, Minghe Yu, Zhiyuan Liu, and Ge~Yu.
\newblock Coast: Enhancing the code debugging ability of llms through communicative agent based data synthesis, 2025{\natexlab{j}}.
\newblock URL \url{https://arxiv.org/abs/2408.05006}.

\bibitem[Yang et~al.(2025{\natexlab{k}})Yang, Stice, Payani, and Mirzasoleiman]{yang2025bootstrapping}
Wenhan Yang, Spencer Stice, Ali Payani, and Baharan Mirzasoleiman.
\newblock Bootstrapping llm robustness for vlm safety via reducing the pretraining modality gap.
\newblock \emph{arXiv preprint arXiv:2505.24208}, 2025{\natexlab{k}}.

\bibitem[Yang et~al.(2024{\natexlab{r}})Yang, Wu, Zhu, Tang, and Luo]{yang2024askchart}
Xudong Yang, Yifan Wu, Yizhang Zhu, Nan Tang, and Yuyu Luo.
\newblock Askchart: Universal chart understanding through textual enhancement, 2024{\natexlab{r}}.
\newblock URL \url{https://arxiv.org/abs/2412.19146}.

\bibitem[Yang et~al.(2023{\natexlab{e}})Yang, Li, Wang, Lin, Azarnasab, Ahmed, Liu, Liu, Zeng, and Wang]{yang2023mmreact}
Zhengyuan Yang, Linjie Li, Jianfeng Wang, Kevin Lin, Ehsan Azarnasab, Faisal Ahmed, Zicheng Liu, Ce~Liu, Michael Zeng, and Lijuan Wang.
\newblock Mm-react: Prompting chatgpt for multimodal reasoning and action, 2023{\natexlab{e}}.
\newblock URL \url{https://arxiv.org/abs/2303.11381}.

\bibitem[Yang et~al.(2025{\natexlab{l}})Yang, Kuang, Xia, and Zhao]{yang_can_2025}
Zheyuan Yang, Zexi Kuang, Xue Xia, and Yilun Zhao.
\newblock Can {LLMs} {Generate} {High}-{Quality} {Test} {Cases} for {Algorithm} {Problems}? {TestCase}-{Eval}: {A} {Systematic} {Evaluation} of {Fault} {Coverage} and {Exposure}, June 2025{\natexlab{l}}.
\newblock URL \url{http://arxiv.org/abs/2506.12278}.
\newblock arXiv:2506.12278 [cs].

\bibitem[Yang et~al.(2024{\natexlab{s}})Yang, Sun, Yue, Devanbu, and Lo]{yang2024robustness}
Zhou Yang, Zhensu Sun, Terry~Zhuo Yue, Premkumar Devanbu, and David Lo.
\newblock Robustness, security, privacy, explainability, efficiency, and usability of large language models for code.
\newblock \emph{arXiv preprint arXiv:2403.07506}, 2024{\natexlab{s}}.

\bibitem[Yang et~al.(2025{\natexlab{m}})Yang, Wang, Fu, He, Xiong, Liu, Miao, Gao, Wang, Ma, et~al.]{kimi_dev}
Zonghan Yang, Shengjie Wang, Kelin Fu, Wenyang He, Weimin Xiong, Yibo Liu, Yibo Miao, Bofei Gao, Yejie Wang, Yingwei Ma, et~al.
\newblock Kimi-dev: Agentless training as skill prior for swe-agents.
\newblock \emph{arXiv preprint arXiv:2509.23045}, 2025{\natexlab{m}}.

\bibitem[Yao et~al.(2022{\natexlab{a}})Yao, Chen, Yang, and Narasimhan]{yao2022webshop}
Shunyu Yao, Howard Chen, John Yang, and Karthik Narasimhan.
\newblock Webshop: Towards scalable real-world web interaction with grounded language agents.
\newblock \emph{Advances in Neural Information Processing Systems}, 35:\penalty0 20744--20757, 2022{\natexlab{a}}.

\bibitem[Yao et~al.(2022{\natexlab{b}})Yao, Zhao, Yu, Du, Shafran, Narasimhan, and Cao]{yao2022react}
Shunyu Yao, Jeffrey Zhao, Dian Yu, Nan Du, Izhak Shafran, Karthik Narasimhan, and Yuan Cao.
\newblock React: Synergizing reasoning and acting in language models, 2022{\natexlab{b}}.

\bibitem[{Yao} et~al.(2023){Yao}, {Chen}, {Hanjie}, {Yang}, and {Narasimhan}]{Colli}
Shunyu {Yao}, Howard {Chen}, Austin~W. {Hanjie}, Runzhe {Yang}, and Karthik {Narasimhan}.
\newblock {COLLIE: Systematic Construction of Constrained Text Generation Tasks}.
\newblock \emph{arXiv e-prints}, art. arXiv:2307.08689, July 2023.
\newblock \doi{10.48550/arXiv.2307.08689}.

\bibitem[Yao et~al.(2023{\natexlab{a}})Yao, Yu, Zhao, Shafran, Griffiths, Cao, and Narasimhan]{yao2023tot}
Shunyu Yao, Dian Yu, Jeffrey Zhao, Izhak Shafran, Thomas~L. Griffiths, Yuan Cao, and Karthik Narasimhan.
\newblock Tree of thoughts: Deliberate problem solving with large language models, 2023{\natexlab{a}}.
\newblock URL \url{https://arxiv.org/abs/2305.10601}.

\bibitem[Yao et~al.(2023{\natexlab{b}})Yao, Zhao, Yu, Du, Shafran, Narasimhan, and Cao]{yao2023react}
Shunyu Yao, Jeffrey Zhao, Dian Yu, Nan Du, Izhak Shafran, Karthik Narasimhan, and Yuan Cao.
\newblock React: Synergizing reasoning and acting in language models.
\newblock In \emph{International Conference on Learning Representations (ICLR)}, 2023{\natexlab{b}}.

\bibitem[Yao et~al.(2024)Yao, Shinn, Razavi, and Narasimhan]{yao2024tau}
Shunyu Yao, Noah Shinn, Pedram Razavi, and Karthik Narasimhan.
\newblock taubench: A benchmark for tool-agent-user interaction in real-world domains.
\newblock \emph{arXiv preprint arXiv:2406.12045}, 2024.

\bibitem[Yao et~al.(2023{\natexlab{c}})Yao, Aminabadi, Ruwase, Rajbhandari, Wu, Awan, Rasley, Zhang, Li, Holmes, et~al.]{yao2023deepspeed}
Zhewei Yao, Reza~Yazdani Aminabadi, Olatunji Ruwase, Samyam Rajbhandari, Xiaoxia Wu, Ammar~Ahmad Awan, Jeff Rasley, Minjia Zhang, Conglong Li, Connor Holmes, et~al.
\newblock Deepspeed-chat: Easy, fast and affordable rlhf training of chatgpt-like models at all scales.
\newblock \emph{arXiv preprint arXiv:2308.01320}, 2023{\natexlab{c}}.

\bibitem[Yao et~al.(2019)Yao, Peddamail, and Sun]{yao2019coacor}
Ziyu Yao, Jayavardhan~Reddy Peddamail, and Huan Sun.
\newblock Coacor: Code annotation for code retrieval with reinforcement learning.
\newblock In \emph{The world wide web conference}, pages 2203--2214, 2019.

\bibitem[Ye and Monperrus(2024)]{ye2024iter}
He~Ye and Martin Monperrus.
\newblock Iter: Iterative neural repair for multi-location patches.
\newblock In \emph{Proceedings of the 46th IEEE/ACM international conference on software engineering}, pages 1--13, 2024.

\bibitem[Ye et~al.(2022{\natexlab{a}})Ye, Martinez, Luo, Zhang, and Monperrus]{ye2022selfapr}
He~Ye, Matias Martinez, Xiapu Luo, Tao Zhang, and Martin Monperrus.
\newblock Selfapr: Self-supervised program repair with test execution diagnostics.
\newblock In \emph{Proceedings of the 37th IEEE/ACM International Conference on Automated Software Engineering}, pages 1--13, 2022{\natexlab{a}}.

\bibitem[Ye et~al.(2022{\natexlab{b}})Ye, Martinez, and Monperrus]{ye2022neural}
He~Ye, Matias Martinez, and Martin Monperrus.
\newblock Neural program repair with execution-based backpropagation.
\newblock In \emph{Proceedings of the 44th international conference on software engineering}, pages 1506--1518, 2022{\natexlab{b}}.

\bibitem[Ye et~al.(2025)Ye, Huang, Xiao, Chern, Xia, and Liu]{ye2025limo}
Yixin Ye, Zhen Huang, Yang Xiao, Ethan Chern, Shijie Xia, and Pengfei Liu.
\newblock Limo: Less is more for reasoning.
\newblock \emph{arXiv preprint arXiv:2502.03387}, 2025.

\bibitem[Yin et~al.(2024{\natexlab{a}})Yin, Ni, Nguyen, Wang, and Yang]{yin2024rectifier0}
Xin Yin, Chao Ni, Tien~N. Nguyen, Shaohua Wang, and Xiaohu Yang.
\newblock Rectifier: Code translation with corrector via llms, 2024{\natexlab{a}}.
\newblock URL \url{https://arxiv.org/abs/2407.07472}.

\bibitem[Yin et~al.(2024{\natexlab{b}})Yin, Ni, Wang, Li, Zeng, and Yang]{yin2024thinkrepair}
Xin Yin, Chao Ni, Shaohua Wang, Zhenhao Li, Limin Zeng, and Xiaohu Yang.
\newblock Thinkrepair: Self-directed automated program repair.
\newblock In \emph{Proceedings of the 33rd ACM SIGSOFT International Symposium on Software Testing and Analysis}, pages 1274--1286, 2024{\natexlab{b}}.

\bibitem[Yong et~al.(2023)Yong, Menghini, and Bach]{yong2023lowresource}
Zheng-Xin Yong, Cristina Menghini, and Stephen~H. Bach.
\newblock Low-resource languages jailbreak gpt-4, 2023.

\bibitem[Young et~al.(2013)Young, Gašić, Thomson, and Williams]{young2013pomdp}
Steve Young, Milica Gašić, Blaise Thomson, and Jason~D. Williams.
\newblock Pomdp-based statistical spoken dialog systems: A review.
\newblock \emph{Proceedings of the IEEE}, 101\penalty0 (5):\penalty0 1160--1179, 2013.
\newblock \doi{10.1109/JPROC.2012.2225812}.

\bibitem[Yu et~al.(2025{\natexlab{a}})Yu, Yan, Li, Xiao, He, Yu, Zhang, Rong, and Huang]{synthcoder}
Dongjun Yu, Xiao Yan, Zhenrui Li, Jipeng Xiao, Haochuan He, Yongda Yu, Hao Zhang, Guoping Rong, and Xiaobo Huang.
\newblock Synthcoder: A synthetical strategy to tune llms for code completion, 2025{\natexlab{a}}.
\newblock URL \url{https://arxiv.org/abs/2508.15495}.

\bibitem[Yu et~al.(2025{\natexlab{b}})Yu, Yan, Li, Xiao, He, Yu, Zhang, Rong, and Huang]{yu2025synthcoder}
Dongjun Yu, Xiao Yan, Zhenrui Li, Jipeng Xiao, Haochuan He, Yongda Yu, Hao Zhang, Guoping Rong, and Xiaobo Huang.
\newblock Synthcoder: A synthetical strategy to tune llms for code completion.
\newblock \emph{arXiv preprint arXiv:2508.15495}, 2025{\natexlab{b}}.

\bibitem[Yu et~al.(2024{\natexlab{a}})Yu, Shen, Ran, Zhang, Zhang, Ma, Liang, Li, Wang, and Xie]{yu2024codereval}
Hao Yu, Bo~Shen, Dezhi Ran, Jiaxin Zhang, Qi~Zhang, Yuchi Ma, Guangtai Liang, Ying Li, Qianxiang Wang, and Tao Xie.
\newblock Codereval: A benchmark of pragmatic code generation with generative pre-trained models.
\newblock In \emph{Proceedings of the 46th IEEE/ACM International Conference on Software Engineering}, pages 1--12, 2024{\natexlab{a}}.

\bibitem[Yu et~al.(2022)Yu, Xu, Tan, Du, Zhang, and Li]{yu2022towards}
Haonan Yu, Zihan Xu, Wei Tan, Yixuan Du, Kun Zhang, and Zhaoran Li.
\newblock Towards safe reinforcement learning with a safety editor policy.
\newblock \emph{arXiv preprint arXiv:2205.15283}, 2022.

\bibitem[Yu et~al.(2025{\natexlab{c}})Yu, Zhang, Zhu, Yuan, Zuo, Yue, Dai, Fan, Liu, Liu, et~al.]{yu2025dapo}
Qiying Yu, Zheng Zhang, Ruofei Zhu, Yufeng Yuan, Xiaochen Zuo, Yu~Yue, Weinan Dai, Tiantian Fan, Gaohong Liu, Lingjun Liu, et~al.
\newblock Dapo: An open-source llm reinforcement learning system at scale.
\newblock \emph{arXiv preprint arXiv:2503.14476}, 2025{\natexlab{c}}.

\bibitem[{Yu} et~al.(2019){Yu}, {Quillen}, {He}, {Julian}, {Narayan}, {Shively}, {Bellathur}, {Hausman}, {Finn}, and {Levine}]{MetaWorld}
Tianhe {Yu}, Deirdre {Quillen}, Zhanpeng {He}, Ryan {Julian}, Avnish {Narayan}, Hayden {Shively}, Adithya {Bellathur}, Karol {Hausman}, Chelsea {Finn}, and Sergey {Levine}.
\newblock {Meta-World: A Benchmark and Evaluation for Multi-Task and Meta Reinforcement Learning}.
\newblock \emph{arXiv e-prints}, art. arXiv:1910.10897, October 2019.
\newblock \doi{10.48550/arXiv.1910.10897}.

\bibitem[Yu et~al.(2025{\natexlab{d}})Yu, Mangal, Zhuo, Fredrikson, and Pasareanu]{yu2025mixture}
Weichen Yu, Ravi Mangal, Terry Zhuo, Matt Fredrikson, and Corina~S Pasareanu.
\newblock A mixture of linear corrections generates secure code.
\newblock \emph{arXiv preprint arXiv:2507.09508}, 2025{\natexlab{d}}.

\bibitem[Yu et~al.(2024{\natexlab{b}})Yu, Rong, Shen, Zhang, Shao, Wang, Wei, Xu, and Wang]{yu2024fine}
Yongda Yu, Guoping Rong, Haifeng Shen, He~Zhang, Dong Shao, Min Wang, Zhao Wei, Yong Xu, and Juhong Wang.
\newblock Fine-tuning large language models to improve accuracy and comprehensibility of automated code review.
\newblock \emph{ACM transactions on software engineering and methodology}, 34\penalty0 (1):\penalty0 1--26, 2024{\natexlab{b}}.

\bibitem[Yu et~al.(2024{\natexlab{c}})Yu, Zhang, Shang, Huang, Xu, Zhao, Hu, and Yin]{yu2024WaveCoderWidespreadVersatile}
Zhaojian Yu, Xin Zhang, Ning Shang, Yangyu Huang, Can Xu, Yishujie Zhao, Wenxiang Hu, and Qiufeng Yin.
\newblock {{WaveCoder}}: {{Widespread And Versatile Enhancement For Code Large Language Models By Instruction Tuning}}, 2024{\natexlab{c}}.

\bibitem[Yu et~al.(2024{\natexlab{d}})Yu, Zhao, Cohan, and Zhang]{humanevalpro}
Zhaojian Yu, Yilun Zhao, Arman Cohan, and Xiao-Ping Zhang.
\newblock Humaneval pro and mbpp pro: Evaluating large language models on self-invoking code generation, 2024{\natexlab{d}}.
\newblock URL \url{https://arxiv.org/abs/2412.21199}.

\bibitem[Yuan et~al.(2024{\natexlab{a}})Yuan, Cui, Wang, Ding, Wang, Deng, Shan, Chen, Xie, Lin, Liu, Zhou, Peng, Liu, and Sun]{yuan2024advancing}
Lifan Yuan, Ganqu Cui, Hanbin Wang, Ning Ding, Xingyao Wang, Jia Deng, Boji Shan, Huimin Chen, Ruobing Xie, Yankai Lin, Zhenghao Liu, Bowen Zhou, Hao Peng, Zhiyuan Liu, and Maosong Sun.
\newblock Advancing llm reasoning generalists with preference trees, 2024{\natexlab{a}}.
\newblock URL \url{https://arxiv.org/abs/2404.02078}.

\bibitem[Yuan et~al.(2024{\natexlab{b}})Yuan, Chen, Hu, Feng, Xie, Mohammadi, Xing, and Quigley]{PrototypeFlow}
Mingyue Yuan, Jieshan Chen, Yongquan Hu, Sidong Feng, Mulong Xie, Gelareh Mohammadi, Zhenchang Xing, and Aaron Quigley.
\newblock Towards human-ai synergy in ui design: Enhancing multi-agent based ui generation with intent clarification and alignment.
\newblock \emph{arXiv preprint arXiv:2412.20071}, 2024{\natexlab{b}}.

\bibitem[Yuan et~al.(2024{\natexlab{c}})Yuan, Song, Chen, Tan, Shen, Kan, Li, and Yang]{yuan2024easytoolenhancingllmbasedagents}
Siyu Yuan, Kaitao Song, Jiangjie Chen, Xu~Tan, Yongliang Shen, Ren Kan, Dongsheng Li, and Deqing Yang.
\newblock Easytool: Enhancing llm-based agents with concise tool instruction, 2024{\natexlab{c}}.
\newblock URL \url{https://arxiv.org/abs/2401.06201}.

\bibitem[Yuan et~al.(2022)Yuan, Zhang, He, Fang, Hung, Hao, and Yin]{yuan2022circle}
Wei Yuan, Quanjun Zhang, Tieke He, Chunrong Fang, Nguyen Quoc~Viet Hung, Xiaodong Hao, and Hongzhi Yin.
\newblock Circle: Continual repair across programming languages.
\newblock In \emph{Proceedings of the 31st ACM SIGSOFT international symposium on software testing and analysis}, pages 678--690, 2022.

\bibitem[Yuan et~al.(2025{\natexlab{a}})Yuan, Moss, Feghali, Singh, Moldavskaya, MacPhee, Caccia, Pereira, Kim, Sordoni, et~al.]{debuggym}
Xingdi Yuan, Morgane~M Moss, Charbel~El Feghali, Chinmay Singh, Darya Moldavskaya, Drew MacPhee, Lucas Caccia, Matheus Pereira, Minseon Kim, Alessandro Sordoni, et~al.
\newblock debug-gym: A text-based environment for interactive debugging.
\newblock \emph{arXiv preprint arXiv:2503.21557}, 2025{\natexlab{a}}.

\bibitem[Yuan et~al.(2025{\natexlab{b}})Yuan, Yue, Zhu, Fan, and Yan]{yuan2025s}
Yufeng Yuan, Yu~Yue, Ruofei Zhu, Tiantian Fan, and Lin Yan.
\newblock What's behind ppo's collapse in long-cot? value optimization holds the secret.
\newblock \emph{arXiv preprint arXiv:2503.01491}, 2025{\natexlab{b}}.

\bibitem[Yuan et~al.(2023)Yuan, Yuan, Li, Dong, Lu, Tan, Zhou, and Zhou]{rft}
Zheng Yuan, Hongyi Yuan, Chengpeng Li, Guanting Dong, Keming Lu, Chuanqi Tan, Chang Zhou, and Jingren Zhou.
\newblock Scaling relationship on learning mathematical reasoning with large language models, 2023.
\newblock URL \url{https://arxiv.org/abs/2308.01825}.

\bibitem[Yuan et~al.(2024{\natexlab{d}})Yuan, Chen, Wang, Yu, Peng, and Lou]{yuan2024semantic}
Zhiqiang Yuan, Weitong Chen, Hanlin Wang, Kai Yu, Xin Peng, and Yiling Lou.
\newblock Semantic alignment-enhanced code translation via an llm-based multi-agent system.
\newblock \emph{arXiv preprint arXiv: 2409.19894}, 2024{\natexlab{d}}.

\bibitem[Yuan et~al.(2024{\natexlab{e}})Yuan, Lou, Liu, Ding, Wang, Chen, and Peng]{yuan2024manualtestsevaluatingimproving}
Zhiqiang Yuan, Yiling Lou, Mingwei Liu, Shiji Ding, Kaixin Wang, Yixuan Chen, and Xin Peng.
\newblock No more manual tests? evaluating and improving chatgpt for unit test generation, 2024{\natexlab{e}}.
\newblock URL \url{https://arxiv.org/abs/2305.04207}.

\bibitem[{Yuchen Lin} et~al.(2025){Yuchen Lin}, {Le Bras}, {Richardson}, {Sabharwal}, {Poovendran}, {Clark}, and {Choi}]{ZebraLogic}
Bill {Yuchen Lin}, Ronan {Le Bras}, Kyle {Richardson}, Ashish {Sabharwal}, Radha {Poovendran}, Peter {Clark}, and Yejin {Choi}.
\newblock {ZebraLogic: On the Scaling Limits of LLMs for Logical Reasoning}.
\newblock \emph{arXiv e-prints}, art. arXiv:2502.01100, February 2025.
\newblock \doi{10.48550/arXiv.2502.01100}.

\bibitem[Yue et~al.(2025)Yue, Yuan, Yu, Zuo, Zhu, Xu, Chen, Wang, Fan, Du, et~al.]{yue2025vapo}
Yu~Yue, Yufeng Yuan, Qiying Yu, Xiaochen Zuo, Ruofei Zhu, Wenyuan Xu, Jiaze Chen, Chengyi Wang, TianTian Fan, Zhengyin Du, et~al.
\newblock Vapo: Efficient and reliable reinforcement learning for advanced reasoning tasks.
\newblock \emph{arXiv preprint arXiv:2504.05118}, 2025.

\bibitem[Yun et~al.(2024)Yun, Thushara, Bhat, Wang, Deng, Wang, Tao, Li, Li, Nakov, et~al.]{yun2024web2code}
Sukmin Yun, Rusiru Thushara, Mohammad Bhat, Yongxin Wang, Mingkai Deng, Jinhong Wang, Tianhua Tao, Junbo Li, Haonan Li, Preslav Nakov, et~al.
\newblock Web2code: A large-scale webpage-to-code dataset and evaluation framework for multimodal llms.
\newblock \emph{Advances in neural information processing systems}, 37:\penalty0 112134--112157, 2024.

\bibitem[Zala et~al.(2023)Zala, Lin, Cho, and Bansal]{zala2023diagrammergpt}
Abhay Zala, Han Lin, Jaemin Cho, and Mohit Bansal.
\newblock Diagrammergpt: Generating open-domain, open-platform diagrams via llm planning.
\newblock \emph{arXiv preprint arXiv:2310.12128}, 2023.

\bibitem[Zan et~al.(2022)Zan, Chen, Yang, Lin, Kim, Guan, Wang, Chen, and Lou]{codegpt}
Daoguang Zan, Bei Chen, Dejian Yang, Zeqi Lin, Minsu Kim, Bei Guan, Yongji Wang, Weizhu Chen, and Jian-Guang Lou.
\newblock {CERT}: Continual pre-training on sketches for library-oriented code generation.
\newblock In \emph{The 2022 International Joint Conference on Artificial Intelligence}, 2022.

\bibitem[Zan et~al.(2024{\natexlab{a}})Zan, Huang, Yu, Lin, Shi, Liu, Chen, Qi, Yu, Yu, Ran, Zeng, Shen, Bian, Liang, Guan, Huang, Xie, Wang, and Wang]{zan2024swebenchjavagithubissueresolving}
Daoguang Zan, Zhirong Huang, Ailun Yu, Shaoxin Lin, Yifan Shi, Wei Liu, Dong Chen, Zongshuai Qi, Hao Yu, Lei Yu, Dezhi Ran, Muhan Zeng, Bo~Shen, Pan Bian, Guangtai Liang, Bei Guan, Pengjie Huang, Tao Xie, Yongji Wang, and Qianxiang Wang.
\newblock Swe-bench-java: A github issue resolving benchmark for java, 2024{\natexlab{a}}.
\newblock URL \url{https://arxiv.org/abs/2408.14354}.

\bibitem[Zan et~al.(2024{\natexlab{b}})Zan, Yu, Liu, Chen, Shen, Li, Yao, Gong, Chen, Guan, Yang, Wang, Wang, and Cui]{zan2024codes}
Daoguang Zan, Ailun Yu, Wei Liu, Dong Chen, Bo~Shen, Wei Li, Yafen Yao, Yongshun Gong, Xiaolin Chen, Bei Guan, Zhiguang Yang, Yongji Wang, Qianxiang Wang, and Lizhen Cui.
\newblock Codes: Natural language to code repository via multi-layer sketch.
\newblock \emph{CoRR abs/2403.16443}, 2024{\natexlab{b}}.
\newblock URL \url{https://arxiv.org/abs/2403.16443}.

\bibitem[Zan et~al.(2025)Zan, Huang, Liu, Chen, Zhang, Xin, Chen, Liu, Zhong, Li, Liu, Xiao, Chen, Zhang, Su, Liu, Long, Shen, and Xiang]{zan2025multiswebenchmultilingualbenchmarkissue}
Daoguang Zan, Zhirong Huang, Wei Liu, Hanwu Chen, Linhao Zhang, Shulin Xin, Lu~Chen, Qi~Liu, Xiaojian Zhong, Aoyan Li, Siyao Liu, Yongsheng Xiao, Liangqiang Chen, Yuyu Zhang, Jing Su, Tianyu Liu, Rui Long, Kai Shen, and Liang Xiang.
\newblock Multi-swe-bench: A multilingual benchmark for issue resolving, 2025.
\newblock URL \url{https://arxiv.org/abs/2504.02605}.

\bibitem[Zeng et~al.(2025{\natexlab{a}})Zeng, Lv, Zheng, Hou, Chen, Xie, Wang, Yin, Zeng, Zhang, Wang, Zhong, Liu, Lu, Cao, Zhang, Huang, Wei, Cheng, An, Niu, Wen, Bai, Du, Wang, Zhu, Zhang, Wen, Wu, Xu, Huang, Zhao, Cai, Yu, Li, Ge, Huang, Zhang, Xu, Zhu, Li, Yin, Lin, Yang, Jiang, Ai, Zhu, Wang, Pan, Wang, Sun, Li, Li, Hu, Zhang, Peng, Tai, Zhang, Wang, Yang, Liu, Zhao, Liu, Yan, Liu, Chen, Li, Zhao, Ren, Jiao, Zhao, Yan, Wang, Gui, Zhao, Liu, Li, Li, Lu, Wang, Yuan, Li, Du, Du, Liu, Zhi, Gao, Wang, Yang, Xu, Fan, Wu, Ding, Wang, Zhang, Li, Xu, Zhao, Zhai, Du, Dong, Lei, Tu, Yang, Lu, Li, Li, Shuang-Li, Yang, Yi, Yu, Tian, Wang, Yu, Tam, Liang, Liu, Wang, Jia, Gu, Ling, Wang, Fan, Pan, Zhang, Zhang, Fu, Zhang, Xu, Wu, Lu, Wang, Zhou, Pan, Zhang, Wang, Li, Su, Geng, Zhu, Yang, Li, Wu, Li, Liu, Wang, Li, Zhang, Liu, Yang, Zhou, Qiao, Feng, Liu, Zhang, Wang, Yao, Wang, Liu, Chai, Li, Zhao, Chen, Zhai, Xu, Huang, Wang, Li, Dong, and Tang]{glm2025glm4_5}
Aohan Zeng, Xin Lv, Qinkai Zheng, Zhenyu Hou, Bin Chen, Chengxing Xie, Cunxiang Wang, Da~Yin, Hao Zeng, Jiajie Zhang, Kedong Wang, Lucen Zhong, Mingdao Liu, Rui Lu, Shulin Cao, Xiaohan Zhang, Xuancheng Huang, Yao Wei, Yean Cheng, Yifan An, Yilin Niu, Yuanhao Wen, Yushi Bai, Zhengxiao Du, Zihan Wang, Zilin Zhu, Bohan Zhang, Bosi Wen, Bowen Wu, Bowen Xu, Can Huang, Casey Zhao, Changpeng Cai, Chao Yu, Chen Li, Chendi Ge, Chenghua Huang, Chenhui Zhang, Chenxi Xu, Chenzheng Zhu, Chuang Li, Congfeng Yin, Daoyan Lin, Dayong Yang, Dazhi Jiang, Ding Ai, Erle Zhu, Fei Wang, Gengzheng Pan, Guo Wang, Hailong Sun, Haitao Li, Haiyang Li, Haiyi Hu, Hanyu Zhang, Hao Peng, Hao Tai, Haoke Zhang, Haoran Wang, Haoyu Yang, He~Liu, He~Zhao, Hongwei Liu, Hongxi Yan, Huan Liu, Huilong Chen, Ji~Li, Jiajing Zhao, Jiamin Ren, Jian Jiao, Jiani Zhao, Jianyang Yan, Jiaqi Wang, Jiayi Gui, Jiayue Zhao, Jie Liu, Jijie Li, Jing Li, Jing Lu, Jingsen Wang, Jingwei Yuan, Jingxuan Li, Jingzhao Du, Jinhua Du, Jinxin Liu, Junkai Zhi, Junli Gao,
  Ke~Wang, Lekang Yang, Liang Xu, Lin Fan, Lindong Wu, Lintao Ding, Lu~Wang, Man Zhang, Minghao Li, Minghuan Xu, Mingming Zhao, Mingshu Zhai, Pengfan Du, Qian Dong, Shangde Lei, Shangqing Tu, Shangtong Yang, Shaoyou Lu, Shijie Li, Shuang Li, Shuang-Li, Shuxun Yang, Sibo Yi, Tianshu Yu, Wei Tian, Weihan Wang, Wenbo Yu, Weng~Lam Tam, Wenjie Liang, Wentao Liu, Xiao Wang, Xiaohan Jia, Xiaotao Gu, Xiaoying Ling, Xin Wang, Xing Fan, Xingru Pan, Xinyuan Zhang, Xinze Zhang, Xiuqing Fu, Xunkai Zhang, Yabo Xu, Yandong Wu, Yida Lu, Yidong Wang, Yilin Zhou, Yiming Pan, Ying Zhang, Yingli Wang, Yingru Li, Yinpei Su, Yipeng Geng, Yitong Zhu, Yongkun Yang, Yuhang Li, Yuhao Wu, Yujiang Li, Yunan Liu, Yunqing Wang, Yuntao Li, Yuxuan Zhang, Zezhen Liu, Zhen Yang, Zhengda Zhou, Zhongpei Qiao, Zhuoer Feng, Zhuorui Liu, Zichen Zhang, Zihan Wang, Zijun Yao, Zikang Wang, Ziqiang Liu, Ziwei Chai, Zixuan Li, Zuodong Zhao, Wenguang Chen, Jidong Zhai, Bin Xu, Minlie Huang, Hongning Wang, Juanzi Li, Yuxiao Dong, and Jie Tang.
\newblock Glm-4.5: Agentic, reasoning, and coding (arc) foundation models, 2025{\natexlab{a}}.
\newblock URL \url{https://arxiv.org/abs/2508.06471}.

\bibitem[Zeng et~al.(2025{\natexlab{b}})Zeng, Shen, Chen, Qi, Das, Gutfreund, Cox, Wornell, Lu, Hong, et~al.]{zeng2025satori}
Guangtao Zeng, Maohao Shen, Delin Chen, Zhenting Qi, Subhro Das, Dan Gutfreund, David Cox, Gregory Wornell, Wei Lu, Zhang-Wei Hong, et~al.
\newblock Satori-swe: Evolutionary test-time scaling for sample-efficient software engineering.
\newblock \emph{arXiv preprint arXiv:2505.23604}, 2025{\natexlab{b}}.

\bibitem[Zeng et~al.(2024{\natexlab{a}})Zeng, Jiang, Wang, Nie, Chen, and Chen]{zeng2024acecoder}
Huaye Zeng, Dongfu Jiang, Haozhe Wang, Ping Nie, Xiaotong Chen, and Wenhu Chen.
\newblock Acecoder: Acing coder rl via automated test-case synthesis.
\newblock \emph{arXiv preprint arXiv:2408.06733}, 2024{\natexlab{a}}.

\bibitem[Zeng et~al.(2025{\natexlab{c}})Zeng, Jiang, Wang, Nie, Chen, and Chen]{zeng2502acecoder}
Huaye Zeng, Dongfu Jiang, Haozhe Wang, Ping Nie, Xiaotong Chen, and Wenhu Chen.
\newblock Acecoder: Acing coder rl via automated test-case synthesis, 2025a.
\newblock \emph{arXiv preprint arXiv:2502.01718}, 2025{\natexlab{c}}.
\newblock URL \url{https://arxiv.org/abs/2502.01718}.

\bibitem[Zeng et~al.(2025{\natexlab{d}})Zeng, Li, Xiao, Li, Liu, Yan, Wei, He, Song, Liu, and Zhou]{zeng2025skyworkswe}
Liang Zeng, Yongcong Li, Yuzhen Xiao, Changshi Li, Chris~Yuhao Liu, Rui Yan, Tianwen Wei, Jujie He, Xuchen Song, Yang Liu, and Yahui Zhou.
\newblock Skywork-swe: Unveiling data scaling laws for software engineering in llms, 2025{\natexlab{d}}.

\bibitem[Zeng et~al.(2024{\natexlab{b}})Zeng, Liu, Fan, Guo, and Zhang]{zeng2024shieldagent}
Q-F Zeng, F~Liu, Y~Fan, Z~Guo, and J.~M. Zhang.
\newblock Shieldagent: Shielding agents via verifiable safety policy reasoning.
\newblock \emph{arXiv preprint arXiv:2405.19253}, 2024{\natexlab{b}}.

\bibitem[Zeng et~al.(2025{\natexlab{e}})Zeng, Tian, Zhang, Wang, Gao, Liu, Yang, Li, Long, Ma, et~al.]{reviewrl}
Sihang Zeng, Kai Tian, Kaiyan Zhang, Yuru Wang, Junqi Gao, Runze Liu, Sa~Yang, Jingxuan Li, Xinwei Long, Jiaheng Ma, et~al.
\newblock Reviewrl: Towards automated scientific review with rl.
\newblock In \emph{Proceedings of the 2025 Conference on Empirical Methods in Natural Language Processing}, pages 16942--16954, 2025{\natexlab{e}}.

\bibitem[{Zha} et~al.(2025){Zha}, {Fan}, {Yang}, {Gao}, and {Chen}]{3DCapacity}
Jirong {Zha}, Yuxuan {Fan}, Xiao {Yang}, Chen {Gao}, and Xinlei {Chen}.
\newblock {How to Enable LLM with 3D Capacity? A Survey of Spatial Reasoning in LLM}.
\newblock \emph{arXiv e-prints}, art. arXiv:2504.05786, April 2025.
\newblock \doi{10.48550/arXiv.2504.05786}.

\bibitem[Zhan et~al.(2024)Zhan, Liu, Liu, Peng, He, and Guo]{zhan2024safe}
Jiaming Zhan, Yantao Liu, Yiran Liu, Fuchun Peng, Zhaofeng He, and Wenbo Guo.
\newblock Safe lora: the silver lining of reducing safety risks when fine-tuning large language models, 2024.

\bibitem[Zhan et~al.(2025)Zhan, Deng, Zhang, Wang, Tang, Lai, Huang, Xiang, Wu, Zhuang, et~al.]{katcoder}
Zizheng Zhan, Ken Deng, Xiaojiang Zhang, Jinghui Wang, Huaixi Tang, Zhiyi Lai, Haoyang Huang, Wen Xiang, Kun Wu, Wenhao Zhuang, et~al.
\newblock Kat-coder technical report.
\newblock \emph{arXiv preprint arXiv:2510.18779}, 2025.

\bibitem[Zhang et~al.(2025{\natexlab{a}})Zhang, Dong, Liu, Zhang, Wang, Yang, Zhang, Liu, Peng, Tan, Zhang, Wang, Wang, He, Deng, Zhou, Huang, and Zhang]{zhang2025codecriticbenchholisticcodecritique}
Alexander Zhang, Marcus Dong, Jiaheng Liu, Wei Zhang, Yejie Wang, Jian Yang, Ge~Zhang, Tianyu Liu, Zhongyuan Peng, Yingshui Tan, Yuanxing Zhang, Zhexu Wang, Weixun Wang, Yancheng He, Ken Deng, Wangchunshu Zhou, Wenhao Huang, and Zhaoxiang Zhang.
\newblock Codecriticbench: A holistic code critique benchmark for large language models, 2025{\natexlab{a}}.
\newblock URL \url{https://arxiv.org/abs/2502.16614}.

\bibitem[Zhang et~al.(2025{\natexlab{b}})Zhang, Li, Xu, Liu, Liu, Hu, Wu, Huang, Li, Yi, et~al.]{zhang2025artifactsbench}
Chenchen Zhang, Yuhang Li, Can Xu, Jiaheng Liu, Ao~Liu, Shihui Hu, Dengpeng Wu, Guanhua Huang, Kejiao Li, Qi~Yi, et~al.
\newblock Artifactsbench: Bridging the visual-interactive gap in llm code generation evaluation.
\newblock \emph{arXiv preprint arXiv:2507.04952}, 2025{\natexlab{b}}.

\bibitem[Zhang et~al.(2025{\natexlab{c}})Zhang, Cardenas, Rezatofighi, Vered, and Say]{zhang2025probabilistic}
Chenyuan Zhang, Cristian~Rojas Cardenas, Hamid Rezatofighi, Mor Vered, and Buser Say.
\newblock Probabilistic active goal recognition.
\newblock \emph{arXiv preprint arXiv:2507.21846}, 2025{\natexlab{c}}.

\bibitem[Zhang et~al.(2024{\natexlab{a}})Zhang, Yang, Zhang, Wu, and Li]{zhang2024human}
D~Zhang, J~Yang, T~Zhang, R~Wu, and Z~Li.
\newblock Human-in-the-loop imitation learning for advanced robotics.
\newblock \emph{Nature Machine Intelligence}, 2024{\natexlab{a}}.

\bibitem[Zhang et~al.(2024{\natexlab{b}})Zhang, Ahmad, Tan, Ding, Nallapati, Roth, Ma, and Xiang]{codesage}
Dejiao Zhang, Wasi~Uddin Ahmad, Ming Tan, Hantian Ding, Ramesh Nallapati, Dan Roth, Xiaofei Ma, and Bing Xiang.
\newblock Code representation learning at scale.
\newblock In \emph{The Twelfth International Conference on Learning Representations}, 2024{\natexlab{b}}.

\bibitem[Zhang et~al.(2023{\natexlab{a}})Zhang, Chen, Zhang, Keung, Liu, Zan, Mao, Lou, and Chen]{zhang2023repocoderrepositorylevelcodecompletion}
Fengji Zhang, Bei Chen, Yue Zhang, Jacky Keung, Jin Liu, Daoguang Zan, Yi~Mao, Jian-Guang Lou, and Weizhu Chen.
\newblock Repocoder: Repository-level code completion through iterative retrieval and generation, 2023{\natexlab{a}}.
\newblock URL \url{https://arxiv.org/abs/2303.12570}.

\bibitem[Zhang et~al.(2024{\natexlab{c}})Zhang, Wu, Bai, Lin, Li, Yu, Wang, Chen, and Keung]{zhang2024humaneval}
Fengji Zhang, Linquan Wu, Huiyu Bai, Guancheng Lin, Xiao Li, Xiao Yu, Yue Wang, Bei Chen, and Jacky Keung.
\newblock Humaneval-v: Benchmarking high-level visual reasoning with complex diagrams in coding tasks.
\newblock \emph{arXiv preprint arXiv:2410.12381}, 2024{\natexlab{c}}.

\bibitem[Zhang et~al.(2025{\natexlab{d}})Zhang, Shen, Wang, Shi, Du, Tao, and Shen]{zhang2025comparativeanalysislargelanguage}
Hang Zhang, Yanxin Shen, Lun Wang, Chuanqi Shi, Shaoshuai Du, Yiyi Tao, and Yixian Shen.
\newblock Comparative analysis of large language models for context-aware code completion using safim framework, 2025{\natexlab{d}}.
\newblock URL \url{https://arxiv.org/abs/2502.15243}.

\bibitem[Zhang and et~al.(2024)]{zhang2024hyperagent}
J.~Zhang and et~al.
\newblock Hyperagent: Generalist multi-agent system for repository-level code generation.
\newblock \emph{arXiv preprint arXiv:2406.11223}, 2024.
\newblock URL \url{https://arxiv.org/abs/2406.11223}.

\bibitem[Zhang et~al.(2025{\natexlab{e}})Zhang, Liu, Qian, Zhang, Liu, Qiao, and Shao]{reef}
Jie Zhang, Dongrui Liu, Chen Qian, Linfeng Zhang, Yong Liu, Yu~Qiao, and Jing Shao.
\newblock {REEF:} representation encoding fingerprints for large language models.
\newblock In \emph{The Thirteenth International Conference on Learning Representations, {ICLR} 2025, Singapore, April 24-28, 2025}. OpenReview.net, 2025{\natexlab{e}}.
\newblock URL \url{https://openreview.net/forum?id=SnDmPkOJ0T}.

\bibitem[Zhang et~al.(2024{\natexlab{d}})Zhang, Zhang, and Jiang]{zhang2024guard}
Jing Zhang, Jian Zhang, and Zhao Jiang.
\newblock Guard: Dual-agent based backdoor defense on chain-of-thought in neural code generation.
\newblock \emph{arXiv preprint arXiv:2405.16723}, 2024{\natexlab{d}}.

\bibitem[Zhang et~al.(2025{\natexlab{f}})Zhang, Zuo, He, Sun, Liu, Jiang, Fan, Tian, Jia, Li, et~al.]{rl_survey}
Kaiyan Zhang, Yuxin Zuo, Bingxiang He, Youbang Sun, Runze Liu, Che Jiang, Yuchen Fan, Kai Tian, Guoli Jia, Pengfei Li, et~al.
\newblock A survey of reinforcement learning for large reasoning models.
\newblock \emph{arXiv preprint arXiv:2509.08827}, 2025{\natexlab{f}}.

\bibitem[Zhang et~al.(2024{\natexlab{e}})Zhang, Li, Dong, Xu, Zhang, Su, Liu, and Jin]{zhang2024codedpo}
Kechi Zhang, Ge~Li, Yihong Dong, Jingjing Xu, Jun Zhang, Jing Su, Yongfei Liu, and Zhi Jin.
\newblock Codedpo: Aligning code models with self generated and verified source code.
\newblock \emph{arXiv preprint arXiv:2410.05605}, 2024{\natexlab{e}}.

\bibitem[Zhang et~al.(2025{\natexlab{g}})Zhang, Li, Li, Dong, and Jin]{zhang2025focuseddpo}
Kechi Zhang, Ge~Li, Jia Li, Yihong Dong, and Zhi Jin.
\newblock Focused-dpo: Enhancing code generation through focused preference optimization on error-prone points.
\newblock \emph{arXiv preprint arXiv:2502.11475}, 2025{\natexlab{g}}.

\bibitem[Zhang et~al.(2025{\natexlab{h}})Zhang, Yang, Yang, Yang, Chen, Zhang, Cui, Hui, and Lin]{zhang2025swe}
Lei Zhang, Jiaxi Yang, Min Yang, Jian Yang, Mouxiang Chen, Jiajun Zhang, Zeyu Cui, Binyuan Hui, and Junyang Lin.
\newblock Swe-flow: Synthesizing software engineering data in a test-driven manner.
\newblock \emph{arXiv preprint arXiv:2506.09003}, 2025{\natexlab{h}}.

\bibitem[Zhang et~al.(2025{\natexlab{i}})Zhang, He, Zhang, Kang, Li, Xie, Wang, Wang, Huang, Fu, Nallipogu, Lin, Dang, Rajmohan, and Zhang]{zhang2025swebenchgoeslive}
Linghao Zhang, Shilin He, Chaoyun Zhang, Yu~Kang, Bowen Li, Chengxing Xie, Junhao Wang, Maoquan Wang, Yufan Huang, Shengyu Fu, Elsie Nallipogu, Qingwei Lin, Yingnong Dang, Saravan Rajmohan, and Dongmei Zhang.
\newblock Swe-bench goes live!
\newblock \emph{arXiv preprint arXiv:2505.23419}, 2025{\natexlab{i}}.

\bibitem[Zhang et~al.(2025{\natexlab{j}})Zhang, Wang, He, Zhang, Kang, Li, Wen, Xie, Wang, Huang, Nallipogu, Lin, Dang, Rajmohan, Zhang, and Zhang]{zhang2025dibenchbenchmarkinglargelanguage}
Linghao Zhang, Junhao Wang, Shilin He, Chaoyun Zhang, Yu~Kang, Bowen Li, Jiaheng Wen, Chengxing Xie, Maoquan Wang, Yufan Huang, Elsie Nallipogu, Qingwei Lin, Yingnong Dang, Saravan Rajmohan, Dongmei Zhang, and Qi~Zhang.
\newblock Di-bench: Benchmarking large language models on dependency inference with testable repositories at scale, 2025{\natexlab{j}}.
\newblock URL \url{https://arxiv.org/abs/2501.13699}.

\bibitem[Zhang et~al.(2024{\natexlab{f}})Zhang, Zan, Yang, Huang, Chen, Shen, Liu, Gong, Huang, Lu, et~al.]{zhang2024codev}
Linhao Zhang, Daoguang Zan, Quanshun Yang, Zhirong Huang, Dong Chen, Bo~Shen, Tianyu Liu, Yongshun Gong, Pengjie Huang, Xudong Lu, et~al.
\newblock Codev: Issue resolving with visual data.
\newblock \emph{arXiv preprint arXiv:2412.17315}, 2024{\natexlab{f}}.

\bibitem[Zhang et~al.(2023{\natexlab{b}})Zhang, Fang, Zhang, Yu, Sun, and Chen]{zhang2023gamma}
Quanjun Zhang, Chunrong Fang, Tongke Zhang, Bowen Yu, Weisong Sun, and Zhenyu Chen.
\newblock Gamma: Revisiting template-based automated program repair via mask prediction.
\newblock In \emph{2023 38th IEEE/ACM International Conference on Automated Software Engineering (ASE)}, pages 535--547. IEEE, 2023{\natexlab{b}}.

\bibitem[Zhang et~al.(2024{\natexlab{g}})Zhang, Shang, Fang, Gu, Zhou, and Chen]{zhang_testbench_2024}
Quanjun Zhang, Ye~Shang, Chunrong Fang, Siqi Gu, Jianyi Zhou, and Zhenyu Chen.
\newblock {TestBench}: {Evaluating} {Class}-{Level} {Test} {Case} {Generation} {Capability} of {Large} {Language} {Models}, September 2024{\natexlab{g}}.
\newblock URL \url{http://arxiv.org/abs/2409.17561}.
\newblock arXiv:2409.17561 [cs].

\bibitem[Zhang et~al.(2017)Zhang, Meng, Bu, Yang, Liu, Pei, Xu, Chen, Dong, Qu, and Song]{DBLP:conf/iwqos/ZhangMBYLPXCDQS17}
Shenglin Zhang, Weibin Meng, Jiahao Bu, Sen Yang, Ying Liu, Dan Pei, Jun Xu, Yu~Chen, Hui Dong, Xianping Qu, and Lei Song.
\newblock Syslog processing for switch failure diagnosis and prediction in datacenter networks.
\newblock In \emph{25th {IEEE/ACM} International Symposium on Quality of Service, IWQoS 2017, Vilanova i la Geltr{\'{u}}, Spain, June 14-16, 2017}, pages 1--10. {IEEE}, 2017.

\bibitem[Zhang et~al.(2024{\natexlab{h}})Zhang, Dong, Li, Zhang, Sun, Wang, Li, Hu, Zhang, Wu, and Wang]{zhang2024instructiontuninglargelanguage}
Shengyu Zhang, Linfeng Dong, Xiaoya Li, Sen Zhang, Xiaofei Sun, Shuhe Wang, Jiwei Li, Runyi Hu, Tianwei Zhang, Fei Wu, and Guoyin Wang.
\newblock Instruction tuning for large language models: A survey, 2024{\natexlab{h}}.
\newblock URL \url{https://arxiv.org/abs/2308.10792}.

\bibitem[Zhang et~al.(2024{\natexlab{i}})Zhang, Wang, Dong, Sun, Zhang, and Pu]{zhang2024experimenting}
Simiao Zhang, Jiaping Wang, Guoliang Dong, Jun Sun, Yueling Zhang, and Geguang Pu.
\newblock Experimenting a new programming practice with llms.
\newblock \emph{CoRR abs/2401.01062}, 2024{\natexlab{i}}.
\newblock URL \url{https://arxiv.org/abs/2401.01062}.

\bibitem[Zhang et~al.(2025{\natexlab{k}})Zhang, Yang, Tao, Chai, Guo, Wu, Chen, Cui, Ding, Xu, Wei, and Zhou]{zhang2025vgamegym}
Wei Zhang, Jack Yang, Renshuai Tao, Lingzheng Chai, Shawn Guo, Jiajun Wu, Xiaoming Chen, Ganqu Cui, Ning Ding, Xander Xu, Hu~Wei, and Bowen Zhou.
\newblock V-gamegym: Visual game generation for code large language models, 2025{\natexlab{k}}.
\newblock URL \url{https://arxiv.org/abs/2509.20136}.

\bibitem[Zhang et~al.(2025{\natexlab{l}})Zhang, Yang, Yang, Wang, Li, Cui, Hui, and Lin]{zhang2025LiveRepoReflection}
Wei Zhang, Jian Yang, Jiaxi Yang, Ya~Wang, Zhoujun Li, Zeyu Cui, Binyuan Hui, and Junyang Lin.
\newblock Turning the tide: Repository-based code reflection, 2025{\natexlab{l}}.
\newblock URL \url{https://arxiv.org/abs/2507.09866}.

\bibitem[Zhang et~al.(2025{\natexlab{m}})Zhang, Chen, Ye, Yang, Chen, Wang, and Petzold]{zhang2025unveiling}
Xinlu Zhang, Zhiyu~Zoey Chen, Xi~Ye, Xianjun Yang, Lichang Chen, William~Yang Wang, and Linda~Ruth Petzold.
\newblock Unveiling the impact of coding data instruction fine-tuning on large language models reasoning.
\newblock In \emph{Proceedings of the AAAI Conference on Artificial Intelligence}, volume~39, pages 25949--25957, 2025{\natexlab{m}}.

\bibitem[Zhang et~al.(2023{\natexlab{c}})Zhang, Li, Zhang, et~al.]{zhang2023deep}
Yifei Zhang, Huan Li, Zhiqiang Zhang, et~al.
\newblock Deep learning for code generation: A survey.
\newblock \emph{arXiv preprint arXiv:2302.06791}, 2023{\natexlab{c}}.
\newblock URL \url{https://arxiv.org/abs/2302.06791}.

\bibitem[Zhang et~al.(2023{\natexlab{d}})Zhang, Liu, Chen, et~al.]{zhang2023copilot}
Yiming Zhang, Zhengzi Liu, Xiang Chen, et~al.
\newblock Practices and challenges of using {GitHub Copilot}: An empirical study.
\newblock In \emph{Proceedings of the 45th International Conference on Software Engineering}, pages 2435--2447, 2023{\natexlab{d}}.

\bibitem[Zhang et~al.(2024{\natexlab{j}})Zhang, Cai, Song, Chen, Sun, and Zheng]{zhang2024reversechaingenericrulellms}
Yinger Zhang, Hui Cai, Xeirui Song, Yicheng Chen, Rui Sun, and Jing Zheng.
\newblock Reverse chain: A generic-rule for llms to master multi-api planning, 2024{\natexlab{j}}.
\newblock URL \url{https://arxiv.org/abs/2310.04474}.

\bibitem[Zhang et~al.(2024{\natexlab{k}})]{zhang2024vulrepair}
Yue Zhang et~al.
\newblock Vulrepair: Vulnerability patching with llms.
\newblock \url{https://github.com/VulRepair/vulrepair}, 2024{\natexlab{k}}.

\bibitem[Zhang et~al.(2025{\natexlab{n}})Zhang, Zhang, Leach, and Huang]{zhang2025formalgrad}
Yueke Zhang, Yifan Zhang, Kevin Leach, and Yu~Huang.
\newblock Formalgrad: Integrating formal methods with gradient-based llm refinement.
\newblock \emph{arXiv preprint arXiv:2508.10059}, 2025{\natexlab{n}}.

\bibitem[Zhang et~al.(2024{\natexlab{l}})Zhang, Ruan, Fan, and Roychoudhury]{zhang2024autocoderover}
Yuntong Zhang, Haifeng Ruan, Zhiyu Fan, and Abhik Roychoudhury.
\newblock Autocoderover: Autonomous program improvement.
\newblock In \emph{Proceedings of the 33rd ACM SIGSOFT International Symposium on Software Testing and Analysis}, pages 1592--1604, 2024{\natexlab{l}}.

\bibitem[Zhang et~al.(2024{\natexlab{m}})Zhang, Chen, Wang, Liu, Yang, Shi, Zhu, Lin, Wan, Yang, Sakai, Feng, and Yamana]{zhang2024toolbehonest}
Yuxiang Zhang, Jing Chen, Junjie Wang, Yaxin Liu, Cheng Yang, Chufan Shi, Xinyu Zhu, Zihao Lin, Hanwen Wan, Yujiu Yang, Tetsuya Sakai, Tian Feng, and Hayato Yamana.
\newblock {T}ool{B}e{H}onest: A multi-level hallucination diagnostic benchmark for tool-augmented large language models.
\newblock In Yaser Al-Onaizan, Mohit Bansal, and Yun-Nung Chen, editors, \emph{Proceedings of the 2024 Conference on Empirical Methods in Natural Language Processing}, pages 11388--11422, Miami, Florida, USA, November 2024{\natexlab{m}}. Association for Computational Linguistics.
\newblock \doi{10.18653/v1/2024.emnlp-main.637}.
\newblock URL \url{https://aclanthology.org/2024.emnlp-main.637/}.

\bibitem[Zhang et~al.(2023{\natexlab{e}})Zhang, Winn, Zhao, Yu, and Halfond]{zhang2023automatically_bug_reports}
Zhaoxu Zhang, Robert Winn, Yu~Zhao, Tingting Yu, and William~GJ Halfond.
\newblock Automatically reproducing android bug reports using natural language processing and reinforcement learning.
\newblock In \emph{Proceedings of the 32nd ACM SIGSOFT International Symposium on Software Testing and Analysis}, pages 411--422, 2023{\natexlab{e}}.

\bibitem[Zhang et~al.(2023{\natexlab{f}})Zhang, Chen, Liu, Liao, Gong, Yu, Li, and Wang]{code_survey_ant}
Ziyin Zhang, Chaoyu Chen, Bingchang Liu, Cong Liao, Zi~Gong, Hang Yu, Jianguo Li, and Rui Wang.
\newblock Unifying the perspectives of nlp and software engineering: A survey on language models for code.
\newblock \emph{arXiv preprint arXiv:2311.07989}, 2023{\natexlab{f}}.

\bibitem[Zhang et~al.(2024{\natexlab{n}})Zhang, Chen, Liu, Liao, Gong, Yu, Li, and Wang]{DBLP:journals/tmlr/ZhangCLLG0L024}
Ziyin Zhang, Chaoyu Chen, Bingchang Liu, Cong Liao, Zi~Gong, Hang Yu, Jianguo Li, and Rui Wang.
\newblock Unifying the perspectives of {NLP} and software engineering: {A} survey on language models for code.
\newblock \emph{Trans. Mach. Learn. Res.}, 2024, 2024{\natexlab{n}}.

\bibitem[Zhang et~al.(2024{\natexlab{o}})Zhang, Zhao, Chen, et~al.]{zhang2024llmcode}
Ziyin Zhang, Chaoyu Zhao, Bingchang Chen, et~al.
\newblock Large language models for code: A survey.
\newblock \emph{arXiv preprint arXiv:2410.01241}, 2024{\natexlab{o}}.

\bibitem[Zhao et~al.(2024{\natexlab{a}})Zhao, Hui, Howland, Nguyen, Zuo, Hu, Choquette{-}Choo, Shen, Kelley, Bansal, Vilnis, Wirth, Michel, Choy, Joshi, Kumar, Hashmi, Agrawal, Gong, Fine, Warkentin, Hartman, Ni, Korevec, Schaefer, and Huffman]{codegemma2024}
Heri Zhao, Jeffrey Hui, Joshua Howland, Nam Nguyen, Siqi Zuo, Andrea Hu, Christopher~A. Choquette{-}Choo, Jingyue Shen, Joe Kelley, Kshitij Bansal, Luke Vilnis, Mateo Wirth, Paul Michel, Peter Choy, Pratik Joshi, Ravin Kumar, Sarmad Hashmi, Shubham Agrawal, Zhitao Gong, Jane Fine, Tris Warkentin, Ale~Jakse Hartman, Bin Ni, Kathy Korevec, Kelly Schaefer, and Scott Huffman.
\newblock Codegemma: Open code models based on gemma.
\newblock \emph{CoRR}, abs/2406.11409, 2024{\natexlab{a}}.
\newblock \doi{10.48550/ARXIV.2406.11409}.
\newblock URL \url{https://doi.org/10.48550/arXiv.2406.11409}.

\bibitem[Zhao et~al.(2023{\natexlab{a}})Zhao, Rong, Guo, He, and Chen]{understanding_fuzz_test_cases}
Jianyu Zhao, Yuyang Rong, Yiwen Guo, Yifeng He, and Hao Chen.
\newblock Understanding programs by exploiting (fuzzing) test cases.
\newblock In Anna Rogers, Jordan~L. Boyd{-}Graber, and Naoaki Okazaki, editors, \emph{Findings of the Association for Computational Linguistics: {ACL} 2023, Toronto, Canada, July 9-14, 2023}, pages 10667--10679. Association for Computational Linguistics, 2023{\natexlab{a}}.
\newblock \doi{10.18653/V1/2023.FINDINGS-ACL.678}.
\newblock URL \url{https://doi.org/10.18653/v1/2023.findings-acl.678}.

\bibitem[Zhao et~al.(2024{\natexlab{b}})Zhao, Chen, Yang, and Shen]{SCCLLM}
Junjie Zhao, Xiang Chen, Guang Yang, and Yiheng Shen.
\newblock Automatic smart contract comment generation via large language models and in-context learning.
\newblock \emph{Information and Software Technology}, 168:\penalty0 107405, 2024{\natexlab{b}}.
\newblock ISSN 0950-5849.
\newblock \doi{https://doi.org/10.1016/j.infsof.2024.107405}.
\newblock URL \url{https://www.sciencedirect.com/science/article/pii/S0950584924000107}.

\bibitem[Zhao et~al.(2023{\natexlab{b}})Zhao, Li, Chia, Ding, and Bing]{zhao2023can}
Ruochen Zhao, Xingxuan Li, Yew~Ken Chia, Bosheng Ding, and Lidong Bing.
\newblock Can chatgpt-like generative models guarantee factual accuracy? on the mistakes of new generation search engines.
\newblock \emph{arXiv preprint arXiv:2304.11076}, 2023{\natexlab{b}}.

\bibitem[Zhao et~al.(2025{\natexlab{a}})Zhao, Zhu, Mozannar, Sontag, Talwalkar, and Chen]{codinggenie}
Sebastian Zhao, Alan Zhu, Hussein Mozannar, David Sontag, Ameet Talwalkar, and Valerie Chen.
\newblock Codinggenie: A proactive llm-powered programming assistant.
\newblock In \emph{Proceedings of the 33rd ACM International Conference on the Foundations of Software Engineering}, pages 1168--1172, 2025{\natexlab{a}}.

\bibitem[Zhao and et~al.(2024)]{zhao2024commit0}
W.~Zhao and et~al.
\newblock Commit0: Library generation from scratch.
\newblock \emph{arXiv preprint arXiv:2412.01769}, 2024.
\newblock URL \url{https://arxiv.org/abs/2412.01769}.

\bibitem[Zhao et~al.(2025{\natexlab{b}})Zhao, Zhou, Li, Tang, Wang, Hou, Min, Zhang, Zhang, Dong, Du, Yang, Chen, Chen, Jiang, Ren, Li, Tang, Liu, Liu, Nie, and Wen]{zhao2025surveylargelanguagemodels}
Wayne~Xin Zhao, Kun Zhou, Junyi Li, Tianyi Tang, Xiaolei Wang, Yupeng Hou, Yingqian Min, Beichen Zhang, Junjie Zhang, Zican Dong, Yifan Du, Chen Yang, Yushuo Chen, Zhipeng Chen, Jinhao Jiang, Ruiyang Ren, Yifan Li, Xinyu Tang, Zikang Liu, Peiyu Liu, Jian-Yun Nie, and Ji-Rong Wen.
\newblock A survey of large language models, 2025{\natexlab{b}}.
\newblock URL \url{https://arxiv.org/abs/2303.18223}.

\bibitem[Zhao et~al.(2025{\natexlab{c}})Zhao, Luo, Shi, Chen, Wang, Liu, and Sun]{zhao2025chartcoder}
Xuanle Zhao, Xianzhen Luo, Qi~Shi, Chi Chen, Shuo Wang, Zhiyuan Liu, and Maosong Sun.
\newblock Chartcoder: Advancing multimodal large language model for chart-to-code generation.
\newblock \emph{arXiv preprint arXiv:2501.06598}, 2025{\natexlab{c}}.

\bibitem[Zhao et~al.(2023{\natexlab{c}})Zhao, Gu, Varma, Luo, Huang, Xu, Wright, Shojanazeri, Ott, Shleifer, Desmaison, Balioglu, Damania, Nguyen, Chauhan, Hao, Mathews, and Li]{zhao2023pytorchfsdpexperiencesscaling}
Yanli Zhao, Andrew Gu, Rohan Varma, Liang Luo, Chien-Chin Huang, Min Xu, Less Wright, Hamid Shojanazeri, Myle Ott, Sam Shleifer, Alban Desmaison, Can Balioglu, Pritam Damania, Bernard Nguyen, Geeta Chauhan, Yuchen Hao, Ajit Mathews, and Shen Li.
\newblock Pytorch fsdp: Experiences on scaling fully sharded data parallel, 2023{\natexlab{c}}.
\newblock URL \url{https://arxiv.org/abs/2304.11277}.

\bibitem[Zhao et~al.(2023{\natexlab{d}})Zhao, Gu, Varma, Luo, Huang, Xu, Wright, Shojanazeri, Ott, Shleifer, et~al.]{zhao2023pytorch}
Yanli Zhao, Andrew Gu, Rohan Varma, Liang Luo, Chien-Chin Huang, Min Xu, Less Wright, Hamid Shojanazeri, Myle Ott, Sam Shleifer, et~al.
\newblock Pytorch fsdp: Experiences on scaling fully sharded data parallel.
\newblock \emph{Proceedings of the VLDB Endowment}, 16\penalty0 (12):\penalty0 3848--3860, 2023{\natexlab{d}}.

\bibitem[Zhao et~al.(2024{\natexlab{c}})Zhao, Huang, Ma, Li, Zhang, Jiang, Liu, Zhu, and Su]{zhao2024repair}
Yuze Zhao, Zhenya Huang, Yixiao Ma, Rui Li, Kai Zhang, Hao Jiang, Qi~Liu, Linbo Zhu, and Yu~Su.
\newblock Repair: Automated program repair with process-based feedback.
\newblock In \emph{Findings of the Association for Computational Linguistics: ACL 2024}, pages 16415--16429, 2024{\natexlab{c}}.

\bibitem[Zhauo et~al.(2025)Zhauo, Liu, Wang, Ahmad, Hui, and Allal]{zhuo2025nlp+}
Terry~Yue Zhauo, Qian Liu, Zijian Wang, Wasi~U Ahmad, Binuian Hui, and Loubna~Ben Allal.
\newblock Nlp+ code: Code intelligence in language models.
\newblock In \emph{Proceedings of the 2025 Conference on Empirical Methods in Natural Language Processing: Tutorial Abstracts}, pages 9--11, 2025.

\bibitem[Zheng et~al.(2024{\natexlab{a}})Zheng, Yuan, Song, Chen, and Liu]{zheng2024seal}
Chiaming Zheng, Cao Yuan, Yang Song, Pin-Yu Chen, and Sijia Liu.
\newblock Seal: Safety-enhanced aligned llm fine-tuning via bilevel data selection.
\newblock \emph{arXiv preprint arXiv:2405.18835}, 2024{\natexlab{a}}.

\bibitem[Zheng et~al.(2025{\natexlab{a}})Zheng, Wang, Shi, Zhang, Ma, Zhang, and Zheng]{zheng2025humanevoevolutionawarebenchmarkrealistic}
Dewu Zheng, Yanlin Wang, Ensheng Shi, Ruikai Zhang, Yuchi Ma, Hongyu Zhang, and Zibin Zheng.
\newblock Humanevo: An evolution-aware benchmark for more realistic evaluation of repository-level code generation, 2025{\natexlab{a}}.
\newblock URL \url{https://arxiv.org/abs/2406.06918}.

\bibitem[Zheng et~al.(2023)Zheng, Xia, Zou, et~al.]{zheng2023codegeex}
Qinkai Zheng, Xiao Xia, Xu~Zou, et~al.
\newblock {CodeGeeX}: A pre-trained model for code generation with multilingual evaluations on {HumanEval-X}.
\newblock \emph{KDD}, 2023.

\bibitem[Zheng et~al.(2024{\natexlab{b}})Zheng, Zhu, Lin, et~al.]{zheng2024codegeex4}
Qinkai Zheng, Tianyu Zhu, Boxuan Lin, et~al.
\newblock {CodeGeeX4-ALL-9B}: Open multilingual code generation model.
\newblock \emph{arXiv preprint arXiv:2407.10845}, 2024{\natexlab{b}}.

\bibitem[Zheng et~al.(2024{\natexlab{c}})Zheng, Zhang, Shen, Liu, Lin, Fu, Chen, and Yue]{opencodeinterpreter}
Tianyu Zheng, Ge~Zhang, Tianhao Shen, Xueling Liu, Bill~Yuchen Lin, Jie Fu, Wenhu Chen, and Xiang Yue.
\newblock Opencodeinterpreter: Integrating code generation with execution and refinement.
\newblock \emph{arXiv preprint arXiv:2402.14658}, 2024{\natexlab{c}}.

\bibitem[Zheng et~al.(2025{\natexlab{b}})Zheng, Xing, Gu, Liang, Qu, Zhou, Li, Wen, Lin, Huang, Liu, Zhang, and Ma]{zheng2025fr3e}
Tianyu Zheng, Tianshun Xing, Qingshui Gu, Taoran Liang, Xingwei Qu, Xin Zhou, Yizhi Li, Zhoufutu Wen, Chenghua Lin, Wenhao Huang, Qian Liu, Ge~Zhang, and Zejun Ma.
\newblock First return, entropy-eliciting explore, 2025{\natexlab{b}}.
\newblock URL \url{https://arxiv.org/abs/2507.07017}.

\bibitem[Zheng et~al.(2024{\natexlab{d}})Zheng, Lou, Cao, Wen, Ji, Lin, Lu, Han, Zhang, and Sun]{zheng2024critic}
Xin Zheng, Jie Lou, Boxi Cao, Xueru Wen, Yuqiu Ji, Hongyu Lin, Yaojie Lu, Xianpei Han, Debing Zhang, and Le~Sun.
\newblock Critic-cot: Boosting the reasoning abilities of large language model via chain-of-thoughts critic.
\newblock \emph{arXiv preprint arXiv:2408.16326}, 2024{\natexlab{d}}.

\bibitem[Zheng et~al.(2022)Zheng, Wang, Dong, Wang, and Li]{Zheng}
Yanzhao Zheng, Haibin Wang, Baohua Dong, Xingjun Wang, and Changshan Li.
\newblock {HIE-SQL:} history information enhanced network for context-dependent text-to-sql semantic parsing.
\newblock In Smaranda Muresan, Preslav Nakov, and Aline Villavicencio, editors, \emph{Findings of the Association for Computational Linguistics: {ACL} 2022,}, pages 2997--3007. Association for Computational Linguistics, 2022.

\bibitem[Zheng et~al.(2024{\natexlab{e}})Zheng, Zhang, Zhang, Ye, and Luo]{zheng-etal-2024-llamafactory}
Yaowei Zheng, Richong Zhang, Junhao Zhang, Yanhan Ye, and Zheyan Luo.
\newblock {L}lama{F}actory: Unified efficient fine-tuning of 100+ language models.
\newblock In Yixin Cao, Yang Feng, and Deyi Xiong, editors, \emph{Proceedings of the 62nd Annual Meeting of the Association for Computational Linguistics (Volume 3: System Demonstrations)}, pages 400--410, Bangkok, Thailand, August 2024{\natexlab{e}}. Association for Computational Linguistics.
\newblock \doi{10.18653/v1/2024.acl-demos.38}.
\newblock URL \url{https://aclanthology.org/2024.acl-demos.38/}.

\bibitem[Zheng et~al.(2024{\natexlab{f}})Zheng, Li, Wang, Shi, Liu, Pu, and Ma]{zheng2024sandboxeval}
Yongwei Zheng, Y-t Li, P~Wang, T~Shi, Y~Liu, G~Pu, and J~Ma.
\newblock Sandboxeval: Towards securing test environment for untrusted code.
\newblock \emph{arXiv preprint arXiv:2405.08375}, 2024{\natexlab{f}}.

\bibitem[Zheng et~al.(2025{\natexlab{c}})Zheng, Cheng, Shen, Zhou, Liu, He, Li, Wei, Hao, Yao, Sheng, Wang, Chai, Korolova, Henderson, Arora, Viswanath, Shang, and Xie]{livecodebenchpro}
Zihan Zheng, Zerui Cheng, Zeyu Shen, Shang Zhou, Kaiyuan Liu, Hansen He, Dongruixuan Li, Stanley Wei, Hangyi Hao, Jianzhu Yao, Peiyao Sheng, Zixuan Wang, Wenhao Chai, Aleksandra Korolova, Peter Henderson, Sanjeev Arora, Pramod Viswanath, Jingbo Shang, and Saining Xie.
\newblock Livecodebench pro: How do olympiad medalists judge llms in competitive programming?, 2025{\natexlab{c}}.
\newblock URL \url{https://arxiv.org/abs/2506.11928}.

\bibitem[Zhong et~al.(2024{\natexlab{a}})Zhong, Wang, and Shang]{debug_like_human}
Li~Zhong, Zilong Wang, and Jingbo Shang.
\newblock Debug like a human: {A} large language model debugger via verifying runtime execution step by step.
\newblock In Lun{-}Wei Ku, Andre Martins, and Vivek Srikumar, editors, \emph{Findings of the Association for Computational Linguistics, {ACL} 2024, Bangkok, Thailand and virtual meeting, August 11-16, 2024}, pages 851--870. Association for Computational Linguistics, 2024{\natexlab{a}}.
\newblock \doi{10.18653/V1/2024.FINDINGS-ACL.49}.
\newblock URL \url{https://doi.org/10.18653/v1/2024.findings-acl.49}.

\bibitem[Zhong et~al.(2024{\natexlab{b}})Zhong, Wang, Wang, Wen, Guan, Tao, and Liu]{zhong2024advancing_bug_detction}
Zhiyuan Zhong, Sinan Wang, Hailong Wang, Shaojin Wen, Hao Guan, Yida Tao, and Yepang Liu.
\newblock Advancing bug detection in fastjson2 with large language models driven unit test generation.
\newblock \emph{arXiv preprint arXiv:2410.09414}, 2024{\natexlab{b}}.

\bibitem[Zhou et~al.(2023{\natexlab{a}})Zhou, Wang, Lu, Shi, Luo, Qin, Lu, Jia, Song, Zhan, et~al.]{zhou2023solving}
Aojun Zhou, Ke~Wang, Zimu Lu, Weikang Shi, Sichun Luo, Zipeng Qin, Shaoqing Lu, Anya Jia, Linqi Song, Mingjie Zhan, et~al.
\newblock Solving challenging math word problems using gpt-4 code interpreter with code-based self-verification.
\newblock \emph{arXiv preprint arXiv:2308.07921}, 2023{\natexlab{a}}.

\bibitem[Zhou et~al.(2024{\natexlab{a}})Zhou, Pujara, Ren, Chen, Cheng, Le, Chi, Zhou, Mishra, and Zheng]{zhou2024self}
Pei Zhou, Jay Pujara, Xiang Ren, Xinyun Chen, Heng-Tze Cheng, Quoc~V Le, Ed~Chi, Denny Zhou, Swaroop Mishra, and Huaixiu~Steven Zheng.
\newblock Self-discover: Large language models self-compose reasoning structures.
\newblock \emph{Advances in Neural Information Processing Systems}, 37:\penalty0 126032--126058, 2024{\natexlab{a}}.

\bibitem[Zhou et~al.(2025{\natexlab{a}})Zhou, Leon, Ying, Zhang, Shao, Ye, Chong, Jin, Xie, Cao, et~al.]{zhou2025browsecomp}
Peilin Zhou, Bruce Leon, Xiang Ying, Can Zhang, Yifan Shao, Qichen Ye, Dading Chong, Zhiling Jin, Chenxuan Xie, Meng Cao, et~al.
\newblock Browsecomp-zh: Benchmarking web browsing ability of large language models in chinese.
\newblock \emph{arXiv preprint arXiv:2504.19314}, 2025{\natexlab{a}}.

\bibitem[Zhou et~al.(2023{\natexlab{b}})Zhou, Alon, Agarwal, and Neubig]{zhou2023codebertscoreevaluatingcodegeneration}
Shuyan Zhou, Uri Alon, Sumit Agarwal, and Graham Neubig.
\newblock Codebertscore: Evaluating code generation with pretrained models of code, 2023{\natexlab{b}}.
\newblock URL \url{https://arxiv.org/abs/2302.05527}.

\bibitem[Zhou et~al.(2023{\natexlab{c}})Zhou, Xu, Zhu, Zhou, Lo, Sridhar, Cheng, Ou, Bisk, Fried, et~al.]{zhou2023webarena2}
Shuyan Zhou, Frank~F Xu, Hao Zhu, Xuhui Zhou, Robert Lo, Abishek Sridhar, Xianyi Cheng, Tianyue Ou, Yonatan Bisk, Daniel Fried, et~al.
\newblock Webarena: A realistic web environment for building autonomous agents.
\newblock \emph{arXiv preprint arXiv:2307.13854}, 2023{\natexlab{c}}.

\bibitem[Zhou et~al.(2024{\natexlab{b}})Zhou, Xu, Zhu, Zhou, Lo, Sridhar, Cheng, Ou, Bisk, Fried, Alon, and Neubig]{zhou2023webarena}
Shuyan Zhou, Frank~F. Xu, Hao Zhu, Xuhui Zhou, Robert Lo, Abishek Sridhar, Xianyi Cheng, Tianyue Ou, Yonatan Bisk, Daniel Fried, Uri Alon, and Graham Neubig.
\newblock Webarena: A realistic web environment for building autonomous agents, 2024{\natexlab{b}}.
\newblock URL \url{https://arxiv.org/abs/2307.13854}.

\bibitem[Zhou et~al.(2025{\natexlab{b}})Zhou, Lin, Jha, Christodorescu, Levchenko, and Chandrasekaran]{zhou2025llmdrivenmultisteptranslationc}
Tianyang Zhou, Haowen Lin, Somesh Jha, Mihai Christodorescu, Kirill Levchenko, and Varun Chandrasekaran.
\newblock Llm-driven multi-step translation from c to rust using static analysis, 2025{\natexlab{b}}.
\newblock URL \url{https://arxiv.org/abs/2503.12511}.

\bibitem[Zhu et~al.(2024{\natexlab{a}})Zhu, Frick, Wu, Zhu, Ganesan, Chiang, Zhang, and Jiao]{zhu2024starling}
Banghua Zhu, Evan Frick, Tianhao Wu, Hanlin Zhu, Karthik Ganesan, Wei-Lin Chiang, Jian Zhang, and Jiantao Jiao.
\newblock Starling-7b: Improving helpfulness and harmlessness with rlaif.
\newblock In \emph{First Conference on Language Modeling}, 2024{\natexlab{a}}.

\bibitem[Zhu et~al.(2025)Zhu, Wang, Chen, Liu, Ye, Gu, Tian, Duan, Su, Shao, et~al.]{zhu2025internvl3}
Jinguo Zhu, Weiyun Wang, Zhe Chen, Zhaoyang Liu, Shenglong Ye, Lixin Gu, Hao Tian, Yuchen Duan, Weijie Su, Jie Shao, et~al.
\newblock Internvl3: Exploring advanced training and test-time recipes for open-source multimodal models.
\newblock \emph{arXiv preprint arXiv:2504.10479}, 2025.

\bibitem[Zhu et~al.()Zhu, Jain, Suresh, Ravindran, Tipirneni, and Reddy]{xlcost}
Ming Zhu, Aneesh Jain, Karthik Suresh, Roshan Ravindran, Sindhu Tipirneni, and Chandan~K Reddy.
\newblock Xlcost: A benchmark dataset for cross-lingual code intelligence, 2022.
\newblock \emph{URL https://arxiv. org/abs/2206.08474}, 88.

\bibitem[Zhu et~al.(2022)Zhu, Suresh, and Reddy]{zhu2022multilingual}
Ming Zhu, Karthik Suresh, and Chandan~K Reddy.
\newblock Multilingual code snippets training for program translation.
\newblock In \emph{Proceedings of the AAAI conference on artificial intelligence}, volume~36, pages 11783--11790, 2022.

\bibitem[Zhu et~al.(2021)Zhu, Sun, Xiao, Zhang, Yuan, Xiong, and Zhang]{zhu2021syntax}
Qihao Zhu, Zeyu Sun, Yuan-an Xiao, Wenjie Zhang, Kang Yuan, Yingfei Xiong, and Lu~Zhang.
\newblock A syntax-guided edit decoder for neural program repair.
\newblock In \emph{Proceedings of the 29th ACM joint meeting on European software engineering conference and symposium on the foundations of software engineering}, pages 341--353, 2021.

\bibitem[Zhu et~al.(2024{\natexlab{b}})Zhu, Guo, Shao, Yang, Wang, Xu, Wu, Li, Gao, Ma, et~al.]{deepseek_coder_v2}
Qihao Zhu, Daya Guo, Zhihong Shao, Dejian Yang, Peiyi Wang, Runxin Xu, Y~Wu, Yukun Li, Huazuo Gao, Shirong Ma, et~al.
\newblock Deepseek-coder-v2: Breaking the barrier of closed-source models in code intelligence.
\newblock \emph{arXiv preprint arXiv:2406.11931}, 2024{\natexlab{b}}.

\bibitem[Zhu et~al.(2024{\natexlab{c}})Zhu, Lau, and Qi]{zhu2024factual}
Rongxin Zhu, Jey~Han Lau, and Jianzhong Qi.
\newblock Factual dialogue summarization via learning from large language models.
\newblock \emph{arXiv preprint arXiv:2406.14709}, 2024{\natexlab{c}}.

\bibitem[Zhuo(2024)]{zhuo2024ice}
Terry~Yue Zhuo.
\newblock Ice-score: Instructing large language models to evaluate code.
\newblock In \emph{Findings of the Association for Computational Linguistics: EACL 2024}, pages 2232--2242, 2024.

\bibitem[Zhuo et~al.()Zhuo, Zebaze, Von~Werra, de~Vries, Liu, and Muennighoff]{zhuo2025parameter}
Terry~Yue Zhuo, Armel~Randy Zebaze, Leandro Von~Werra, Harm de~Vries, Qian Liu, and Niklas Muennighoff.
\newblock Parameter-efficient instruction tuning code large language models: An empirical study.
\newblock In \emph{ICLR 2025 Third Workshop on Deep Learning for Code}.

\bibitem[Zhuo et~al.(2023{\natexlab{a}})Zhuo, Li, Huang, Shiri, Wang, Haffari, and Li]{zhuo2023robustness}
Terry~Yue Zhuo, Zhuang Li, Yujin Huang, Fatemeh Shiri, Weiqing Wang, Gholamreza Haffari, and Yuan-Fang Li.
\newblock On robustness of prompt-based semantic parsing with large pre-trained language model: An empirical study on codex.
\newblock \emph{arXiv preprint arXiv:2301.12868}, 2023{\natexlab{a}}.

\bibitem[Zhuo et~al.(2023{\natexlab{b}})Zhuo, Yang, Sun, Wang, Li, Du, Xing, and Lo]{zhuo2023source}
Terry~Yue Zhuo, Zhou Yang, Zhensu Sun, Yufei Wang, Li~Li, Xiaoning Du, Zhenchang Xing, and David Lo.
\newblock Source code data augmentation for deep learning: A survey.
\newblock \emph{arXiv preprint arXiv:2305.19915}, 2023{\natexlab{b}}.

\bibitem[Zhuo et~al.(2024)Zhuo, Vu, Chim, Hu, Yu, Widyasari, Yusuf, Zhan, He, Paul, et~al.]{zhuo2024bigcodebench}
Terry~Yue Zhuo, Minh~Chien Vu, Jenny Chim, Han Hu, Wenhao Yu, Ratnadira Widyasari, Imam Nur~Bani Yusuf, Haolan Zhan, Junda He, Indraneil Paul, et~al.
\newblock Bigcodebench: Benchmarking code generation with diverse function calls and complex instructions.
\newblock \emph{arXiv preprint arXiv:2406.15877}, 2024.

\bibitem[Zhuo et~al.(2025{\natexlab{a}})Zhuo, Huang, Chen, Du, and Xing]{zhuo2025bypassing}
Terry~Yue Zhuo, Yujin Huang, Chunyang Chen, Xiaoning Du, and Zhenchang Xing.
\newblock Bypassing guardrails: Lessons learned from red teaming chatgpt.
\newblock \emph{ACM Transactions on Software Engineering and Methodology}, 2025{\natexlab{a}}.

\bibitem[Zhuo et~al.(2025{\natexlab{b}})Zhuo, Jin, Liu, Jiang, Liu, Gong, Bishnoi, Mishra, Suppa, Ziems, et~al.]{zhuo2025bigcodearena}
Terry~Yue Zhuo, Xiaolong Jin, Hange Liu, Juyong Jiang, Tianyang Liu, Chen Gong, Bhupesh Bishnoi, Vaisakhi Mishra, Marek Suppa, Noah Ziems, et~al.
\newblock Bigcodearena: Unveiling more reliable human preferences in code generation via execution.
\newblock \emph{arXiv preprint arXiv:2510.08697}, 2025{\natexlab{b}}.

\bibitem[Zhuo et~al.(2025{\natexlab{c}})Zhuo, Wang, Ding, Kumar, and Wang]{zhuo2025cyber}
Terry~Yue Zhuo, Dingmin Wang, Hantian Ding, Varun Kumar, and Zijian Wang.
\newblock Cyber-zero: training cybersecurity agents without runtime.
\newblock \emph{arXiv preprint arXiv:2508.00910}, 2025{\natexlab{c}}.

\bibitem[Zhuo et~al.(2025{\natexlab{d}})Zhuo, Wang, Ding, Kumar, and Wang]{zhuo2025training}
Terry~Yue Zhuo, Dingmin Wang, Hantian Ding, Varun Kumar, and Zijian Wang.
\newblock Training language model agents to find vulnerabilities with ctf-dojo.
\newblock \emph{arXiv preprint arXiv:2508.18370}, 2025{\natexlab{d}}.

\bibitem[Zou et~al.(2023)Zou, Wang, Kolter, and Fredrikson]{zou2023universal}
Andy Zou, Zifan Wang, J.~Zico Kolter, and Matt Fredrikson.
\newblock Universal and transferable adversarial attacks on aligned language models, 2023.

\bibitem[Zuccotto et~al.(2024)Zuccotto, Castellini, Torre, Mola, and Farinelli]{zuccotto2024reinforcement}
Maddalena Zuccotto, Alberto Castellini, Davide~La Torre, Lapo Mola, and Alessandro Farinelli.
\newblock Reinforcement learning applications in environmental sustainability: a review.
\newblock \emph{Artificial Intelligence Review}, 57\penalty0 (4):\penalty0 88, 2024.

\end{thebibliography}

\end{document}